Sujet de la thèse:

# Gamma-ray pulsar physics: gap-model populations and light-curve analyses in the FERMI era.

Devant le jury composé de:

| | |
|---|---|
| Prof. Marcello Fulchignoni | Président de jury |
| Prof. Patrizia Caraveo | Rapporteur |
| Prof. David Smith | Rapporteur |
| Dr. Gilles Theureau | Examinateur |
| Dr. Andrea Possenti | Examinateur |
| Prof. Isabelle Grenier | Directeur de thèse |

# Contents

# Avant-propos

Ce travail de thèse a été fait dans le laboratoire AIM/Service d'Astrophysique du Commissariat à l'Energie Atomique, CEA Saclay. Il concerne l'étude du mécanisme d'émission de rayons gamma des pulsars à la lumière des nouvelles observations fournies par le *FERMI LAT γ-ray space telescope*. Ce nouveau télescope a identifié ce qui est actuellement reconnu comme une nouvelle population d'objets jeunes qui sont caractérisés par une haute perte d'énergie rotationnelle, de grands champs magnétiques et de hautes fréquences de rotation. Le thème de mon projet de doctorat est l'étude de cette population de jeune pulsars isolés détectés par l'instrument LAT, ainsi que leur première phase évolutive étant donné qu'il est maintenant possible de les observer avec une grande statistique.

La première partie de mon projet de doctorat a consisté à synthétiser plusieurs populations de pulsars basées sur les modèles d'émission actuellement disponibles dans le but d'étudier leurs propriétés collectives dans le contexte des nouvelles données FERMI. J'ai implémenté 4 différents modèles d'émission, les 3 premiers étant: le *Polar Cap* (PC, Muslimov & Harding, 2003), le *Slot Gap* (SG, Muslimov & Harding, 2004) et le *Outer Gap* (OG, Cheng et al., 2000). J'ai par ailleurs implémenté une variation du OG proposée par Romani & Watters (2010), le *One pole caustic model* (OPC). Les modèles d'émission théoriques ont été calculés pour une population de $10^7$ objets distribuée dans la Voie lactée en tenant compte de leurs caractéristiques de naissance, leurs positions, leurs vitesses d'éjection initiale lors de l'explosion de supernova, le champ magnétique, la période et la première dérivée de la période, et optimisés pour reproduire les distributions de la population radio observée. J'ai fait évolué cet échantillon d'étoiles à neutrons (NS) a évolué dans le potentiel gravitationnel Galactique jusqu'au temps présent. En utilisant un ensemble de modèles d'émission pour chaque modèle (le diagramme de phase), j'ai assigné un profil de courbe de lumière à chaque NS de l'échantillon. Finalement, j'ai utilisé les cartes de sensibilité gamma et radio pour sélectionner tous les pulsars de l'échantillon qui auraient été découverts par le LAT pendant sa première année d'activité et visibles pour le LAT pendant un an d'activité et visibles pour les 10 relevés radio les plus importants.

Dans la deuxième partie de mon projet de doctorat, j'ai effectué un ajustement simultané des courbes de lumière observées en gamma et radio
d'un échantillon de pulsars détectés par le LAT en utilisant les différents
modèles d'émission. Ceci m'a permis d'estimer l'obliquité magnétique $\alpha$ et
l'angle de visée de l'observateur $\zeta$ qui décrivent au mieux l'orientation dans
l'espace du pulsar ainsi que son rayonnement gamma et radio. Ces angles ont
une importance fondamentale. La largeur de la région d'émission du pulsar
(*gap region*) est une de ses caractéristiques les plus discutées mais qui n'est
pas directement observable. Dû à la différence entre la densité de charge réelle
et celle de Goldreich-Julian (Goldreich & Julian, 1969), c'est dans cette région
de la magnétosphère qu'un fort champ électrique se développe et accélère les
particules. Sa localisation et sa largeur déterminent tant la géométrie que
l'intensité de son émission. Puisque la largeur de cette région est fonction de $\alpha$
et $\zeta$, le fait d'obtenir des estimations exactes de ces angles aide à contraindre
les différents modèles d'émission. La largeur de la *gap region* définit aussi la
luminosité totale des particules.

La troisième partie de mon projet de doctorat a concerné l'évaluation
des caractéristiques morphologiques des courbes de lumière. Je les ai classées
selon des paramètres structurels comme la largeur entre pics et le nombre de
maxima et minima présents dans la courbe de lumière. Cette classification
a produit l'histogramme des formes les plus fréquentes dans l'échantillon
visible pour chaque modèle. Ces résultats sont à comparer aux profils
observés par le LAT afin de définir lequel des modèles décrit au mieux
les observations. Une autre analyse structurelle et sa comparaison avec les
observations LAT se sont intéressées à la séparation des pics et au décalage
radio. Ces caractéristiques des courbes de lumière sont particulièrement
importantes parce qu'elles peuvent être bien mesurées et leur comparaison
avec les observations du LAT représente une des rares contraintes directes
de la structure de champ magnétique. Etant donné que j'ai considéré des
modèles d'émission de la magnétosphère à basse et haute altitude (PC et
SG/OG/OPC, respectivement) et des configurations de champ magnétique
avec un ou deux pôles d'émission (PC–SG ou OG/OPC, respectivement), la
séparation entre pics et la comparaison du décalage radio-gamma permettent
de discriminer entre différentes positions de l'accélérateur. J'ai effectué une
analyse morphologique des courbes de lumière en utilisant les coefficients
d'asymétries et d'aplatissement qui décrivent la symétrie et le contraste des
courbes. Cette méthode représente une approche originale dans l'analyse
des courbes de lumière des pulsars. La comparaison avec les détections du
LAT aide à contraindre la magnétosphère du pulsar et sa disposition et,
conjointement avec la séparation entre pics et l'étude du décalage radio, elle
aidera à comprendre la nature de la configuration du champ magnétique du
pulsar.

# Foreword

This thesis work has been carried out in the laboratory AIM/Service d'Astrophysique of the Commissariat à l'Energie Atomique, CEA, in Saclay. It concerns the study of the $\gamma$-ray emission mechanism from pulsars in the light of the new observations provided by the FERMI LAT $\gamma$-ray space telescope.

The new telescope has identified what is by this time known as a new population of young object, characterised by high spin-down energy loss, high magnetic fields and high spin frequency values. The topic of my PhD project is the study of the young LAT isolated pulsar population, and the fact that it is now possible to observe, with high statistics, its early evolutionary phase.

The first part of my PhD project consisted in synthesising different populations of pulsars based on the emission models currently available to study their collective properties in the light of the FERMI data. I have implemented 4 different emission models, the Polar Cap (PC) (Muslimov & Harding, 2003), Slot Gap (SG) (Muslimov & Harding, 2004), and Outer Gap (OG) (Cheng et al., 2000). I have also implemented a variation of the OG proposed by Romani & Watters (2010), the One Pole Caustic model (OPC). The theoretical emission models have been implemented for a population of $10^7$ objects distributed in the Milky Way with birth characteristics, including position, initial kick velocity, magnetic field, period and period first time derivative, optimised to reproduce the distributions of the observed radio population. This NS sample has been evolved in the Galactic gravitational potential up to the present time. By using a set of emission patterns for each model (phase-plot) I assigned a light curve profile to each NS of our sample. Finally I used gamma and radio sensitivity maps to select all the pulsars of the sample that would have been detected by the LAT during the first year and would be visible for the LAT in one year and would be visible in the 10 major radio surveys.

In the second part of my PhD project I have performed a joint fit of the observed gamma and radio light curves of a sample of LAT detected pulsars by using the emission pattern derived from each model. The goal of the fitting was to give an estimate of the magnetic obliquity $\alpha$ and the observer line of sight $\zeta$ angle that best describes the pulsar space orientation to explain both the $\gamma$-ray and radio emission. These angles are of fundamental importance. The width of the pulsar emission region (gap region) is one of the most debated pulsar characteristics that is not directly observable. It is the region of the magnetosphere in which, due to a difference between the real charge density and the Goldreich-Julian one (Goldreich & Julian, 1969), a strong electric field develops and accelerates particles. Its location and size determine both the emission geometry and intensity. Since the width of the gap is a function of $\alpha$ and $\zeta$, having accurate estimates of these angles helps to constrain the different emission models. The gap width also sets the total particle luminosity.

The third part of my PhD project concerns the evaluation of the light curve morphological characteristics. I have classified the light curve shapes according to structural parameters, like the peak width and the number of maxima and minima in the light curve. The classification produced the histogram of the most frequent shapes in the visible sample for each model. These results can be compared with the observed profiles to define which model best describes the observations. Another structural analysis and comparison with the LAT observations focussed on peak separation and radio to gamma-ray lag. These light curve characteristics are particularly important because they can be well measured and their comparison with the LAT observations represents one of the rare direct constraints on the magnetic field structure. Since I have assumed low and high altitude emission models (PC and SG,OG/OPC respectively) and one pole emission or two pole emission configurations (PC, SG or OG/OPC respectively) the peak separation and radio lag comparison helps to discriminate between different gap locations. I have performed a light curve morphological analysis using the skewness and kurtosis. These characteristics describe the curve symmetry and sharpness. This method represents a novel approach in the pulsar light curve analysis.The comparison with the LAT detections helps to constrain the pulsar magnetosphere layout and jointly with the peak separation and radio-gamma lag study, it will help to understand the nature of the pulsar magnetic field configuration.

# Abstract

This thesis research focusses on the study of the young and energetic isolated ordinary pulsar population detected by the Fermi gamma-ray space telescope.

We compared the model expectations of four emission models and the LAT data. We found that all the models fail to reproduce the LAT detections, in particular the large number of high $\dot{E}$ objects observed. This inconsistency is not model dependent. A discrepancy between the radio-loud/radio-quiet objects ratio was also found between the observed and predicted samples. The $L_\gamma \propto \dot{E}^{0.5}$ relation is robustly confirmed by all the assumed models with particular agreement in the slot gap (SG) case. On luminosity bases, the intermediate altitude emission of the two pole caustic SG model is favoured. The beaming factor $f_\Omega$ shows an $\dot{E}$ dependency that is slightly visible in the SG case.

Estimates of the pulsar orientations have been obtained to explain the simultaneous gamma and radio light-curves. By analysing the solutions we found a relation between the observed energy cutoff and the width of the emission slot gap. This relation has been theoretically predicted. A possible magnetic obliquity $\alpha$ alignment with time is rejected -for all the models- on timescale of the order of $10^6$ years.

The light-curve morphology study shows that the outer magnetosphere gap emission (OGs) are favoured to explain the observed radio-gamma lag. The light curve moment studies (symmetry and sharpness) on the contrary favour a two pole caustic SG emission.

All the model predictions suggest a different magnetic field layout with an hybrid two pole caustic and intermediate altitude emission to explain both the pulsar luminosity and light curve morphology . The low magnetosphere emission mechanism of the polar cap model, is systematically rejected by all the tests done.

# Chapter 1

# Pulsar physics and origin

In this chapter I will introduce the concept of a pulsar in order to create the basis that is needed to understand the contents of this research work. I will start by describing the history of pulsar research, how they were discovered and the connection between the first empirical observations and the theoretical interpretation of the new objects. After this brief historical overview I will describe the classical and accepted pulsar theory, how it is generated, and the physical structure that sustains the pulsar engine, with particular attention to the main intrinsic pulsar characteristics. In the last part of the chapter, these concepts will be followed by a description of the first theoretical models formulated to explain the pulsar radio emission mechanism. I will close this section by describing how pulsars evolve and the different populations we have detected after forty-three years of observations.

## 1.1 History

Pulsars are relatively young astrophysical objects. Their discovery dates from 1967 when the PhD student Jocelyn Bell-Burnell and her PhD supervisor Antony Hewish, observing at a frequency of 81.5 MHz with a self-made radio telescope built in the Cambridge countryside, detected the first high-accuracy periodic signal (Hewish et al. 1968 ) ever seen from a celestial body (figure 1.1). At that time, at least from the observational point of view, no astrophysical object was known that could emit a so accurate periodic signal and the first interpretations of the new phenomenon were in the direction of an artificial emission coming from some man-made device. Nevertheless there were some other elements that focused Bell's attention on the possibility that the periodic signal was coming from outer space. The main one was that plotting the observations over several days, it was evident that the 1.337 second periodic signal (Hewish et al. 1968 ) reappeared on the detector exactly one sidereal day (23 hours and 56 minutes) after the start of the previous detection. This was a key element that pushed Jocelyn Bell to investigate in depth the nature of the new signal, to the extent that she asked for a confirmation by another radio telescope situated in a completely different position on the Earth. When also the second telescope confirmed the presence of a periodic signal coming from the same direction of the sky, she had the confirmation that the signal was not Earth made. At first, these objects were classified by using the prefix LGM (little green man). But the LGM interpretation was not to persist for a long time and when a survey of the sky showed the presence of other periodic signals from different directions Jocelyn Bell had to exclude a synchronised alien effort to contact the Earth, and the scientific community had to accept that they were confronted with a new class of astrophysical objects.

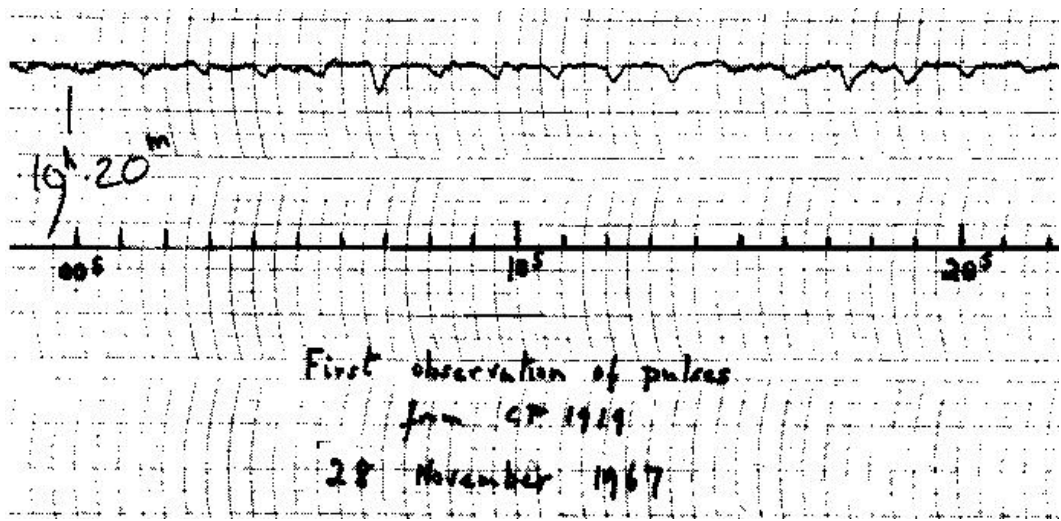

Figure 1.1: One of the first plots produced by the radio telescope detector of the first pulsar's signal ever observed.

## 1.2 The pulsar nature

A pulsar is a neutron star (NS) that spins rapidly and is characterised by a very strong magnetic field, often above $10^{12} - 10^{13}$ Gauss, probably one of the most intense in the known universe. The possibility that a NS could exist in a stable configuration was first formulated, in 1934, by a Swiss astrophysicist and a German astronomer, Fritz Zwicki and Walter Baade of Mount Wilson Observatory, just two years after the discovery of the neutron (thus much longer before the discovery of the first pulsar). They were the same two astronomers that introduced the term "*supernova*" and suggested that a supernova represents the intermediate stage between an ordinary star and a NS.

### 1.2.1 How a pulsar is born

The universally accepted theory about how a pulsar is born assumes that the NS formation is a consequence of a *core-collapse supernova explosion*. It is the final evolution of an intermediate-high mass star, with $M_* \geq 8M_\odot$. During the AGB phase (Asymptotic Giant Branch phase in the colour-magnitude diagram) such a star undergoes a double shell burning: an inner one where Helium fusion produces C that accretes on a inert Carbon-Oxygen (C-O) core and a more external one where Hydrogen produces Helium. In the post-AGB phase the C-O core has contracted and heated enough to trigger the fusion of the Carbon via $\alpha$-capture ($T < 10^8 \ °K$)

$$^{12}\mathrm{C} + {}^4\mathrm{He} \longrightarrow {}^{16}\mathrm{O} + \gamma$$

and via direct fusion (T$\sim 6 \times 10^8 \ °K$)

$$^{12}\mathrm{C} + {}^{12}\mathrm{C} \longrightarrow \left\{ \begin{array}{l} {}^{16}\mathrm{O} + 2{}^4\mathrm{He} \\ {}^{20}\mathrm{Ne} + {}^4\mathrm{He} \\ {}^{23}\mathrm{Na} + p^+ \\ {}^{23}\mathrm{Mg} + n \\ {}^{24}\mathrm{Mg} + \gamma \ . \end{array} \right.$$

In the last equation and hereafter, $\gamma$, $n$, and $p$ respectively indicate photon, neutron, and proton. This last fusion process will lead to the formation of an initially inert Oxygen core that will contract and heat until the Oxygen fusion is triggered. It will start a cascade fusion mechanism that, beginning from the external stellar regions up to the core, will trigger the nuclear fusion of heavier elements; first there will be the formation of an inert core composed by the heaviest elements of the star, later this core will increase in mass, contract and heat enough to start to generate heavier elements by nuclear fusion. After the fusion of the Carbon in Oxygen, for $T_* \sim 10^9 \ °K$, the Oxygen starts to burn via direct fusion:

$$^{16}\mathrm{O} + {}^{16}\mathrm{O} \longrightarrow \left\{ \begin{array}{l} {}^{24}\mathrm{Mg} + 2{}^4\mathrm{He} \\ {}^{28}\mathrm{Si} + {}^4\mathrm{He} \\ {}^{31}\mathrm{P} + p^+ \\ {}^{31}\mathrm{S} + n \\ {}^{32}\mathrm{S} + \gamma \end{array} \right. \ .$$

This process will lead to the formation of a $^{28}\mathrm{Si}$ core. After the formation of the $^{28}\mathrm{Si}$, the production of heavier elements will be mainly caused by $\alpha$ capture process (T$\sim 6 \times 10^9 \ °K$)

$$^{28}\mathrm{Si} + {}^4\mathrm{He} \longrightarrow {}^{32}\mathrm{S} + \gamma$$
$$^{32}\mathrm{S} + {}^4\mathrm{He} \longrightarrow {}^{36}\mathrm{Ar} + \gamma$$
$$... \ ... \ ... \ ...$$

But this cascade-burning-shells process will not go on indefinitely up the highest known atomic weight elements. In fact, when reactions such as

$$^{52}\mathrm{Cr} + {}^4\mathrm{He} \longrightarrow {}^{56}\mathrm{Ni} + \gamma$$

start to generate elements of the chemical group of Iron, any other fusion reaction to produce heavier elements will be endothermic and so cannot be triggered. An inert core will be generated, composed of $^{56}_{26}$Fe, $^{54}_{26}$Fe e $^{56}_{28}$Ni, with an excess of $^{56}_{26}$Fe, and the star will have assumed the classical onion structure in the static equilibrium picture Figure 1.2. Because it is impossible for the Iron core to trigger other nuclear reactions, it will contract and heat up to temperatures much bigger than $10^9$ °K at which photo-disintegration processes such as

$$^{56}\text{Fe} + \gamma \longrightarrow 13\,^4\text{He} + 4n$$
$$^4\text{He} + \gamma \longrightarrow 2p + 2n$$

will start. At this point, if the mass of the star is equal to or bigger than $8M_\odot$, the $^{56}_{26}$Fe inert core will pass the Chandrasekhar limit ($\sim\ 1.44M_\odot$). Beyond this limit the pressure of the non degenerate iron nuclei gas that composes the core is no longer strong enough to prevent the gravitational collapse. There will be a collapse of the iron core that will stop just when the atomic structure of the iron nuclei will be destroyed through a reaction like

$$p + e^- \longrightarrow n + \nu$$

and the core will be composed of neutrons. At this point the degeneracy pressure of the neutrons,

$$p_d = \frac{1}{20} \left(\frac{3}{\pi}\right)^{2/3} h^2 m_N^{-1} \left(\frac{\rho}{2m_N}\right)^{5/3} \tag{1.1}$$

much higher than the bigger iron nuclei one, will balance the force of the gravitational collapse and the nucleus will reach a new equilibrium.

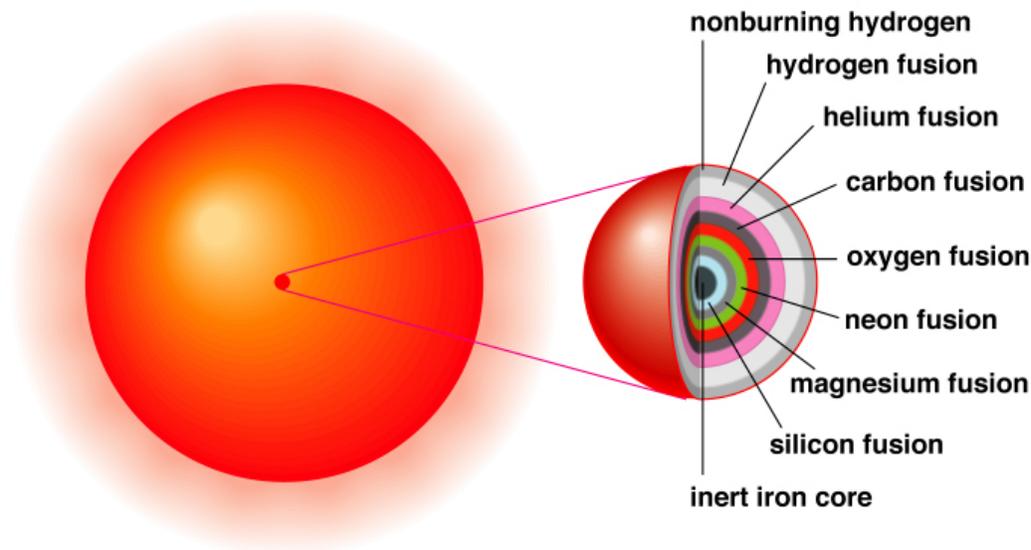

Figure 1.2: An example of typical onion structure of a star with M$\gtrsim 8M_\odot$ in the last phase of its life. From the surface to the core is triggered the fusion of elements gradually heavier, from the hydrogen until the formation of an inert iron core.

Nevertheless, when the core collapse reaches the new equilibrium configuration, the star envelope is still collapsing toward it. It is not clear which kind of process will act in this phase but the interaction of the collapsing star envelope and the infinitely rigid neutron core will generate a shock wave that, propagating from the core to the external star region, will violently eject the envelope in a supernova explosion. The amount of energy the neutrinos generated in the neutron formation in the core will convey to the envelope is unclear. What will remain from the explosion is a star largely composed of neutrons that will spin very rapidly spin motion (due to the conservation of the angular momentum during the core collapse phase) and will be characterised by a very intense magnetic field (trapped in the collapse). This object will be surrounded by the remnant of the star envelope ejected during the explosion.

### 1.2.2   The oblique rotator model (Gold 1968)

The oblique rotator model was formulated in 1968, before the discovery of the first pulsars, by T. Gold. This model was based on the assumption that a rotating, highly magnetised neutron star, whose rotational and magnetic axes are not aligned, should radiate a big amount of electromagnetic radiation at the NS's spin frequency. Moreover, it was predicted that the electromagnetic emission should have been energetically supplied by the rotational kinetic energy, causing a progressive slowdown of the NS.

Calling $\mathbf{m}_0$ the magnetic moment of the NS and $\alpha$ the angle between magnetic and rotational axis, the magnetic moment component on the equatorial plane is $\mathbf{m} = \mathbf{m}_0 \sin\alpha$. This component will co-rotate with the NS with the same angular velocity $\Omega_{NS}$. In this case, the power irradiated by a magnetic dipole is given by the Larmor formula:

$$\frac{d\epsilon}{dt} = \frac{2}{3}\frac{1}{c^3}\left(\frac{d^2\mathbf{m}}{dt^2}\right)^2 = \frac{2}{3}\frac{1}{c^3}\Omega_{NS}^4(\mathbf{m_0}\sin\alpha)^{\mathbf{2}} \tag{1.2}$$

and is emitted at the frequency $\nu = \Omega_{NS}/2\pi$. If this emission is energetically supplied by the rotational slowdown, we can write

$$\frac{2}{3}\frac{1}{c^3}\Omega_{NS}^4(\mathbf{m}_0\sin\alpha)^2 = \frac{d\epsilon}{dt} = -\frac{dT_{NS}}{dt} = -I_{NS}\Omega_{NS}\dot{\Omega}_{NS} \tag{1.3}$$

where $I_{NS}$ is the moment of inertia and $T_{NS}$ the rotational kinetic energy.

From this equation we obtain

$$\dot{\Omega} = -\frac{2}{3}\frac{1}{c^3}\Omega_{NS}^3\frac{(\mathbf{m}_0\sin\alpha)^2}{I_{NS}} \qquad \text{or} \qquad P\dot{P} = \frac{8\pi^2}{3c^3}\frac{(\mathbf{m}_0\sin\alpha)^2}{I_{NS}}. \qquad (1.4)$$

For $\alpha = 1$ we write the dipole moment as a function of the magnetic field on the NS surface: $|\mathbf{m}_0| \sim B_{NS}R_{NS}^3$, and from the latter, one obtains

$$\frac{2}{3}\frac{1}{c^3}\Omega_{NS}^4 B_{NS}^2 R_{NS}^6 = -I_{NS}\Omega_{NS}\dot{\Omega}_{NS}$$

or, written with respect to the period,

$$4\pi^2 I_{NS}\frac{\dot{P}}{P^3} = \frac{2}{3}\frac{1}{c^3}(2\pi)^4 R_{NS}^6\frac{B_{NS}^2}{P^4}. \qquad (1.5)$$

Equation 1.5 can be written in the classical form, with respect to the magnetic field, as

$$B_{NS} = \left(\frac{3c^3}{8\pi^2}\frac{I_{NS}}{R_{NS}^6}P\dot{P}\right)^{1/2} \qquad \text{or more synthetically} \qquad B_{NS}^2 \propto P\dot{P}. \qquad (1.6)$$

These last equations have a big importance in the pulsar study because they define a well defined dependency between fundamental parameters like magnetic field on the surface ($B_{NS}$), period ($P$), and period's first derivative ($\dot{P}$).

**Pulsar age**

By taking into account what has been defined in the previous sections, it is possible to estimate the pulsar age. Equation 1.4 can be written in a more general form:

$$\dot{\nu} = -K\nu^n \qquad (1.7)$$

where $n$ is called *braking index* and depends on the radiation mechanism we assume. Writing the same equation with respect to the period, we have $\dot{P} = KP^{2-n}$, that integrated assuming $K = constant$ and $n \neq 1$ becomes

$$T = \frac{P}{(n-1)\dot{P}}\left[1 - \left(\frac{P_0}{P}\right)^{n-1}\right] \qquad (1.8)$$

Where $P_0$ is the born pulsar period. Now, assuming that the initial spin period is much lower than the observed one and that the slow down ($\dot{P}$) is due to the dipole radiation (that implies $n = 3$, equation 1.7), the previous equation becomes

$$\tau_c \equiv \frac{P}{2\dot{P}} \simeq 15.8 \text{ Myr} \left(\frac{P}{s}\right)\left(\frac{\dot{P}}{10^{-15}}\right)^{-1}. \qquad (1.9)$$

The latter defines what is called *characteristic age of a pulsar*.

Another age constraint can be expressed from a pure kinematic and geometrical assumption. If we assume that all the pulsars were born on the galactic plane and their velocity distribution at the observation time it is consistent with that at birth, the kinematic age is defined as

$$\tau_k = \frac{z}{v_z}. \tag{1.10}$$

In this equation $z$ is the vertical distance of the pulsar from the galactic plane, and $v_z$ is the vertical component of the pulsar velocity, always with respect to the galactic plane.

These two different age estimates yield consistent values just up to a *characteristic age* $\lesssim 10^7$ years, while for older objects, the age evaluated with the equation (1.9) are clearly overestimate. The reasons of such an inconsistency are essentially two. When we evaluate the pulsar age on the basis of its distance from the galactic plane we should take into account that the pulsar could be in a descending orbit in the galactic potential. In fact, $\sim 10^7$ years is an important fraction of the characteristic orbital period of an object in the galactic gravitational field. In this case the *kinematic age* will be an under-estimation of the true pulsar age. The second inconsistency reason is in the way we evaluate the pulsar distance that is based on the distribution of ionised gas in the galaxy. Since the ionised gas is practically absent above a certain distance from the galactic plane, all the pulsars above this limit and under the same observer line of sight angle will appear to have the same distance. In this case, the *kinematic age* will give an underestimation of the true age.

### 1.2.3 Intrinsic properties

**Mass**

Since NSs are supposed to be generated from a collapsed star core that reached the Chandrasekhar limit (section 1.2.1), it is reasonable to assume a typical value for their mass not so far from the Chandrasekhar mass of $\sim 1.44\ M_\odot$. This mass value has been experimentally constrained by using the masses estimates obtained from the NS-NS and NS-white dwarf binary systems (Lattimer & Prakash (2007), figure 1.3). In this thesis work, the typical NS mass is assumed to be $1.5 M_\odot$. As shown in figure 1.3, this assumption is consistent with the observations.

In general relativity, it is possible to define a maximum value for the NS mass. Rhoades & Ruffini (1974) derived an upper limit of 3.2 $M_\odot$ while Hartle & Sabbadini (1977) showed that this upper limit scales with the total mass energy density $\epsilon_f$, like:

$$M_{max} = 4.2\sqrt{\epsilon_f/\epsilon_s}M_\odot \tag{1.11}$$

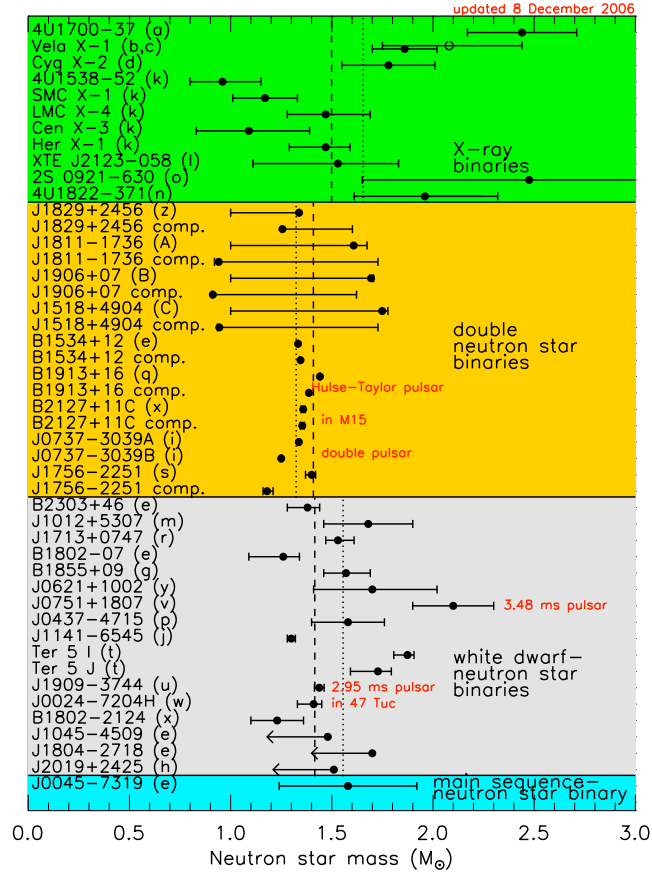

Figure 1.3: Masses of four different typology of NS: X-ray binaries (green), double NS (yellow), white dwarf-NS binary system (grey) and a main sequence NS binaries (cyan). From Lattimer & Prakash 2007.

where $\epsilon_f = \rho_f c^2$, $\rho_f = 4.6 \times 10^{14}$ g cm$^{-3}$ is a fiducial mass energy density and $\epsilon_s$ is the star energy density (Lattimer & Prakash 2007 and references there in ). For a mass value and density higher that this limit, no physical theory can describe a state of equilibrium between gravity and internal pressure: the star will collapse into a black hole singularity.

**Size**

Concerning the dimension of a NS, it is possible to obtain an estimate of its radius lower and upper value. By assuming that the sound speed inside a NS has to be lower than the speed of the light and a soft transition between high and low density region (Lattimer el al.1990; Glendenning 1992) the equation that define a minimum value for the NS radius expressed in terms of the Schwarzschild radius, is:

$$R_{min} \simeq 1.5 R_S = \frac{3GM}{c^2} = 6.2 \text{ km.} \left( \frac{M}{1.4 M_\odot} \right) \qquad (1.12)$$

This result agrees with the one previously obtained by Lindblom (Lattimer & Prakash 2007 and references there in ) based on a completely different assumption. Starting from a maximum compactness assumption, for $\rho_f = 3 \times 10^{14}$ g cm$^{-3}$, and assuming causality, the maximum value for the gravitational redshift is:

$$z = \frac{1}{\sqrt{1 - 2GM/Rc^2}} - 1 \leq 0.863 \ \rightarrow \ R \geq 2.83 \frac{GM}{c^2}. \tag{1.13}$$

Turning now to the upper limit of the NS radius, it is possible to obtain an estimate including the centrifugal force due to the rotational motion in the equilibrium relation:

$$R_{max} \simeq \left(\frac{GMP^2}{4\pi^2}\right)^{1/3} = 16.8 \text{ km} \left(\frac{M}{1.4M_\odot}\right)^{1/3} \left(\frac{P}{1 \text{ ms}}\right)^{2/3} \tag{1.14}$$

In the previous and following equations $G$ is the universal gravitational constant, $c$ the speed of light, $M$ and $P$ the mass and spin period of the NS, and $R_S$ the Schwarzschild radius, defined as:

$$R_S = \frac{2GM}{c^2} \simeq 4.2 \text{ km} \left(\frac{M}{1.4M_\odot}\right). \tag{1.15}$$

In this thesis work, the typical NS radius is assumed to be 13 km.

Figure 1.4 shows the mass and radius choices with respect to different solutions of the equation of state for the neutron star interior (from (Lattimer & Prakash, 2007)).

**Density and moment of inertia**

Starting from the previous considerations about dimensions and mass, it is possible to estimate two important parameters, the moment of inertia $I = kMR^2$ and the density. For a sphere the k parameter is 0.4. In this thesis work, the typical moment of inertia of a pulsar is the one evaluated by Lattimer & Prakash 2007:

$$I \simeq 0.237MR^2(1 + 2.84\beta + 18.9\beta^4)M_\odot \ km^2 \tag{1.16}$$

where $M$ and $R$ are the mass and radius of the pulsar and $\beta = GM/Rc^2$. Taking into account the 13 km radius and the $1.5M_\odot$ mass of our standard star, we obtain, from equation 1.16, $I = 1.79 \times 10^{38}$ kg m$^2$ = $1.79 \times 10^{45}$ g cm$^2$. Because of the errors on the mass and radius estimates, this value has an uncertainty of about the 70%. The present choice of moment of inertia is slightly larger than the $10^{38}$ kg m2 value often quoted in the literature. This choice is driven by the need for high luminosities to match the LAT data, as will be seen in the next chapters.

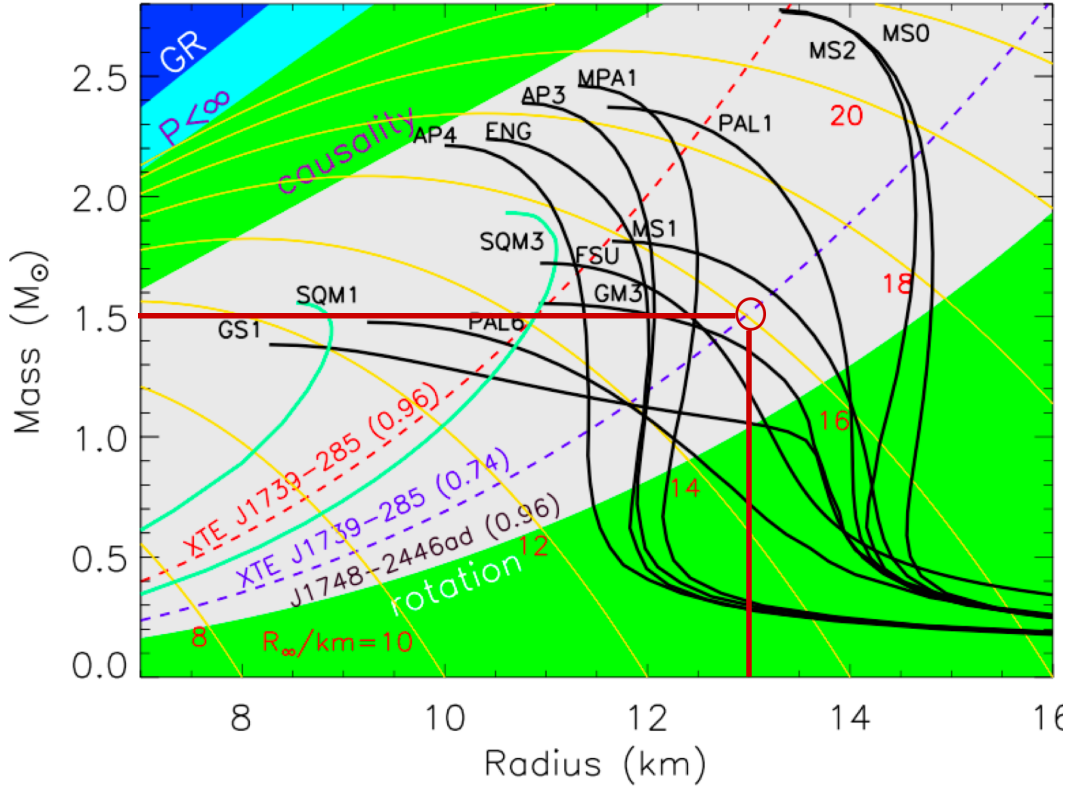

Figure 1.4: Neutron star Mass-radius plane. The black lines indicate typical equation of state (EOSs) solutions. Blue, cyan, green, and light green zones respectively indicate non-allowed EOS solutions regions for General relativity (GR), finite pressure constraint, causality, and rotation of the 716 Hz pulsar J1748-2446ad (Hessels et al., 2006). The orange lines show contour of radiation. The upper red dashed curve is the corresponding rotational limit for the 1122 Hz X-ray source XTE J1739-285 (Kaaret et al., 2007); the lower blue dashed curve is the rigorous causal minimum period limit. The red lines and circle indicate the mass-radius solution choose in this thesis (Plot from Lattimer & Prakash (2007)).

As for the density, the value obtained from the mass and radius assumed in the previous section is $\langle \rho \rangle = 3.2 \times 10^{17}$ kg m$^{-3}$ = $3.2 \times 10^{14}$ g cm$^{-3}$ that is even higher than the the atomic nucleus one $\langle \rho \rangle = 2.7 \times 10^{14}$ g cm$^{-3}$.

**Structure**

Figure (1.5) shows a schematic interpretation of the commonly accepted NS structure. It is possible to distinguish at least 5 different concentric regions:

- **Atmosphere**, composed by a super-hot plasma (T $10^5 - 10^6$ K)

- **Outer crust**, ~200 meters deep composed by a crystal lattice of nuclei plus electrons

- **Inner crust**, ∼1000 meters deep composed by a crystal lattice of nuclei plus electrons and neutron drip

- **Outer core**, composed by a fluid of atomic particles

- **Inner core**, composed by exotic matter. For the nucleus composition pions or quark have been proposed.

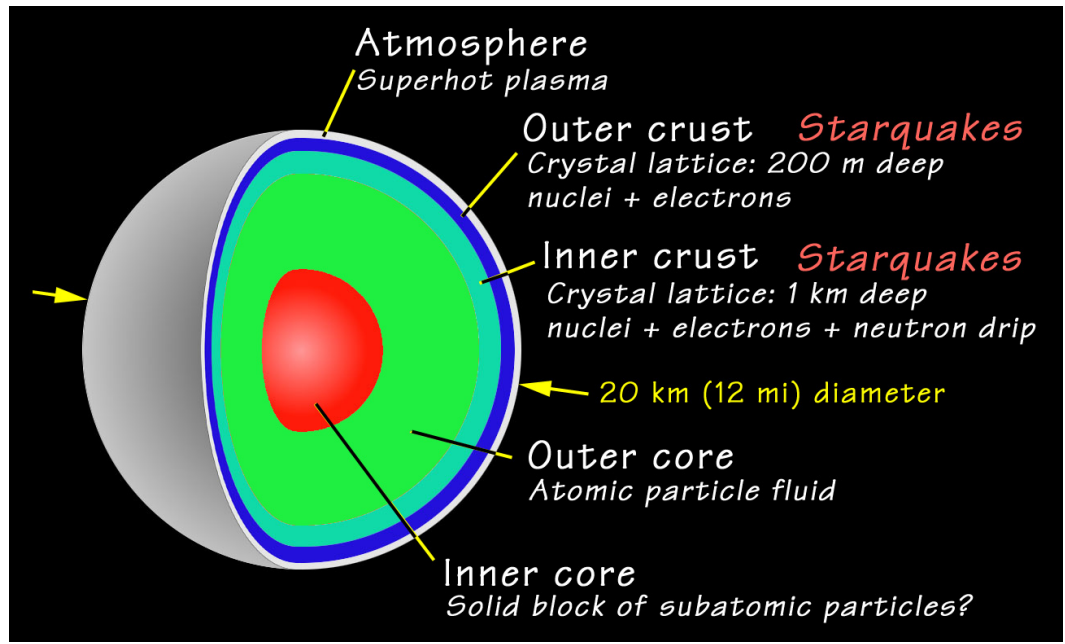

Figure 1.5: Classical schema of the shell structure of a neutron star. In this figure (Max-Plank-Institute für Radioastronomie http://www.mpifr-bonn.mpg.de/div/fundamental/research.html) the neutron star radius is assumed to be 10 km.

Going through the solid crust, composed by iron nuclei and degenerate electrons with density ∼ $10^6$ g cm$^{-3}$, the density increases so much that electrons and protons combine into neutrons. An inner crust, composed by neutron rich nuclei, will be generated. Going deeper through the "neutron drip point", several hundred meters below the surface, the density reaches values around ∼ $4 \times 10^{11}$ g cm$^{-3}$ and the rate of neutrons released by the nuclei rapidly increases. When the density reaches the value of ∼ $2 \times 10^{14}$ g cm$^{-3}$ matter consists of a neutron superfluid with ∼ 5% of protons and electrons. Many different theories try to address the particles composition of the inner NS core. One of the most extreme interpretation, from Shapiro & Teukolsky, was given in 1983 and supposes the core to be composed of pions or quarks.

**Spin & spin-down**

As it will be discussed in section 1.4, a pulsar can have a spin period spanning a wide range, from a value of the order of a millisecond to few seconds (figure 1.8). The majority of the pulsars known so far ($\sim 2000$) have periods $P \sim 0.5$ s that increase at a rate $\dot{P} \sim 10^{-15}$ s/s, with an important fraction characterised by much lower periods, $1.0 \text{ ms} \lesssim P \lesssim 30$ ms, increasing at a much smaller rate, $\dot{P} \lesssim 10^{-19}$ $s/s$ (figure 1.8). The spin-down can be related to the NS rotational energy loss through the *spin-down luminosity*, equation 1.3. Assuming the moment of inertia evaluated in the previous *Density and moment of inertia* section, a typical value of the spin-down luminosity is:

$$\dot{E} \simeq 7.07 \times 10^{31} \text{erg s}^{-1} \left( \frac{\dot{P}}{10^{-15} \text{s/s}} \right) \left( \frac{P}{1\text{s}} \right)^{-3} \tag{1.17}$$

Just a fraction of the whole spin-down luminosity will be converted in electromagnetic radiation: a small amount will be converted in radio pulsed emission but the biggest part of $\dot{E}$ will be converted in high energy electromagnetic radiation and will be spent to accelerate particles out of the magnetosphere and generate the pulsar wind.

**Magnetic field**

The magnetic field represents, without any doubts, the most impressive characteristic of a pulsar. Observational measurements of NS magnetic fields could be performed taking advantage of the cyclotron radiation from X-ray binary systems (Trümper et al. 1978 ; Wheaton et al. 1979) and also from isolated NS (Bignami et al. 2003). These measurements show a typical dipole magnetic field value of $\sim 10^{11} - 10^{13}$ Gauss. The magnetic momentum $\mathbf{m}$, is connected with the magnetic field intensity $B$ by the relation $B \approx |\mathbf{m}| / r^3$ where $r$ is the distance from the NS centre. A theoretical estimate of the magnetic field on the NS surface could be done by taking into account the *oblique rotator model* (section 1.2.2). From the equality of the spin-down power, Equation 1.3, with Equation 1.2 that gives the radiation power of a rotating dipole in the CGS system, we can obtain Equation 1.4 that define the spin frequency evolution that, written in the magnetic field form gives

$$B_S \equiv B(r = R) = \sqrt{\frac{3c^3}{8\pi^2} \frac{I}{R^6 \sin^2 \alpha} P \dot{P}}. \tag{1.18}$$

In these last equations, $c$ is the speed of light, $\Omega$ the NS angular velocity, $\alpha$ the magnetic obliquity defined as the angle between the rotational and magnetic axes, $I$ the NS moment of inertia, and $R$ the NS radius. Equation 1.18 is only valid for a vacuum dipole magnetic field. If the neutron star is surrounded by a particle flow, the term $\sin^2 \alpha$ can be ignored or at least there should be another term that makes $B_S$ finite at $\alpha=0$.

Assuming the oblique rotator model (section 1.2.2) for a 13 km radius pulsar, with a moment of inertia $I= 1.79\times10^{45}$ g cm$^2$, and an $\alpha$ value of 90°, we obtain a standard value for the surface magnetic field of

$$B_S = 1.95 \times 10^{19} [\text{Gauss}] \sqrt{P\dot{P}} \qquad (1.19)$$

The equation (1.19) defines the "characteristic magnetic field" and should be considered like a first order estimate of the NS magnetic field.

## 1.3 Dynamics & electrodynamics: the classical models

The main and most studied characteristic of a Pulsars is the capability to emit periodic signals in a very broad spectral range, from the radio wavelengths to photons of several GeV (Giga elettronvolt, $10^9$ eV). The pulsed nature of the signal has to be ascribed mainly to the geometrical layout of the magnetic an rotational axes with respect to the observer line of sight and the positions of the emitting regions in the magnetosphere (section 1.3.1). The positions of the emitting regions are presumed to change as a function of the observed wavelength.

At radio frequencies, since the emission is likely to be generated in a region close to the magnetic polar cap of the pulsar, and it is collimated in roughly conical beam, coaxial with the magnetic axis (figure 1.6), we will see the pulsed emission just if the emission beam crosses our line of sight during the pulsar rotation. In this case we will talk about a *lighthouse effect* (bottom right panel of figure 1.6). If the magnetic and rotational axes are aligned, the signal is not going to be modulated by the spin period and the possibility to detect it as pulsed is very low.

The situation is a bit more complicated at high energy. More than one model has been proposed to explain where the radiation is generated. In some model we can detect pulsed emission at large $\zeta$ angles, if the magnetic and rotational axes are aligned. This is the case for example of the outer gap and slot gap models (sections 2.2 & 2.1) for which the emission regions are located in the external magnetosphere and the wide beams shine far off the magnetic axis. The emission gap position for the most studied high energy emission models, is indicated, in red (PC), yellow (SG), and cyan (OG), in figure 1.6.

The first formulation of a model that was able to describe the radio emission mechanism from pulsar dates from 1969 and has been proposed by Goldreich & Julian on the basis of a previous work by Deutsch (1955). Even if the Goldreich-Julian model is still one of the best theoretical description of the pulsar electrodynamics, it has been formulated starting from an unreliable assumption and so, it does not describe a real situation. In fact, the first and most important assumption of the model is the condition of parallelism between magnetic and rotational axes, condition that is not consistent with the lighthouse effect that is an important condition to have a radio emission pulsed enough to be detected[1].

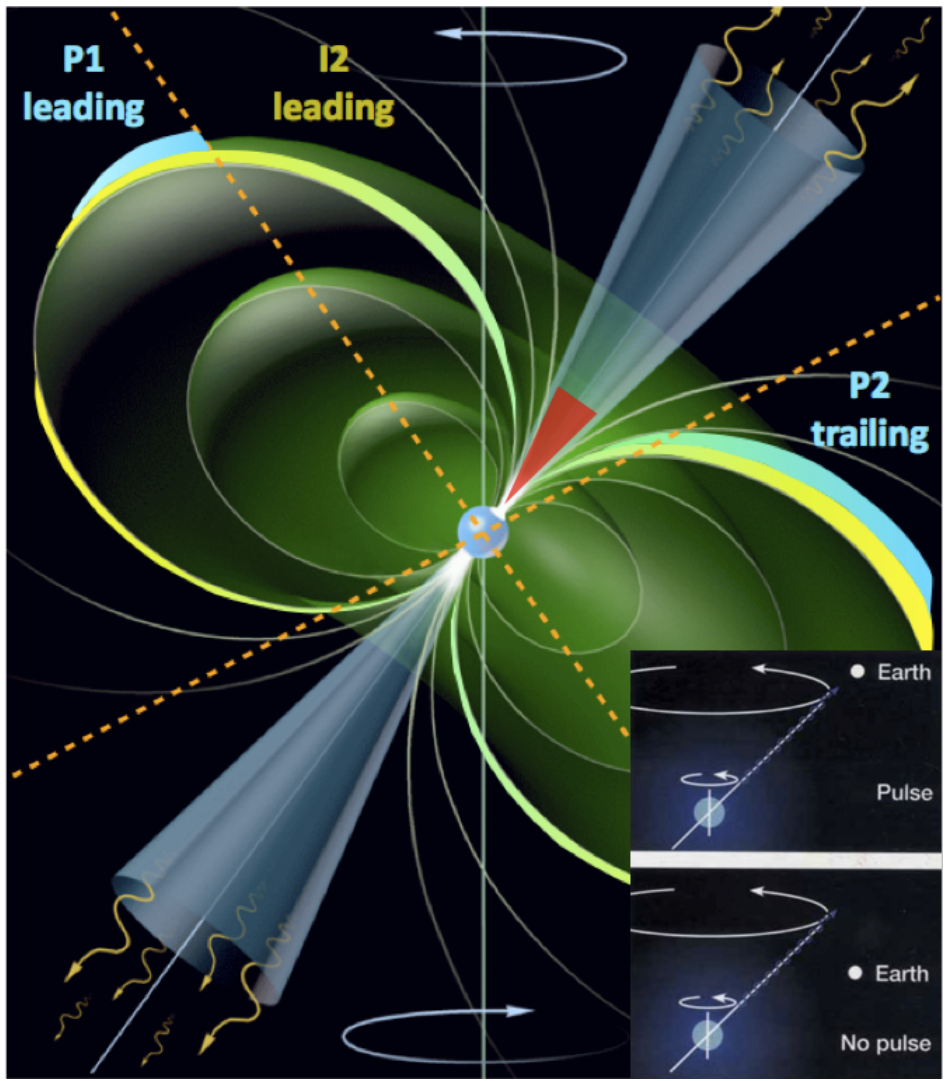

Figure 1.6: Structure of the pulsar magnetosphere according to the most accepted emission models. The radio and PC high-energy emissions, indicated in red, are collimated in conical beams, coaxial with the magnetic axis. In yellow and cyan are respectively indicated the $\gamma$-ray emission region for the slot gap (SG) and outer gap (OG) models. Two insert panels illustrate the *lighthouse effect*, responsible of the pulsation.

---

[1] When Goldreich & Julian formulated the model the only known pulsated emission was at radio wavelengths so they did not take into account any possible pulsated high energy emission.

### 1.3.1 The Goldreich-Julian model

The first assumptions of the Goldreich-Julian model is that inside the pulsar, the Lorentz force is much more intense than the gravitational one and the NS matter, in both degenerate interior and non degenerate atmosphere, could be considered an excellent conductor. At any point within a rotating magnetised sphere with spin frequency $\Omega$ there will be an induced electric field $(\mathbf{\Omega} \times \mathbf{r}) \times \mathbf{B}$ due to the presence of the magnetic field $\mathbf{B}$. If this sphere is a perfect conductor as well, the induced electric field will be balanced by a charge segregation that will generate an electric field $\mathbf{E}$. Under these conditions, for every point $\mathbf{r}$ inside the sphere, a *force-free* state is obtained

$$\mathbf{E} + \frac{1}{c}(\mathbf{\Omega} \times \mathbf{r}) \times \mathbf{B} = 0. \tag{1.20}$$

Now, by assuming an empty space surrounding the NS, the electrical charges on the surface will induce a quadrupole external field (unipolar induction)

$$\Phi(r, \theta) = \frac{B_S \Omega R^5}{6cr^3}(3\cos^2 \theta - 1) \tag{1.21}$$

where $(r, \theta)$ are the polar coordinates of a system centred on the star. From this induced field an electric field will be generated

$$E_\parallel = \frac{\mathbf{E} \cdot \mathbf{B}}{B}\bigg|_{r=R} = -\frac{\Omega B_S R}{c}\cos^3 \theta \tag{1.22}$$

causing an electric force $F = qE_\parallel$ on the surface charges that, for the typical pulsar parameters, will exceed the gravity force by more than 10 orders of magnitude. Charged particles will be extracted from the NS surface and will fill the space region surrounding the star. This demonstrates that the NS surrounding cannot be empty but should be filled with charges. The initial condition of a vacuum region surrounding the NS cannot be maintained anymore and the NS will be surrounded by a plasma of density

$$\rho_e(r, \theta) = \frac{1}{4\pi}\nabla \mathbf{E} = -\frac{\mathbf{\Omega B}}{2\pi c} = -\frac{B_S \Omega R^3}{4\pi cr^3}\left(3\cos^2 \theta - 1\right). \tag{1.23}$$

This plasma will generate, with the magnetic field lines, the *pulsar's magnetosphere*. The charge density above the NS's polar caps and the equatorial region will have opposite sign with the null charge surface corresponding to $\cos \theta = \sqrt{1/3}$ (figure 1.7). The *force free state* condition, defined from the equation (1.20), will hold also in the magnetosphere. With such a charge density, the particle density $n = \rho_e/e$ is

$$n_{GJ} = \frac{\Omega B_S}{2\pi ce} \simeq \frac{B_S}{ceP} = 7 \times 10^{10}\text{cm}^{-3}\left(\frac{P}{1\text{s}}\right)^{-1/2}\left(\frac{\dot{P}}{10^{-15}\text{s/s}}\right)^{1/2} \tag{1.24}$$

and it is called *Goldreich-Julian density.*

Since the magnetosphere and the NS interior are permeated by the same field $\mathbf{E} \times \mathbf{B}$, the magnetosphere is forced to co-rotate with the NS. This co-rotation must break at very high distances from the NS surface where the linear velocity $v = \Omega_{NS} r$ approaches the speed of light. The distance $R_{lc}$, for which $v = \Omega_{NS} r = c$, is called light cylinder radius, and represents the radius of a cylindrical surface that contains the co-rotating magnetosphere. It is defined as:

$$R_{CL} = \frac{c}{\Omega} = \frac{cP}{2\pi} \simeq 1.77 \times 10^4 \text{km} \left( \frac{P}{s} \right) \tag{1.25}$$

and the magnetic field computed at a distance $R_{lc}$ from the star surface is

$$B_{LC} = B_S \left( \frac{\Omega R}{c} \right)^3. \tag{1.26}$$

The existence of the light cylinder implies the existence of two kind of magnetic field lines:

**closed** that co-rotate with the pulsar, inside the light cylinder and connect the two magnetic poles

**opened** that do not co-rotate with the pulsars, do not connect the magnetic poles and are free to escape the pulsar

This distinction is a direct consequence of the dipolar magnetic field structure and the light cylinder radius. In fact, the closer we get to the magnetic pole, the longer is the way the magnetic field line has to cover to reach the other pole and the bigger will be the distance reached from the line to the NS surface. So there will be a magnetic latitude on the pulsar surface above which all the magnetic field lines at higher latitude are opened and free, while for lower latitude they are closed and co-rotating. Since the quantity $(\sin^2 \theta / r)$ is constant for a dipolar field, for the last closed magnetic field line we have

$$\frac{\sin^2 \theta}{r} = \frac{1}{R_{LC}} = \frac{2\pi}{cP} = \frac{\sin^2 \theta_P}{R}. \tag{1.27}$$

Under the assumption that the radius of the polar cap $R_p$ (from which the open lines come out) is not too big, we can use the equation (1.27) to estimate it. For a 13 km radius pulsar we have

$$R_P \simeq R \sin \theta_P = \sqrt{\frac{2\pi R^3}{cP}} = 215 \text{ m} \left( \frac{P}{s} \right)^{-1/2}. \tag{1.28}$$

This model is a very nice and powerful instrument to have an idea of how a pulsar works, what is the basic physics involved, and to have a rough estimation of the value of some of the most important pulsar parameters. All the same, the Goldreich-Julian model cannot be applied when magnetic and rotational axes are misaligned.

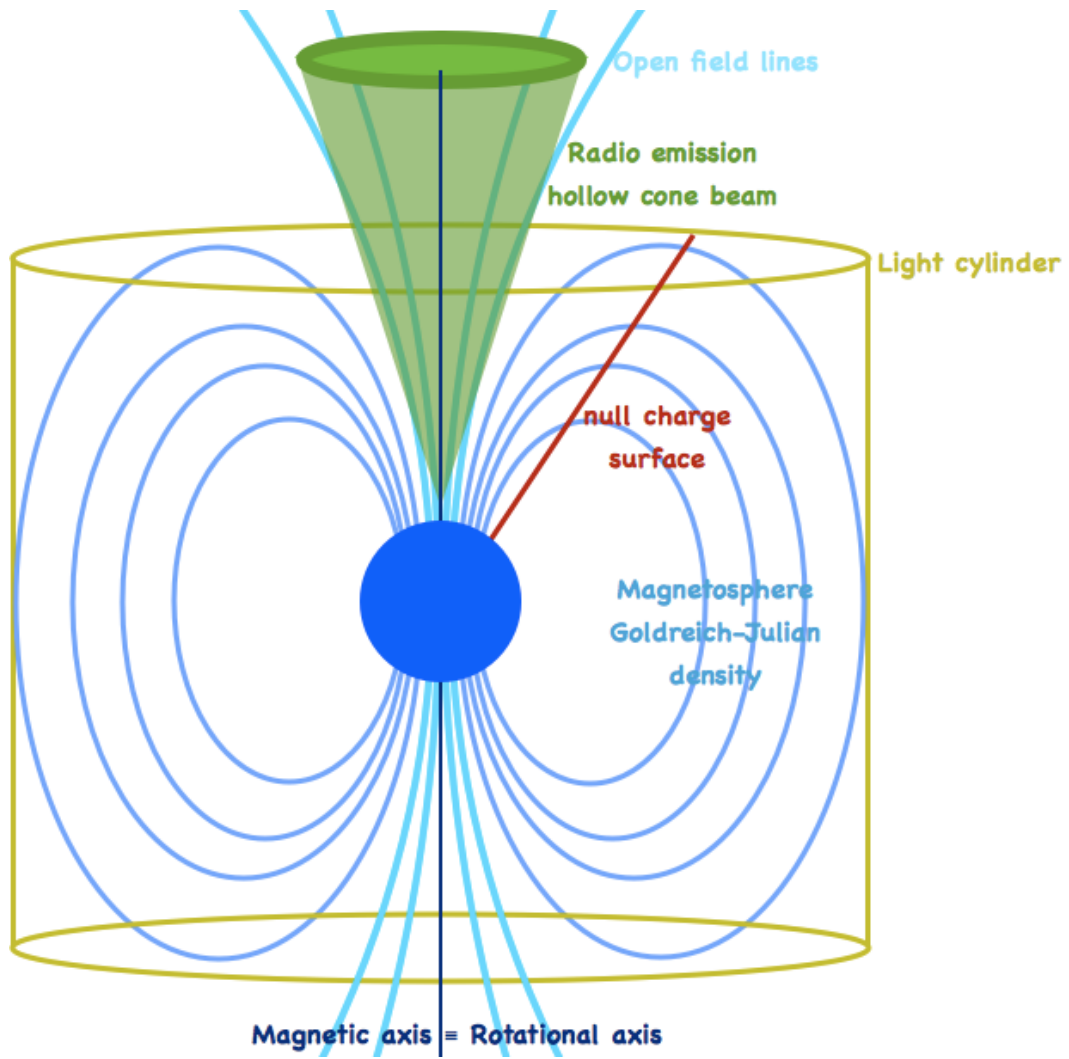

Figure 1.7: The Goldreich-Julian model. The magnetic and rotational pulsar axes are aligned with the light cylinder axis. The conical emission beam is indicated in green, starting from the bases of the opened field line while the closed ones, with the the Goldreich-Julian charge density, form the magnetosphere. In red it is indicated the limit at which the magnetosphere charge density changes sign.

### 1.3.2  The emission mechanism, first approach

The nature of the electromagnetic pulsed signal is strictly connected with the intrinsic characteristics of the pulsar magnetosphere. After the discovery of the first pulsar, several theoretical models were proposed to explain the emission mechanism (at that time just the radio emission were known), for example Smith 1969-1970 . The pure geometrical model that was accepted and also today represents one of the best interpretations of the radio emission was formulated by Radhakrishnan & Cooke in 1969 and from Komesaroff in 1970. In this model the emission is collimated in conical beams that are coaxial with the magnetic field axis. Taking advantage of the Goldreich-Julian model, Radhakrishnan & Cooke proposed that the particles can escape the pulsar along the opened magnetic field lines. In this way, escaping the pulsar, each particle will experience an acceleration due to the curvature of the magnetic field line, curvature that will gradually decrease on lines closer to the magnetic pole. These particles, accelerated in the pulsar magnetic field will emit a radiation, called *curvature radiation* in a direction that is tangent to the magnetic field line at the moment of emission. It is analogous to the synchrotron radiation and it will be characterised by frequencies of the order of

$$\nu = \frac{3}{2}\gamma^3 \frac{c}{R_c} \tag{1.29}$$

where $R_c$ is the magnetic field line curvature radius. Let us note that for $R_c \approx R_{NS}$ and for a maximum Lorentz factor of $\gamma_{max} \sim 10^7$, we get emission frequencies of $10^{25}$ Hz, corresponding to $\gamma$ photons of energy $10^{-2}$ erg. These super energetic photons, in the intense magnetic field of the pulsar, will generate a $e^+ - e^-$ pair (pair production first proposed by Sturrock 1971), which, for the curvature principle, will generate other photons. It will start an *electromagnetic cascade* of $e^+ - e^-$ pairs and photons of decreasing energies that will degrade the energy until the radio wavelength (this is one of the several possible explanation of the pulsar radio emission). The radiation produced in this process will be collimated in a widening conical beam with aperture defined by the direction of the tangent to the last opened magnetic field line at the altitude of the emission. If the cone amplitude reach the very last open magnetic field line, its amplitude could be expressed, in polar coordinates, as

$$\tan\theta = -\frac{3}{2\tan\rho} \pm \sqrt{2 + \left(\frac{3}{2\tan\rho}\right)^2}, \tag{1.30}$$

Where $\rho$ is the opening angle of the emission cone.

In this model the radiation density inside the emission beam it is not supposed to be uniform. In fact, from the equation (1.29) it is clear that the emission frequency decrease near like $R_c^{-1}$ and tend to zero in correspondence to the magnetic pole, where $R_c = \infty$. In this model the emission beam structure assumes the shape of an *hollow cone*. This *hollow cone model* is able to explain some pulsar characteristics observed in the radio, like double peaks. However, a number of pulsed profiles show more complex structures that are not explicable just by the *hollow cone* beam model. To be able to explain these structures, the *hollow cone model* has been extended to include something like a core *pencil like beam* (Beker 1976).

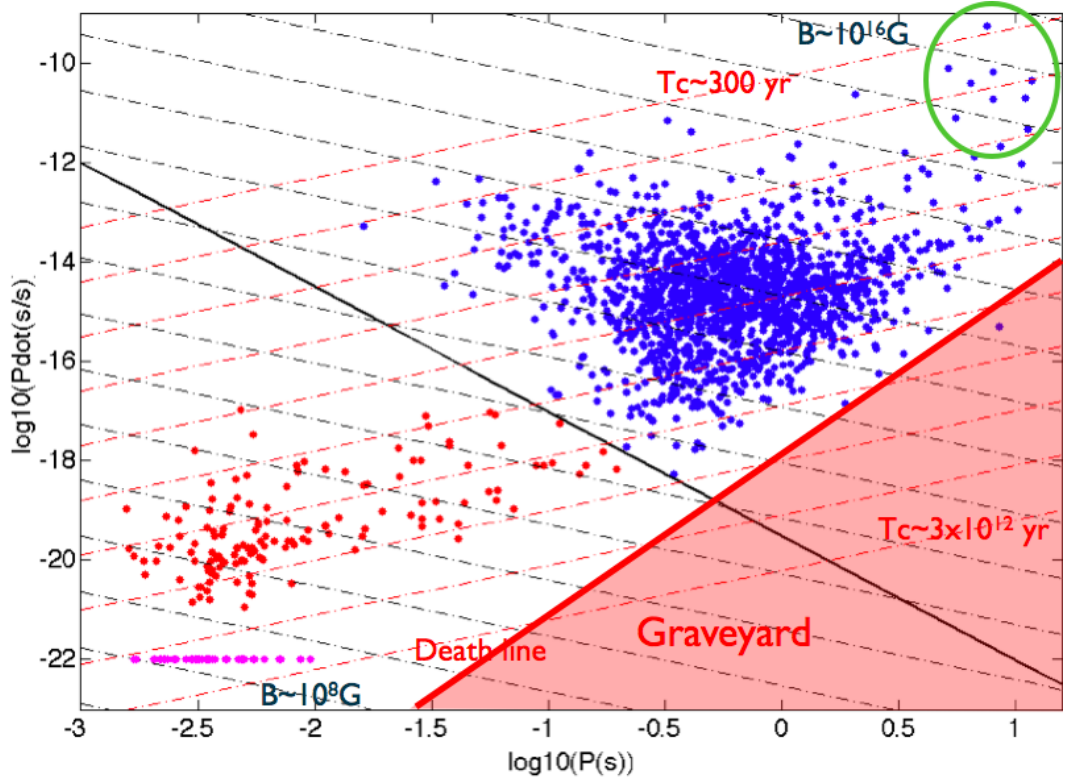

Figure 1.8: $P - \dot{P}$ diagram of all the ATNF database pulsars as of the end of October 2009. The plot is divided in two regions: the active pulsars one, and the death pulsars one (Graveyard), respectively to the left and to the right of the thick red death line (equation 1.32). The ordinary pulsars are plotted in blue and the millisecond pulsars in red. These two pulsar populations are separated in the plane by the black thick line, defined by equation (1.31). In the top right part of the diagram is, circled in green, the region occupied by a third pulsar population, the magnetars. The MSP with no $\dot{P}$ estimate have been plotted with $\dot{P} = 10^{-22}$ in magenta.

## 1.4   $P - \dot{P}$ diagram

The $P - \dot{P}$ diagram is a cartesian plane with a period scale on the abscissa and a period first time derivative one on the ordinate. It allows to recognise the existence of different pulsar populations and to study the possible evolutive scenario of each pulsar population. In figure (1.8) is showed the $P - \dot{P}$ diagram for all the pulsars in the ATNF[2] (Australian Telescope National Facility) catalogue (Manchester, R. N., Hobbs, G. B., Teoh, A. & Hobbs, 1993-2006), update as of October 2009.

In the plot are drawn all the pulsar of the ATNF database: the blue ones are normal pulsars (NPs) while the red ones are millisecond pulsars (MSPs).

The distinction between these two main pulsar populations is defined by the condition

$$\log \dot{P} < -19.5 - 2.5 \log P \qquad (1.31)$$

that, in figure 1.8 is indicated by a thick black line that separates blue NPs from red MSPs. If the Equation 1.32 is satisfied, the pulsar is classified as a MSP, otherwise it is classified as a NP. In figure (1.8) is shown one other pulsar group, coloured in magenta. These are the pulsar with no acceptable $\dot{P}$ estimation. In figure (1.8) is indicated one other pulsar population, circled in green, at the top right side of the diagram. The pulsars in this region are called *magnetars* and are characterised by the most intense magnetic field.

From equation (1.6), given $P$ and $\dot{P}$ and assuming a value for the radius and the moment of inertia of the pulsar, it is possible draw constant magnetic field lines in figure (1.8) as dotted black lines. From equation (1.9), it is possible to draw lines of constant characteristic age as red dotted lines.

Another very important characteristic that it is possible to show in the $P - \dot{P}$ diagram is the region in which the emission pulsar mechanism is no longer efficient. That region is called *pulsar graveyard* and is indicated like the red $P - \dot{P}$ diagram side in figure (1.8). The thick red line that marks the graveyard is called *pulsar death line*. All the pulsars with a too low angular frequency or with a too low $\dot{P}$ value will cross the death line and will stop to produce pulsed emission. Mathematically the death line could be obtained for the radio or $\gamma$-ray emission mechanism from the following conditions:

- The mean free path of the $\gamma$-ray should be lower that the magnetic field scale

- The pairs $e^+ - e^-$ should be energetic enough to produce $\gamma$-ray and start the electromagnetic cascade.

These two conditions require that

$$\frac{B}{10^{12}} P^{-2} \gtrsim 0.2 \qquad (1.32)$$

that defines the death line.

Turning now to pulsar evolution, in the most accepted evolutive scenario a pulsar is born with a rapid rotation and an intense magnetic field, in the left top region of figure (1.8). It will slow down and shift toward higher period and lower $\dot{P}$ regions in the $P - \dot{P}$ diagram.

## 1.5   Pulsar populations and evolution

One of the most accepted model to explain the formation of different types of systems that include a pulsar is roughly illustrated in figure (1.9). In a binary system, a NS is formed as a consequence of a SN explosion. In this case, assuming a symmetric explosion, if the mass ejected by the SN is more than half the total mass of the system, this one will be destroyed. The NS and the companion will move away in more or less opposite directions. In the other case we will have a binary system composed by a NS and by an ordinary star. After the explosion,for $10^{7-8}$ years , the NS could be observed like a pulsar (it could develop a pulsed periodic signal), gradually increasing its rotational period as a function of time. After this time interval, the pulsar will have converted a considerable fraction of its kinetic energy into electromagnetic radiation and it will no longer be able to produce pulsed emission. The pulsar will cross the *death line* (figure 1.8). After the "pulsar death", just for the system that survived the SN explosion, the ordinary star companion will evolve until the red giant phase. It will fill the *Roche lobe* and mass will start to flow from the younger giant to be accreted by the NS. The accretion process may transfer angular momentum to the pulsar that will spin up until its frequency will be able to reactivate the pulsed emission mechanism. The results of this process will be a recycled pulsar that will have spun up to rotational frequencies of the order of a few milliseconds. During the accreting phase the system will emit X-rays via gravitational energy conversion during the mass transfer. The system could be observed as an *X-ray binary*. If the system encompasses a high mass star (*HMXB, High Mass X-ray Binary*), such star can also explode. In this case, if the system survives the explosion, a double pulsar system forms. With a young one and a recycled one that rapidly move away from each other. If the companion star is a low mass one (*LMXB, Low Mass X-ray Binary*), the spin up phase will go on until the companion will have lost all of its envelope to the pulsar. At the end of the accretion process we will have a binary system composed by a very fast millisecond pulsar ($P \sim 1$ ms) in a very rapid orbit around a white dwarf.

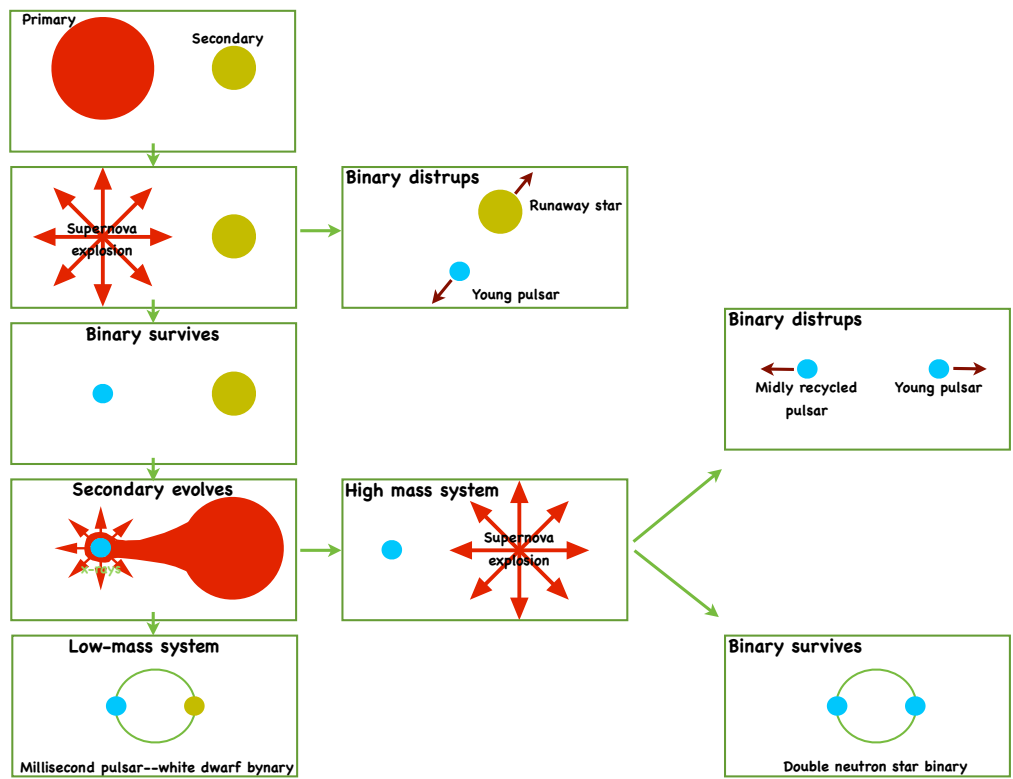

Figure 1.9: Diagram that schematically shows one of the most accepted theory about pulsar evolution. This model also provides an explanation for the presence of MSPs on the galactic plane, classifying them like the results of a HMXB system destroyed by the SN explosion of the companion.

Some millisecond pulsars have not been observed in a binary system. While the existence of these objects could be understood in a globular cluster (GC) it is much more difficult to explain their presence in the galactic plane. In fact, because of the very high star density in GC, the probability of gravitational interaction is much higher than that in the galactic plane, where the gravitational interaction probability for two masses is negligible. The simplest explanation of the existence of isolated MSPs in the galactic plane, is based on the distruption of the binary system after the SN explosion of the massive companion. This will generate an isolated MSP and an isolated young pulsar rapidly moving away from each other. This last assumption could be confirmed by the fact that plotting these galactic plane MSPs on the $P - \dot{P}$ diagram, they will occupy the binary NS region. Some other MSPs are isolated because they were *Black Widows* (King et al., 2003) that evaporated their companion.

# Chapter 2

# $\gamma$-ray emission gap models & radio model

In this chapter I will describe the $\gamma$-ray emission models and the radio model I have developed in this thesis work. The implemented $\gamma$-ray emission models are the low and high altitude slot gap models, referred to here as Polar Cap (PC) and Slot Gap (SG) (Muslimov & Harding 2003 & Muslimov & Harding 2004), then the Outer Gap model (OG) (Cheng et al. 2000), and a variation of the OG, proposed by Watters et al. (2009) & Romani & Watters (2010), the One Pole Caustic model (OPC). The last section of the chapter is dedicated to the description of the implemented radio model (Gonthier et al., 2004; Harding et al., 2007).

## 2.1  The low and high altitude slot gaps: *PC & SG models*

PC & SG emission could be considered as two separate energetic and geometrical components although they originate from the very same physical process: a polar cap $e^+$-$e^-$ pair cascade. As we will see, this physical phenomenon implies two distinct $\gamma$-ray emission components: one from the cascade radiation above the *pair formation front*, PFF, at low altitude, near the polar cap, and the other one from primary electrons in a narrow gap, the slot gap, that extends from low altitude near the NS surface up to the external magnetosphere.

### 2.1.1  PC & SG: emission region and structure

**Polar cap**

In the *polar cap model* the emission comes from a region close to the NS surface and well confined above the NS magnetic polar cap. Primary charged particles above the surface are accelerated in the strong electrostatic field generated by the difference between the actual space charge density and the Goldreich-Julian one, and by the inertial frame dragging. Escaping the NS along the open magnetic field lines, they emit high energy photons by curvature radiation (CR) and inverse Compton scattering (ICS) radiation. The primary γ-rays are absorbed and produce $e^+$-$e^-$ pairs in the intense magnetic field near the surface. It is important to note that the particles are not accelerated during the pair cascades. Above the PFF the parallel electric field is screened within a very short distance. The energy for the cascade comes from the primary particles acceleration.

The PFF (figure 2.1) is the region in which the first $e^+$-$e^-$ pairs are generated. The further interactions of these pairs via $e_{acc}^{-/+} + B \rightarrow \gamma$, $\gamma + B \rightarrow e^+ - e^-$ ... will generate a pair cascade.

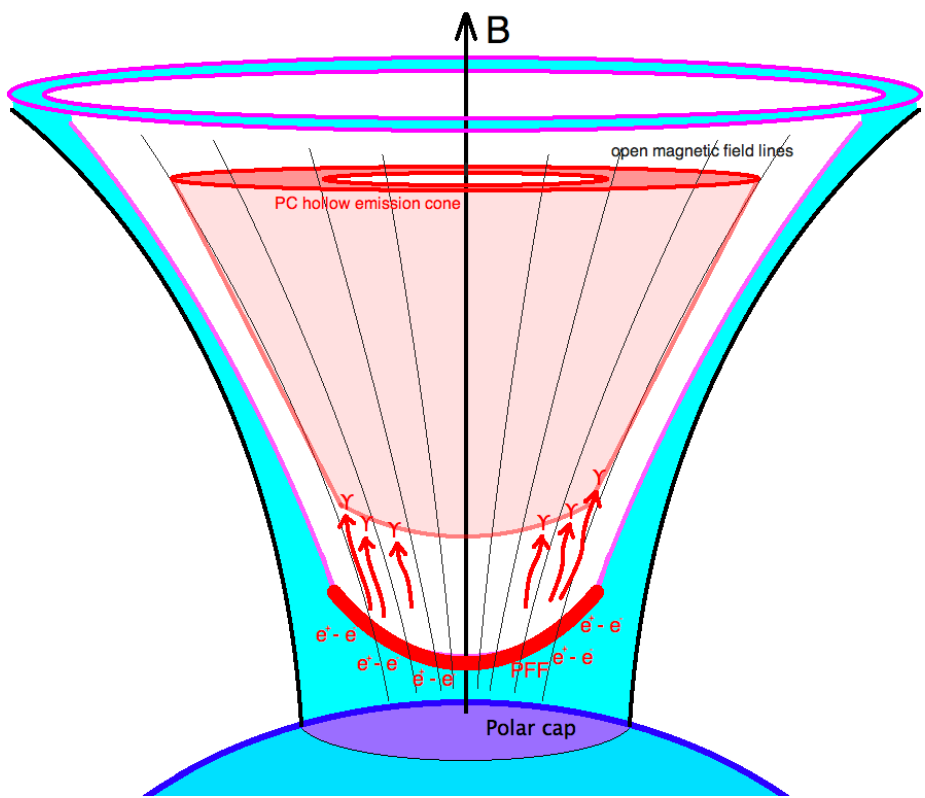

Figure 2.1: Schematic representation of the PC model emission. The charged particles generated in the pair formation front, interacting with the NS magnetic field, will start an electromagnetic cascade that will lead the γ-ray PC emission. Because the proximity of the emission region to the NS surface and since the charged particles emit along the magnetic field lines, the emission beam is conical and well collimated with the polar cap edges. The hollow cone structure is due to the fact the intensity of the emission is proportional to the magnetic lines curvature that decreases going from the polar cap edges toward the magnetic pole.

The pair cascade may be initiated by high energy photons by curvature radiation or inverse Compton (IC) of thermal X-rays from the neutron star surface by the primary electrons. CR-initiated cascade can screen the parallel electric field, ICS-initiated cascade cannot. The CR-initiated pairs are produced in abundance only by the younger pulsars having ages $\tau \lesssim 10^7$ yr and by some millisecond pulsars (Harding & Muslimov, 2001). The pair cascade process will lead to a high charge multiplicity that will generate a charge density that will screen the parallel electric field within a short distance above the PFF. The pair plasma will likely establish force-free conditions along the magnetic field lines above the PFF, as well as radiating a spectrum of $\gamma$-rays via synchrotron emission. Over most of the polar cap the PFF and $\gamma$-ray emission occur well within a stellar radius of the surface, and are well aligned with the polar cap. This alignment makes the PC emission beam very collimated to the polar cap edges, forming a narrow conical $\gamma$-ray beam. The main contribution to the $\gamma$-ray emission comes from the curvature radiation (CR). Because of the curvature of the magnetic field lines, the charges will feel a centripetal acceleration that will imply emission in a direction that is tangent to the field line at the emission point. Since the curvature of the magnetic field lines decreases from the edges of the polar cap toward the magnetic axis, the CR will confer to the beam the structure of a hollow cone (fig. 2.1).

**Slot gap**

The *slot gap* emission is generated exactly from the same physical process. Going toward the polar cap edges, near the boundary of the open magnetic field lines region, $E_\parallel \to 0$ because force-free conditiona apply to the closed magnetosphere The PFF rises to higher altitude (figure 2.2), and the electrons must accelerate over a longer distance to initiate the pairs cascade. A narrow gap, *the slot gap*, is formed along the first close magnetic field line, where the PFF is never established, and electrons can continue accelerating and radiating into the outer magnetosphere (figure 2.2). The SG region is defined between the last open magnetic field line and the magnetic field line with a colatitude value $(1 - \Delta\xi)$ times smaller, with $\Delta\xi$, the SG width, expressed in unit of the dimensionless colatitude of a PC magnetic field line $\xi$. As it will be accurately described in section (2.1.1), we have:

$$\xi = \frac{\theta}{\theta_0}, \;\; with \;\; \theta_0 \sim \left[ \frac{\Omega R}{cf(1)} \right]^{0.5}$$

where $\theta_0$ (see figure 2.2) is PC half angle, $f(1)$ is a correction factor for the dipole component of the magnetic field in a Schwarzschild metric, $R$ is the NS radius, and $c$ is the speed of light. The SG high energy emission beam assumes the shape of a hollow cone. Its origin from the high altitude cascades above the interior edge of the slot gap, makes it much broader compared to the PC one and well of the magnetic axis. This beam structure implies that the SG emission pattern considerably changes for different observer lines of sight.

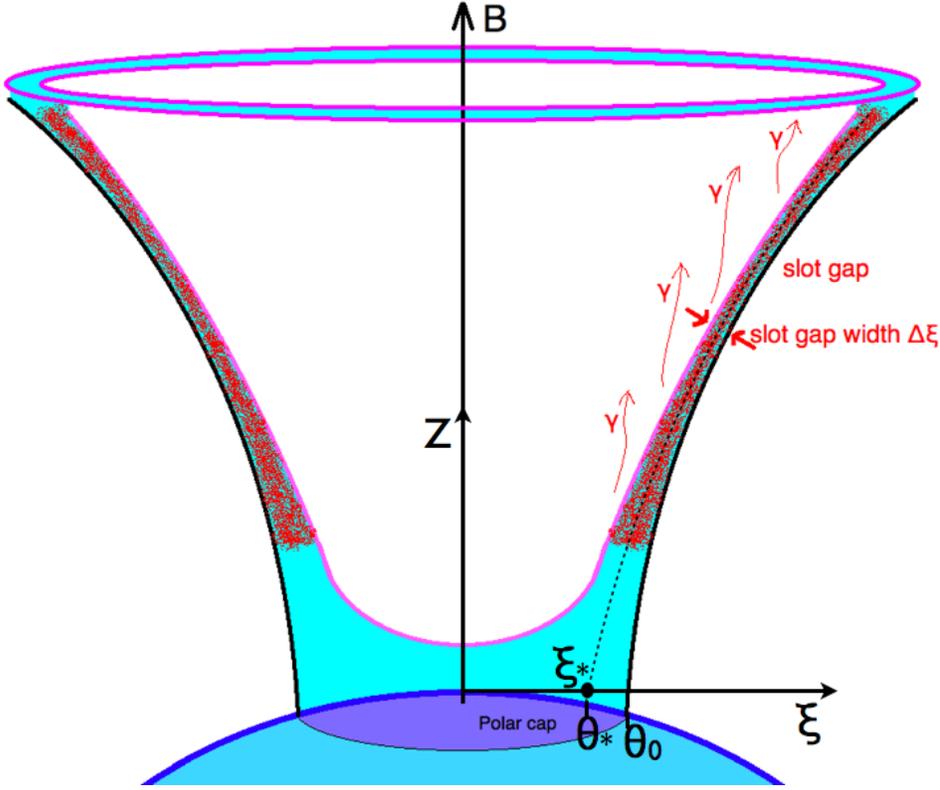

Figure 2.2: Schematic representation of the SG model emission. The polar cap PFF asymptotically approaches the outer boundary of the open field lines region, indicated by the black thick magnetic field line . The narrow SG region is formed and the charges will accelerate and emit along the slot gap up to the outer magnetosphere. A thin black line midway in the slot gap, intersects the polar cap in $\theta_*$, the colatitude at the centre of the slot gap.

**Luminosity**

It is possible to estimate the emission from the PC pair cascades along the low-altitude part of the gap, above the PC, by assuming that mono-energetic radiation is emitted tangent to field lines, with a distribution

$$N_\gamma = \begin{cases} N_0 \exp[-(s_{fade} - s)/\Delta_{in}], & s < s_{fade} \\ N_0 \exp[-(s - s_{fade})/\Delta_{out}], & s > s_{fade} \end{cases} \qquad (2.1)$$

where $s$ is distance above the neutron star surface, $s_{fade} = 2.5R$, $\Delta_{in} = 1.0R$ and $\Delta_{out} = 2.0R$. Such an emission distribution approximates that of simulated pair cascades (Muslimov & Harding 2003). The width of the slot gap, $\Delta\xi_{SG}$, is a function of pulsar period, $P$, and surface magnetic field, $B_{12} \equiv B_0/10^{12}$ G (Muslimov & Harding 2003), and can be expressed as a fraction in colatitude $\xi \equiv \theta/\theta_0$ of the polar cap opening angle, $\theta_0 \simeq (\Omega R/c)^{1/2}$ (figure 2.2). In open-volume coordinates (*ovc*), the photons from the low-altitude SG pair cascades are emitted at $r_{ovc}^{min} = r_{ovc}^{max} = 1 - \Delta\xi_{SG}$. The luminosity of the SG from each pole is

$$L_\gamma^{SG} = \varepsilon_\gamma \alpha c \int_0^{2\pi} d\phi_{PC} \int_{\theta_0(1-\Delta\xi_{SG}/2)}^{\theta_0} \rho(\eta)\Phi(\eta)r^2 \sin\theta d\theta \qquad (2.2)$$

where $\rho(\eta)$ and $\Phi(\eta)$ are the primary charge density and potential as a function of the emission altitude $\eta \equiv r/R$, in units of NS radius, and $\varepsilon_\gamma$ is the radiation efficiency. Using the expressions for $\Phi$ and for $\rho$ from (Muslimov & Harding 2003)

$$\Phi(\eta, \xi_*, \phi_{PC}) = \phi_0\theta_0^2 \left\{ \nu_{SG}\mathscr{A}(1-\xi_*^2)\cos\alpha + 2\mathscr{B} \times \right.$$
$$\left. \times \left[ 1 - \frac{\cos h(\sqrt{\nu_{SG}}\xi_*)}{\cos h(\sqrt{\nu_{SG}})} \right] \sin\alpha\cos\phi_{PC} \right\} \qquad (2.3)$$

$$\rho = -\frac{\Omega B_0}{2\pi c\alpha\eta^3}\frac{f(\eta)}{f(1)} \left[ (1-\kappa)\cos\alpha + \frac{3}{2}\theta_{SG}(1)H(1)\sin\alpha\cos\phi_{PC} \right] \qquad (2.4)$$

we have

$$L_\gamma^{SG,low} = \varepsilon_\gamma \dot{E}_{sd}\Delta\xi_{SG}^3(1-\frac{\Delta\xi_{SG}}{2})[\kappa(1-\kappa)(1-\frac{1}{\eta^3})\cos^2\alpha +$$
$$+ \frac{9}{8}\theta_0^2(1-\frac{\Delta\xi_{SG}}{2})H^2(1)\left[ \frac{H(\eta)}{H(1)}\sqrt{\eta\frac{f(\eta)}{f(1)}-1} \right]\sin^2\alpha] \qquad (2.5)$$

where $\dot{E}_{sd} = \Omega^4 B_0^2 R^6/6c^3 f(1)^2$ is the spin-down power, $\kappa = 0.15I_{45}/R_6^3$, $I_{45}$ is the NS moment of inertia in unit of $10^{45}$ g cm$^2$, $H$ is a relativistic correction factor of order 1, $f$ is the correction factor for the dipole component of the magnetic field in a Schwarzschild metric, $\alpha$ is the pulsar obliquity and $\phi_{PC}$ is the magnetic azimuthal angle (Muslimov & Tsygan 1992 and Harding & Muslimov 1998). The symbols $\mathscr{A}$ & $\mathscr{B}$, and $\nu_{SG}$ (Muslimov & Harding 2003) are defined as:

$$\mathscr{A} = \kappa \left( 1 - \frac{1}{\eta^3} \right) \qquad (2.6)$$

$$\mathscr{B} = \frac{3}{2} \left[ H(\eta)\theta_{SG}(\eta) - H(1)\theta_{SG}(1) \right] \qquad (2.7)$$

$$\nu_{SG} = \frac{1}{4}\Delta\xi_{SG}^2 \qquad (2.8)$$

where $\theta_{SG}(\eta)$ and $\theta_{SG}(1)$ are respectively the slot gap colatitudes at the fractional height $\eta$ and on the surface.

To model the emission component from primary electrons in the high-altitude SG, in other words in the rising part of the gap along the last closed field line, we assume that radiation is emitted along the field lines in the SG, up to altitude $\eta = \eta_{max}$. We assume a distribution of emissivity across the SG,

$$N(\xi_*) = (1-\xi_*^2) \qquad (2.9)$$

from $r_{ovc}^{min} = (1 - \Delta\xi_{SG})$ to $r_{ovc}^{max} = \Delta\xi_{SG}$, where $\xi_* = 0$ (Figure 2.2) at the center of the SG and

$$\xi_* = 1 - \frac{2(1 - \xi)}{\Delta\xi_{SG}}. \tag{2.10}$$

Such a distribution follows from the $\xi_*$ distribution of the SG potential (Muslimov & Harding 2004). In the high SG model, the width of the slot gap $\Delta\xi$, could be estimated as the magnetic colatitude where the variation in height of the curvature radiation PFF with colatitude becomes comparable to a fraction $\lambda$ of the stellar radius R:

$$\left(\frac{\partial z_0}{\partial\xi}\right)_{\xi=\xi_{SG}} = \lambda. \tag{2.11}$$

The equation 2.11 defines the SG condition: the SG acceleration sets in when the characteristic variation of the PFF height becomes comparable to $\lambda$ times the stellar radius. In equation 2.11, $z_0$ represents the dimensionless altitude, of the PFF due to curvature radiation (Figure 2.1)

$$z_0 = 7 \times 10^{-2}\frac{P_{0.1}^{7/4}}{B_{12}I_{45}^3/4}\frac{1}{\xi^{1/2}(1 - \xi^2)^{3/4}} \tag{2.12}$$

where $I_{45}$ is the moment of inertia in unit of $10^{45}$ g cm$^{-2}$, $P_{0.1} = P/0.1$ s, and $B_{12}$ is the magnetic fiels in unit of $10^{12}$ Gauss. By deriving equation 2.12, as indicated in equation 2.11, it is possible to have an estimation of the SG width $\Delta\xi$

$$\frac{\partial z_0}{\partial\xi} = \frac{z_a}{2}\xi^{-3/2}(1 - \xi^2)^{-7/4}(4\xi^2 - 1) = \lambda \tag{2.13}$$

where $z_a = 0.07P_{0.1}^{7/4}B_{12}^{-1}I_{45}^{-3/4}$. By solving numerically equation 2.13 one gets $\xi_{SG}$ for a specific pulsar, and the $\Delta\xi$ gap width value is then obtained as

$$\Delta\xi = 1 - \xi_{SG}. \tag{2.14}$$

As we will see in section 5.4.2, the choice of the $\lambda$ parameter is the most important assumption from which depends the luminosity and the emission geometry of the SG model.

Using the equations for $\Phi$ and for $\rho$ from Muslimov & Harding 2004

$$\Phi = \left(\frac{\Omega R}{c}\right)^2\frac{B_0}{f(1)}R\nu_{SG}\left\{\left[\kappa\left(\beta - \frac{1}{\eta_c^3}\right) + 1 - \beta\right]\left(1 + \frac{\eta}{\eta_{lc}}\right)\cos\alpha + \right.$$
$$\left. + \frac{3}{2}\theta_0 H(1)\left[\frac{H(\eta_c)}{H(1)}\sqrt{\eta_c\frac{f(1)}{f(\eta_c)}} - \beta\right]\sin\alpha\cos\phi_{PC}\right\}(1 - \xi_*^2) \tag{2.15}$$

$$\rho \approx -\rho_0\frac{f(\eta)}{f(1)}\frac{1}{\alpha\eta^3}\beta(\eta)\left\{[\alpha_0(\xi) + \alpha_1(1,\xi)]\cos\alpha + [b_0(\xi) + \right.$$
$$\left. + b_1(1,\xi)]\sin\alpha\cos\phi_{PC}\right\} \tag{2.16}$$

the high-altitude SG luminosity from each pole can also be determined from equation (2.2)

$$L_\gamma^{SG,high} = \varepsilon_\gamma \dot{E}_{sd} \Delta \xi_{SG}^3 \beta (1 - \frac{\Delta \xi_{SG}}{2}) \left\{ [\kappa(\beta - \kappa)(1 - \frac{1}{\eta_c^3}) + 1 - \beta] \times \right.$$

$$\left. \times (1 + \frac{\eta}{\eta_{lc}}) \cos^2 \alpha + \frac{9}{8} \theta^2 H^2(1) \left[ \frac{H(\eta_c)}{H(1)} \sqrt{\eta_c \frac{f(1)}{f(\eta_c)} - \beta} \right] \sin^2 \alpha \right\} \quad (2.17)$$

where $\beta = (1 - 3\eta/4\eta_{lc})^{1/2}$ and $\eta_{lc} = r_{lc}/R = c/\Omega R$. According to Muslimov & Harding 2004, the energies of the primary electrons in the SG will quickly become radiation-reaction limited, with the rate of acceleration balancing the curvature radiation loss rate, so that one expects 100% radiative efficiency in this case ($\varepsilon_\gamma = 1$).

## 2.2 The outer gap : *OG & one pole caustic (OPC) models*

In this section two different versions of the outer gap model will be described: the classic OG, formulated by Cheng et al. 2000 and its OPC variation, that takes into account a different gap width dependence, assumed from Watters et al. (2009) and Romani & Watters (2010). The OPC model assumes exactly the same emission pattern as the classical OG version from Cheng et al. 2000. The differences between the two formulations are the dependence of the total luminosity with the gap width and the gap width evolution with respect to the pulsar magnetic inclination angle $\alpha$ and age.

### 2.2.1   OG & OPC emission region and structure

The outer gap (OG) model has not only a different geometry from the slot gap but also a completely different electrodynamics. The outer gaps are vacuum regions characterised by a strong electric field along the magnetic field lines, located in the outer magnetosphere of a pulsar (Holloway 1973; Cheng et al 1976), near the null charge surface where $\mathbf{B} \cdot \mathbf{\Omega} = 0$. Here $\mathbf{\Omega}$ & $\mathbf{B}$ are the NS angular velocity and the NS magnetic field (figure 2.3). Four outer gap regions (Cheng et al. 1976 ) can exist in the $(\mathbf{\Omega}, \mu)$ plane ($\mu$ is magnetic momentum): a long and a short one for each pole. Generally it is assumed that for large magnetic obliquities, the $\gamma$-ray emission is generated just in the long gap region while for a nearly aligned rotator all the regions could be active. However, also in this last case the emission from the long gap region should be strong enough to quench the short gap one.

The $\gamma$-ray emission mechanism is a consequence of a variation of the Goldreich & Julian density due to a charges segregation owing to the change of sign of the charged particles going through the null charge surface (figure 2.3). The magnetospheric charge density, expressed by the Goldreich & Julian density equation

$$\rho_0 = -\frac{\mathbf{\Omega} \cdot \mathbf{B}}{2\pi c}\left[1 + O\left(\frac{|\mathbf{\Omega} \times \mathbf{r}|}{c}\right)\right]$$

changes sign at the null surface $\mathbf{B} \cdot \mathbf{\Omega} = 0$. A charge-deficient region, the outer gap, will form between the null surface and the light cylinder in the open magnetosphere if a charge-separated flow is formed. The difference between the charge density $\rho$ and the Goldreich & Julian one $\rho_0$ will generate an electric field along the magnetic field lines. If this induced $\mathbf{E}$ is strong enough, it will be able to accelerate $e^+ - e^-$ pairs to ultra-relativistic energies to radiate γ-rays in a direction tangent to the $\mathbf{B}$ lines. The gamma-ray photons interact with thermal X-rays from the NS surface to produce pairs on field lines interior to the last open field line because of field line curvature. Since the pairs screen the electric field, the gap inner surface is defined by the surface of pair formation. The bulk of the gamma rays are radiated by pairs on the gap inner edge. A full calculation of the width of the OG radiating layer is complicated (Hirotani 2006, 2008) since the pair production, screening, and radiation occur in the same location.

**Luminosity**

It is possible to treat the OG radiation approximately by assuming that the radiation occurs in an infinitely thin layer along the gap inner edge (Watters et al. (2009) & Romani & Watters (2010)). To determine the gap width, we consider two different prescriptions. In the first one (Watters et al. (2009)) it is simply assumed that the gap width is equal to the gamma radiation efficiency:

$$L_{\gamma,OPC} = w_{OPC}\dot{E}_{sd}. \tag{2.18}$$

This assumed gap width, we call the OPC model, is not based on any physical prescription and is very different from the usual dependence luminosity $\propto$(gap width)³ (both SG and OG) based on the electrodynamics. Because of the $L_\gamma \propto \dot{E}^{0.5}$ relation observed among the LAT pulsars, the gap with should follow as

$$w_{OPC} = Const \times (\dot{E})^{0.5}. \tag{2.19}$$

Our second prescription for determining the width of the outer gap is to follow the calculations presented in Zhang et al. (2004), who determine the gap width by computing the location of the pair formation surface. The outer gap extends from the last open field line, the light cylinder, to its inner boundary, and the null surface. From Kapoor & Shukre 1998, the polar angle $\theta_c$ corresponding to the magnetic field line tangent to the light cylinder is:

$$\tan\theta_c = -\frac{3}{4\tan\alpha}\left[1 + (1 + \frac{8}{9}\tan^2\alpha)^{0.5}\right] \tag{2.20}$$

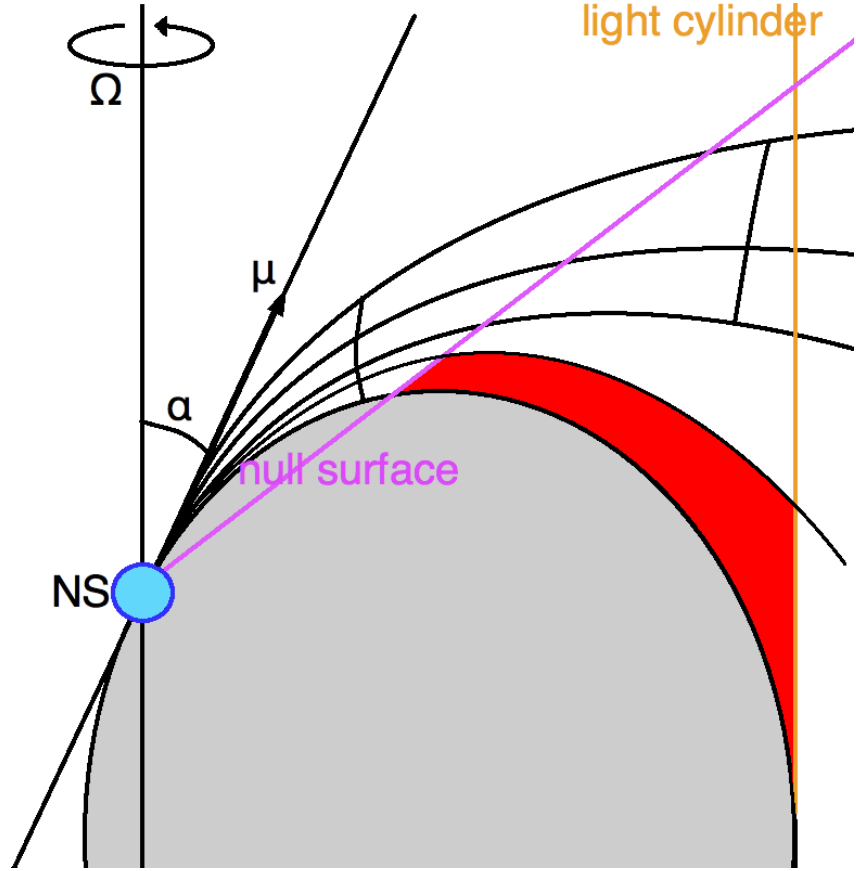

Figure 2.3: Schematic representation of the OG model emission. The OG region lies in the $(\boldsymbol{\Omega}, \mu)$ plane. The inner and outer edges that define the gap extension are defined by the emission regions that go thought the null charge surface, indicated in magenta, up to the light cylinder.

with the light cylinder radius given by

$$R_L = \frac{r_c}{\sin \theta_c}. \tag{2.21}$$

The lower boundary of the outer gap is estimated from the null-charge surface, $\boldsymbol{\Omega} \cdot \mathbf{B} = 0$, that in two dimensions is described by $(r_{in}, \theta_{in})$. By definition, the polar angle at the inner edge of the outer gap is

$$\theta_{in} = \frac{1}{2} \left( 3 \tan \alpha + \sqrt{9 \tan^2 \alpha + 8} \right). \tag{2.22}$$

The computation of $r_{in}$ is done starting from the fact that, along the last open field line, the equation

$$\frac{\sin^2(\theta - \alpha)}{r} = \frac{\sin^2(\theta_c - \alpha)}{r_c} \tag{2.23}$$

is valid. That implies

$$r_{in} = \frac{R_L \sin^2(\theta_{in} - \alpha)}{\sin \theta_c \sin^2(\theta_c - \alpha)} \tag{2.24}$$

with $\tan \theta_{in} = \frac{1}{2} \left( 3 \tan \alpha + \sqrt{9 \tan^2 \alpha + 8} \right)$.

In Zhang et al. (2004), as in previous papers dealing with OG gap geometry, the total gamma-ray luminosity is

$$L_{\gamma,OG} = w_{OG}^3 (\langle r \rangle) \dot{E}_{sd} \qquad (2.25)$$

where $w_{\mathrm{OG}}$ is the fractional width of the gap at the average gap radius $\langle r \rangle$. Zhang et al. (2004) include the dependence of gap geometry on pulsar inclination angle $\alpha$, so that the gap width is a function of $\alpha$, $P$ and $B_0$,

$$w_{\mathrm{OG}}(\langle r \rangle, P, B_0) = \eta(\alpha, P, B_0) \, w_0(P, B_0) \qquad (2.26)$$

where, $w_0(P, B_0) = 5.5 P^{26/21} B_{12}^{-4/7}$ is the fractional gap size obtained by ignoring the dependency on the inclination angle $\alpha$. Pair production too close the gap uses X-ray emission from the NS. X-rays are produced by pulsar cooling and PC-heating. In this thesis, the X-rays come from the bombardment of the NS surface by the full return current from the OG (the self-sustaining OG model). The bright X-ray luminosity allows OGs and gamma-ray emission for many old pulsars. The relation that define the fractional OG size in this case is:

$$w_{OG} = 5.2 B_{12}^{-4/7} P^{26/21} R_6^{-10/7} G(r, \alpha) \qquad (2.27)$$

where $B_{12}$ and $R_6$ are respectively the pulsar magnetic field in units of $10^{12}$ Gauss and the distance to the pulsar in unit of $10^6$ m. $G(r, \alpha)$ is a factor that is numerically solved for each pulsar by calculating the average distance $\langle r \rangle$ at which primary $\gamma$-rays are produced and along which magnetic field line they would pair produce when they interact with an X-ray coming from the NS surface (Zhang et al., 2004).

## 2.3   Radio emission model

The empirical radio emission model implemented in the simulations, has been described in detail in Gonthier et al. (2004) and Harding, Grenier & Gonthier (2007). We assume that the radio beam is composed of a core component originating relatively near the neutron star surface and a conical component radiated at higher altitude, both centered on the magnetic axis in the co-rotating frame. The adopted form of this model is similar to that proposed by Arzoumanian et al. (2002), based on the work of Rankin (1983) and Mitra & Despande (1999), and modified to include frequency dependence $\nu$ by (Harding et al., 2007). The summed flux from the two components seen at angle $\theta$ to the magnetic field axis is

$$S(\theta, \nu) = F_{core} e^{-\theta^2 / \rho_{core}^2} + F_{cone} e^{-(\theta - \bar{\theta})^2 / \omega_e^2} \qquad (2.28)$$

where

$$F_i(\nu) = \frac{-(1 + \alpha_i)}{\nu} \left( \frac{\nu}{50 MHz} \right)^{\alpha_i + 1} \frac{L_i}{\Omega_i D^2}. \quad (2.29)$$

The index $i$ refers to the core or cone, $\alpha_i$ is the spectral index of the total angle-integrated flux, $L_i$ is the component luminosity, and $D$ is the distance to the pulsar. The total solid angles of the gaussian beam profiles describing the core and cone components are

$$\Omega_{core} = \pi \rho_{core}^2 \quad (2.30)$$

$$\Omega_{cone} = 0.8\pi \rho_{cone}^2. \quad (2.31)$$

The width of the Gaussian describing the core beam is

$$\rho_{core} = 1.5° \left( \frac{P}{1s} \right)^{-0.5} \quad (2.32)$$

where $P$ is the pulsar period in seconds. The annulus and width of the cone beam are

$$\bar{\theta} = (1. - 2.63\,\delta_w)\rho_{cone} \quad (2.33)$$

$$w_e = \delta_w \rho_{cone} \quad (2.34)$$

where $\delta_w = 0.18$ (Gonthier et al. 2006), and

$$\rho_{cone} = 1.24° r_{KG}^{0.5} \left( \frac{P}{1s} \right)^{-0.5} \quad (2.35)$$

is the radius of the open field volume at the emission altitude derived by Kijak & Gil (2003)

$$r_{KG} \approx 40 \left( \frac{\dot{P}}{10^{-15} s\,s^{-1}} \right)^{0.07} P^{0.3} \left( \frac{\nu}{\nu_{1000}} \right)^{-0.26} \quad (2.36)$$

where $\nu_{1000} = 1000$ MHz. $r_{KG}$ is in units of stellar radius. The core to cone peak fluxes ratio $r$, is expressed as

$$r = r_1 \left( \frac{\nu}{\nu_1} \right)^{\alpha_{core} - \alpha_{cone} - 0.26} \quad (2.37)$$

and requires $\alpha_{core} - \alpha_{cone} - 0.26 = 0.9$, $\alpha_{core} - \alpha_{cone} = 0.64$. Gonthier et al. (2006), who have carried out a study of 20 pulsars having three peaks in their average-pulse profiles, at three frequencies, 400, 600 and 1400 MHz, find a core-to-cone peak flux ratio

$$r = \frac{F_{core}}{F_{cone}} = \begin{cases} 10^{4.1} P^{1.3} (\frac{\nu}{\nu_1})^{-0.9}, & P < 0.7s \\ 10^{3.3} P^{-1.8} (\frac{\nu}{\nu_1})^{-0.9}, & P > 0.7s \end{cases} \quad (2.38)$$

that is consistent with the core-to-cone peak flux ratio of Arzoumanian et al. (2002), at periods above about 1 s, but predicts that pulsars with $P \lesssim 0.05$ s are cone dominated. In the latter equation, $\nu_1 =$1 MHz. Such a picture is supported by polarization observations of young pulsars. Crawford et al. (2001) measured polarization of a number of pulsars younger than 100 kyr, finding that they possess a high degree of linear polarization and very little circular polarization. The luminosities of the core and cone components are

$$L_{cone} = \frac{L_{radio}}{1 + (1/r_0)}, \qquad L_{core} = \frac{L_{radio}}{1 + r_0}, \qquad (2.39)$$

where

$$r_0 = \frac{\Omega_{cone}}{\Omega_{core}} \frac{(\alpha_{core} + 1)}{(\alpha_{cone} + 1)} \frac{1}{r} \left(\frac{\nu}{\nu_0}\right)^{\alpha_{core} - \alpha_{cone}}, \qquad (2.40)$$

$\alpha_{core} = -1.96$, $\alpha_{cone} = -1.32$, and

$$L_{radio} = 2.805 \times 10^9 \, P^{-1} \dot{P}^{0.35} \, mJy \ kpc^2 \ MHz \qquad (2.41)$$

as modified from Arzoumanian et al. (2002),.

To incorporate this radio emission geometry in the retarded dipole magnetic field that I am using to simulate the high energy emission, I apply the flux $S(\theta, \varepsilon_R)$ given by Eqn (2.28) to the field lines of the *ovc*. The differential flux radiated from a bundle of field lines centred at *ovc* coordinates $(r_{ovc}, l_{ovc})$ is

$$dS(\theta, \nu) = S_i(\theta, \nu) \sin\theta \, d_{ovc} \, \theta_0 \, r_{ovc}^{max} \frac{2\pi}{N_l} \, d\nu \qquad (2.42)$$

where $N_l$ is the number of azimuthal divisions of each ring. The flux is assumed to be emitted at altitude $1.8R$ for the core component and at altitude given by Eq (2.36) for the cone component.

# Chapter 3

# The FERMI $\gamma$-ray space telescope

The Fermi Gamma-ray Space Telescope (Fermi) mission, formerly the Gamma-ray Large Area Space Telescope (GLAST), is a new generation space telescope for $\gamma$-ray observations. Successfully launched on June 11, 2008 the *FERMI satellite* hosts two instruments: the Large Area Telescope (LAT), the main one, which uses a pair-conversion technique to detect photons in an energy range from 20 MeV to above 300 GeV and a NaI and GBO scintillation detector, the Gamma-ray burst monitor (GMB), that records transient phenomena in the sky in the energy range from 8 keV to 40 MeV. Compared with its predecessor, the Energetic Gamma Ray Experiment Telescope (EGRET), the LAT has a sensitivity that is at least an order of magnitude greater and, unlike EGRET, is able to observe the whole sky several times per day allowing a much deeper and dynamic monitoring of the transient high-energy phenomena in the sky.

## 3.1   The Large Area Telescope

The Large Area Telescope (LAT) has been designed to measure the directions, the energies, and the exact time of arrivals (TOA) of the $\gamma$-rays incident on the detector. The instrument collecting surface is able to detect photons from a very wide field of view. The effective area after event reconstruction and background rejection is >8000 cm$^2$ above 1 GeV and for normal incident photons, as compared to 1200 cm$^2$ for EGRET. With such an effective area, the field of view of the LAT, $FoV = \int A_{eff}(\theta, \phi)d\Omega/A_{eff}(0,0)$, reaches 2.5 sr above 1 GeV, that corresponds to nearly 20% of the sky at a given time. ($A_{eff}$ is the effective area of the LAT). Another very important characteristic that presents a considerable improvement compared to EGRET is the angular resolution. For the LAT telescope, a typical on-axis single-photon 68% containment radius is of the order of 0.1° for a photon energy > 10 GeV. Such a big effective area and $FoV$ allow a systematic survey observation strategy. During the orbital motion, the LAT scans the sky up to 50° away from the orbital plane and so covering 75% of the sky. After one orbit the satellite rocks up to 50° on the other side of the orbital plane and it continues to scan. In this way, it is able to scan the whole sky, with a good uniformity, every 3 hours. On average, the effective area, the $FoV$, and the spatial resolution give the LAT a steady point-source sensitivity $\sim 30$ times better than EGRET.

The LAT telescope is equipped with: (1) a precision converter tracker (TKR) that provides the direction of the incoming photon, (2) a calorimeter (CAL) that measures the energy of the incident photons, (3) an anti-coincidence detector (ACD) that provides charged-particle background rejection, and (4) a programmable trigger and data acquisition system (DAQ) that utilises the signals from the other three instruments to produce a trigger and record data. The global structure of the on board instruments, and the choice of the detector, together define one of the most important features of the LAT telescope, the self-triggering capability. The latter is possible because of the choice of the silicon-strip detector that do not require an external trigger. Another important characteristic of the LAT, that makes it more stable and durable compared with other space telescopes, is that all the instrument subsystems on board utilise technologies that do not use consumable materials (like gas). To give an overview on how the different subsystems work together the acquisition process could be described as follow: upon triggering, the DAQ initiates the read out of the three subsystems, ACD, TKR, and CAL, and utilises on-board event processing to reduce the rate of events transmitted to the ground to a rate compatible with the 1 Mbps average downlink available to the LAT. The onboard processing is optimised for rejecting events triggered by cosmic-ray background particles while maximising the number of events triggered by $\gamma$-rays. Heat produced by the tracker, calorimeter, and DAQ electronics is transferred to radiators through heat pipes in the grid. The ACD, the TKR and the CAL are physically assembled as shown in figure 3.1.

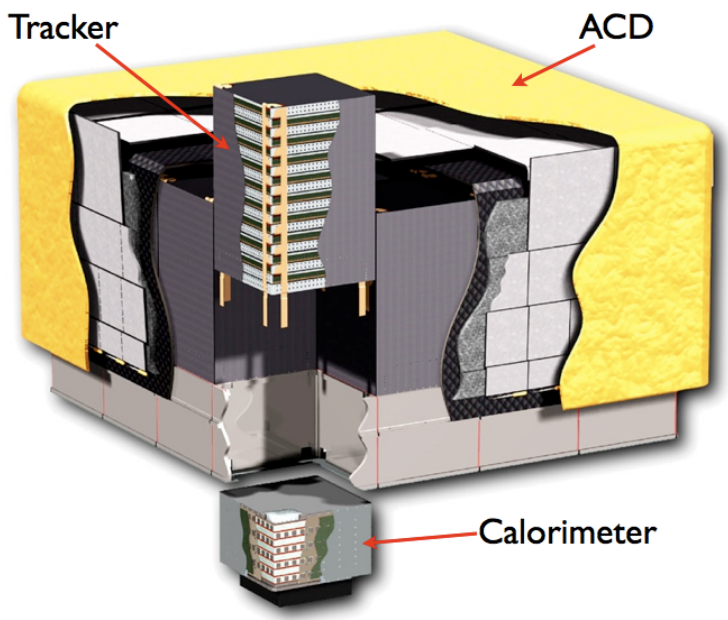

Figure 3.1: Built disposition of the three main LAT systems, ACD, TKR, and CAL.

### 3.1.1   The precision converter tracker

The converter tracker (Atwood et al. (2009)) is composed by 16 foils of heavy material (tungsten, w) in which the incident $\gamma$-rays can produce an $e^+-e^-$ pair. The converter planes are interleaved with position-sensitive silicon detectors that record the passage of the particles and allow to reconstruct the direction of the incident $\gamma$ rays. Each tracker module has 18 (x, y) tracking planes, consisting of two layers (x and y) of single-sided silicon strip detectors.

The support structure for the detectors and converter foil planes is a stack of 19 composite panels, or "trays", supported by carbon-composite sidewalls that also serve to conduct heat to the base of the tracker array. The tray structure is a low-mass, carbon-composite assembly made of a carbon-carbon closeout, carbon-composite face sheets, and a vented aluminium honeycomb core. Carbon was chosen for its long radiation length, high modulus (stiffness)-to-density ratio, good thermal conductivity, and thermal stability.

The probability distribution for the reconstructed direction of incident $\gamma$-rays from a point source is referred to as the point-spread function (PSF). Multiple scattering of the $e^+$ and $e^-$ and bremsstrahlung production limit the obtainable resolution. To get optimal results requires that the $e^+$ and $e^-$ directions be measured immediately following the conversion. At 100 MeV, losing one of the first detections implies a loss in resolution of about a factor 2, resulting in long tails in the PSF. To minimise the missing events on the first layer following a conversion, the tungsten foil in each plane covers only the active areas of the silicon strip detector.

One of the most complex problems in the building of the LAT telescope is the compromise between the necessity to have a thin layer detector, to optimise the PSF at low energy, and to have thick converter, to maximise the probability of detection and so the effective area at high energy. In fact the PSF at low energy is mainly determined by the $\sim$Energy$^{-1}$ dependence of multiple scattering while the high-energy photons require a thick layer to increase the detection probability. The solution was to divide the tracker into two regions, *front* and *back*. The front region (first 12 (x,y) tracking planes) has thin converters, each 0.03 radiation length thick, to optimise the PSF at low energy, while the converters in the back (four (x, y)-planes after the front tracker section) are $\sim$6 times thicker, to maximise the effective area at the expense of less than a factor of 2 in angular resolution (at 1 GeV), for photons converting in that region.

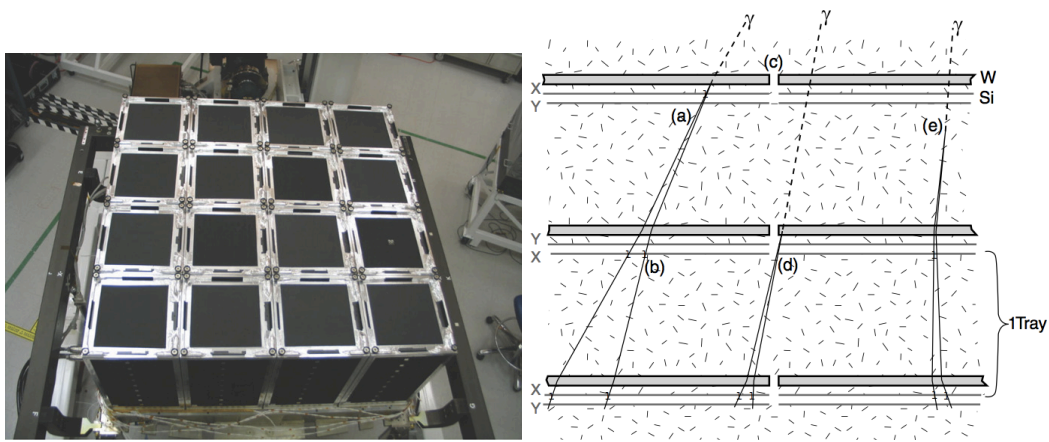

Figure 3.2: *Left*: Completed tracker array before integration with the ACD. *Right*: Illustration of tracker design principles. The first two points dominate the measurement of the photon direction, especially at low energy. (Note that in this projection only the x hits can be displayed.) (a) Ideal conversion in W: Si detectors are located as close as possible to the W foils, to minimise the lever arm for multiple scattering. Therefore, scattering in the second W layer has very little impact on the measurement. (b) Fine detector segmentation can separately detect the two particles in many cases, enhancing both the PSF and the background rejection. (c) Converter foils cover only the active area of the Si, to minimise conversions for which a close-by measurement is not possible. (d) A missed hit in the first or second layer can degrade the PSF by up to a factor of 2, so it is important to have such inefficiencies well localised and identifiable, rather than spread across the active area. (e) A conversion in the structural material or Si can give long lever arms for multiple scattering, so such material is minimised. Good two-hit resolution can help identify such conversions.

### 3.1.2 Calorimeter

The main purposes of the FERMI LAT calorimeter (Atwood et al. (2009)) are two: (1) to measure the energy deposition due to the electromagnetic $e^+ - e^-$ pairs generated by the incident $\gamma$-ray photon (2) generate an image of the cascade profile and measure the shown maximum to provide a very important background discriminator and an estimator of the electromagnetic cascade fluctuation. Each calorimeter module is composed of 96 CsI(Tl) scintillation crystal, providing 8.6 radiation lengths. The crystals are optically isolated from each other and are disposed horizontally in 8 layers of 12 crystals each. Each calorimeter module layer is perpendicular to its neighbours forming an (x,y) hodoscopic array. Each crystal element is read out by PIN photodiodes, mounted on both ends of the crystal, which measure the scintillation light that is transmitted to each end. There are two photodiodes at each end of the crystal, a large photodiode with area 147 $mm^2$ and a small photodiode with area 25 $mm^2$, providing two readout channels to cover the large dynamic range of energy deposition in the crystal. The large photodiodes cover the range 2 MeV-1.6 GeV, while the small photodiodes cover the range 100 MeV-70 GeV. An illustration of the LAT calorimeter is shown in figure 3.3.

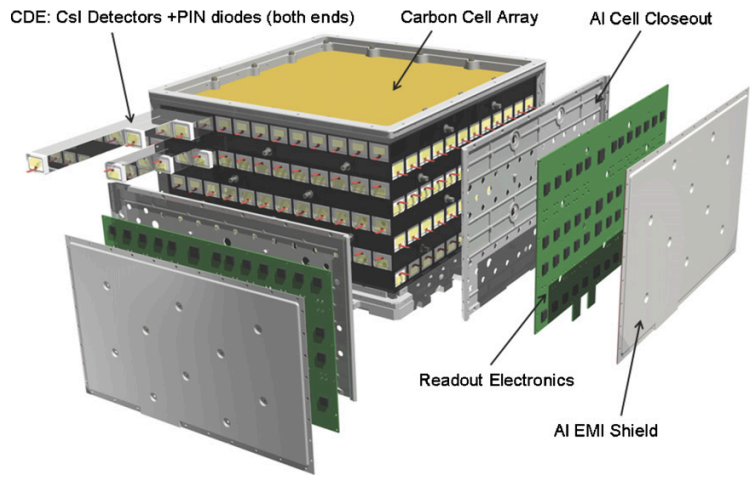

Figure 3.3: LAT calorimeter module. The 96 CsI(Tl) scintillator crystal detector elements are arranged in eight layers, with the orientation of the crystals in adjacent layers rotated by $\pi/2$. The total calorimeter depth (at normal incidence) is 8.6 radiation lengths.

### 3.1.3   Anti coincidence detector

The Anti coincidence detector (Atwood et al. (2009)) consists of an array of plastic scintillator tiles and wavelength shifting fibres (WLS). The scintillation light from each tile is collected by the WLS and coupled to two photomultiplier tubes for redundancy. This arrangement provides uniformity of light collection that is typically better than 95%. The purpose of the ACD is to provide charged-particle background rejection; therefore its main requirement is to have high detection efficiency for charged particles. The ACD is required to provide at least 0.9997 efficiency (averaged over the ACD area) for detection of single charged particles entering the FoV of the LAT. Since the LAT is designed to measure $\gamma$-rays with energies up to at least 300 GeV, a heavy calorimeter ($\sim$1800 kg) is needed to absorb enough of the photon-induced shower energy. However, a calorimeter of such a big mass creates a problem called the backsplash effect. This effect is generated by the isotropically distributed secondary particles (mostly 100-1000 keV photons) from the electromagnetic shower created by the incident high-energy photon that can Compton scatter in the ACD and thereby create false signals from the recoil electrons. The same effect was present in EGRET, where the instrument detection efficiency above 10 GeV was a factor of at least 2 or lower than at 1GeV due to false vetoes caused by backsplash. The resolution of the backsplash effect was found in a structural modification of the ACD. To minimise the veto signals the ACD has been segmented so that only the ACD segment near to the incident candidate photon detection may be considered. This choice reduces dramatically the area of ACD that can contribute to backsplash. To minimise the chance of light leaks due to penetrations of the light-tight wrapping by micrometeoroids and space debris, the ACD is completely surrounded by a low-mass micrometeoroid shield (0.39 g cm$^{-2}$).

### 3.1.4   Trigger and Data Acquisition System

The Data Acquisition System (DAQ, Atwood et al. (2009)) collects the data from the other subsystems, implements the multilevel event trigger, provides onboard event processing to run filter algorithms to reduce the number of downlinked events, and provides an onboard science analysis platform to rapidly search for transients. In figure (3.4) is shown the DAQ hierarchical architecture. The 16 Tower Electronics Modules (TEM) provide the interface to tracker and calorimeter pair. Each TEM generates instrument trigger primitives from combinations of tower subsystem triggers, provides event buffering to support event readout and communicates with the EMB, Event Builder Module, that is part of the GASU, Global-trigger/ACD-module/Signal distribution unit. The GASU consists of (1) the Command Response Unit (CRU) that sends and receives commands and distributes the DAQ clock signal, (2) the Global-Trigger Electronics Module (GEM) that generates LAT-wide readout decision signals based on trigger primitives from the TEMs and the ACD, (3) the ACD Electronics Module (AEM) that performs tasks, much like a TEM, for the ACD, and (4) the EBM (Event Builder Module) that builds complete LAT events out of the information provided by the TEMs and the AEM, and sends them to dynamically selected target Event Processor Units (EPUs).

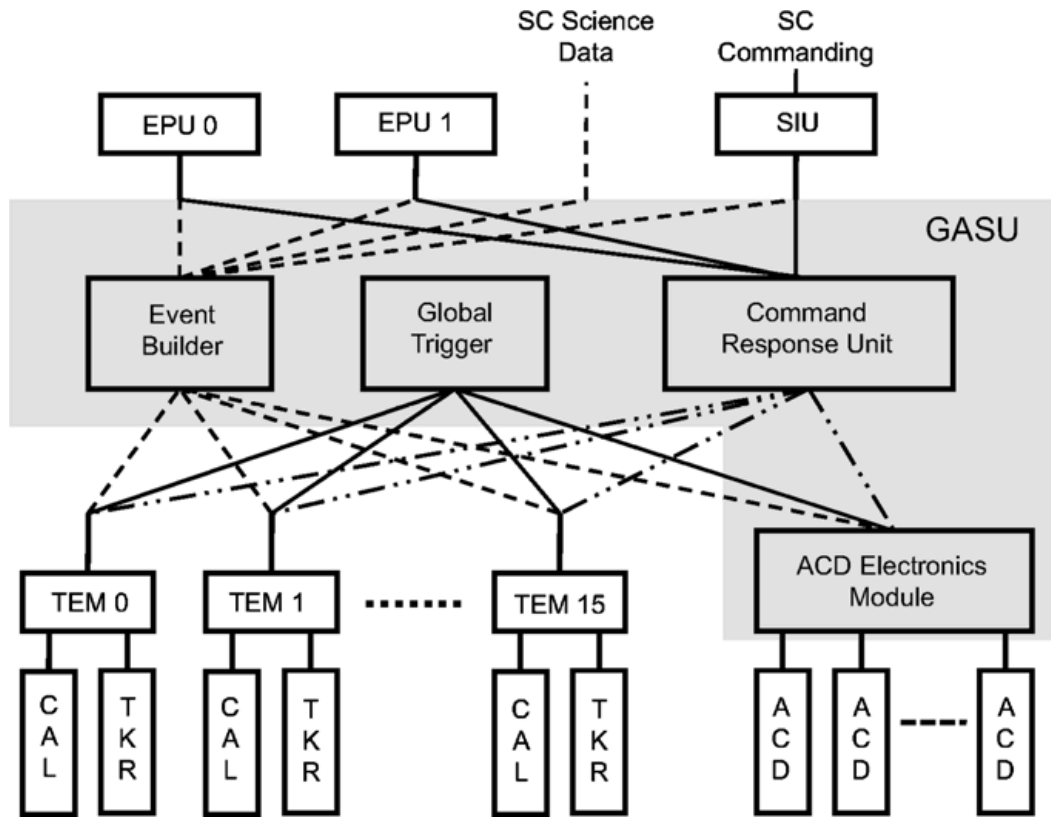

Figure 3.4: Data acquisition system architecture.

### 3.1.5 Event classification

It is possible to define three analysis classes, based on different characteristics like *(1)* the backgrounds expected in orbit *(2)* our current knowledge of the $\gamma$-ray emission from the universe and *(3)* on the LAT performances. The difference between the classes is defined by an increasingly tighter condition that the incoming photon in the tracker and calorimeter behaves as expected for a $\gamma$-ray induced electromagnetic cascade. The classified event classes are:

- *Transient class*: the background rejection is set to allow a background rate of $< 2$ Hz, estimated using a background model.

- *Source class*: the residual background contamination is similar to the one expected from the extragalactic $\gamma$-ray background flux over the entire field of view. The background rate is of 0.4 Hz.

- *Diffuse class*: has the best background rejection, 0.1 Hz, and was designed such that harsher cuts would not significantly improve the signal to noise ratio.

Such classification has been defined in the pre-launch phase and all the conditions have been optimised during the on-orbit calibration procedure. It's important to note that this classification has a hierarchical structure; all the events contained in the diffuse class are included in the source one and all the events contained in the source class are in the transient one. An overview of the behaviour of each class, expressed as the ratio between the background level and the extragalactic diffuse background versus energy, is shown in figure 3.5.

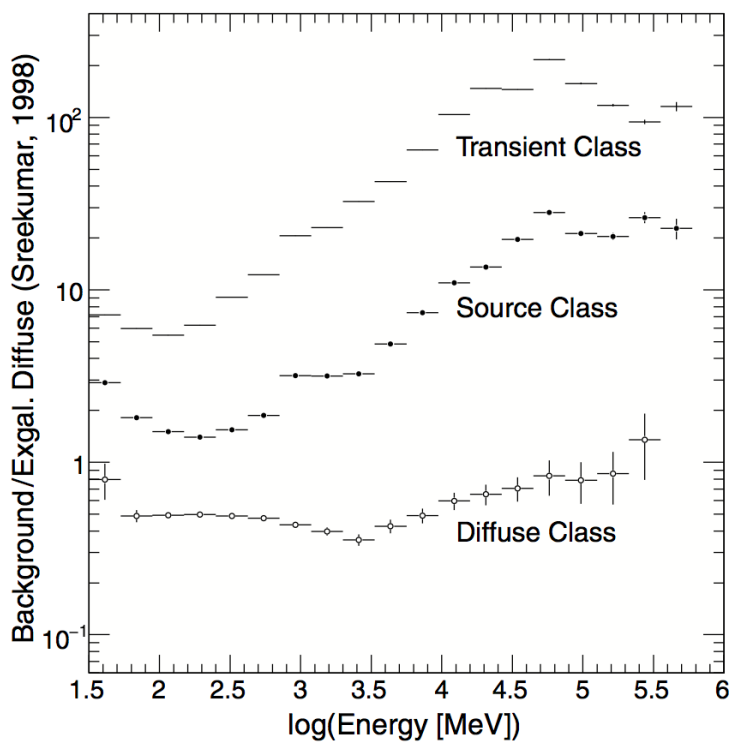

Figure 3.5: Ratio of the residual background to the extragalactic diffuse background inferred from EGRET observations Sreekumar et al. (1998)(Sreekumar et al. 1998) for each of the three analysis classes (from Atwood et al. (2009)).

### 3.1.6    Scientific performances

The performances of the LAT (Atwood et al. (2009); Rando for the Fermi LAT Collaboration (2009)) are essentially defined by the design of the hardware, the event reconstruction algorithms, and the event selection algorithms. The LAT pre-launch response was tuned using Monte Carlo simulations and beam test data. Obviously, during the early part of on orbit operation these performance parameters[1] have changed owing to the optimisation of the selection algorithms. Figures 3.6 to 3.10 summarise the LAT performances after the on-orbit calibration phase and using an optimised event selection algorithm (P6_V3, pass 6 version 3). All plots are given for this selection and for diffuse class events. In all the plots, red and blue curves respectively refer to photon detected in the front and back part of the tracker. The black curve gives the performance for the total population.

#### Effective ara

Figure 3.6 gives the LAT effective area which increases from the minimum value at $\sim$ 30 MeV up to its maximum value for energy bigger than 1 GeV. This behaviour is due to the structure of the tracker (section 3.1.1). The first 12 thin layers optimise the PSF at low energy (see section 3.1.1), but with a low photon interaction probability, they do not optimise the effective area. The last very thick 4 layers, increase a lot the photon interaction probability starting from 1 GeV, maximising the effective area at high energy, without affecting significantly the PSF.

---

[1]All the plots and most of the information presents in this section are periodically updated at the WEB address $http://www-glast.slac.stanford.edu/software/IS/glast\_lat\_performance.htm$

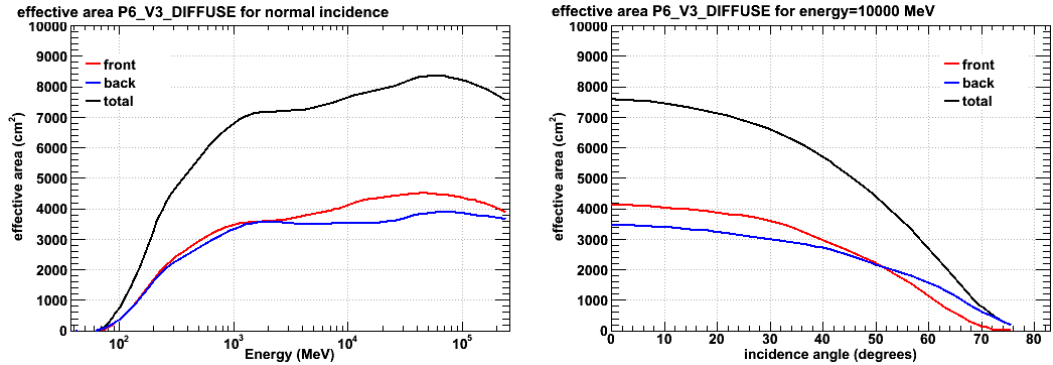

Figure 3.6: *Left*: effective area for normal incidence photons (defined here as cos($\theta$) >0.975); *right* effective area for 10 GeV photons as a function of incidence angle.

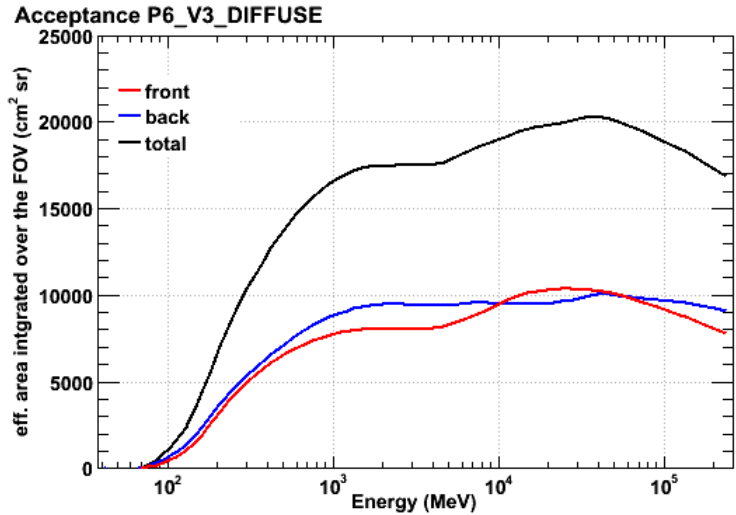

Figure 3.7: Acceptance as a function of energy.

### Acceptance

The LAT acceptance is defined as the effective area integrated over the solid angle. Figure 3.7 shows the intrinsic acceptance that does not take into account the orbital characteristics. To obtain the effective acceptance, each curve has to be scaled by a constant factor which depends on the dead time, the South Atlantic Anomaly (SAA, where the LAT does not take data) and from the observation strategy.

**Point Spread Function (PSF)**

Figures 3.8 & 3.9 show the behaviour of the PSF (see section 3.1.1) with respect to the energy and to the incidence angle of the photon on the tracker.

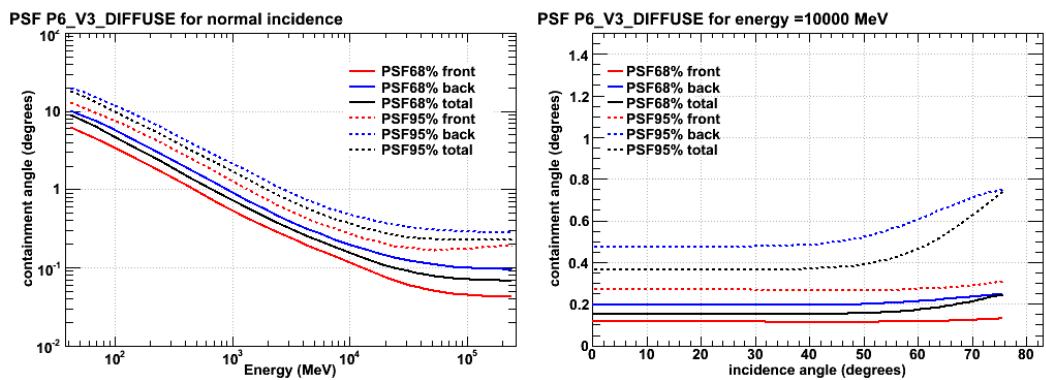

Figure 3.8: 68% and 95% PSF containment ratio versus energy, for a normal incident photon,(*left*) and versus the incidence angle, for a 10 GeV photon (*right*).

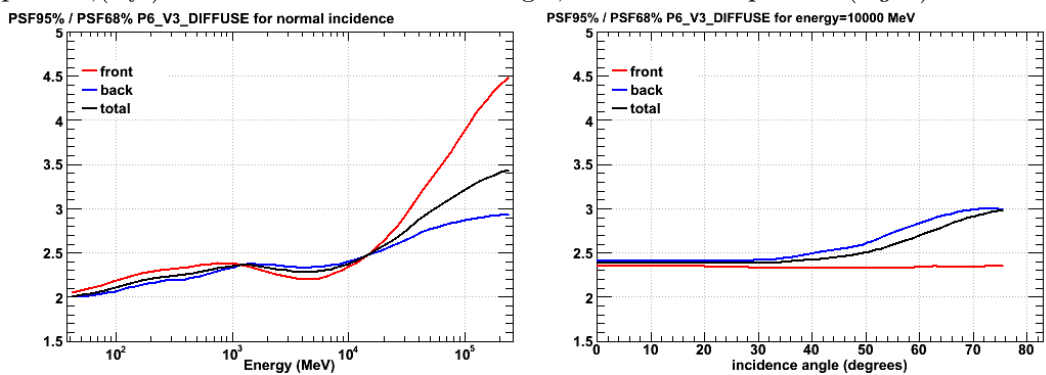

Figure 3.9: Ratio PSF95% / PSF68% versus energy for a normal incident photon (left) and versus the incidence angle for a 10 GeV photon (right).

Figure 3.9 shows the ratio PSF95% / PSF68% which is a useful indicator of the magnitude of the tails of the distribution.

**Energy resolution**

The plot in figure 3.10 shows the behaviour of the energy resolution which does not evolve much across the entire energy band and which is precise enough to study the continuum spectra expected from astrophysics sources.

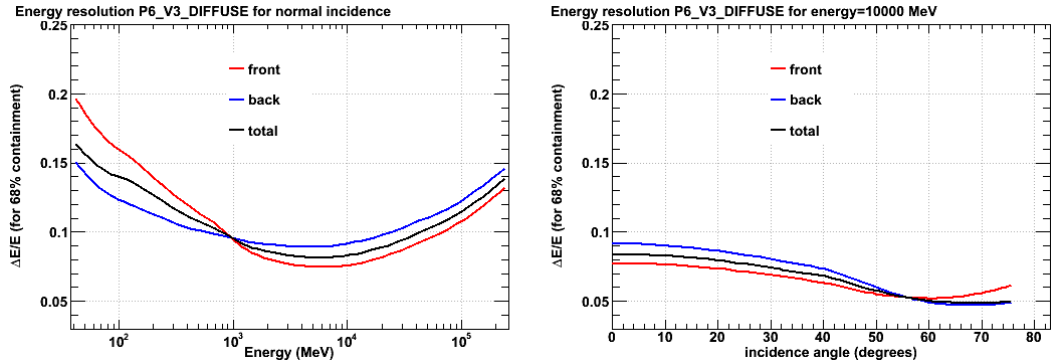

Figure 3.10: Energy resolution versus energy for a normal incident photon *(left)* and versus the incidence angle for a 10 GeV photon *(right)*

**LAT time resolution and timing accuracy**

To have a very good time resolution and timing accuracy is a fundamental requirement to be able to perform pulsars observations and timing. In the case of the LAT this is also one of the most impressive and stable telescope characteristic. The LAT time resolution and accuracy intrinsically depend on the response timescale of the on-board instruments and on the way the different subsystems exchange informations. The time between a particle interaction in the LAT, that causes an event trigger, and the latching of the tracker discriminators, is 2.3-2.4 $\mu$s, mainly due to the analog rise time in the tracker front-end electronics. Similarly, the latching of the analog sample-and-holds for the calorimeter and the ACD are delayed (programmable delay of $\sim 2.5 \mu$s) until the shaped analog signals peak. In addition, the time required to latch the trigger information in the GEM and send it from the GEM (Global-Trigger Electronics Module, section 3.1.4) to the EBM (Event Builder Module, section 3.1.4) is 26.5 us. This limits the minimum instrumental dead-time. The LAT uses event buffering and does not read out channels where there is no signal above some minimum channel, therefore, for events illuminating many channels and/or very high event rates, the true readout time is not much more than the minimum.

Concerning the timing accuracy, when an event is detected by the LAT, the exact time at which this event occurred has to be registered. In the pre-launch phase a ground test has been performed to measure the difference between two GPS systems mounted on the LAT and on another $\gamma$-ray receiver, an atmospheric muon detector. A scintillator pair has been placed close to the LAT and its signal triggered with a GPS system, tested with a ground $\gamma$-telescope on the Crab pulsar. The LAT timestamps have been registered and compared with the reference GPS ones. The results, plotted in figure 3.11, showed that the LAT timestamps agreed with the reference GPS to within 0.3 μs (Abdo et al., 2009a). After the launch, the LAT timing accuracy has been tested again and the on-orbit telemetry showed that the LAT time is still well within 1μs of the GPS time used by the spacecraft that is maintained within 20ns (1σ) of UTC (Abdo et al. (2009a), Astropart. Phys., 32, 193). FERMI is equipped with 4 GPS *Viceroy^{TM} spaceborne receivers*, for a total of 2 antenna for each side of the spacecraft to see all the sky and to be able to detect the signal from as many GPS satellites at a time as possible.

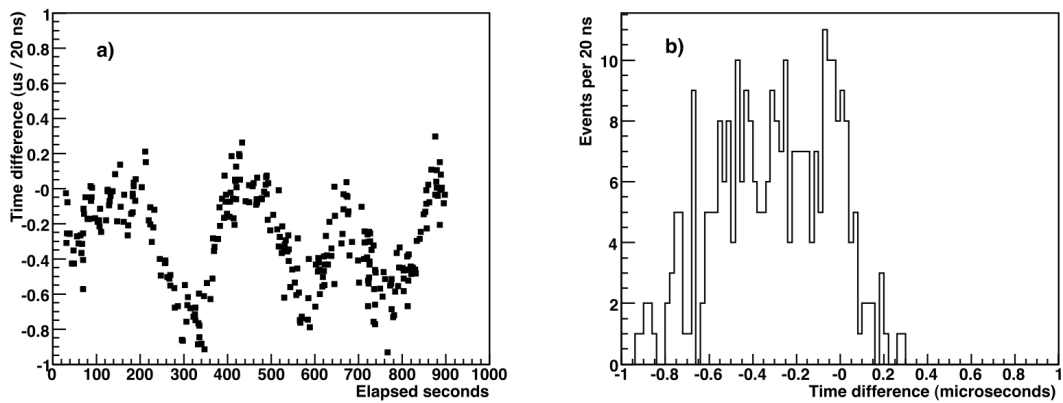

Figure 3.11: *left*: Difference in the time recorded at ground between the LAT and another acquisition system during the tests of the pre-launch phase. *right*: histogram of time differences indicating mean and RMS values around 0.3 μs.

On the spacecraft the time measurement is based on a 20 MHz clock closely tracked and synchronised each second with a one pulse-per-second (PPS) signal from the GPS system of the spacecraft. The system that registers the event timestamp is the GEM. When a γ-ray photon is detected the fractional part of the event timestamps is obtained from the 20 MHz clock, $T_{photon} = UTC_{PPS} + (T_{PPS} - T_{\gamma})/d\nu$ where in average $d\nu = 20$ MHz. In this way, the TOA (Time Of Arrival) of each photon is defined by a reference PPS from the GPS plus a fractional part obtained counting the 20MHz cycles between the reference PPS and the events. The 1 PPS GPS spacecraft clock has an accuracy of ±1.5 μs.

For different reasons, possible occasional short losses of the GPS signal reception could occur and when it happens the time starts to drift. To avoid the complete shift of the time scale an internal oscillator is in charge to maintain the PPS accuracy. The system keeps the drift to a maximum of 1 μs per 100 second of drift time. The main reasons for which a loss of the GPS clock can occur are 3: *(i)* when the two antennae detect the signal from less than 5 GPS satellites, *(ii)* when the Dilution Of Precision (DOP, a parameter that takes into account the geometric strength of the LAT) became too high, usually >6, *(iii)* when a series of other circumstances, connected to the time related LAT parameters, occur simultaneously.

# Chapter 4

# LAT pulsar observations

In this chapter I will review the LAT capability to observe $\gamma$-ray emission coming from pulsars. I will describe the main search techniques used to detect new objects and the different kinds of LAT data analysis proposed to define their characteristics. The last section of the chapter will be dedicated to the description of the results reached by the *FERMI LAT* pulsar observations in the first part of its activity. Particular attention will be given in describing the characteristics of what emerges as a population of $\gamma$-ray neutron stars: *the LAT pulsar population*

## 4.1  Pulsar search, analysis, and observation with the LAT

Before the launch of the Compton Gamma Ray Observatory (CGRO) (Thompson et al. 1999 ), hosting the Energetic Gamma Ray Experiment Telescope (EGRET) on the 5th of April 1991, the high energy $\gamma$-ray emission from pulsars was a largely unexplored field. With the advent of EGRET, just the most powerful pulsars were seen to pulsate at very high $\gamma$-ray energy (Vela, Crab, B1706-44, Geminga, B1055-52, B1905+32) and others, at first detected as $\gamma$-ray sources (Thompson et al 1994 ), were detected as pulsating sources in other experiments, consequently to an improvement of the scientific performances of the detectors. The reason of the poor pulsar detection of the EGRET instrument was mainly due to its scientific performances that: it was not sensitive enough to detect the bulk of what we know to be a much more numerous pulsar population emitting in a wider energy range and/or pulsating in a wider period interval.

To be able to detect pulsars $\gamma$-ray emission in the EGRET energy range, between 20 MeV and 30 GeV, while evaluating with a high accuracy the timing and spectral characteristics of the objects, a detector should be characterised by high sensitivity in a wide energy range, high time accuracy, short instrumental dead time, and high angular resolution. With such a detector, the high sensitivity will ensure the detection of the lower flux sources, the high accuracy of the timing parameters will provide a clear identification (or rejection) of an unidentified object as a pulsar.

With a field of view of $\sim 2.4$ sr, the capability to detect photons in the energy interval 20 MeV-300 GeV, an on-axis effective area of $\sim$8000 cm$^2$ for E >1 GeV, and an accuracy of the time stamps relative to UTC $< 1\mu$s, (section 3.1.6) the LAT space telescope represents one of the best instruments ever to observe and study both the timing and spectral sides of the pulsar physics.

All the detection techniques and analysis described in the next sections refer to the method used in the first observation campaign performed by the LAT in the first 6-8 months of activity. A detailed description can be found in Smith et al. 2008 and in the first pulsar catalogue paper Abdo et al. 2010.

### 4.1.1 Search and detection strategies

The first step of the pulsar analysis consisted in detecting the well known pulsars and define a search strategy to search for $\gamma$-ray emission from other sources. Two different approaches have been pursued. The first one focused on already known radio pulsars for which highly accurate ephemerides come from radio timing (or with much less accuracy from X-rays for Geminga (Jackson & Halpern 2005)). The second one tried to find pulsed $\gamma$-ray emission in *blind period search*, e.g. without any indication of the spin pulsar parameters but mostly performing searches at positions corresponding to bright unidentified sources.

The advantage of the first technique is to be more sensitive to very low fluxes. In general the $\gamma$-ray sources have a very low flux, of the order of one photon emitted every several hundred rotations of the star. For example, the Crab pulsar emits a photon more or less every 500 rotations. A fainter pulsar emits few tens of photons in several months of observation and to know with high precision the pulsar spin period makes it possible to align in phase the few detected events and unveil the periodicity. On the other side, a blind search method will scan the Fourier frequency space checking for periodicity just at some predefined frequency values that will never correspond to the exact spin frequency of the pulsar: they will represent an approximation of the real pulsar pulsation. In this last case, if the pulsar flux is high enough, the periodicity will be detected, otherwise the phase dispersion due to the frequency approximation will be too large and the method will not find any significant signal. The blind search method has different advantages. First it allows to discover new pulsars with selection biases different from those introduced by the radio ephemeris search. The blind search technique favours the discovery of pulsars with larger magnetic obliquities $\alpha$. It also allows the discovery of radio-quiet pulsars.

**Observation of pulsars with known radio ephemeris**

Contemporaneus ephemerides provided by a consortium of radio observatories plus 5 X-ray telescopes (Smith et al. 2008) were considered to search for pulsation. The radio observations to evaluate the ephemerides have been performed by: *Parkes, Nancay, Jodrell Bank, Green Bank Telescope (GBT), Arecibo,* and the *Westerbork Synthesis Radio Telescope.* The need for simultaneous ephemeris is justified by the fact that in the ATNF pulsar catalogue, the best $\gamma$-ray pulsar candidates are the high $\dot{E}$ objects. Unfortunately these objects are characterised by an unstable slow-down trend, known as glitches and timing noise, that often causes the degradation of the timing solution within a few months. The only way to keep stable the timing noise of high $\dot{E}$ pulsar ephemerides is provided by a contemporaneous monitoring.

The consortium provided for 762 ephemeris divided in two groups: a first one encompassing 218 objects with $\dot{E} > 10^{34}$erg s$^{-1}$ (Weltevrede et al. 2010a) and a second one with the remaining candidates, that reduce the selection bias introduced by choosing high $\dot{E}$ pulsars for the first analysis. In the very first analysis, this method yielded the discovery of $\gamma$-ray emission from 46 already known pulsars, of which 5 millisecond pulsars (MSPs) all in binary systems.

**Blind search for $\gamma$-ray pulsars**

The blind search method has been applied to $\sim$100 candidates selected before the FERMI launch and to $\sim$200 sources selected during the first period of the LAT activity. The frequency interval scanned covered the region from 0.5 Hz to 64 Hz (156.26 ms to 2 s) and the frequency first time derivative space ($\dot{f}$) has been scanned from zero to the spin down value of the young Crab pulsars, $\dot{f} = -3.7 \times 10^{-10}$. The $P - \dot{P}$ space tested corresponds to $\sim$86% of that of the entire ATNF pulsar population (Abdo et al. 2009c). The blind search method yielded the discovery of 16 unknown pulsars (Abdo et al. 2008, 2009c). For all the 16 blind search pulsars plus Geminga and J1124-5916, for which ephemeris were lacking, the timing has been performed directly using the LAT data and, for Geminga, this technique yielded the best timing solution ever obtained.

### 4.1.2 Timing analysis

In this section I summarise the timing analyses performed to evaluate ephemeris of the blind search pulsars and of few other objects both radio quiet (Geminga) or particularly radio weak. An accurate description can be found in Smith et al. 2008 and in the first pulsar catalogue paper (Abdo et al. 2010).

The timing analysis is the procedure used to find, with high accuracy, the rotational parameters of a newly discovered pulsar. The first thing to do is to select the dataset to apply the timing procedure. Usually the data collected in the post launch calibration period of the LAT instrument are not used to perform scientific investigation. This is due to the fact that during the first observation period a lot of calibration tests were made, and the configuration of the instruments and of the analysis tools changed several times in a short period. Nevertheless all this changes did not affect the timing characteristics and accuracy of the LAT instruments and so all the events, starting from the first one detected, have been used to perform the blind search timing analysis. Just the *diffuse* class (section 3.1.5) photons, coming from a small region of interest (ROI) around the pulsar position (a circle with radius of 0°.5 or 1°.0) and with an energy above 300 MeV were taken into account. The TOA of each photon has been converted to the solar system barycenter, using the FERMI science tool *GTBARY*.

The timing procedure consists in using a timing model together with the software *TEMPO2* (Hobbes et al. 2006) in its predictive mode to generate polynomial coefficients that could describe the pulsar phase as a function of time in the chosen reference frame. The rotation phase of the pulsar has been calculated by using a truncated Taylor series expansion like

$$\phi_i(t_i) = \phi_0 + \sum_{j=0}^{j=N} \frac{f_j \times (t_i - T_0)^{j+1}}{(j+1)}, \qquad (4.1)$$

where $T_0$ is the reference epoch of the pulsar ephemeris, $\phi_0$ is the pulsar phase at $t = T_0$, and the coefficients $f_i$ are the rotation frequency derivatives of order $j$. The next step consisted in using these reference phases to fold a pulse profile over a discretional LAT data period established on the basis of the pulsar flux. By defining several folded intervals for each pulsars, each folded profile is cross-correlated, in the Fourier domain, with a template (Taylor 1992) to assign a TOA to each data segment. For most of the pulsars analysed, the template to cross-correlate with the profile is obtained by fitting the light curve with several gaussian distributions. Nevertheless not all the pulsar profiles are well fitted by a multi-gaussian distribution and in at least one case the template has been obtained in a different way. This is the case of the Geminga pulsar, that owing to its complex profile, has been cross-correlated using its own light curve. The last step of the timing procedure, with which the ephemeris have been generated, consisted in using *TEMPO2* to fit a timing model to each pulsar. From this fitting procedure, the rms obtained for the residuals are between the 0.5% and the 2.9% of a phase rotation (Abdo et al., 2010b).

### 4.1.3 Spectral analysis

In this section I will briefly describe the spectral analysis performed during the first observation campaign of the LAT. For more details see Abdo et al.

(2010b).

At variance with the timing analysis, which was unhampered by the LAT calibration procedures, the spectral analysis was significantly affected by the fine tuning performed during the calibration period. Thus, in order to avoid inconsistencies, the spectral analyses of the observed LAT pulsars have been performed taking into account the data collected after the end of the calibration and test period, August 4th 2008. All the photons with $E > 100$ MeV and belonging to the *diffuse* class (section 3.1.5) have been taken into account and, to minimise the contribution of the $\gamma$-ray produced by cosmic-ray interactions in the Earth's atmosphere, the whole 10° degrees ROI selected around each source, was within 105° of zenith angle. To fit the pulsar spectra a power law with an exponential cutoff was used:

$$\frac{dN}{dE} = K E_{GeV}^{-\Gamma} e^{(-\frac{E}{E_{cutoff}})}. \tag{4.2}$$

In this equation $N$ is the photon number, $E$ is the photon energy, and the three free parameters are the photon index at low energy $\Gamma$, the energy cutoff $E_{cutoff}$, and the constant $K$, a normalisation factor expressed in $[ph\ cm^{-2}\ s^{-1}\ MeV^{-1}]$.

The fit was performed taking into account both the direction and energy of each photon, by maximisation of an un-binned likelihood (Abdo et al. 2009c). The errors on the parameters have been computed by developing the logarithm of the likelihood around the best fit position. Starting from the best fit parameters, the photon flux above 100 MeV, has been evaluated as

$$F_{100} = \int_{100MeV}^{100GeV} \frac{dN}{dE} dE, \tag{4.3}$$

and the energy flux above 100 MeV as

$$G_{100} = \int_{100MeV}^{100GeV} E \frac{dN}{dE} dE. \tag{4.4}$$

## 4.2 The LAT pulsar population

In the last section of this chapter I will illustrate the public FERMI Large Area Telescope pulsar profiles obtained in more than two years of observation. In figure 4.1 is showed the $P$-$\dot{P}$ diagram for the public LAT pulsars. The isolated ordinary pulsar (OP) LAT population represents the most energetic (higher $\dot{E}$ and magnetic field values) and young component of the global pulsar population. The centre of the radio loud (RL) LAT pulsars distribution (red points), seems to be lightly shifted toward lower $\dot{E}$ values compared to the radio quiet (RQ) LAT one (blue NPs points), and all the LAT millisecond pulsars (MSPs) observed so far are RL objects. In addition to the spin characteristics defined in the timing analysis, a set of intrinsic parameters, both structural and energetic, has been evaluated for the LAT detected pulsars. All such parameters are listed in tables 4.1, 4.2, 4.3, 4.4.

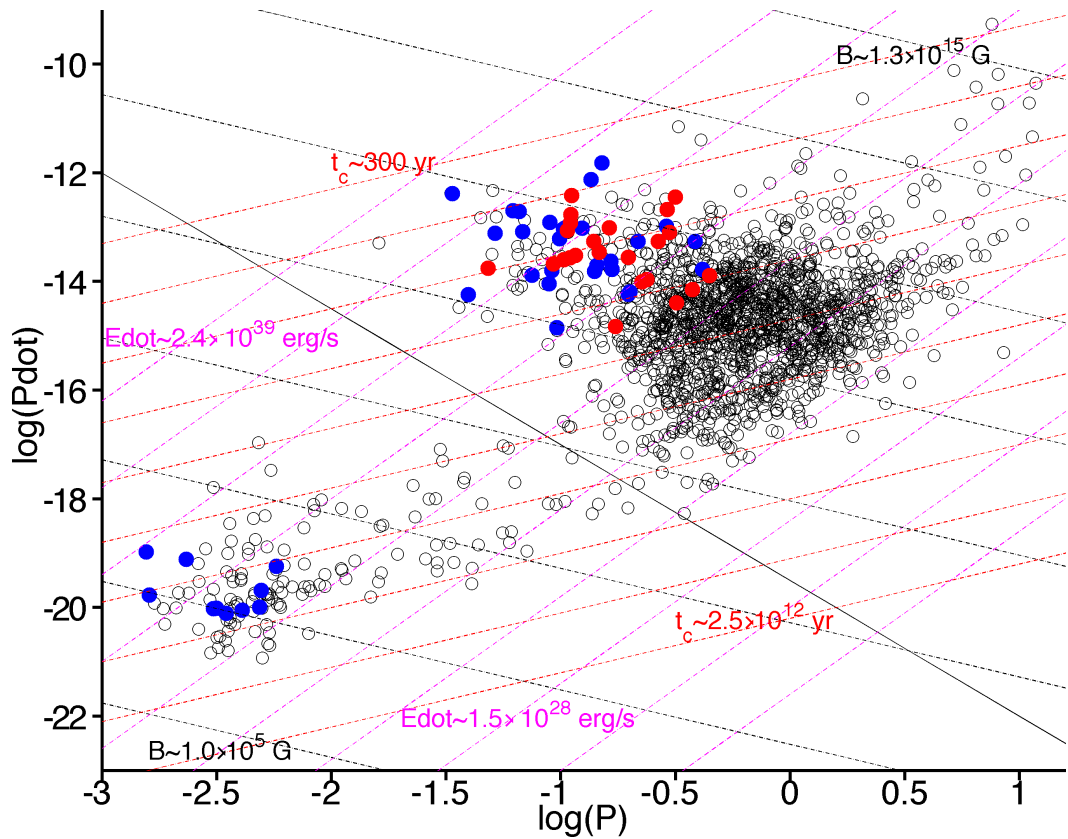

Figure 4.1: $\log_{10}(P)$-$\log_{10}(\dot{P})$ of the known pulsar population. In black are plotted the radio pulsars, in blue and red the LAT radio loud and radio quiet pulsars, respectively. The black line from the top left to the bottom right of the plot define the millisecond pulsar (MSP) condition: all the objets to the left of this line are defined as MSPs. The black, red, and magenta dotted lines respectively mark iso-magnetic field, iso-crone, and iso-$\dot{E}$ $\log_{10}(P)$-$\log_{10}(\dot{P})$ plane regions.

| | $P$ s | $\dot{P}$ s/s | $E_{sd}$ erg/s | $B$ Gauss | $B_{lc}$ Gauss | type | $\zeta_{ng}$ Deg | $\delta\zeta$ Deg |
|---|---|---|---|---|---|---|---|---|
| **J0007+7303** | 0.316 | 3.6e-13 | 8.08e+35 | 2.08e+12 | 5.38 | g | ... | ... |
| **J0205+6449** | 0.0657 | 1.94e-13 | 4.83e+37 | 6.96e+11 | 200 | rg | 91.6 | 2.5 |
| **J0248+6021** | 0.217 | 5.51e-14 | 3.81e+35 | 6.74e+11 | 5.37 | rg | ... | ... |
| **J0357+3205** | 0.444 | 1.2e-14 | 9.69e+33 | 4.5e+11 | 0.419 | g | ... | ... |
| **J0534+2200** | 0.0336 | 4.23e-13 | 7.87e+38 | 7.35e+11 | 1.58e+03 | rg | 61.3 | 1.1 |
| **J0631+1036** | 0.288 | 1.05e-13 | 3.12e+35 | 1.07e+12 | 3.67 | g | ... | ... |
| **J0633+0632** | 0.297 | 7.95e-14 | 2.14e+35 | 9.48e+11 | 2.94 | g | ... | ... |
| **J0633+1746** | 0.237 | 1.1e-14 | 5.84e+34 | 3.15e+11 | 1.93 | g | ... | ... |
| **J0659+1414** | 0.385 | 5.5e-14 | 6.82e+34 | 8.97e+11 | 1.28 | rg | ... | ... |
| **J0742-2822** | 0.167 | 1.68e-14 | 2.56e+35 | 3.26e+11 | 5.74 | rg | ... | ... |
| **J0835-4510** | 0.0894 | 1.24e-13 | 1.23e+37 | 6.49e+11 | 74.2 | rg | 63.6 | 0.6 |
| **J1028-5819** | 0.0914 | 1.61e-14 | 1.49e+36 | 2.37e+11 | 25.3 | rg | ... | ... |
| **J1048-5832** | 0.124 | 9.61e-14 | 3.59e+36 | 6.72e+11 | 29 | rg | ... | ... |
| **J1057-5226** | 0.197 | 5.8e-15 | 5.36e+34 | 2.08e+11 | 2.22 | rg | ... | ... |
| **J1124-5916** | 0.135 | 7.54e-13 | 2.15e+37 | 1.97e+12 | 64.6 | rg | 105 | 7 |
| **J1418-6058** | 0.111 | 1.7e-13 | 8.9e+36 | 8.45e+11 | 51 | g | ... | ... |
| **J1420-6048** | 0.0682 | 8.28e-14 | 1.85e+37 | 4.63e+11 | 119 | rg | ... | ... |
| **J1459-6053** | 0.103 | 2.55e-14 | 1.64e+36 | 3.16e+11 | 23.5 | g | ... | ... |
| **J1509-5850** | 0.0889 | 9.17e-15 | 9.23e+35 | 1.76e+11 | 20.4 | rg | ... | ... |
| **J1709-4429** | 0.102 | 9.26e-14 | 6.08e+36 | 6.01e+11 | 45.5 | rg | 53.3 | 3.3 |
| **J1718-3825** | 0.0747 | 1.32e-14 | 2.24e+36 | 1.94e+11 | 37.9 | rg | ... | ... |
| **J1732-3131** | 0.197 | 2.61e-14 | 2.43e+35 | 4.42e+11 | 4.74 | g | ... | ... |
| **J1741-2054** | 0.414 | 1.68e-14 | 1.68e+34 | 5.14e+11 | 0.592 | rg | ... | ... |
| **J1747-2958** | 0.0988 | 6.13e-14 | 4.49e+36 | 4.8e+11 | 40.5 | rg | ... | ... |
| **J1809-2332** | 0.147 | 3.44e-14 | 7.69e+35 | 4.38e+11 | 11.3 | g | ... | ... |
| **J1813-1246** | 0.0481 | 1.76e-14 | 1.12e+37 | 1.79e+11 | 132 | g | ... | ... |
| **J1826-1256** | 0.11 | 1.21e-13 | 6.39e+36 | 7.12e+11 | 43.4 | g | ... | ... |
| **J1833-1034** | 0.0619 | 2.02e-13 | 6.03e+37 | 6.89e+11 | 237 | rg | 85.4 | 0.3 |
| **J1836+5925** | 0.173 | 1.52e-15 | 2.07e+34 | 1e+11 | 1.57 | g | ... | ... |
| **J1907+0602** | 0.107 | 8.73e-14 | 5.09e+36 | 5.95e+11 | 40 | g | ... | ... |
| **J1952+3252** | 0.0395 | 5.79e-15 | 6.63e+36 | 9.33e+10 | 123 | rg | ... | ... |
| **J1958+2846** | 0.29 | 2.22e-13 | 6.41e+35 | 1.57e+12 | 5.21 | g | ... | ... |
| **J2021+3651** | 0.104 | 9.57e-14 | 6.06e+36 | 6.14e+11 | 44.9 | rg | 79 | 2.2 |
| **J2021+4026** | 0.265 | 5.48e-14 | 2.08e+35 | 7.43e+11 | 3.25 | g | ... | ... |
| **J2032+4127** | 0.143 | 1.96e-14 | 4.72e+35 | 3.27e+11 | 9.06 | rg | ... | ... |
| **J2043+2740** | 0.0961 | 1.3e-15 | 1.04e+35 | 6.89e+10 | 6.33 | rg | ... | ... |
| **J2229+6114** | 0.0516 | 7.79e-14 | 4e+37 | 3.91e+11 | 232 | rg | 46 | 6.3 |
| **J2238+5903** | 0.163 | 9.86e-14 | 1.62e+36 | 7.81e+11 | 14.8 | g | ... | ... |
| **J1023-5746** | 0.111 | 3.84e-13 | 1.96e+37 | 1.28e+12 | 75.1 | g | ... | ... |
| **J1044-5737** | 0.139 | 5.46e-14 | 1.44e+36 | 5.37e+11 | 16.3 | g | ... | ... |
| **J1413-6205** | 0.11 | 2.77e-14 | 1.48e+36 | 3.4e+11 | 21 | g | ... | ... |
| **J1429-5911** | 0.116 | 3.05e-14 | 1.39e+36 | 3.66e+11 | 19.2 | g | ... | ... |
| **J1846+0919** | 0.226 | 9.93e-15 | 6.12e+34 | 2.92e+11 | 2.07 | g | ... | ... |
| **J1954+2836** | 0.0927 | 2.12e-14 | 1.88e+36 | 2.73e+11 | 28 | g | ... | ... |
| **J1957+5033** | 0.375 | 7.08e-15 | 9.51e+33 | 3.18e+11 | 0.492 | g | ... | ... |
| **J2055+2539** | 0.32 | 4.08e-15 | 8.85e+33 | 2.23e+11 | 0.556 | g | ... | ... |
| J1513-5850 | 0.151 | 1.54e-12 | 3.19e+37 | 2.97e+12 | 70.8 | rg | ... | ... |
| J2240+5832 | 0.14 | 1.54e-14 | 3.98e+35 | 2.86e+11 | 8.52 | rg | ... | ... |
| J1648-4611 | 0.165 | 2.37e-14 | 3.74e+35 | 3.85e+11 | 7 | rg | ... | ... |
| **J2030+3641** | 0.2 | 6.5e-15 | 5.75e+34 | 2.22e+11 | 2.27 | g | ... | ... |
| J1119-6127 | 0.408 | 4.02e-12 | 4.2e+36 | 7.89e+12 | 9.49 | rg | ... | ... |
| J0030+0451 | 0.00487 | 1.02e-20 | 6.27e+33 | 4.34e+07 | 30.7 | rg | ... | ... |
| J0218+4232 | 0.00232 | 7.74e-20 | 4.37e+35 | 8.27e+07 | 538 | rg | ... | ... |
| J0437-4715 | 0.00576 | 1.4e-20 | 5.19e+33 | 5.53e+07 | 23.6 | rg | ... | ... |
| J0613-0200 | 0.00306 | 9e-21 | 2.22e+34 | 3.24e+07 | 91.9 | rg | ... | ... |
| J0751+1807 | 0.00348 | 6e-21 | 1.01e+34 | 2.82e+07 | 54.6 | rg | ... | ... |
| J1614-2230 | 0.00315 | 4e-21 | 9.05e+33 | 2.19e+07 | 57.1 | rg | ... | ... |
| J1744-1134 | 0.00407 | 7e-21 | 7.32e+33 | 3.29e+07 | 39.7 | rg | ... | ... |
| J2124-3358 | 0.00493 | 1.2e-20 | 7.08e+33 | 4.74e+07 | 32.3 | rg | ... | ... |
| J0034-0534 | 0.00188 | 4.97e-21 | 5.32e+34 | 1.88e+07 | 232 | rg | ... | ... |
| J1939+2134 | 0.00156 | 1.05e-19 | 1.96e+36 | 7.88e+07 | 1.7e+03 | rg | ... | ... |
| J1959+2048 | 0.00161 | 1.69e-20 | 2.88e+35 | 3.21e+07 | 631 | rg | ... | ... |
| J0614-3329 | 0.00315 | 1.78e-20 | 4.03e+34 | 4.62e+07 | 121 | r | ... | ... |
| J1231-1411 | 0.00368 | 2.12e-20 | 3e+34 | 5.45e+07 | 88.9 | r | ... | ... |
| J2214+3000 | 0.00312 | 1.5e-20 | 3.5e+34 | 4.22e+07 | 113 | r | ... | ... |
| J1823-3021A | 0.00544 | 3.38e-18 | 1.49e+36 | 8.36e+08 | 423 | rg | ... | ... |

Table 4.1: Pulsar spin characteristics. Indicated are the spin period $P$ and its first derivative $\dot{P}$, the magnetic field at the surface and at the light cylinder, a radio loud-quiet flag, and the known line of sight $\zeta$ estimates with the respective errors (Ng & Romani, 2008). In the upper part of the table are listed the ordinary pulsars, in the bottom part the MSP ones. In bold are indicated the pulsars that have been analysed in this thesis. The $\dot{E}$ values have been evaluated using the mass, radius, and moment of inertia values assumed in this thesis.

| | $l$ | $b$ | $D$ | $\delta D^{+}_{-}$ | $\Delta D^{sup}_{inf}$ |
|---|---|---|---|---|---|
| | Deg | Deg | kpc | kpc | kpc |
| **J0007+7303** | 120 | 10.5 | 1.4 | -0.3 + 0.3 | ...- ... |
| **J0205+6449** | 131 | 3.06 | ... | ... + ... | 2.6- 3.2 |
| **J0248+6021** | 137 | 0.7 | 2 | -0.2 + 0.2 | ...- ... |
| **J0357+3205** | 163 | -16 | ... | ... + ... | ...- ... |
| **J0534+2200** | 185 | -5.78 | 2 | -0.5 + 0.5 | ...- ... |
| **J0631+1036** | 201 | 0.38 | ... | ... + ... | 0.75-3.62 |
| **J0633+0632** | 205 | -0.93 | ... | ... + ... | ...- ... |
| **J0633+1746** | 195 | 4.27 | 0.25 | -0.062 + 0.12 | ...- ... |
| **J0659+1414** | 201 | 8.27 | 0.288 | -0.027 + 0.033 | ...- ... |
| **J0742-2822** | 244 | -2.44 | 2.07 | -1.07 + 1.38 | ...- ... |
| **J0835-4510** | 264 | -2.78 | 0.287 | -0.017 + 0.019 | ...- ... |
| **J1028-5819** | 285 | -0.46 | 2.33 | -0.7 + 0.7 | ...- ... |
| **J1048-5832** | 287 | 0.55 | 2.71 | -0.81 + 0.81 | ...- ... |
| **J1057-5226** | 286 | 6.65 | 0.72 | -0.2 + 0.2 | ...- ... |
| **J1124-5916** | 292 | 1.75 | 4.8 | -1.2 + 0.7 | ...- ... |
| **J1418-6058** | 313 | 0.13 | ... | ... + ... | 2- 5 |
| **J1420-6048** | 314 | 0.23 | 5.6 | -1.7 + 1.7 | ...- ... |
| **J1459-6053** | 318 | -1.82 | ... | ... + ... | ...- ... |
| **J1509-5850** | 320 | -0.59 | 2.6 | -0.8 + 0.8 | ...- ... |
| **J1709-4429** | 343 | -2.68 | ... | ... + ... | 1.4- 3.6 |
| **J1718-3825** | 349 | -0.43 | 3.82 | -1.15 + 1.15 | ...- ... |
| **J1732-3131** | 356 | 1 | ... | ... + ... | ...- ... |
| **J1741-2054** | 6.42 | 4.9 | 0.38 | -0.11 + 0.11 | ...- ... |
| **J1747-2958** | 359 | -0.8 | 2 | -0.6 + 0.6 | ...- ... |
| **J1809-2332** | 7.39 | -2 | 1.7 | -1 + 1 | ...- ... |
| **J1813-1246** | 17.2 | 2.43 | ... | ... + ... | ...- ... |
| **J1826-1256** | 18.6 | -0.37 | ... | ... + ... | ...- ... |
| **J1833-1034** | 21.5 | -0.82 | 4.7 | -0.4 + 0.4 | ...- ... |
| **J1836+5925** | 88.9 | 25 | ... | ... + ... | 0- 0.8 |
| **J1907+0602** | 40.2 | -0.87 | 3.21 | -0.3 + 0.3 | ...- ... |
| **J1952+3252** | 68.8 | 2.82 | 2 | -0.5 + 0.5 | ...- ... |
| **J1958+2846** | 65.9 | -0.37 | ... | ... + ... | ...- ... |
| **J2021+3651** | 75.2 | 0.14 | 2.1 | -1 + 2.1 | ...- ... |
| **J2021+4026** | 78.2 | 2.09 | 1.5 | -0.45 + 0.45 | ...- ... |
| **J2032+4127** | 80.2 | 1.03 | 3.65 | -1.08 + 1.08 | ...- ... |
| **J2043+2740** | 70.6 | -9.15 | 1.8 | -0.54 + 0.54 | ...- ... |
| **J2229+6114** | 107 | 2.97 | ... | ... + ... | 0.8- 6.5 |
| **J2238+5903** | 107 | 0.52 | ... | ... + ... | ...- ... |
| **J1023-5746** | 284 | -0.4 | ... | ... + ... | ...- ... |
| **J1044-5737** | 287 | 1.2 | ... | ... + ... | ...- ... |
| **J1413-6205** | 312 | -0.7 | ... | ... + ... | ...- ... |
| **J1429-5911** | 315 | 1.3 | ... | ... + ... | ...- ... |
| **J1846+0919** | 40.7 | 5.3 | ... | ... + ... | ...- ... |
| **J1954+2836** | 65.2 | 0.4 | ... | ... + ... | ...- ... |
| **J1957+5033** | 84.6 | 11 | ... | ... + ... | ...- ... |
| **J2055+2539** | 70.7 | -12.5 | ... | ... + ... | ...- ... |
| J1513-5850 | 320 | -1.16 | 5.2 | -1.4 + 1.4 | ...- ... |
| J2240+5832 | 107 | -0.111 | ... | ... + ... | 3.8- 7.7 |
| J1648-4611 | 339 | -0.79 | 4.89 | -1.5 + 1.5 | ...- ... |
| **J2030+3641** | 76.1 | -1.44 | 8 | -2.4 + 2.4 | ...- ... |
| J1119-6127 | 292 | -0.54 | 8.4 | 0.4 + 0.4 | ...- ... |
| J0030+0451 | 113 | -57.6 | 0.3 | -0.09 + 0.09 | ...- ... |
| J0218+4232 | 140 | -17.6 | ... | ... + ... | 2.5- 4 |
| J0437-4715 | 253 | -41.9 | 0.156 | -0.0013 +0.0013 | ...- ... |
| J0613-0200 | 210 | -9.33 | 0.48 | -0.11 + 0.19 | ...- ... |
| J0751+1807 | 203 | 21.1 | 0.6 | -0.2 + 0.6 | ...- ... |
| J1614-2230 | 353 | 20.3 | 1.27 | -0.39 + 0.39 | ...- ... |
| J1744-1134 | 14.9 | 9.15 | 0.357 | -0.035 + 0.043 | ...- ... |
| J2124-3358 | 10.9 | -45.4 | 0.25 | -0.08 + 0.25 | ...- ... |
| J0034-0534 | 111 | -68.1 | 0.53 | 0.21 + 0.21 | ...- ... |
| J1939+2134 | 57.5 | -0.29 | 2.3 | -0.5 + 0.8 | ...- ... |
| J1959+2048 | 59.2 | -4.7 | 2.49 | -0.75 + 0.75 | ...- ... |
| J0614-3329 | 240 | -21.8 | 1.89 | -0.57 + 0.57 | ...- ... |
| J1231-1411 | 296 | 48.4 | 0.44 | -0.13 + 0.13 | ...- ... |
| J2214+3000 | 86.9 | -21.7 | 1.54 | -0.46 + 0.46 | ...- ... |
| J1823-3021A | 2.79 | -7.91 | 7.9 | ... + ... | ...- ... |

Table 4.2: Pulsar positions in galactic coordinates (longitude and latitude), distance with its uncertainties ($D \pm \delta D^{+}_{-}$), and a distance interval for the objects for which it was not possible to give a distance estimate ($\delta D^{sup}_{inf}$). In the upper part of the table are listed the ordinary pulsars, in the bottom part the MSP ones. In bold are indicated the pulsars that have been analysed in this thesis.

| | $\nu F_{\nu,>100MeV}$ erg cm$^{-2}$ s$^{-1}$ | $\delta\nu F_{\nu,>100MeV}$ erg cm$^{-2}$ s$^{-1}$ | $\Gamma$ | $\delta\Gamma$ | $E_{cut}$ GeV | $\delta E_{cut}$ GeV | $S_{radio,1400}$ mJy | $\delta S$ mJy |
|---|---|---|---|---|---|---|---|---|
| **J0007+7303** | 3.82e-10 | 1.11e-11 | 1.38 | 0.043 | 4.6 | 0.44 | 0 | 0.1 |
| **J0205+6449** | 6.65e-11 | 5.52e-12 | 2.09 | 0.144 | 3.54 | 1.44 | 0.045 | 0 |
| **J0248+6021** | 3.08e-11 | 5.75e-12 | 1.15 | 0.495 | 1.37 | 0.58 | 13.7 | 2.7 |
| **J0357+3205** | 6.4e-11 | 3.74e-12 | 1.29 | 0.179 | 0.94 | 0.17 | ... | ... |
| **J0534+2200** | 1.31e-09 | 1.12e-10 | 1.97 | 0.06 | 5.8 | 1.2 | 14 | 3 |
| **J0631+1036** | 3.04e-11 | 5.08e-12 | 1.38 | 0.352 | 3.57 | 1.82 | 0.8 | 0 |
| **J0633+0632** | 8.01e-11 | 6.37e-12 | 1.29 | 0.18 | 2.22 | 0.56 | 0 | 0.2 |
| **J0633+1746** | 3.39e-09 | 2.85e-11 | 1.08 | 0.016 | 1.94 | 0.05 | 0 | 1 |
| **J0659+1414** | 3.17e-11 | 3.06e-12 | 2.37 | 0.416 | 0.65 | 0.46 | 3.7 | 0.8 |
| **J0742-2822** | 1.83e-11 | 3.56e-12 | 1.76 | 0.399 | 1.99 | 1.44 | 15 | 1.5 |
| **J0835-4510** | 8.81e-09 | 4.55e-11 | 1.57 | 0.008 | 3.23 | 0.06 | 1.1e+03 | 15 |
| **J1028-5819** | 1.77e-10 | 1.24e-11 | 1.25 | 0.175 | 1.86 | 0.43 | 0.36 | 0.06 |
| **J1048-5832** | 1.73e-10 | 1.1e-11 | 1.31 | 0.149 | 1.98 | 0.37 | 6.5 | 0.7 |
| **J1057-5226** | 2.72e-10 | 8.12e-12 | 1.06 | 0.079 | 1.32 | 0.12 | 11 | 0 |
| **J1124-5916** | 3.8e-11 | 5.73e-12 | 1.43 | 0.334 | 1.74 | 0.7 | 0.08 | 0.02 |
| **J1418-6058** | 2.36e-10 | 3.15e-11 | 1.32 | 0.202 | 1.9 | 0.36 | 0 | 0.06 |
| **J1420-6048** | 1.59e-10 | 2.8e-11 | 1.73 | 0.195 | 2.75 | 1 | 0.9 | 0.1 |
| **J1459-6053** | 1.06e-10 | 9.66e-12 | 1.83 | 0.2 | 2.67 | 1.13 | 0 | 0.2 |
| **J1509-5850** | 9.69e-11 | 1.01e-11 | 1.36 | 0.226 | 3.46 | 1.14 | 0.15 | 0.03 |
| **J1709-4429** | 1.24e-09 | 2.21e-11 | 1.7 | 0.027 | 4.88 | 0.39 | 7.3 | 0.7 |
| **J1718-3825** | 6.75e-11 | 1.65e-11 | 1.26 | 0.617 | 1.28 | 0.62 | 1.3 | 0.4 |
| **J1732-3131** | 2.42e-10 | 1.22e-11 | 1.27 | 0.118 | 2.15 | 0.32 | 0 | 0.2 |
| **J1741-2054** | 1.28e-10 | 6.63e-12 | 1.39 | 0.138 | 1.16 | 0.19 | 0.16 | 0 |
| **J1747-2958** | 1.31e-10 | 1.36e-11 | 1.11 | 0.275 | 0.98 | 0.24 | 0.25 | 0.03 |
| **J1809-2332** | 4.13e-10 | 1.29e-11 | 1.52 | 0.063 | 2.86 | 0.34 | 0 | 0.06 |
| **J1813-1246** | 1.69e-10 | 1.07e-11 | 1.83 | 0.12 | 2.88 | 0.77 | 0 | 0.2 |
| **J1826-1256** | 3.34e-10 | 1.46e-11 | 1.49 | 0.093 | 2.39 | 0.34 | 0 | 0.06 |
| **J1833-1034** | 1.02e-10 | 1.22e-11 | 2.25 | 0.154 | 7.69 | 4.79 | 0.07 | 0 |
| **J1836+5925** | 6e-10 | 1.08e-11 | 1.35 | 0.034 | 2.32 | 0.14 | 0 | 0.007 |
| **J1907+0602** | 2.75e-10 | 1.29e-11 | 1.84 | 0.077 | 4.62 | 0.95 | 0 | 0.02 |
| **J1952+3252** | 1.34e-10 | 7.41e-12 | 1.75 | 0.099 | 4.55 | 1.16 | 1 | 0.1 |
| **J1958+2846** | 8.46e-11 | 6.88e-12 | 0.775 | 0.256 | 1.23 | 0.24 | ... | ... |
| **J2021+3651** | 4.7e-10 | 1.46e-11 | 1.65 | 0.06 | 2.63 | 0.29 | 0.1 | 0 |
| **J2021+4026** | 9.77e-10 | 1.7e-11 | 1.79 | 0.034 | 3.03 | 0.24 | ... | ... |
| **J2032+4127** | 1.11e-10 | 1.22e-11 | 0.68 | 0.381 | 2.14 | 0.55 | 0.24 | 0.05 |
| **J2043+2740** | 1.55e-11 | 2.76e-12 | 1.07 | 0.553 | 0.76 | 0.34 | 7 | 3 |
| **J2229+6114** | 2.2e-10 | 8.11e-12 | 1.74 | 0.07 | 3.03 | 0.49 | 0.25 | 0 |
| **J2238+5903** | 5.45e-11 | 5.97e-12 | 1 | 0.356 | 1.02 | 0.31 | ... | ... |
| J1023-5746 | 2.69e-10 | 1.8e-11 | 1.58 | 0.13 | 1.8 | 0.3 | ... | ... |
| J1044-5737 | 1.03e-10 | 6.5e-12 | 1.6 | 0.12 | 2.5 | 0.5 | ... | ... |
| J1413+6205 | 1.29e-10 | 1e-11 | 1.32 | 0.16 | 2.6 | 0.6 | ... | ... |
| J1429-5911 | 9.26e-11 | 8.1e-12 | 1.93 | 0.14 | 3.3 | 1 | ... | ... |
| J1846+0919 | 3.58e-11 | 3.5e-12 | 1.6 | 0.19 | 4.1 | 1.5 | ... | ... |
| J1954+2836 | 9.75e-11 | 6.8e-12 | 1.55 | 0.14 | 2.9 | 0.7 | ... | ... |
| J1957+5033 | 2.27e-11 | 2e-12 | 1.12 | 0.28 | 0.9 | 0.2 | ... | ... |
| J2055+2539 | 1.15e-10 | 7e-12 | 11.5 | 0.77 | 0.71 | 0.19 | ... | ... |
| J1513-5850 | 1e-11 | 0 | 2.9 | 0 | ... | ... | 0.94 | 0.1 |
| J2240+5832 | 1e-11 | 4.5e-12 | 1.8 | 0.61 | 5.7 | 4.5 | 2.7 | 0.7 |
| J1648-4611 | ... | ... | ... | ... | ... | ... | 0.58 | 0.07 |
| **J2030+3641** | ... | ... | ... | ... | ... | ... | ... | ... |
| J1119-6127 | ... | ... | ... | ... | ... | ... | 0.09 | ... |
| J0030+0451 | 5.28e-11 | 3.6e-12 | 1.22 | 0.163 | 1.76 | 0.38 | 0.6 | 0.2 |
| J0218+4232 | 3.62e-11 | 5.21e-12 | 2.02 | 0.228 | 5.12 | 4.17 | 0.9 | 0.2 |
| J0437-4715 | 1.86e-11 | 2.31e-12 | 1.74 | 0.323 | 1.32 | 0.65 | 142 | 53 |
| J0613-0200 | 3.23e-11 | 3.59e-12 | 1.38 | 0.239 | 2.72 | 0.97 | 1.4 | 0.2 |
| J0751+1807 | 1.1e-11 | 3.23e-12 | 1.56 | 0.584 | 3 | 4.31 | 3.2 | 0.7 |
| J1614-2230 | 2.74e-11 | 4.21e-12 | 1.34 | 0.36 | 2.44 | 1.05 | ... | ... |
| J1744-1134 | 2.81e-11 | 4.69e-12 | 1.02 | 0.593 | 0.72 | 0.36 | 3 | 1 |
| J2124-3358 | 2.75e-11 | 3.46e-12 | 1.05 | 0.28 | 2.68 | 0.98 | 1.6 | 0.4 |
| J0034-0534 | 1.9e-11 | 2.2e-12 | 1.5 | 0.22 | 1.7 | 0.61 | 0.61 | 0.09 |
| J1939+2134 | ... | ... | ... | ... | ... | ... | 10 | 1 |
| J1959+2048 | ... | ... | ... | ... | ... | ... | 0.4 | 0.2 |
| J0614-3329 | 1.09e-10 | 1.12e-11 | 1.44 | 0.086 | 4.49 | 1.1 | ... | ... |
| J1231-1411 | 1.03e-10 | 9.29e-12 | 1.4 | 0.086 | 2.98 | 0.48 | ... | ... |
| J2214+3000 | 3.32e-11 | 3.2e-12 | 1.44 | 0.12 | 2.53 | 0.61 | ... | ... |
| J1823-3021A | 2e-11 | 7e-12 | 1.2 | 0.5 | 1.6 | 0.9 | 0.72 | 0.02 |

Table 4.3: Pulsar spectral characteristics. Columns are as follows: the energy flux, the spectral index and error, the high energy cutoff and error, and the radio flux and error. In the upper part of the table are listed the ordinary pulsars, in the bottom part the MSP ones. In bold are indicated the pulsars that have been analysed in this thesis.

| | Peak number | Peak separation phase | δ phase | Radio lag phase | δ phase |
|---|---|---|---|---|---|
| **J0007+7303** | 2 | 0.23 | 0.01 | ... | ... |
| **J0205+6449** | 2 | 0.5 | 0.01 | 0.08 | 0.01 |
| **J0248+6021** | 1 | ... | ... | 0.35 | 0.01 |
| **J0357+3205** | 1 | ... | ... | ... | ... |
| **J0534+2200** | 2 | 0.4 | 0.01 | 0.09 | 0.01 |
| **J0631+1036** | 1 | ... | ... | 0.54 | 0.02 |
| **J0633+0632** | 2 | 0.48 | 0.01 | ... | ... |
| **J0633+1746** | 2 | 0.5 | 0.01 | ... | ... |
| **J0659+1414** | 1 | ... | ... | 0.21 | 0.01 |
| **J0742-2822** | 1 | ... | ... | 0.61 | 0.02 |
| **J0835-4510** | 2 | 0.43 | 0.01 | 0.13 | 0.01 |
| **J1028-5819** | 2 | 0.47 | 0.01 | 0.19 | 0.01 |
| **J1048-5832** | 2 | 0.42 | 0.02 | 0.15 | 0.01 |
| **J1057-5226** | 2 | 0.2 | 0.07 | 0.35 | 0.05 |
| **J1124-5916** | 2 | 0.49 | 0.01 | 0.23 | 0.01 |
| **J1418-6058** | 2 | 0.47 | 0.01 | ... | ... |
| **J1420-6048** | 2 | 0.18 | 0.02 | 0.26 | 0.02 |
| **J1459-6053** | 2 | 0.15 | 0.03 | ... | ... |
| **J1509-5850** | 2 | 0.2 | 0.03 | 0.18 | 0.03 |
| **J1709-4429** | 2 | 0.25 | 0.01 | 0.24 | 0.01 |
| **J1718-3825** | 1 | ... | ... | 0.42 | 0.02 |
| **J1732-3131** | 2 | 0.42 | 0.02 | ... | ... |
| **J1741-2054** | 2 | 0.18 | 0.02 | 0.3 | 0.01 |
| **J1747-2958** | 2 | 0.42 | 0.04 | 0.18 | 0.01 |
| **J1809-2332** | 2 | 0.35 | 0.01 | ... | ... |
| **J1813-1246** | 2 | 0.47 | 0.02 | ... | ... |
| **J1826-1256** | 2 | 0.47 | 0.01 | ... | ... |
| **J1833-1034** | 2 | 0.44 | 0.01 | 0.15 | 0.01 |
| **J1836+5925** | 2 | 0.48 | 0.01 | ... | ... |
| **J1907+0602** | 2 | 0.4 | 0.01 | ... | ... |
| **J1952+3252** | 2 | 0.49 | 0.01 | 0.15 | 0.01 |
| **J1958+2846** | 2 | 0.45 | 0.01 | ... | ... |
| **J2021+3651** | 2 | 0.47 | 0.01 | 0.17 | 0.01 |
| **J2021+4026** | 2 | 0.48 | 0.01 | ... | ... |
| **J2032+4127** | 2 | 0.5 | 0.01 | 0.15 | 0.01 |
| **J2043+2740** | 2 | 0.36 | 0.01 | 0.2 | 0.01 |
| **J2229+6114** | 1 | ... | ... | 0.49 | 0.01 |
| **J2238+5903** | 2 | 0.5 | 0.01 | ... | ... |
| **J1023-5746** | 2 | 0.45 | 0.01 | ... | ... |
| **J1044-5737** | 2 | 0.35 | 0.01 | ... | ... |
| **J1413-6205** | 2 | 0.31 | 0.02 | ... | ... |
| **J1429-5911** | 2 | 0.46 | 0.01 | ... | ... |
| **J1846+0919** | 1 | ... | ... | ... | ... |
| **J1954+2836** | 2 | 0.43 | 0.01 | ... | ... |
| **J1957+5033** | 1 | ... | ... | ... | ... |
| **J2055+2539** | 1 | ... | ... | ... | ... |
| J1513-5850 | 2 | 0.37 | 0.02 | -0.04 | 0.01 |
| J2240+5832 | 1 | ... | ... | 0.58 | 0.01 |
| J1648-4611 | 0 | ... | ... | ... | ... |
| J2030+3641 | 0 | ... | ... | ... | ... |
| J1119-6127 | 1 | ... | ... | ... | 0.8 |
| J0030+0451 | 2 | 0.44 | 0.01 | 0.18 | 0.01 |
| J0218+4232 | 2 | 0.36 | 0.02 | 0.32 | 0.02 |
| J0437-4715 | 1 | ... | ... | 0.43 | 0.02 |
| J0613-0200 | 1 | ... | ... | 0.42 | 0.01 |
| J0751+1807 | 1 | ... | ... | 0.43 | 0.02 |
| J1614-2230 | 2 | 0.51 | 0.01 | 0.19 | 0.01 |
| J1744-1134 | 1 | ... | ... | 0.83 | 0.02 |
| J2124-3358 | 1 | ... | ... | 0.86 | 0.02 |
| J0034-0534 | 2 | 0.274 | 0.015 | -0.027 | 0.008 |
| J1939+2134 | 0 | ... | ... | ... | ... |
| J1959+2048 | 0 | ... | ... | ... | ... |
| J0614-3329 | 0 | ... | ... | ... | ... |
| J1231-1411 | 0 | ... | ... | ... | ... |
| J2214+3000 | 0 | ... | ... | ... | ... |
| J1823-3021A | 2 | 0.642 | 0.008 | 0 | ... |

Table 4.4: Pulsar light curve structural characteristics. Columns are as follows: The numbers of peaks, the peak separation for the double peak light curves, and the radio lag for the radio-loud LAT objects. In the upper part of the table are listed the ordinary pulsar, in the bottom part the MSP ones. In bold are indicated the pulsars that have been analysed in this thesis.

### 4.2.1 Young LAT pulsars

The pulsars detected by the LAT and studied in this thesis belong to the non-millisecond, isolated ordinary pulsar population. All the pulsars for which I have obtained the $\gamma$-ray and radio light curves and that have been analysed in this thesis are indicated in bold in tables 4.1, 4.2, 4.3, 4.4. In the next sections I will describe the data selection and the procedure adopted to obtain their light curves.

### 4.2.2 Data selection and Light curves production

The dataset I selected to build the pulsar light curves includes LAT observations between MET=240202715 seconds and MET=299978038 seconds (where MET is the Mission Elapsed Time starting on January 1st 2001 at 00:00:00.000 UTC). This time interval corresponds to the period between August 8th 2008 and April 7th 2010.

To have the highest background rejection only photons with energy $E_{ph} > 100$ MeV and belonging to the *diffuse* class (section 3.1.5) have been used. To avoid spurious detection due to the $\gamma$-rays scattered from the Earth atmosphere, I used the events collected within a zenith angle of 105 degrees. Each pulsar has been analysed by applying a PSF-wide selection. Just the events inside the LAT containment ratio for their measured energy have been used to build the pulsar light curves (section 3.1.6, figure 3.8 & 3.9), and a lower and upper limit for the ROI radius of $0°.3$ and $1°.0$ has been applied. The ephemerides used to produce the pulsar light curves have been generated by the joint work of the LAT pulsar group and the radio consortium that collaborates and exchanges information with the LAT team. With the exception of the blind search pulsars, for which no radio counterpart has been identified, and few other pulsars, the radio consortium provides the radio ephemeris with the highest accuracy possible.

The light curve generation has been implemented using the LAT tool *gtselect*[1] and the TEMPO2 software (Hobbs et al., 2006). To make the light curve generation process automatic and recursive, I wrote a script package to manage the data selection and the folding. For each pulsar, the script package reads the information in the ephemeris file and runs *gtselect* to select the data corresponding to the time intervals of validity for the pulsar ephemeris. In the same run, *gtselect* applies also the selections for the zenith angle, the photon event class, and the ROI around the pulsar position. The next step was to run a special TEMPO2 plug-in[2], implemented by Lucas Guillemot, on the *gtselect* output data file. The plug-in applies the geocentric photon correction and uses the Period, period time derivatives, and all the informations available in the ephemeris file to fold the $\gamma$-ray light curve. The procedure is automatically run for all the pulsars by the script package.

---

[1] FERMI LAT science tools package http://fermi.gsfc.nasa.gov/ssc/data/analysis/
[2] http://fermi.gsfc.nasa.gov/ssc/data/analysis/user/

The $\gamma$-ray light curves generated in this way are visible, for all the pulsars of our sample, from figure 4.2, to figure 4.9. Two plots are shown for each pulsar, a right one in which a red cross indicates the position of each photon in the phase-time diagram, and the left one, that shows the vertical integration of the phase-time diagram in a 45 bin light curve. The phase-time diagram is particularly important to check the timing solution we used to build our light curves. The timing solution can be trusted if all the photons of a peak are aligned at the same phase to form a vertical, straight band.

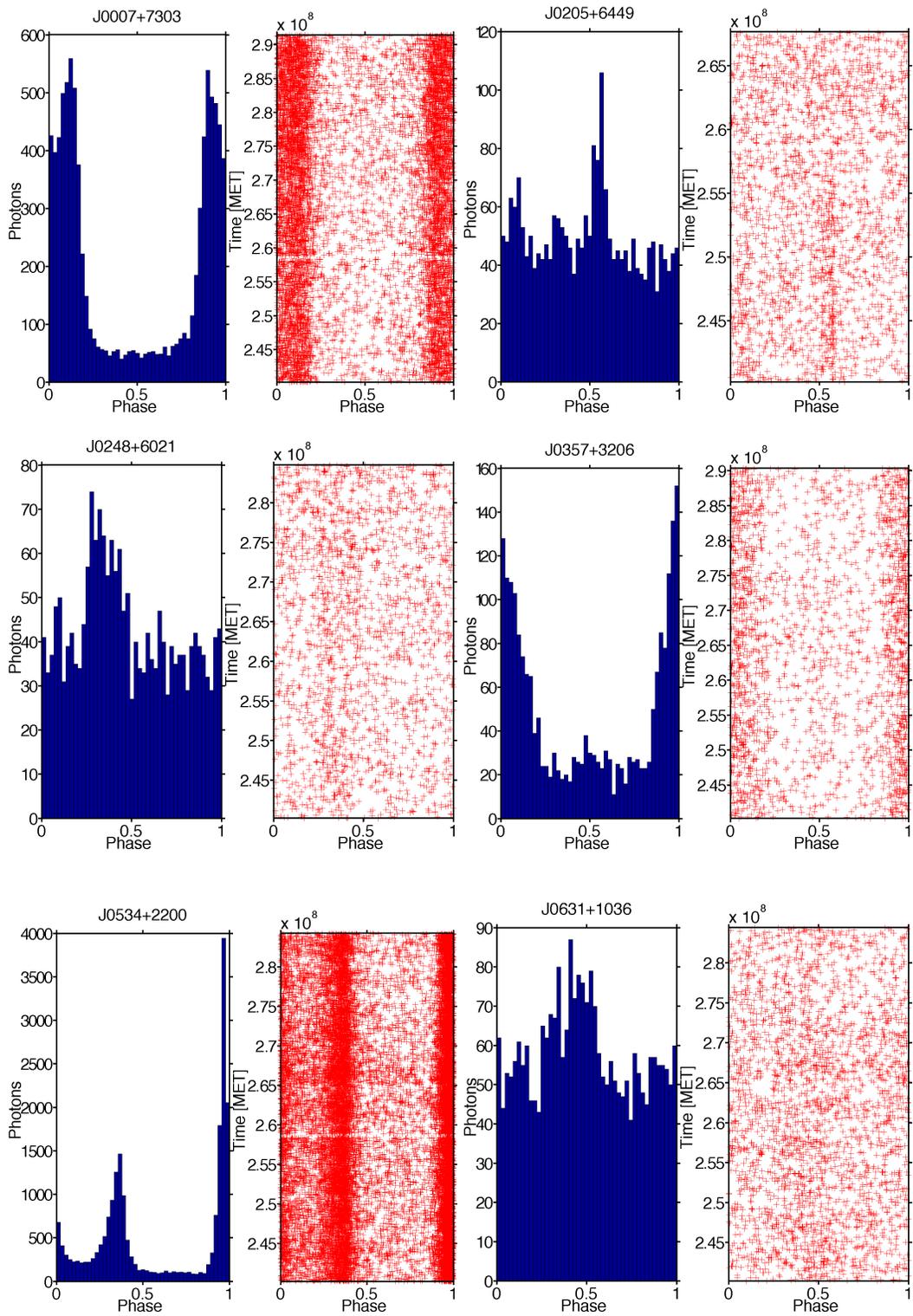

Figure 4.2: Gamma-ray light curve and phase-time diagram for the pulsars J0007+7303, J0205+6449, J0248+6021, J0357+3206, J0534+2200, J0631+1036.

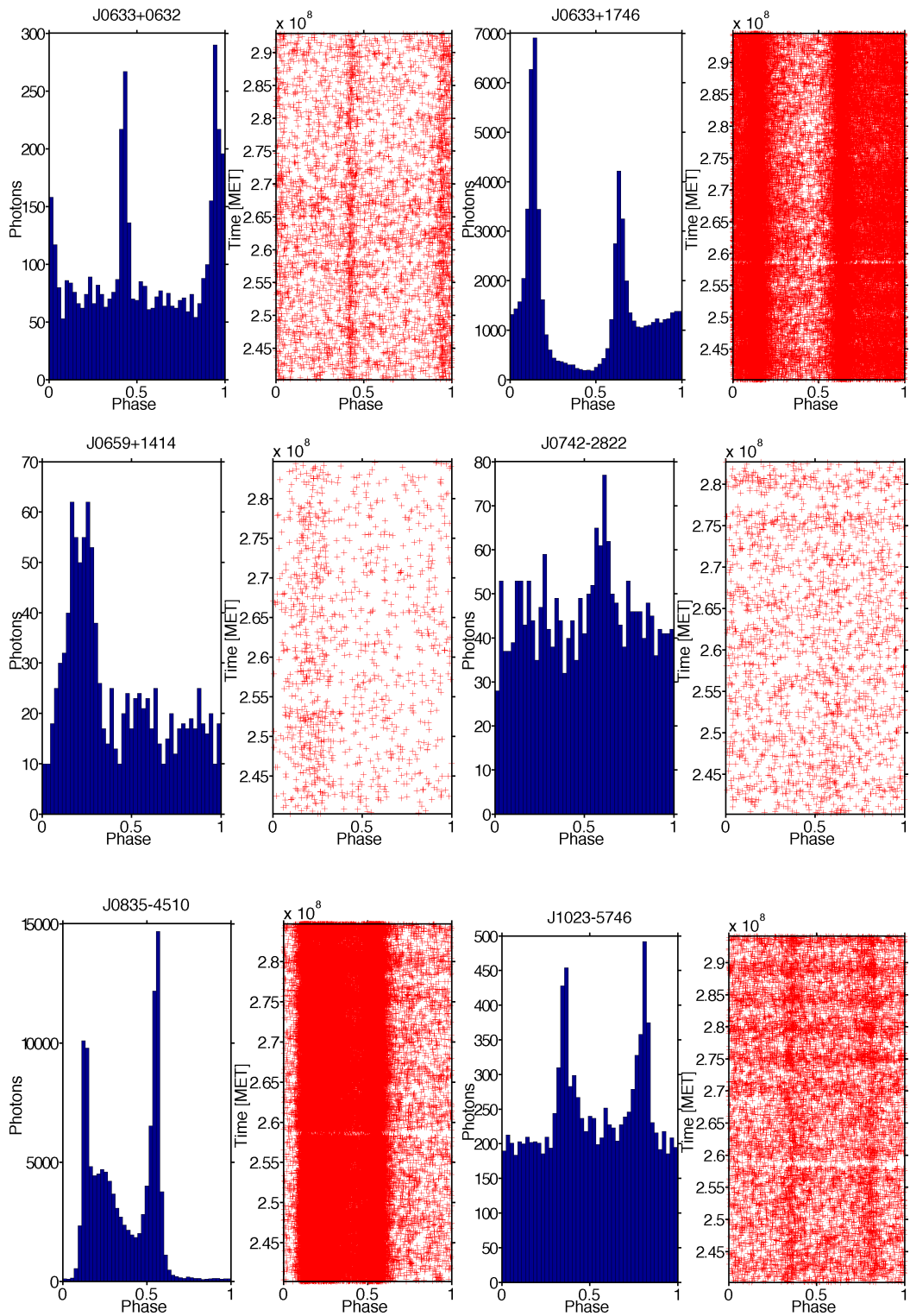

Figure 4.3: Gamma-ray light curve and phase-time diagram for the pulsars
J0633+0632, J0633+1746, J0659+1414, J0742-282, J0835-4510, J1023-5746.

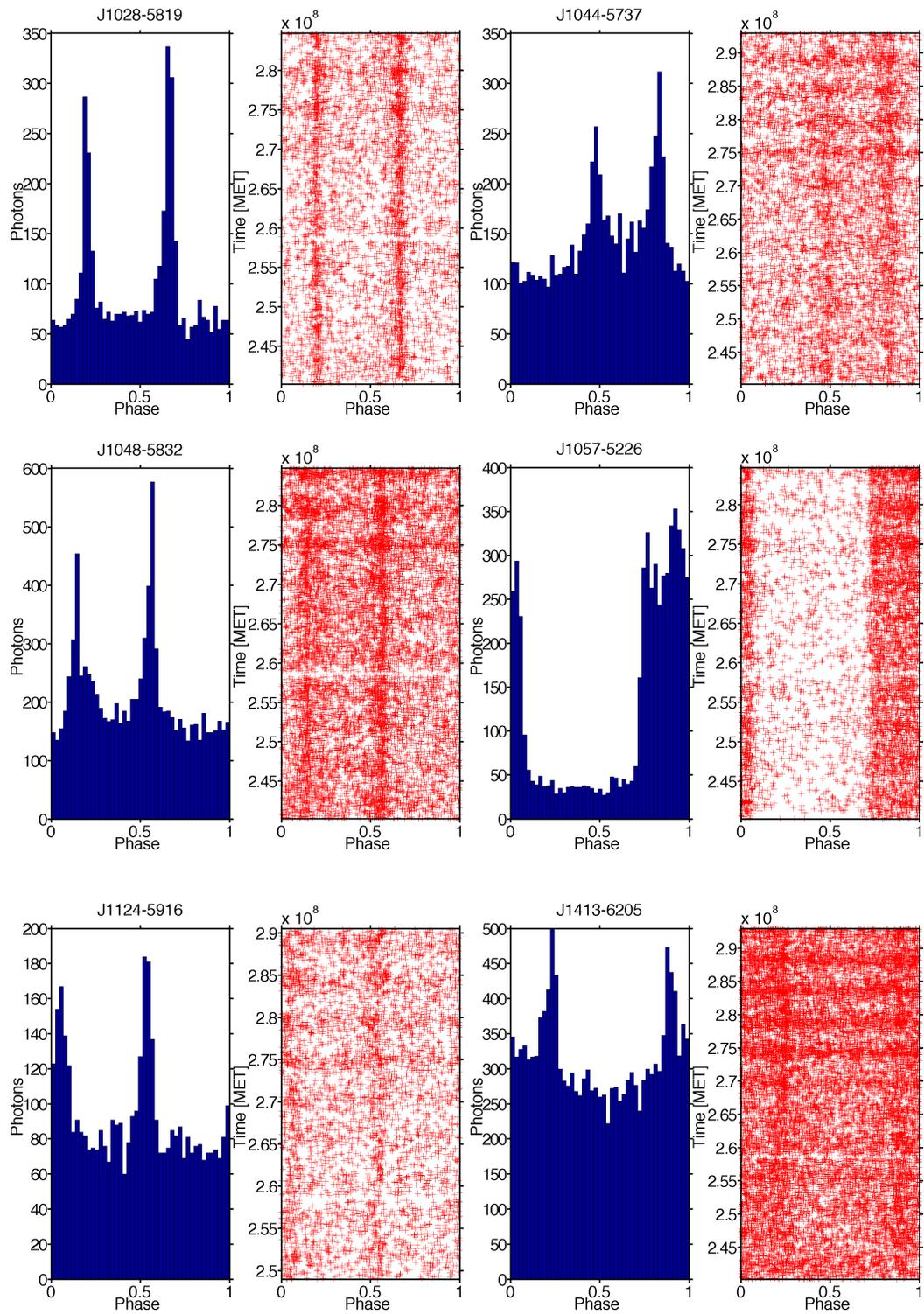

Figure 4.4: Gamma-ray lightcurve and phase-time diagram for the pulsars J1028-5819, J1044-5737, J1048-5832, J1057-5226, J1124-5916, J1413-6205.

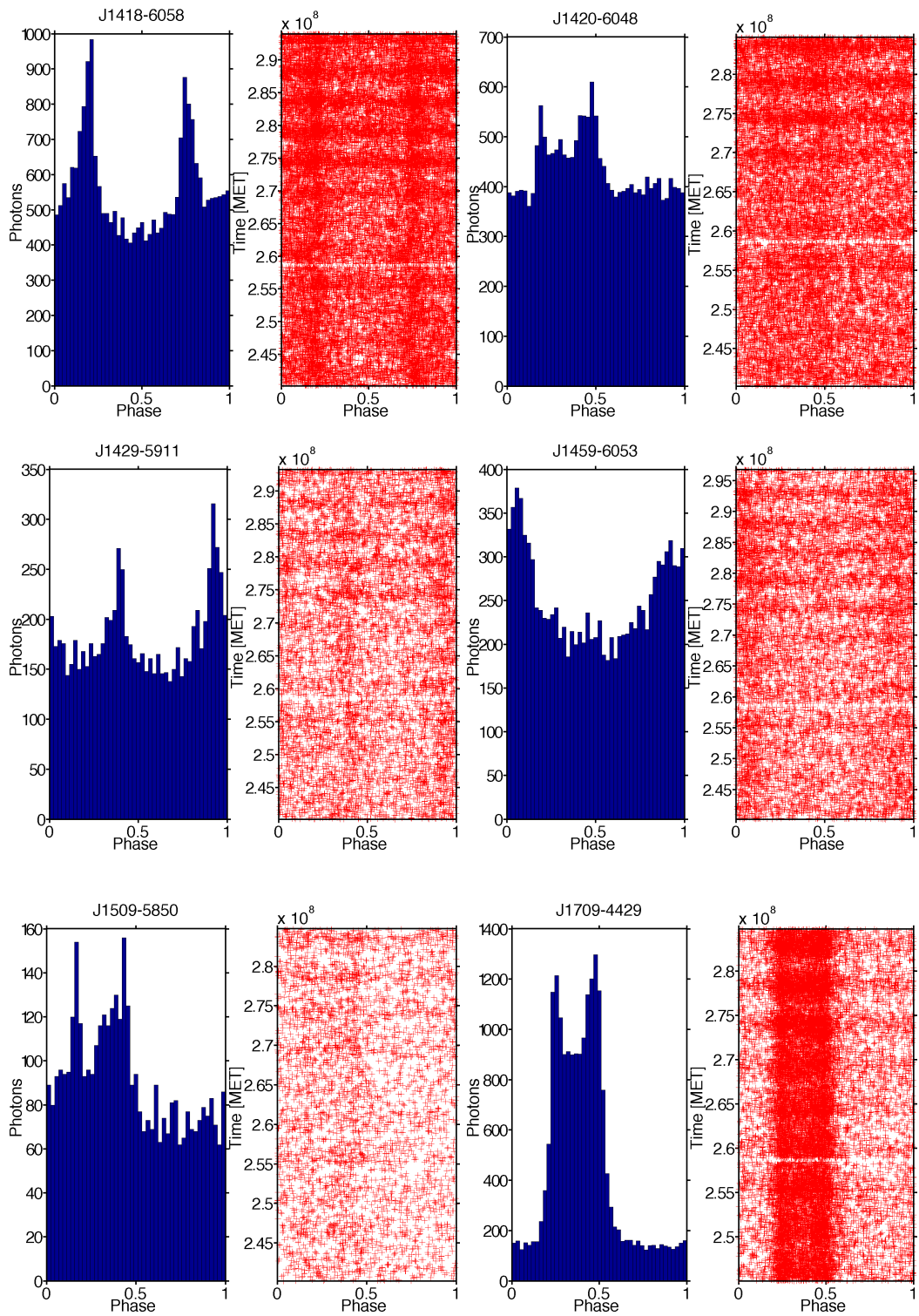

Figure 4.5: Gamma-ray light curve and phase-time diagram for the pulsars J1418-6058, J1420-6048, J1429-5911, J1459-6053, J1509-5850, J1709-4429.

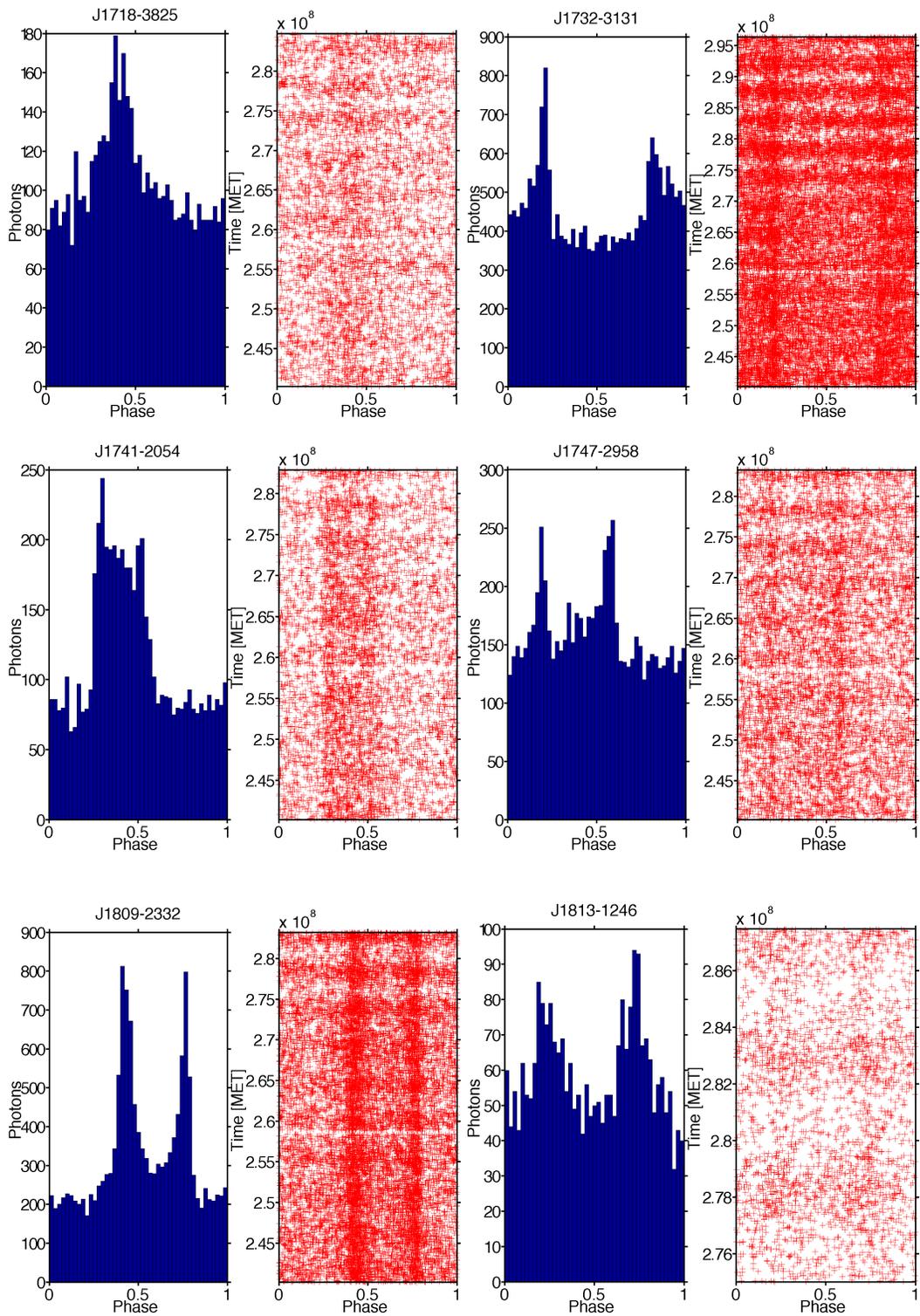

Figure 4.6: Gamma-ray light curve and phase-time diagram for the pulsars J1718-3825, J1732-3131, J1741-2054, J1747-2958, J1809-2332, J1813-1246.

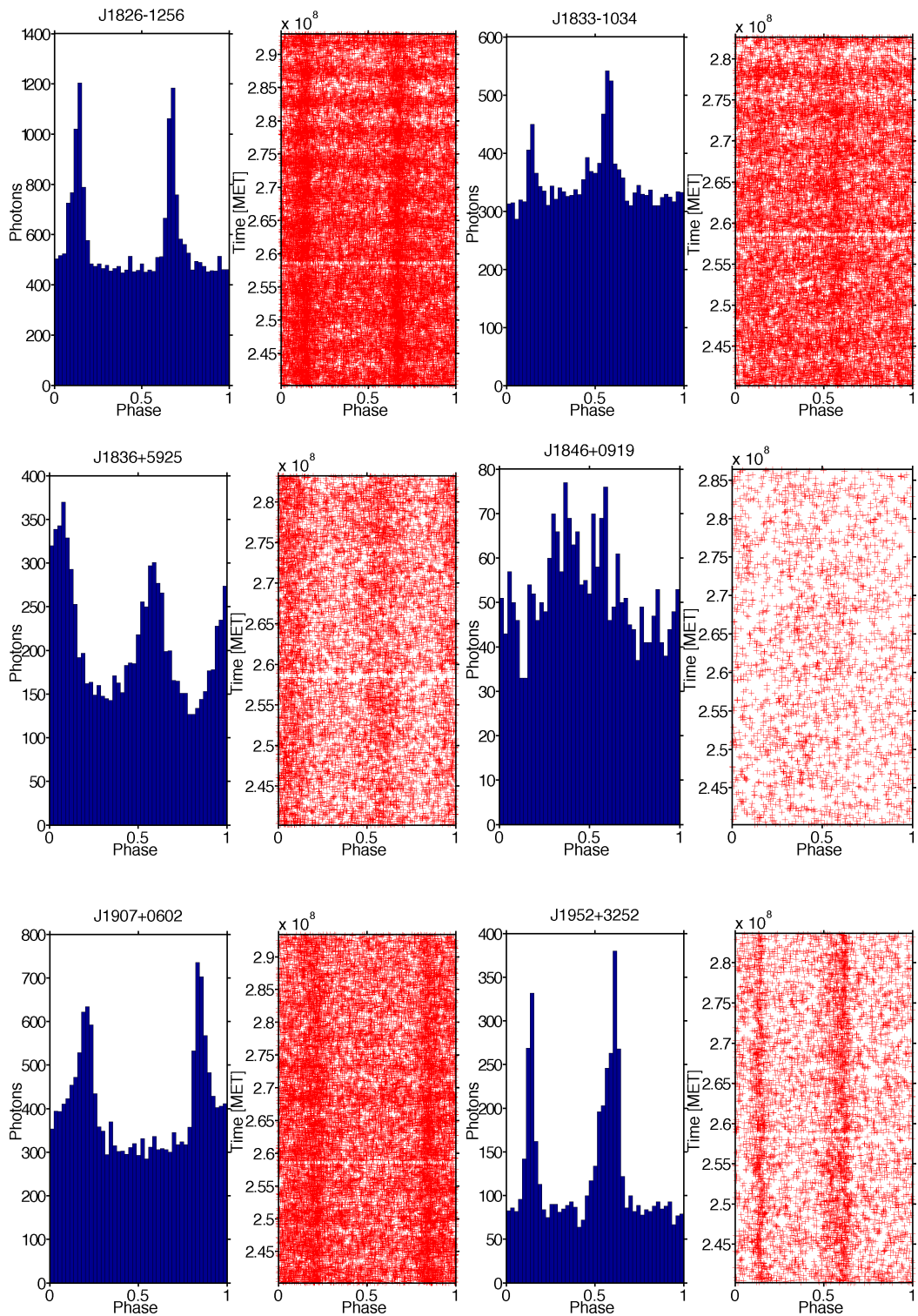

Figure 4.7: Gamma-ray light curve and phase-time diagram for the pulsars J1826-1256, J1833-1034, J1836+5925, J1846+0919, J1907+0602, J1952+3252.

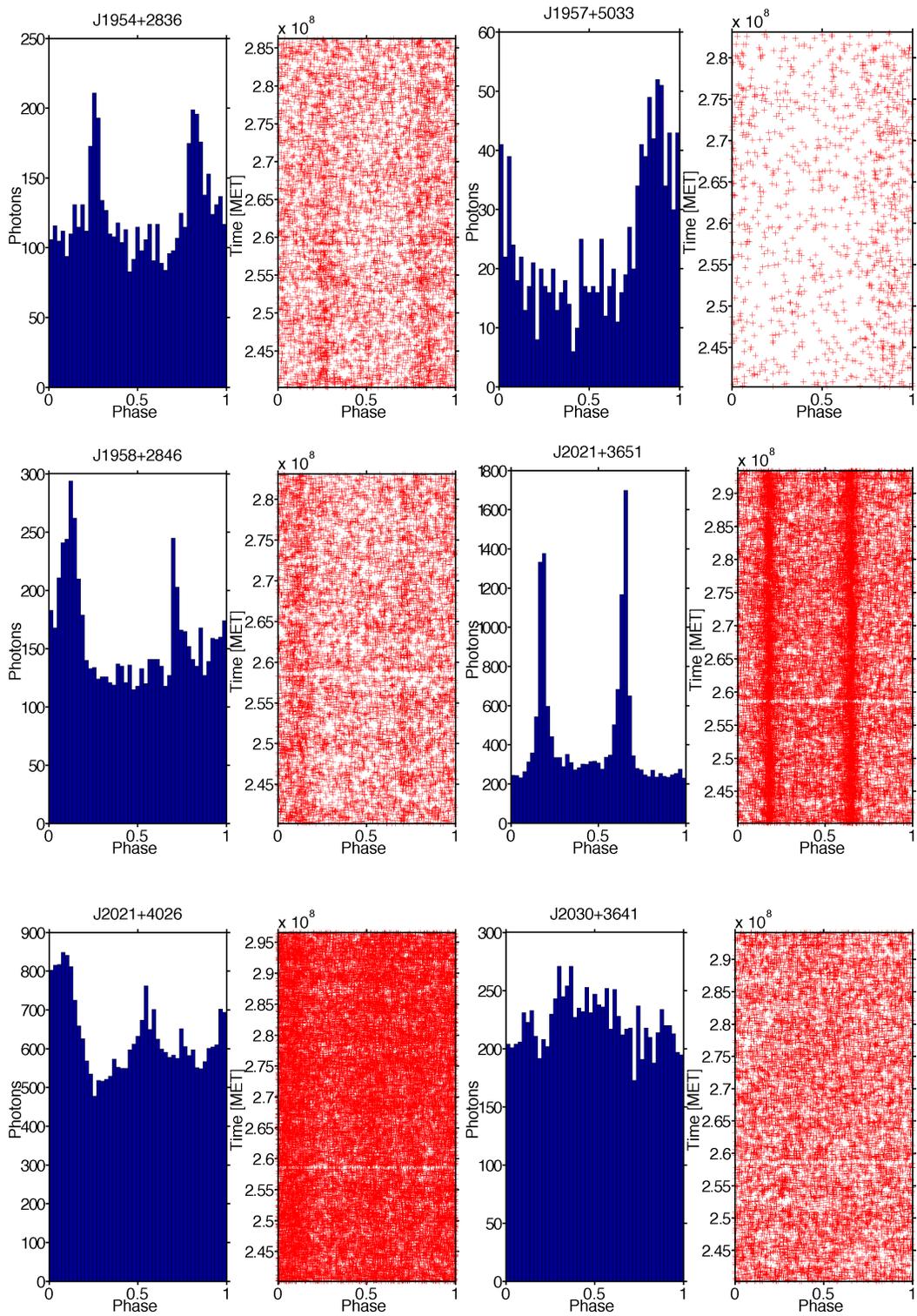

Figure 4.8: Gamma-ray light curve and phase-time diagram for the pulsars J1954+2836, J1957+5033, J1958+2846, J2021+3651, J2021+4026, J2030+3641.

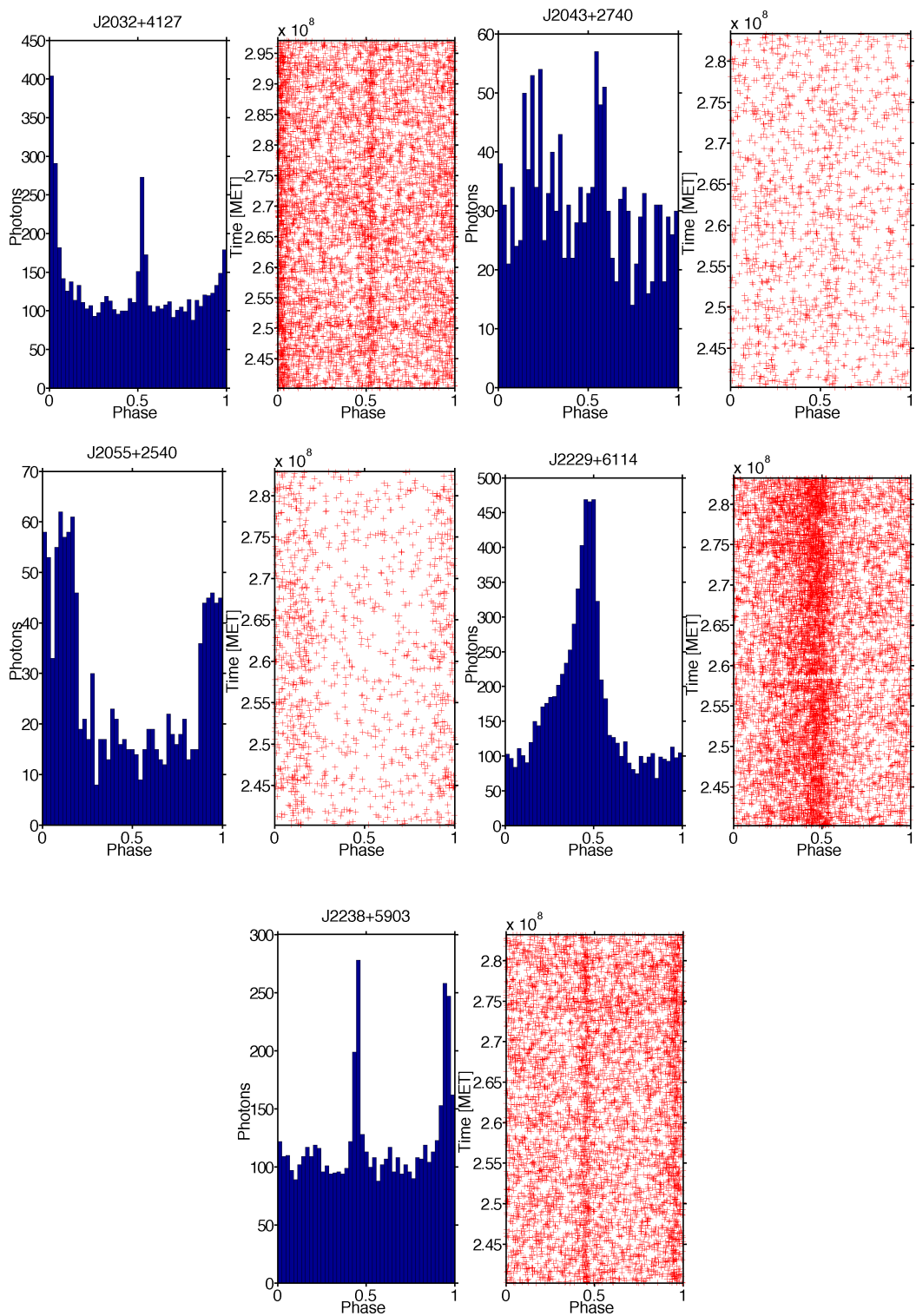

Figure 4.9: Gamma-ray light curve and phase-time diagram for the pulsars J2032+4127, J2043+2740, J2055+2540, J2229+6114, J2238+5903.

Concerning the radio-loud pulsar sample analysed in this thesis, the radio light-curves have been downloaded from the Fermi Science support centre webpage[3]. Each radio profile, resampled to 45 bins, is shown in Figures 4.10 to 4.13.

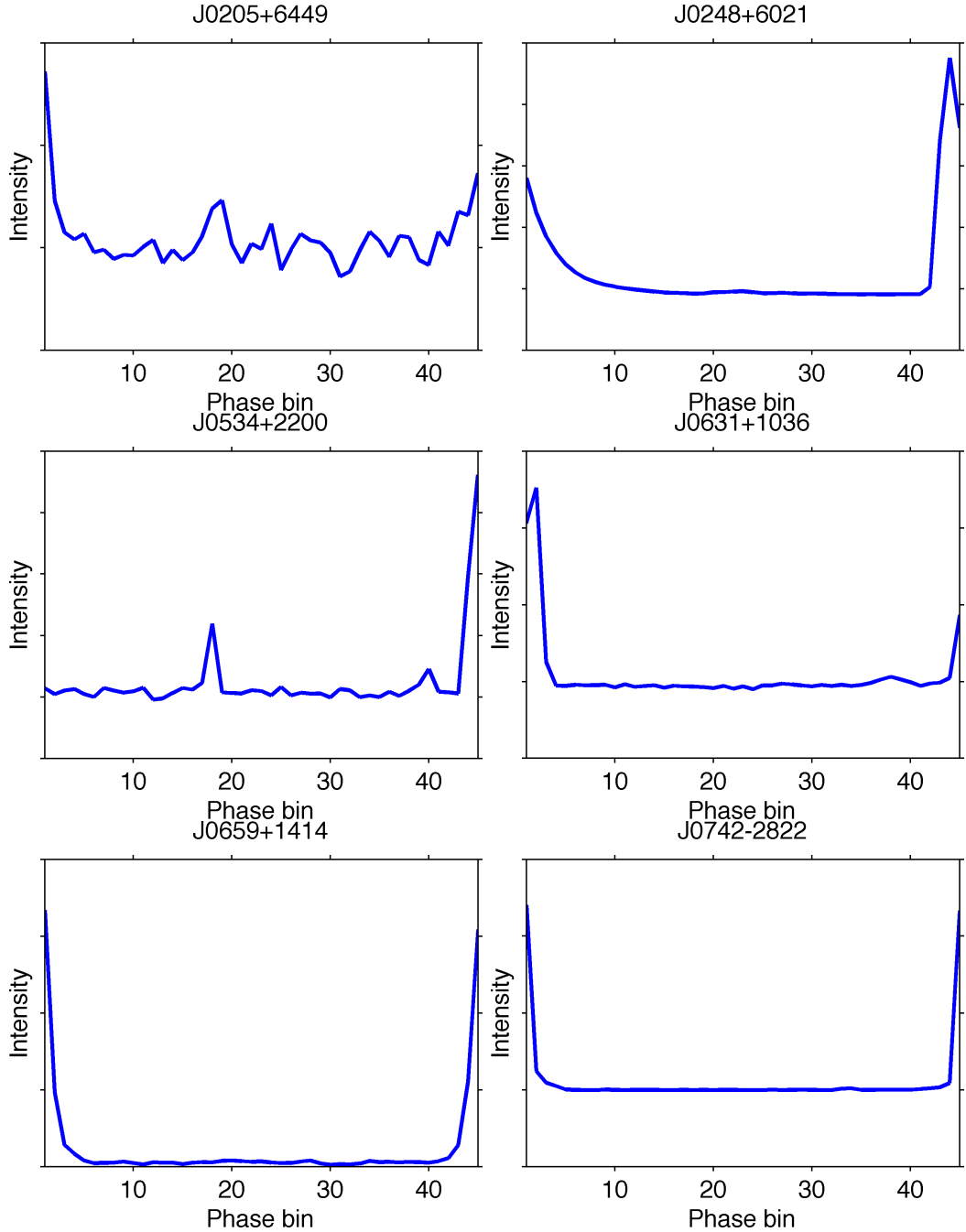

Figure 4.10: Radio light curve for the pulsars J0205+6449, J0248+6021, J0534+2200, J0631+1036, J0659+1414, J0742-2822.

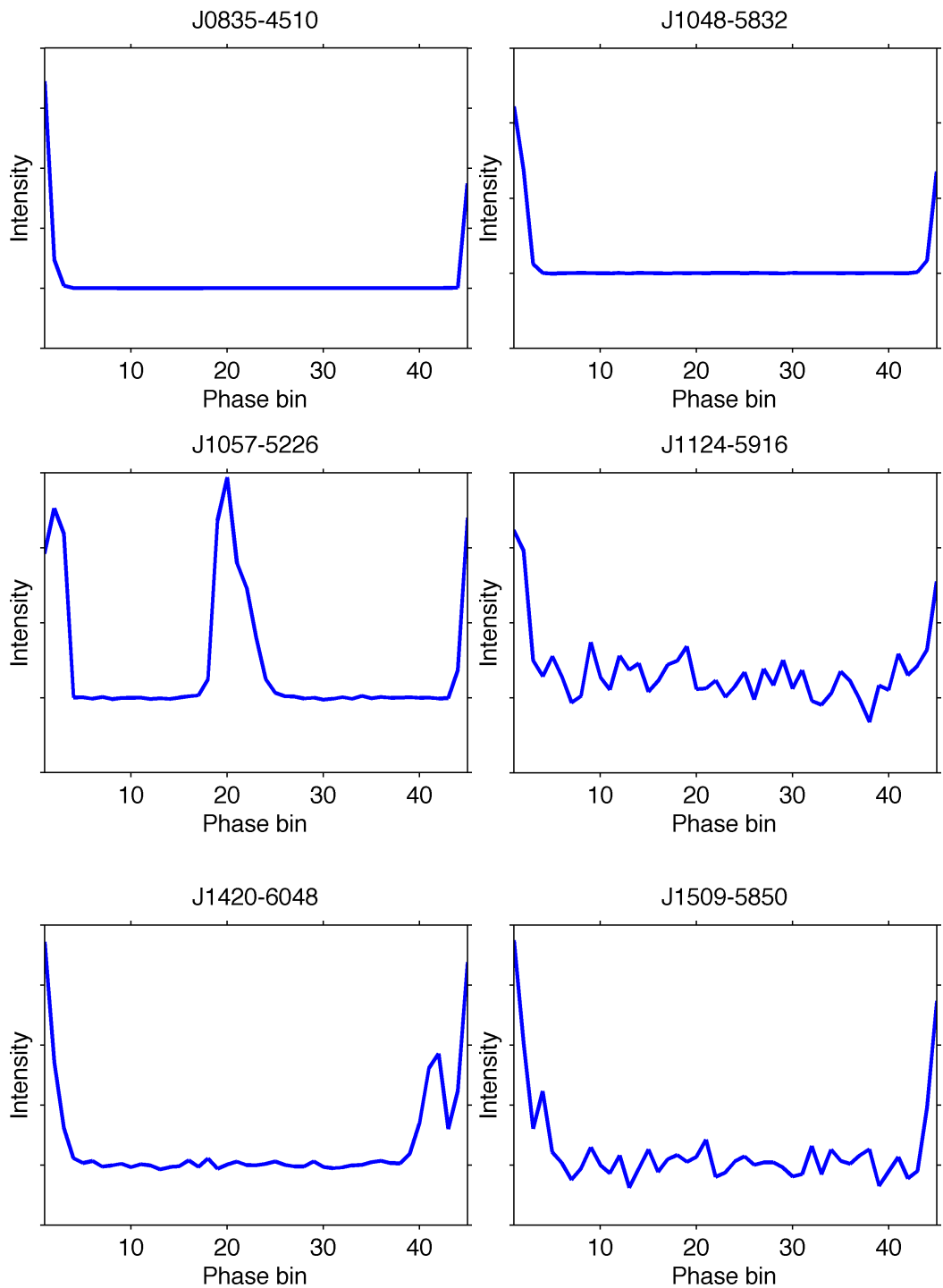

Figure 4.11: Radio light curve for the pulsars J0835-4510, J1048-5832, J1057-5226, J1124-5916, J1420-6048, J1509-5850.

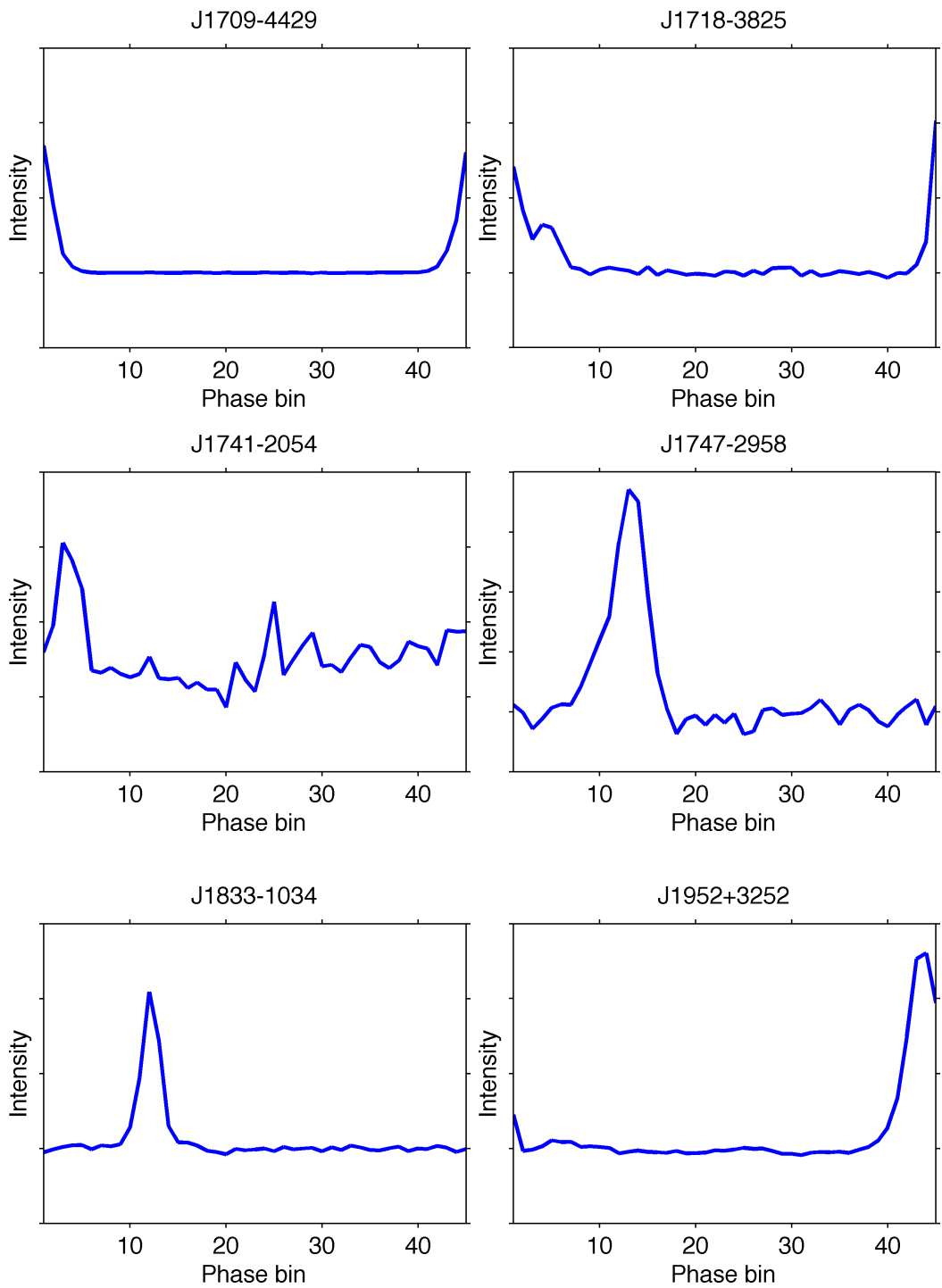

Figure 4.12: Radio light curve for the pulsars J1709-4429, J1718-3825, J1741-2054, J1747-2958, J1833-1034 ,J1952+3252.

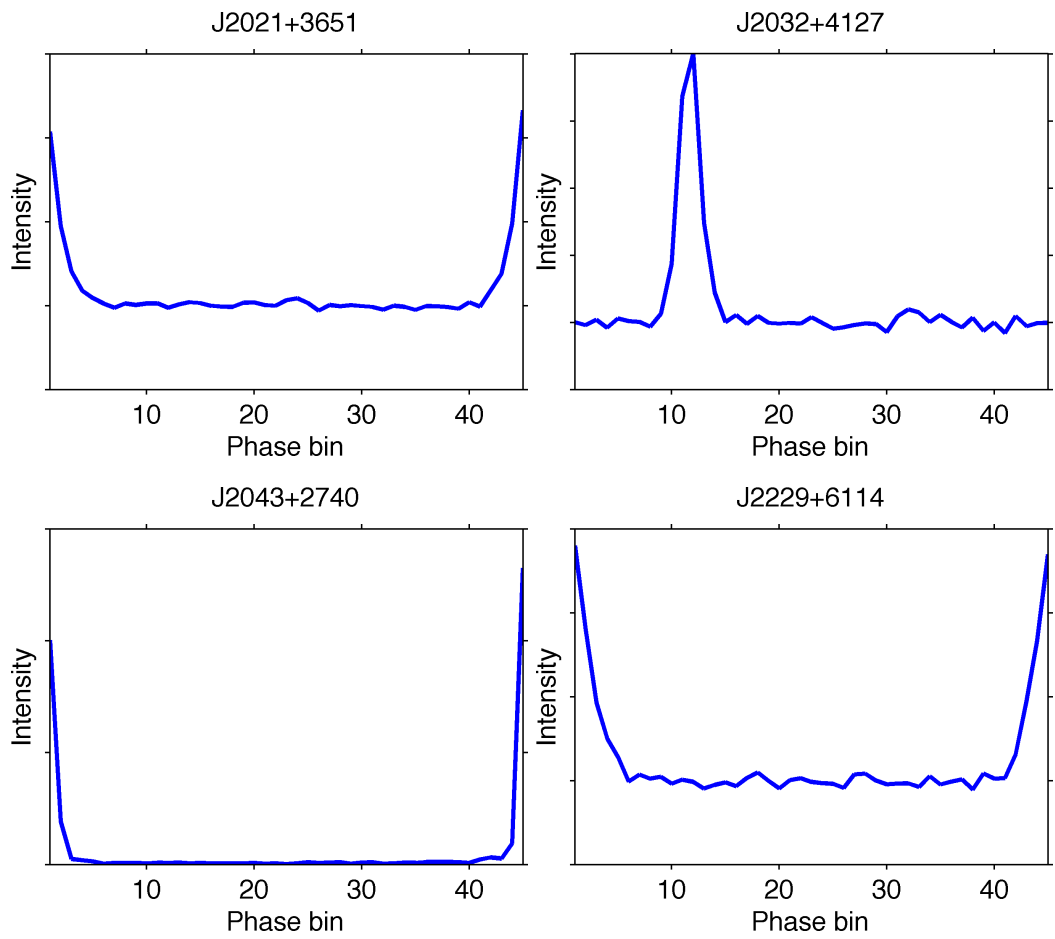

Figure 4.13: Radio light curve for the pulsars J2021+3651, J2032+4127, J2043+2740, J2229+6114.

# Chapter 5

# A simulated $\gamma$-ray pulsar population: comparison with the LAT pulsar sample

In this chapter I will describe the first part of my thesis project, the synthesis of a $\gamma$-ray pulsar population and the comparison with the LAT observations. The synthesis of the NS population has been implemented in collaboration with Peter Gonthier that generated and evolved in the Galactic gravitational potential the neutron star samples. The population has been synthesised taking into account the structure of our Galaxy as well as the known radio pulsar population characteristics and evolved up to the present time. To reproduce and study the population behaviour with respect to different emission mechanisms, each $\gamma$-ray emission model described in Chapter 2 has been considered. The results of the comparison between the LAT observations and each emission model will be discussed in the last section of the chapter.

## 5.1 The simulated pulsar sample: population characteristics

The simulated pulsar population used for this part of my thesis project has been synthesised in collaboration with Isabelle Grenier, Alice Harding, and Peter Gonthier, and is the starting point of the paper in preparation: *'Population synthesis of radio and $\gamma$-ray pulsars: confronting Fermi observations with current emission models'*, by Marco Pierbattista, Isabelle Grenier, Alice Harding, Peter Gonthier.

We synthesised 40 NSs samples for a total of 26442434 objects. In each sample, the NSs to the left of the radio death line are $2.5 \times 10^5$, for a global population of $10^7$ normal radio pulsars. An exponential magnetic field decay with a time scale of $2.8 \times 10^6$ yr has been assumed to match the radio survey data. The radio death line we used is defined from the equation

$$\log \dot{P} < a \times b \log P. \tag{5.1}$$

It is composed by three different segments (Story et al. (2007) & Zhang et al. (2000)), each one refers to a specific period interval characterised from the following $a$ and $b$ values

$$
\begin{aligned}
P \leq 15ms & \qquad a = -19.00 \quad b = 0.814 \\
15ms < P \leq 300ms & \qquad a = -17.60 \quad b = 1.370 \\
P > 300ms & \qquad a = -16.69 \quad b = 2.590
\end{aligned}
\tag{5.2}
$$

A set of birth intrinsic characteristics has been modelled for each pulsar, then we have evolved both the spin characteristics under dipole assumption and the pulsar position and velocity in the Galactic gravitational potential, up to the present time.

### 5.1.1  Period P, period first time derivative $\dot{\mathbf{P}}$, and magnetic field B

By assuming a value for the pulsar mass and radius, the pulsar electrodynamics is completely defined by its period **P** and its period first time derivative $\dot{\mathbf{P}}$ (the Goldreich-Julian model, sections 1.3.1 & the oblique rotator model, section 1.2.2). Since **P** and $\dot{\mathbf{P}}$ are two observable characteristics, the pulsar magnetic field intensity at the surface can be estimated from the relation $B_S^2 \propto P\dot{P}$ (equation 1.18). Starting from these assumptions and considering a purely time-dependent magnetic field and an absolute time reference frame, it is possible to follow the evolution of the whole pulsar electrodynamic characteristics from a time $t_1$ to a time $t_2$.

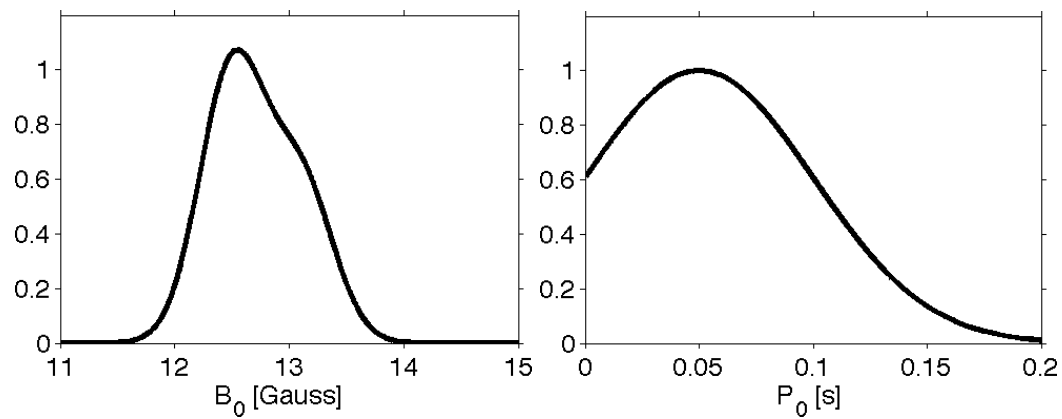

Figure 5.1: *Left:* The assumed surface magnetic field distribution at birth. *Right:* The assumed spin period distribution at birth.

The birth distribution for the magnetic field $\mathbf{B_0}$ (in $\log_{10}$ scale) is shown in figure 5.1. It has been built as the sum of two gaussians, both of 0.4 Gauss in width, respectively centred at $10^{12.5}$ and $10^{13.1}$ Gauss, and with an amplitude ratio $1:7/12$. This has been chosen *a posteriori* to best describe the magnetic field distribution of the observed population. The period $\mathbf{P_0}$ distribution at birth, plotted in the right panel of figure 5.1, follows a single gaussian of width 50 ms centred at 50 ms. The $\mathbf{\dot{P}_0}$ birth distribution has been derived from $\mathbf{P_0}$ and $\mathbf{B_0}$ by using equation 1.18. Moreover, an exponential constant rate magnetic field decay on a time scale of $2.8 \times 10^6$ yr has been assumed. The simulated pulsar population at birth, is shown, in red, in the $\mathbf{P}$-$\mathbf{\dot{P}}$ diagram of figure 5.2. Since our final purpose is to compare the evolved sample with the young LAT one and since we find in section 5.7 that all models underpredict the number of visible $\gamma$-ray pulsars at high $\dot{E}$ we optimised the birth distributions to have a higher fraction of LAT-like evolved objects.

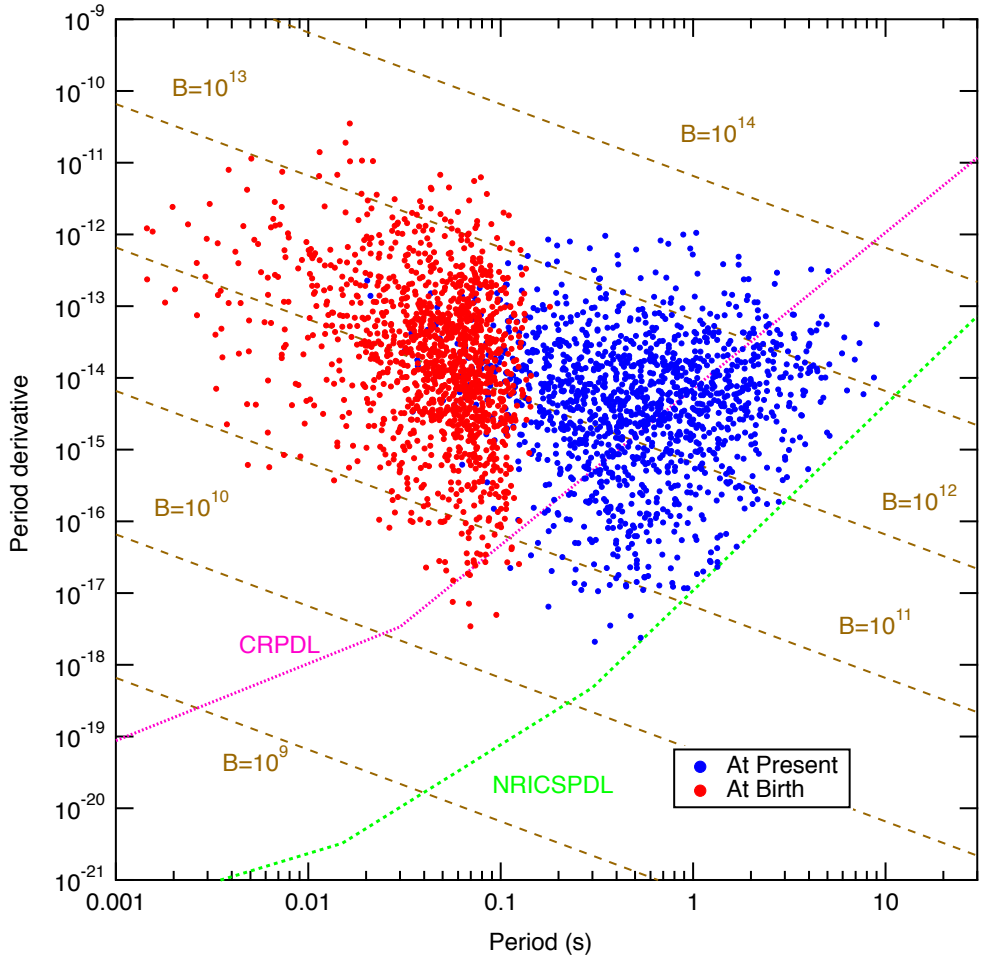

Figure 5.2: $P$-$\dot{P}$ diagram of the pulsar population at birth (red), and the evolved one (blue). The pink and green radio death lines, respectively for the birth and evolved populations, have been obtained from equations 5.1 & 5.2.

From (Gonthier et al., 2002), by knowing the analytical expression for $B(t) = f(B_0, t)$, it is possible to follow the evolution of the population parameters from the birth time $t_0$ to the present time $t_p$. The magnetic decay is described by

$$B(t) = B_{0,12} e^{-t/\tau_D} \tag{5.3}$$

where $\tau_D = 2.8$ Myr is the decay timescale, and $B_{0,12}$ is the birth magnetic field in unit of $10^{12}$ Gauss. Assuming magnetic dipole spin-down and initial period $P_0$, the period and the period first time derivative at the present time can be obtained as

$$P^2 = P_0^2 + K B_{0,12}^2 \tau_D 31557600 (1 - e^{-2t/\tau_D}) \tag{5.4}$$

$$\dot{P} = K e^{-2t/\tau_D} \frac{B_{0,12}^2}{P} \tag{5.5}$$

$$K = \frac{8\pi^2 R_{NS}^6}{3c^3 I} \tag{5.6}$$

where $P$ and $P_0$ are in seconds, and $t$ and $\tau_D$ are in years. Because a magnetic field decay is assumed, the spin-down age of the pulsar should be expressed as

$$Age = \frac{\tau_D}{2} \ln\left(\frac{3.17 \times 10^{-8} P}{\dot{P} \tau_D} + 1\right). \tag{5.7}$$

The last equation asymptotically reaches the classical characteristic age form $P/2\dot{P}$ when $\tau_D$ goes to infinity. The evolved pulsar population is shown, in blue, in the $\mathbf{P}$-$\dot{\mathbf{P}}$ diagram of figure 5.2.

### 5.1.2 Birth distribution in the Galactic plane

To follow the dynamical evolution of the pulsars in the Galactic reference frame, we synthesised their birth position $x$, $y$, $z$ in the Galaxy as well as kick velocity and direction.

One of the most debated and open question in the pulsar population study concerns their distribution within our Galaxy. The vast majority of $\sim 2000$ pulsars known so far have been first observed in radio. Since a pulsar is a weak radio emitter, the whole known population is located within a few thousand parsecs from our Solar System. This bias makes it difficult to extrapolate a global statistic that, on the Galactic scale, could indicate specific regions for neutron star formation.

We emulated the distributions of the O & B star NS's progenitors (section 1.2.1) through the location of the HII regions. The latter are good tracers of massive stars because O-B stars are required to ionise the hydrogen bubbles. For the number density of pulsars at birth as a function of Galactocentric distance, we used the HII region profile recently obtained by Bania et al. 2010 from radio observations that can probe HII regions to large distance with little absorption. Figure 5.3 shows the comparison between the birth distribution used in earlier publications (Paczyński, 1990), in red, and the HII region profile used in this thesis, in blue. Both the distributions extends from the galactic centre up to 40 kpc and have been normalised to have the surface density $\sum P(R) \times R = 1$.

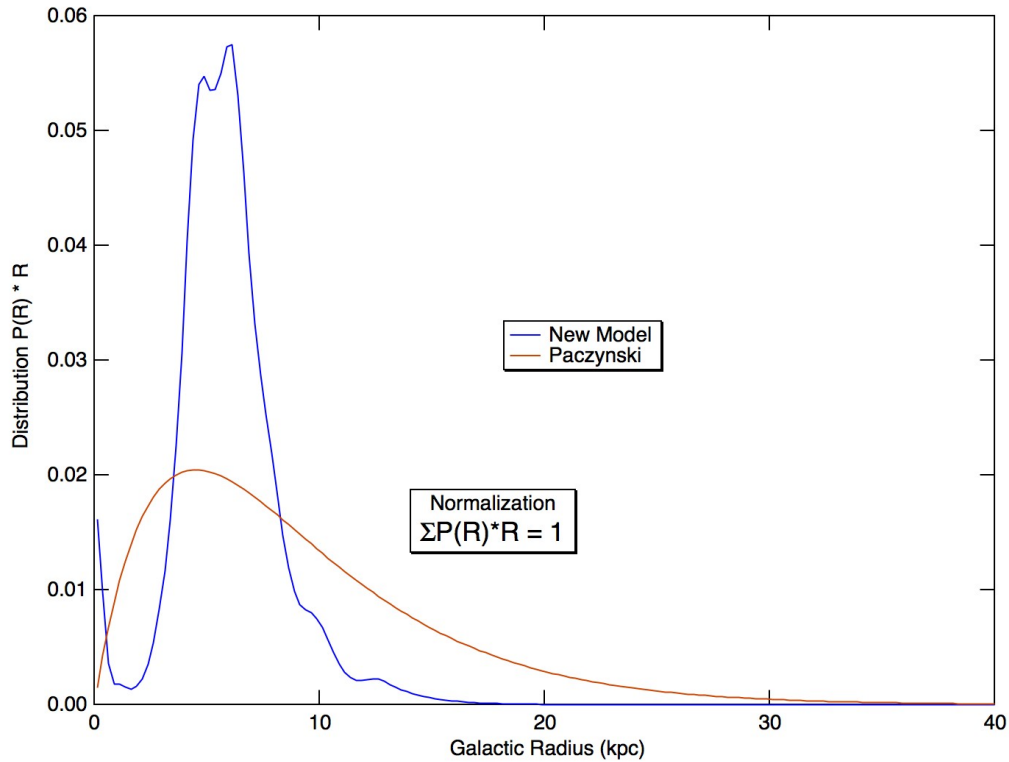

Figure 5.3: Surface density of the new-born neutron stars. In red is plotted the Paczynski distribution (Paczyński, 1990) and in blue the adopted one following the distribution of radio HII regions. Both curves are normalized to an integral of 1 in the Galaxy.

With respect to the Galactic latitude distribution, we assume that all the NSs are born in the Galactic plane and move away because of the large supernova kick velocity, an intrinsic space velocity conferred to the NS in the explosion. The kick velocity assumed in this thesis is shown in figure 5.4. It is defined by a maxwellian distribution, characterised by a mean of 400 km s$^{-1}$ and a width of 256 km s$^{-1}$ (Hobbs et al., 2005). One can follow each pulsar position in the Galactic gravitational potential to the present time. The Galactic gravitational potential adopted in this thesis is described by the negative of the potential funcion as described in Paczyński (1990),

$$\Phi_i(R, z) = \frac{-GM_i}{\{R^2 + [a_i + (z^2 + b_i^2)^{1/2}]^2\}^{1/2}}, \ (5.8)$$

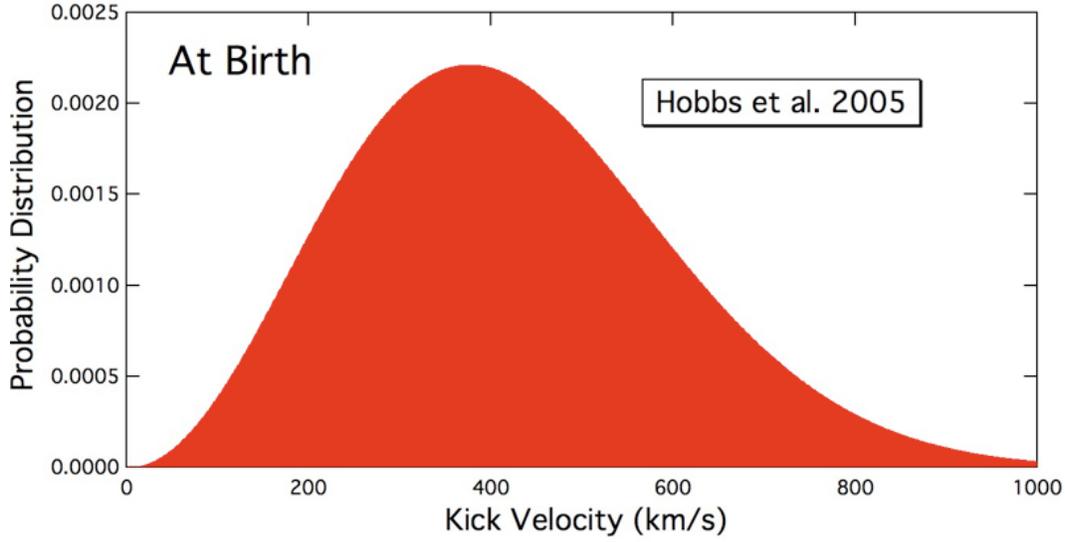

Figure 5.4: Maxwellian probability distribution of the supernova kick velocity assumed for the simulated pulsar sample (Hobbs et al., 2005).

$$\Phi_h(r) = \frac{-GM_c}{r_c} \left[ 1 + \frac{1}{2}\ln\left(1 + \frac{R_h^2}{r_c^2}\right) - \frac{1}{2}\ln\left(1 + \frac{r^2}{r_c^2}\right) - \frac{r_c}{r}\tan^{-1}\left(\frac{r}{r_c}\right) \right], (5.9)$$

where $\Phi_1$, $\Phi_2$, and $\Phi_h$ respectively describe the potential component of spheroid, disk, and halo, $G$ is the gravitational constant and $M$ the mass of the considered component. In equation 5.8, $R$ is the distance from the Galactic centre, $z$ is the altitude on the galactic plane, and $a$ and $b$ scale constants. In equation 5.9, $r$ is the distance from the halo centre, $R_h$ is the halo radius (typical value of 41 kpc, Binney & Tremaine (1987)), and $r_c$ a scale constant. As described in Paczyński (1990), the parameters used to evaluate the Gravitational Potential are

$$a_1 = 0, \quad b_1 = 0.227 \ kpc, \quad M_1 = 1.12 \times 10^{10} \ M_\odot \qquad (5.10)$$

$$a_2 = 3.7 \ kpc, \quad b_2 = 0.20 \ kpc, \quad M_2 = 8.07 \times 10^{10} \ M_\odot \qquad (5.11)$$

$$r_c = 6.0 \ kpc, \quad M_c = 5.0 \times 10^{10} M_\odot. \qquad (5.12)$$

After the time evolution of the spin and dynamical parameters, important characteristics of the evolved population have been derived. In figure 5.5 are illustrated the distributions of some characteristics of the evolved population and their comparison with the observed population. Exception made for the distribution of radio fluxes at 400 MHz, the simulated distributions well describe the observed sample. The inconsistency in radio flux at 400 MHz is probably due to the high uncertainty on pulsar distances. This is suggested from the comparison between the $Sd^2$ parameter, that well describes the observations, and the evaluated flux $S$, that is not consistent with the data. As we will see in section 5.5, the simulated radio flux is used to assess the visibility of pulsar. Since the majority of the pulsars analysed in this thesis have radio

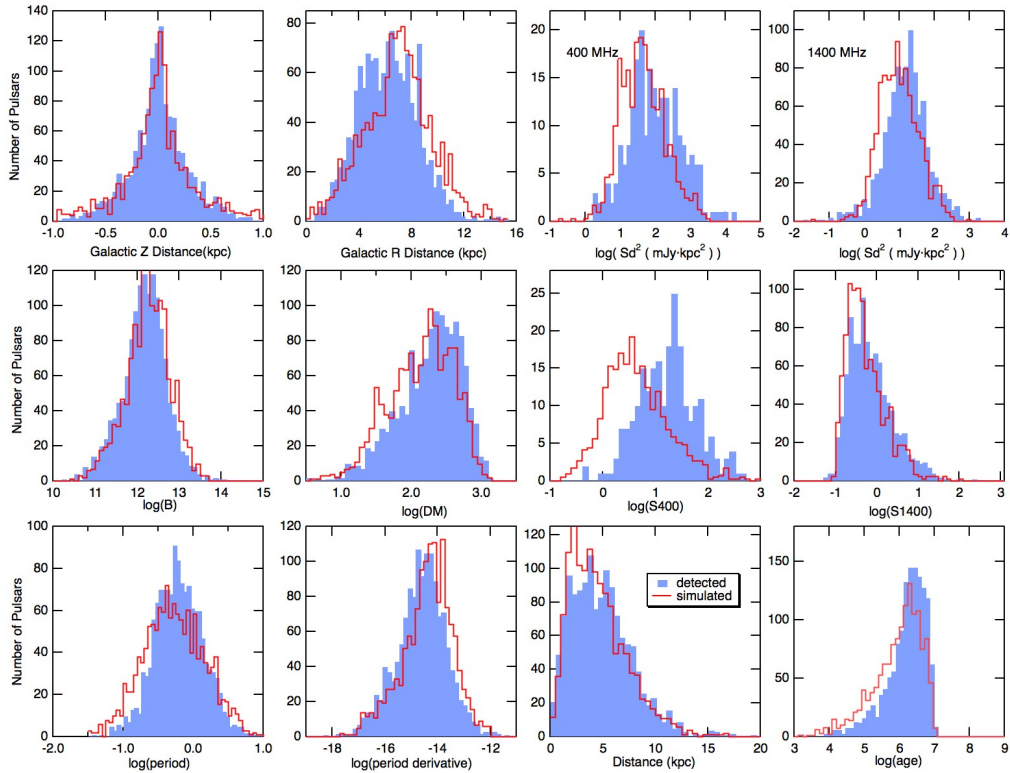

Figure 5.5: Comparison between simulated and observed pulsar characteristics. From the top left to the bottom right are respectively plotted the distribution for: height on the Galactic plane, Galactic distance, brightness $Sd^2$ at 400 & 1400 MHz, magnetic field, dispersion measure, radio flux $S$ at 400 MHz and 1400 MHz, spin period, spin period first time derivative, distance, and age. In blue the ATNF sample (http://www.atnf.csiro.au/research/pulsar/psrcat/), in red the simulated population

flux estimates at 1400 MHz, this inconsistency does not considerably affect the results.

### 5.1.3 Other simulated pulsar characteristics

Following the evolved pulsar positions in the Galactic frame, values of the radio dispersion measure (DM), the radio rotation measure (RM), and the sky temperatures at 408 MHz ($T_{sky,408}$) have been derived using the NE2001 model from (Cordes & Lazio, 2001).

A value of the magnetic obliquity $\alpha$ (angle between the pulsar rotation and magnetic axes) and of the observer line of sight $\zeta$ (angle between the pulsar rotation axis and the observer line of sight) has been randomly assigned to each pulsar of the sample. Since our model does not contain favour pulsar orientation, we have used a random flat distribution both in $\alpha$ and $\zeta$ angles.

To be able to assess the γ-ray visibility of a pulsar (in photon flux) we need a spectral prescription to convert the modelled energy fluxes into photon fluxes. A value of the high energy spectral index ($\Gamma$) and high-energy cutoff ($E_{cut}$) has been generated for each pulsar of our sample. The description of the spectral parameters will be given in section 5.4.1.

## 5.2 Phase-plot calculation and normalisation

### 5.2.1 Definition

The γ-ray emission models discussed in chapter 2 describe the particle luminosity. To provide the γ-ray emission pattern for each emission mechanism, we used the geometric emission model from Dyks et al. 2004, based on the following assumptions: *(i) the pulsar magnetic field is dipolar and swept up by the pulsar rotation (retarded potentials) , (ii) the γ-ray emission is tangent to the magnetic field line and oriented in the direction of the accelerated electron velocity in the star frame.* Relativistic aberration and time of flight delays are taken into account.

In the computation of the emission pattern, the first step consists in localising the position of the magnetic field line from which the radiation is produced. Each field line is then divided into segments and for each segment the tangent direction and height with respect to the NS surface is evaluated. Since the emission gap is located, for each model, in a different magnetospheric region, the emission patterns are obtained by selecting the *segments* corresponding to the gap position in each model. By defining a co-rotating reference frame $CRF$ and an observer's one $ORF$ it is possible to derive the structure of the emission pattern with respect to the pulsar orientation. The observer line of sight angle $\zeta$ will be defined by the choice of $ORF$, and the phase $\phi$ of the pulsar emission is defined by the direction of the emitted photons with respect to $CRF$. The results of this computation will be a two dimensional emission pattern in the plane ($\phi,\zeta$) that is called **phase-plot**.

The light curve of a pulsar characterised by magnetic obliquity and line of sight ($\alpha_{psr}, \zeta_{psr}$) is obtained by cutting horizontally the phase-plot panel evaluated for $\alpha_{psr}$ in correspondence of the line of sight $\zeta_{psr}$.

### 5.2.2 Calculation

Concerning the phase-plots generated in this thesis work, the ($\phi,\zeta$) space has been divided in $180 \times 180$ bins so that each bin contains the number of photons $dN_\gamma/d\Omega$ per solid angle $d\Omega$.

Each $(\phi, \zeta)$ phase-plot will be obtained for a specific set of pulsar parameters that define its magnetospheric structure: the spin period **P**, the surface magnetic field **B**, and the magnetic obliquity $\alpha$. For the model studied in this thesis, the phase-plot $(EM_{pp})$ has the following dependencies:

$$EM_{pp,PC/Radio} = f(P, B, \alpha)$$

$$EM_{pp,SG} = f(\Delta\xi, \alpha) \ \ and \ \ \Delta\xi = f(P, B)$$

$$EM_{pp,OG/OPC} = f(w, \alpha) \ \ and \ \ w = f(P, B).$$

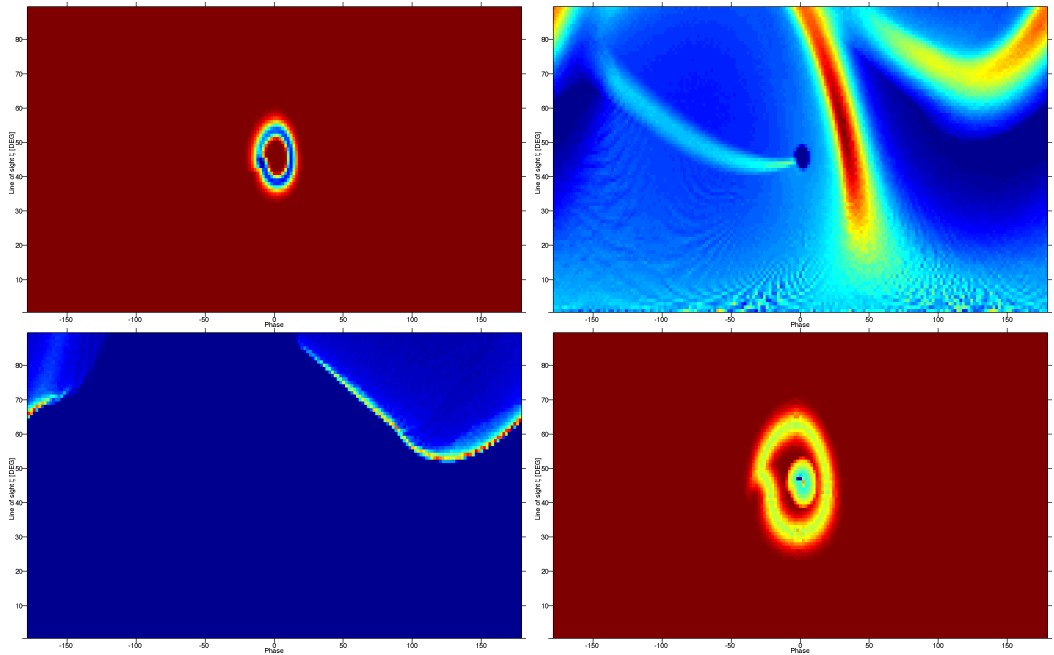

Figure 5.6: From the top left to the bottom right panel are respectively indicated the $\gamma$-ray phase-plot emission pattern for: PC model, obtained for $B = 10^{12}$ G, $P$=75 ms, and $\alpha = 45°$; SG model, obtained for $\Delta\xi = 0.25$ and $\alpha = 45°$; OG/OPC model, obtained for $w = 0.3$ and $\alpha = 45°$; Radio core plus cone model, obtained for $B = 10^{12}$ G, $P$=75 ms, and $\alpha = 45°$. It has to be noted that the radio phase plot shown is just an example of the emission pattern. The central core and the external cone components intensity are not scaled.

For each emission model, I have evaluated phase-plots for 18 $\alpha$ values, from 5° to 90°, with a step of 5°. In the PC and radio cases, for each $\alpha$ value, the phase-plots have been evaluated for 2 magnetic field values and 9 spin period values, for a total of 324 phase-plot panels per model. In the SG and OG/OPC cases, for each $\alpha$ value, the phase-plots have been evaluated for 16 gap width values, for a total of 288 phase-plot panels per model. The complete set of the parameters sample is listed in Table 5.1. An example phase-plot is shown, for each model, in Figure 5.6.

| | **B**<br>Gauss | **P**<br>milliseconds | **α**<br>Degrees | Gap Width values |
|---|---|---|---|---|
| PC/Radio | $10^{12}, 10^{13}$ | 30, 40, 50, 75, 100<br>300, 500, 750, 1000 | 5-90<br>$5°\,step$ | PC: 0.04, 0.06, 0.08, 0.1, 0.13, 0.16, 0.2, 0.225<br>0.25, 0.275, 0.3, 0.34, 0.38, 0.42, 0.46, 0.50 |
| SG | none | none | 5-90<br>$5°\,step$ | 0.04, 0.06, 0.08, 0.1, 0.13, 0.16, 0.2, 0.225<br>0.25, 0.275, 0.3, 0.34, 0.38, 0.42, 0.46, 0.50 |
| OG/OPC | none | none | 5-90<br>$5°\,step$ | 0.01, 0.025, 0.04, 0.05, 0.067, 0.084, 0.1, 0.2<br>0.3, 0.4, 0.5, 0.53, 0.56, 0.59, 0.62, 0.65 |

Table 5.1: Magnetic field, period, and gap width values for which the phase-plots have been evaluated for each emission model. The SG and OG/OPC emission patterns do not depends directly from the pulsar period and magnetic field.

**Emission: location and profile in the gap**

In the phase-plot computation it is important to define the start and stop altitude of the emission and the emission profile assumed across the gap. Since the emission gap has a finite width (the size of which is well defined in chapter 2), to be able to reproduce the gap emission pattern it is fundamental to assume an emission profile across the width of the gap.

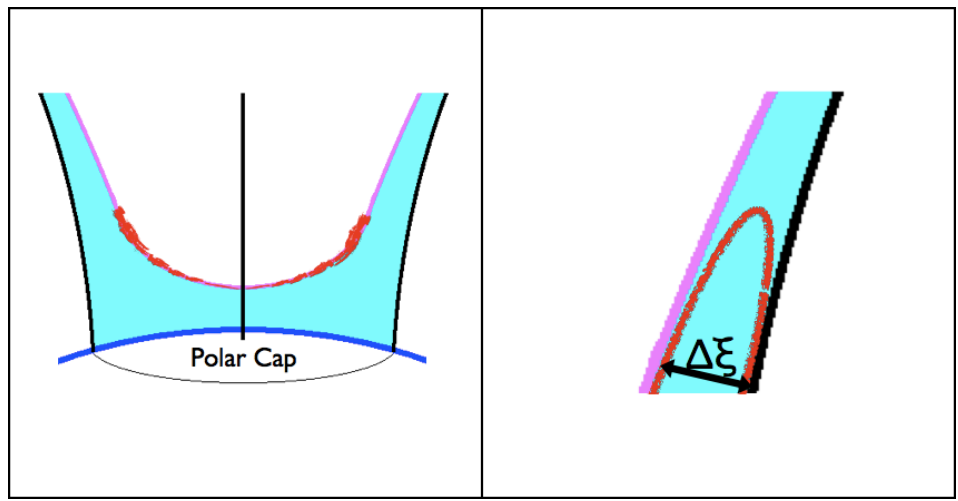

Figure 5.7: *left:* polar cap emission profile inside the gap. *Right:* slot gap emission profile across the gap width $\Delta\xi$ In both the panels, the emission profile is drawn, in red, on the top of the gap.

Within the PC and SG models, the primary emission is mainly generated from curvature radiation. Since the magnetic field line curvature decreases toward the magnetic pole, the emission decline toward the pole toward the pole. In the PC model, the emission profile in colatitude is infinitely thin along the inner edge of the slot gap (where the SG width is defined the same as the high altitude SG) and the emission along the field line is modulated by exponentials. In the SG case, the emission distribution across the gap peaks in the centre of the gap and decreases toward the gap edges. The profile follows the predicted current distribution across the gap. The shape for the PC and SG emission distributions across the gap is schematically shown in figure 5.7.

For the OG/OPC emission model we describe the emitting region, as an infinitely thin layer along the inner surface of the gap.

### 5.2.3 Phase-plot normalisation and connection with the $\gamma$-ray emission gap models

The phase-plot normalisation is a procedure to connect the $\gamma$-ray emission models with the geometrical model by Dyks et al. 2004 in order to transform the modelled luminosity into a pulsar flux.

Each bin $n(\phi, \zeta)$ of the phase-plot gives the number of photons per solid angle per primary particle that can be observed in the $\zeta$ direction at the rotational phase $\phi$:

$$n(\phi, \zeta) = \frac{dN_\gamma}{\sin \zeta d\zeta d\phi} \tag{5.13}$$

Assuming a value for the primary particle production rate $\dot{N}_e$, the energy $E_\gamma$ of each photon, one gets a radiation luminosity per phase-plot bin:

$$dL_\gamma = \dot{N}_e E_\gamma n(\phi, \zeta) \sin \zeta d\zeta d\phi = An(\phi, \zeta) \sin \zeta d\zeta d\phi \tag{5.14}$$

where $A$ is a proportionality constant. Each model gives a total particle luminosity per pole $L_{pole}$. For a radiative efficiency $\epsilon_\gamma$ of the particles one can normalise the phase-plot to the total radiation luminosity over the two poles according to:

$$2\epsilon_\gamma L_{pole} = A \int_0^\pi \sin \zeta d\zeta \int_0^{2\pi} n(\phi, \zeta) d\phi \tag{5.15}$$

We define the specific intensity $I$ as

$$I = \frac{dL_\gamma}{d\Omega} \quad \rightarrow \quad I(\phi, \zeta) = \frac{An(\phi, \zeta) \sin \zeta d\zeta d\phi}{\sin \zeta d\zeta d\phi} = An(\phi, \zeta). \tag{5.16}$$

It is now possible to get the average energy flux observed by an Earth observer for a line of sight $\zeta_{obs}$:

$$\langle \nu F_\nu \rangle = \frac{\int_0^{2\pi} I(\zeta_{obs}, \phi) d\phi}{2\pi D^2}. \tag{5.17}$$

Here, $D$ is the pulsar distance.

From the equations 5.17 and 5.16, we can write the equation for the average energy flux observed at the Earth as:

$$\langle \nu F_\nu \rangle = \frac{\epsilon_\gamma L_{pole}}{\pi D^2} \frac{\int_0^{2\pi} n(\zeta_{obs}, \phi) d\phi}{\int_0^\pi \sin \zeta d\zeta \int_0^{2\pi} n(\phi, \zeta) d\phi} \tag{5.18}$$

This last equation defines the connection between the theoretical emission models described in chapter 2, and the geometrical emission model by Dyks et al. (2004). It establishes the relation between the particle luminosity derived in the framework of a given model, $L_{pole}$, and the integral of the pulsar light curve $\int_0^{2\pi} n(\zeta_{obs}, \phi) d\phi$, obtained, from the phase-plot, for $\zeta = \zeta_{obs}$. It assumes a simple radiative efficiency for all the particles accelerated in the gap.

## 5.3 Phase-plots interpolation

In this section I will describe the procedure used to assign a light curve profile to each pulsar. Apart from the light curve integral used to obtain the average energy flux observed at Earth (equation 5.18), the complete light curves will be extensively used in the second and third part of my thesis project to fit the observed pulsar profiles and to study the modulation characteristics of each
emission model.

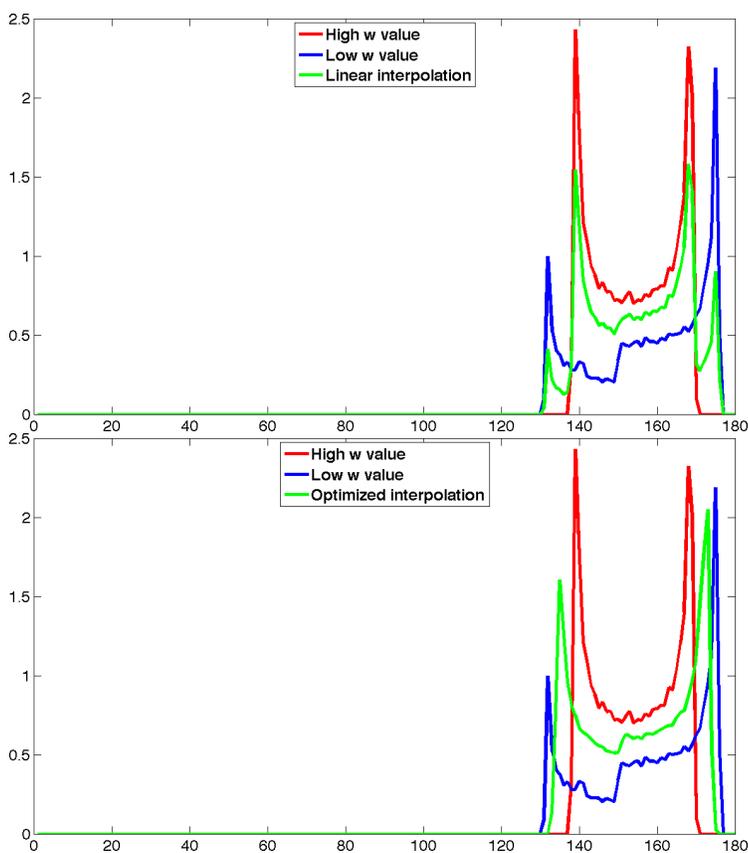

Figure 5.8: Top panel: linear interpolation between two OG light curves, obtained from two contiguous gap width values. The interpolated curve has a completely different structure compared to the parent ones and twice as many peaks. Bottom panel: Non linear interpolation of the same curves. The interpolated curve has retained the original structure of the parent curves. In both panels, the blue and red curves are the parent ones and the green is the result of the interpolation.

Since the phase-plots have been evaluated for given sets of **P**, **B**, and $\alpha$,
or Gap width and $\alpha$ (see table 5.1), to compute each pulsar light curve I had
to interpolate phase-plots and preserve the structures of the parent profiles in
the interpolated curve. The curve characteristics I tried to preserve were the
peak number and the size of the emission region. A first interpolation trial
was made by using a classical linear interpolation between the parent curves.
The result, shown in the upper panel of figure 5.8 indicates how this kind of
interpolation fails. In fact, if one considers two phase-plot panels obtained
for two contiguous values of a specific parameter, one realizes that the most
conspicuous change in the light curve structure is the phase extension of the
emission. Even though the structure of the curve is kept stable in the transition
between the two phase-plots, the phase extension is always different. Applying
a linear interpolation to these curve will yield an interpolated profile with a
distorted structure and a number of peaks doubled with respect to the parent
curves (upper panel, figure 5.8).

Thus, we adopted a non linear interpolation which expands the curve
which covers the smallest phase range up to the extent of the most extended
one, then applies a linear interpolation, and contracts the resulting curve
down to the interpolated phase size between the parent profiles. Looking at
figure 5.8 it is evident that the new interpolation method solves the phase-plot
interpolation problem. It is possible to retain the light curve structure, to
preserve the peak number, to follow the evolution of the peak separation, and
to preserve the smallest structures present in both the parent curves.

## 5.4 Flux calculations

In the next sections I will describe the spectral and NS structure assumptions
made to evaluate the flux of the simulated pulsars.

### 5.4.1 Spectral parameters

To be able to evaluate the photon flux of each pulsar of the sample, each pulsar
should be characterised by a set of spectral parameters, like *spectral index* $\Gamma$,
and *energy cutoff value* $E_{cut}$. The LAT pulsar spectra are well fitted by a
power-law with an exponential cutoff, like

$$\frac{dN}{dE} = k \left( \frac{E}{E_0} \right)^{-\alpha} e^{-E/E_c}. \tag{5.19}$$

Both the spectral index and the energy cutoff distributions have been generated
starting from two gaussian distributions. Their parameters have been chosen to obtain a spectral index and high energy cutoff distributions as close as possible to the observed LAT pulsar population ones. By defining the first and second gaussian parameters as

$$Gaussian\ 1,\ X: \quad mean = 1.97; \quad variance = 0.18$$

$$(5.20)$$

$$Gaussian\ 2,\ Y: \quad mean = 3.06; \quad variance = 0.37$$

The spectral index $\Gamma$ and the $\log_{10}(E_{cut})$ are defined like:

$$\theta = 0.5982 \quad [rad]$$
$$\Gamma = X \cos \theta - Y \sin \theta$$
$$\log_{10}(E_{cut}) = X \sin \theta + Y \cos \theta.$$

$$(5.21)$$

In figure 5.9 are indicated the distributions for the spectral index and energy cutoff generated by applying the parameters 5.20 to the equations 5.21. The gaussian widths, centroids, and correlation had been derived from the analysis of the spectral parameters measured for the 1st LAT pulsar catalogue. We took here the very same values.

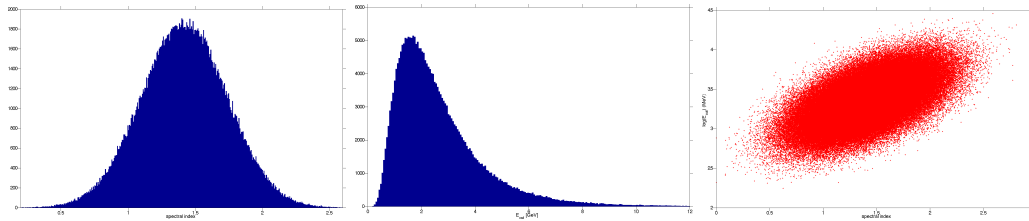

Figure 5.9: From the left to the right panel: spectral index distribution, cutoff energy distribution, and correlation between spectral index and cutoff energy for the simulated pulsar population as drawn from the observed characteristics of the LAT sample.

Once $\Gamma$ and $E_{cut}$ values have been assigned to each simulated pulsar, it is possible to convert the energy flux to a photon flux. Let us define the ratio between the photon flux and the energy flux, above 100 MeV, as

$$R_{flux} = \frac{Ph_{flux,100}}{En_{flux,100}}.$$

The photon flux is obtained as

$$F_{\gamma} = \int_{E_1}^{\infty} \frac{dN}{dE} dE = k \left( \frac{E}{E_0} \right)^{-\alpha} \int_{E_1}^{\infty} \left( \frac{E}{E_c} \right)^{-\alpha} e^{-E/E_c} dE$$

$$(5.22)$$

that by writing $x = E/E_c$ and so $dx = dE/E_c$, can be written as

$$F_{\gamma} = k \left( \frac{E}{E_0} \right)^{-\alpha} \int_{x_1}^{\infty} x^{-\alpha} e^{-x} dx.$$

$$(5.23)$$

Let us remember the *gamma function* and the *incomplete gamma function*, respectively as

$$\Gamma(a) = \int_0^\infty e^{-t} t^{a-1} dt \qquad (5.24)$$

$$\Gamma_i(x,a) = \frac{1}{\Gamma(a)} \int_0^x e^{-t} t^{a-1} dt. \qquad (5.25)$$

that imply

$$\lim_{x \to 0} \Gamma_i(x,a) = x^a \qquad (5.26)$$

$$\lim_{x \to \infty} \Gamma_i(x,a) = 1. \qquad (5.27)$$

The pulsar photon flux can then be written as

$$F_\gamma = k E_c \left(\frac{E}{E_c}\right)^{-\alpha} \Gamma(1-\alpha) \left[\Gamma_i(\infty, 1-\alpha) - \Gamma_i(x_1, 1-\alpha)\right] \qquad (5.28)$$

that by using equation 5.26, for $E \geq E_1$, becomes

$$F_\gamma = k E_c \left(\frac{E}{E_c}\right)^{-\alpha} \Gamma(1-\alpha) \left[1 - \Gamma_i\left(\frac{E_1}{E_c}, 1-\alpha\right)\right]. \qquad (5.29)$$

By applying the same equation and definition, the energy flux for $E \geq E_1$, results as

$$\nu F_\nu = k E_c^2 \left(\frac{E_c}{E_0}\right)^{-\alpha} \Gamma(2-\alpha) \left[1 - \Gamma_i\left(\frac{E_1}{E_c}, 2-\alpha\right)\right]. \qquad (5.30)$$

From equations 5.29 & 5.30, the flux ratio $R_{flux}$ is defined by

$$R_{flux} = \frac{F_\gamma}{\nu F_\nu} = \frac{1}{1.602 \times 10^{-6} E_c} \frac{\Gamma(1-\alpha)}{\Gamma(2-\alpha)} \frac{\left[1 - \Gamma_i\left(\frac{E_1}{E_c}, 1-\alpha\right)\right]}{\left[1 - \Gamma_i\left(\frac{E_1}{E_c}, 2-\alpha\right)\right]}. \qquad (5.31)$$

### 5.4.2 The gap width calculations: the slot gap

Because of the direct dependence of each model luminosity on the gap width (PC: equation 2.5; SG: equation 2.17; OG: equation 2.25; OPC: equation 2.18), particular attention has to be paid to the evaluation of the size of the emission region for each pulsar. In the OG, and OPC models the gap width computation is quite direct, since it does not need we to formulate any pulsar structure assumption to be evaluated. For these models, the gap width has been directly evaluated by using Equation 2.27 for the OG and 2.19 for the OPC models.

Concerning the SG the parameter $\lambda$ is defined in section 2.1.1 as altitude gradient at which the PC pair formation front curves up and which marks the inner edge of the SG. This important parameter constrains both the energetics and emission pattern of the SG emission. To choose a large $\lambda_{SG}$ value implies thin gaps with narrow peaks as observed in most of the LAT light curves. However assuming too large a $\lambda_{SG}$ value implies luminosities too low compared to the LAT fluxes. We found a compromise between the narrow light curve structures and the $\gamma$ luminosity through a reasonable radiation efficiency $\epsilon_\gamma$. We tried two different approaches to constrain $\lambda_{SG}$: a first one based on energetic arguments and a second one, based on the optimisation of the expected profile for some of the LAT observed pulsars.

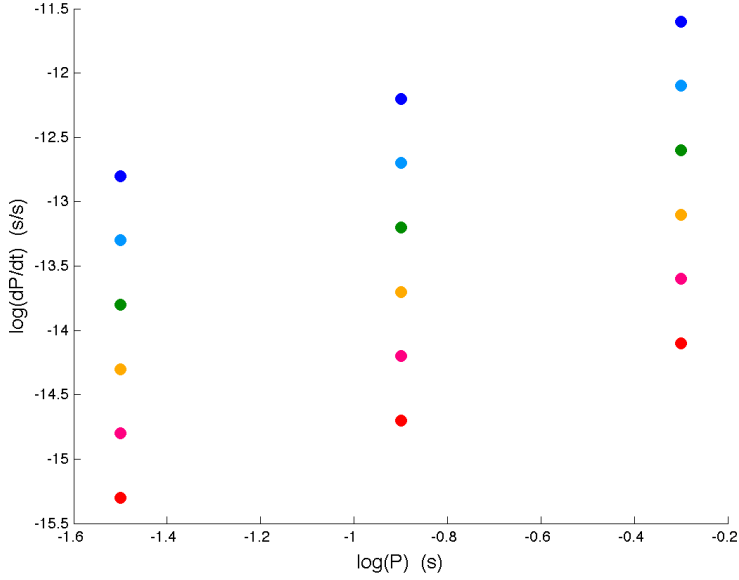

Figure 5.10: $P-\dot{P}$ diagram for 18 pulsars with period and period first time derivative values that span the LAT pulsars population ones. The points with the same colour indicate objects with the same age.

In figure 5.10 is shown a $P-\dot{P}$ diagram of 18 neutron stars with $P$ and $\dot{P}$ values that span the parameters of the LAT pulsars with age increasing from $10^{3.5}$ (blue) to $10^6$ (red). In figures 5.11 & 5.12 are shown the slot gap width and the luminosity as a function of $\dot{E}$ for 4 different values of $\lambda$: 0.10, 0.35, 0.5, and 1.0. Since $L_\gamma$ scales as $\Delta\xi^3 \times \dot{E}$ and since we want the luminosity to be close to $L_\gamma \propto \dot{E}^{1/2}$ we need to get $\Delta\xi \propto \dot{E}^{-1/6}$ to get a reasonable agreement with the LAT data. The luminosity remains close to $\dot{E}^{0.5}$ for all the tested lambda values, but favours $\lambda < 0.4$ to explain the bright LAT pulsars. It is possible to see that we loose a lot of power when we increase lambda (figure 5.12) because the gap thins out. On the other hand, the choice of a too small lambda ($\lambda = 0.1$) will generate a too large gap that will not be able to reproduce the LAT sharp profiles. A good compromise is found for $\lambda = 0.35$ (figure 5.11). One can calculate the pair formation front shape for the P and B values of some of the best known pulsars, Crab, Vela, CTA1, and Geminga, and to obtain an approximate $\Delta\xi$ value. The results yield $\Delta\xi_{Crab}$=0.03, $\Delta\xi_{Vela}$=0.1, $\Delta\xi_{CTA1}$=0.16, $\Delta\xi_{Geminga}$=0.3 for $\lambda_{SG}$ values between 0.02-0.6.

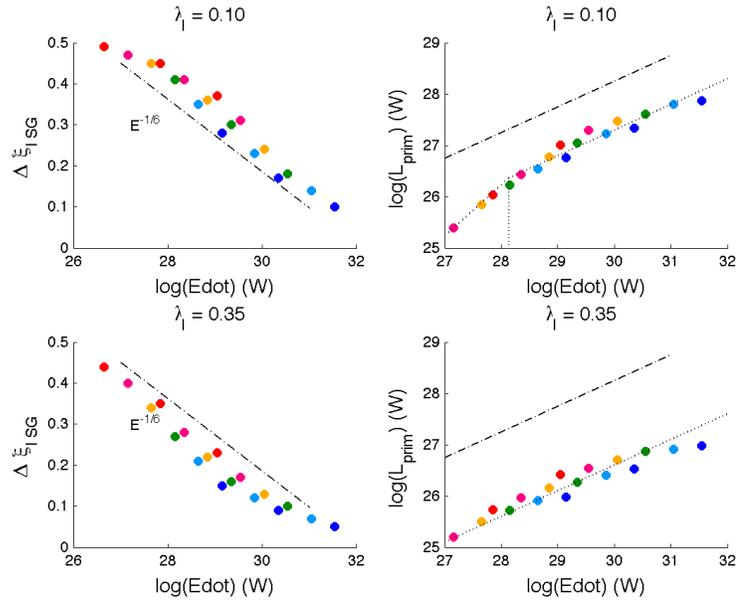

Figure 5.11: Slot gap with $\Delta\xi$ (left) and primary particle luminosity (right) as a function of $\dot{E}$ for $\lambda = 0.1$ and $0.35$.

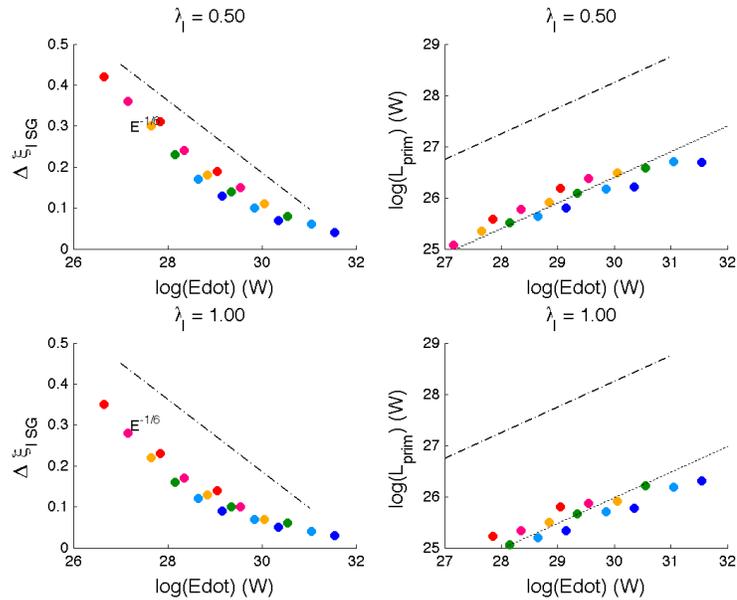

Figure 5.12: Slot gap with $\Delta\xi$ (left) and primary particle luminosity (right) as a function of $\dot{E}$ for $\lambda = 0.5$ and $1.0$.

In order to investigate how the pulsar profile changes as a function of $\lambda$, I performed a fit of some LAT light curves with the SG phase-plots, evaluated for a set of $\Delta\xi$ values obtained for different $\lambda_{SG}$ values (equation 2.14). Figure 5.13 shows the behaviour of the best fit likelihood value as a function of the $\lambda_{SG}$ parameter, for Vela, Crab, J1028-5820, J1048-5832, J2021+3651, and J2229+6114. When $\lambda$ increases the peak sharpness increases too. So, we expect that for the very sharp LAT pulsars, like the Crab, the Vela, J1028-5820, J1048-5832, and J2021+3651 the likelihood increase continuously with increasing $\lambda$ (the simulated curve sharpness get closer to the real one) while for a broader profile like the J2229+6114 one, the likelihood should decrease with increasing $\lambda$ (for low $\lambda$ the simulated curve is as broad as the real one).This is observed in figure 5.13. In the $\lambda < 0.4$ range that allow bright enough pulsars, the maximum likelihood value presents a local maximum corresponding between 0.2 and 0.4. By increasing $\lambda$, the monotone expected trend is broken in correspondence of a $\lambda$ value that better describes the real pulsar profile. This result is in good agreement with the $\lambda_{SG}$ estimate obtained from the luminosity study and the pair formation front evaluation from Crab, Vela, CTA1, and Geminga.

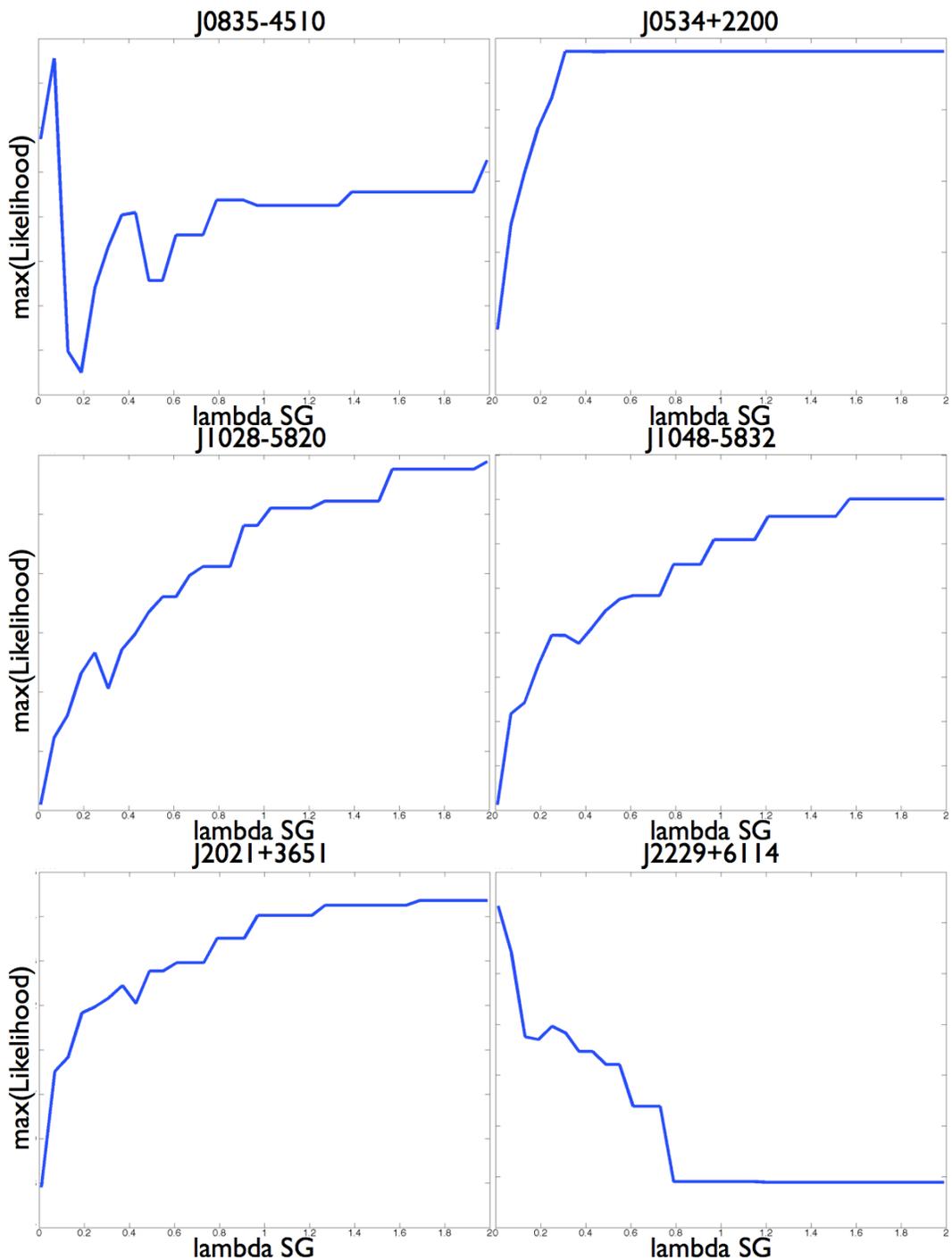

Figure 5.13: Best-fit likelihood value versus $\lambda_{SG}$ for some LAT pulsars. The fit has been done, for each pulsar, comparing the real pulsar profile and all the phase-plot light curves generated for different $\lambda_{SG}$ values chosen in the interval 0-2.

For the population study we therefore set $\lambda_{SG}$=0.35 as a good compromise to reproduce the bulk of the observed light curves and to yield reasonable SG luminosities.

### 5.4.3   Computation

By using the non linear interpolation method described in section 5.3, I assigned a $\gamma$-ray and radio light curve to each pulsar of the sample, storing the value of the curve integral $\int_0^{2\pi} n(\zeta_{obs}, \phi)d\phi$ for the flux computation. The width of the emission gaps is computed using equation 2.14 for the SG, and equations 2.27 and 2.19 for the OG & OPC, while each pulsar luminosity has been obtained through equations 2.5, 2.17, 2.25, and 2.18, respectively for the PC, SG, OG and OPC models. A different gap width upper limit has been set for the PC & SG models, and for the OG & OPC models. Because the PC and SG models do not apply when the gap becomes too large (pair-starved gaps should then be used), gap widths larger than 0.5 have been set to 0. Because no emission remains visible from the thin inner edge of OG/OPC gaps when the gap width exceeds 0.7 we have set larger gap widths to 0. So all the pulsars with a gap width above these threshold levels cannot produce any $\gamma$-ray and have been *"killed"*.

For the radio luminosity computation, I have used Equation 2.41 to evaluate the total radio luminosity and the equations 2.39 to evaluate the luminosities of each core and cone component.

With each pulsar luminosity and light curve integral, the energy flux and photon flux computation have been done using Equation 5.18 and Equation 5.31.

## 5.5 Radio pulsar visibility

The synthesis of the population used in this thesis is not based on any NS birth rate assumption. The criterion we followed in the choice of the simulated pulsar number was to generate a large enough population to reduce the Poisson Montecarlo fluctuations and to have good statistics in the analysis results. The simulated sample had to be scaled to the real number of pulsars detected in the Galaxy. The scaling factor has been evaluated by selecting all the ATNF radio pulsars present in several surveys and comparing this number with simulated pulsars visible in the same regions.

We used 10 radio surveys: the ratio between the number of simulated pulsars meeting any of the surveys visibility criteria and the total number of objects detected by the 10 surveys defines the scaling factor:

$$S_f = \frac{N_{obs,10\ surv}}{N_{sim,10\ surv}} \tag{5.32}$$

where $S_f$ is the scaling factor, $N_{obs,10\ surv}$ is the number of objects detected in the 10 selected surveys, and $N_{sim,10\ surv}$ is the number of simulated pulsar that meet the visibility criteria of at least one of the 10 surveys. In the next two sections I will describe the characteristics of the selected surveys and the visibility criteria optimisation we used to select the simulated and observed radio pulsars.

### 5.5.1 Radio surveys

We considered the 10 radio surveys from the ATNF database[1] that are well parametrized and that cover the largest possible sky surface while minimising the overlapping regions. These surveys are: Molonglo2 (Manchester et al., 1978), Green Bank 2 & 3 (Dewey et al., 1985; Stokes et al., 1985), Parkes 2 (70 cm) (Lyne et al., 1998), Arecibo 2 & 3 (Stokes et al., 1986; Nice et al., 1995), Parkes 1 (Johnston et al., 1992), Jodrell Bank 2 , Parkes Multi-beam (Manchester et al., 2001) and the extended Swinburne surveys (Edwards et al., 2001; Jacoby et al., 2009).

### 5.5.2 Radio pulsar selection

Prior to apply a selection to the simulated pulsars that fall inside the visibility criteria of the selected survey, it was necessary to re-define some of the survey parameters.

During a radio survey, the edges of the survey region are defined by the position most distant from the centre of the radio-telescope beam. Nevertheless, because of the solid angle extension and complexity of the beam, it is possible to observe a pulsar out of the declared survey region. Thus, to say that all the pulsars observed during a survey fall inside the declared survey coordinates edges is not totally correct. The first parameter I re-evaluated for each survey is the number of pulsar insides a given region.

---

[1]http://www.atnf.csiro.au/research/pulsar/psrcat/

The second important parameter is the survey efficiency $\epsilon_{surv}$. It is defined as a filling factor, e.g. the ratio between the actual solid angle covered by the radio telescope beam during the observations, and the area within the declared survey boundaries. The survey efficiency can be considered as the probability of observing a pulsar present in the survey region if the parent spatial distribution is uniform. However, a parameter defining the observed percentage of all the observable pulsars present in the survey region is essential to evaluate the number of simulated pulsars within the radio survey visibility criteria. To evaluate the boundaries of the survey region and to define the survey efficiency we decided:

1. to slightly extend the sky region covered by the survey in order to include the largest number of pulsars actually detected by a survey, without changing too much the original boundaries

2. to evaluate the detection flux threshold for each pulsar within a survey by scaling the Dewey formula (Dewey et al., 1985) with a free parameter, $\epsilon_{Dewey}$, to match the observations.

$$S_{min} = \epsilon_{Dewey} \times S_{threshold} \tag{5.33}$$

where the threshold flux $S_{threshold}$ is expressed by the Dewey formula

$$S_{threshold} = \frac{\sigma_{S/N}[T_{rec} + T_{sky}(l,b)]}{G\sqrt{N_p B t}}\sqrt{\frac{W}{P-W}}. \tag{5.34}$$

The Dewey formula, or radiometer formula, is a relation that takes into account the characteristics of a given radio telescope and detector as well as pulsar period and direction to give the minimum flux the survey would be able to detect. In equation 5.34 $\sigma_{S/N}$ is the minimum signal to noise ratio taken into account, $T_{rec}$ is the receiver temperature, $T_{sky}$ is the sky temperature at 408 MHz, $G = Gain/\beta$ is the ratio between the radio telescope gain and the dimensionless factor $\beta$ that accounts for system losses, $N_p$ is the number of measured polarisations, $B$ is the total receiver bandwidth, $t$ is the integration time, and $P$ is the pulsar period. $W$ is the effective pulse broadening, defined as

$$W^2 = W_0^2 + \tau_{samp}^2 + \tau_{DM}^2 + \tau_{scat}^2 + \tau_{trailDM}. \tag{5.35}$$

Here, $W_0$ is the intrinsic pulse width, $t_{samp}$ is a low-pass filter time constant applied before sampling (when this parameter is unknown, a value equal to twice the sampling time has been used), $\tau_{DM}$ is the pulse smearing due to the DM over one frequency interval $\Delta\nu$, and $\tau_{scat}$ is the pulse broadening due to interstellar scattering. The dispersion broadening time, $\tau_{DM}$ (ms), across one frequency channel, $\Delta\nu$, is related to the dispersion measure (DM) as

$$\tau_{DM} = 8.3 \times 10^6 \frac{\Delta\nu DM}{\nu^3}. \tag{5.36}$$

The dispersion measure, DM (pc cm$^{-3}$), is obtained using the Cordes & Lazio NE2001 model. The same model provides the scattering measure, SM (kpc m$^{20/3}$), which allows to estimate the broadening time due to interstellar scattering as

$$\tau_{scat} = 1000 \left(\frac{SM}{292}\right)^{0.5} d \left(\frac{\nu}{1000}\right)^{-4.4} \tag{5.37}$$

The last term of equation 5.35, $\tau_{trailDM}$, is an additional time broadening added when the sampling is performed for a DM value different from the real one. This term becomes important just for low period pulsars.

The sky temperature at frequencies other than 408 MHz is obtained as:

$$T_{sky}(\nu) = T_{sky.408} \left(\frac{408\ MHz}{\nu}\right)^{2.6}. \tag{5.38}$$

Table 5.2 lists all the radio telescope and detector characteristics of the surveys I took into account. Survey parameters that in the literature were listed as average values, have been re-evaluated using the exact dependency (table 5.2, $G$ & $T_{rec}$ for the surveys Arecibo 2 & 3 (Stokes et al., 1986; Nice et al., 1995))

The scaling of the radiometer equation (equation 5.33) was motivated by the intrinsic uncertainties related to the Dewey formula, mainly mainly because of the flux oscillation due to scintillation. The scintillation is a phenomenon generated by the turbulent variations in the interstellar medium that the pulsar light has to cross before reaching the observer. The consequence is an oscillation (scintillation) of the pulsar flux that, during a survey, could introduce spurious detections of pulsars with a flux generally lower than the survey threshold or could cause the non-detection of pulsars with a flux higher than the survey threshold. So we scaled the $S_{threshold}$ level in order to take into account possible spurious detections or missed detections due to scintillation.

$S_{threshold}$ should not be lower than the flux of the weakest pulsar of the survey. A reasonable estimate is to take the average of the low-flux tail of the pulsars of the survey. We scaled the Dewey formula so that the average of the lowest fluxes divided by the $S_{threshold}$ values for the pulsars is 1. The ratios are displayed in Figures 5.14 to 5.18 and the scaling $\epsilon_{Dewey}$ values are given in Table 5.3.

In the ATNF database we can count how many pulsars fall within a survey boundary, how many would match the survey visibility criterion (flux $> F_{threshold}$), and how many of these pulsars have really been observed by the survey. The comparison of the ratios of the radio flux recorded for each pulsar to the minimum $S_{min}$ visible flux in its direction provides an estimate of the Dewey scaling factor. The distribution of the flux ratios is shown in Figures

| | $\frac{Gain}{\beta}$ (K Jy$^{-1}$) | $\sigma_{S/N}$ | $T_{rec}$ (K) | $\nu_{surv}$ (MHz) | $T_{int}$ (s) | $t_{samp}$ (ms) | $\Delta\nu_b$ (MHz) | $\Delta\nu_{ch}$ (MHz) |
|---|---|---|---|---|---|---|---|---|
| Mol2 | 5.100 | 5.4 | 225 | 408 | 40.96 | 40.0 | 4.0 | 4.0 |
| GrB2 | 0.886 | 7.5 | 30.0 | 390 | 136 | 33.5 | 16.0 | 2.0 |
| GrB3 | 0.950 | 8.0 | 30.0 | 390 | 132 | 2.0 | 8.0 | 0.25 |
| Pks2 | 0.430 | 8.0 | 50.0 | 436 | 157.3 | 0.6 | 32.0 | 0.125 |
| Are2 | 10.91[1] | 8.0 | 100[2] | 430 | 39.3 | 0.4 | 0.96 | 0.06 |
| Are3 | 13.35[3] | 8.5 | 70.0[4] | 430 | 67.7 | 0.5 | 10.0 | 0.078 |
| Pks1 | 0.256 | 8.0 | 45.0 | 1520 | 157.3 | 2.4 | 320.0 | 5.0 |
| JBk2 | 0.400 | 6.0 | 40.0 | 1400 | 524.0 | 4.0 | 40.0 | 5.0 |
| PMBM | 0.460 | 8.0 | 21.0 | 1374 | 2100.0 | 0.25 | 288.0 | 3.0 |
| Swnb | 0.427 | 10.0 | 21.0 | 1374 | 265.0 | 0.25 | 288.0 | 3.0 |

[1] Computed using $\frac{Gain}{\beta} = \frac{19-(0.42\times|19-dec|)}{1.1375}$.

[2] Computed using $T_{rec} = 90 + 2.083 \times |19 - dec|$)

[3] Computed using $\frac{Gain}{\beta} = \frac{19.7-(0.42\times|19-dec|)}{1.2236}$.

[4] Computed using $T_{rec} = 65 + 2.083 \times |19 - dec|$).

Table 5.2: Instrumental parameters of the radio surveys. For the Arecibo surveys we chose to adopt a more accurate definition for the gain and the receiver temperature that are function of the declination $\delta$. Respectively from the left to the right column, are indicated: telescope gain divided by a system losses factor, minimum signal to noise detected, receiver temperature, central observation frequency, integration time, sampling time, total bandwidth, and channel bandwidth.

5.14 to 5.18 (right plots) for each survey. Only the ratios below 10 are displayed to focus near the visibility threshold. Then, for each survey, we derived the ratio between the number of pulsars really detected by the survey and the total number of observable ATNF pulsars (the sum of the detected ones plus those that match the position and flux survey criteria but were not detected). We consider this last ratio as the new survey efficiency, $\epsilon_{surv}$, e.g. the percentage of pulsars detected by the survey with respect to all the detectable pulsars in the survey region.

By using the newly estimated survey parameters listed in tables 5.3 & 5.2, and by using the radiometer equation 5.34, I have re-evaluated the number of pulsars detected by each survey. The new selection results have been used to evaluate the new efficiency value, indicated, for each survey, in table 5.3, and led to a total number of 1429 radio pulsars to be used to scale the number of simulated pulsars that match the same selection criteria in position and flux.

|         | $l_{st}$ (°) | $l_{ed}$ (°) | $b_{st}$ (°) | $b_{ed}$ (°) | $dec_{st}$ (°) | $dec_{ed}$ (°) | $\epsilon_{surv}$ | $\epsilon_{Dewey}$ | *Duty cycle* |
|---------|------|------|-------|------|--------|--------|-------|-------|------|
| Mol2    | -    | -    | -     | -    | -85.0  | 20.0   | 0.59  | 0.4   | 0.03 |
| GrB2    | -    | -    | -     | -    | -18.0  | 90.0   | 0.31  | 0.7   | 0.03 |
| GrB3    | 15.0 | 230  | -15   | 15   | -      | -      | 0.41  | 0.75  | 0.03 |
| Pks2    | -    | -    | -     | -    | -90.0  | 0.00   | 0.89  | 0.75  | 0.03 |
| Are2    | 40   | 66   | -10   | 10   | 9.50   | 25.0   | 0.52  | 1.0   | 0.05 |
| Are3    | 38   | 66   | -8.1  | 8.2  | 5.00   | 26.5   | 0.66  | 0.7   | 0.05 |
| Pks1    | -92  | 20   | -4    | 4    | -      | -      | 0.40  | 0.6   | 0.03 |
| JBk2    | -5   | 105  | -1.3  | 1.3  | -      | -      | 0.47  | 0.8   | 0.03 |
| PkMB    | -105 | 52   | -6.03 | 6.35 | -      | -      | 0.99  | 0.9   | 0.05 |
| Swnb    | -100 | 50   | 4.5   | 30   | -      | -      | 0.87  | 1.0   | 0.05 |

Table 5.3: Estimated survey parameters. Respectively, from the left to the right
column, are indicated: longitude start & end, latitude start & end, declination start
& end, new survey efficiency, Dewey scaling factor, and pulsar duty cycle.

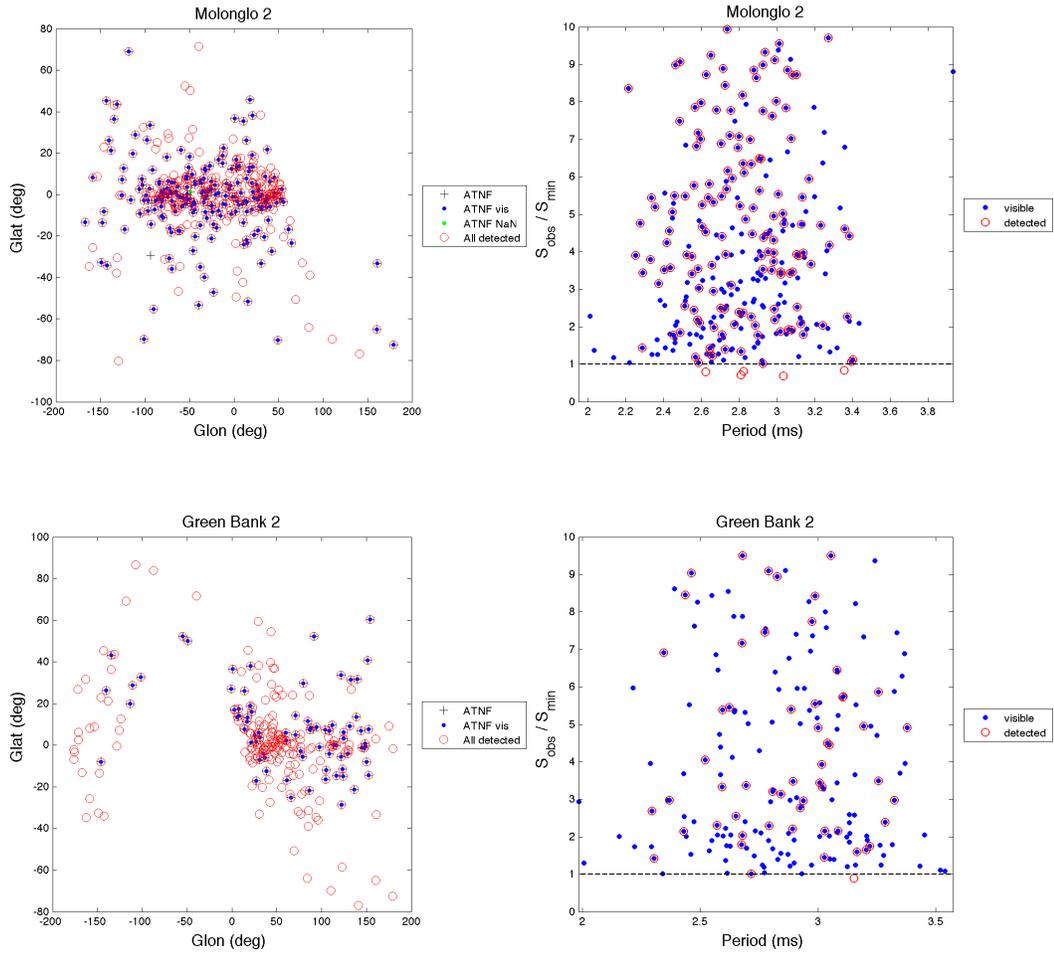

Figure 5.14: Definition of the radio visibility criteria for the surveys Molonglo 2, Green Bank 2, Green Bank 3. For each survey we show the pulsar selection, classification, and counting in the redefined survey coordinates region (table 5.3), and the ratio between the pulsar fluxes and the threshold ones re-evaluated by taking into account the fudge factor defined in table 5.3. In the sky position figures (left), the pulsars classified as *NaN* are the ATNF catalogue objects for which, because of some undefined parameter, it was not possible evaluate the threshold flux. ATNF, detected, and visible, indicate pulsars that are in the new survey coordinates region (Table 5.2) and, respectively, belong to the ATNF database, have been detected by applying the new survey parameters (Table 5.3), and that are visible but non necessarily detected (by the original survey or by applying our new survey parameters).

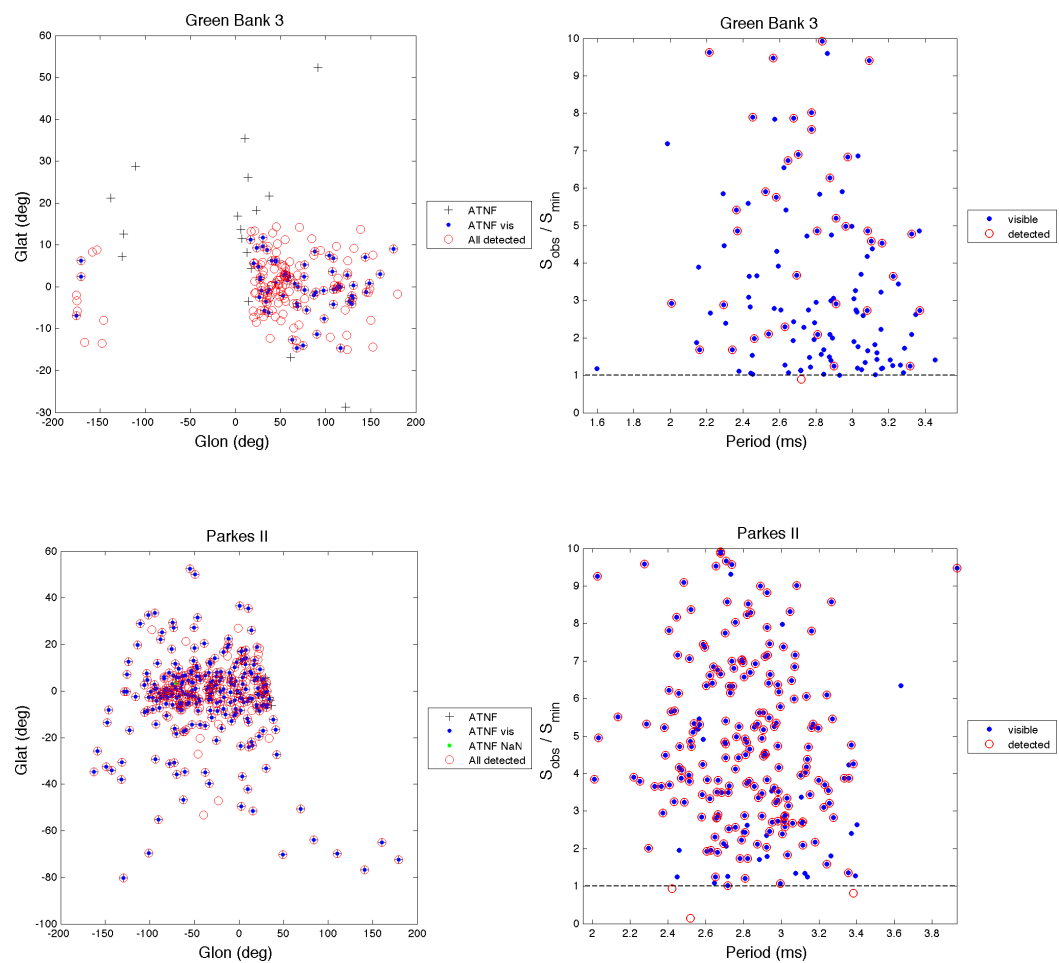

Figure 5.15: Definition of the radio visibility criteria for the surveys Green Bank 3,Parkes 2. For each survey we show the pulsar selection, classification, and counting in the redefined survey coordinates region (table 5.3), and the ratio between the pulsar fluxes and the threshold ones re-evaluated by taking into account the fudge factor defined in table 5.3. In the sky position figures (left), the pulsars classified as *NaN* are the ATNF catalogue objects for which, because of some undefined parameter, it was not possible evaluate the threshold flux. ATNF, detected, and visible, indicate pulsars that are in the new survey coordinates region (Table 5.2) and, respectively, belong to the ATNF database, have been detected by applying the new survey parameters (Table 5.3), and that are visible but non necessarily detected (by the original survey or by applying our new survey parameters).

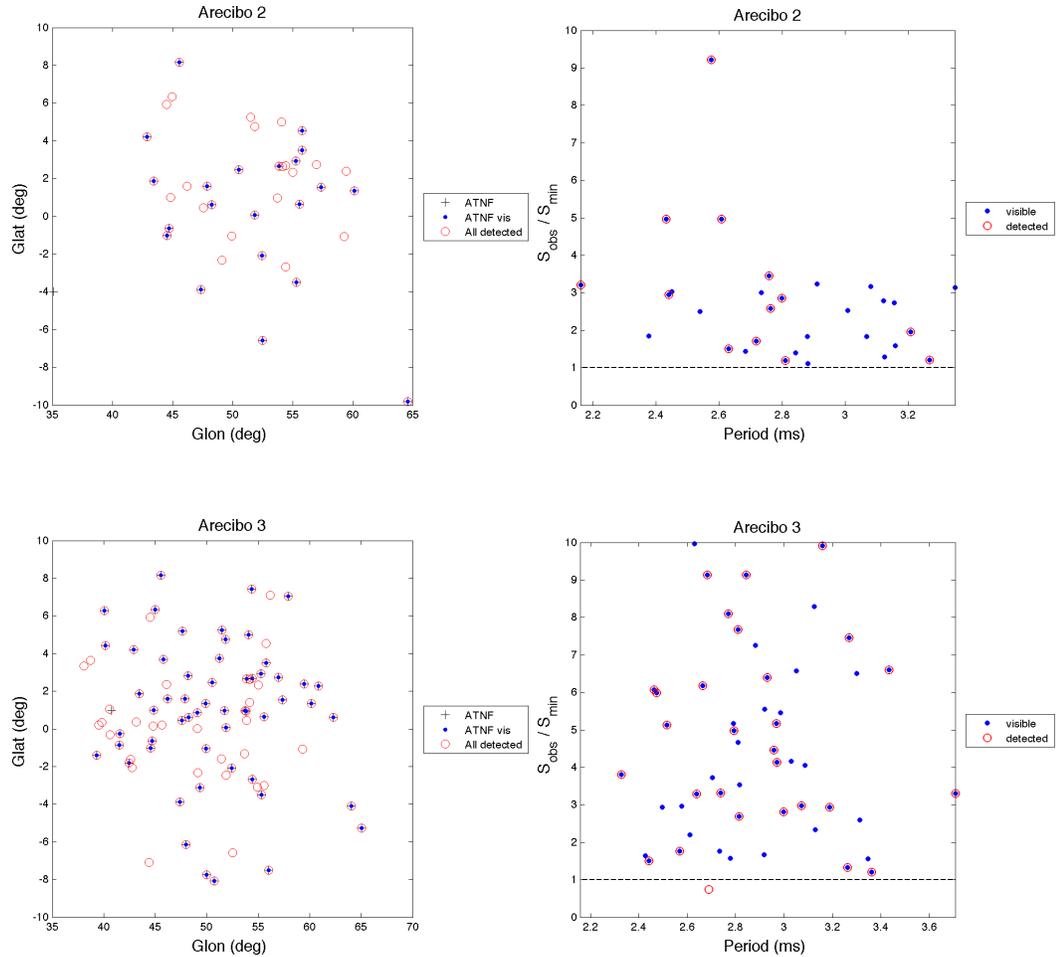

Figure 5.16: Definition of the radio visibility criteria for the surveys Arecibo 2, Arecibo3. For each survey we show the pulsar selection, classification, and counting in the redefined survey sky region (table 5.3), and the ratio between the pulsar fluxes and the threshold ones re-evaluated by taking into account the fudge factor defined in table 5.3. In the sky position figures (left), the pulsars classified as *NaN* are the ATNF catalogue objects for which, because of some undefined parameter, it was not possible evaluate the threshold flux. ATNF, detected, and visible, indicate pulsars that are in the new survey coordinates region (Table 5.2) and, respectively, belong to the ATNF database, have been detected by applying the new survey parameters (Table 5.3), and that are visible but non necessarily detected (by the original survey or by applying our new survey parameters).

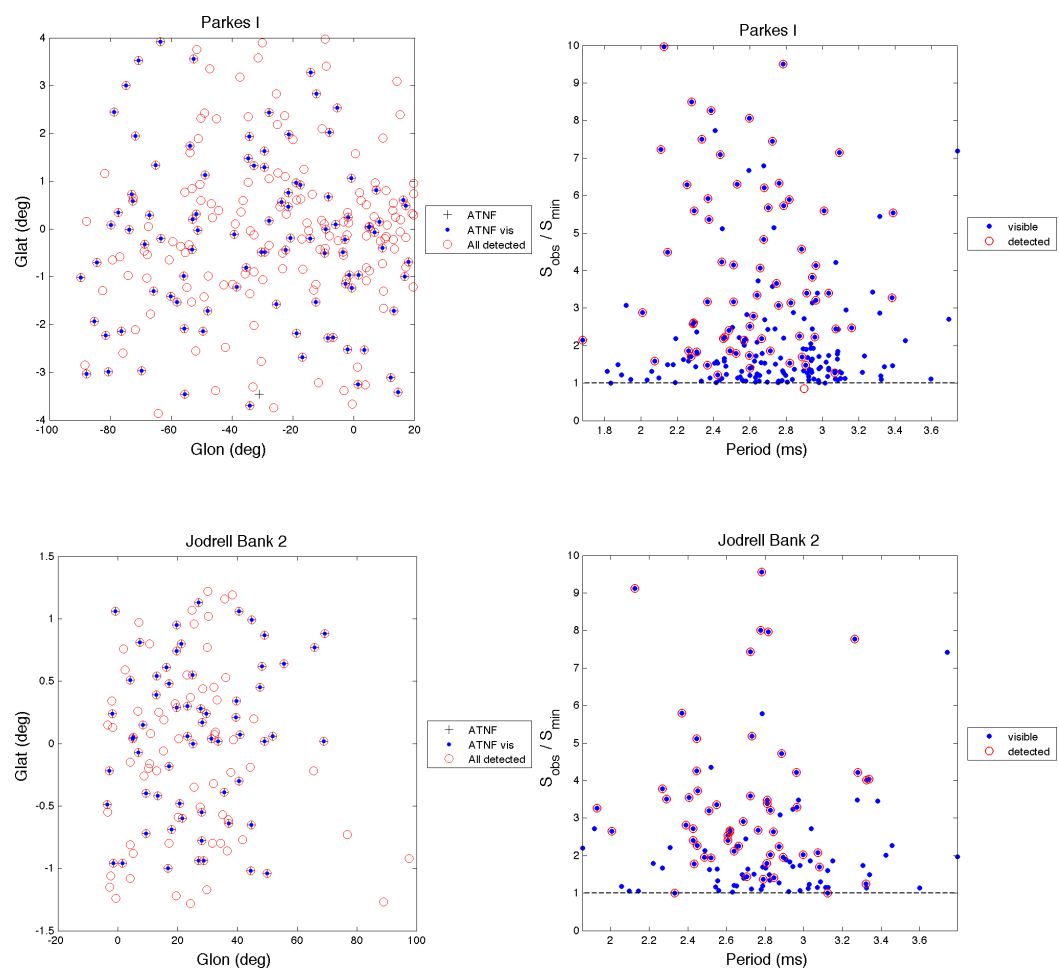

Figure 5.17: Definition of the radio visibility criteria for the surveys Parkes 1, Jodrell Bank 2. For each survey we show the pulsar selection, classification, and counting in the redefined survey sky region (table 5.3), and the ratio between the pulsar fluxes and the threshold ones re-evaluated by taking into account the fudge factor defined in table 5.3. In the sky position figures (left), the pulsars classified as *NaN* are the ATNF catalogue objects for which, because of some undefined parameter, it was not possible evaluate the threshold flux. ATNF, detected, and visible, indicate pulsars that are in the new survey coordinates region (Table 5.2) and, respectively, belong to the ATNF database, have been detected by applying the new survey parameters (Table 5.3), and that are visible but non necessarily detected (by the original survey or by applying our new survey parameters).

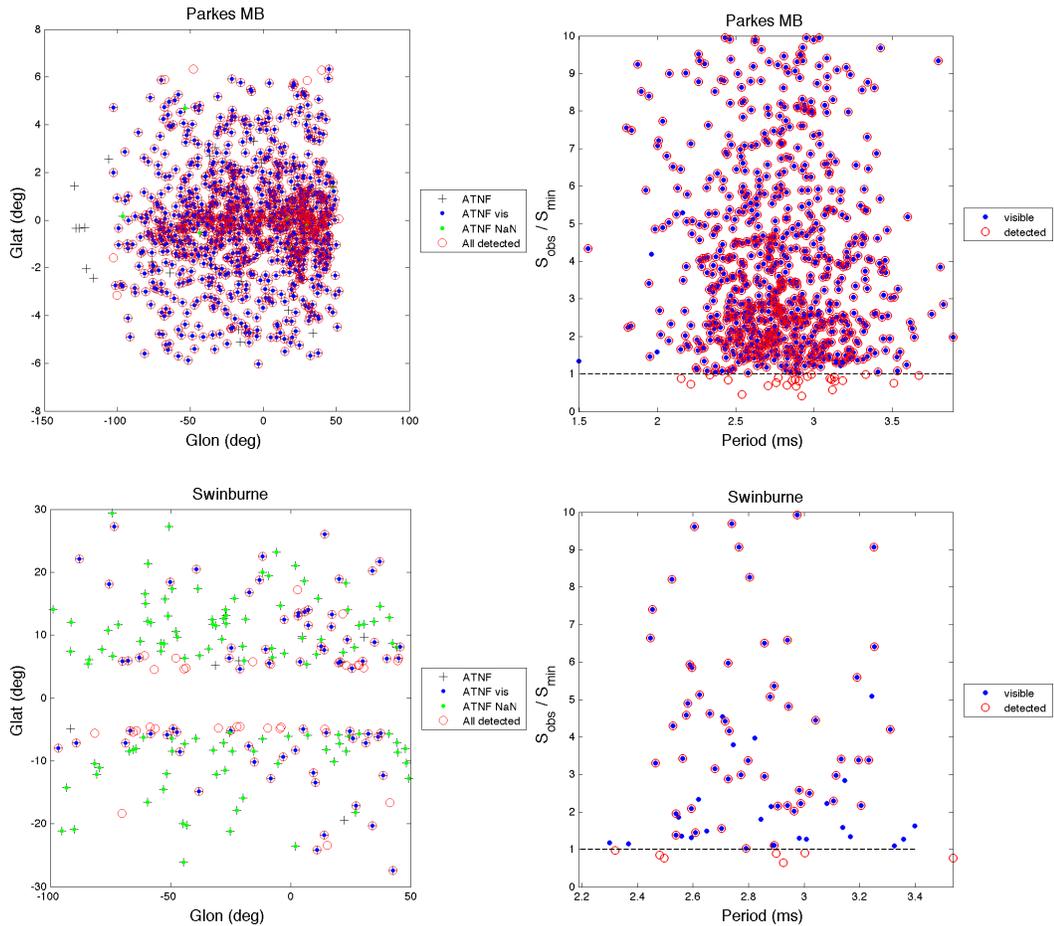

Figure 5.18: Definition of the radio visibility criteria for the surveys Parkes Multibeam and Swinburne. It is shown the pulsar selection, classification, and counting in the redefined survey sky region (table 5.3), and the ratio between the pulsar fluxes and the threshold ones re-evaluated by taking into account the fudge factor defined in table 5.3.In the sky position figure (left), the pulsars classified as *NaN* are the ATNF catalogue objects for which, because of some undefined parameter, it was not possible evaluate the threshold flux.ATNF, detected, and visible, indicate pulsars that are in the new survey coordinates region (Table 5.2) and, respectively, belong to the ATNF database, have been detected by applying the new survey parameters (Table 5.3), and that are visible but non necessarily detected (by the original survey or by applying our new survey parameters).

## 5.6 γ-ray pulsar visibility

In order to select the simulated pulsars that could be observed by the LAT during one year of observation, I applied a selection by using the visibility criteria of the LAT telescope. I made use of the 6 months LAT visibility map published for the 1st LAT pulsar catalogue. The map, figure 5.19, gives the minimum visible flux in [photon×second$^{-1}$×cm$^{-2}$]. This map has been obtained taking into account the real LAT observation time for each sky position as well as the incidence angle and energy of the detected photons, and the effective photon collection area corrected for the different incidence directions. The 1 year sensitivity map of the LAT, has been obtained by scaling to 1 year the 6 months sensitivity map of figure 5.19 as:

$$F_{th,1y} = F_{th,6m} \times \sqrt{\frac{181.3518}{365.25}}$$

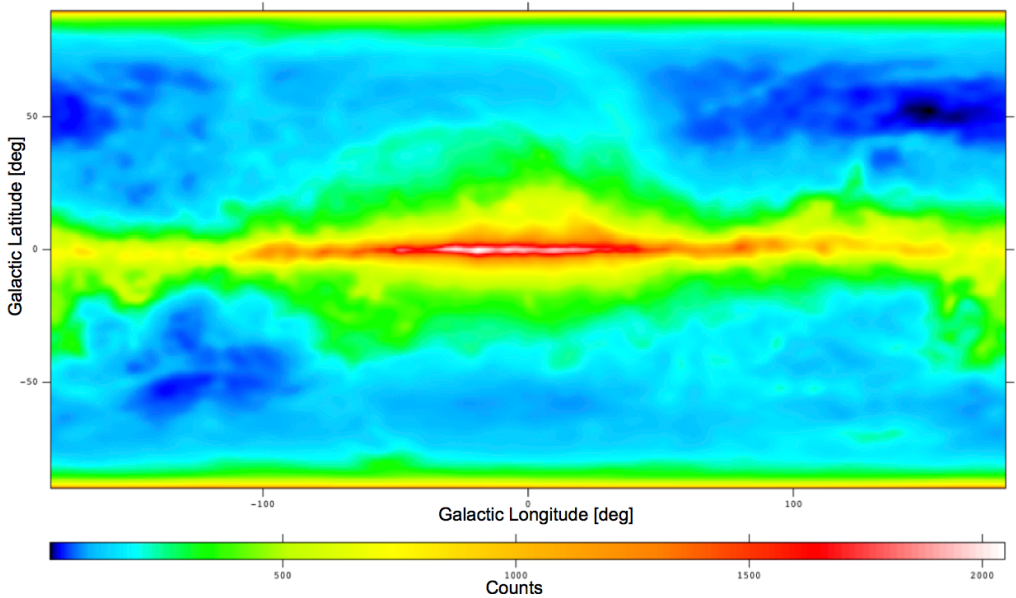

Figure 5.19: Pulsar sensitivity map for 6 months of observation of the LAT telescope. Each two dimensional bin contains the value of the minimum flux that would have been observed by the LAT after 6 months of observation. The colour scale refers to the photon number per second and square centimeter.

## 5.7 Population synthesis results and comparison with the observed pulsar sample

In this section I will show the results of the population synthesis. For each emission model, namely PC, SG, OG, and OPC, I have applied the radio and γ visibility criteria described in sections 5.5.2 & 5.6. The most important characteristics of the resulting $\gamma$-ray pulsar populations will be compared to the FERMI LAT pulsar observations in order to decide which of the emission models best describes the FERMI LAT $\gamma$-ray pulsar population.

**Observed flux and radiative efficiency**

To assess the photon luminosity predicted by each model from the particle luminosity calculation, it was necessary to assign a radiative efficiency to each model. Such an efficiency, $\epsilon_\gamma$, is defined as the percentage of the primary particle energy that will be converted into pulsed luminosity. This is clearly a very important parameter, that determines the whole flux outcome of the pulsar.

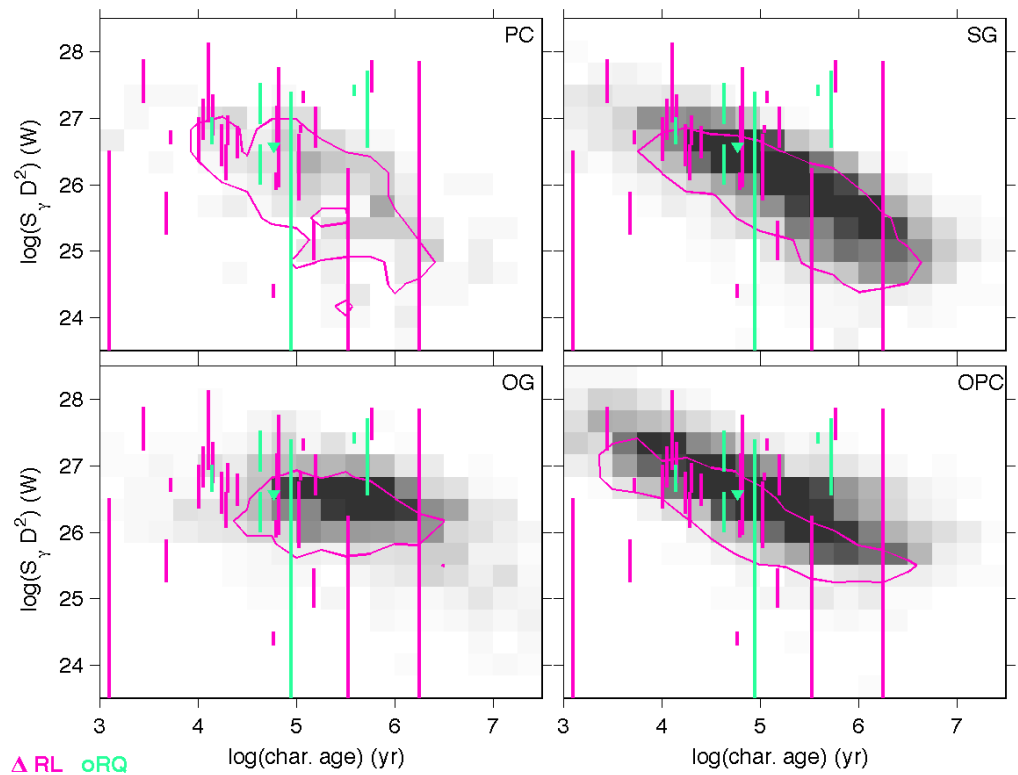

Figure 5.20: Number density of the visible gamma-ray pulsars obtained for each model as a function of characteristic age and energy flux times the square of the distance. These parameters can all be measured from the observations. The linear grey scale saturates at 1 star/bin for the polar cap and 2.5 star/bin for the other models. The pink contours outline the density obtained for the simulated radio-loud gamma-ray sub-sample (at 20% of the maximum density). The pink and green lines show the data for the radio-loud and radio-quiet LAT pulsars, respectively. The triangle marks an upper-limit.

In this thesis $\epsilon_\gamma$ has been chosen to provide a good agreement between the observed and simulated $S_\gamma D^2$ distributions as a function of characteristic age, with $S_\gamma$ the photon flux and $D$ the pulsar distance. This distribution involves only readily observable quantities. The solution found for each model is shown in figure 5.20. The choice of radiative efficiencies are: $\epsilon_{PC} = 1.0$, $\epsilon_{SG} = 7.0$, $\epsilon_{OG} = 1.0$, and $\epsilon_{OPC} = 0.5$. The very high value of $\epsilon_{SG} = 700\%$ needed for the SG requires either a *super Goldreich-Julian current* or a stronger value of the accelerating electric field in the gap than the original calculation by Muslimov and Harding (2004). This is quite possible if the polar cap is slightly offset, i.e. non symmetrical around the magnetic axis as one expects from the shape of the magnetic field lines distorted by the stellar rotation. Harding and Muslimov (2011) show that this distortion leads to a larger pair multiplicity, thus larger current, as well as an increased electrical field along the field lines, thus an enhanced potential of the pairs. Offset polar caps can sustain the modest increase in particle energy that is required in the present population study to account for the flux and pulsar counts observed by the LAT without invoking a radiation efficiency larger than 1. The offset polar cap prediction was not available at the time of the population synthesis work, so we keep here the original polar cap luminosity and $\epsilon_{SG} = 700\%$.

**The beaming factor $f_\Omega$**

One important and most debated pulsar characteristic is the *beaming fraction*, defined as

$$f_\Omega = \frac{L_\gamma}{4\pi D^2 \langle \nu F_\nu \rangle} \tag{5.39}$$

where $L_\gamma$, $D$, and $\langle \nu F_\nu \rangle$ are respectively the pulsar luminosity, distance, and observed average energy flux. Physically, the beaming factor describes the energy flux contained in the fraction of the $4\pi$ steradian solid angle swept by the pulsar beam after one complete rotation. So, its value depends both on the global solid angle of the emission beam, on the beam inclination so the spin axis, and on the amount of energy that it contains.

From equations 5.39 & 5.18, the *beaming fraction* with respect to the phase-plot is

$$L_{2\ pole} = 4\pi D^2 f_\Omega \langle \nu F_\nu \rangle \quad if \quad f_\Omega = \frac{1}{2\int_0^{2\pi} n(\zeta_{obs}, \phi) d\phi}. \tag{5.40}$$

To be able to compare the $\gamma$-ray luminosity evaluated using a radiative efficiency for each model (previous paragraph) to the LAT pulsar, I had to assign a value of $f_\Omega$ to each observed pulsar. In figure 5.21 is shown the behaviour of the beaming factor as a function of $\dot{E}$. For each model I have fitted the radio-loud and radio-quiet components of the population. The best fit functions $f_{\Omega,RL/RQ} = f(\dot{E})$ are indicated as solid and dotted lines in figure 5.21.

The decreasing of $f_\Omega$ with age (decreasing $\dot{E}$) is a trend observed in all the emission models and is due to the shrinking of the polar cap as a consequence of the pulsar slow-down. Since the pulsar period increases with age, the light cylinder radius also increases, implying a reduction of the polar cap radius. As the polar cap shrinks, the solid angle of the emission beam decreases for all the emission models, so that also the beaming factor follows a decreasing trend that is different for each model. In the PC case, both the radio-quiet and radio-loud $f_\Omega$ values are very dispersed and very low because of the extremely collimated PC beam (characterised by low solid angle values).

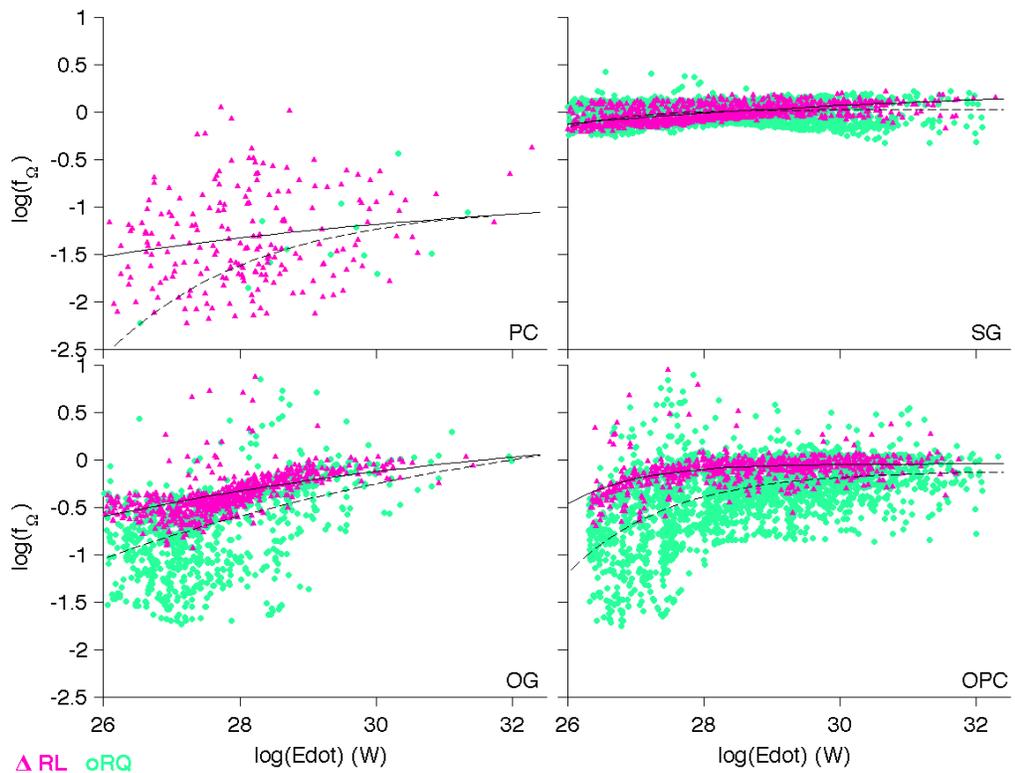

Figure 5.21: Distribution of the gamma-ray beaming factors obtained for each model as a function of the spin-down power. Pink and green dots refer to radio-loud and radio-quiet gamma-ray visible objects, respectively. The solid and dotted lines give the best fit to the radio-loud and radio-quiet data points, respectively.

For the SG, OG, and OPC, the $f_\Omega$ distribution of the radio-quiet population component appears always more dispersed than the radio-loud one. The dispersion for the radio-quiet objects is always toward the small $f_\Omega$ values, while the radio-loud objects exhibit high beaming factors and their distribution is more collimated. A pulsar characterised by a large $\gamma$-ray solid angle has a higher probability to have overlapping radio and $\gamma$ beams, thus to be radio-loud. This large solid angle objects will be characterised by high value of $f_\Omega$.

On the other hand, when the beam solid angle decreases, the probability for the radio and $\gamma$ beam to overlap decreases too. Such objects are characterised by low $f_\Omega$ values and will be mainly radio-quiet.

The SG case shows a minimal change in beaming factor with age for both the radio-loud and radio quiet objects. Since there is emission in nearly all the direction of the sky from the bright caustic, the radio-quiet and radio-loud beaming factor distributions are centred around $f_\Omega$=1.

In the OG and OPC cases, we observe the most pronounced radio-quiet $f_\Omega$ dispersion. The OG and OPC models share the same emission pattern (phase-plot), thus the dispersion covers the same range of values. The largest spread is observed in the OG at low $\dot{E}$, typically below $10^{28}$ W where the beaming factor seems to assume a constant value.

In the OPC case the trend is exactly the opposite, it increases up to $10^{28}$ W and then it stays stable around 1. Since all the objects at a given $\dot{E}$ have the same $\gamma$-ray luminosity (by construction), the observed spread in the $f_\Omega$ values of the radio-quiet objects reflect the spread in beam flux as sum from different perspectives. It amounts typically to one order of magnitude and can reach two in rarer case.

**Luminosity**

In Figure 5.22 is shown, for each emission model, the evolution of the $\gamma$-ray luminosity with the spin-down power $\dot{E}$ compared with the LAT results. The luminosity of the LAT pulsars has been computed from the measured pulsed flux using equation 5.40 with a beaming factor $f_\Omega(\dot{E})$ coming from the best-fit trend plotted in figure 5.21, according to the radio-loud or radio-quiet quality of the LAT pulsars.

The observed gamma luminosity evolution $L_\gamma \propto \dot{E}^{0.5}$ is roughly predicted by all the models and so the luminosity trend can not be used to discriminate between the emission mechanisms.

In the PC case, the $\gamma$-ray emission comes from the same region as the radio one and this implies a high probability for the $\gamma$ and radio beams to overlap and so, a low number of radio-quiet objects. Anyway the radio-quiet objects populate just the high $\dot{E}$ and $L_\gamma$ side of the PC $\gamma$ visible population. The comparison with the LAT population indicates that the PC population is not luminous enough to account for most of the observed pulsars.

In the SG case the luminosity of the high $\dot{E}$ objects is mainly sustained by the radio-quiet component of the population. Because of the large radiative efficiency (increased current power in the gap), the SG luminosity reasonably follows the LAT data: the SG population is the one that best describes the trend of the observed one. The fraction of radio-quiet objects moderately increases at low $\dot{E}$ and its distribution get less dispersed. For both the PC and SG models, since the $\gamma$-ray emission is sustained by the particles generated by the same polar cap electromagnetic cascade (section 2.1), $L_\gamma$ follows the same trend steepening from $L_\gamma \propto E^{1/2}$ to $L_\gamma \propto E$ with decreasing $\dot{E}$ when the pulsar pours out most of its pin-down power into $\gamma$-rays. This trend is predicted but not yet observed because of the large dispersion in the LAT data points and because of the large uncertainties in LAT pulsar distances.

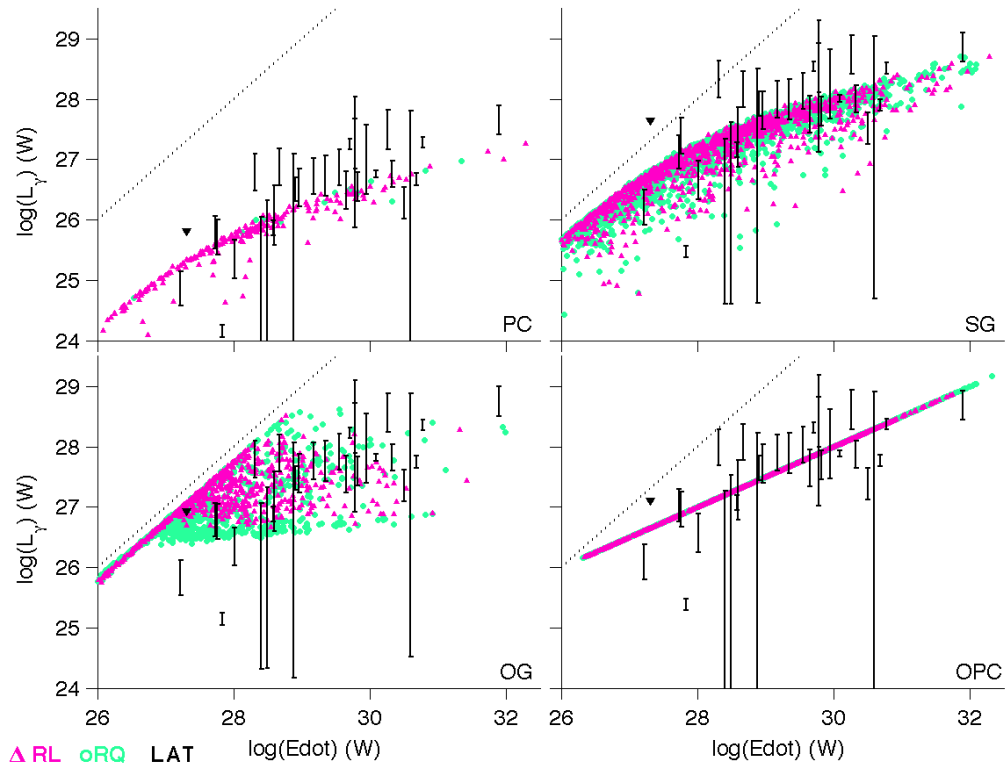

Figure 5.22: Distribution of the gamma-ray luminosities obtained for each model as a function of the spin-down power. Pink and green dots refer to radio-loud and radio-quiet gamma-ray visible objects, respectively. The observed LAT luminosities (black lines and upper limit) have been derived using the pulsar energy-flux measurement and the $f_\Omega(\dot{E})$ value estimated from the fit to the simulated data for the particular spin-down power and radio-loud or quiet state of the LAT pulsar. The dotted line shows the 100% efficiency boundary.

The OG luminosity evolution shows a different behaviour for hight and low spin-down values. At $\dot{E} > 10^{26.5}$ W for radio-quiet pulsars and $\dot{E} > 10^{27}$ for radio-laud ones, the spread in $\gamma$-ray luminosity results primarily from the evolution of the gap width with $\dot{E}$ and $\alpha$. Going towards the low $\dot{E}$ values, the luminosity decreases along the 100% efficiency boundary and the low dispersion is due to the fact that the gap width assumes an approximatively constant value. The LAT pulsars fall within the range of predicted luminosities but they exhibit less dispersion of high $\dot{E}$ than predicted from the model.

In the OPC case the $L_\gamma(\dot{E})$ evolution is a built-in assumption of the model. Since $L_\gamma \propto \Delta w_{OPC} \dot{E}$, this implies the complete absence of dispersion around a perfect power law. As in the SG case, the luminosity of the high $\dot{E}$ objects is mainly sustained by the radio-quiet component of the population. Energetically the OPC well describes the power of the observed object, however, this strong agreement can not be a physical constraint of the validity of the model since the constant $\Delta w_{OPC}$ has not been derived but rather chosen to fit the observations.

**Detection statistics**

Table 5.4 indicates, for each model, the numbers of NSs that passed the radio and/or $\gamma$ visibility criteria and their comparison with the LAT pulsar detection after 1 year of observation. The number of radio visible pulsars in the simulation has been scaled to the 1429 ATNF radio pulsars that passed the same selection criteria. The scale factor of **0.03952** that allows to match the simulated and observed radio sample has been applied to all star counts quoted hereafter, in particular to the $\gamma$-ray simulated samples. The choice of radiative efficiencies driven by a reasonable agreement in the $S_\gamma D^2$ evolution with characteristic age yields the right number of pulsar detections with respect to the LAT findings, except for the under-luminous PC model.

| | $PSR_{radio,vis}$ | $PSR_{\gamma RL,vis}$ | $PSR_{\gamma RQ,vis}$ | $RL/RQ_{(mod,LAT)}$ | $N_{mod}/N_{LAT,A7}$ |
|---|---|---|---|---|---|
| PC | 1420.82 | 8.18 | 0.51 | (16.04,1.043) | 0.18 |
| SG | 1390.39 | 38.61 | 72.79 | (0.53,1.043) | 2.37 |
| OG | 1402.60 | 26.40 | 37.70 | (0.70,1.043) | 1.36 |
| OPC | 1402.76 | 26.24 | 83.98 | (0.31,1.043) | 2.34 |

Table 5.4: Are respectively indicated, for each model and from the left to the right column: the number of just radio visible pulsars, the scaled $\gamma$-ray visible radio-loud and radio-quiet pulsars, the radio-loud/radio-quiet pulsar number ratio predicted ($mod$) and observed ($LAT$), and the ratio of the total number of $\gamma$-ray detections over the LAT detection. All the values refer to 1 year of LAT observation.

## 5.7.1 The visible component of the sample

**The radio visible sample: comparison with the observed population**

Figure 5.23 shows the comparison between the $P - \dot{P}$ diagram of the radio visible component of the simulated population (left panel) and the observed one (right panel). The real sample has been derived from the ATNF catalogue and both populations have been filtered with the new radio visibility criteria defined in section 5.5.2.

The simulated population is able to describe the $P - \dot{P}$ distribution of the observed sample but just for the slowest rotators. This discrepancy is more evident in figure 5.24 where the distributions for the total visible samples (both radio or gamma visible) are plotted. In these histograms, the distributions are totally dominated by the radio sample since the $\gamma$-ray pulsars contribution is much smaller. The simulated spin period distribution is too broad to be able to describe the observed proportion between the number of the intermediate period objects ($\sim 50$ ms) and the ones that populate the wings of the distribution. Even so, the range of the spin periods is well-covered by the simulated sample and well centred, but we lack objects in the 0.3-1.0 second range. Whereas this would be problematic to study radio beam models, the good representation of the period distribution at $P < 500$ ms (and the slight excess of lower period objects) where most of the LAT pulsars are found, supports the study of $\gamma$-ray models.

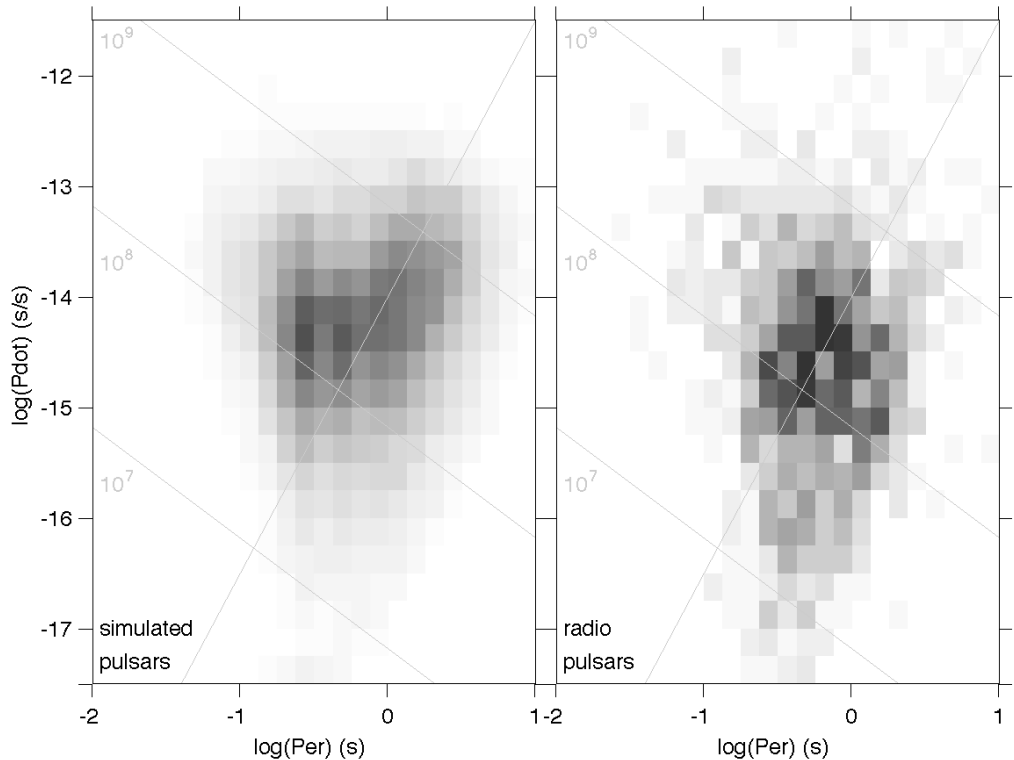

Figure 5.23: Number density of the radio visible pulsars as a function of period and period derivative. The left and right plots respectively show the simulation and observed data with the same grey scale saturating at 25 star/bin and the same visibility criteria. The rising grey line marks the slot-gap death line. The declining grey lines mark the iso-magnetic lines at $10^7$, $10^8$, and $10^9$ (T) Tesla.

The simulated distributions in $\dot{P}$ and $B$, are all shifted to an excess of young and energetic pulsars compared to the observed ones. There is also a clear excess of nearby ones. The obtained visible distributions result from the choice of the birth characteristics that emphasised nearby and high $\dot{E}$ objects while preserving the bulk of the radio distributions.

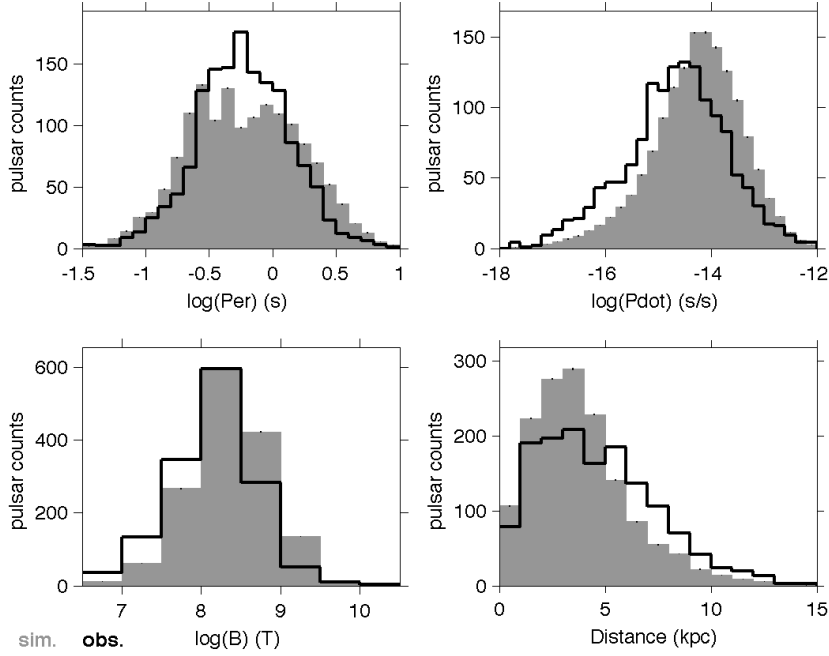

Figure 5.24: Number distributions in period, period derivative, magnetic field strength, and distance obtained for the total visible sample (radio or gamma). The shaded histograms from the simulation are compared to the LAT plus radio-pulsar distribution (thick line). The ATNF radio pulsar sample has been restricted to the objects that pass the same position and flux selection criteria as in the simulation. The slot-gap visible gamma-ray simulated objects have been used in this Figure. The other models yield similar pictures because of the overwhelming population of radio objects that dominate the curves.

**The γ-ray visible samples: comparison with the LAT objects**

Figure 5.25 shows the $P - \dot{P}$ diagram of the γ-ray visible population for each model. The grey scale indicates the modelled number density of the model population while the LAT pulsars are plotted as magenta triangles, for radio-loud objects, and green dots, for radio-quiet ones.

The PC model reproduces very poorly the observed population. The simulated pulsars are characterised by too high $P$ and too low $\dot{P}$ when compared with the much more energetic LAT population. Moreover the PC model predicts too few ($\sim 9$) γ visible pulsars, $\sim 6\%$ of which are radio-quiet, when LAT find a much larger proportion.

The SG $P - \dot{P}$ distribution of the γ-visible objects is much wider compared with the PC one and covers the $P - \dot{P}$ region occupied by the LAT objects. Nevertheless, the core of the simulated population is too squeezed toward the SG death-line, in the low energetic object region of the diagram.

Although the OG model predicts a number of radio-quiet & radio-loud objects (table 5.4) consistent with the LAT detections, the $P - \dot{P}$ distribution of the simulated population is different from the observed one. The OG $P - \dot{P}$ diagram is closer to the *ordinary older* radio pulsar population shown in figure 5.23. The core of the distribution is not as extended as the observed one and it is too close to the pulsar death line. The most energetic component of the $P$ and $\dot{P}$ distributions of the LAT objects is not well represented by the OG model.

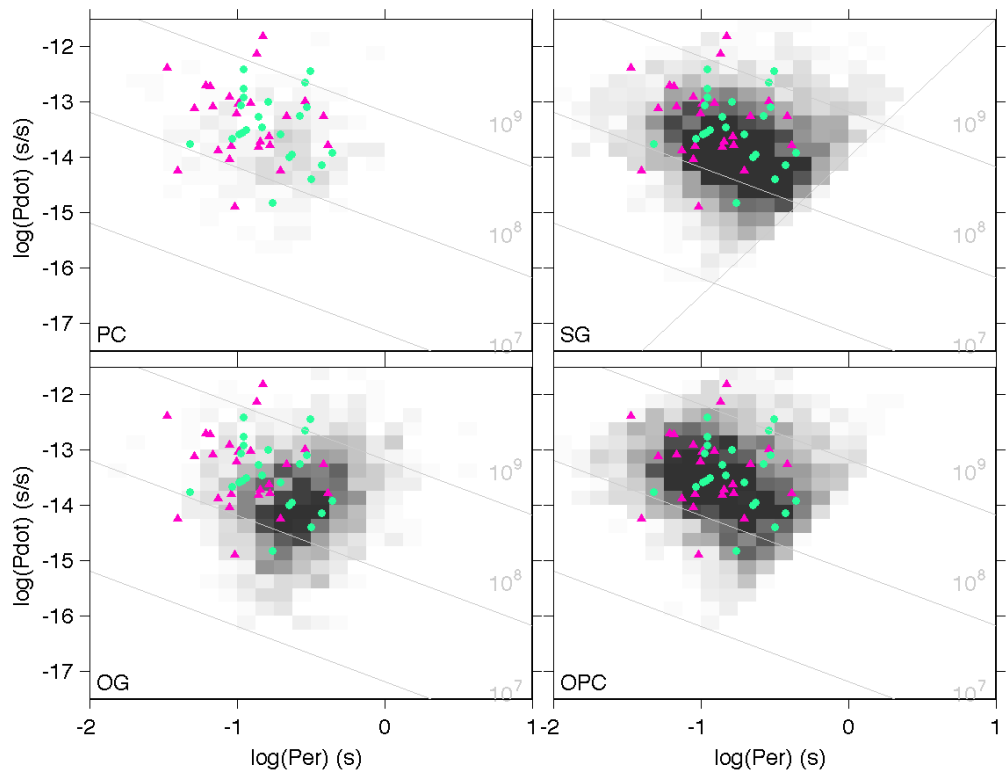

Figure 5.25: Number density of the visible gamma-ray pulsars obtained for each model as a function of period and period derivative. The linear grey scale saturates at 1.5 star/bin. The pink triangles and green dots show the radio-loud and radio-quiet LAT pulsars, respectively. The rising grey line in the slot-gap subplot marks the slot-gap death line. The declining grey lines mark the iso-magnetic lines at $10^7$, $10^8$, and $10^9$ T.

The OPC $\gamma$-visible population best describes the observed $P$ and $\dot{P}$ of the LAT population: a core centred on the observed objects, and tails that cover the overall dispersion. As has already been discussed in the previous sections, this model is not based on any physical assumptions but just contains a gap width definition optimised to fit the data.

**Light cylinder magnetic field & age**

The B strength at the light cylinder (LC) magnetic field is useful to study the outer magnetosphere emission mechanism. Figure 5.26 shows the comparison between the model predictions and the observed population as a function of LC magnetic field strength and age. For both the simulated and observed populations, the magnetic field at the light cylinder $B_{LC}$, has been evaluated by using equation 1.26

$$B_{CL} = B_S \left( \frac{\Omega R}{c} \right)^3$$

where $B_S$ is the surface magnetic field defined by equation 1.18. One should remember that $B_S$ is a derived pulsar parameter and its computation is based on the assumption of a dipolar field structure as well as stellar moment of inertia. A few stellar radii off the surface the quadrupole components would be negligible. The radius and mass values are those defined in the thesis: $R_* = 1.3 \times 10^6$ m, and $M_* = 1.5 M_\odot$.

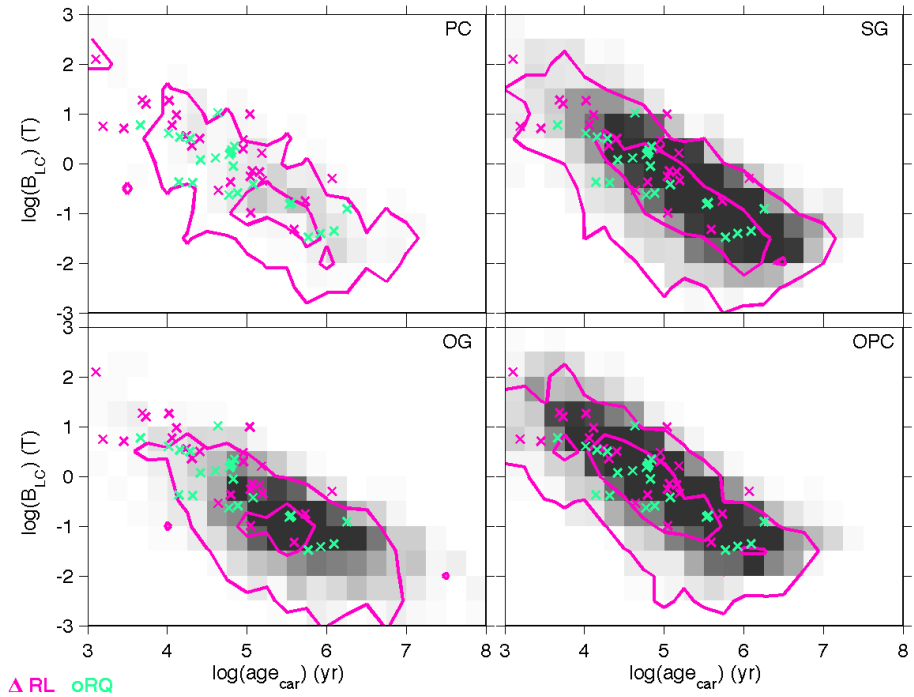

Figure 5.26: Number density of the visible gamma-ray pulsars obtained for each model as a function of characteristic age and magnetic field strength at the light cylinder. The linear gray scale saturates at 2 star/bin. The pink contours outline the density obtained for the simulated radio-loud gamma-ray sub-sample (at 5% and 50% of the maximum density). The pink and green crosses show the radio-loud and radio-quiet LAT pulsars, respectively.

The $B_{LC}$ evolution is not relevant for the inner polar cap PC model. In the OG model, the simulated population is not able to describe the $B_{LC}$ evolution of the very young and energetic objects. Since the OG is one of the most extreme external magnetosphere emission models, this inconsistency with the observed trend suggests a lack of energetic pulsars predicted by this model. The SG and OPC results are able to describe properly the $B_{LC}$ evolution found for the LAT pulsars. The different beam geometries between SG and OPC gaps, reaching below the null surface or not, yield indistinguishable distributions in the $(B_{LC}, age)$ plane.

**The spin-down power**

One of the most important pulsar characteristics, and for sure the one that best describes the pulsar energetics, is its spin-down power $\dot{E}$. It is defined (equation 1.3) as the rate with which the pulsar loses rotational kinetic energy, as

$$\dot{E} \equiv \frac{-\mathrm{d}E_{rot}}{\mathrm{d}t} = 4\pi^2 I \dot{P} P^{-3}. \qquad (5.41)$$

The latter equation could be considered as the best approximation available to evaluate the pulsar $\dot{E}$ but it remains an approximation. The main assumption is related to the NS structure through the moment of inertia $I$. The $I$ computation depends on the NS radius and mass and these parameters can be chosen inside an interval of allowed of values that could affect the final $\dot{E}$ estimate by more than a factor 3.

The pulsar spin-down power discussed in this section has been evaluated, for both the LAT population and the simulated one, using equation 5.41, and a moment of inertia according to equation 1.16 (Lattimer & Prakash, 2007), for $M_* = 1.5 M_\odot$, and $R_* = 1.3 \times 10^6$ m.

Figure 5.27 shows the comparison between the spin-down power present in the observed population and those of the $\gamma$-visible simulated samples. A significant discrepancy is seen for all the models. The LAT distribution has the shape of a reverse Maxwell-Boltzmann function with a steep fall around $10^{30} - 10^{31}$ W, and a peak around $10^{29}$ W. The first inconsistency between the simulated and observed samples is in the shapes of the distributions. All the models show a rapidly increasing histogram at low $\dot{E}$ and slowly decreasing at high $\dot{E}$, opposite to the observed one. The difference in shape suggests that the $\dot{E}$ inconsistency is not due to a simple scale problem but to a gap evolution problem: even by introducing a fudge factor in the spin-down power computation, none of the models would be able to describe the observed distribution.

As already hinted in previous plots the PC model is not luminous enough to reproduce the large number of energetic objects found by the LAT. The reason of this lack in luminosity is explained in the PC panel of figure 5.27. In the SG case we see the same problem, even though the high $\dot{E}$ region is more populated.

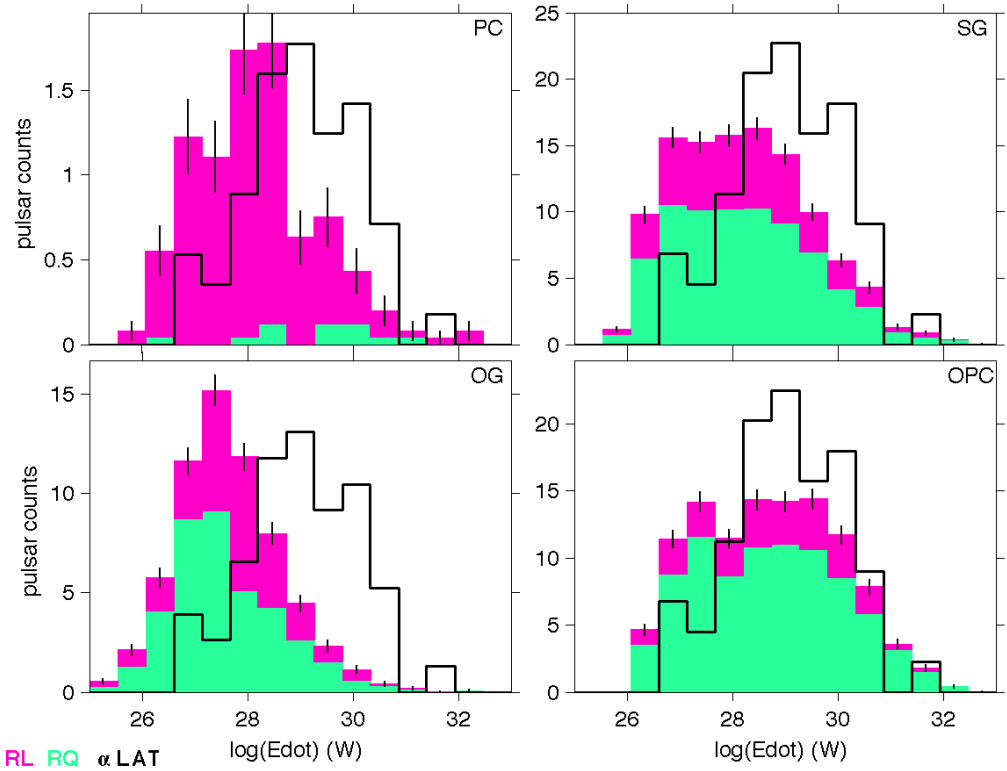

Figure 5.27: Spin-down power distributions obtained for each model for the visible gamma-ray objects. The stacked pink and green histograms refer to radio-loud and radio-quiet gamma-ray objects, respectively. The observed LAT pulsar distribution (in black) has been scaled to the total number of visible objects for each model to ease the comparison and show the relative lack of young energetic objects with $\dot{E} > 10^{29}$ W.

The OG model provides the poorest description of the observed spin-down power distribution. The observed and simulated distribution are antithetic: the first peaking at high $\dot{E}$, the second peaking at low $\dot{E}$. Energetically they describe two different pulsar populations: high and low $\dot{E}$ ones.

The OPC model gives a better results. The inconsistency between observed and simulated distribution is still present but appears less dramatic with respect to the other models. The simulated distribution is well centred on the observed one and well contained inside the LAT histogram but it is not able to reproduce the abundance of energetic LAT objects. Since this model was made to fit the luminosity evolution of the LAT data the lack of visible high $\dot{E}$ objects relates to the evolution of the beam pattern and its visibility.

The lack of hight $\dot{E}$ visible pulsars is rather puzzling since they are the intrinsically brightest objects (high particle power and large $f_\Omega$) with the widest beams (large open magnetosphere) swapping across the sky. The problem affects all the models, so its origin does not depend much on the emission pattern or the luminosity trend with $\dot{E}$. It is also insensitive to the relative orientation of radio and $\gamma$-ray beams since both radio-loud and radio-quiet are missing at high $\dot{E}$. Nor is the problem related to a visibility selection since all the models match reasonably the low $\dot{E}$ region of the LAT histogram.

By testing different population configurations I have tried to figure out which pulsar parameter has the largest impact on the high $\dot{E}$ object number. A set of different birth distributions for spin period, magnetic field and age has been tested. Even decreasing as much as possible the birth spin period to have, after evolution, a very young and energetic pulsar population (section 5.1.1), the gain in the number of $\gamma$-visible energetic objects was very small. The case illustrated here is significantly biased to young, energetic pulsars, within tolerable limits with respect to the total radio and $\gamma$ population (Figure 5.24). Another solution was searched by scanning the allowed domain of the intrinsic luminosity (e.g. SG $\lambda$ parameter, section 5.4.2), but without success

The parameter that reduced the discrepancy was the newly assumed birth location profile and the spatial density of younger supernovae in the inner Galaxy (see sections 5.1.2). So, the $\dot{E}$ problem is totally model independent and seems to have one of its causes in birth location and population evolution.

A speculative explanation could be given on the basis of a possible evolution of the magnetic obliquity $\alpha$ with age. That would modify the swept-up solid angle and the visibility. When the pulsar gets older (the $\dot{E}$ decreases), if we assume that $\alpha$ gradually increasing with age, originally undetected objects will start to be detected as time passes. Another speculative explanation could be found in a slower evolution of the dipole radiation for very young and energetic objects. This would be a strong assumption that requires strong theoretical basis to be confirmed.

Figure 5.28 shows the comparison of the characteristic age of the simulated and observed populations with an obvious lack of young (this high $\dot{E}$) visible pulsars. Since in the $P - \dot{P}$ diagram the region characterised by the same characteristic age

$$\tau_c = \frac{P}{2\dot{P}}$$

defines rising isochrone lines, the $\tau_c$ could be used as a tracker of the grey scale $P - \dot{P}$ diagrams in figure 5.25.

The observed $\tau_c$ distribution (Figure 5.28) shows a low significance double peaked structure, peaked at $10^{4.2}$ and $10^{4.8}$ years. A double structure is marginally seen just in the PC case but for much older objects. The SG and OG show the same trend described for the $P - \dot{P}$ of figure 5.25. For all the model it is possible to appreciate how the core of the simulated distribution is shifted with respect to the observed one. Again the OPC seems to give the best results. Considering the nearby objects, it describes better than the other models the LAT population, but it shows a clear excess of objects older than $10^{5.1}$ years, where the LAT number falls down.

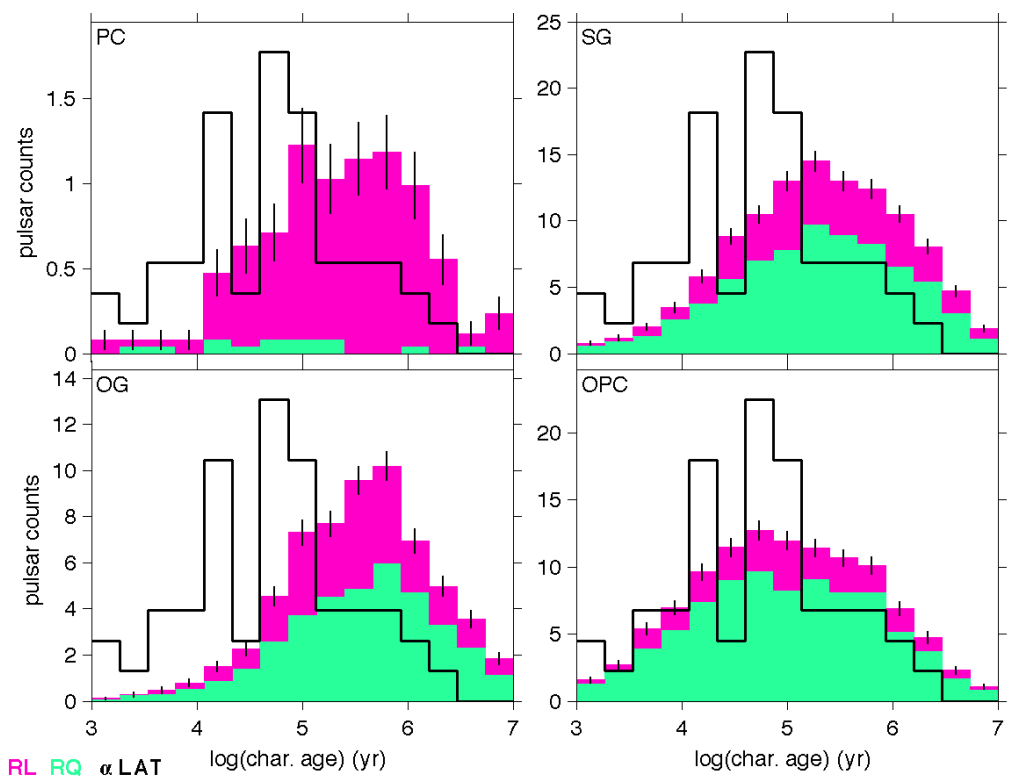

Figure 5.28: Age distributions obtained for each model for the visible gamma-ray objects. Pink and green dots refer to radio-loud and radio-quiet gamma-ray objects, respectively. The observed LAT pulsar distribution (in black) has been scaled to the total number of visible objects for each model to ease the comparison and show the relative lack of visible objects with age <100 kyr.

**Spatial distribution in the Milky Way**

The number of high $\dot{E}$ objects in the $\gamma$-visible population is particularly sensitive to the Galactic distribution of the NS at birth. Figure 5.29 shows, as a polar view of the Galaxy, the visible radio or $\gamma$ objects distribution, resulting from of the birth location profile described in section 5.1.2. The gamma-ray visibility contour matches well the Galactic region where the LAT pulsars have been detected. This excludes a distacnce related visibility problem as the cause of the lack of high $\dot{E}$ object discussed in the previous section. The Galactic latitude distribution for the $\gamma$ visible component derived from each model is shown in figure 5.30. The PC, SG, and OG models show an inconsistency with the observed population close to the Galactic disk. The OPC prediction nicely describes the observed latitude distribution because it is the model that predicts the largest number of young visible pulsars, young therefore still at low altitude. The fact that the modelled Galactic latitude matches the observed distribution is meaningless without a comparison with the distances of the same objects. Comparing figure 5.30 with figure 5.31 it is possible to note that the observed consistency between the OPC Galactic latitude distribution and the observed one, does not correspond to a consistency between the object distances. Figure 5.31 gives the distance distributions of the visible $\gamma$-ray pulsars. All the distribution are skewed to short distances as in the initial birth distribution which strongly peaks in the inner Galaxy.

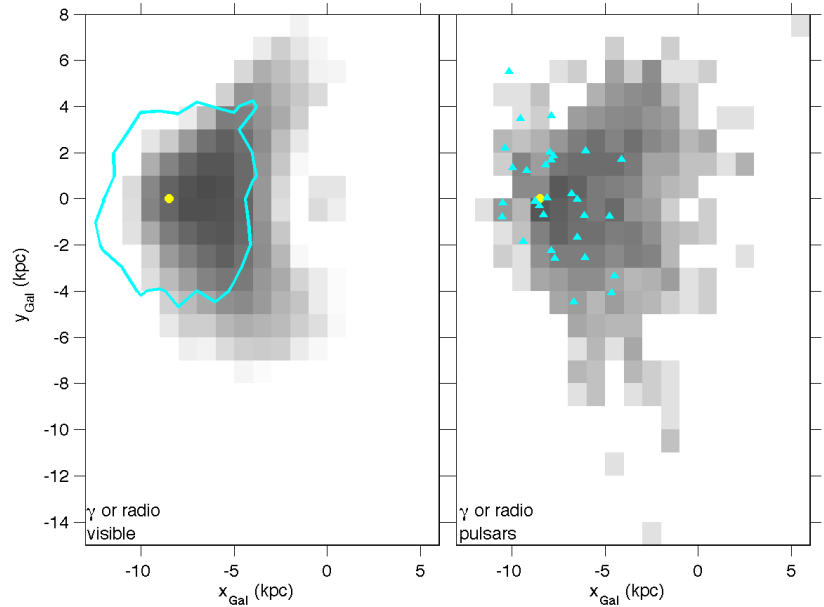

Figure 5.29: Number density of the visible radio or gamma-ray pulsars in the Milky Way (polar view). The left and right plots respectively show the simulation and observed data with the same logarithmic gray scale saturating at 100 star/bin and the same visibility criteria. The cyan contour outlines the region where simulated gamma-ray pulsars are detectable. The cyan triangles show the location of the LAT pulsars. The yellow dot marks the Sun.

An interesting trend, observed in all the models except for the PC, is in the evolution of the RL/RQ ratio with distance. It is observed that it tends to 1 and this could implying that all the far away objects tend to lose the radio emission. The simulated star position evolution has been done by assuming that all the pulsars are born in the Galactic plane with a supernova kick velocity distribution, and they spread in the Galaxy as time goes by. So, the $RL/RQ \rightarrow 1$ for high distances is mainly due to the decreasing of the pulsar radio visibility.

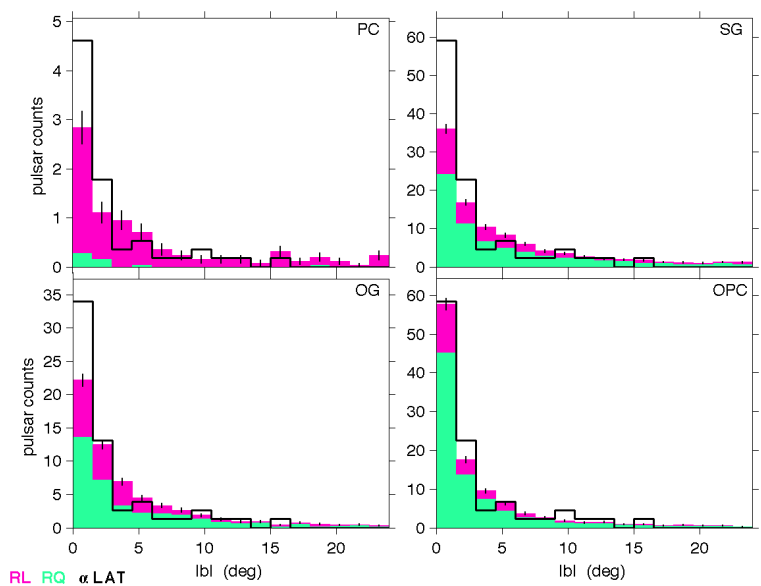

Figure 5.30: Latitude distributions obtained for each model for the visible gamma-ray objects. Pink and green dots refer to radio-loud and radio-quiet gamma-ray objects, respectively. The observed LAT pulsar distribution (in black) has been scaled to the total number of visible objects for each model to ease the comparison.

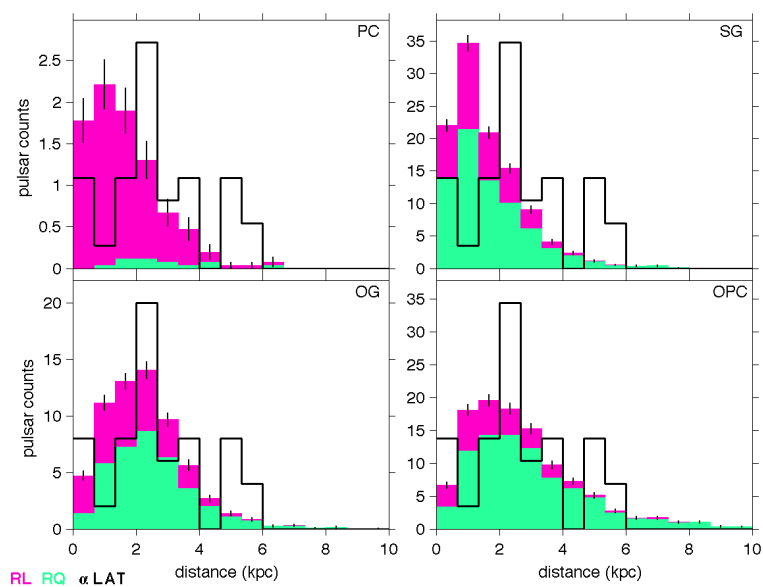

Figure 5.31: Distance distributions obtained for each model for the visible gamma-ray objects. The stacked pink and green histograms refer to radio-loud and radio-quiet gamma-ray objects, respectively. The observed LAT pulsar distribution (in black) has been scaled to the total number of visible objects for each model to ease the comparison and show the relative overabundance of nearby objects and lack of distant ones.

Since the measurement of the LAT pulsar distances is often affected by huge uncertainties, the distances and Galactic coordinates comparison between the models and the observed population could lead to uncertain conclusions.

**The radio $\gamma$-loud population**

Another interesting comparison between the LAT observed pulsars and the model predictions can be made by plotting the radio flux of the *gamma*-visible pulsars in figure 5.32. In the PC case, the radio flux of the $\gamma$-loud objects is clearly over-predicted and this could be one of the causes of radio-loud pulsars over-prediction. Considering the $\gamma$-loud objects, the radio emission model is not able to describe the observed population: it completely misses the majority of the observed distribution at low radio fluxes. The SG model shows a much broader distribution that, like in the PC case, over-predicts the radio flux. On the other side, the low flux tail of the $\gamma$-loud objects flux distribution covers the whole range of the observed values and is able to well describe the second peak in the observed flux distribution.

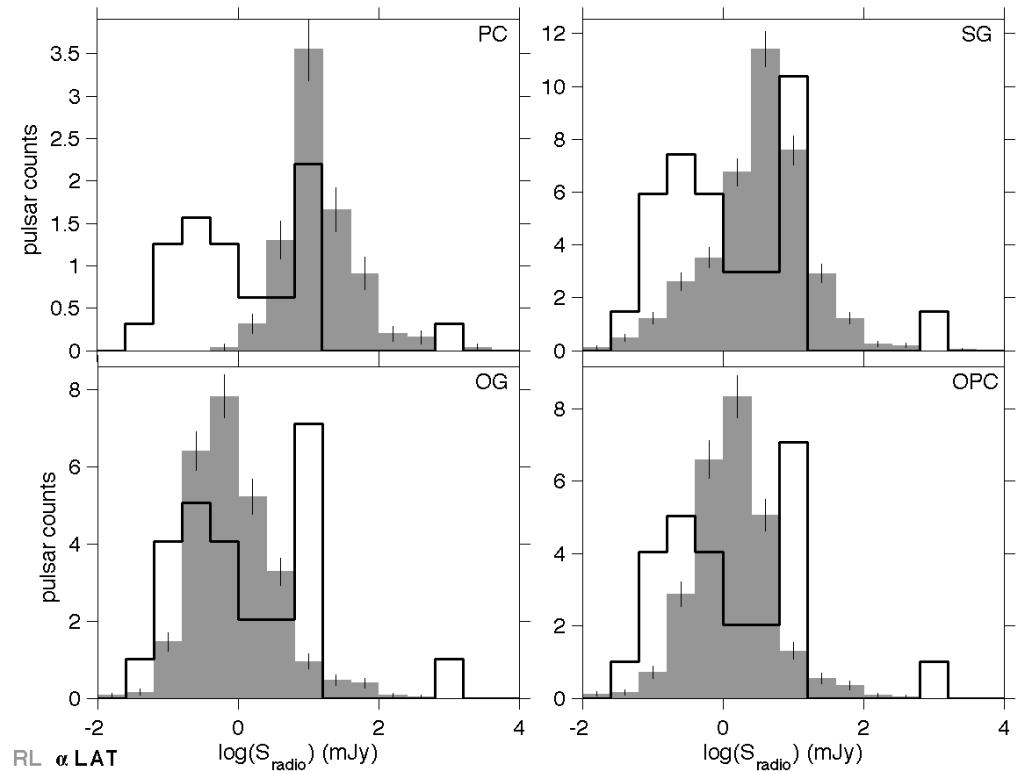

Figure 5.32: Radio-flux distributions obtained for the visible radio-loud gamma-ray objects for each model. The observed LAT pulsar distribution (in black) has been scaled to the total number of visible objects for each model to ease the comparison.

The OG and OPC γ-loud objects show more or less the same radio flux distributions. In both cases, the radio emission is predicted in the whole range of the observed values with the OG γ-loud population that best describes the radio emission for the first peak of the observed flux distribution.

**Evolution of the radio-loud and radio-quiet populations**

We studied the evolution of the radio-visible and gamma-visible samples of the simulated and real populations with respect to the spin-down power $\dot{E}$.

Figure 5.33 shows the evolution of the ratio $N_{rad+\gamma}/N_\gamma$ versus $\dot{E}$. For the LAT pulsars we see a decrease of the number of radio-loud objects with age (decreasing $\dot{E}$): the LAT pulsars are born as radio and γ sources and gradually loose the radio component as they age. None of the models can describe the observed trend. The predicted evolution of the $N_{rad+\gamma}/N_\gamma$ ratio with $\dot{E}$ is in fact opposite to the observed one. Moreover there is a precise interval of $\dot{E}$ values (timescale) where the most important change occurs in the LAT sample for $\dot{E} > 10^{29.5} W$. After this first steep decrease, $N_{rad+\gamma}/N_\gamma$ remains stable around $\sim 55\%$. This trend is difficult to explain by the shrinking of the radio emission beam with age that should generate a constant decrease of the ratio with decreasing $\dot{E}$. It could just be responsible of the slight decrease of the $N_{rad+\gamma}/N_\gamma$ ratio below $10^{29}$ W. The timescale that is sampled by the LAT pulsars is of the order of $10^5$ years. The loss of radio emission in such a short timescale could be due to a fast variation of the magnetic obliquity $\alpha$ or it could depend on different evolution of the dipole radiation for very high spin-down objects.

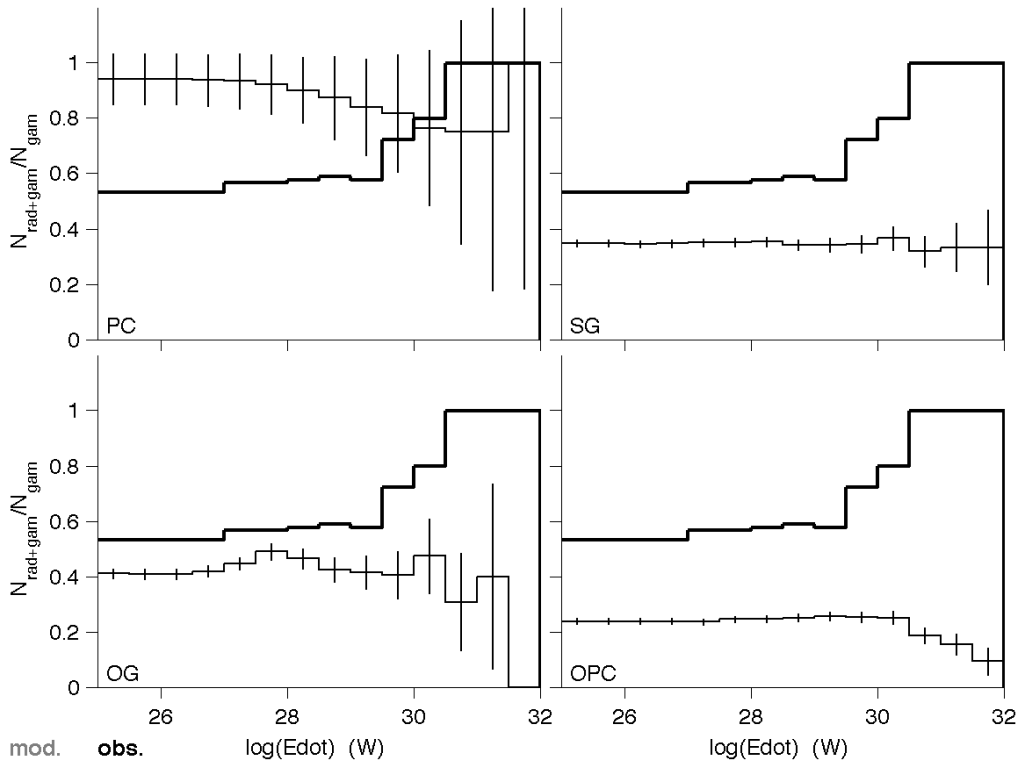

Figure 5.33: Evolution with spin-down power of the fraction of radio-loud objects among the gamma-ray loud ones. The thin lines give the simulation results for each model. The thick line gives the fraction evolution in the LAT sample.

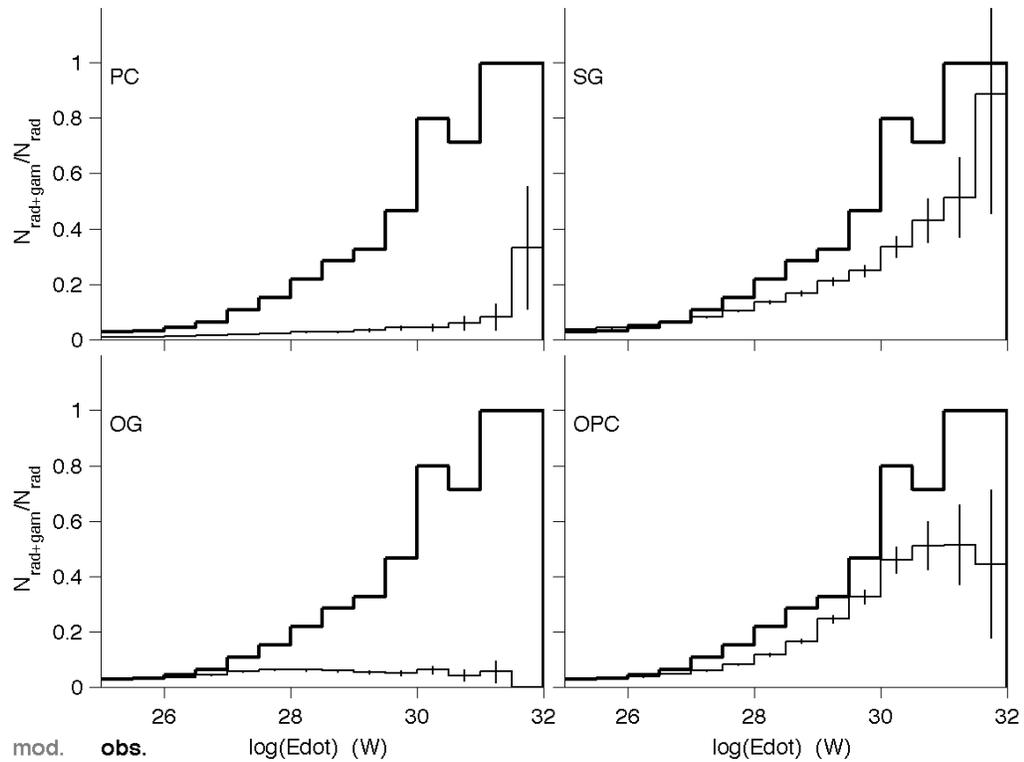

Figure 5.34: Evolution with spindown power of the fraction of gamma-loud objects among the radio loud ones. The thin lines give the simulation results for each model. The thick line gives the fraction evolution in the LAT sample.

Figure 5.34 shows the evolution of the ratio $N_{rad+\gamma}/N_{rad}$ versus $\dot{E}$. The LAT population is characterised by a relative loss of gamma emission among the radio loud objects with decreasing $\dot{E}$. Figures 5.33 and 5.34 jointly illustrate the very high probability of detecting both the radio and $\gamma$-ray beams in LAT objects with $\dot{E} > 10^{30}$ W, as opposed to all model predictions. The regular decrease of the $N_{rad+\gamma}/N_{rad}$ ratio for $\dot{E}$ lower than $10^{30}$ W is an argument in favour of a short timescale process that enhances the probability of detecting both beams at high $\dot{E}$. The constant decrease of the $N_{rad+\gamma}/N_{rad}$ ratio is due to the fact that the population is getting older, the polar cap shrinks and we loose the radio emission (the probability for the observer line of sight to cross the radio beam decreases).

While the PC and OG hardly manage to describe the observed increasing trend with $\dot{E}$, the SG and OPC do describe the correct trend especially when considering the lack of high $\dot{E}$ detections in these models.

In summary, all models studied here under-predict the number of visible $\gamma$-ray pulsars at high $\dot{E}$, as well as the number of $\gamma$-loud and radio-loud objects at high $\dot{E}$. The discrepancy with the observations is significant despite our choice of birth distributions skewed to young energetic objects at slight variance with the constraints imposed by the total radio and gamma pulsar sample observed. The use of different $\gamma$-ray luminosity evolutions and different beam patterns in the 4 models suggests a different origin of the discrepancy. Concentrating the birth location in the inner Galaxy helped, but did not solve the discrepancy. Increasing the number of energetic pulsars near the sun would conflict with the observed pulsar distances even more than in Figure 5.31. The estimate of the visibility threshold in radio or gamma flux is not at stake since all models over-predict the number of older, fainter, visible objects. The impact of a magnetic alignment with age or the choice of different braking indices for the pulsar spin-down will be studied in the future.

# Chapter 6

# Evaluation of the magnetic obliquity $\alpha$ and line of sight $\zeta$ of the observed *Fermi-LAT* $\gamma$-ray pulsar

This chapter describes the second part of my thesis project, the evaluation of the magnetic obliquity $\alpha$ and line of sight angle $\zeta$ for some of the $\gamma$-ray pulsars observed by the *Fermi-LAT* $\gamma$-ray telescope. The adopted strategy is based on fitting the observed pulsar light-curves with the high-energy and radio phase-plots (section 5.2) that describe the emission pattern of the $\gamma$-ray and radio emission models (chapter 2).

At first, I performed a fit between the LAT light curves, obtained as described in section 4.2.2, and the $\gamma$-ray phase-plot light curves simulated following the models we have studied. This $\gamma$-ray fit led to a first $\alpha$-$\zeta$ estimate based just on the high-energy emission geometry.

Next, I performed a fit of some of the LAT pulsar radio profiles with the radio model phase-plot light curves. This radio fit has been used to implement a joint $\gamma$-radio estimate of the pulsar orientation parameters that better constrains $\alpha$ and $\zeta$.

In the following sections I will give a detailed description of the adopted fit strategy and of the joint $\gamma$-radio estimate of the LAT pulsar $\alpha$ & $\zeta$ angles. They represent the starting point of the paper in preparation: *'Observational constraints on the obliquities and aspect angles of the Fermi pulsars*, by Marco Pierbattista, Isabelle Grenier, Alice Harding, Peter Gonthier.

## 6.1 Individual $\gamma$-ray fit

A first estimate of the $\alpha$ & $\zeta$ angles obtained by fitting the LAT pulsar profiles and the phase-plot light curves concerns 47 LAT pulsars. The names of the analysed pulsars have been printed in bold in the tables from 4.1 to 4.4.

### 6.1.1   LAT pulsar, phase-plots & probability distribution

Once I obtained the light-curves of the LAT pulsars as described in section 4.2.1, I used the emission models to generate the phase-plots for the specific gap width of the pulsar (evaluated with equations 2.14, 2.27, and 2.19 for SGs, OG, and OPC), the real period and magnetic field values (just for the PC), and by applying the same interpolation technique described in section 5.3, with an $\alpha$ step of 1 degree. The emission pattern of each pulsar was described by 90 phase-plot panels, from $\alpha$=1 to 90 degrees, obtained for the real pulsar characteristics.

Since the LAT pulsar light curves are computed by time integration of a phase-time diagram (figures from 4.2 to 4.9) containing discrete, independent, and rare events, the probability distribution that best describes the pulsar $\gamma$-ray emission follows the Poisson statistics.

**Poisson probability distribution**

The Poisson probability distribution is expressed as

$$P_j(N_{obs,j}, N_{sim,j}) = \frac{e^{-N_{sim,j}} N_{sim,j}^{N_{obs,j}}}{N_{obs,j}!}. \tag{6.1}$$

This equation expresses the probability that the measurement in the *j-bin* of the observed light curve (*obs*) is compatible with that expected in the the *j-bin* of the simulated phase-plot one (*sim*). The total likelihood between the observed and simulated distributions (light curves, LTCs) is

$$L_{tot} = P_{tot}(LTC_{obs}, LTC_{sim}) = \prod_j \frac{e^{-N_{sim,j}} N_{sim,j}^{N_{obs,j}}}{N_{obs,j}!} \tag{6.2}$$

that, written in logarithmic form, becomes

$$\ln L_{tot} = -\sum_j N_{sim,j} + \sum_j N_{obs,j} \ln N_{sim,j} - \sum_j \ln(N_{obs,j}!). \tag{6.3}$$

Since the best-fit solution is obtained by maximising the probability distribution with respect to the fit parameters, the constant additive term $\ln(N_{obs,j}!)$ is not going to change the slope of the likelihood distribution, so it can be dropped. The final likelihood probability distribution I used to fit the LAT observed pulsar light curves can be written as

$$\ln L_{tot} = -\sum_j N_{sim,j} + \sum_j N_{obs,j} \ln N_{sim,j}. \tag{6.4}$$

**Gaussian probability distribution**

I have implemented a different $\gamma$-ray fit using Regular Binned (RB) curves, and a Gaussian ($\chi^2$) probability distribution. The Gaussian probability distribution is defined as

$$P_j(LTC_{obs,j}, LTC_{sim,j}) = \frac{1}{\sigma_{sim,j}\sqrt{2\pi}} e^{-\frac{(LTC_{obs,j} - LTC_{sim,j})^2}{2\sigma_{sim,j}^2}} \qquad (6.5)$$

that written as a logarithmic likelihood of the total distribution (light curve) is

$$\ln L_{tot} = -\sum_j \frac{(LTC_{obs,j} - LTC_{sim,j})^2}{2\sigma_{sim,j}^2} - \sum_j \ln(\sigma_{sim,j}\sqrt{2\pi}). \qquad (6.6)$$

Since the errors $\sigma_{sim}$ do not depend on the model free parameters, the last constant term could be dropped, and it is possible to write

$$\ln L_{tot} = -\sum_j \frac{(LTC_{obs,j} - LTC_{sim,j})^2}{2\sigma_{sim,j}^2} = -\frac{1}{2}\chi^2. \qquad (6.7)$$

Equation 6.7 defines the $\chi^2$ probability distribution I used for the second fit estimate.

### 6.1.2   Fitting method

**Poisson fit between fixed-count binned light curves**

First, I have implemented a 5 free parameter fit, by using the Poisson distribution described in equation 6.4 and re-binning both the simulated and observed light curves by using the Fixed Count Binning (FCB) technique.

The FCB is a re-binning method to increase the resolution of the peaks in a binned distribution. It consists in re-defining the size of each bin in order to have the same number of counts per bin. By dividing the total photon number contained in the light curve for the total bin number, one obtains the photon number to assign to each bin

$$\langle n_{ph} \rangle = \frac{1}{N_{bin}} \sum_{i=1}^{N_{bin}} n_i$$

Starting from phase $\phi = 0$ and proceeding toward phase $\phi = 1$, the new bin size $\Delta\phi_i = \phi_i - \phi_{i-1}$ will have to contain $\langle n_{ph} \rangle$ photons, and so the boundaries $\phi_i$ and $\phi_{i-1}$ will be defined by using the relation

$$\int_{\phi_{i-1}}^{\phi_i} n(\phi) d\phi = \langle n_{ph} \rangle. \qquad (6.8)$$

By re-binning the curve in this way, we will have few bins in the light curve region with a low photon number and many small bins in the light curve regions with a lot of events e.g. the peaks of the light curve. The advantage
in performing a fit between two FCB curves is that, since the peak regions are
described by higher bin number, the statistical weight of the peak is higher
than the rest of the curve and the best fit solution will be more driven by the
peak structure.

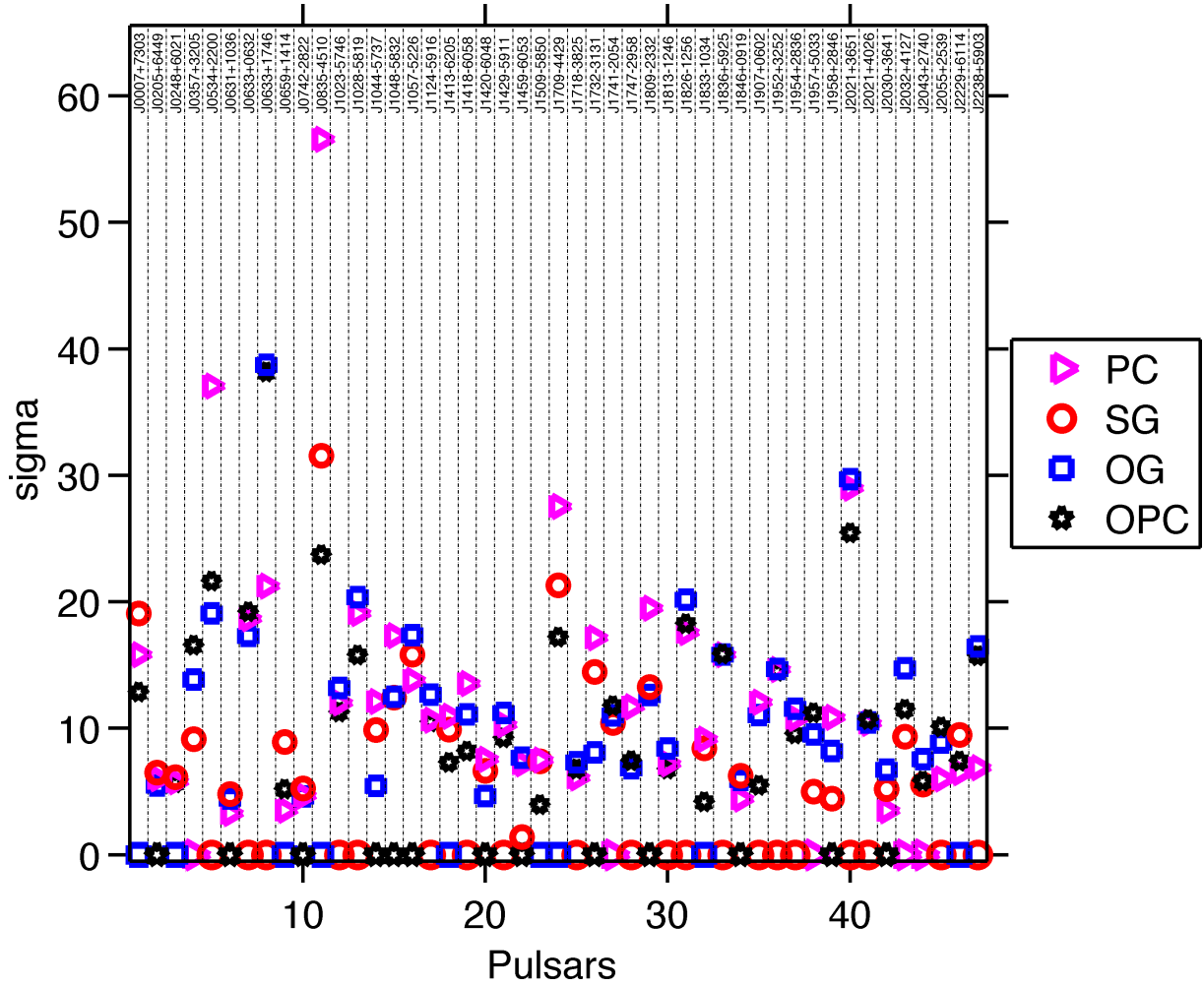

Figure 6.1: Relative significance of the Poisson likelihood fit solutions obtained for
each pulsar of tables 6.1 & 6.2 as derived from the log-likelihood ratio found between
two models. The rms significance between the best-fit solution for one model and
the very best fit model is given by $\sigma = \sqrt{2\ln(L_{bestfitmodel}) - \ln(L_{model,max})}$. The
very best solution is plotted at 0 sigma. A 10 sigma value means that the given
model yields a 10 sigma worse solution than the very best one.

The free parameters are: the $\alpha$ angle, the $\zeta$ angle, a background flat level,
a light-curve normalisation factor, and a phase shift. Computationally, for
each LAT pulsar of the sample and each $\gamma$-ray model, the logarithmic Poisson
likelihood, defined in equation 6.4, has been evaluated for all the possible
combinations of the 5 free parameters. We used phase shift from 0 to 1 in 45 phase steps, for each one we tested 15 flat background steps, and, for each background value, 10 different normalisation values have been applied. As a result, for each pulsar a total of $5.4675 \times 10^6$ likelihood values have been stored, corresponding to a likelihood matrix of $90_\alpha \times 90_\zeta \times 45_{\phi~steps} \times 15_{flat~bkg} \times 10_{norm}$. This matrix has then been maximised with respect to the last 3 dimensions and the likelihood map and and the best-fit value of each parameter have been stored.

| | $\alpha_{PC}$ | $\alpha_{SG}$ | $\alpha_{OG}$ | $\alpha_{OPC}$ | $\zeta_{PC}$ | $\zeta_{SG}$ | $\zeta_{OG}$ | $\zeta_{OPC}$ |
|---|---|---|---|---|---|---|---|---|
| J0007+7303 | $6^4_6$ | $40^{0.02}_{0.04}$ | $19^{0.03}_{0.02}$ | $17^{0.02}_{0.05}$ | $3^{0.005}_{0.005}$ | $69^{0.02}_{0.02}$ | $87^{0.02}_{0.01}$ | $73^{0.04}_{0.01}$ |
| J0205+6449 | $62^{0.1}_{0.07}$ | $51^{0.1}_{0.1}$ | $54^{0.2}_{0.1}$ | $60^{0.3}_{0.1}$ | $57^{0.1}_{0.07}$ | $48^{0.04}_{0.1}$ | $47^{0.07}_{0.1}$ | $90^{0.0005}_{0.01}$ |
| J0248+6021 | $7^{0.1}_{0.1}$ | $39^{0.5}_{0.3}$ | $10^{0.2}_{0.7}$ | $7^{0.09}_{0.3}$ | $14^{0.09}_{0.3}$ | $68^{0.2}_{0.07}$ | $82^{0.2}_{0.07}$ | $78^{0.03}_{0.1}$ |
| J0357+3205 | $6^4_6$ | $82^{0.1}_{0.1}$ | $82^{0.03}_{0.03}$ | $70^{0.03}_{0.02}$ | $3^{0.03}_{0.02}$ | $21^{0.1}_{0.2}$ | $86^{0.03}_{0.04}$ | $71^{0.07}_{0.04}$ |
| J0534+2200 | $16^{0.005}_{0.005}$ | $42^{0.07}_{0.05}$ | $35^{0.008}_{0.02}$ | $36^{0.05}_{0.01}$ | $15^{0.01}_{0.004}$ | $74^{0.009}_{0.02}$ | $73^{0.007}_{0.007}$ | $72^{0.01}_{0.007}$ |
| J0631+1036 | $10^{0.1}_{0.2}$ | $60^{0.2}_{0.3}$ | $28^{0.5}_{0.1}$ | $80^{0.3}_{0.6}$ | $7^{0.2}_{0.2}$ | $14^{0.2}_{0.1}$ | $83^{0.1}_{0.1}$ | $23^{0.3}_{0.3}$ |
| J0633+0632 | $90^{0.0005}_{0.002}$ | $57^{0.3}_{0.2}$ | $30^{0.3}_{0.03}$ | $9^{0.3}_{0.02}$ | $90^{0.0005}_{0.001}$ | $87^{0.1}_{0.1}$ | $88^{0.02}_{0.02}$ | $82^{0.02}_{0.01}$ |
| J0633+1746 | $87^{0.02}_{0.003}$ | $60^{0.3}_{0.1}$ | $56^{0.005}_{0.007}$ | $58^{0.006}_{0.005}$ | $87^{0.009}_{0.02}$ | $89^{0.03}_{0.03}$ | $78^{0.005}_{0.002}$ | $60^{0.004}_{0.002}$ |
| J0659+1414 | $9^{0.3}_{0.1}$ | $33^{0.2}_{0.2}$ | $22^{0.8}_{0.07}$ | $22^{0.09}_{0.1}$ | $9^{0.05}_{0.1}$ | $37^1_{0.3}$ | $89^{0.2}_{0.09}$ | $85^{0.06}_{0.1}$ |
| J0742-2822 | $16^{0.3}_{0.005}$ | $62^{0.3}_{0.01}$ | $64^{0.7}_{0.2}$ | $70^{0.3}_{0.4}$ | $4^{0.2}_{0.04}$ | $36^{0.5}_{0.3}$ | $63^{0.2}_{0.3}$ | $29^{0.09}_{0.4}$ |
| J0835-4510 | $7^{0.005}_{0.3}$ | $78^{0.01}_{0.2}$ | $48^{0.005}_{0.006}$ | $56^{0.003}_{0.002}$ | $7^{0.008}_{0.002}$ | $47^{0.01}_{0.02}$ | $83^{0.004}_{0.01}$ | $77^{0.004}_{0.006}$ |
| J1023-5746 | $7^{0.04}_{0.03}$ | $85^{0.07}_{0.6}$ | $13^{0.03}_{0.05}$ | $11^{0.2}_{0.01}$ | $6^{0.02}_{0.02}$ | $35^{0.2}_{0.2}$ | $78^{0.02}_{0.02}$ | $74^{0.02}_{0.02}$ |
| J1028-5819 | $87^{0.1}_{0.05}$ | $86^{0.05}_{0.03}$ | $31^{0.04}_{0.02}$ | $60^{0.04}_{0.02}$ | $89^{0.6}_{0.1}$ | $43^{0.08}_{0.09}$ | $84^{0.04}_{0.06}$ | $87^{0.02}_{0.01}$ |
| J1044-5737 | $4^2_2$ | $66^{0.1}_{0.8}$ | $70^{0.05}_{0.2}$ | $57^{0.07}_{0.03}$ | $9^{0.9}_{0.01}$ | $50^{0.08}_{0.08}$ | $81^{0.06}_{0.1}$ | $75^{0.03}_{0.04}$ |
| J1048-5832 | $4^2_4$ | $80^{0.2}_{0.04}$ | $39^{0.02}_{0.03}$ | $60^{0.06}_{0.08}$ | $8^{0.01}_{0.02}$ | $38^{0.3}_{0.2}$ | $81^{0.1}_{0.02}$ | $76^{0.04}_{0.01}$ |
| J1057-5226 | $10^{0.02}_{0.03}$ | $76^{0.08}_{0.06}$ | $28^{0.02}_{0.01}$ | $15^{0.02}_{0.02}$ | $5^{0.01}_{0.01}$ | $19^{0.07}_{0.04}$ | $90^{0.04}_{0.02}$ | $88^{0.03}_{0.02}$ |
| J1124-5916 | $90^{0.0005}_{0.01}$ | $89^{0.2}_{0.2}$ | $2^2_2$ | $61^{0.05}_{0.09}$ | $88^{0.02}_{0.08}$ | $35^{0.05}_{0.2}$ | $78^{0.04}_{0.01}$ | $83^{0.03}_{0.08}$ |
| J1413-6205 | $7^{0.06}_{0.04}$ | $70^{0.1}_{0.09}$ | $24^1_{0.03}$ | $51^{0.05}_{0.04}$ | $9^{0.1}_{0.02}$ | $42^{0.2}_{0.1}$ | $80^{0.2}_{0.05}$ | $76^{0.04}_{0.06}$ |
| J1418-6058 | $7^{0.03}_{0.02}$ | $84^{0.05}_{0.1}$ | $43^{0.02}_{0.02}$ | $60^{0.04}_{0.03}$ | $8^{0.01}_{0.03}$ | $34^{0.05}_{0.08}$ | $85^{0.2}_{0.02}$ | $84^{0.2}_{0.1}$ |
| J1420-6048 | $14^{0.07}_{0.08}$ | $64^{0.1}_{0.2}$ | $25^{0.07}_{0.07}$ | $17^{0.1}_{0.2}$ | $10^{0.09}_{0.2}$ | $37^{0.2}_{0.1}$ | $74^{0.07}_{0.07}$ | $71^{0.2}_{0.2}$ |
| J1429-5911 | $7^{0.03}_{0.05}$ | $47^{0.6}_{1.6}$ | $13^{0.04}_{0.09}$ | $10^{0.02}_{0.06}$ | $7^{0.02}_{0.05}$ | $85^{1.1}_{0.9}$ | $83^{0.05}_{0.04}$ | $77^{0.03}_{0.02}$ |
| J1459-6053 | $13^{2.1}_2$ | $22^{1.4}_{0.7}$ | $88^{0.2}_{0.1}$ | $78^{0.06}_{0.2}$ | $3^{0.01}_{0.04}$ | $52^{2.3}_{1.3}$ | $51^{0.2}_{0.02}$ | $12^{0.08}_{0.03}$ |
| J1509-5850 | $16^{0.02}_{0.05}$ | $41^{0.1}_{0.2}$ | $27^{0.6}_{0.09}$ | $29^{0.1}_{0.1}$ | $2^2_{0.05}$ | $37^{0.05}_{0.03}$ | $87^{0.2}_{0.04}$ | $71^{0.2}_{0.05}$ |
| J1709-4429 | $13^2_2$ | $38^{0.02}_{0.02}$ | $16^{0.01}_{0.01}$ | $6^{0.01}_{0.009}$ | $3^{0.006}_{0.005}$ | $68^{0.01}_{0.008}$ | $76^{0.01}_{0.01}$ | $73^{0.007}_{0.01}$ |

Table 6.1: $\alpha$ and $\zeta$ best fit solutions (in degrees )resulting from the $\gamma$-ray fit for the first 24 pulsars of the analysed sample. At each value is associated the statistical error.

For each pulsar, such fitting process produced a bi-dimensional $\alpha$-$\zeta$ likelihood map and a best fit value for the flat level, normalisation factor, and the phase shift. The best fit $\alpha$ and $\zeta$ estimate, obtained by maximisation of the $\alpha - \zeta$ likelihood map, are listed in Tables 6.1 & 6.2.

The best-fit flat level, normalisation factor, and phase shift have been used to plot the $\gamma$-ray best-fit light curves (Figure 6.3 to Figure 6.49). Figure 6.1 compares the significance obtained between models as measured by the decrease in log-likelihood between the best-fit solutions found for each model.

| | $\alpha_{PC}$ | $\alpha_{SG}$ | $\alpha_{OG}$ | $\alpha_{OPC}$ | $\zeta_{PC}$ | $\zeta_{SG}$ | $\zeta_{OG}$ | $\zeta_{OPC}$ |
|---|---|---|---|---|---|---|---|---|
| J1718-3825 | $16^{0.1}_{0.04}$ | $32^{0.2}_{0.2}$ | $11^{0.4}_{0.8}$ | $7^{0.4}_{0.1}$ | $8^{0.2}_{0.5}$ | $36^{0.2}_{0.1}$ | $79^{0.06}_{0.06}$ | $79^{0.07}_{0.2}$ |
| J1732-3131 | $59^{0.1}_{0.1}$ | $47^{0.07}_{0.07}$ | $29^{0.04}_{0.5}$ | $46^{0.02}_{0.02}$ | $57^{0.2}_{0.06}$ | $51^{0.06}_{0.03}$ | $87^{0.5}_{0.3}$ | $82^{0.02}_{0.03}$ |
| J1741-2054 | $3^{3}_{3}$ | $82^{0.2}_{0.2}$ | $84^{0.1}_{0.08}$ | $30^{0.03}_{0.08}$ | $40^{0.01}_{0.02}$ | $15^{0.09}_{0.04}$ | $90^{0.0005}_{0.0005}$ | $90^{0.0005}_{0.0005}$ |
| J1747-2958 | $7^{0.07}_{0.06}$ | $70^{0.2}_{0.4}$ | $39^{0.5}_{0.06}$ | $57^{0.1}_{0.1}$ | $8^{0.06}_{0.2}$ | $49^{0.4}_{0.2}$ | $81^{0.1}_{0.04}$ | $79^{0.1}_{0.1}$ |
| J1809-2332 | $4^{2}_{4}$ | $32^{0.2}_{0.1}$ | $70^{0.01}_{0.05}$ | $37^{0.06}_{0.01}$ | $90^{0.005}_{0.01}$ | $78^{0.02}_{0.1}$ | $78^{0.02}_{0.02}$ | $72^{0.02}_{0.02}$ |
| J1813-1246 | $8^{0.1}_{0.1}$ | $90^{0.0005}_{0.1}$ | $2^{3}_{2}$ | $11^{0.08}_{0.04}$ | $10^{0.3}_{0.09}$ | $28^{0.5}_{0.3}$ | $78^{0.1}_{0.03}$ | $75^{0.04}_{0.05}$ |
| J1826-1256 | $7^{0.02}_{0.01}$ | $77^{0.08}_{0.4}$ | $20^{0.009}_{0.05}$ | $65^{0.007}_{0.03}$ | $70^{0.008}_{0.01}$ | $82^{0.06}_{0.05}$ | $79^{0.03}_{0.004}$ | $84^{0.02}_{0.01}$ |
| J1833-1034 | $17^{0.08}_{0.09}$ | $80^{0.2}_{0.2}$ | $66^{0.2}_{0.2}$ | $87^{0.2}_{0.3}$ | $5^{0.1}_{0.1}$ | $38^{0.1}_{0.1}$ | $81^{0.1}_{0.1}$ | $74^{0.1}_{0.2}$ |
| J1836+5925 | $15^{0.2}_{0.09}$ | $89^{0.2}_{0.04}$ | $37^{0.05}_{0.3}$ | $32^{0.09}_{0.04}$ | $8^{0.06}_{0.2}$ | $20^{0.04}_{0.05}$ | $90^{0.0005}_{0.3}$ | $87^{0.04}_{0.2}$ |
| J1846+0919 | $10^{0.05}_{0.07}$ | $24^{0.1}_{0.2}$ | $32^{0.1}_{1.1}$ | $13^{0.2}_{0.3}$ | $5^{0.08}_{0.3}$ | $20^{0.2}_{0.3}$ | $90^{0.0005}_{0.3}$ | $90^{0.0005}_{0.02}$ |
| J1907+0602 | $7^{0.03}_{0.02}$ | $77^{0.05}_{0.05}$ | $35^{0.04}_{0.1}$ | $21^{0.1}_{0.02}$ | $90^{0.008}_{0.02}$ | $31^{0.1}_{0.02}$ | $81^{0.04}_{0.03}$ | $73^{0.05}_{0.04}$ |
| J1952+3252 | $12^{0.01}_{0.02}$ | $85^{0.04}_{0.03}$ | $45^{0.08}_{0.1}$ | $8^{0.1}_{0.3}$ | $7^{0.04}_{0.02}$ | $47^{0.08}_{0.03}$ | $84^{0.04}_{0.3}$ | $86^{0.3}_{0.1}$ |
| J1954-2836 | $7^{0.08}_{0.03}$ | $57^{0.1}_{0.4}$ | $40^{0.05}_{0.1}$ | $54^{0.2}_{0.2}$ | $70^{0.04}_{0.03}$ | $80^{0.08}_{0.02}$ | $87^{0.1}_{0.07}$ | $80^{0.3}_{0.06}$ |
| J1957+5033 | $5^{5}_{5}$ | $83^{0.2}_{0.3}$ | $88^{0.2}_{0.1}$ | $84^{0.06}_{0.07}$ | $3^{0.06}_{0.06}$ | $12^{0.2}_{0.1}$ | $81^{0.06}_{0.06}$ | $60^{0.08}_{0.08}$ |
| J1958+2846 | $31^{0.03}_{0.2}$ | $41^{0.06}_{0.09}$ | $63^{0.2}_{0.03}$ | $47^{0.3}_{0.2}$ | $25^{0.08}_{0.07}$ | $46^{0.4}_{0.07}$ | $90^{0.0005}_{0.1}$ | $85^{0.08}_{0.09}$ |
| J2021+3651 | $87^{0.06}_{0.03}$ | $66^{0.07}_{0.005}$ | $20^{0.003}_{0.009}$ | $65^{0.003}_{0.01}$ | $88^{0.04}_{0.04}$ | $84^{0.02}_{0.03}$ | $79^{0.007}_{0.001}$ | $84^{0.06}_{0.004}$ |
| J2021+4026 | $6^{0.05}_{0.1}$ | $51^{0.2}_{0.2}$ | $68^{0.04}_{0.04}$ | $77^{0.09}_{0.04}$ | $13^{0.2}_{0.1}$ | $30^{0.07}_{0.04}$ | $64^{0.03}_{0.04}$ | $26^{0.04}_{0.02}$ |
| J2030+3641 | $10^{0.08}_{0.04}$ | $82^{0.4}_{0.5}$ | $85^{0.1}_{0.1}$ | $13^{0.2}_{0.3}$ | $4^{0.09}_{0.2}$ | $9^{0.1}_{0.2}$ | $90^{0.0005}_{0.03}$ | $90^{0.0005}_{0.03}$ |
| J2032+4127 | $89^{0.1}_{0.03}$ | $64^{0.3}_{0.6}$ | $48^{0.06}_{0.07}$ | $61^{0.01}_{0.01}$ | $83^{0.06}_{0.08}$ | $90^{0.0005}_{0.005}$ | $63^{0.06}_{0.03}$ | $90^{0.0005}_{0.03}$ |
| J2043+2740 | $4^{2.2}_{4}$ | $46^{0.1}_{0.2}$ | $61^{0.08}_{0.2}$ | $66^{0.3}_{0.1}$ | $9^{0.05}_{0.3}$ | $52^{0.2}_{0.7}$ | $86^{0.09}_{0.08}$ | $89^{0.2}_{0.1}$ |
| J2055+2539 | $6^{4}_{6}$ | $75^{0.2}_{0.2}$ | $85^{0.06}_{0.06}$ | $87^{0.06}_{0.04}$ | $3^{0.04}_{0.04}$ | $25^{0.2}_{0.3}$ | $90^{0.0005}_{0.02}$ | $69^{0.05}_{0.05}$ |
| J2229+6114 | $15^{0.02}_{0.009}$ | $36^{0.04}_{0.04}$ | $83^{0.9}_{0.4}$ | $68^{0.03}_{0.02}$ | $70^{0.03}_{0.03}$ | $58^{0.01}_{0.03}$ | $25^{0.04}_{0.04}$ | $26^{0.02}_{0.02}$ |
| J2238+5903 | $90^{0.0005}_{0.004}$ | $65^{0.04}_{0.04}$ | $14^{0.04}_{0.03}$ | $10^{0.01}_{0.03}$ | $88^{0.03}_{0.09}$ | $89^{0.3}_{0.06}$ | $83^{0.04}_{0.02}$ | $77^{0.02}_{0.02}$ |

Table 6.2: $\alpha$ and $\zeta$ best fit solutions (in degrees) resulting from the $\gamma$-ray fit for the last 23 pulsars of the analysed sample. At each value is associated the statistical error.

## $\chi^2$ fit between regular binned light curves and systematic errors on $(\alpha, \zeta)$ estimates

In order to give an estimate of the systematic errors that affect the $\alpha$ & $\zeta$ measures from the statistical analysis, a second fit has been implemented by using the logarithmic Gaussian $\chi^2$ distribution described in equation 6.7. For this second fit the simulated and observed light-curves have been produced in regular binning. To be consistent with the Poisson fit, I have used the same free parameters interval, equally stepped. A $\chi^2$ likelihood requires the knowledge of one more parameter, the *variance $\sigma$* of the simulated distribution. Since the calculation of our model phase-plots does not yield any error, each simulated light curve variance has been evaluated by setting the reduced $\chi^2$ to 1 for the best fit light-curve obtained from the Poisson FCB likelihood.

Let us define, for each pulsar, the best-fit light curve obtained by Poisson

FCB likelihood as $f^*(x)$. The reduced $\chi^2$ criterion gives

$$\frac{1}{n_{free}} \sum_j \frac{[y_j - f^*(x_j)]^2}{\sigma_*^2} = 1 \qquad (6.9)$$

that, solved for $\sigma_*^2$, gives

$$\sigma_*^2 = \frac{1}{n_{free}} \sum [y_j - f^*(x_j)]^2 \qquad (6.10)$$

where $y$ is the observed light curve. $\sigma_*$ has then been used to fit the light-curves with $\sigma_{sim} = \sigma_*$ in Equation 6.7. The $\alpha$-$\zeta$ best fit solutions obtained by using the $\chi^2$ statistic and regular binning slightly differ from those found with the Poisson likelihood fixed count binning.

|        |          | PC       | SG   | OG   | OPC  |
|--------|----------|----------|------|------|------|
| $1\sigma$ | $\alpha$ | $\pm 1$  | $\pm 1$ | $\pm 2$ | $\pm 4$ |
|        | $\zeta$  | $\pm < 1$ | $\pm 1$ | $\pm 1$ | $\pm 1$ |
| $1.5\sigma$ | $\alpha$ | $\pm 3$ | $\pm 3$ | $\pm 5$ | $\pm 10$ |
|        | $\zeta$  | $\pm 2$  | $\pm 3$ | $\pm 2$ | $\pm 4$ |
| $2\sigma$ | $\alpha$ | $\pm 27$ | $\pm 24$ | $\pm 30$ | $\pm 58$ |
|        | $\zeta$  | $\pm 27$ | $\pm 24$ | $\pm 10$ | $\pm 46$ |

Table 6.3: Estimate of the systematic errors on $\alpha$ and $\zeta$ obtained from the comparison of the fixed-count binning Poisson fit and the regular binning gaussian fit.

We have studied the distribution of the discrepancy between the solutions obtained with the two different methods. Figure 6.2 shows the function $1 - f_{cum}(x)$, where $f_{cum}(x)$ is the cumulative distribution of the differences between the two sets of solutions, and the confidence region that can be estimated at the 1, 1.5, and $2\sigma$ levels displayed in Table 6.3. They show that the formal fitting errors in Tables 6.1 and 6.2 are clearly under-estimated and that the fitting method itself yields an uncertainties of a few degrees at least on $\alpha$ and $\zeta$.

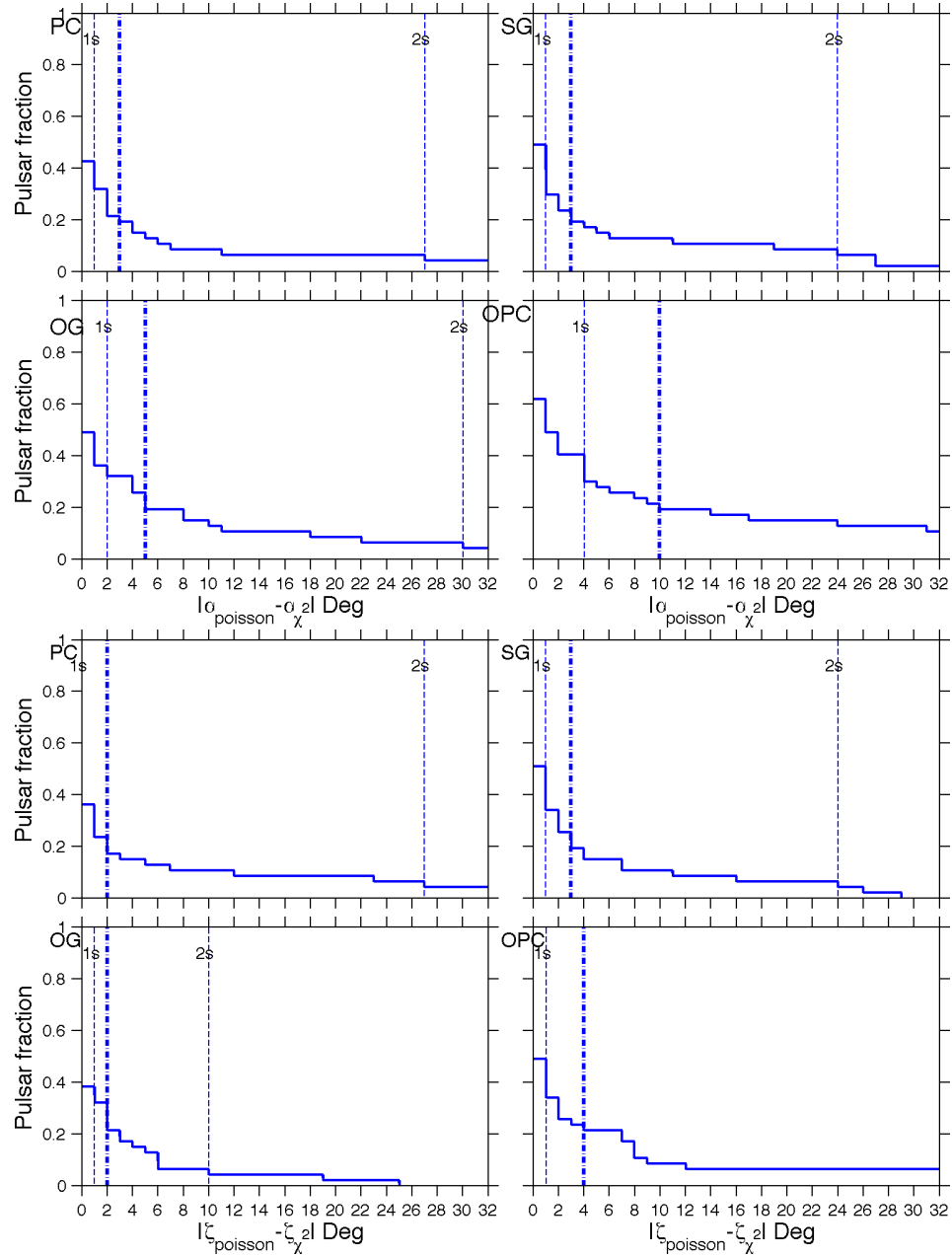

Figure 6.2: One minus the cumulative function $(1 - f_{cum}(x))$ for the quantity $|\alpha_{poisson} - \alpha_{\chi^2}|$ (top) and $|\zeta_{poisson} - \zeta_{\chi^2}|$ (bottom) evaluated between the $\alpha$ & $\zeta$ estimates obtained with Poisson (fixed count binning) and $\chi^2$ (regular binning) fit for each $\gamma$-ray model. The dashed lines indicate the 68.2%, 80%, and 95.4% containment region that correspond to 1, 1.5, and $2\sigma$.

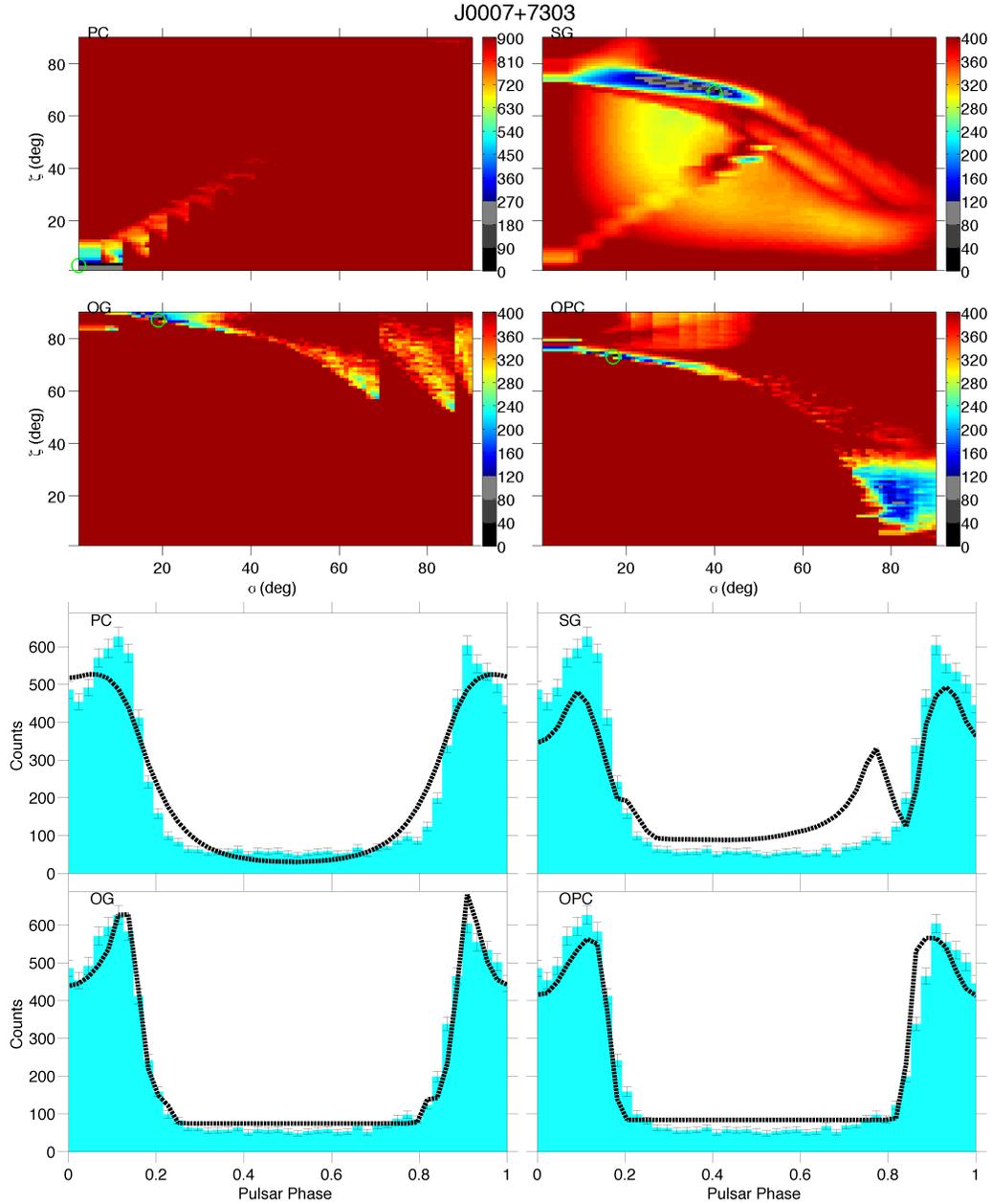

Figure 6.3: PSR J0007+7303. *Top*: for each model is shown the $\alpha$ & $\zeta$ likelihood map obtained with the Poisson FCB γ-ray fit. The color-bar is in $\sigma$ units, zero corresponds to the best fit solution. *Bottom*: the best γ-ray light curve (black dotted line) obtained, for each model, by maximising each likelihood map, superimposed to the FERMI pulsar light curve (in blue).

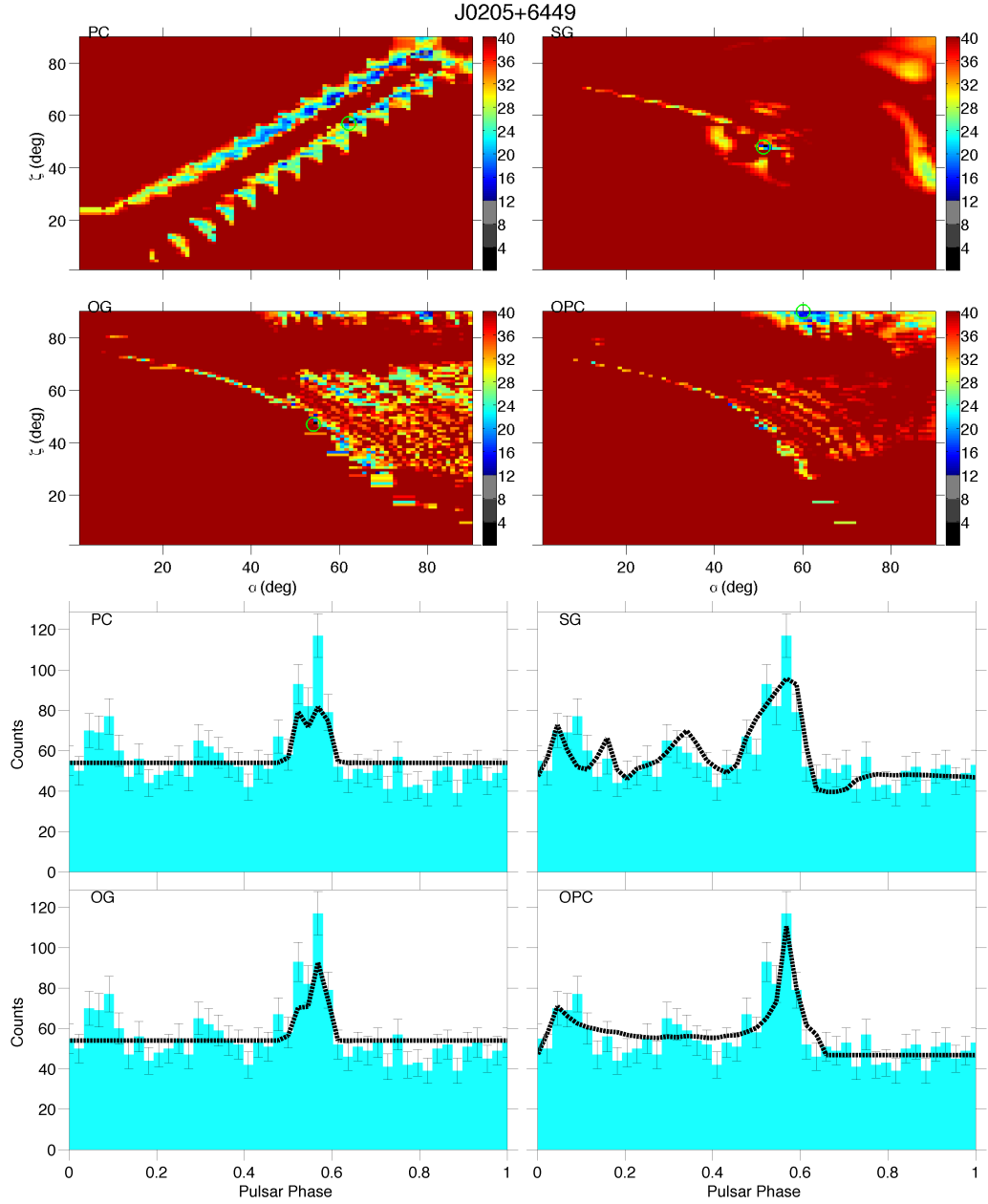

Figure 6.4: PSR J0205+6449. *Top*: for each model is shown the $\alpha$ & $\zeta$ likelihood map obtained with the Poisson FCB $\gamma$-ray fit. The color-bar is in $\sigma$ units, zero corresponds to the best fit solution.*Bottom*: the best $\gamma$-ray light curve (black dotted line) obtained, for each model, by maximising each likelihood map, superimposed to the FERMI pulsar light curve (in blue).

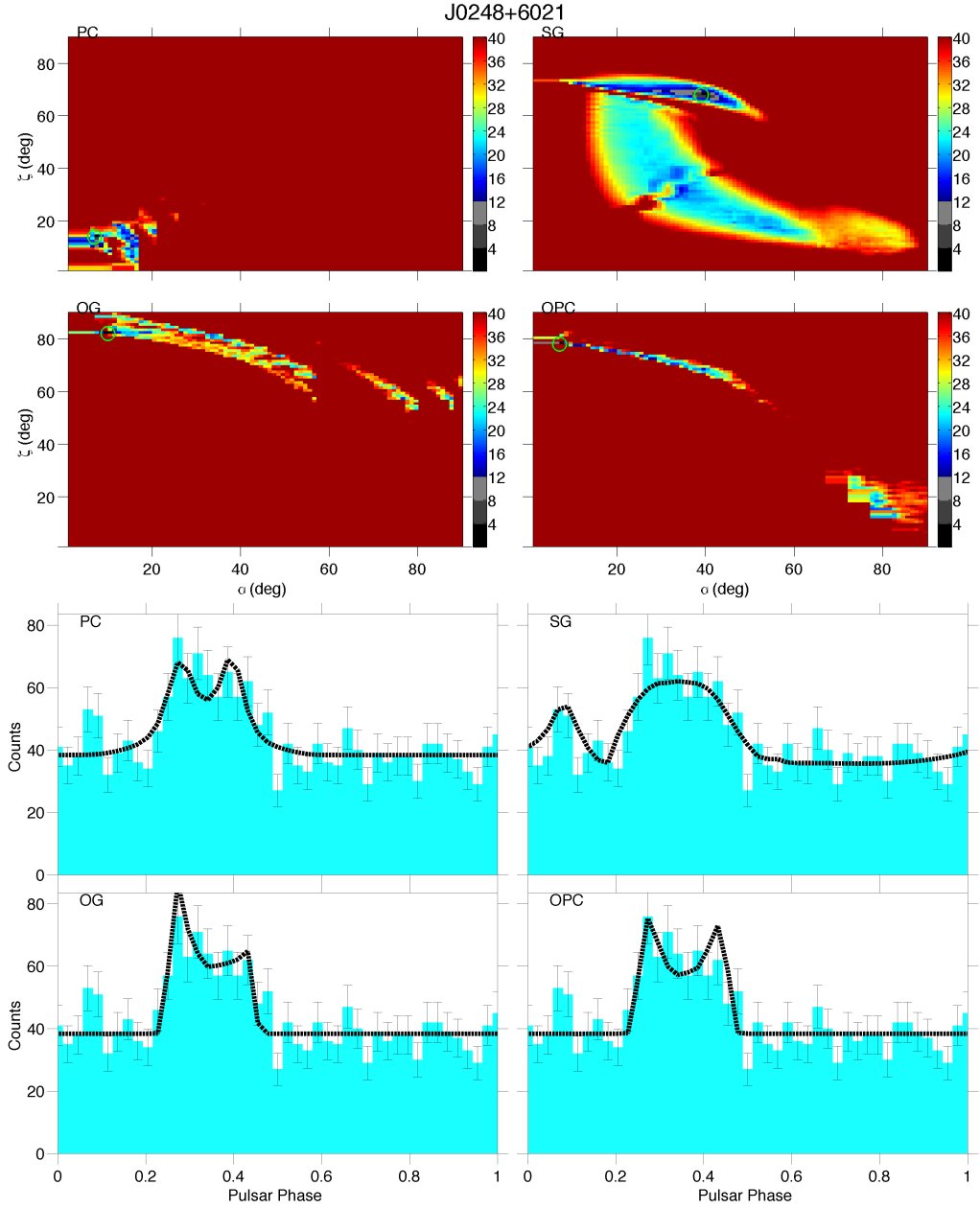

Figure 6.5: PSR J0248+6021. *Top*: for each model is shown the α & ζ likelihood map obtained with the Poisson FCB γ-ray fit. The color-bar is in σ units, zero corresponds to the best fit solution.*Bottom*: the best γ-ray light curve (black dotted line) obtained, for each model, by maximising each likelihood map, superimposed to the FERMI pulsar light curve (in blue).

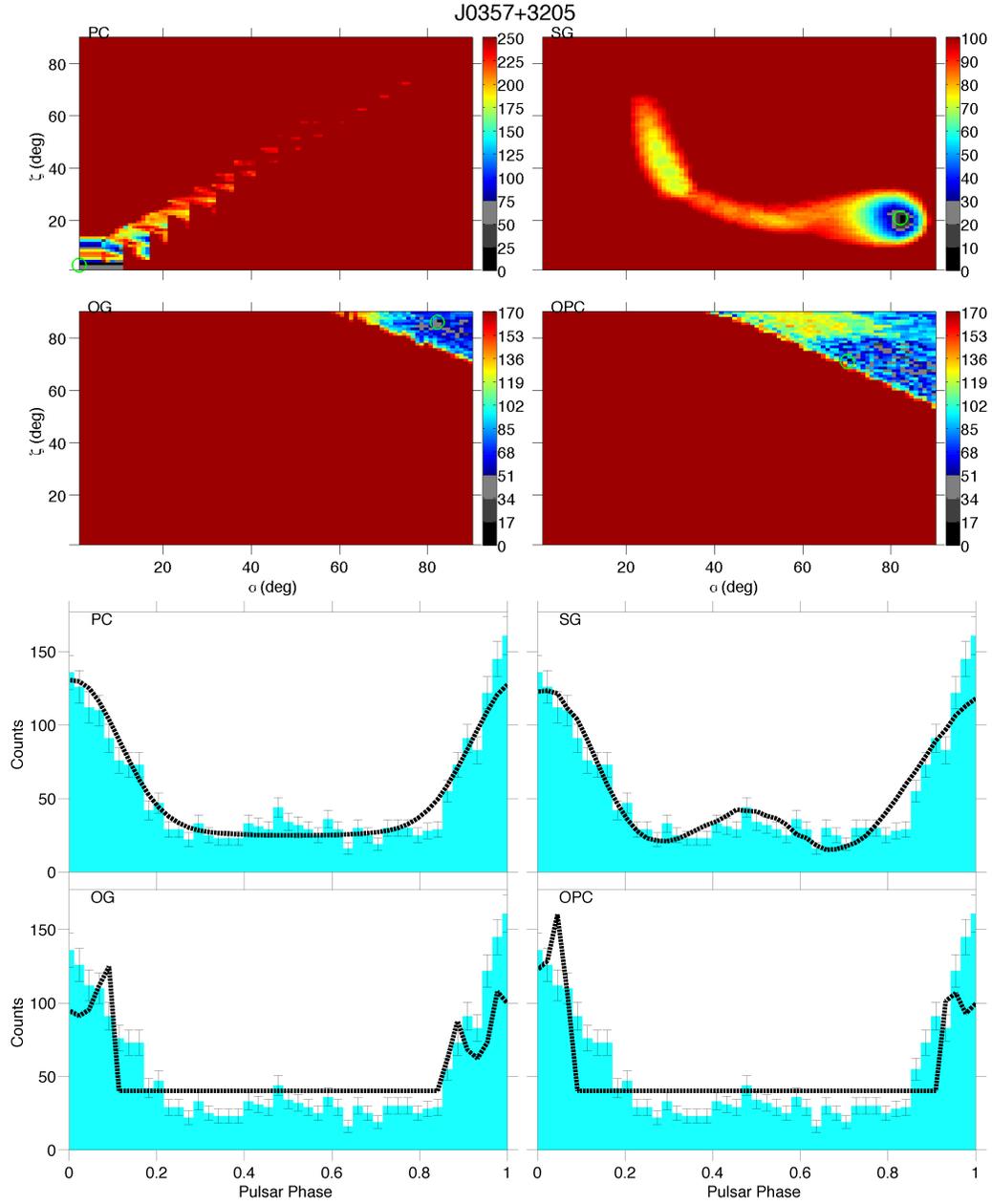

Figure 6.6: PSR J0357+3205. *Top*: for each model is shown the $\alpha$ & $\zeta$ likelihood map obtained with the Poisson FCB $\gamma$-ray fit. The color-bar is in $\sigma$ units, zero corresponds to the best fit solution.*Bottom*: the best $\gamma$-ray light curve (black dotted line) obtained, for each model, by maximising each likelihood map, superimposed to the FERMI pulsar light curve (in blue).

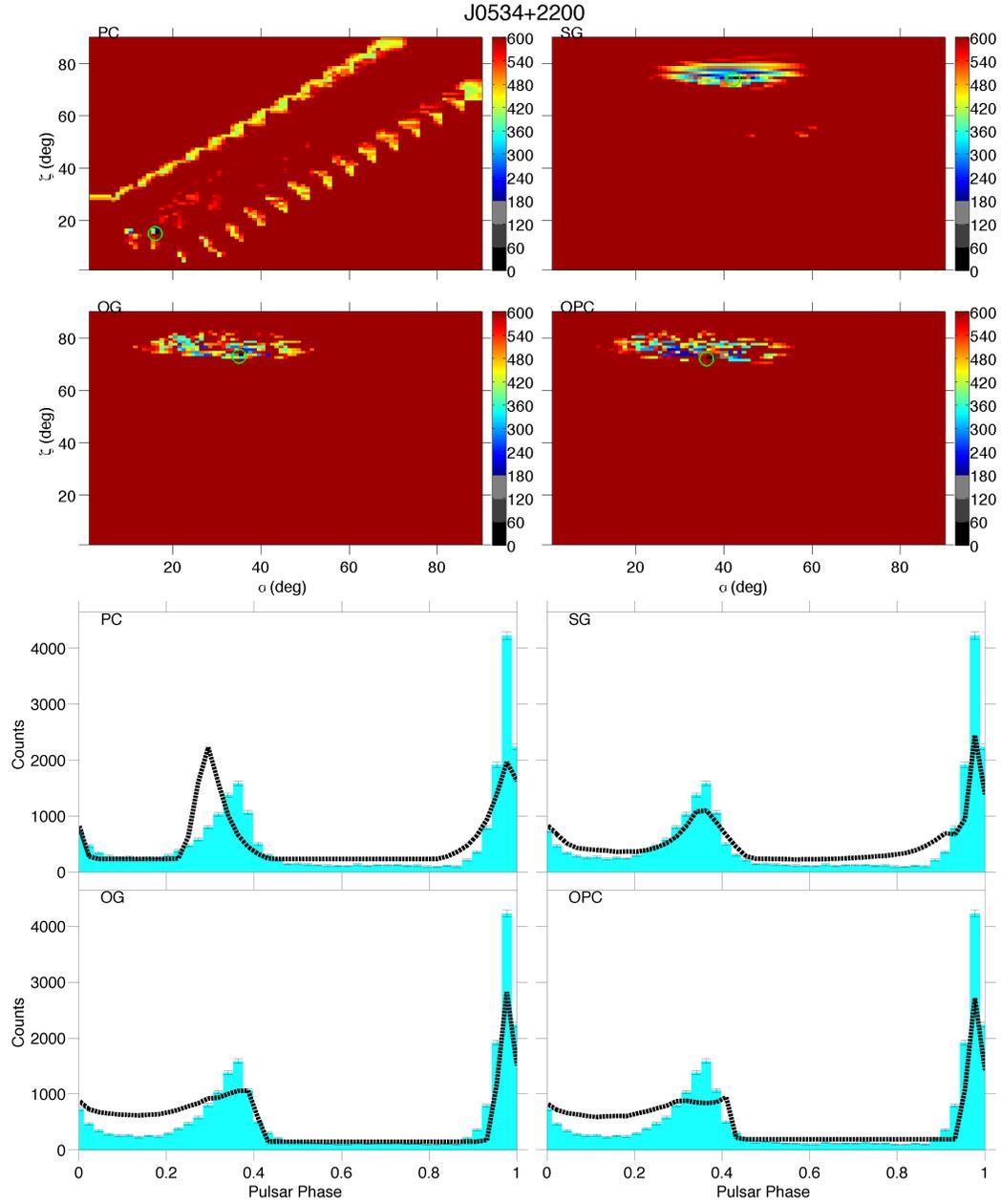

Figure 6.7: PSR J0534+2200. *Top*: for each model is shown the α & ζ likelihood map obtained with the Poisson FCB γ-ray fit. The color-bar is in σ units, zero corresponds to the best fit solution. *Bottom*: the best γ-ray light curve (black dotted line) obtained, for each model, by maximising each likelihood map, superimposed to the FERMI pulsar light curve (in blue).

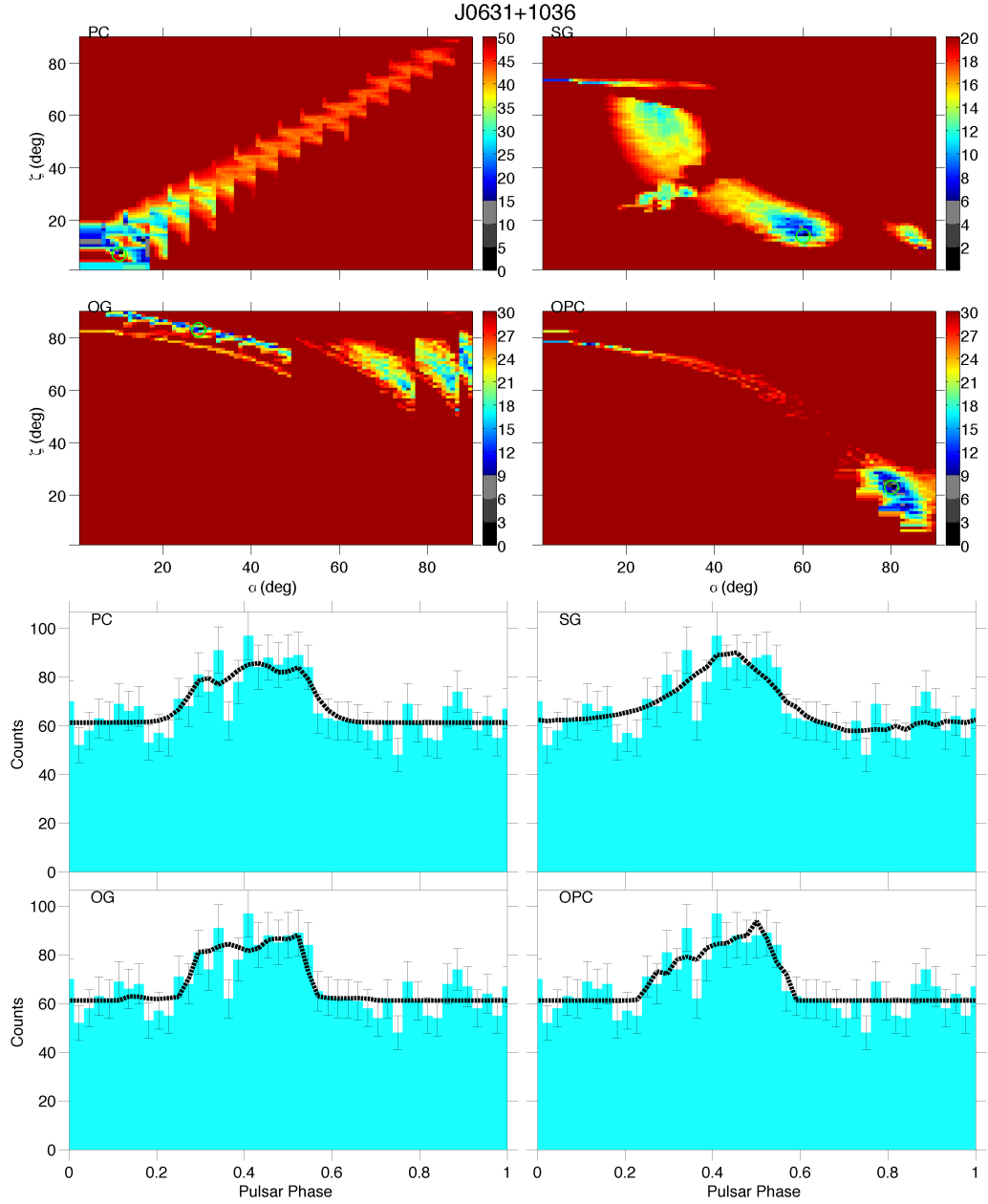

Figure 6.8: PSR J0631+1036. *Top*: for each model is shown the $\alpha$ & $\zeta$ likelihood map obtained with the Poisson FCB $\gamma$-ray fit. The color-bar is in $\sigma$ units, zero corresponds to the best fit solution.*Bottom*: the best $\gamma$-ray light curve (black dotted line) obtained, for each model, by maximising each likelihood map, superimposed to the FERMI pulsar light curve (in blue).

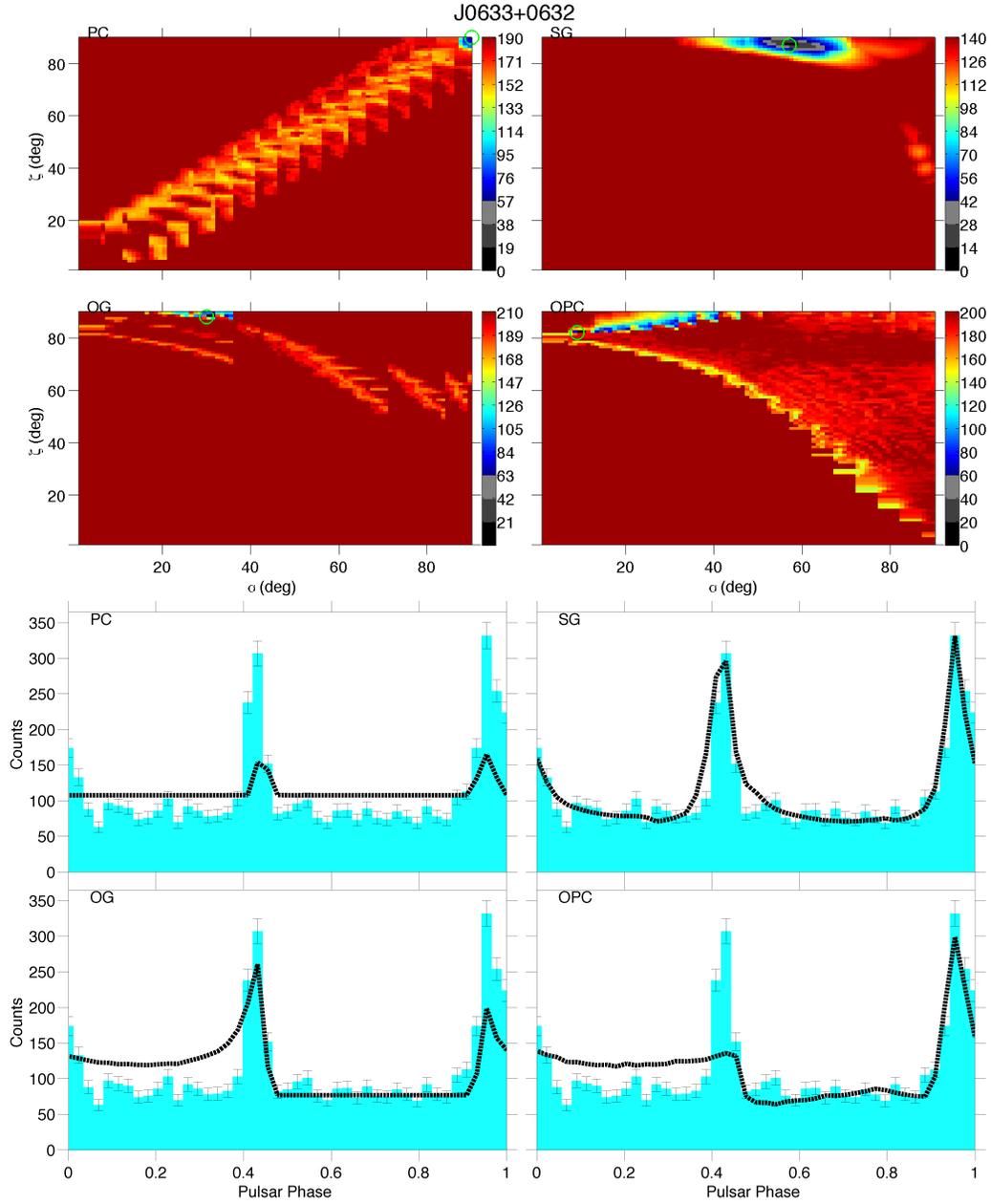

Figure 6.9: PSR J0633+0632. *Top*: for each model is shown the α & ζ likelihood map obtained with the Poisson FCB γ-ray fit. The color-bar is in σ units, zero corresponds to the best fit solution.*Bottom*: the best γ-ray light curve (black dotted line) obtained, for each model, by maximising each likelihood map, superimposed to the FERMI pulsar light curve (in blue).

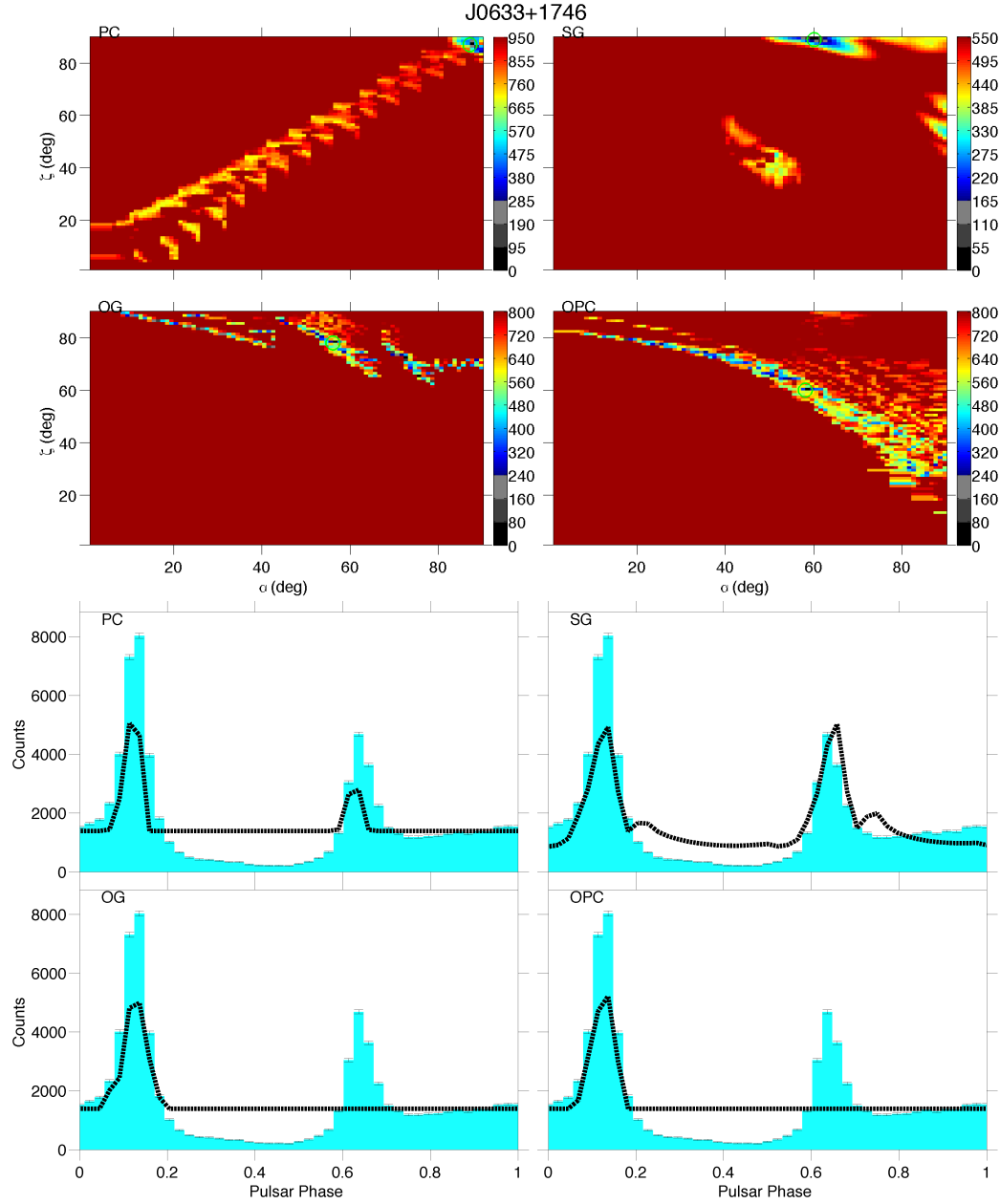

Figure 6.10: PSR J0633+1746. *Top*: for each model is shown the α & ζ likelihood map obtained with the Poisson FCB γ-ray fit. The color-bar is in σ units, zero corresponds to the best fit solution.*Bottom*: the best γ-ray light curve (black dotted line) obtained, for each model, by maximising each likelihood map, superimposed to the FERMI pulsar light curve (in blue).

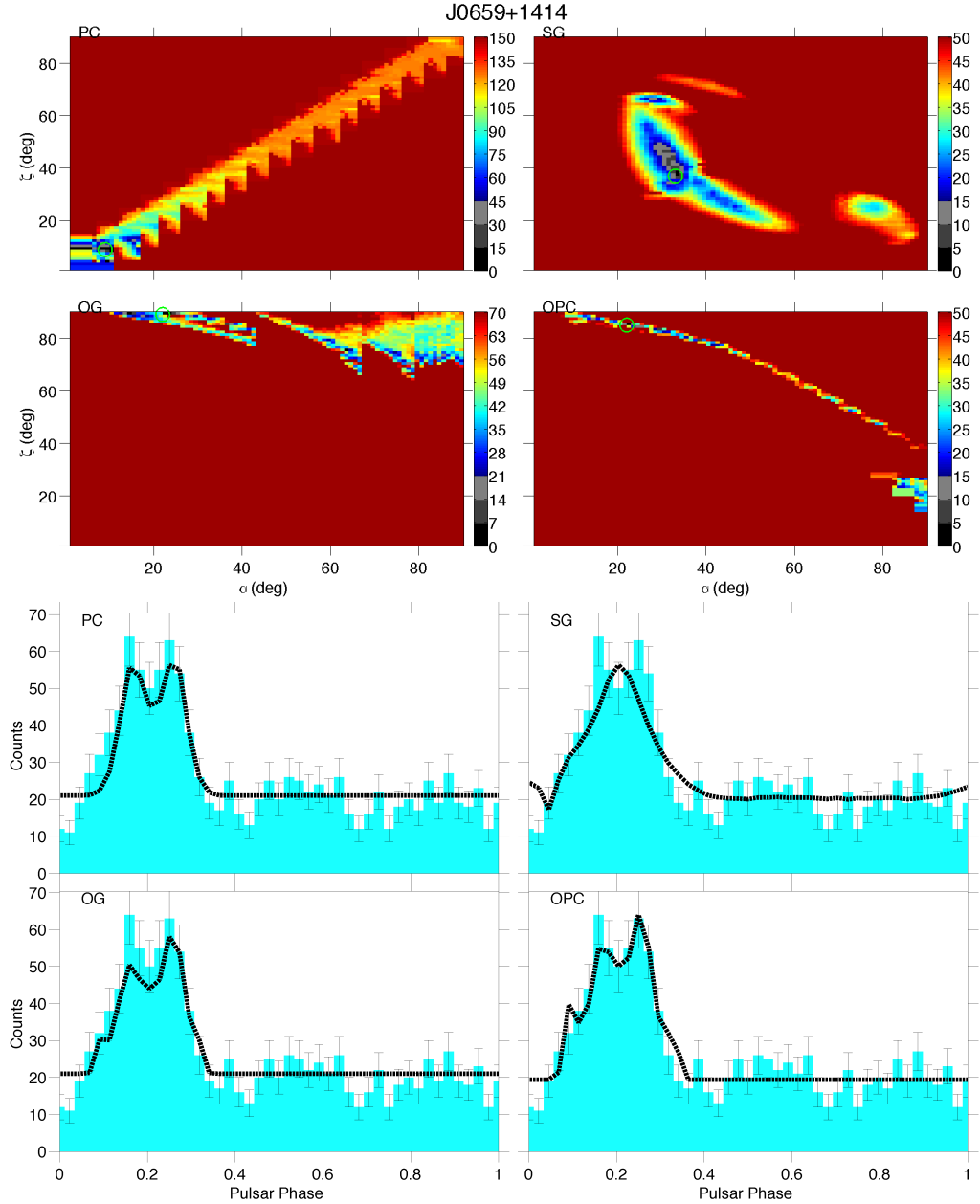

Figure 6.11: PSR J0659+1414. *Top*: for each model is shown the $\alpha$ & $\zeta$ likelihood map obtained with the Poisson FCB γ-ray fit. The color-bar is in $\sigma$ units, zero corresponds to the best fit solution.*Bottom*: the best γ-ray light curve (black dotted line) obtained, for each model, by maximising each likelihood map, superimposed to the FERMI pulsar light curve (in blue).

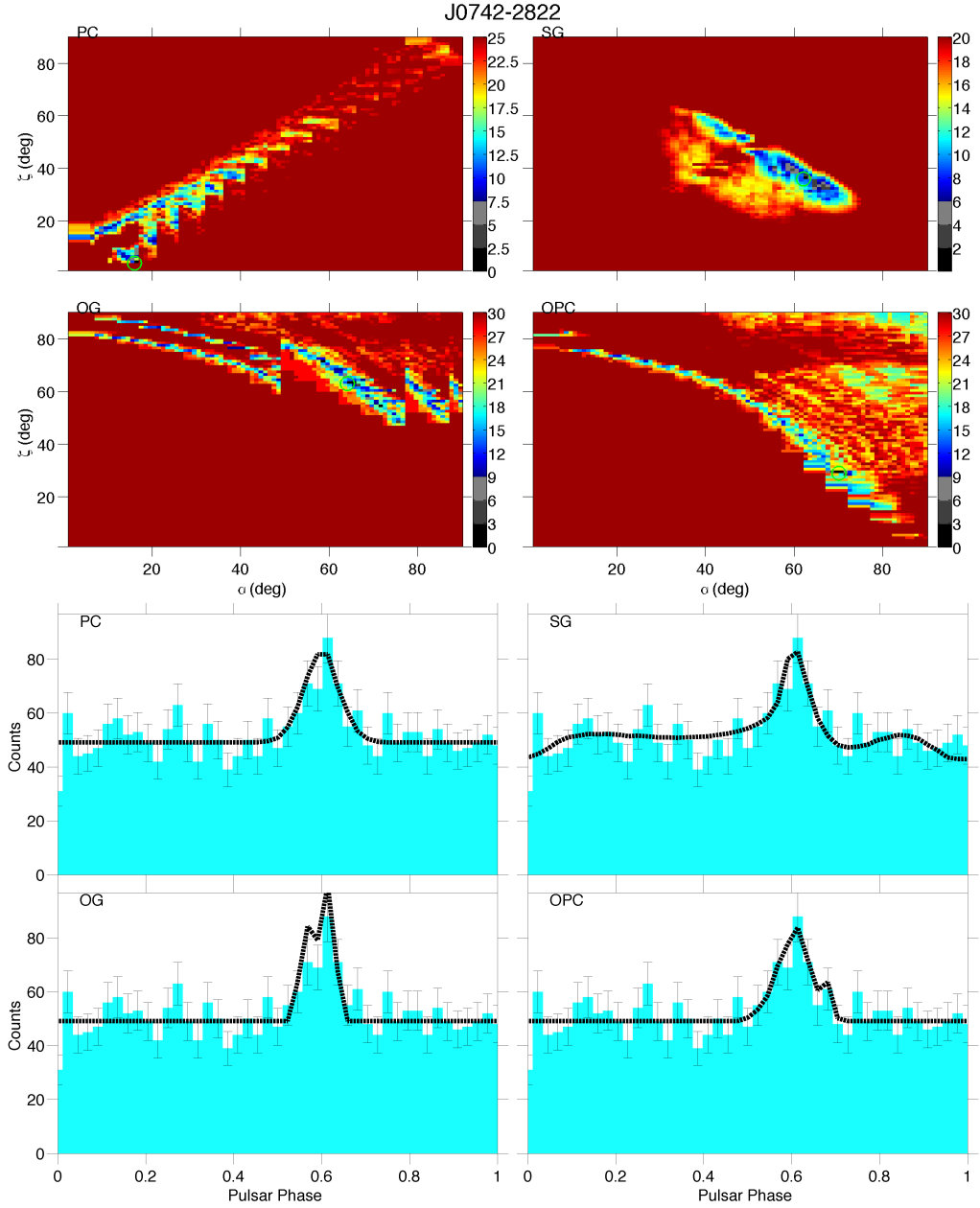

Figure 6.12: PSR J0742-2822. *Top*: for each model is shown the α & ζ likelihood map obtained with the Poisson FCB γ-ray fit. The color-bar is in σ units, zero corresponds to the best fit solution.*Bottom*: the best γ-ray light curve (black dotted line) obtained, for each model, by maximising each likelihood map, superimposed to the FERMI pulsar light curve (in blue).

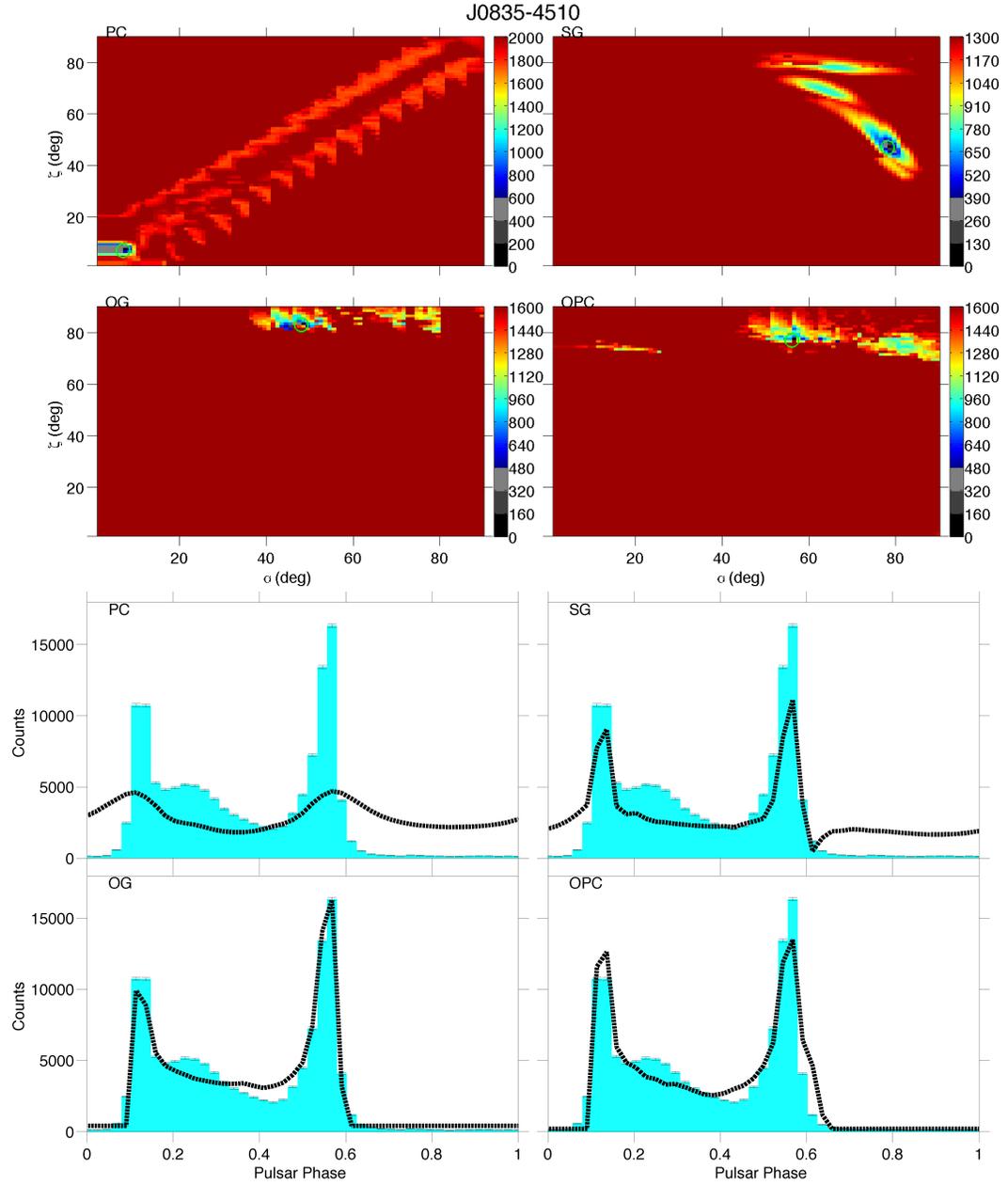

Figure 6.13: PSR J0835-4510. *Top*: for each model is shown the $\alpha$ & $\zeta$ likelihood map obtained with the Poisson FCB $\gamma$-ray fit. The color-bar is in $\sigma$ units, zero corresponds to the best fit solution.*Bottom*: the best $\gamma$-ray light curve (black dotted line) obtained, for each model, by maximising each likelihood map, superimposed to the FERMI pulsar light curve (in blue).

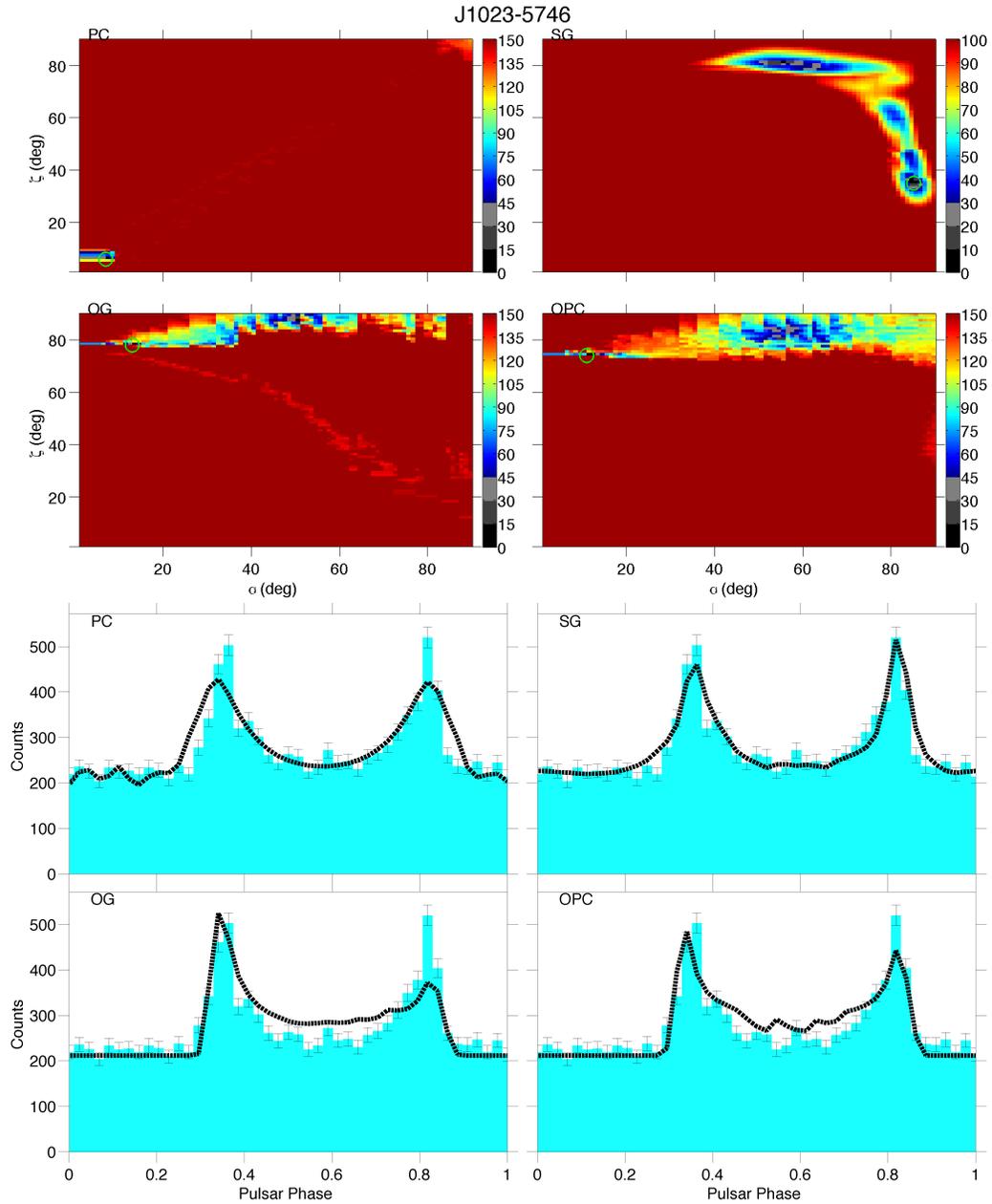

Figure 6.14: PSR J1023-5746. *Top*: for each model is shown the α & ζ likelihood
map obtained with the Poisson FCB γ-ray fit. The color-bar is in σ units, zero
corresponds to the best fit solution.*Bottom*: the best γ-ray light curve (black dotted
line) obtained, for each model, by maximising each likelihood map, superimposed
to the FERMI pulsar light curve (in blue).

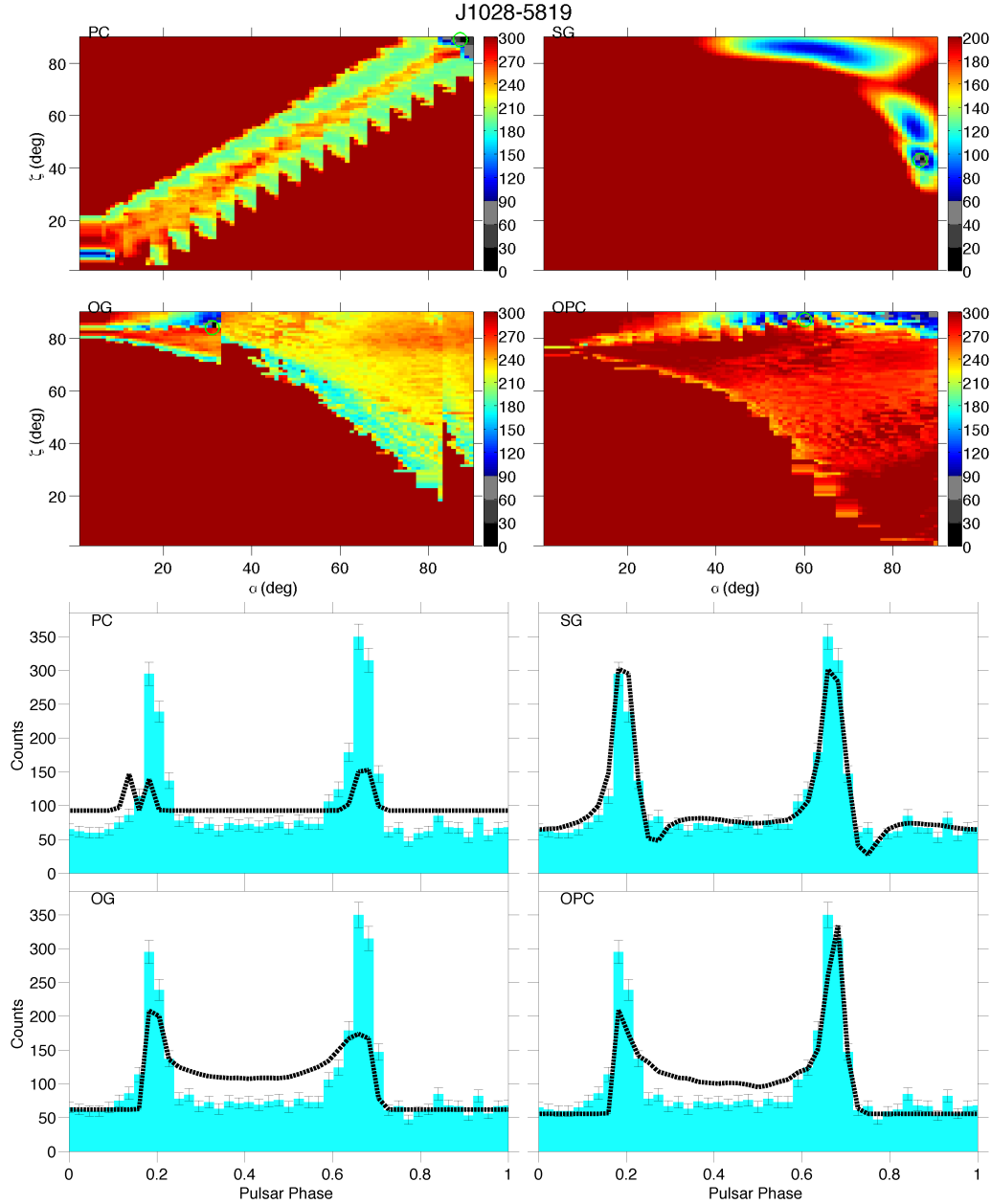

Figure 6.15: PSR J1028-5819. *Top*: for each model is shown the $\alpha$ & $\zeta$ likelihood map obtained with the Poisson FCB γ-ray fit. The color-bar is in $\sigma$ units, zero corresponds to the best fit solution.*Bottom*: the best γ-ray light curve (black dotted line) obtained, for each model, by maximising each likelihood map, superimposed to the FERMI pulsar light curve (in blue).

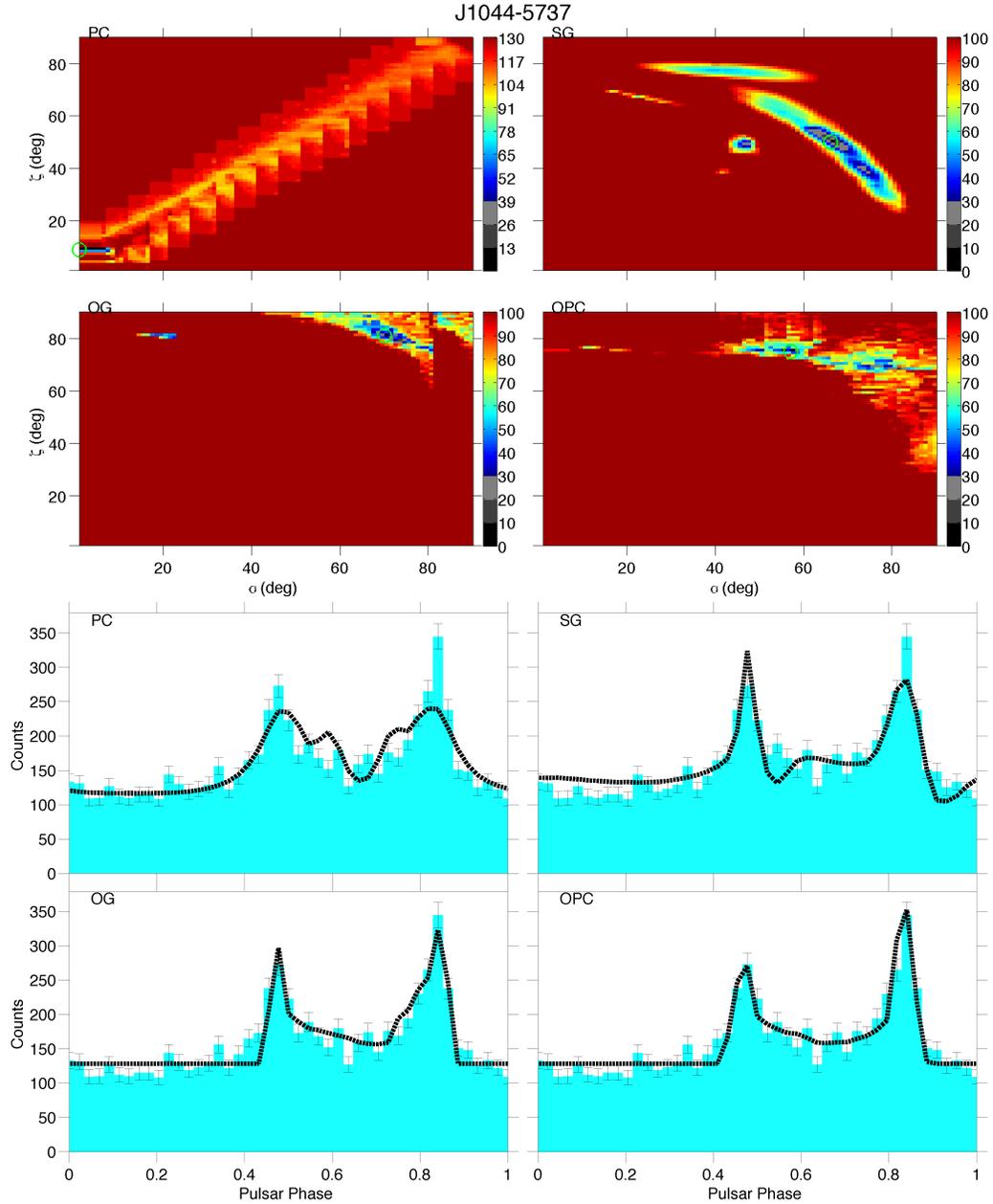

Figure 6.16: PSR J1044-5737. *Top*: for each model is shown the $\alpha$ & $\zeta$ likelihood map obtained with the Poisson FCB $\gamma$-ray fit. The color-bar is in $\sigma$ units, zero corresponds to the best fit solution. *Bottom*: the best $\gamma$-ray light curve (black dotted line) obtained, for each model, by maximising each likelihood map, superimposed to the FERMI pulsar light curve (in blue).

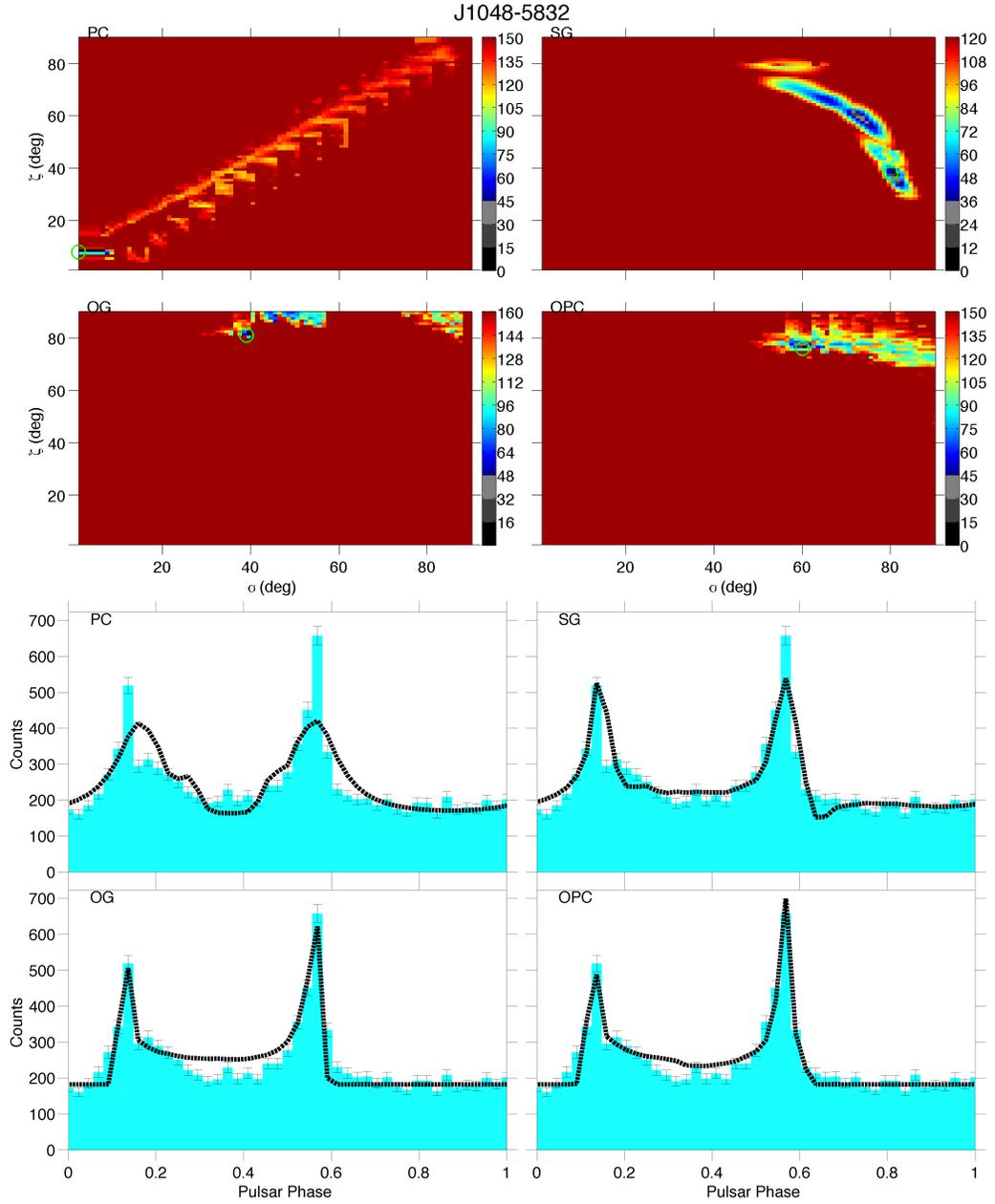

Figure 6.17: PSR J1048-5832. *Top*: for each model is shown the $\alpha$ & $\zeta$ likelihood map obtained with the Poisson FCB $\gamma$-ray fit. The color-bar is in $\sigma$ units, zero corresponds to the best fit solution. *Bottom*: the best $\gamma$-ray light curve (black dotted line) obtained, for each model, by maximising each likelihood map, superimposed to the FERMI pulsar light curve (in blue).

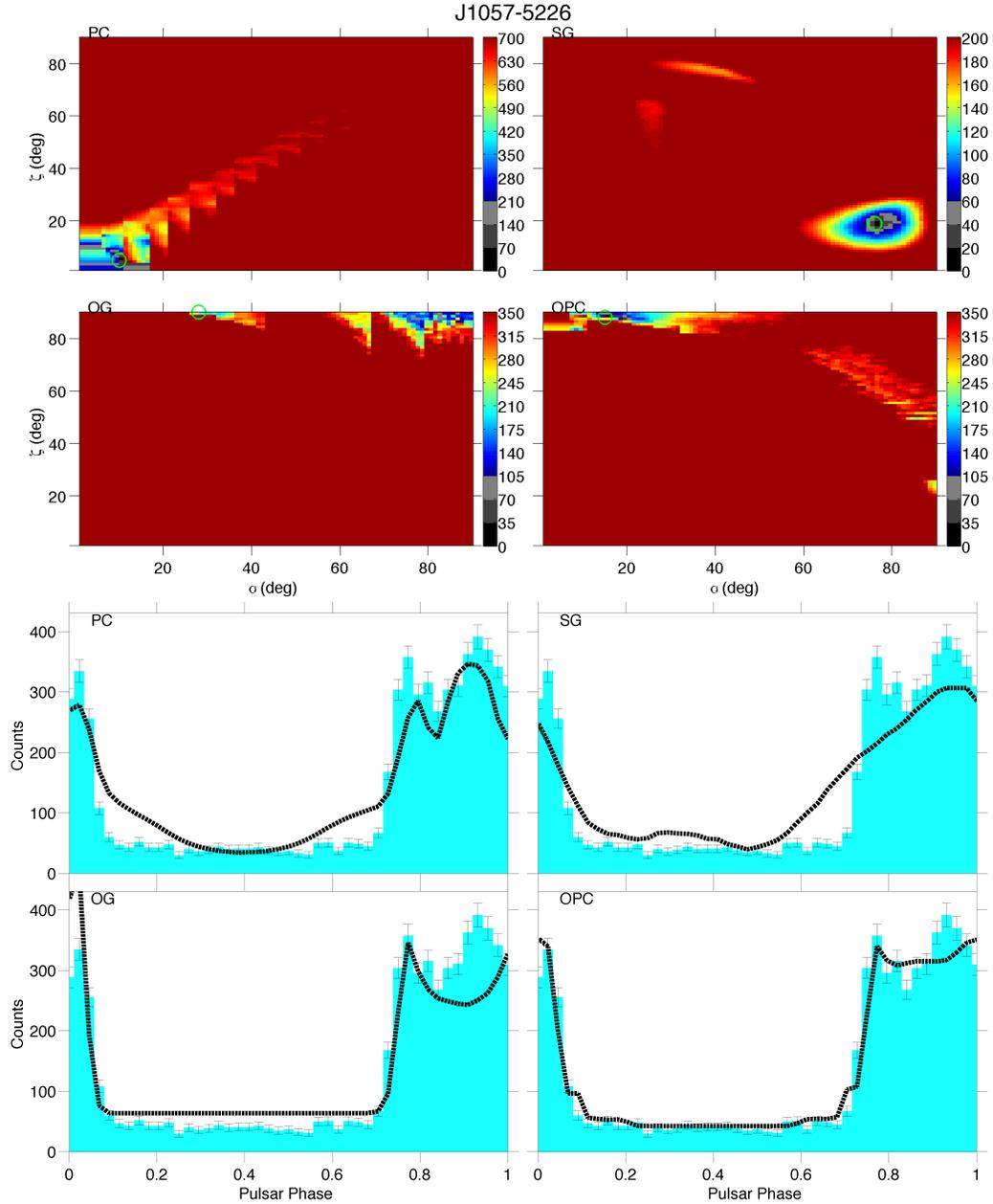

Figure 6.18: PSR J1057-5226. *Top*: for each model is shown the $\alpha$ & $\zeta$ likelihood
map obtained with the Poisson FCB $\gamma$-ray fit. The color-bar is in $\sigma$ units, zero
corresponds to the best fit solution. *Bottom*: the best $\gamma$-ray light curve (black dotted
line) obtained, for each model, by maximising each likelihood map, superimposed
to the FERMI pulsar light curve (in blue).

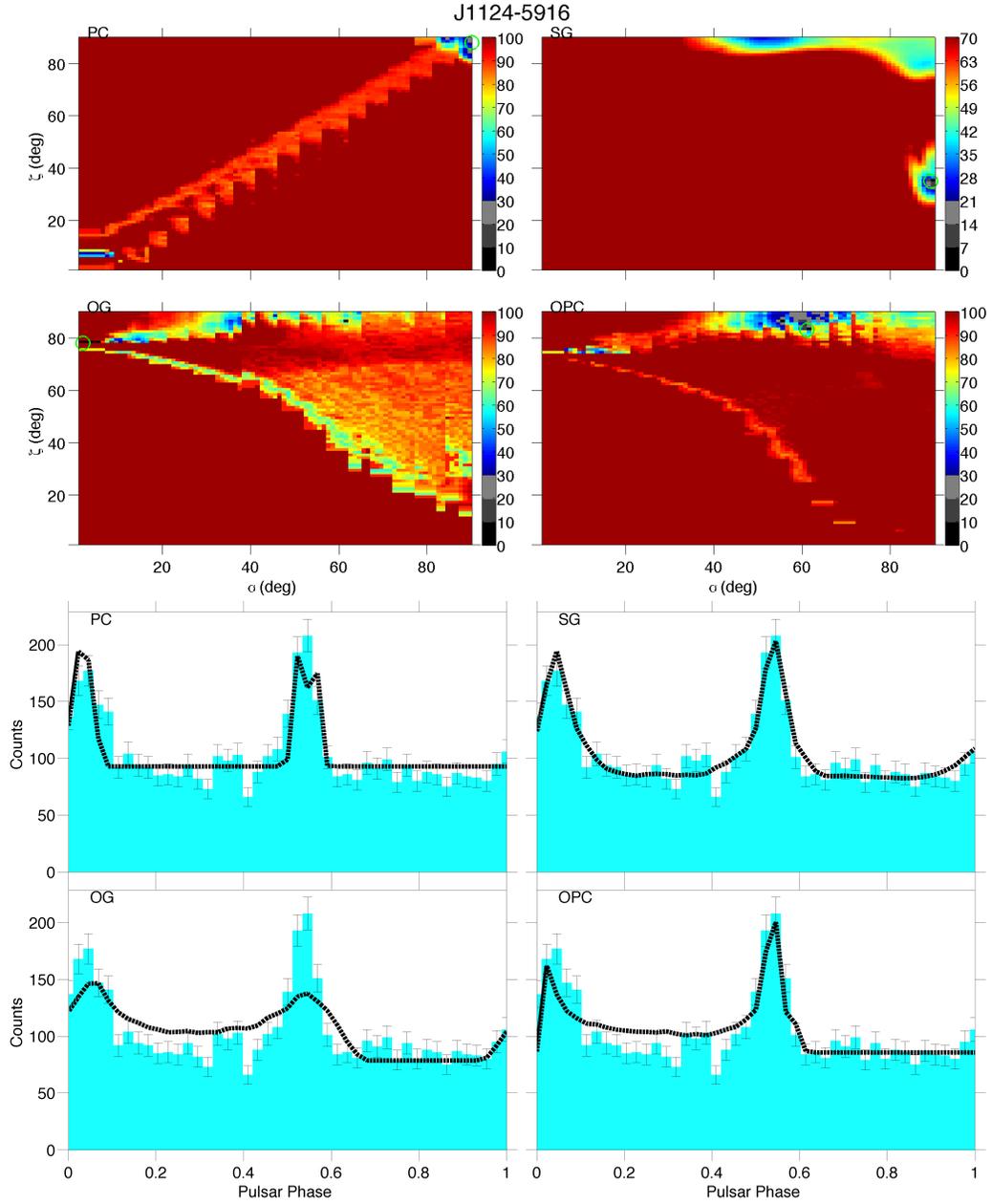

Figure 6.19: PSR J1124-5916. *Top*: for each model is shown the α & ζ likelihood map obtained with the Poisson FCB γ-ray fit. The color-bar is in σ units, zero corresponds to the best fit solution.*Bottom*: the best γ-ray light curve (black dotted line) obtained, for each model, by maximising each likelihood map, superimposed to the FERMI pulsar light curve (in blue).

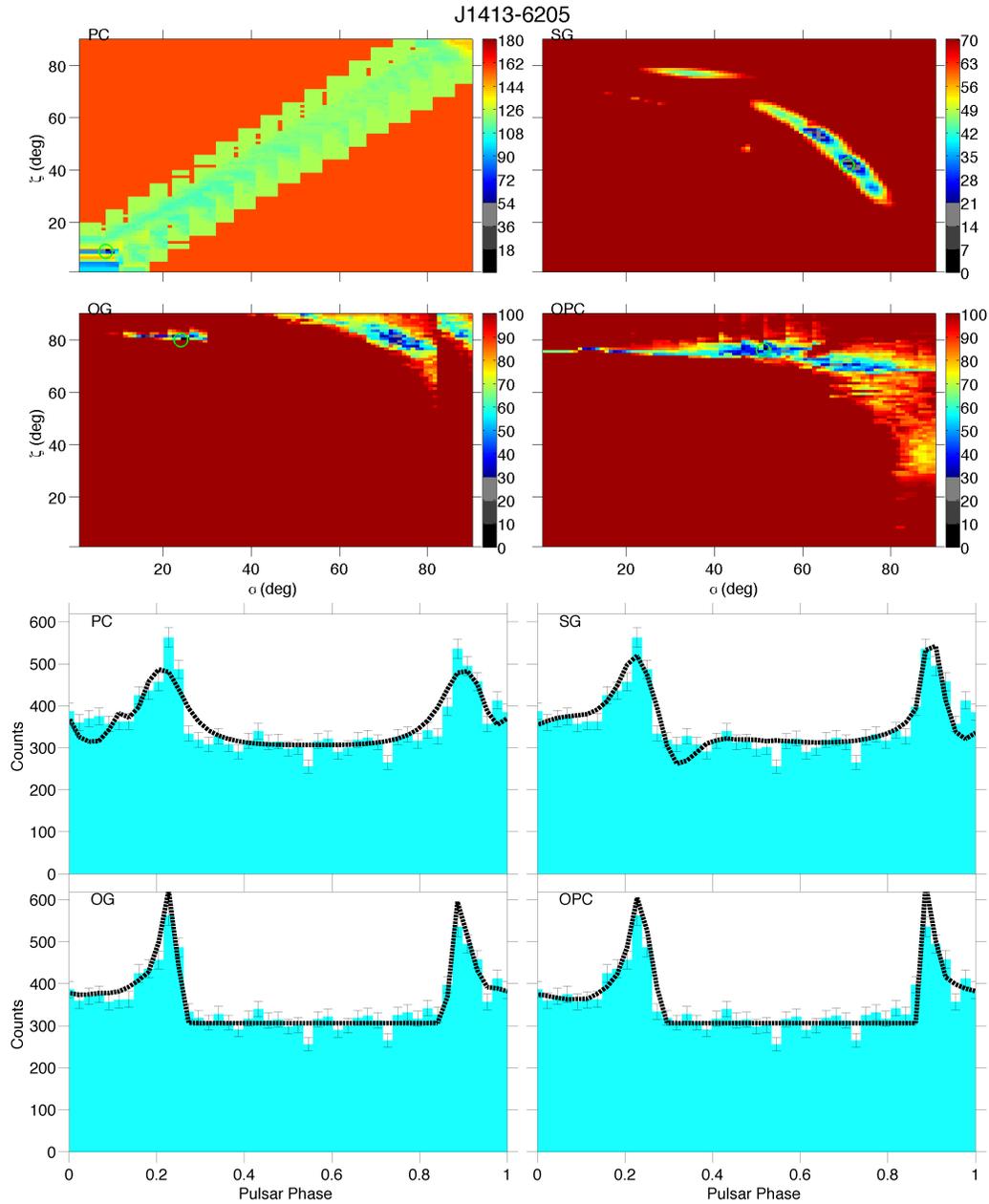

Figure 6.20: PSR J1413-6205. *Top*: for each model is shown the $\alpha$ & $\zeta$ likelihood map obtained with the Poisson FCB $\gamma$-ray fit. The color-bar is in $\sigma$ units, zero corresponds to the best fit solution. *Bottom*: the best $\gamma$-ray light curve (black dotted line) obtained, for each model, by maximising each likelihood map, superimposed to the FERMI pulsar light curve (in blue).

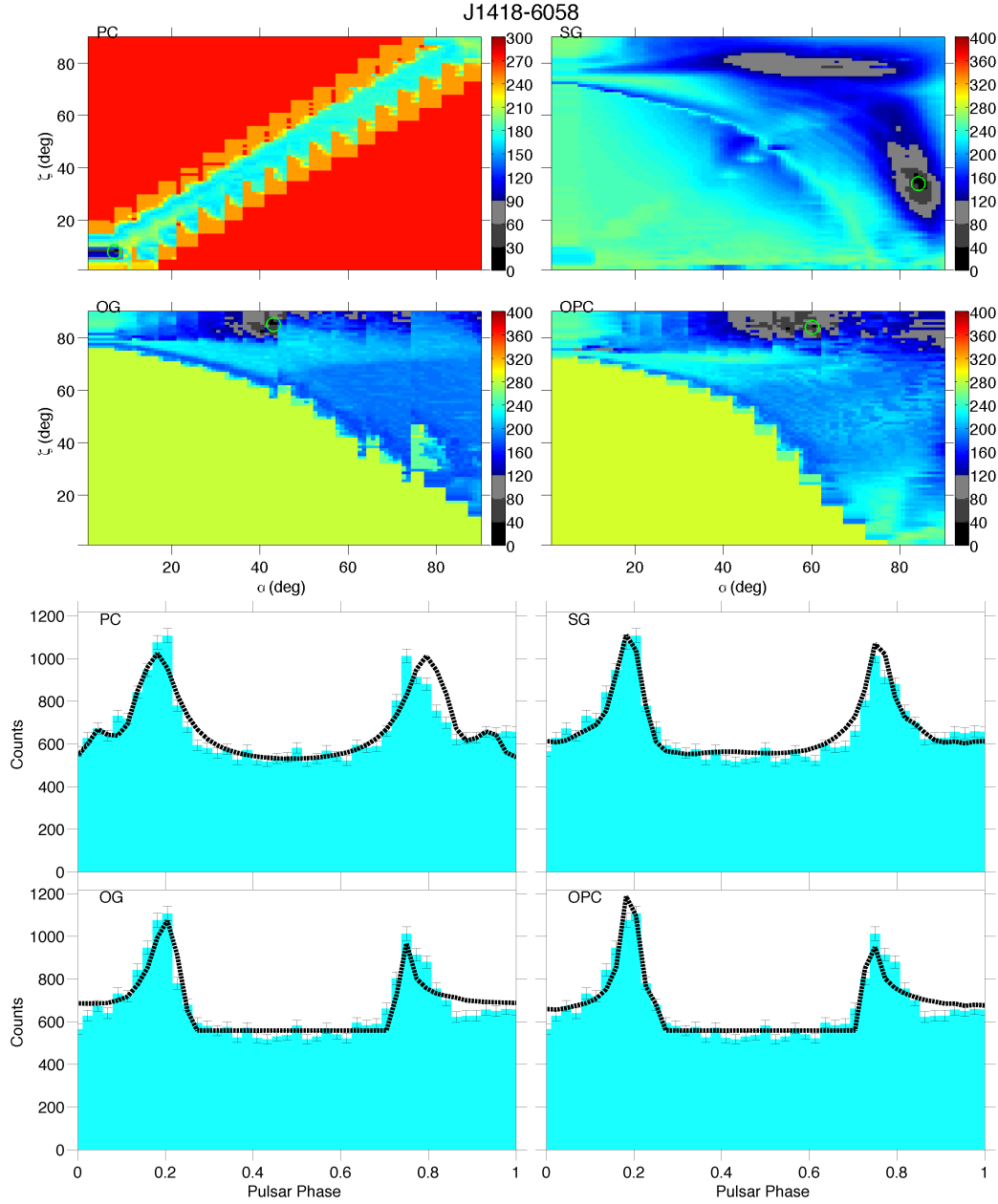

Figure 6.21: PSR J1418-6058. *Top*: for each model is shown the α & ζ likelihood map obtained with the Poisson FCB γ-ray fit. The color-bar is in σ units, zero corresponds to the best fit solution.*Bottom*: the best γ-ray light curve (black dotted line) obtained, for each model, by maximising each likelihood map, superimposed to the FERMI pulsar light curve (in blue).

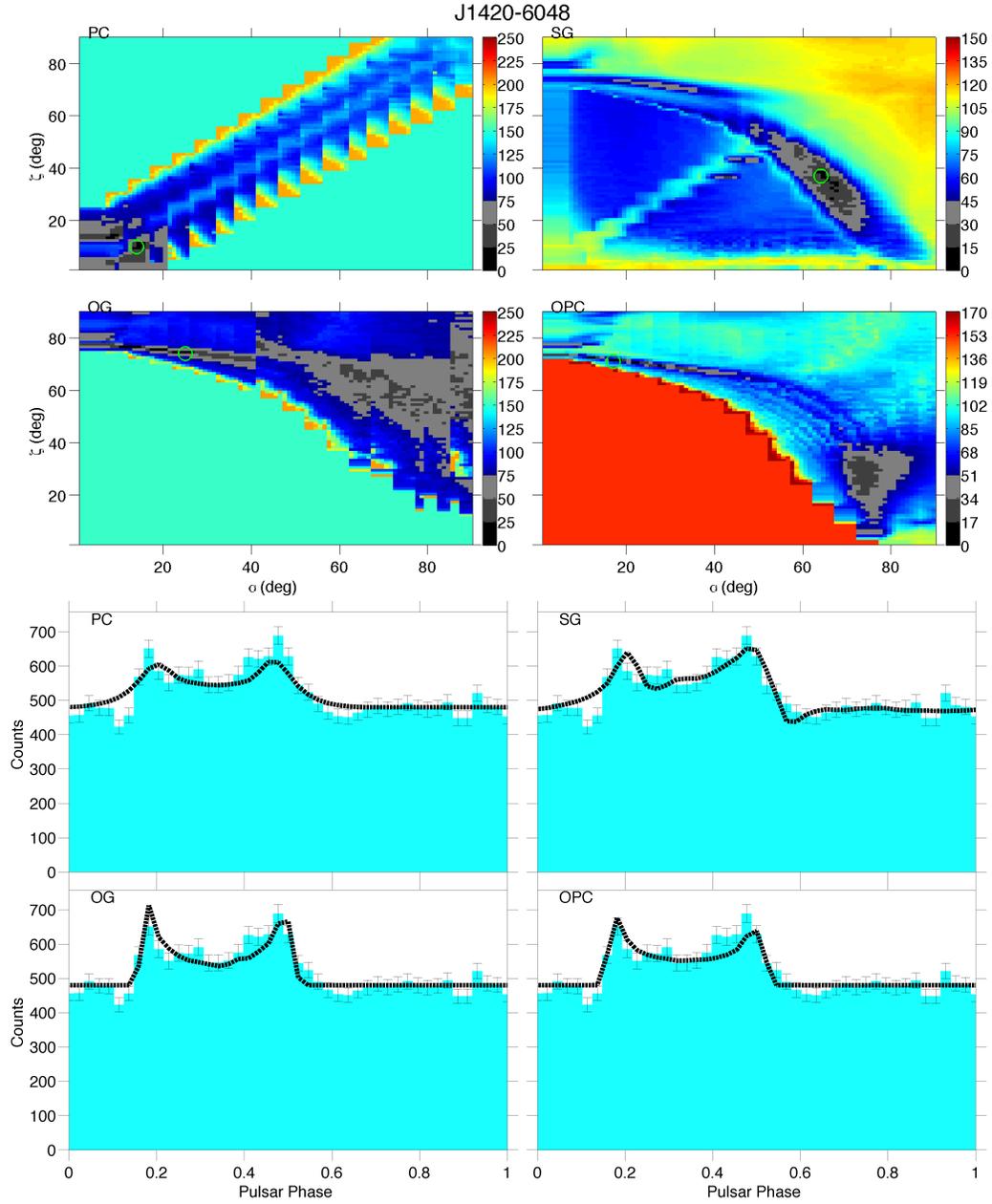

Figure 6.22: PSR J1420-6048. *Top*: for each model is shown the $\alpha$ & $\zeta$ likelihood map obtained with the Poisson FCB $\gamma$-ray fit. The color-bar is in $\sigma$ units, zero corresponds to the best fit solution. *Bottom*: the best $\gamma$-ray light curve (black dotted line) obtained, for each model, by maximising each likelihood map, superimposed to the FERMI pulsar light curve (in blue).

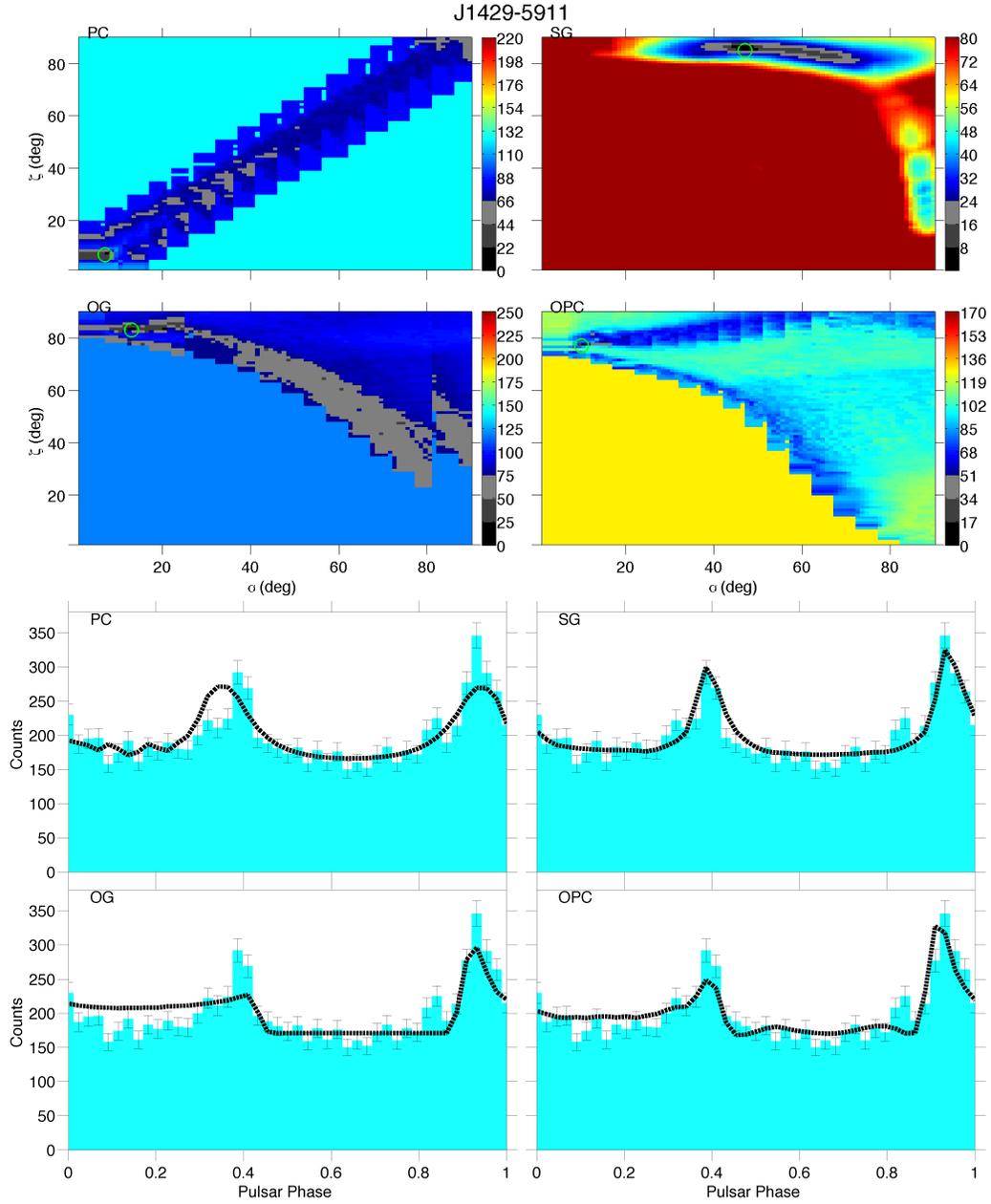

Figure 6.23: PSR J1429-5911. *Top*: for each model is shown the $\alpha$ & $\zeta$ likelihood map obtained with the Poisson FCB $\gamma$-ray fit. The color-bar is in $\sigma$ units, zero corresponds to the best fit solution.*Bottom*: the best $\gamma$-ray light curve (black dotted line) obtained, for each model, by maximising each likelihood map, superimposed to the FERMI pulsar light curve (in blue).

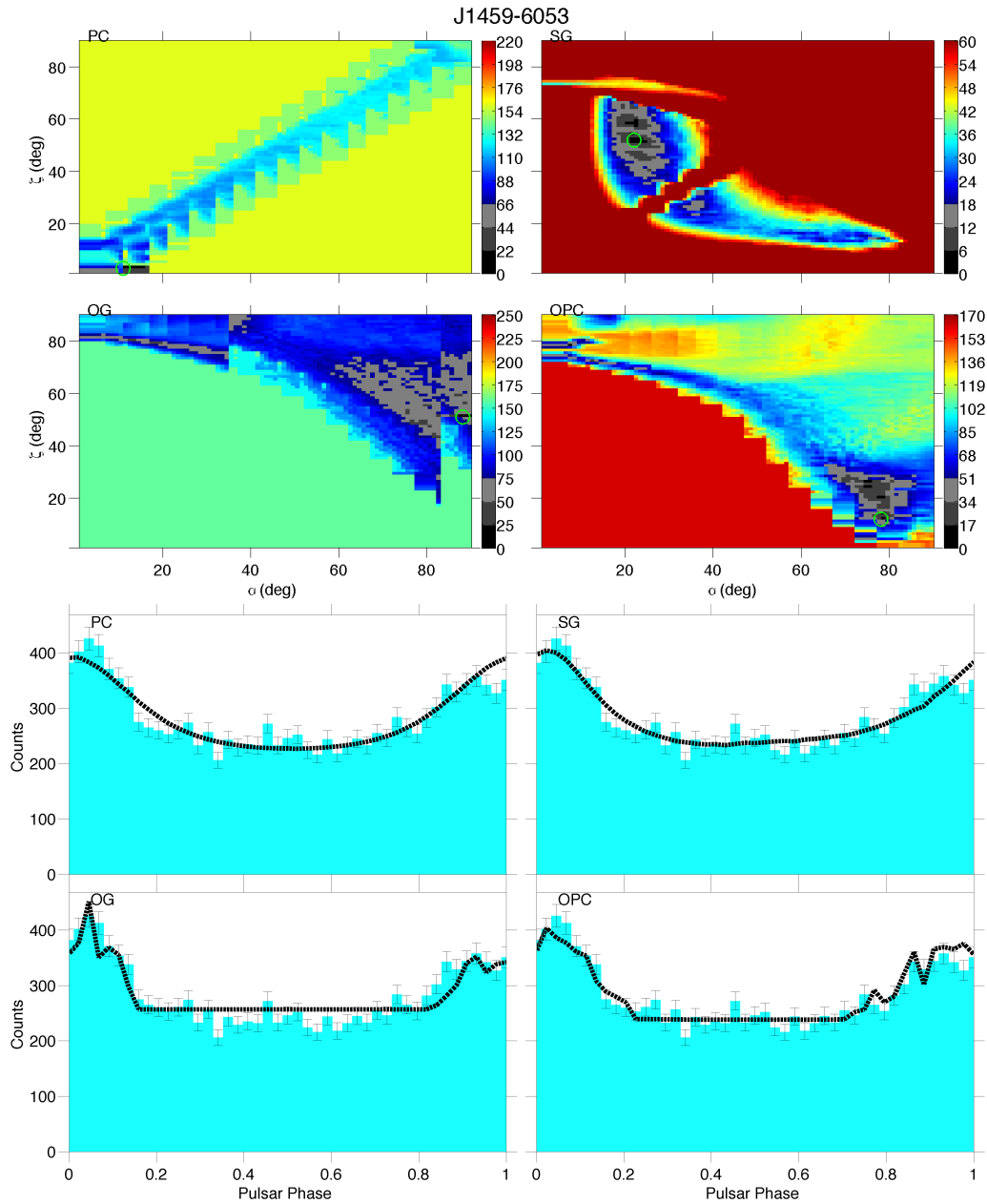

Figure 6.24: PSR J1459-6053. *Top*: for each model is shown the $\alpha$ & $\zeta$ likelihood map obtained with the Poisson FCB $\gamma$-ray fit. The color-bar is in $\sigma$ units, zero corresponds to the best fit solution.*Bottom*: the best $\gamma$-ray light curve (black dotted line) obtained, for each model, by maximising each likelihood map, superimposed to the FERMI pulsar light curve (in blue).

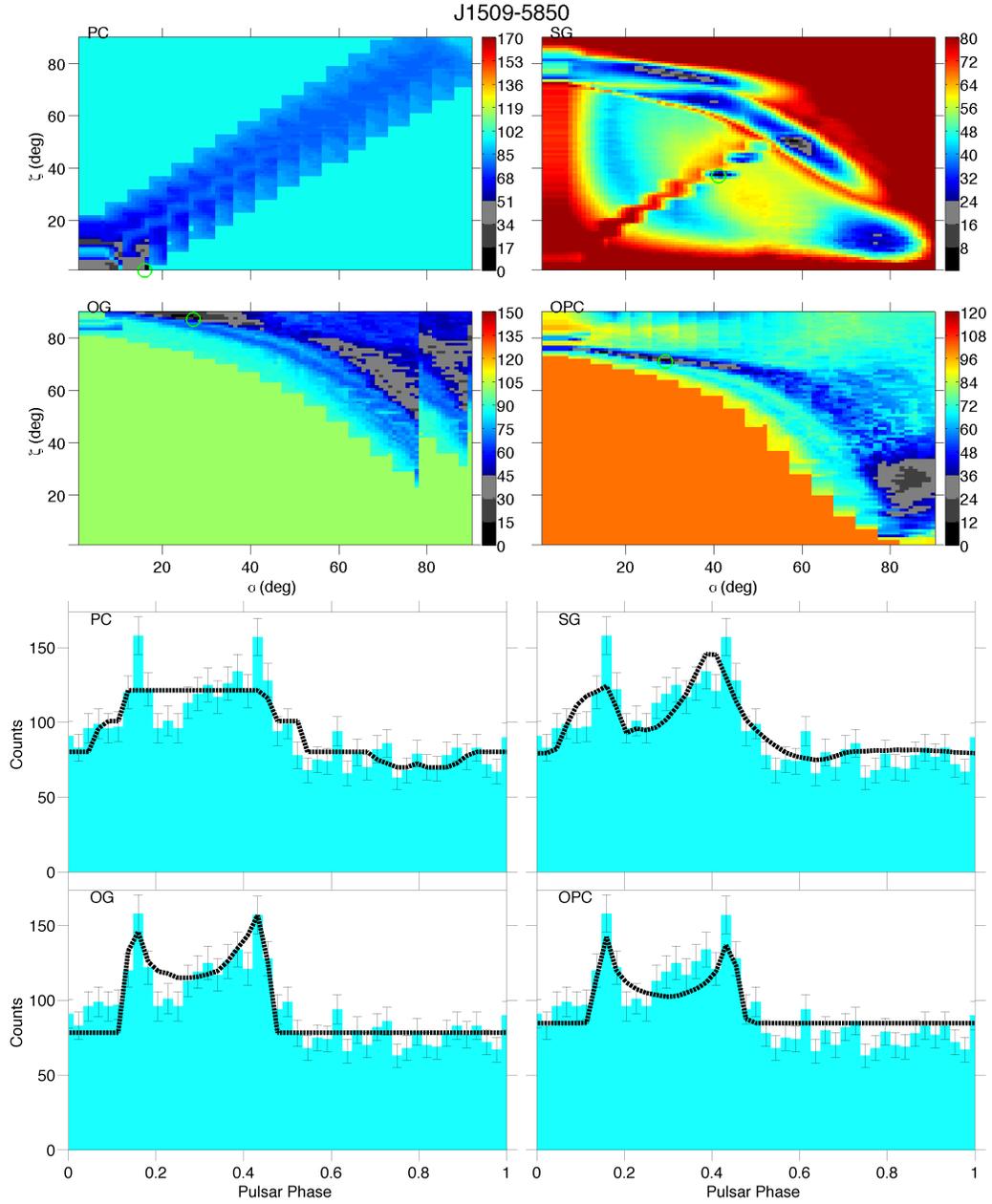

Figure 6.25: PSR J1509-5850. *Top*: for each model is shown the $\alpha$ & $\zeta$ likelihood map obtained with the Poisson FCB $\gamma$-ray fit. The color-bar is in $\sigma$ units, zero corresponds to the best fit solution. *Bottom*: the best $\gamma$-ray light curve (black dotted line) obtained, for each model, by maximising each likelihood map, superimposed to the FERMI pulsar light curve (in blue).

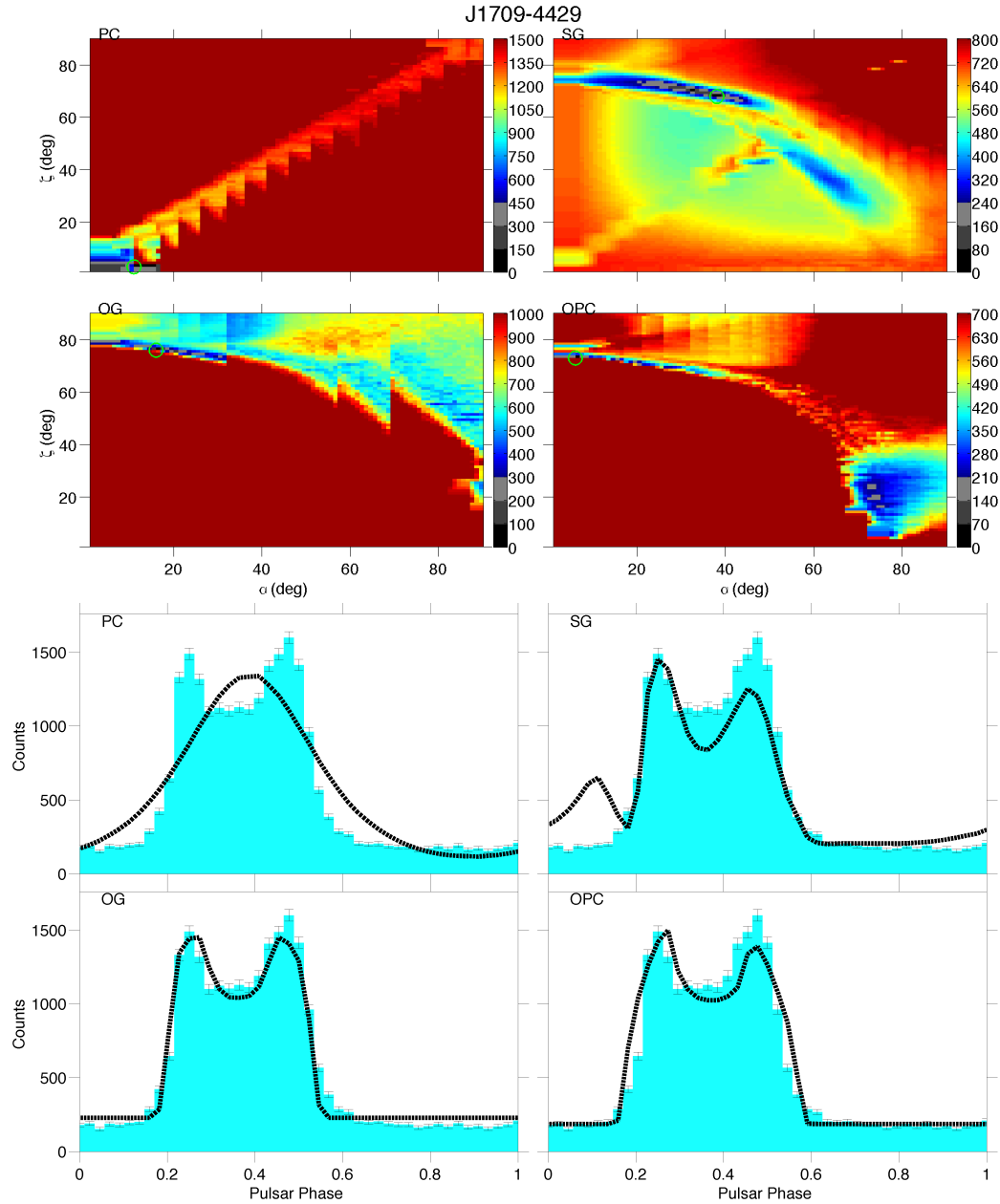

Figure 6.26: PSR J1709-4429. *Top*: for each model is shown the α & ζ likelihood map obtained with the Poisson FCB γ-ray fit. The color-bar is in σ units, zero corresponds to the best fit solution.*Bottom*: the best γ-ray light curve (black dotted line) obtained, for each model, by maximising each likelihood map, superimposed to the FERMI pulsar light curve (in blue).

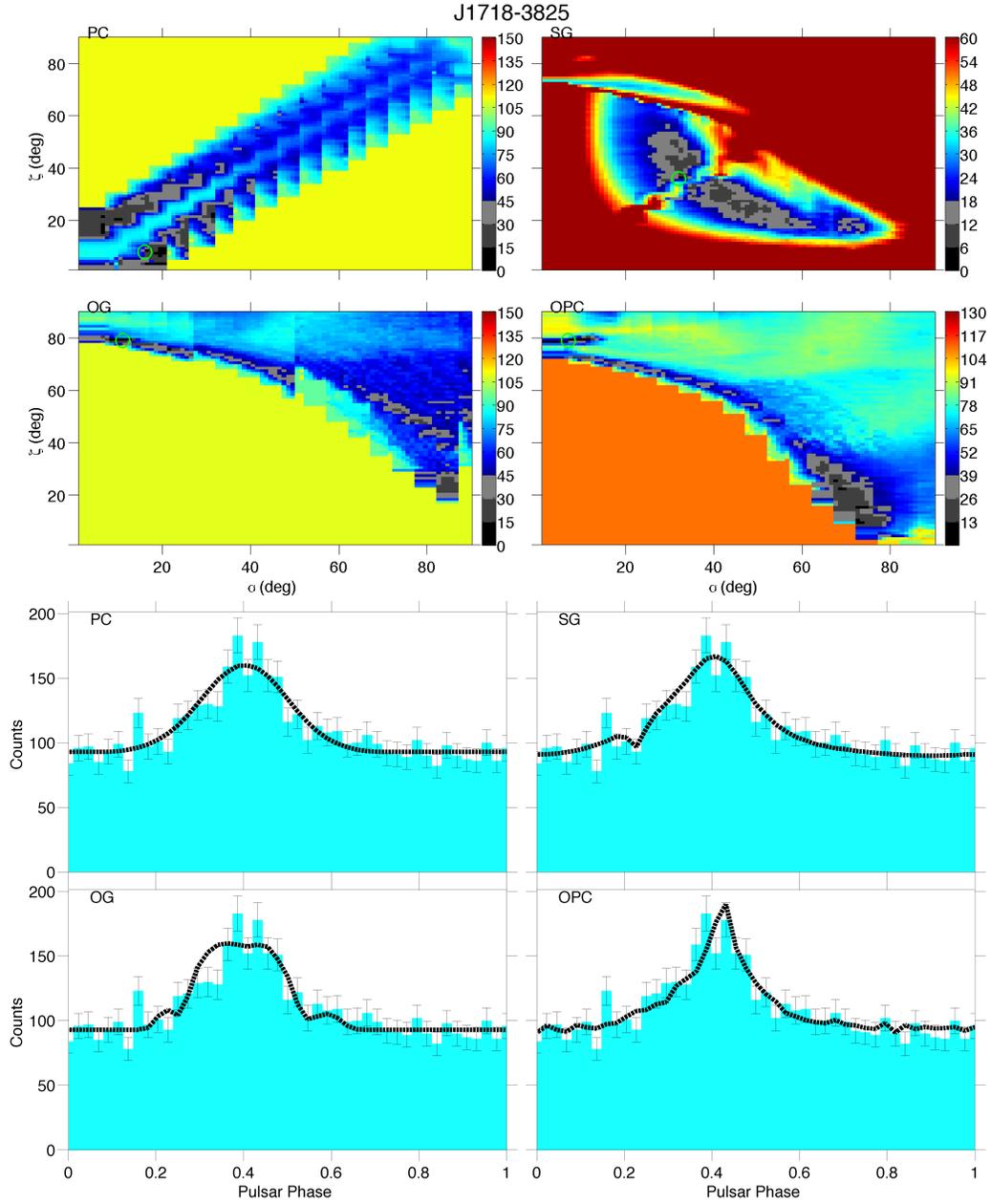

Figure 6.27: PSR J1718-3825. *Top*: for each model is shown the $\alpha$ & $\zeta$ likelihood map obtained with the Poisson FCB γ-ray fit. The color-bar is in $\sigma$ units, zero corresponds to the best fit solution. *Bottom*: the best γ-ray light curve (black dotted line) obtained, for each model, by maximising each likelihood map, superimposed to the FERMI pulsar light curve (in blue).

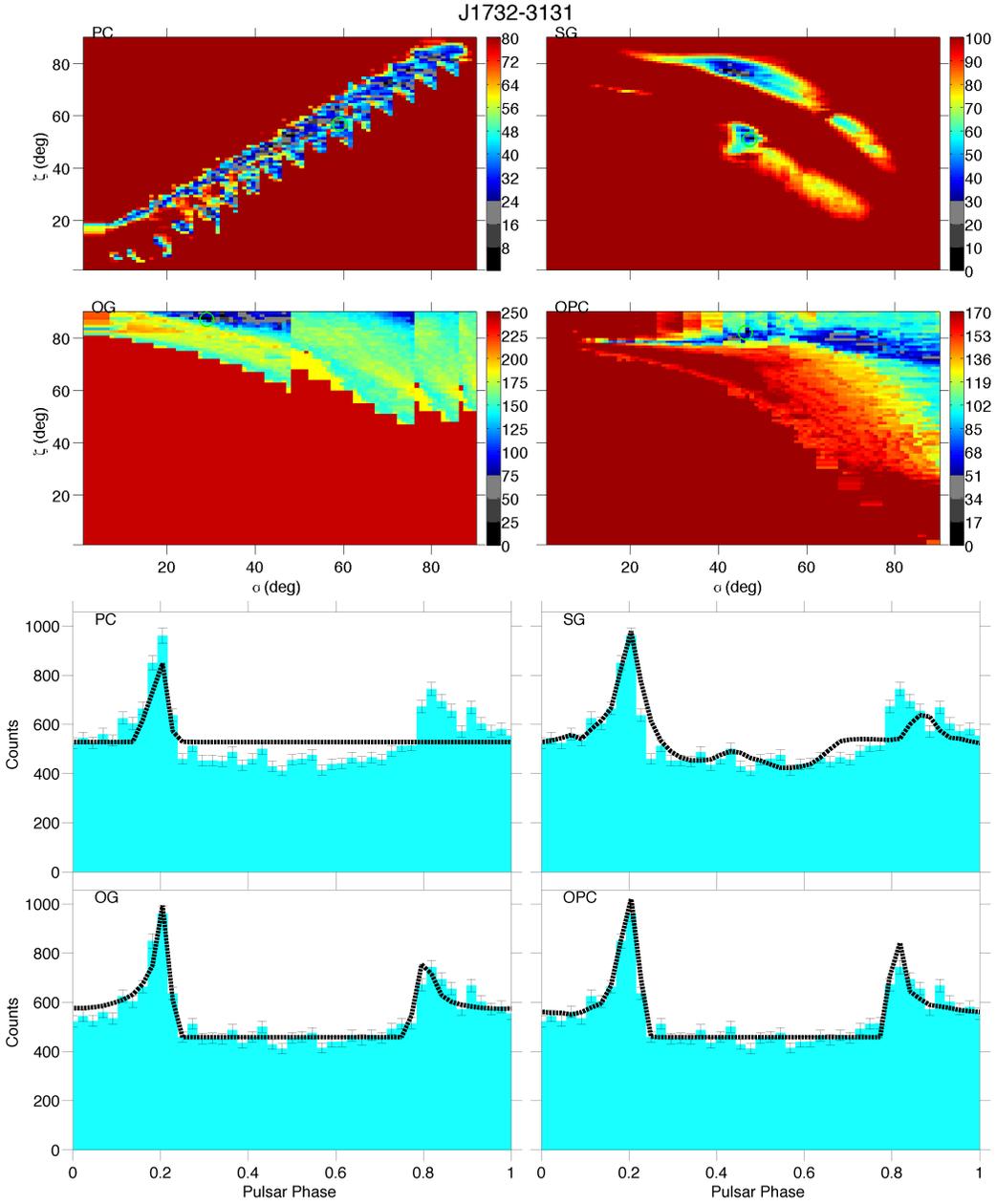

Figure 6.28: PSR J1732-3131. *Top*: for each model is shown the α & ζ likelihood map obtained with the Poisson FCB γ-ray fit. The color-bar is in σ units, zero corresponds to the best fit solution.*Bottom*: the best γ-ray light curve (black dotted line) obtained, for each model, by maximising each likelihood map, superimposed to the FERMI pulsar light curve (in blue).

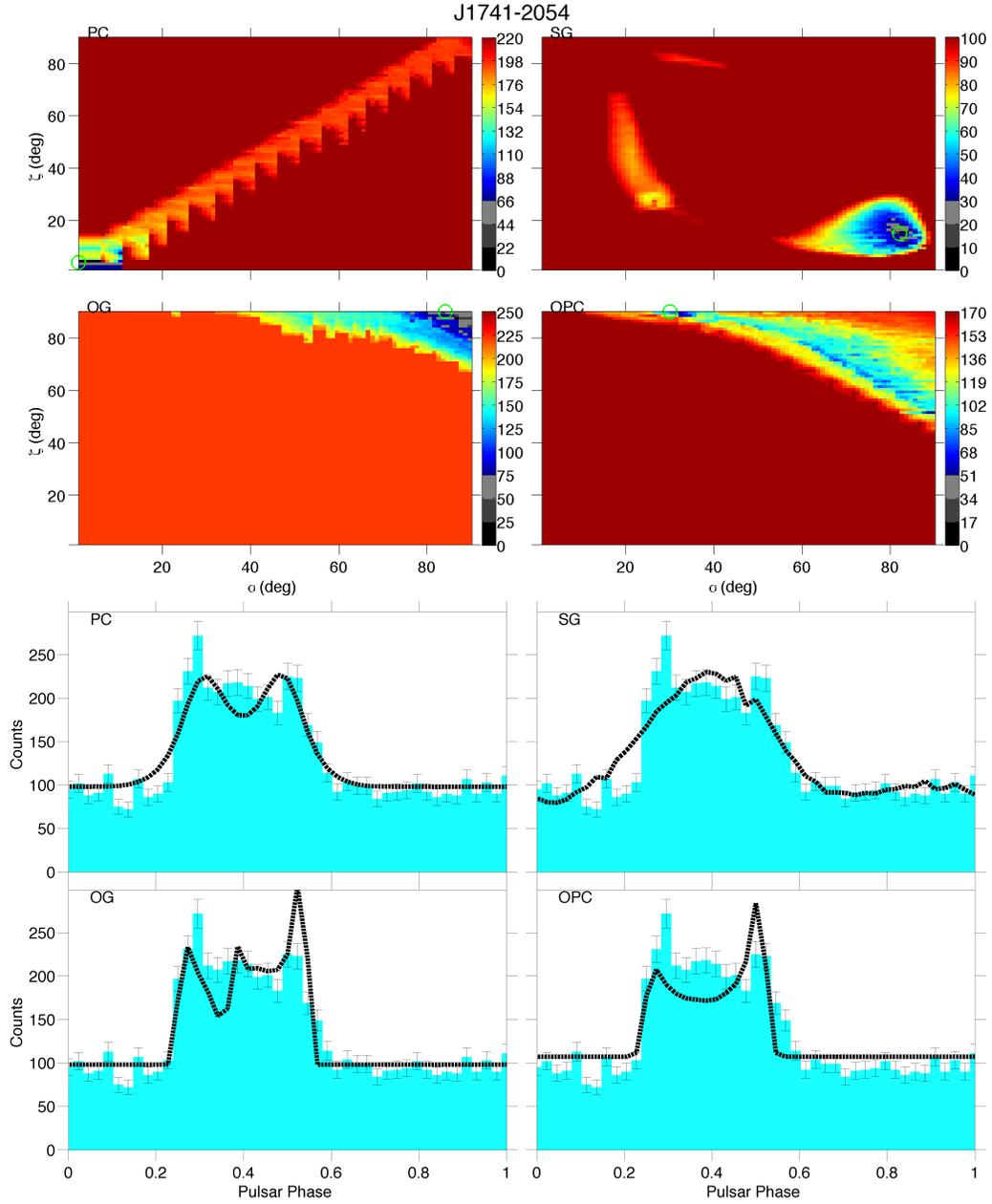

Figure 6.29: PSR J1741-2054. *Top*: for each model is shown the α & ζ likelihood map obtained with the Poisson FCB γ-ray fit. The color-bar is in σ units, zero corresponds to the best fit solution.*Bottom*: the best γ-ray light curve (black dotted line) obtained, for each model, by maximising each likelihood map, superimposed to the FERMI pulsar light curve (in blue).

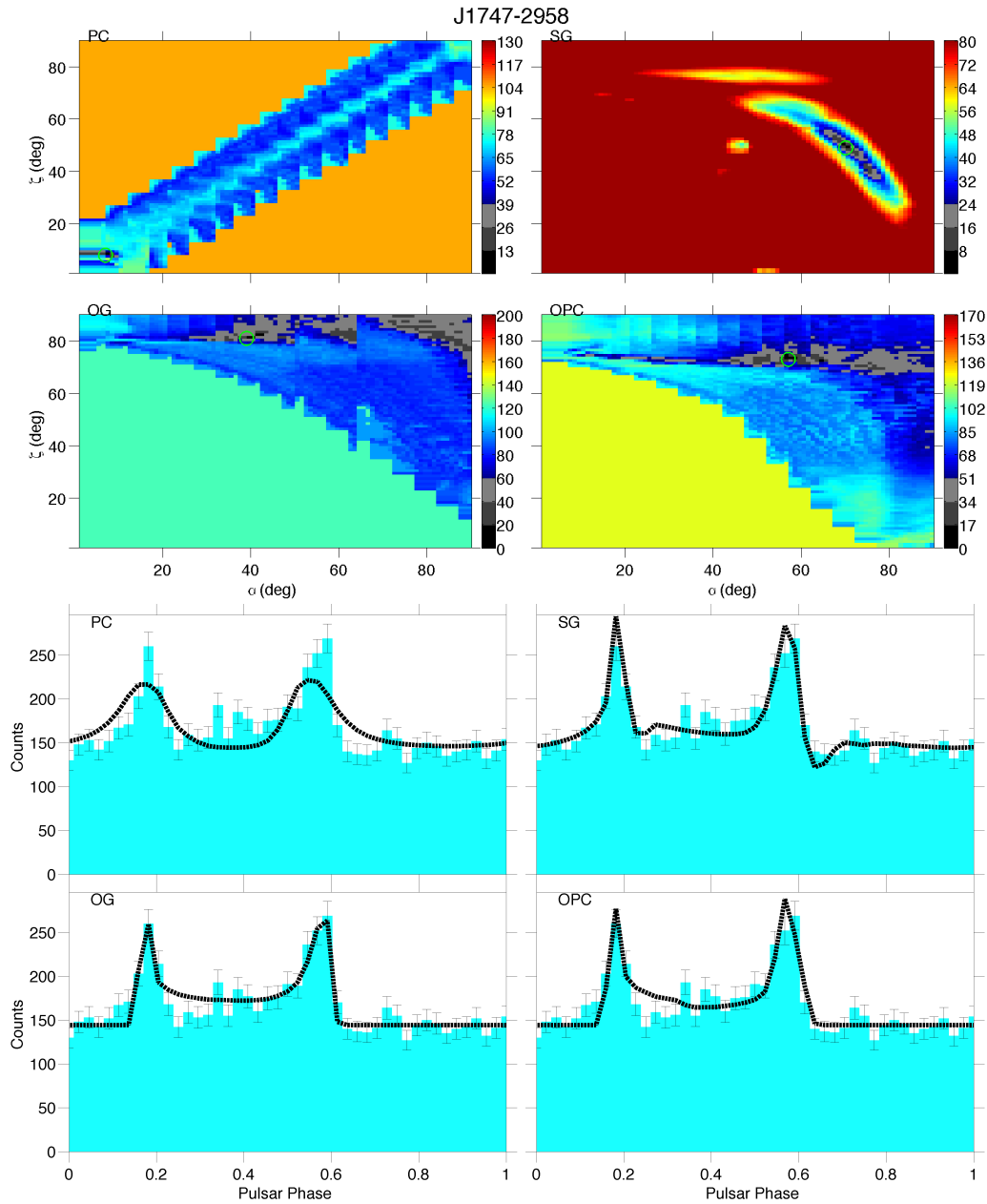

Figure 6.30: PSR J1747-2958. *Top*: for each model is shown the α & ζ likelihood map obtained with the Poisson FCB γ-ray fit. The color-bar is in σ units, zero corresponds to the best fit solution.*Bottom*: the best γ-ray light curve (black dotted line) obtained, for each model, by maximising each likelihood map, superimposed to the FERMI pulsar light curve (in blue).

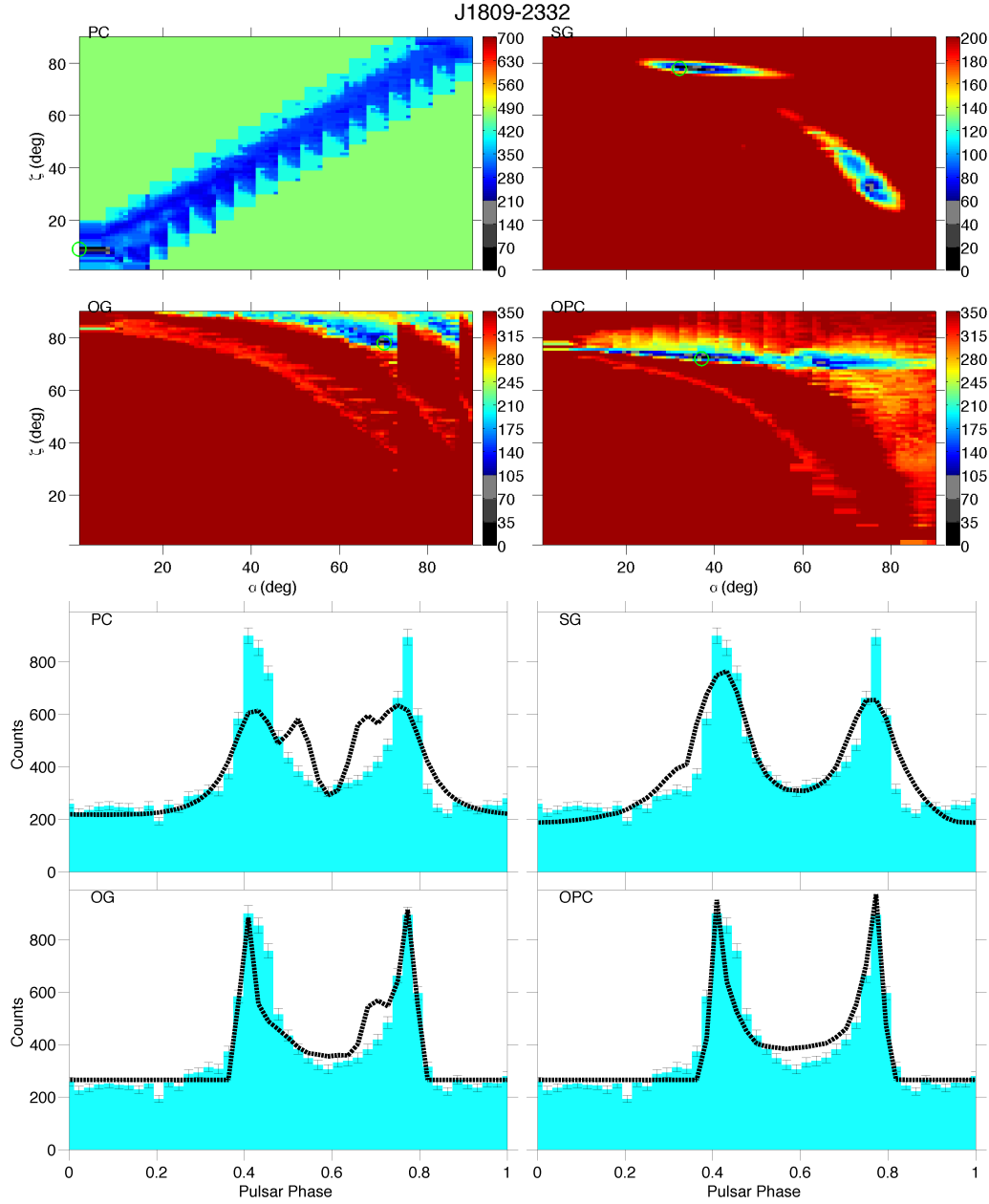

Figure 6.31: PSR J1809-2332. *Top*: for each model is shown the $\alpha$ & $\zeta$ likelihood map obtained with the Poisson FCB $\gamma$-ray fit. The color-bar is in $\sigma$ units, zero corresponds to the best fit solution. *Bottom*: the best $\gamma$-ray light curve (black dotted line) obtained, for each model, by maximising each likelihood map, superimposed to the FERMI pulsar light curve (in blue).

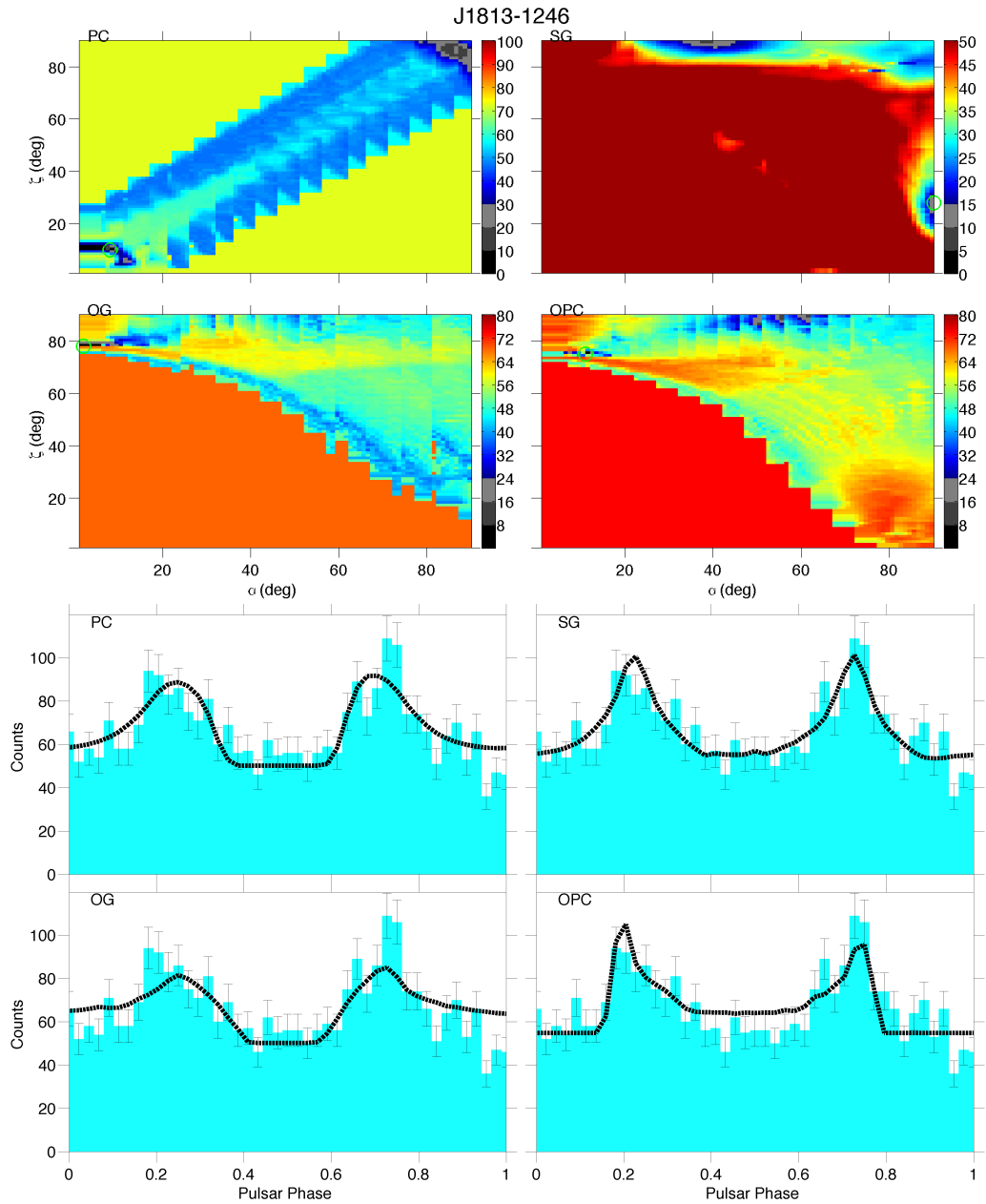

Figure 6.32: PSR J1813-1246. *Top*: for each model is shown the α & ζ likelihood
map obtained with the Poisson FCB γ-ray fit. The color-bar is in σ units, zero
corresponds to the best fit solution.*Bottom*: the best γ-ray light curve (black dotted
line) obtained, for each model, by maximising each likelihood map, superimposed
to the FERMI pulsar light curve (in blue).

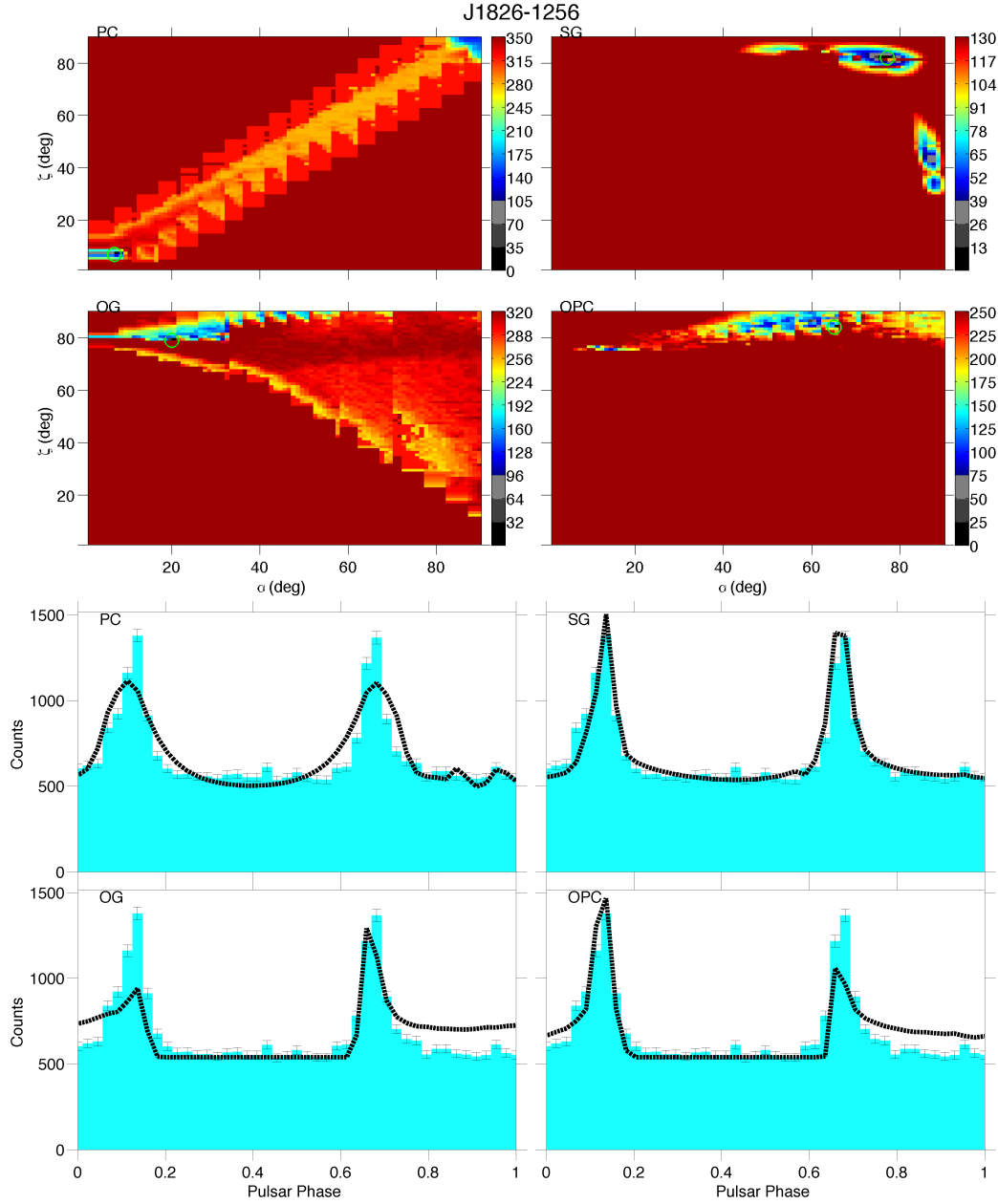

Figure 6.33: PSR J1826-1256. *Top*: for each model is shown the $\alpha$ & $\zeta$ likelihood map obtained with the Poisson FCB $\gamma$-ray fit. The color-bar is in $\sigma$ units, zero corresponds to the best fit solution.*Bottom*: the best $\gamma$-ray light curve (black dotted line) obtained, for each model, by maximising each likelihood map, superimposed to the FERMI pulsar light curve (in blue).

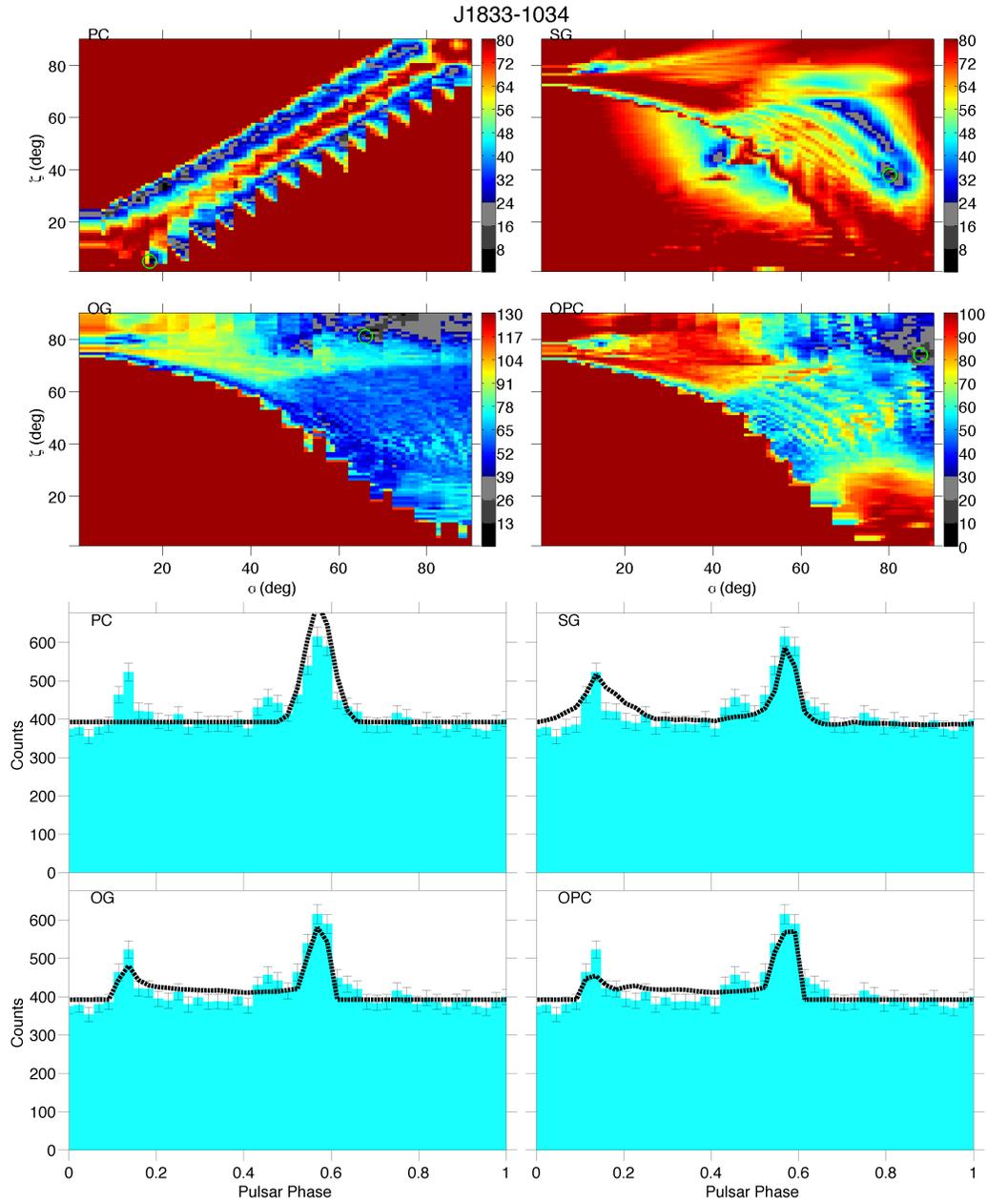

Figure 6.34: PSR J1833-1034. *Top*: for each model is shown the α & ζ likelihood
map obtained with the Poisson FCB γ-ray fit. The color-bar is in σ units, zero
corresponds to the best fit solution.*Bottom*: the best γ-ray light curve (black dotted
line) obtained, for each model, by maximising each likelihood map, superimposed
to the FERMI pulsar light curve (in blue).

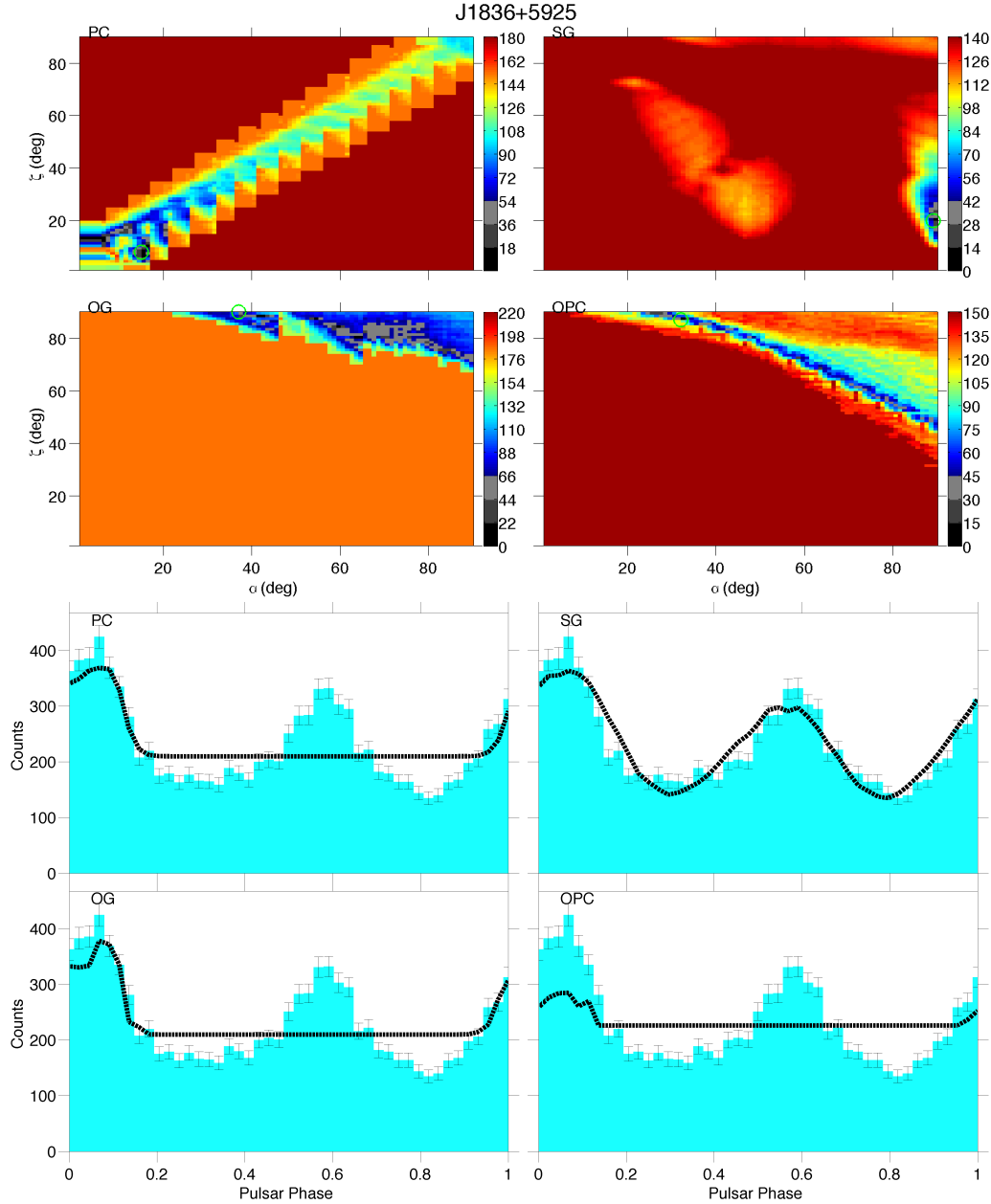

Figure 6.35: PSR J1836+5925. *Top*: for each model is shown the $\alpha$ & $\zeta$ likelihood map obtained with the Poisson FCB $\gamma$-ray fit. The color-bar is in $\sigma$ units, zero corresponds to the best fit solution. *Bottom*: the best $\gamma$-ray light curve (black dotted line) obtained, for each model, by maximising each likelihood map, superimposed to the FERMI pulsar light curve (in blue).

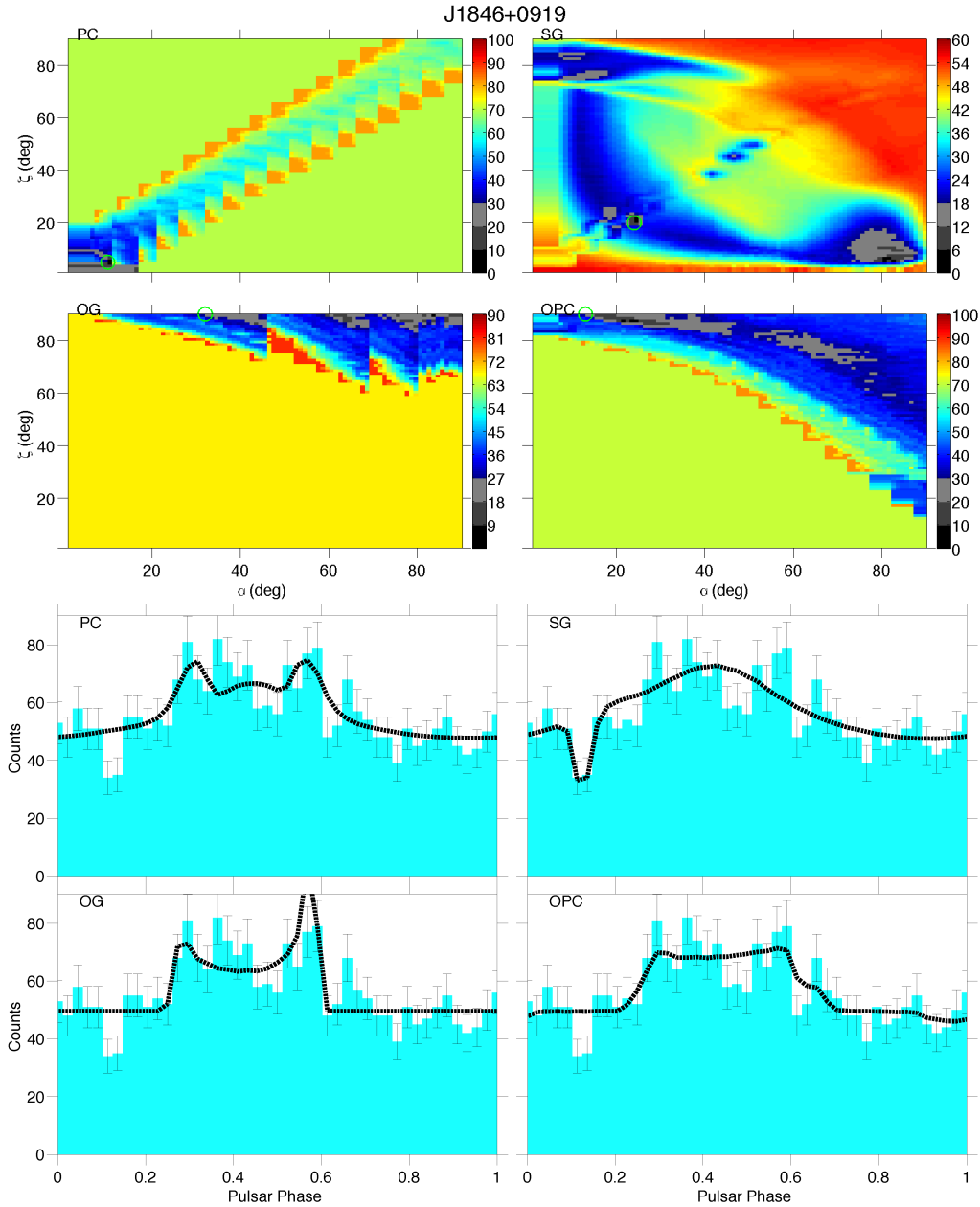

Figure 6.36: PSR J1846+0919. *Top*: for each model is shown the α & ζ likelihood map obtained with the Poisson FCB γ-ray fit. The color-bar is in σ units, zero corresponds to the best fit solution. *Bottom*: the best γ-ray light curve (black dotted line) obtained, for each model, by maximising each likelihood map, superimposed to the FERMI pulsar light curve (in blue).

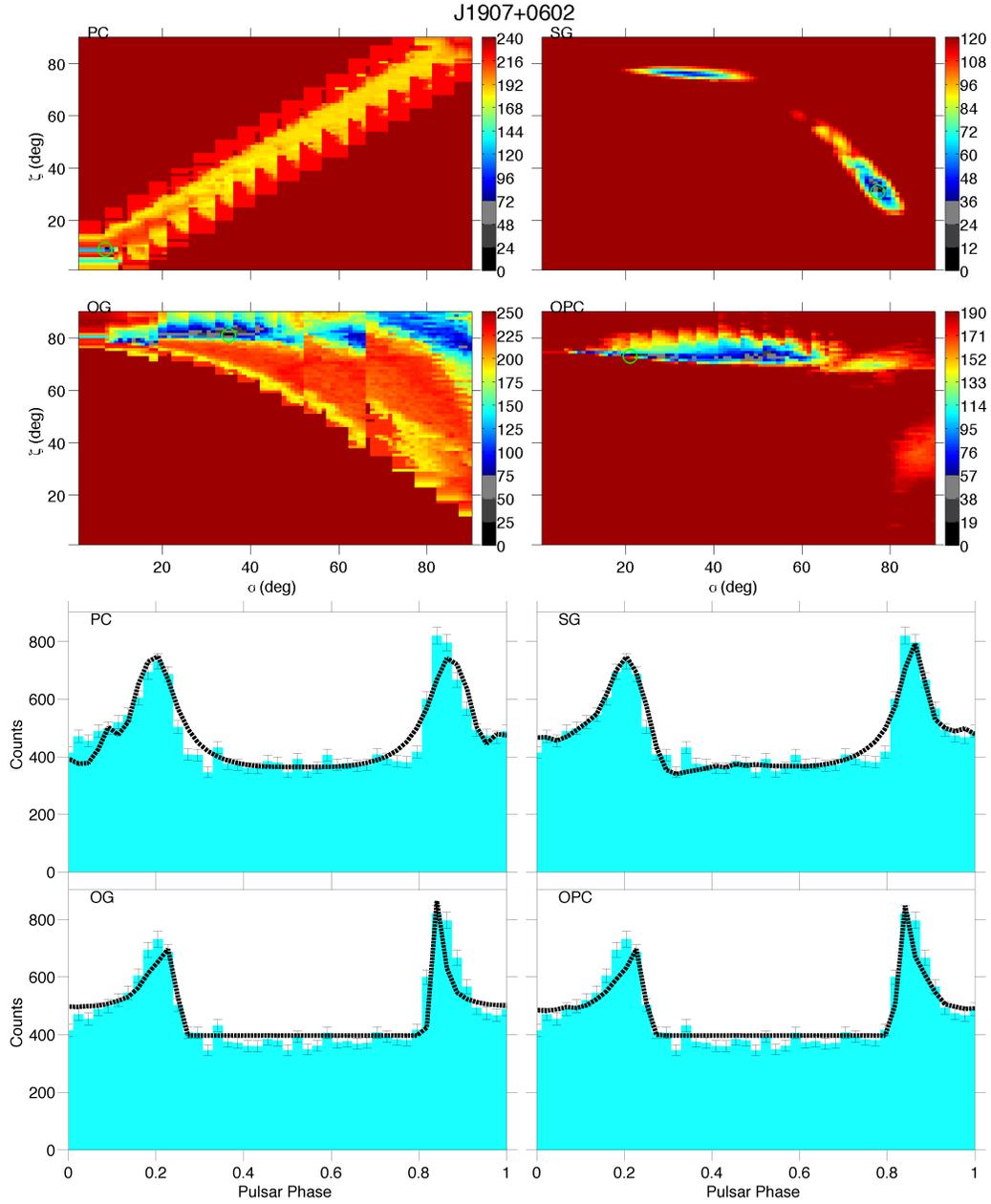

Figure 6.37: PSR J1907+0602. *Top*: for each model is shown the $\alpha$ & $\zeta$ likelihood map obtained with the Poisson FCB $\gamma$-ray fit. The color-bar is in $\sigma$ units, zero corresponds to the best fit solution. *Bottom*: the best $\gamma$-ray light curve (black dotted line) obtained, for each model, by maximising each likelihood map, superimposed to the FERMI pulsar light curve (in blue).

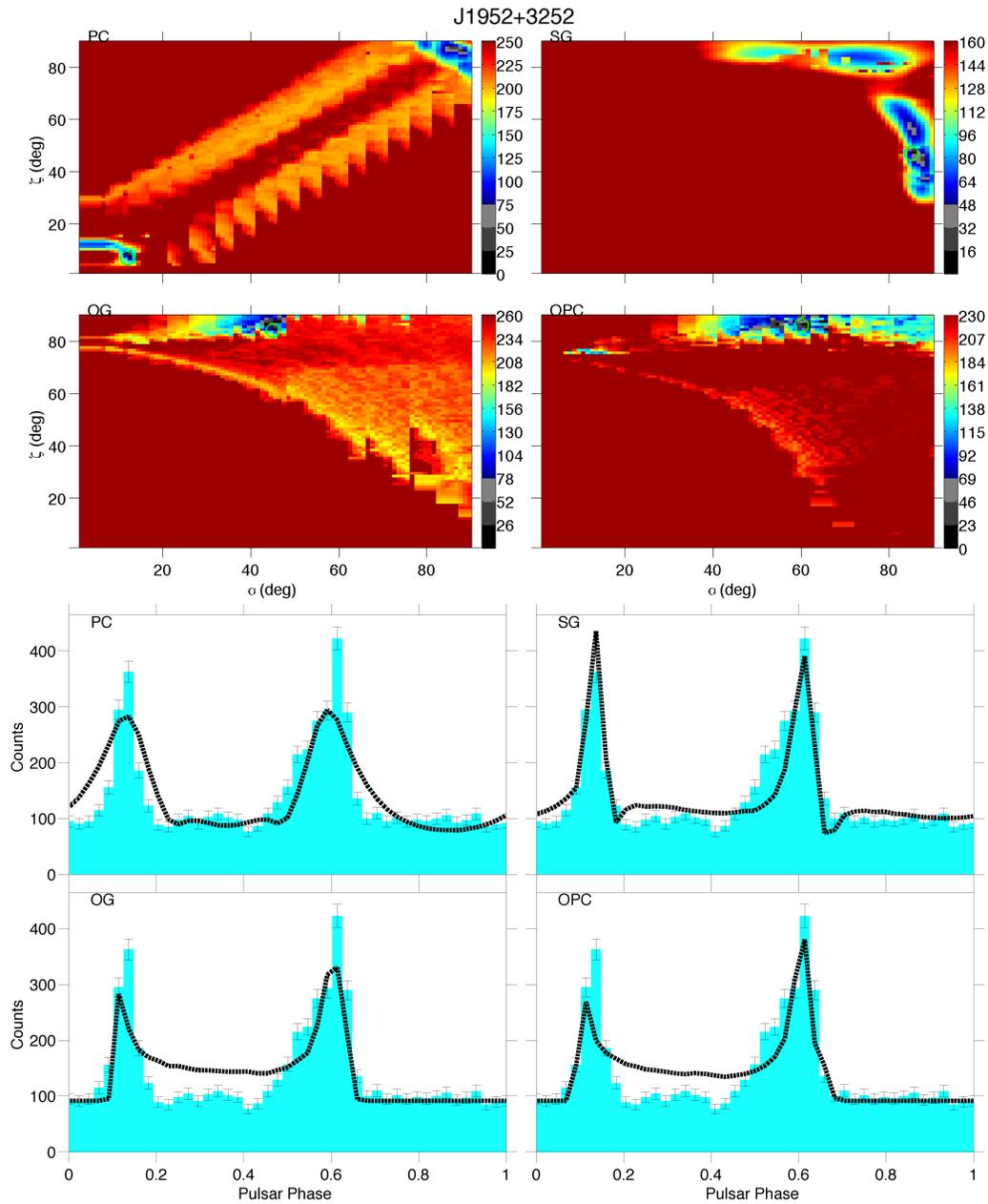

Figure 6.38: PSR J1952+3252. *Top*: for each model is shown the α & ζ likelihood map obtained with the Poisson FCB γ-ray fit. The color-bar is in σ units, zero corresponds to the best fit solution.*Bottom*: the best γ-ray light curve (black dotted line) obtained, for each model, by maximising each likelihood map, superimposed to the FERMI pulsar light curve (in blue).

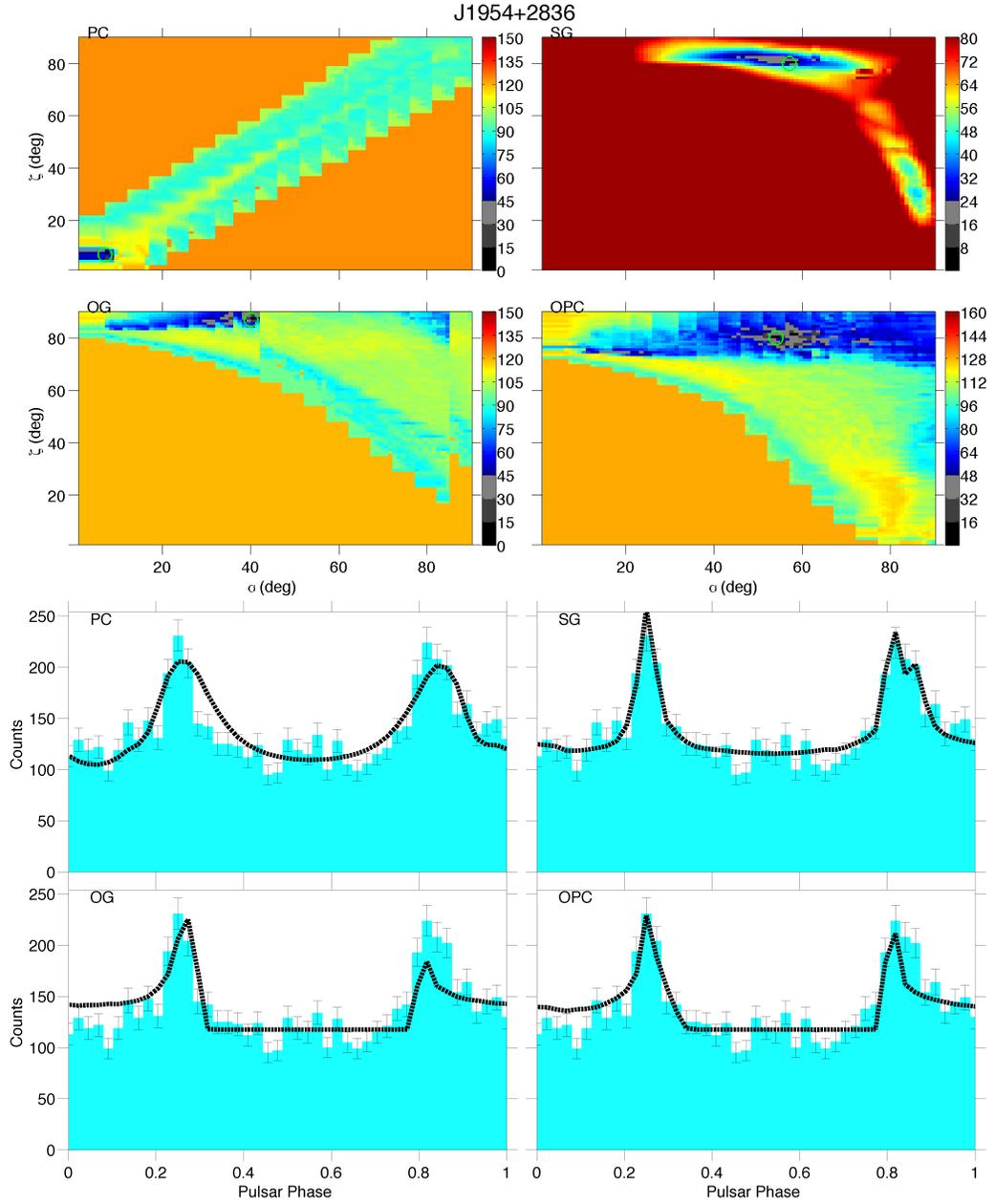

Figure 6.39: PSR J1954+2836. *Top*: for each model is shown the α & ζ likelihood map obtained with the Poisson FCB γ-ray fit. The color-bar is in σ units, zero corresponds to the best fit solution.*Bottom*: the best γ-ray light curve (black dotted line) obtained, for each model, by maximising each likelihood map, superimposed to the FERMI pulsar light curve (in blue).

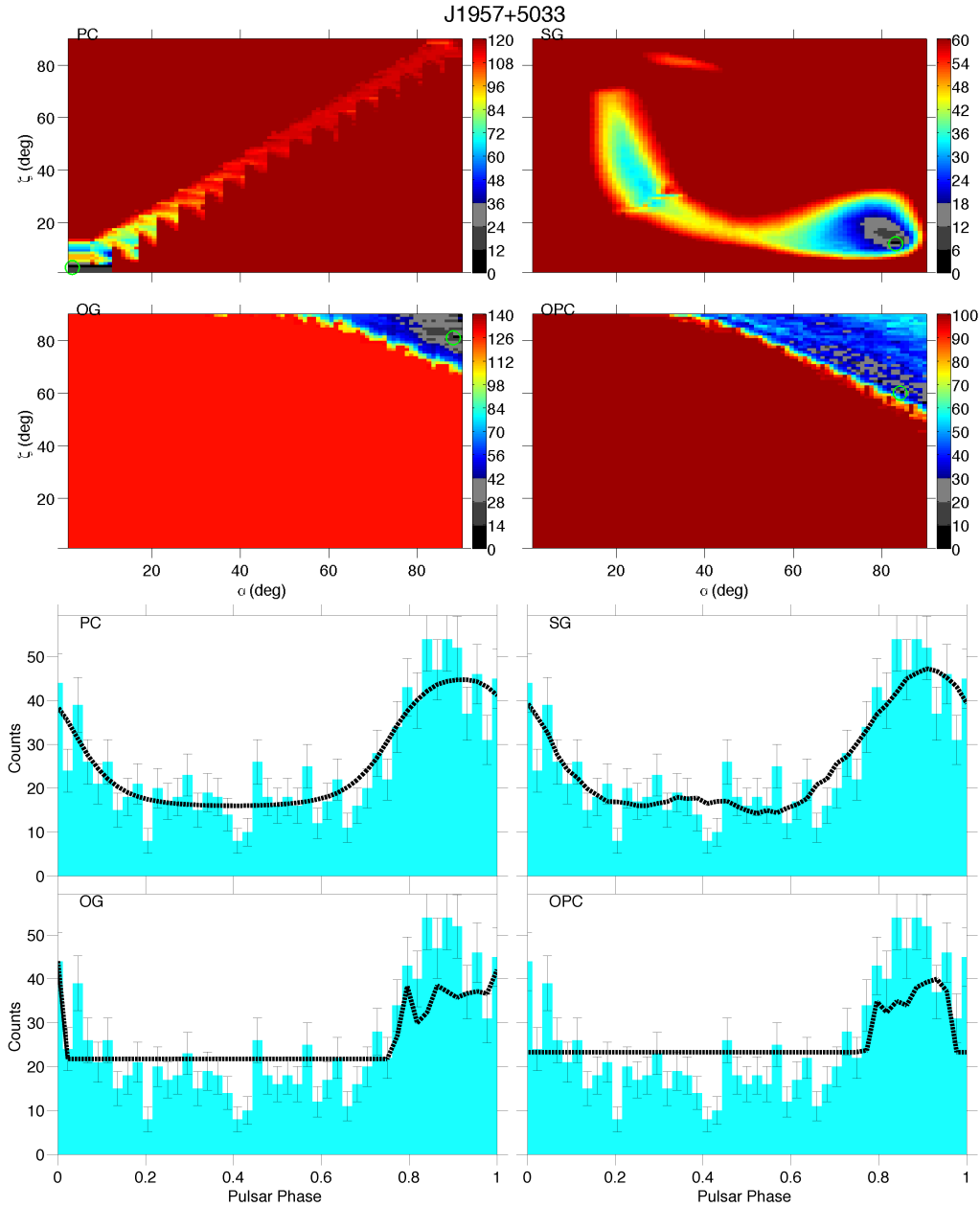

Figure 6.40: PSR J1957+5033. *Top*: for each model is shown the $\alpha$ & $\zeta$ likelihood map obtained with the Poisson FCB $\gamma$-ray fit. The color-bar is in $\sigma$ units, zero corresponds to the best fit solution. *Bottom*: the best $\gamma$-ray light curve (black dotted line) obtained, for each model, by maximising each likelihood map, superimposed to the FERMI pulsar light curve (in blue).

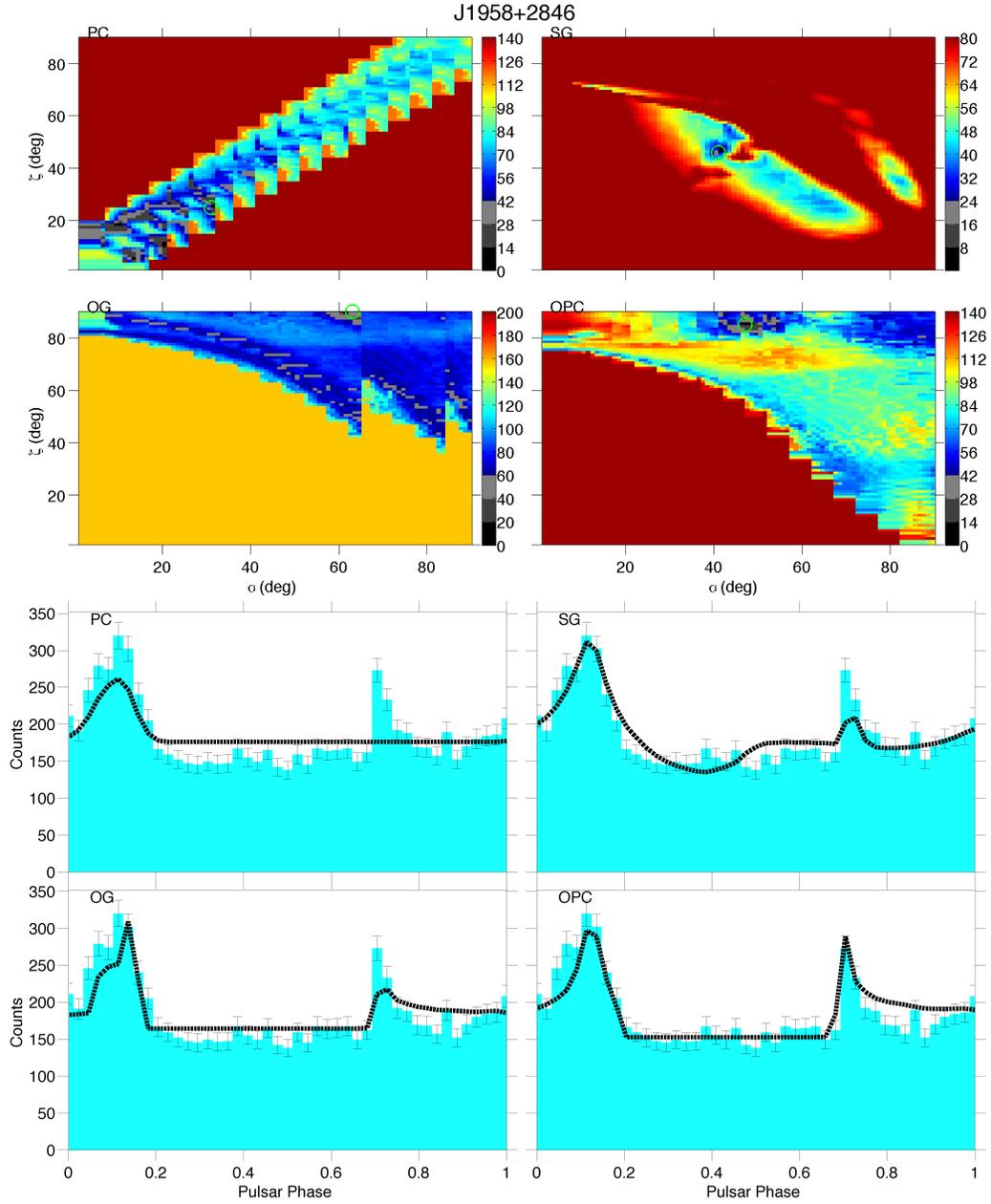

Figure 6.41: PSR J1958+2846. *Top*: for each model is shown the α & ζ likelihood map obtained with the Poisson FCB γ-ray fit. The color-bar is in σ units, zero corresponds to the best fit solution.*Bottom*: the best γ-ray light curve (black dotted line) obtained, for each model, by maximising each likelihood map, superimposed to the FERMI pulsar light curve (in blue).

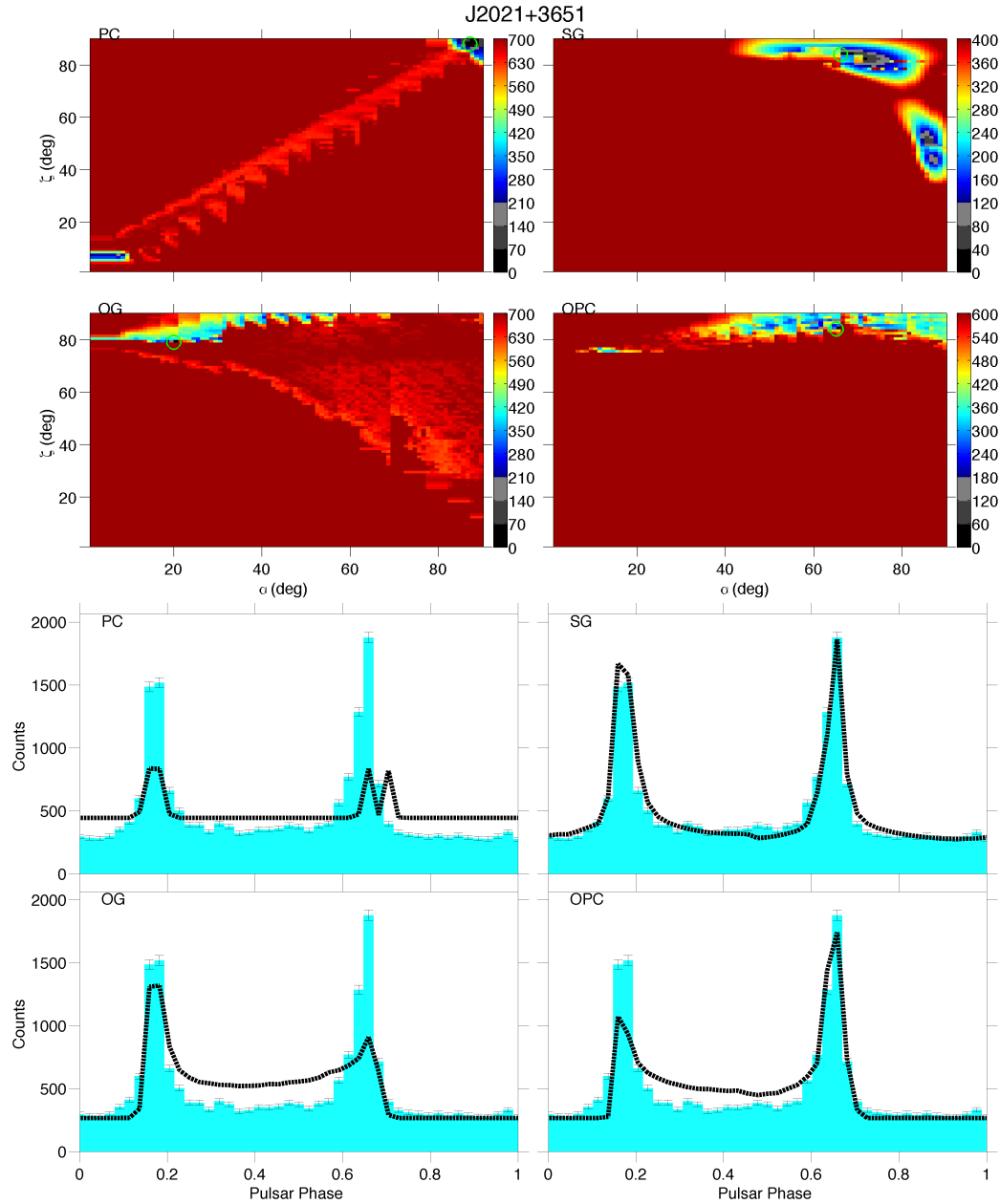

Figure 6.42: PSR J2021+3651. *Top*: for each model is shown the $\alpha$ & $\zeta$ likelihood map obtained with the Poisson FCB $\gamma$-ray fit. The color-bar is in $\sigma$ units, zero corresponds to the best fit solution.*Bottom*: the best $\gamma$-ray light curve (black dotted line) obtained, for each model, by maximising each likelihood map, superimposed to the FERMI pulsar light curve (in blue).

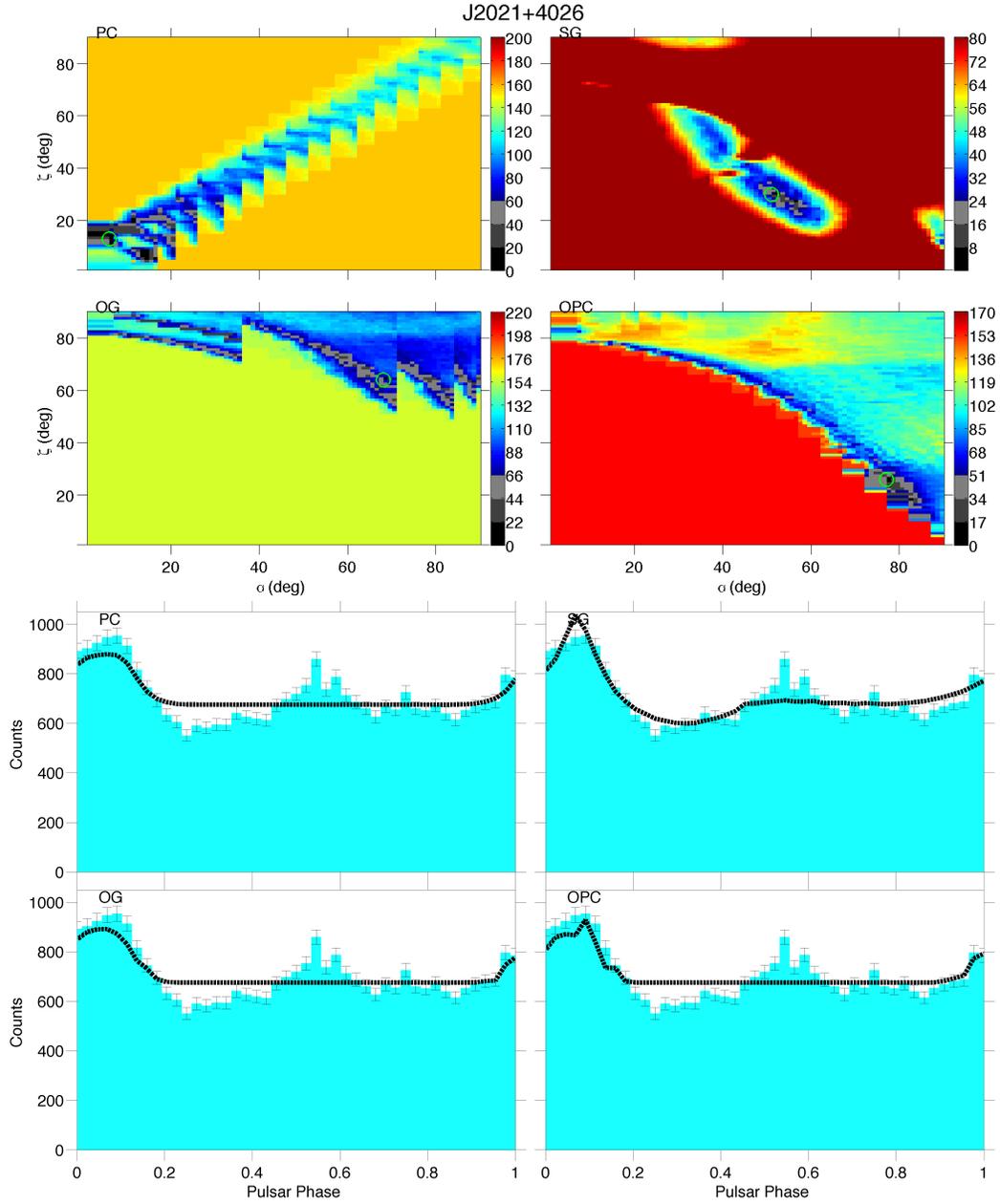

Figure 6.43: PSR J2021+4026. *Top*: for each model is shown the α & ζ likelihood map obtained with the Poisson FCB γ-ray fit. The color-bar is in σ units, zero corresponds to the best fit solution.*Bottom*: the best γ-ray light curve (black dotted line) obtained, for each model, by maximising each likelihood map, superimposed to the FERMI pulsar light curve (in blue).

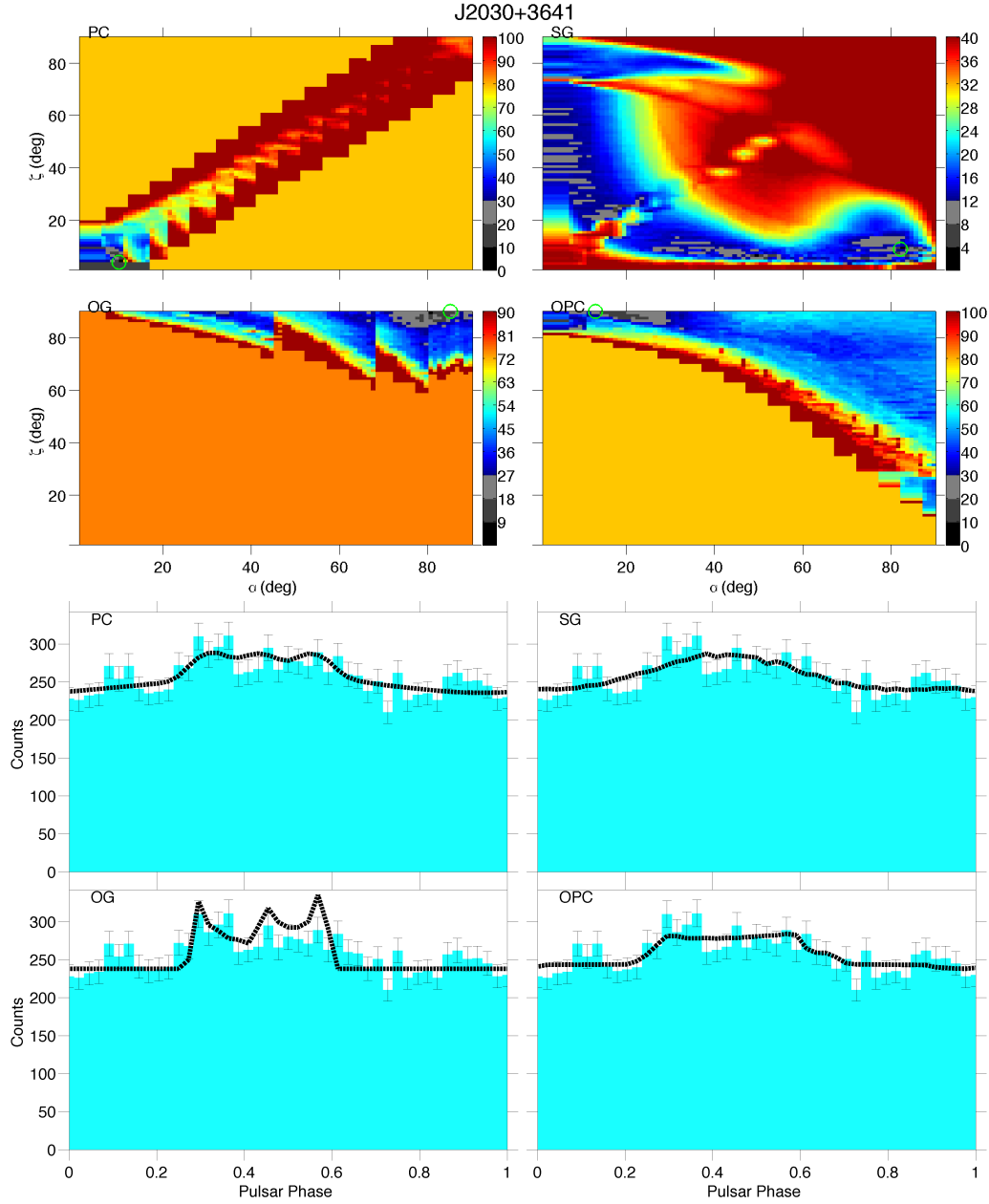

Figure 6.44: PSR J2030+3641. *Top*: for each model is shown the $\alpha$ & $\zeta$ likelihood map obtained with the Poisson FCB $\gamma$-ray fit. The color-bar is in $\sigma$ units, zero corresponds to the best fit solution.*Bottom*: the best $\gamma$-ray light curve (black dotted line) obtained, for each model, by maximising each likelihood map, superimposed to the FERMI pulsar light curve (in blue).

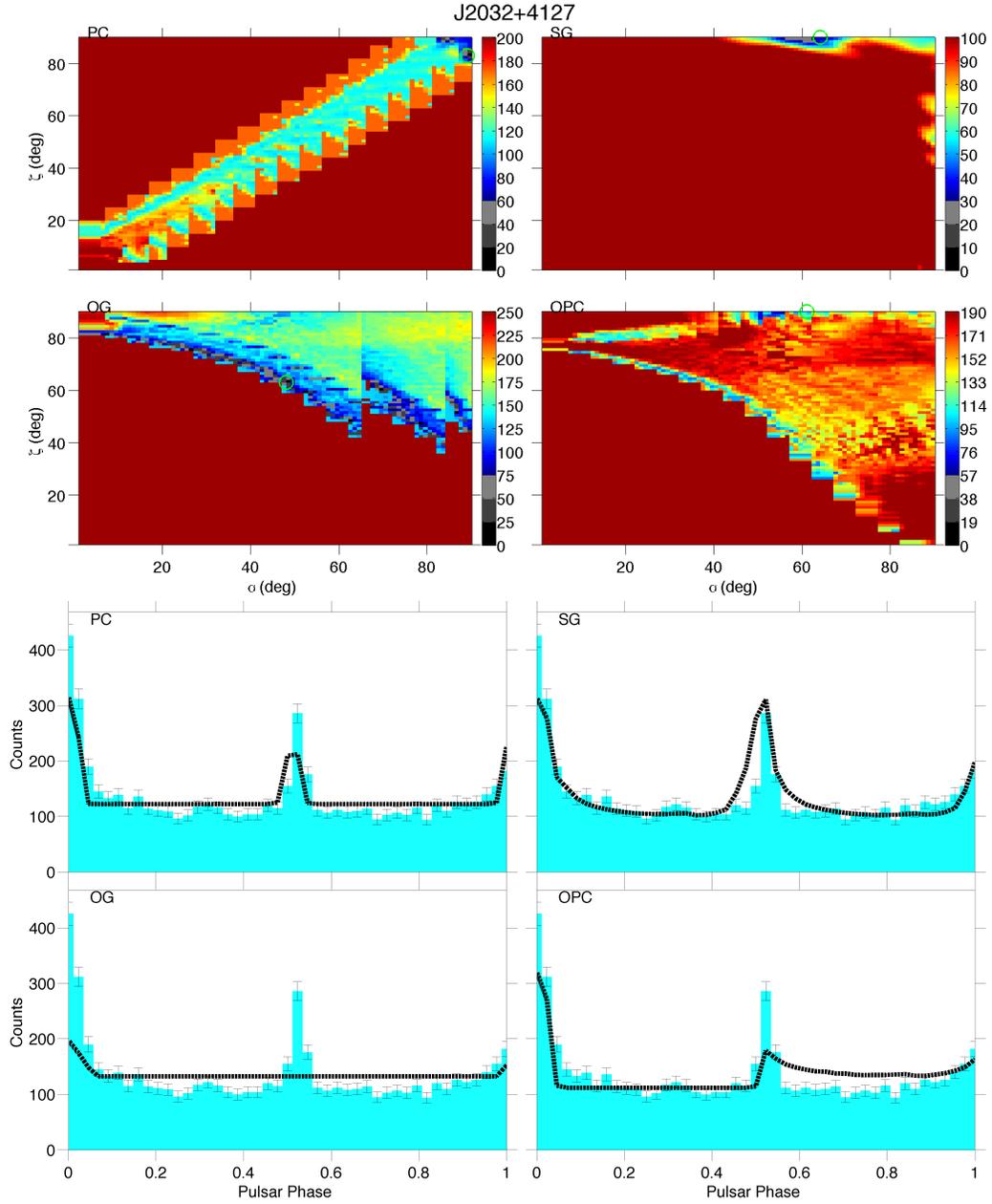

Figure 6.45: PSR J2032+4127. *Top*: for each model is shown the $\alpha$ & $\zeta$ likelihood map obtained with the Poisson FCB γ-ray fit. The color-bar is in $\sigma$ units, zero corresponds to the best fit solution.*Bottom*: the best γ-ray light curve (black dotted line) obtained, for each model, by maximising each likelihood map, superimposed to the FERMI pulsar light curve (in blue).

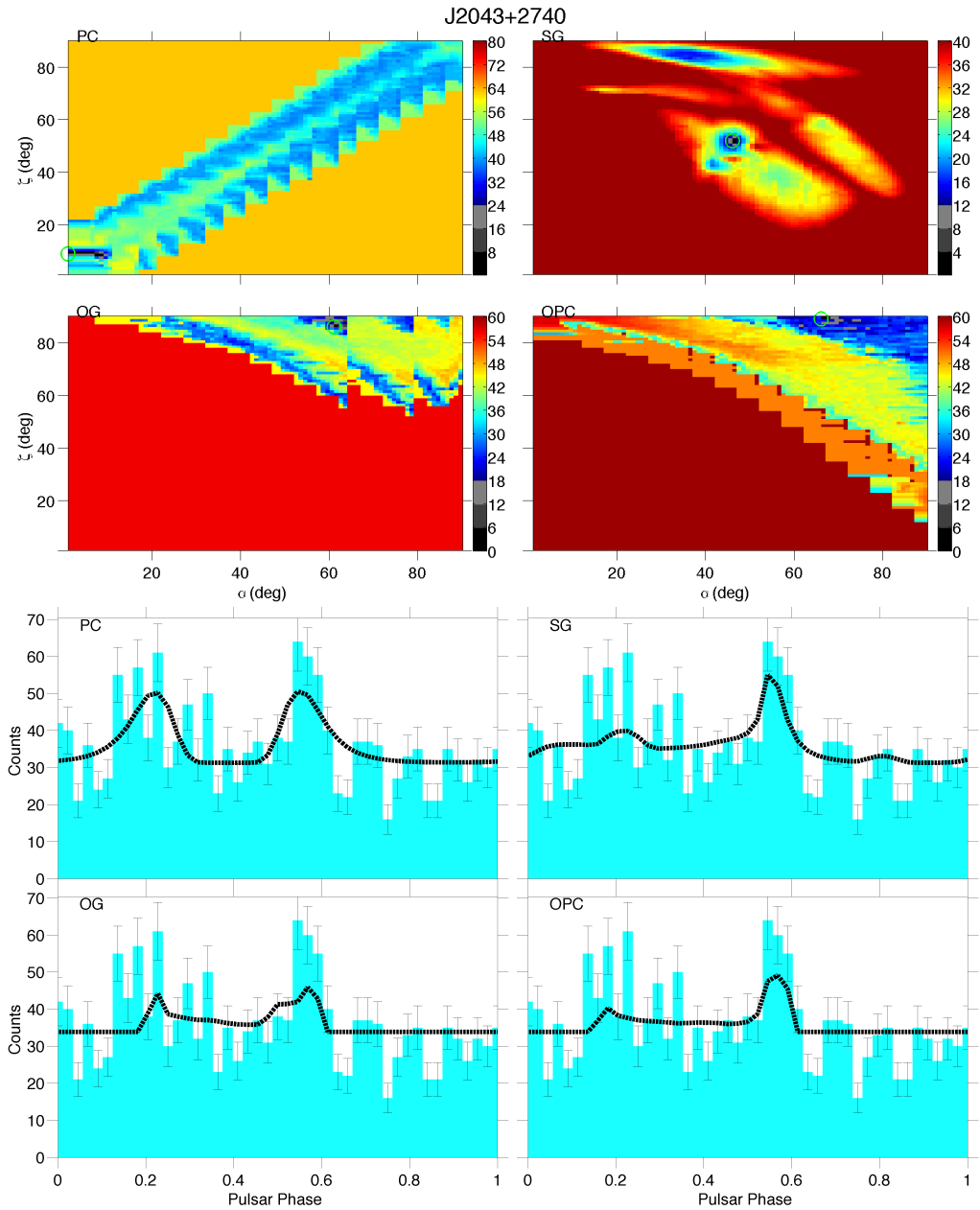

Figure 6.46: PSR J2043+2740. *Top*: for each model is shown the α & ζ likelihood map obtained with the Poisson FCB γ-ray fit. The color-bar is in σ units, zero corresponds to the best fit solution. *Bottom*: the best γ-ray light curve (black dotted line) obtained, for each model, by maximising each likelihood map, superimposed to the FERMI pulsar light curve (in blue).

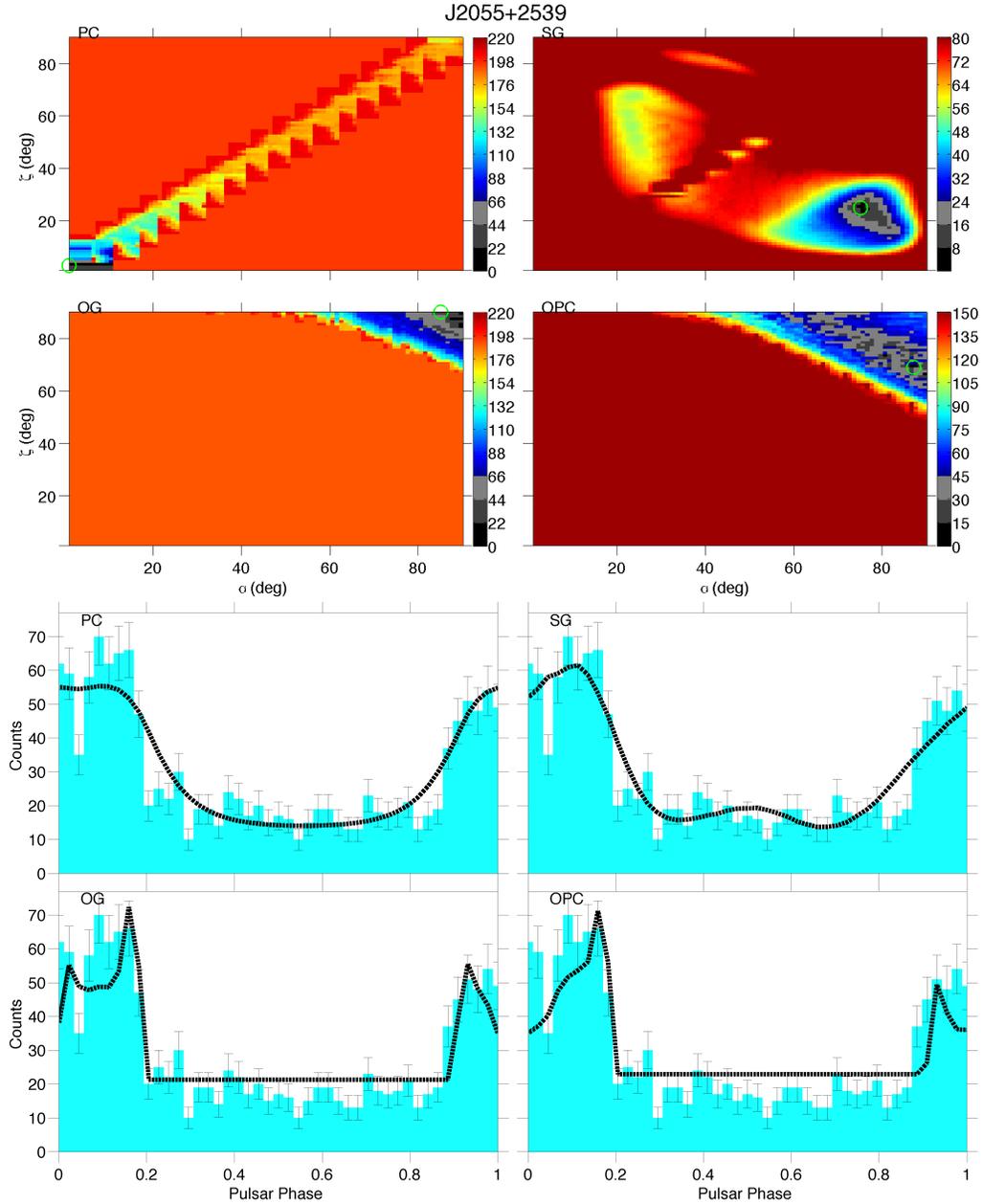

Figure 6.47: PSR J2055+2539. *Top*: for each model is shown the α & ζ likelihood map obtained with the Poisson FCB γ-ray fit. The color-bar is in σ units, zero corresponds to the best fit solution. *Bottom*: the best γ-ray light curve (black dotted line) obtained, for each model, by maximising each likelihood map, superimposed to the FERMI pulsar light curve (in blue).

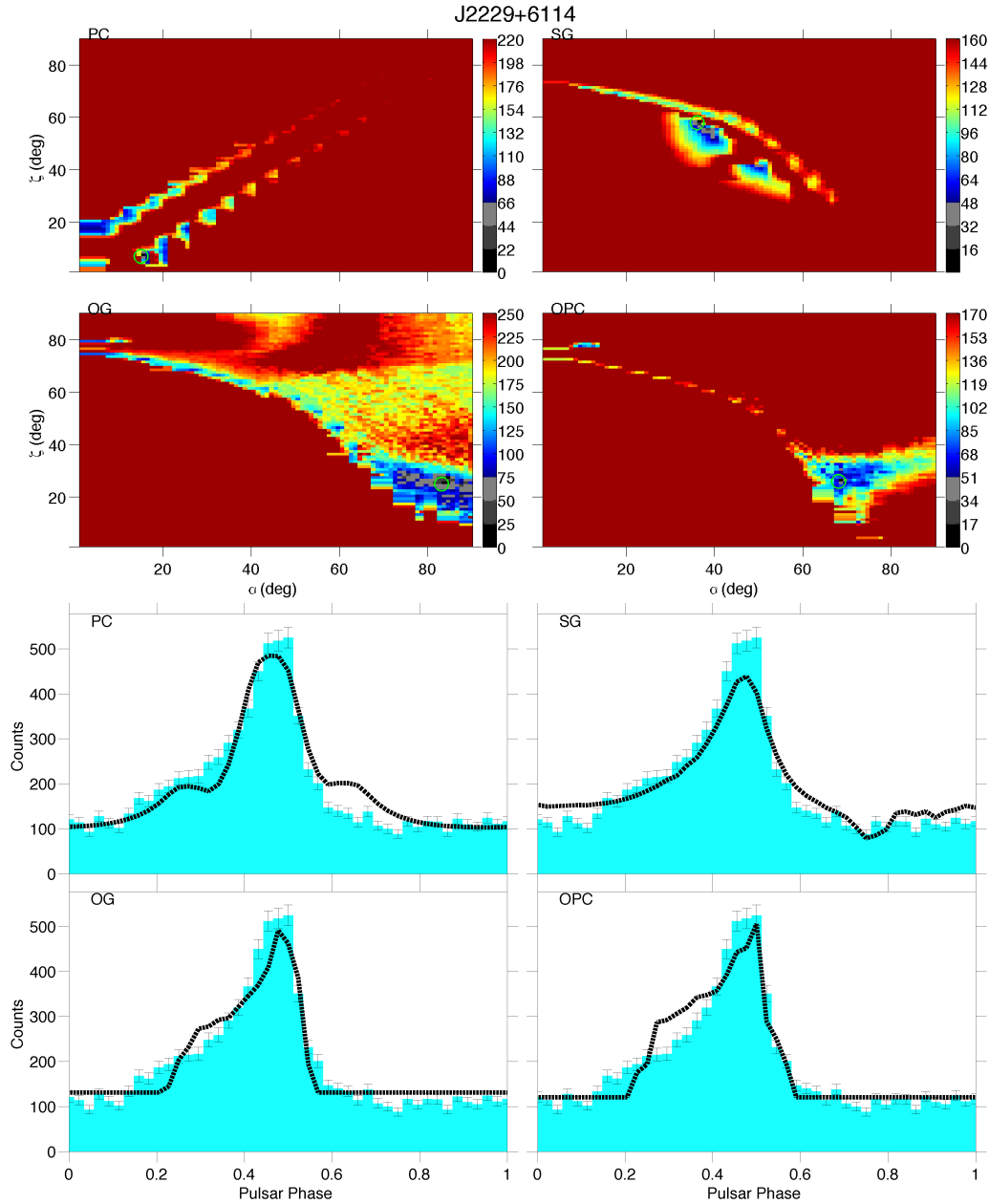

Figure 6.48: PSR J2229+6114. *Top*: for each model is shown the $\alpha$ & $\zeta$ likelihood map obtained with the Poisson FCB $\gamma$-ray fit. The color-bar is in $\sigma$ units, zero corresponds to the best fit solution.*Bottom*: the best $\gamma$-ray light curve (black dotted line) obtained, for each model, by maximising each likelihood map, superimposed to the FERMI pulsar light curve (in blue).

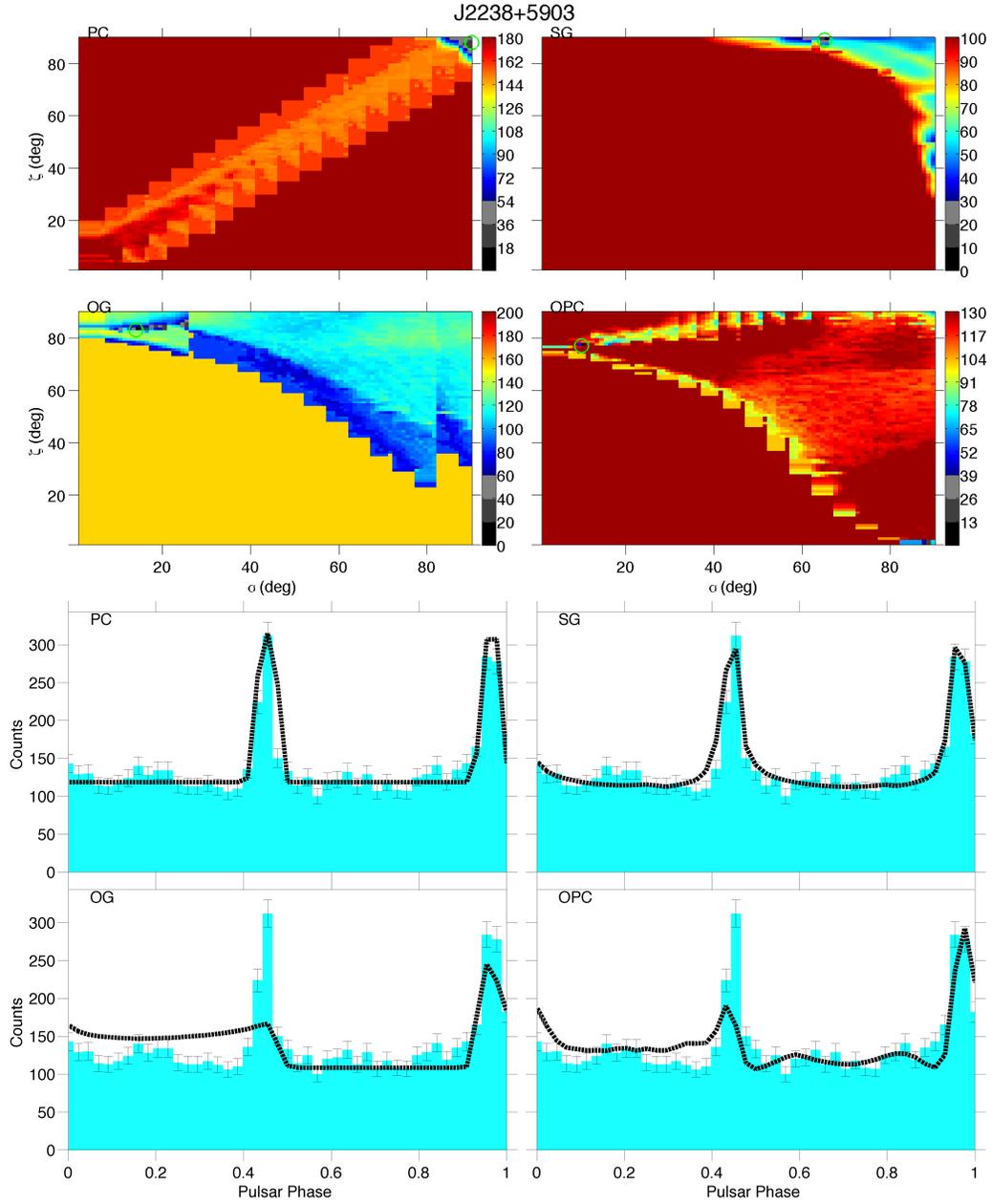

Figure 6.49: PSR J2238+5903. *Top*: for each model is shown the $\alpha$ & $\zeta$ likelihood map obtained with the Poisson FCB γ-ray fit. The color-bar is in $\sigma$ units, zero corresponds to the best fit solution. *Bottom*: the best γ-ray light curve (black dotted line) obtained, for each model, by maximising each likelihood map, superimposed to the FERMI pulsar light curve (in blue).

## 6.2 Fitting both the $\gamma$-ray and radio emission

The joint $\gamma$-radio estimation of the $\alpha$ and $\zeta$ angles has been performed for
22 of the radio pulsars detected in $\gamma$-ray by the LAT. The radio light curve
profiles have been downloaded from the NASA science support centre FERMI
web page[1]. The $\gamma$-ray light curves used to perform all the fits described in
this chapter have been evaluated by using the same criteria and data selection as
described in section 4.2.1.

The individual $\gamma$-ray fit gives an estimate of the pulsar orientation just
based on the high energy emission patterns. The radio light curve profile is
used against the radio phase-plot model to yield an independent estimate of the
pulsar orientation.

I have first selected the radio light curve profiles of the LAT pulsars and
I have fitted them by using the phase-plot light curves of the radio emission
model implemented in this thesis. Second, the likelihood maps have been
summed with the previously obtained $\gamma$-ray ones to obtain a pulsar orientation
estimate based on both the radio and high-energy emission mechanisms. In the
next sections I will give a detailed description of the combination of individual
$\gamma$-ray and radio fits to obtain a joint evaluation of the $\alpha$ and $\zeta$ pulsar orientation
angles.

### 6.2.1 Individual radio fit

I have implemented a 5 free parameters $\chi^2$ likelihood fit for some of the radio-
loud LAT pulsars. No light curve re-binning has been applied; each likelihood
value has been evaluated between regular binned (RB) light curves. To be able
to compare the fit results with the previously obtained $\gamma$-ray ones, the same 5
free parameters defined in section 6.1.2, equally stepped in the same intervals,
have been chosen. The evaluation of the fit likelihood values has been done by
using equation 6.7 with the variance $\sigma$ directly evaluated on the observed radio
light-curves. The maximisation of the $90_\alpha \times 90_\zeta \times 45_{\phi\ steps} \times 15_{flat\ bkg} \times 10_{norm}$
likelihood matrix, has been done evaluating, for each pulsar, the best fit
normalisation factor, flat level, and shift. An $\alpha$-$\zeta$ likelihood map has been
generated for each pulsar, and the $\alpha$-$\zeta$ best fit values have been evaluated by
maximising the obtained map.

This first fit has been implemented under the assumption that $\sigma_{sim} = \sigma_{obs}$.
Since the observed radio light curve have a very high accuracy (very low
variance) the fit is very highly constrained. It is in fact so highly constrained
compared to the $\gamma$-ray one, rendering the joint radio-$\gamma$ fit useless because it is
entirely driven by the radio likelihood map. To lower the weight of the radio
fit, I have used these first best-fit $f^*(x)$ light-curves to derive a new $\sigma_{sim}$ value based on the reduced $\chi^2 = 1$ criterion.

From the first fit with with $\sigma_{obs}$ of the observed light-curve $y$ we get

$$\ln L_{max} = -\frac{1}{2\sigma_{obs}^2} \sum_j [y_j - f^*(x_j)]^2 \quad with \quad \sigma_{obs} = \sigma_*. \qquad (6.11)$$

Now, from the reduced $\chi^2=1$ criterion defined in equation 6.9 we can define the optimised $\sigma_*$ as

$$\sigma_*^2 = -\frac{2\ln L_{max}}{n_{free}} \sigma_{obs}^2. \qquad (6.12)$$

The new $\sigma_*$ optimised fit has been applied to generate the radio $\alpha$-$\zeta$ likelihood map to be used for the joint $\gamma$-radio estimation of the pulsar orientation.

### 6.2.2   Joint γ-radio estimation of the LAT pulsar orientations

The total probability that 2 independent phenomena occur simultaneously is given by the product of the probability that each single phenomenon occurs. Since the $\gamma$ & radio likelihood maps have been evaluated in logarithmic scale and since the radio and $\gamma$ ray emissions occur simultaneously, to evaluate the best joint $\alpha$-$\zeta$ solution it is enough to maximise the summed $\alpha$-$\zeta$ likelihood maps evaluated for each pulsar.

During the first trials made to evaluate a common solution, the real difficulty was that the joint solution was always dominated by the radio fit e.g. the range of the radio likelihood values was huge compared with the $\gamma$ one ($\sim 10^6$ times). Because of this, the shape of the summed $\gamma$+radio likelihood map was driven by the radio one and the best fit solution found, identical to the radio one. The reason lied in the use of the very small variance $\sigma_{obs}$ in the radio fit compared with the very low counts statistics and large errors in $\gamma$-rays. The huge likelihood drops found across an $\alpha, \zeta$ map in the radio case cannot therefore be compared to the modest likelihood changes across the map in the gamma-ray case.

One solution is to use the same Gaussian probability distribution and to use the same reduced $\chi^2 = 1$ criterion to derive comparable variances for the radio and gamma-ray curves. I have summed the $\gamma$ & radio likelihood maps evaluated, in regular binning with the $\sigma$ optimised $\chi^2$ distribution described in sections 6.1.2 & 6.2.1. The best joint $\alpha$-$\zeta$ solution has been evaluated, for each analysed pulsar, by maximising the summed maps. The complete set of the $\alpha$ and $\zeta$ angles, estimated by a joint $\gamma$-radio fit, are listed, for each analysed pulsar, in table 6.4. Figure 6.51 to figure 6.72 illustrate the results of the joint $\gamma$-radio fit. In each figure we provide, for each model, and from the top to the bottom panel:

- The gamma and radio $\alpha$-$\zeta$ likelihood maps, optimised for $\sigma_{\gamma,*}$ and $\sigma_{radio,*}$ and their summed maps.

- The best $\gamma$-ray light curves corresponding to the maximum likelihood solution of the summed $\alpha$-$\zeta$ likelihood map, compared with the observed LAT profile

- The best radio light curves corresponding to the maximum likelihood solution of the summed $\alpha$-$\zeta$ likelihood map, compared with the observed radio profile.

| | $\alpha_{PC}$ | $\alpha_{SG}$ | $\alpha_{OG}$ | $\alpha_{OPC}$ | $\zeta_{PC}$ | $\zeta_{SG}$ | $\zeta_{OG}$ | $\zeta_{OPC}$ |
|---|---|---|---|---|---|---|---|---|
| J0205+6449 | $87^{1.1}_{0.9}$ | $85^{2.7}_{1.4}$ | $74^{1.2}_{0.8}$ | $90^{0.0005}_{2.1}$ | $81^{0.6}_{1.6}$ | $75^{1.4}_{1.4}$ | $87^{1.4}_{0.9}$ | $90^{0.0005}_{4.2}$ |
| J0248+6021 | $60^{0.2}_{0.3}$ | $58^{0.6}_{0.3}$ | $67^{0.6}_{0.4}$ | $60^{0.2}_{0.4}$ | $59^{0.2}_{0.2}$ | $54^{1.3}_{0.8}$ | $65^{0.4}_{0.5}$ | $59^{0.2}_{0.6}$ |
| J0534+2200 | $75^{0.1}_{0.3}$ | $27^{0.2}_{0.3}$ | $64^{0.3}_{0.2}$ | $65^{0.09}_{0.2}$ | $60^{0.4}_{0.2}$ | $63^{1.4}_{0.4}$ | $30^{0.5}_{0.7}$ | $27^{0.2}_{0.2}$ |
| J0631+1036 | $53^{0.4}_{0.2}$ | $50^{0.08}_{1}$ | $87^{1}_{1.1}$ | $50^{0.08}_{0.5}$ | $51^{0.3}_{0.5}$ | $67^{0.4}_{0.2}$ | $72^{0.3}_{0.2}$ | $67^{0.5}_{0.2}$ |
| J0659+1414 | $40^{0.03}_{0.09}$ | $37^{0.3}_{0.6}$ | $75^{0.07}_{0.03}$ | $75^{0.07}_{0.03}$ | $42^{0.4}_{0.3}$ | $32^{0.2}_{0.1}$ | $75^{0.04}_{0.04}$ | $75^{0.04}_{0.04}$ |
| J0742-2822 | $61^{0.02}_{0.03}$ | $67^{0.9}_{0.02}$ | $67^{1}_{0.02}$ | $67^{0.8}_{0.02}$ | $46^{0.4}_{0.03}$ | $87^{0.3}_{0.4}$ | $87^{0.3}_{0.4}$ | $87^{0.3}_{0.5}$ |
| J0835-4510 | $51^{0.03}_{0.05}$ | $54^{0.2}_{0.1}$ | $54^{0.2}_{0.06}$ | $54^{0.2}_{0.1}$ | $63^{0.02}_{0.05}$ | $77^{0.1}_{0.07}$ | $77^{0.1}_{0.06}$ | $77^{0.1}_{0.08}$ |
| J1048-5832 | $33^{0.04}_{0.03}$ | $37^{0.06}_{0.08}$ | $75^{0.01}_{0.4}$ | $68^{0.04}_{0.03}$ | $24^{0.03}_{0.04}$ | $50^{0.02}_{0.07}$ | $63^{0.03}_{0.05}$ | $78^{0.04}_{0.05}$ |
| J1057-5226 | $11^{4.5}_{0.2}$ | $75^{1}_{0.0005}$ | $85^{0.3}_{0.3}$ | $67^{0.2}_{0.3}$ | $30^{0.07}_{0.1}$ | $13^{2.4}_{0.2}$ | $90^{0.0005}_{0.7}$ | $71^{0.1}_{0.3}$ |
| J1124-5916 | $7^{0.7}_{0.4}$ | $84^{1.7}_{0.7}$ | $77^{0.9}_{0.4}$ | $81^{0.4}_{0.3}$ | $6^{1.2}_{0.1}$ | $79^{1.3}_{0.6}$ | $88^{0.6}_{2.6}$ | $84^{0.9}_{0.3}$ |
| J1420-6048 | $60^{0.2}_{1.2}$ | $43^{1.4}_{2}$ | $70^{0.2}_{0.4}$ | $65^{0.1}_{0.1}$ | $20^{0.1}_{0.1}$ | $65^{0.5}_{0.6}$ | $70^{0.2}_{0.2}$ | $62^{0.3}_{0.8}$ |
| J1509-5850 | $40^{5.2}_{0.1}$ | $34^{1}_{0.9}$ | $56^{1.1}_{0.7}$ | $61^{0.2}_{0.6}$ | $26^{0.7}_{0.5}$ | $17^{1.5}_{0.5}$ | $68^{1.6}_{0.3}$ | $47^{1}_{1.2}$ |
| J1709-4429 | $35^{0.03}_{0.1}$ | $15^{0.1}_{0.1}$ | $59^{0.1}_{0.1}$ | $53^{0.2}_{0.4}$ | $29^{0.08}_{0.08}$ | $28^{0.1}_{0.1}$ | $70^{0.09}_{0.08}$ | $59^{0.7}_{0.2}$ |
| J1718-3825 | $10^{0.4}_{0.1}$ | $40^{0.3}_{0.7}$ | $75^{0.09}_{0.9}$ | $38^{0.8}_{0.7}$ | $3^{0.5}_{0.2}$ | $65^{0.3}_{0.3}$ | $52^{0.7}_{0.4}$ | $62^{0.3}_{0.3}$ |
| J1741-2054 | $5^{5.07}_{0.9}$ | $83^{2}_{3.5}$ | $84^{1.3}_{3.3}$ | $89^{1}_{4.2}$ | $4^{0.1}_{0.3}$ | $15^{3.6}_{2.8}$ | $90^{0.0005}_{0.5}$ | $51^{0.2}_{0.2}$ |
| J1747-2958 | $29^{1.9}_{0.1}$ | $26^{0.8}_{0.1}$ | $61^{0.3}_{0.3}$ | $50^{0.07}_{0.5}$ | $13^{0.4}_{6.4}$ | $16^{0.5}_{0.7}$ | $46^{0.2}_{0.1}$ | $43^{0.3}_{0.2}$ |
| J1833-1034 | $45^{0.1}_{0.1}$ | $61^{0.04}_{0.1}$ | $61^{0.04}_{0.1}$ | $53^{0.3}_{0.4}$ | $36^{0.3}_{0.3}$ | $36^{0.2}_{0.2}$ | $36^{0.2}_{0.1}$ | $40^{0.1}_{0.2}$ |
| J1952+3252 | $36^{0.5}_{0.3}$ | $51^{0.1}_{0.1}$ | $46^{0.09}_{0.1}$ | $52^{0.6}_{0.1}$ | $59^{0.3}_{0.3}$ | $80^{0.8}_{0.8}$ | $76^{0.1}_{0.2}$ | $77^{0.1}_{0.2}$ |
| J2021+3651 | $31^{0.05}_{0.2}$ | $84^{0.3}_{0.3}$ | $72^{0.4}_{0.3}$ | $73^{0.5}_{0.5}$ | $22^{0.3}_{0.1}$ | $81^{0.05}_{0.05}$ | $84^{0.5}_{0.6}$ | $86^{0.8}_{1.1}$ |
| J2032+4127 | $75^{1.2}_{0.02}$ | $75^{1.2}_{0.02}$ | $75^{1.2}_{2.3}$ | $75^{1.2}_{2.3}$ | $87^{0.8}_{0.5}$ | $87^{0.8}_{0.5}$ | $87^{0.8}_{0.5}$ | $87^{0.8}_{0.5}$ |
| J2043+2740 | $51^{0.04}_{0.3}$ | $51^{0.04}_{0.02}$ | $63^{0.5}_{1}$ | $86^{4}_{0.07}$ | $58^{0.08}_{0.2}$ | $58^{0.08}_{0.2}$ | $90^{0.0005}_{0.0005}$ | $66^{0.2}_{0.1}$ |
| J2229+6114 | $31^{0.05}_{0.3}$ | $45^{0.2}_{1.1}$ | $75^{0.03}_{0.2}$ | $44^{1}_{0.2}$ | $19^{0.2}_{0.1}$ | $62^{0.2}_{0.3}$ | $56^{0.1}_{0.2}$ | $61^{0.3}_{0.2}$ |

Table 6.4: $\alpha$ and $\zeta$ best fit solutions (in degrees) resulting from the joint radio plus $\gamma$ fit for the 22 pulsars of the analysed sample. At each value is associated the statistical error.

In figure 6.50 is plotted, for each pulsar, the relative significance of the best fits for the different $\gamma$-ray models.

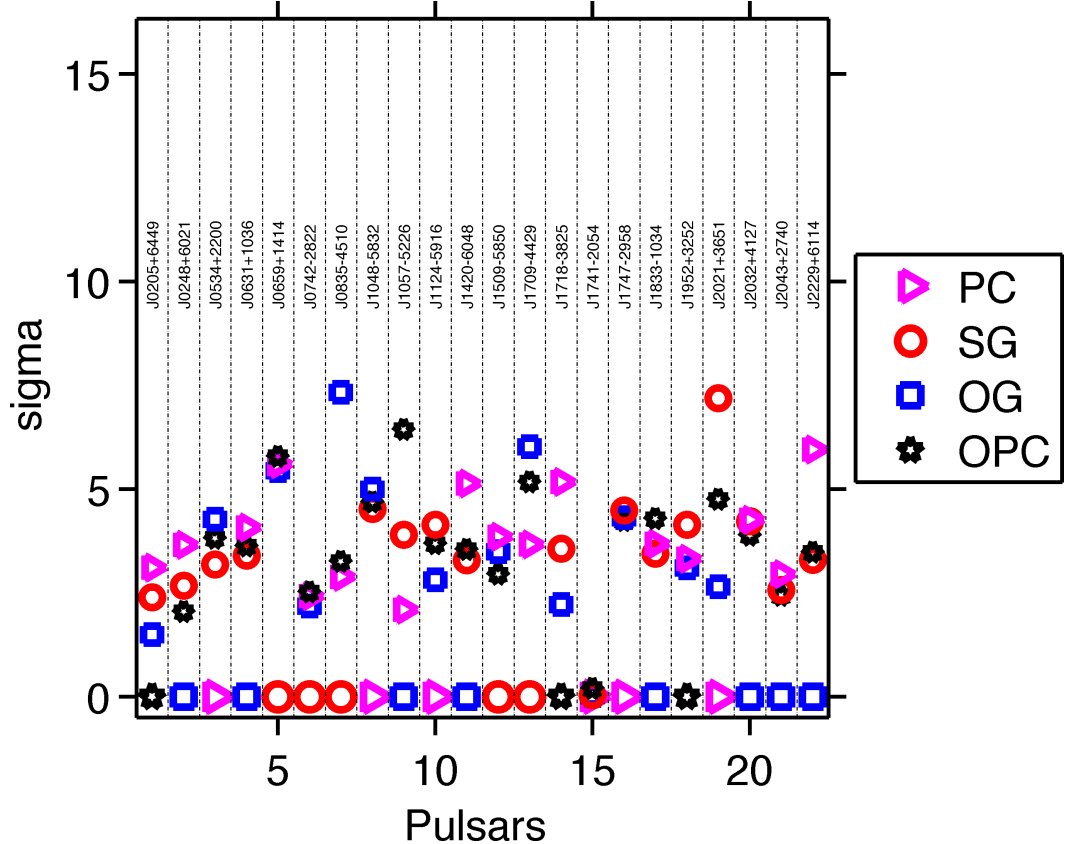

Figure 6.50: Relative significance of the joint radio-γ solutions obtained for each pulsar of table 6.4 as derived from the log-likelihood ratio found between two γ-ray models. The rms significance between the best-fit solution for one model and the very best fit model is given by $\sigma = \sqrt{2\ln(L_{bestfitmodel}) - \ln(L_{model,max})}$ The very best solution is plotted at 0 sigma. A 10 sigma value means that the given model yields a 10 sigma worse solution than the very best one.

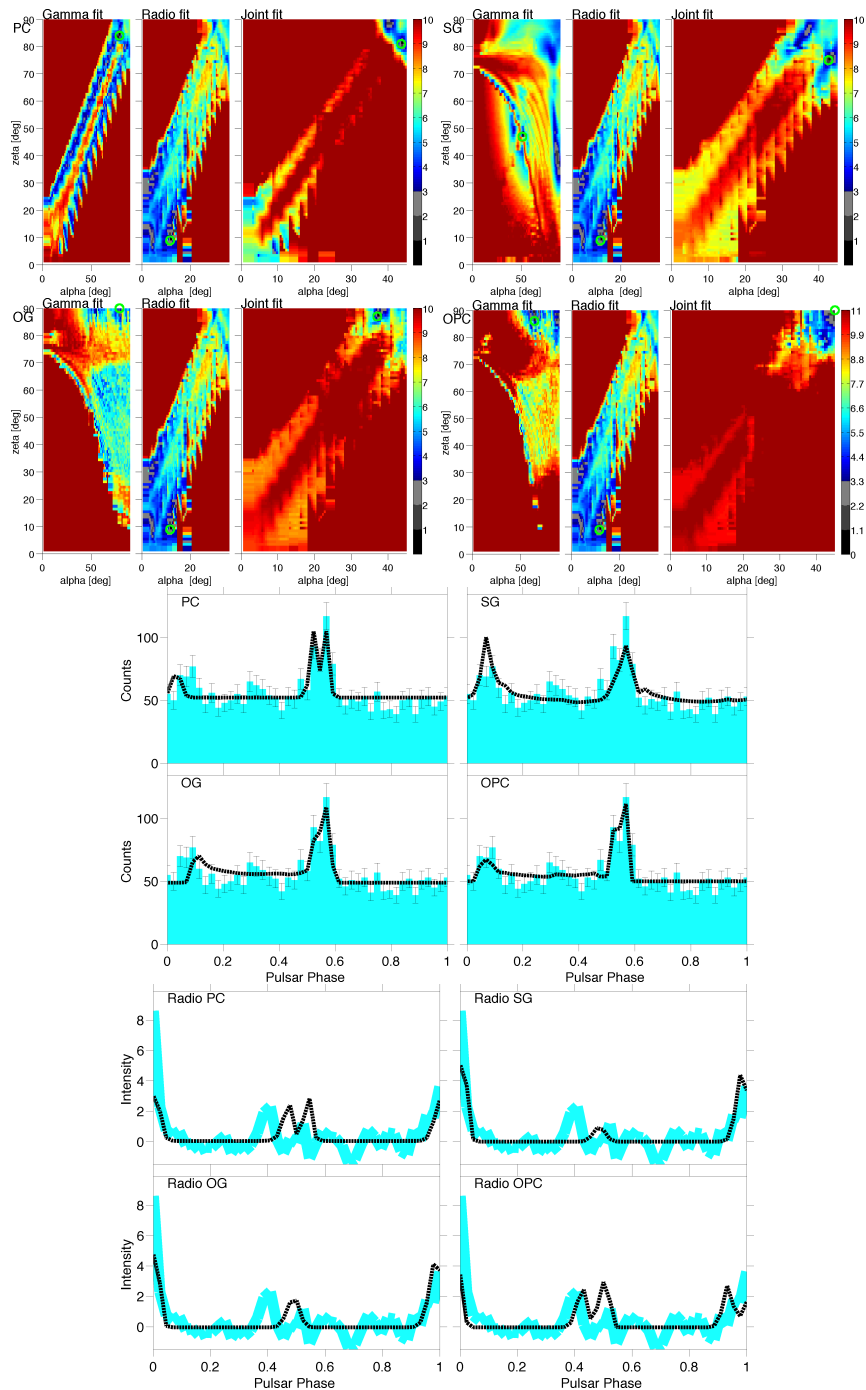

Figure 6.51: PSR J0205+6449. *Top*: for each model is shown the α-ζ likelihood map for the γ-ray χ² fit, the radio χ² fit, each one optimised variance, and the sum of the maps. The color-bar is in σ units, zero corresponds to the best fit solution. *Middle*: the LAT light-curve (in blue) is compared to the γ-ray light curve obtained, for each model, by maximising the joint likelihood map. *Bottom*: the radio profile (thick blue line) is compared to the radio light curve obtained, for each model, by maximising the joint likelihood map. The radio model is unique, but the (α, ζ) solutions vary for each γ-ray model.

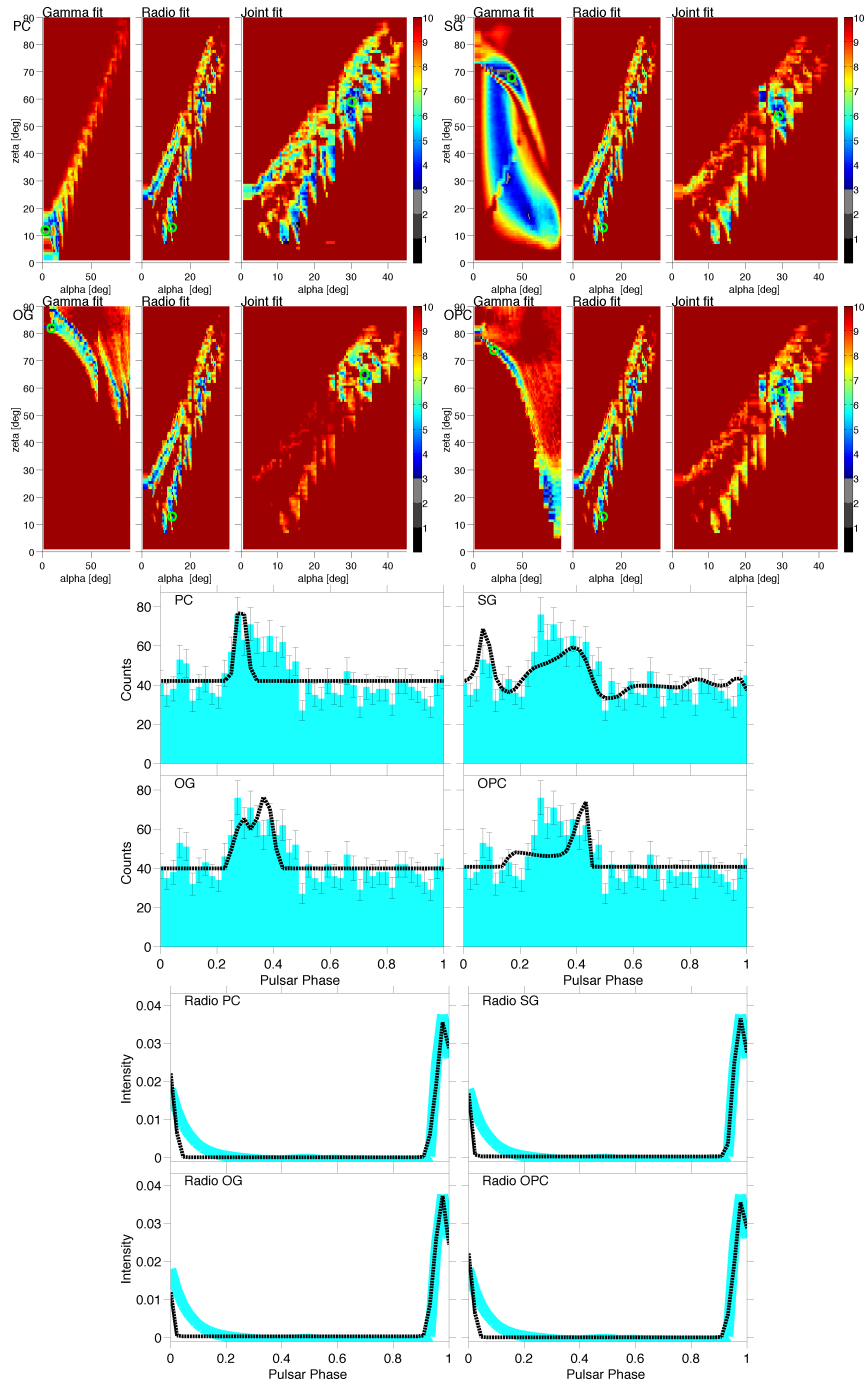

Figure 6.52: PSR J0248+6021. *Top*: for each model is shown the $\alpha$-$\zeta$ likelihood map for the $\gamma$-ray $\chi^2$ fit, the radio $\chi^2$ fit, each one optimised variance, and the sum of the maps. The color-bar is in $\sigma$ units, zero corresponds to the best fit solution. *Middle*: the LAT light-curve (in blue) is compared to the $\gamma$-ray light curve obtained, for each model, by maximising the joint likelihood map. *Bottom*: the radio profile (thick blue line) is compared to the radio light curve obtained, for each model, by maximising the joint likelihood map. The radio model is unique, but the $(\alpha, \zeta)$ solutions vary for each $\gamma$-ray model.

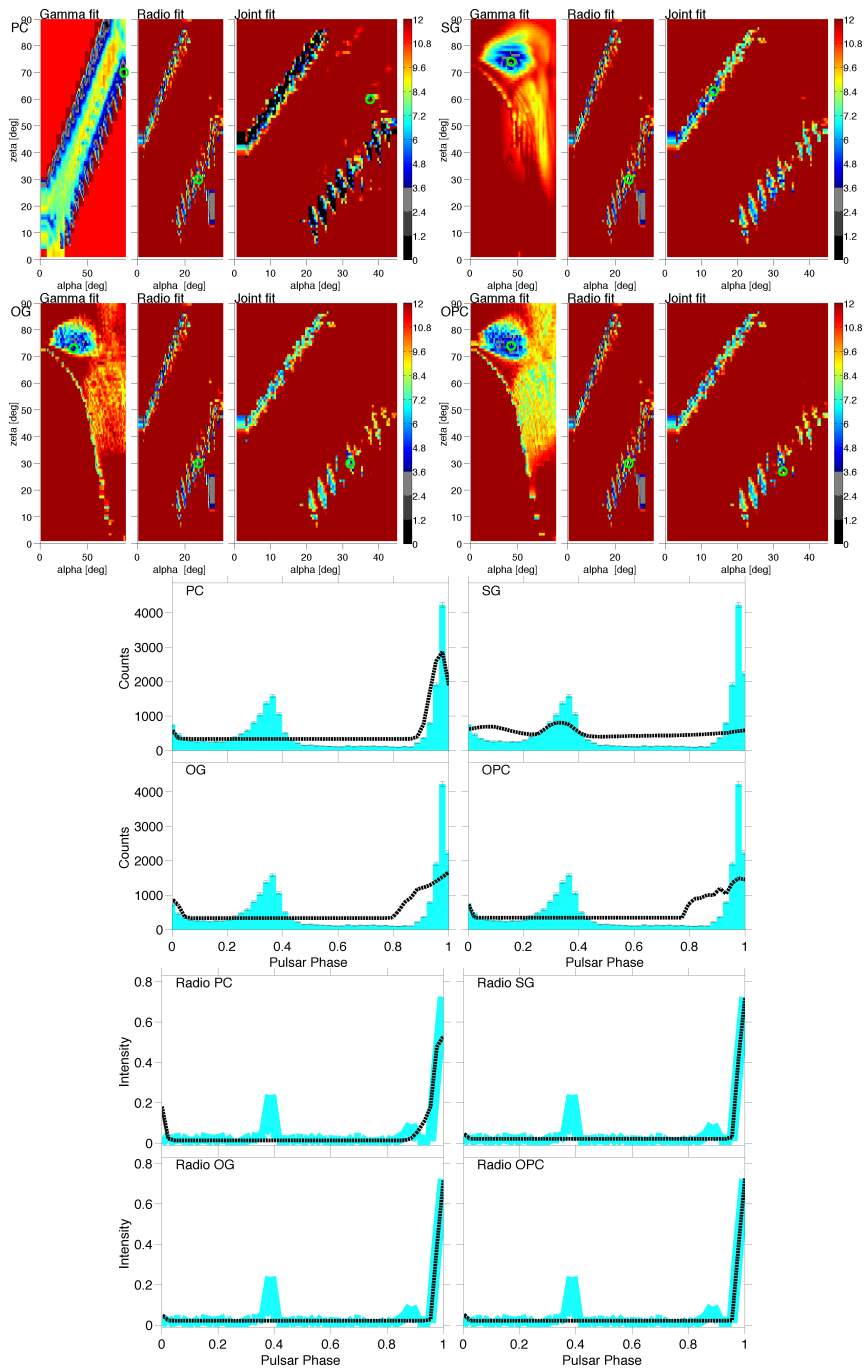

Figure 6.53: PSR J0534+2200. *Top*: for each model is shown the $\alpha$-$\zeta$ likelihood map for the $\gamma$-ray $\chi^2$ fit, the radio $\chi^2$ fit, each one optimised variance, and the sum of the maps. The color-bar is in $\sigma$ units, zero corresponds to the best fit solution. *Middle*: the LAT light-curve (in blue) is compared to the $\gamma$-ray light curve obtained, for each model, by maximising the joint likelihood map. *Bottom*: the radio profile (thick blue line) is compared to the radio light curve obtained, for each model, by maximising the joint likelihood map. The radio model is unique, but the $(\alpha, \zeta)$ solutions vary for each $\gamma$-ray model.

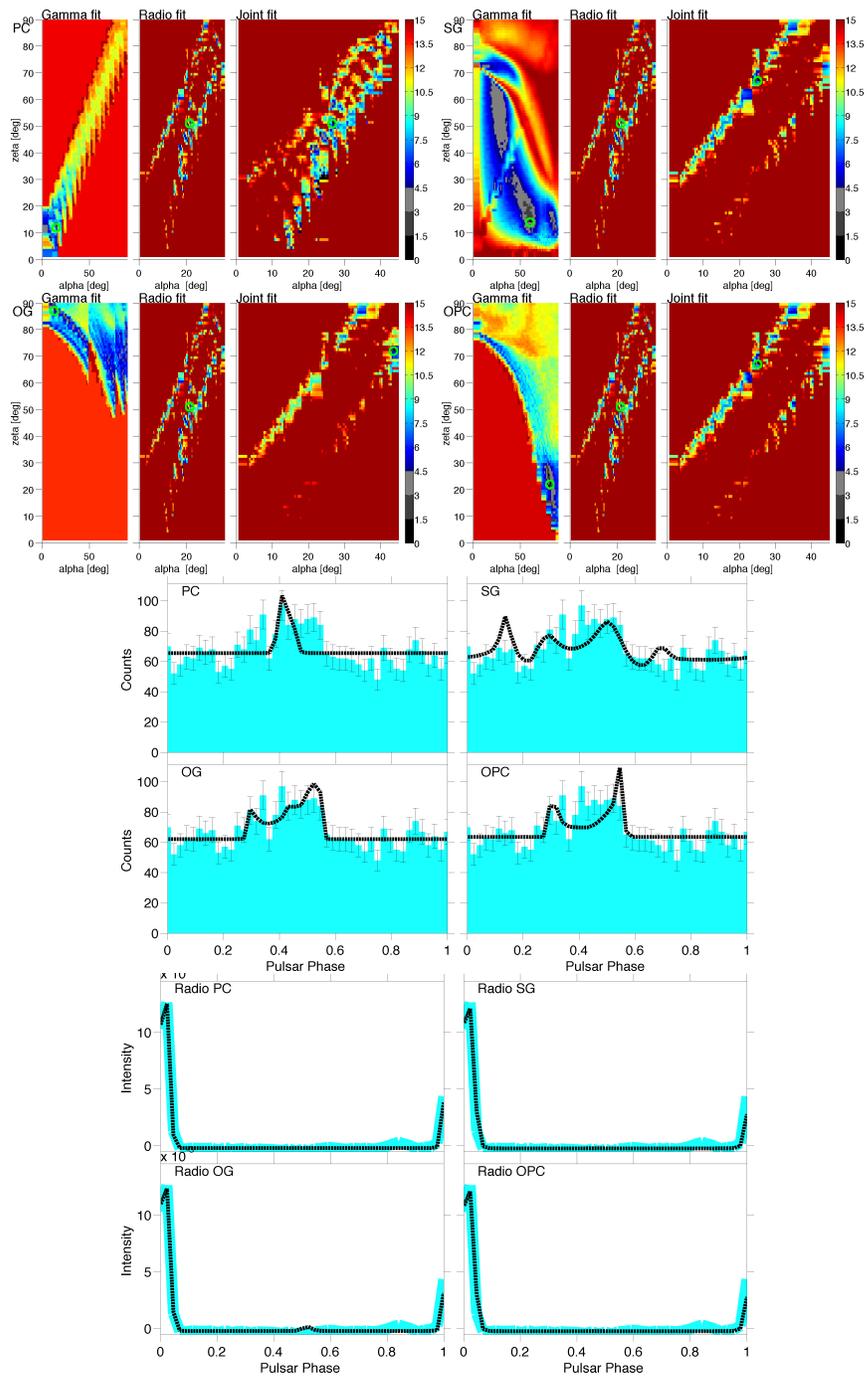

Figure 6.54: PSR J0631+1036. *Top*: for each model is shown the $\alpha$-$\zeta$ likelihood map for the $\gamma$-ray $\chi^2$ fit, the radio $\chi^2$ fit, each one optimised variance, and the sum of the maps. The color-bar is in $\sigma$ units, zero corresponds to the best fit solution. *Middle*: the LAT light-curve (in blue) is compared to the $\gamma$-ray light curve obtained, for each model, by maximising the joint likelihood map. *Bottom*: the radio profile (thick blue line) is compared to the radio light curve obtained, for each model, by maximising the joint likelihood map. The radio model is unique, but the $(\alpha, \zeta)$ solutions vary for each $\gamma$-ray model.

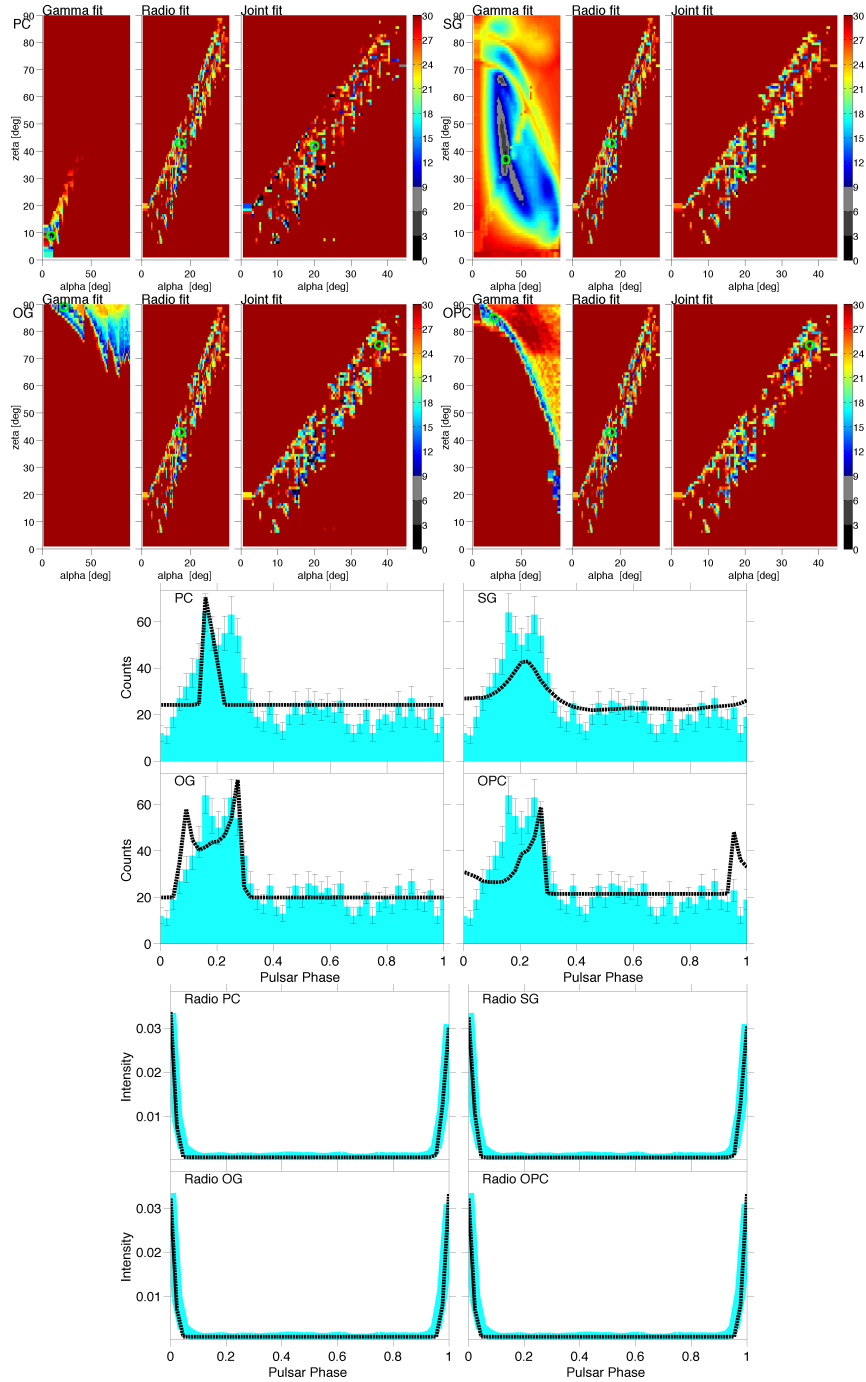

Figure 6.55: PSR J0659+1414. *Top*: for each model is shown the α-ζ likelihood map for
the γ-ray χ² fit, the radio χ² fit, each one optimised variance, and the sum of the maps. The
color-bar is in σ units, zero corresponds to the best fit solution. *Middle*: the LAT light-curve
(in blue) is compared to the γ-ray light curve obtained, for each model, by maximising the joint
likelihood map. *Bottom*: the radio profile (thick blue line) is compared to the radio
light curve obtained, for each model, by maximising the joint likelihood map. The radio
model is unique, but the (α, ζ) solutions vary for each γ-ray model.

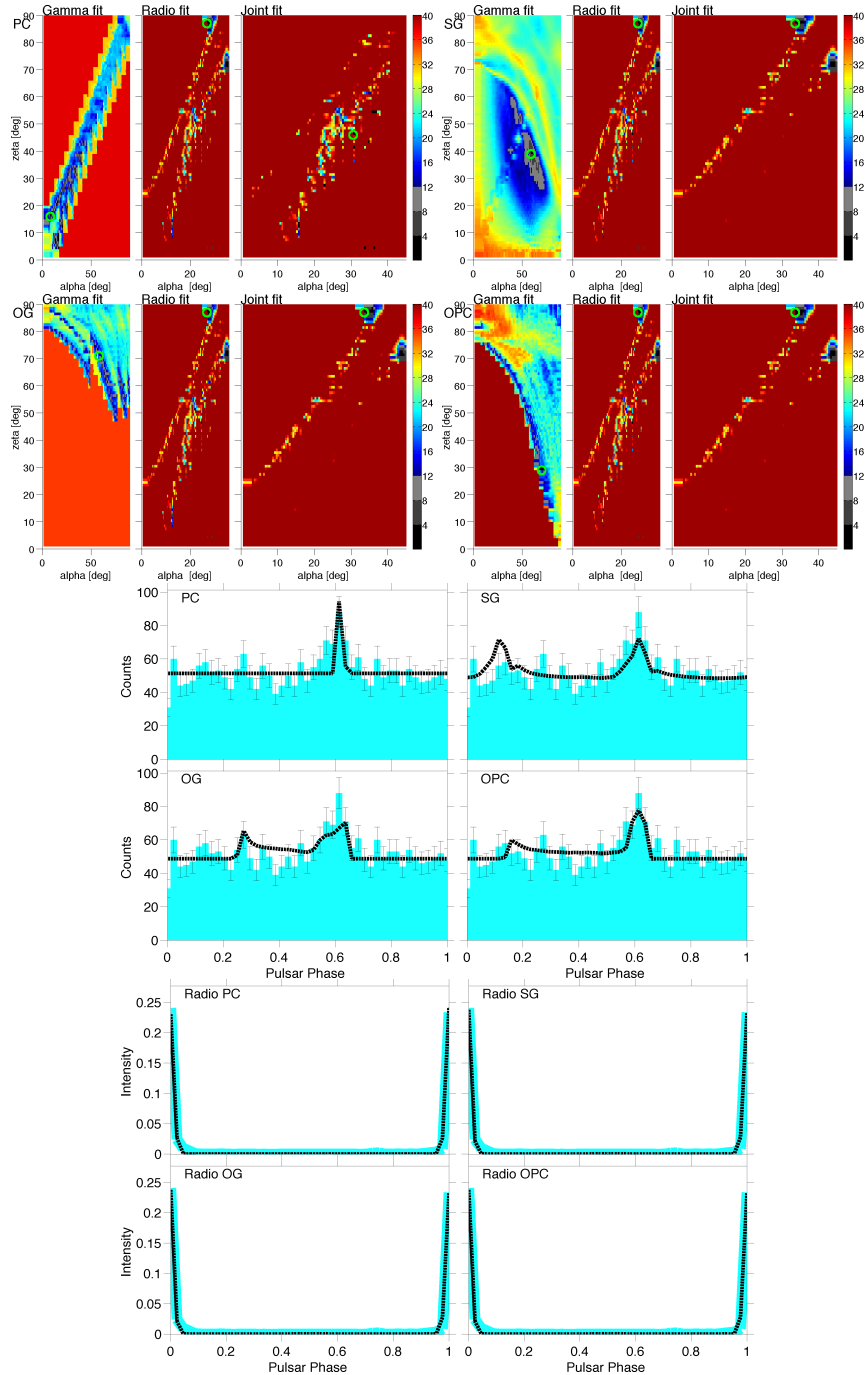

Figure 6.56: PSR J0742-2822. *Top*: for each model is shown the $\alpha$-$\zeta$ likelihood map for the γ-ray $\chi^2$ fit, the radio $\chi^2$ fit, each one optimised variance, and the sum of the maps. The color-bar is in $\sigma$ units, zero corresponds to the best fit solution. *Middle*: the LAT light-curve (in blue) is compared to the γ-ray light curve obtained, for each model, by maximising the joint likelihood map. *Bottom*: the radio profile (thick blue line) is compared to the radio light curve obtained, for each model, by maximising the joint likelihood map. The radio model is unique, but the $(\alpha, \zeta)$ solutions vary for each γ-ray model.

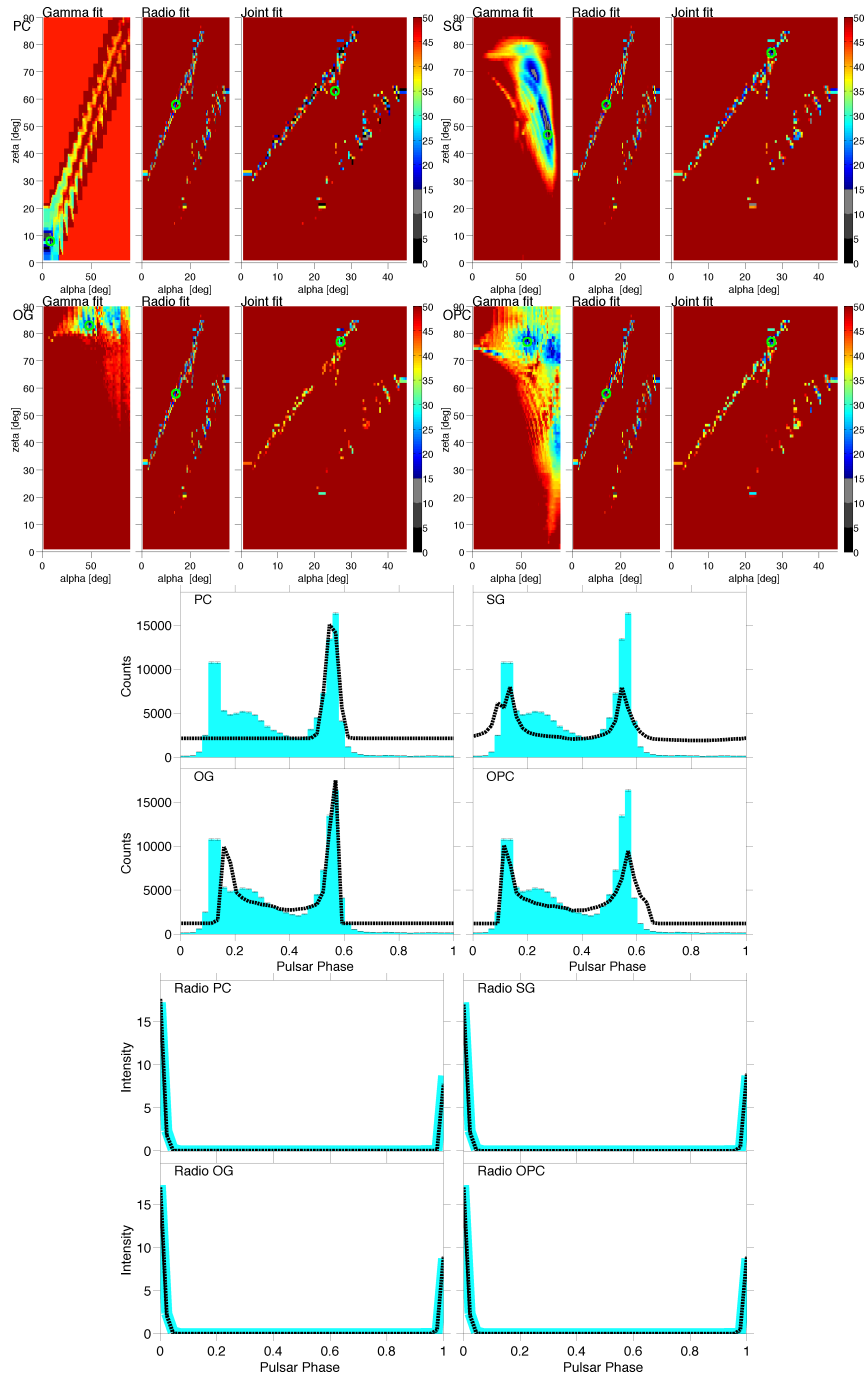

Figure 6.57: PSR J0835-4510. *Top*: for each model is shown the $\alpha$-$\zeta$ likelihood map for the $\gamma$-ray $\chi^2$ fit, the radio $\chi^2$ fit, each one optimised variance, and the sum of the maps. The color-bar is in $\sigma$ units, zero corresponds to the best fit solution. *Middle*: the LAT light-curve (in blue) is compared to the $\gamma$-ray light curve obtained, for each model, by maximising the joint likelihood map. *Bottom*: the radio profile (thick blue line) is compared to the radio light curve obtained, for each model, by maximising the joint likelihood map. The radio model is unique, but the $(\alpha, \zeta)$ solutions vary for each $\gamma$-ray model.

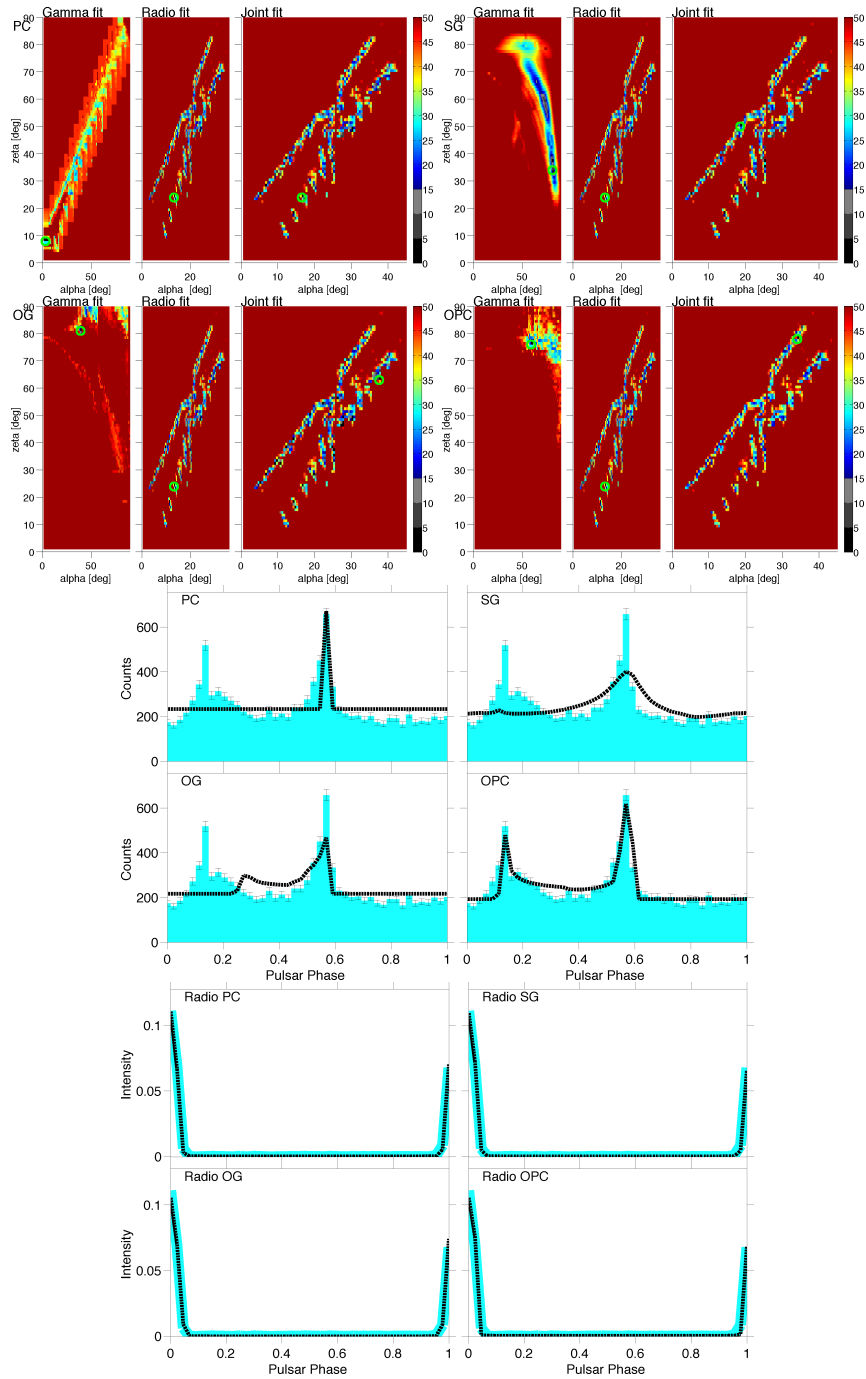

Figure 6.58: PSR J1048-5832. *Top*: for each model is shown the $\alpha$-$\zeta$ likelihood map for the γ-ray $\chi^2$ fit, the radio $\chi^2$ fit, each one optimised variance, and the sum of the maps. The color-bar is in $\sigma$ units, zero corresponds to the best fit solution. *Middle*: the LAT light-curve (in blue) is compared to the γ-ray light curve obtained, for each model, by maximising the joint likelihood map. *Bottom*: the radio profile (thick blue line) is compared to the radio light curve obtained, for each model, by maximising the joint likelihood map. The radio model is unique, but the $(\alpha, \zeta)$ solutions vary for each γ-ray model.

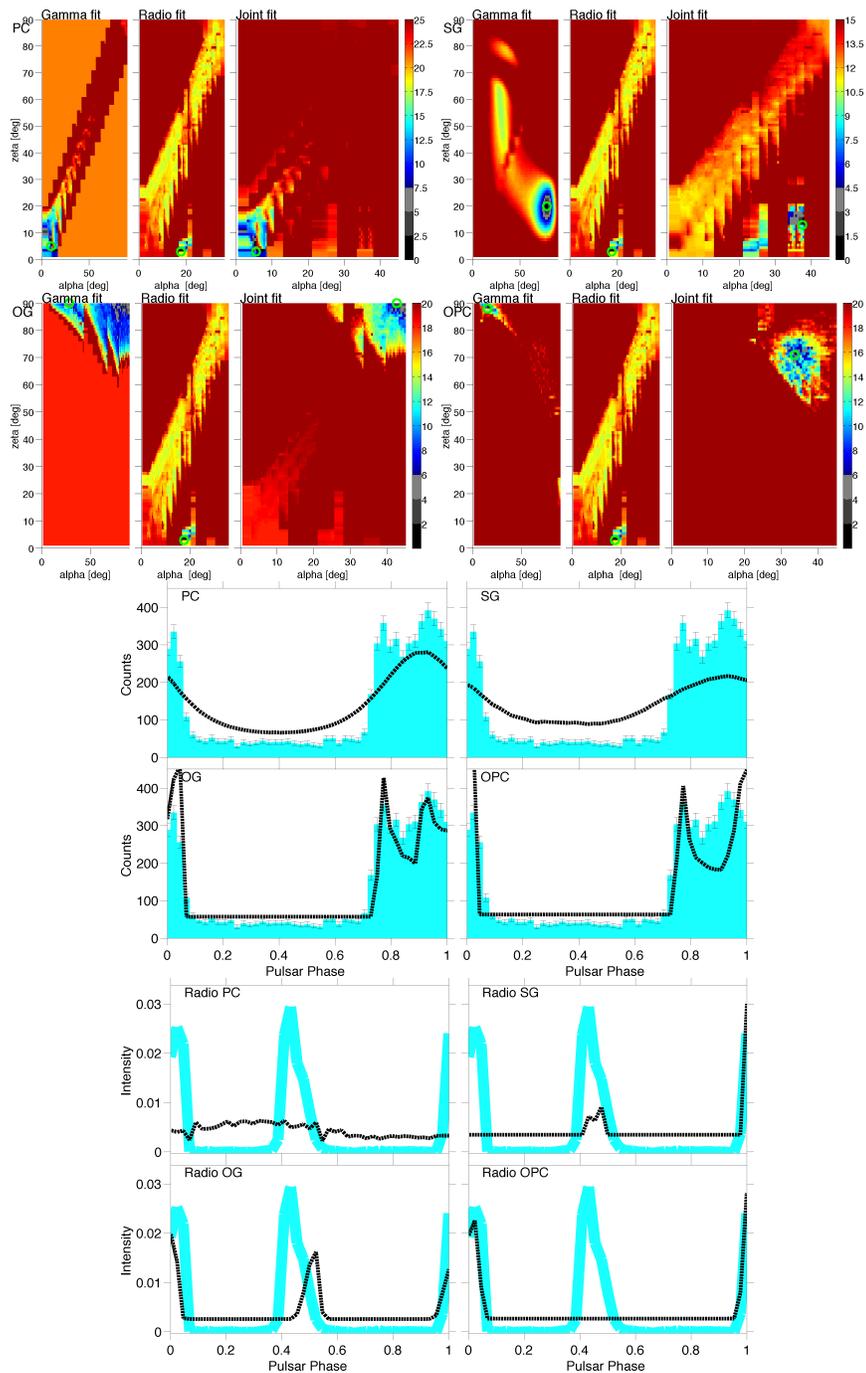

Figure 6.59: PSR J1057-5226. *Top*: for each model is shown the α-ζ likelihood map for the γ-ray $\chi^2$ fit, the radio $\chi^2$ fit, each one optimised variance, and the sum of the maps. The color-bar is in σ units, zero corresponds to the best fit solution. *Middle*: the LAT light-curve (in blue) is compared to the γ-ray light curve obtained, for each model, by maximising the joint likelihood map. *Bottom*: the radio profile (thick blue line) is compared to the radio light curve obtained, for each model, by maximising the joint likelihood map. The radio model is unique, but the (α, ζ) solutions vary for each γ-ray model.

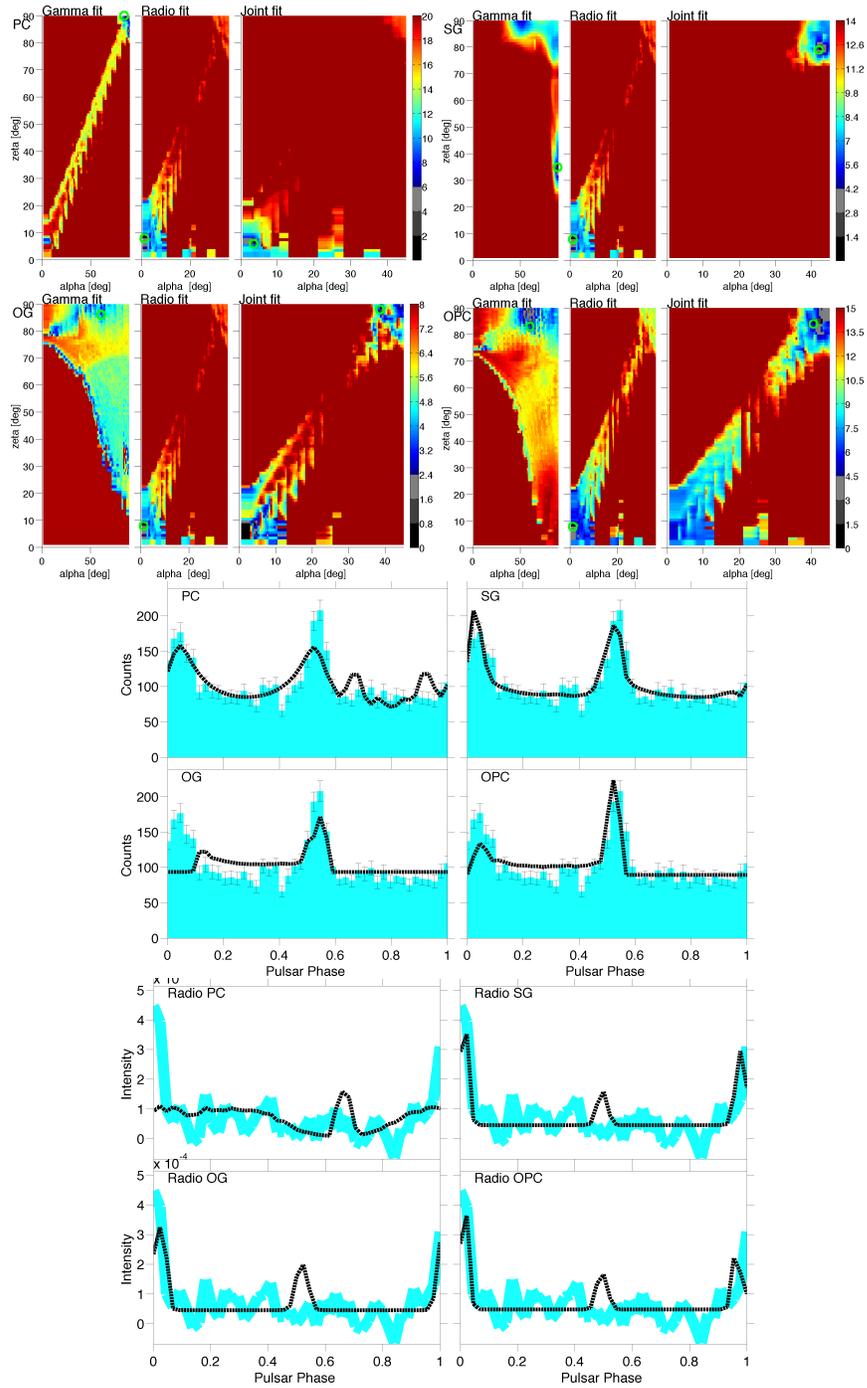

Figure 6.60: PSR J1124-5916. *Top*: for each model is shown the α-ζ likelihood map for the γ-ray χ² fit, the radio χ² fit, each one optimised variance, and the sum of the maps. The color-bar is in σ units, zero corresponds to the best fit solution. *Middle*: the LAT light-curve (in blue) is compared to the γ-ray light curve obtained, for each model, by maximising the joint likelihood map. *Bottom*: the radio profile (thick blue line) is compared to the radio light curve obtained, for each model, by maximising the joint likelihood map. The radio model is unique, but the (α, ζ) solutions vary for each γ-ray model.

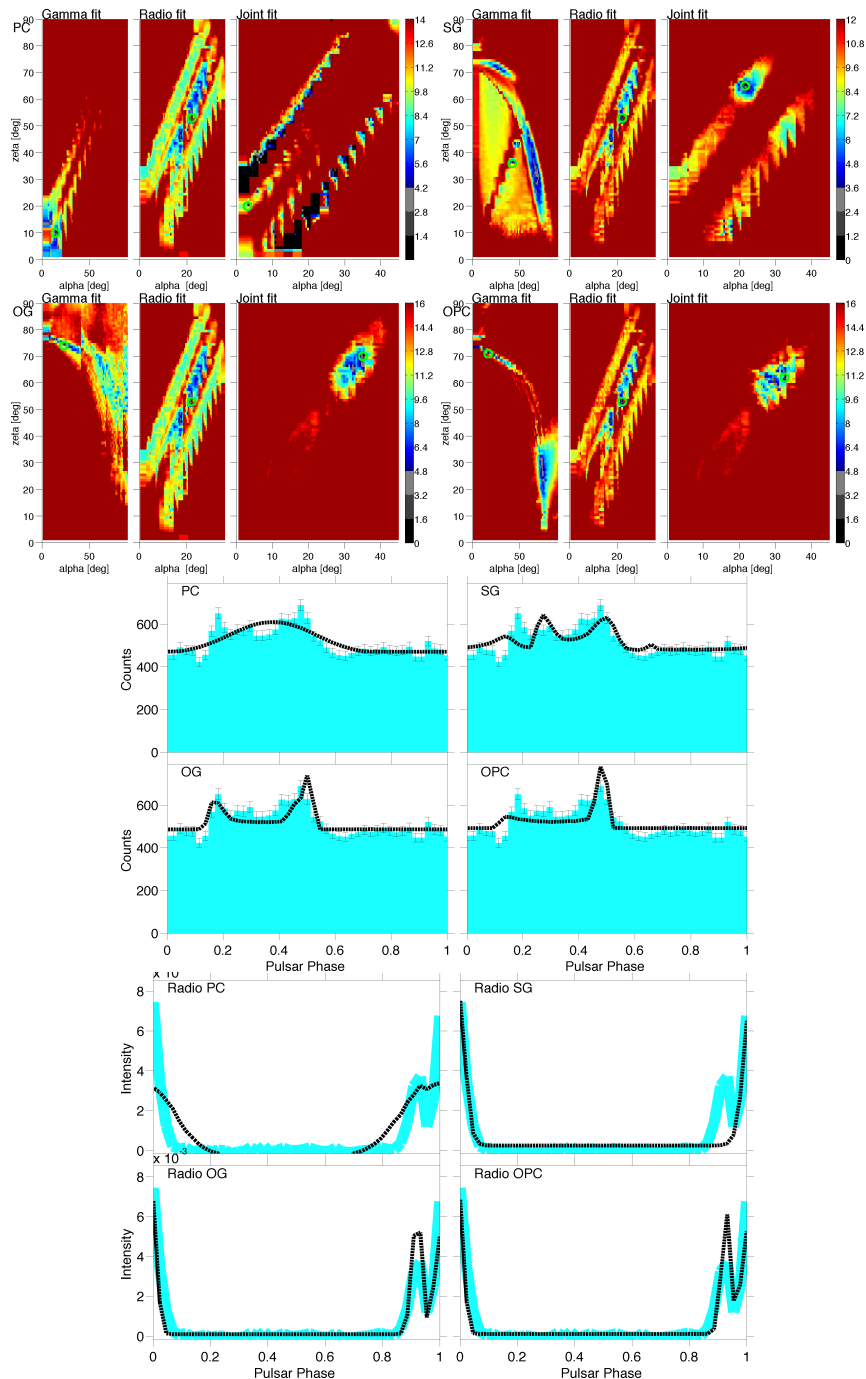

Figure 6.61: PSR J1420-6048. *Top*: for each model is shown the α-ζ likelihood map for the
γ-ray χ² fit, the radio χ² fit, each one optimised variance, and the sum of the maps. The
color-bar is in σ units, zero corresponds to the best fit solution. *Middle*: the LAT light-curve
(in blue) is compared to the γ-ray light curve obtained, for each model, by maximising the
joint likelihood map. *Bottom*: the radio profile (thick blue line) is compared to the radio
light curve obtained, for each model, by maximising the joint likelihood map. The radio
model is unique, but the (α, ζ) solutions vary for each γ-ray model.

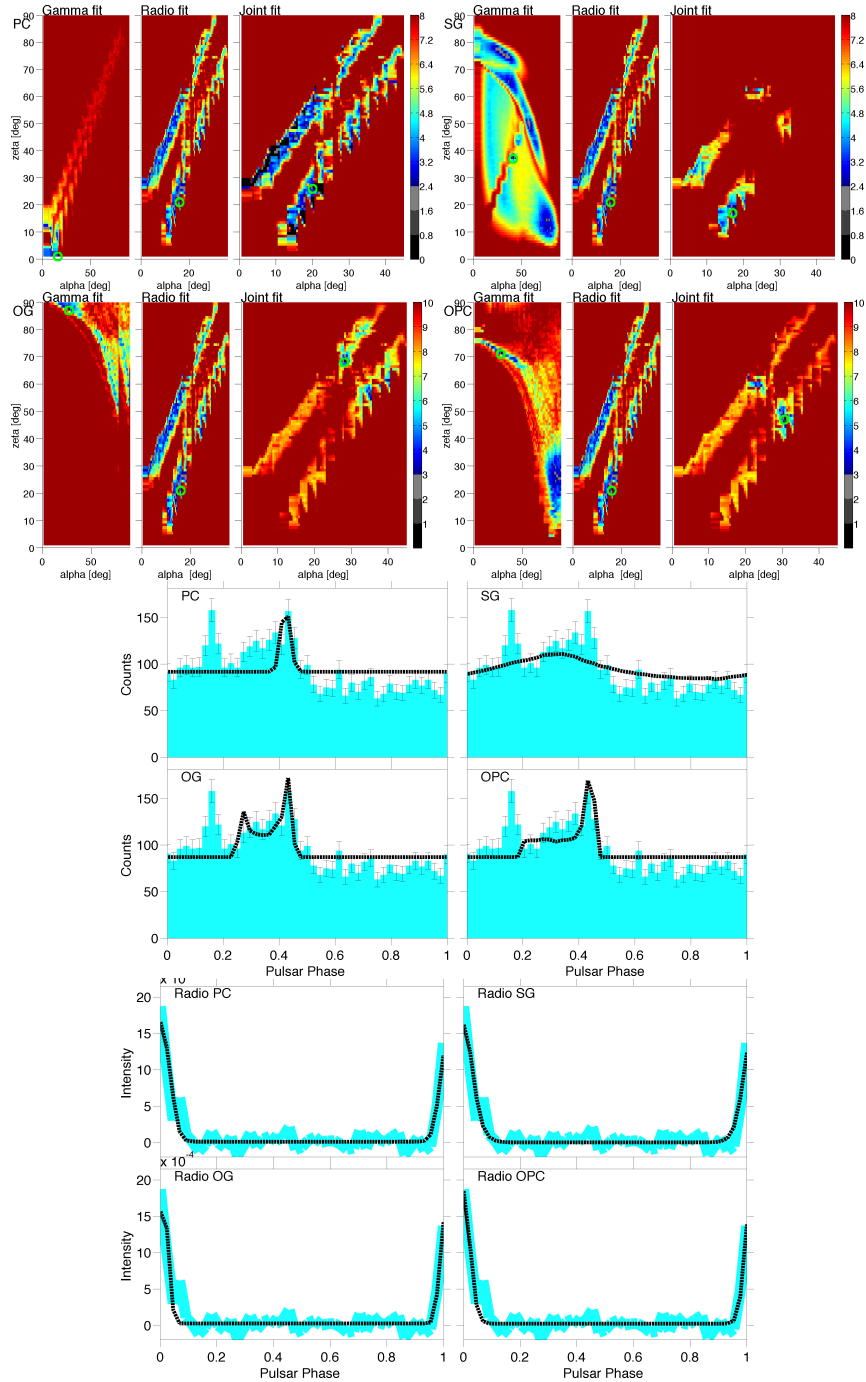

Figure 6.62: PSR J1509-5850. *Top*: for each model is shown the $\alpha$-$\zeta$ likelihood map for the $\gamma$-ray $\chi^2$ fit, the radio $\chi^2$ fit, each one optimised variance, and the sum of the maps. The color-bar is in $\sigma$ units, zero corresponds to the best fit solution. *Middle*: the LAT light-curve (in blue) is compared to the $\gamma$-ray light curve obtained, for each model, by maximising the joint likelihood map. *Bottom*: the radio profile (thick blue line) is compared to the radio light curve obtained, for each model, by maximising the joint likelihood map. The radio model is unique, but the $(\alpha, \zeta)$ solutions vary for each $\gamma$-ray model.

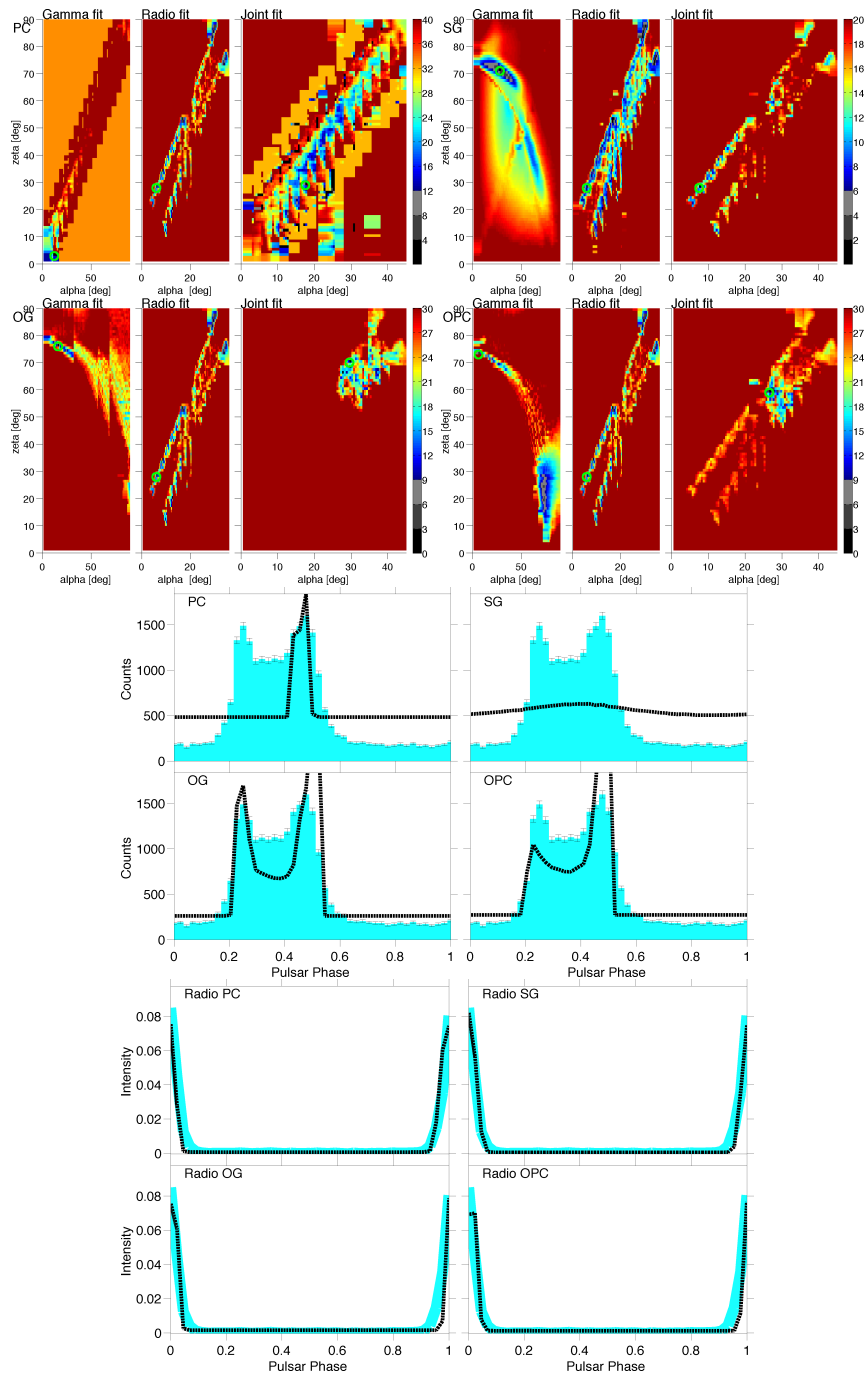

Figure 6.63: PSR J1709-4429. *Top*: for each model is shown the $\alpha$-$\zeta$ likelihood map for the $\gamma$-ray $\chi^2$ fit, the radio $\chi^2$ fit, each one optimised variance, and the sum of the maps. The color-bar is in $\sigma$ units, zero corresponds to the best fit solution. *Middle*: the LAT light-curve (in blue) is compared to the $\gamma$-ray light curve obtained, for each model, by maximising the joint likelihood map. *Bottom*: the radio profile (thick blue line) is compared to the radio light curve obtained, for each model, by maximising the joint likelihood map. The radio model is unique, but the $(\alpha, \zeta)$ solutions vary for each $\gamma$-ray model.

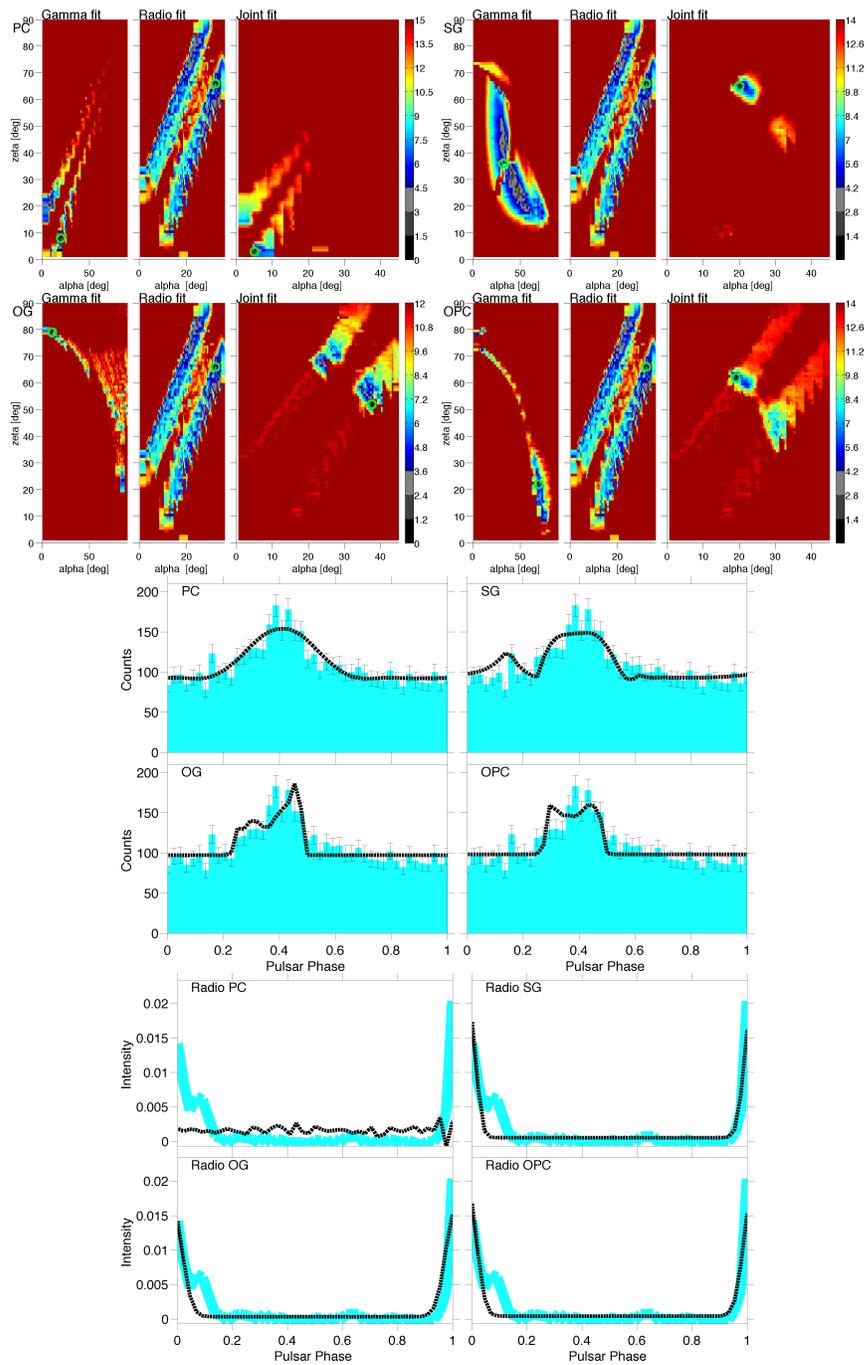

Figure 6.64: PSR J1718-3825. *Top*: for each model is shown the $\alpha$-$\zeta$ likelihood map for the $\gamma$-ray $\chi^2$ fit, the radio $\chi^2$ fit, each one optimised variance, and the sum of the maps. The color-bar is in $\sigma$ units, zero corresponds to the best fit solution. *Middle*: the LAT light-curve (in blue) is compared to the $\gamma$-ray light curve obtained, for each model, by maximising the joint likelihood map. *Bottom*: the radio profile (thick blue line) is compared to the radio light curve obtained, for each model, by maximising the joint likelihood map. The radio model is unique, but the $(\alpha, \zeta)$ solutions vary for each $\gamma$-ray model.

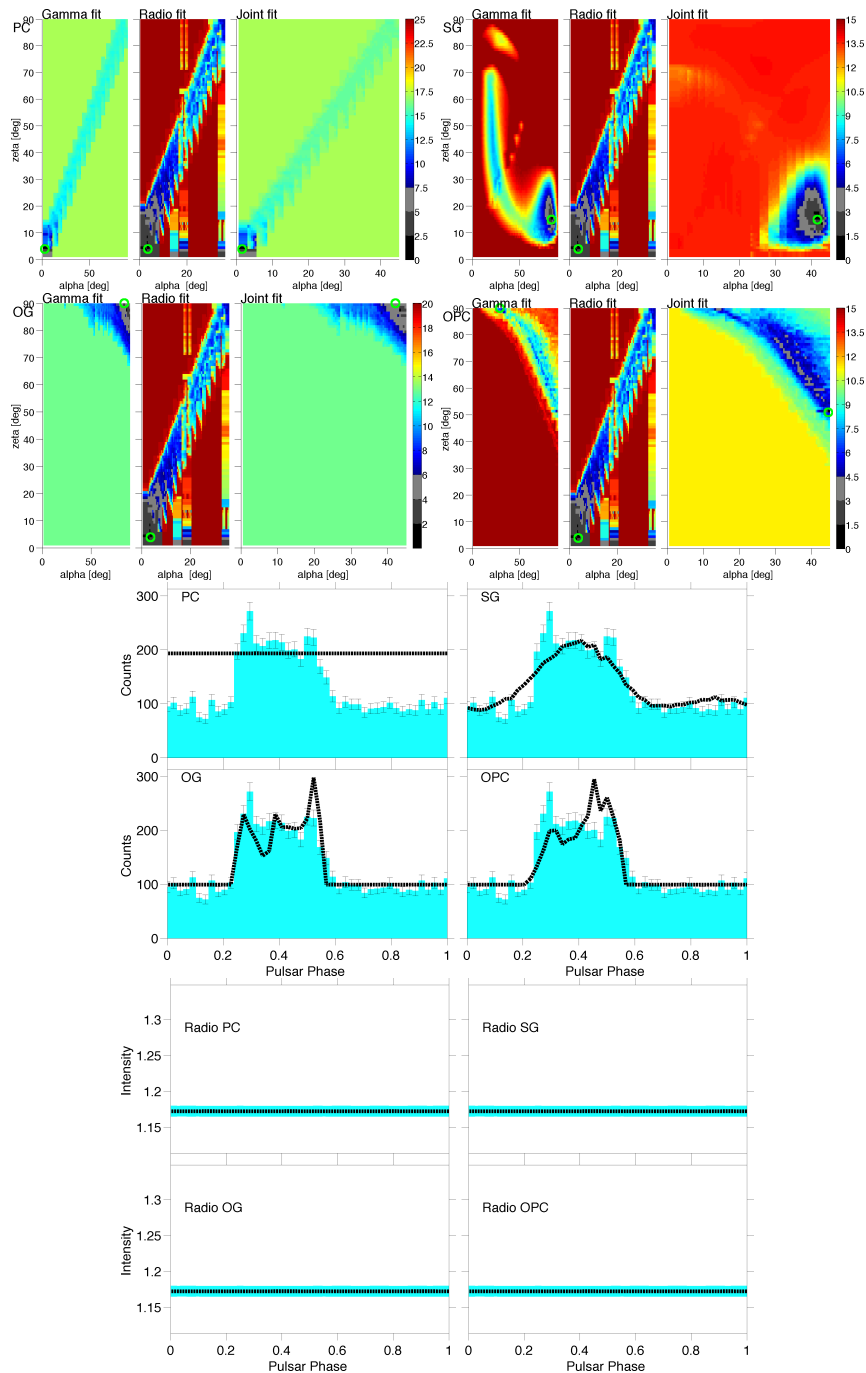

Figure 6.65: PSR J1741-2054. *Top*: for each model is shown the $\alpha$-$\zeta$ likelihood map for the
$\gamma$-ray $\chi^2$ fit, the radio $\chi^2$ fit, each one optimised variance, and the sum of the maps. The
color-bar is in $\sigma$ units, zero corresponds to the best fit solution. *Middle*: the LAT light-curve
(in blue) is compared to the $\gamma$-ray light curve obtained, for each model, by maximising the
joint likelihood map. *Bottom*: the radio profile (thick blue line) is compared to the radio
light curve obtained, for each model, by maximising the joint likelihood map. The radio
model is unique, but the $(\alpha, \zeta)$ solutions vary for each $\gamma$-ray model.

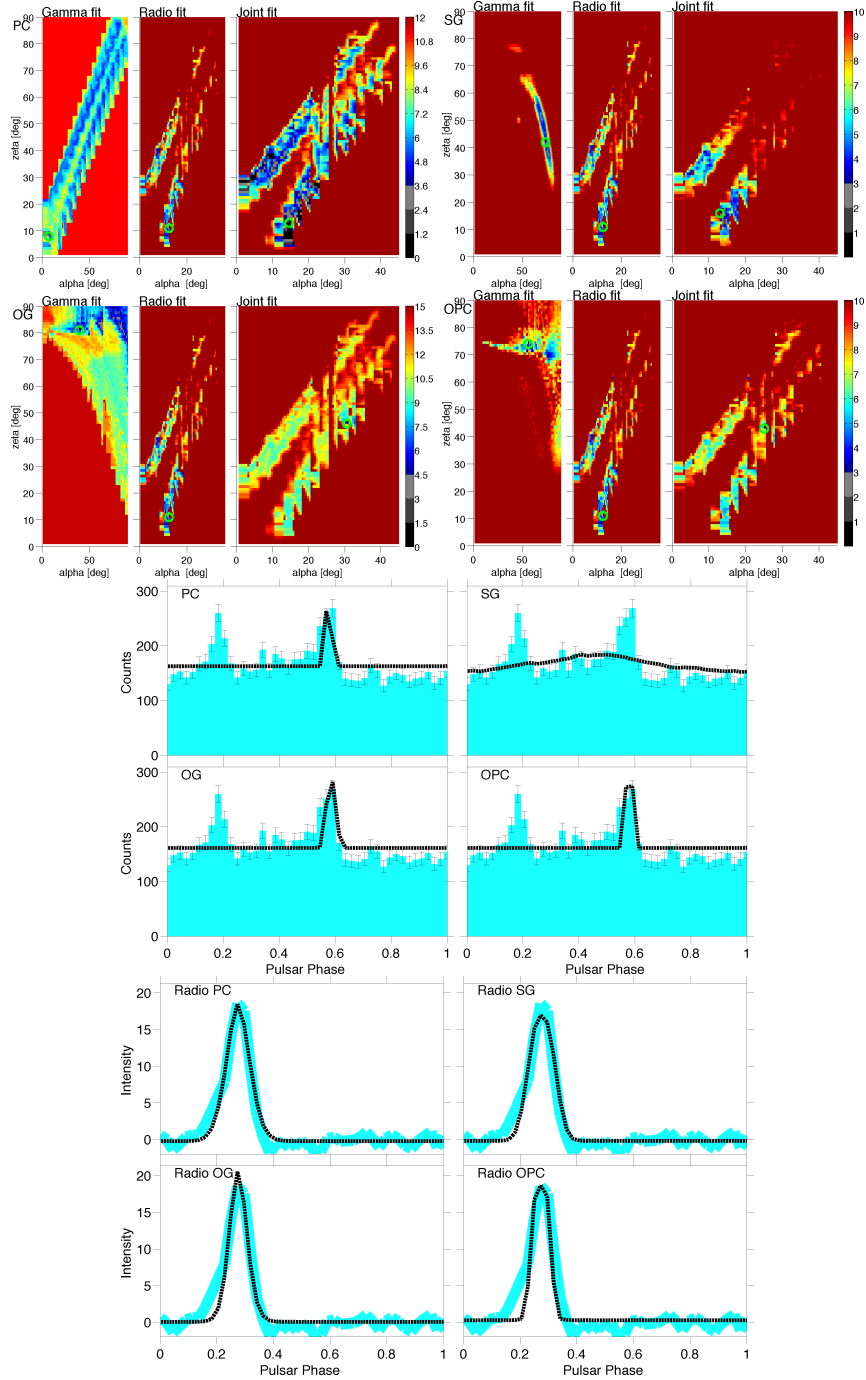

Figure 6.66: PSR J1747-2958. *Top*: for each model is shown the $\alpha$-$\zeta$ likelihood map for the $\gamma$-ray $\chi^2$ fit, the radio $\chi^2$ fit, each one optimised variance, and the sum of the maps. The color-bar is in $\sigma$ units, zero corresponds to the best fit solution. *Middle*: the LAT light-curve (in blue) is compared to the $\gamma$-ray light curve obtained, for each model, by maximising the joint likelihood map. *Bottom*: the radio profile (thick blue line) is compared to the radio light curve obtained, for each model, by maximising the joint likelihood map. The radio model is unique, but the $(\alpha, \zeta)$ solutions vary for each $\gamma$-ray model.

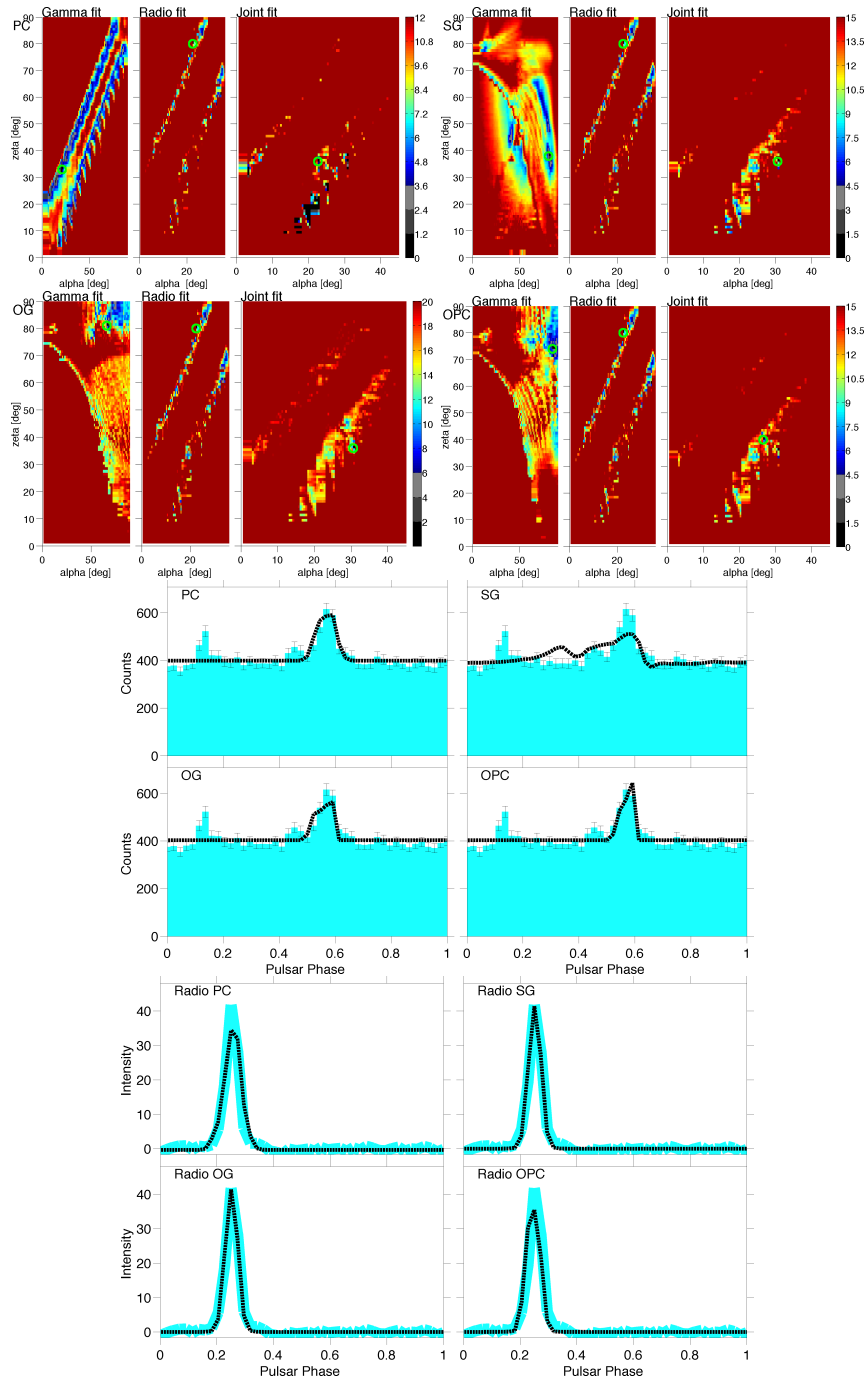

Figure 6.67: PSR J1833-1034. *Top*: for each model is shown the α-ζ likelihood map for the γ-ray χ² fit, the radio χ² fit, each one optimised variance, and the sum of the maps. The color-bar is in σ units, zero corresponds to the best fit solution. *Middle*: the LAT light-curve (in blue) is compared to the γ-ray light curve obtained, for each model, by maximising the joint likelihood map. *Bottom*: the radio profile (thick blue line) is compared to the radio light curve obtained, for each model, by maximising the joint likelihood map. The radio model is unique, but the (α,ζ) solutions vary for each γ-ray model.

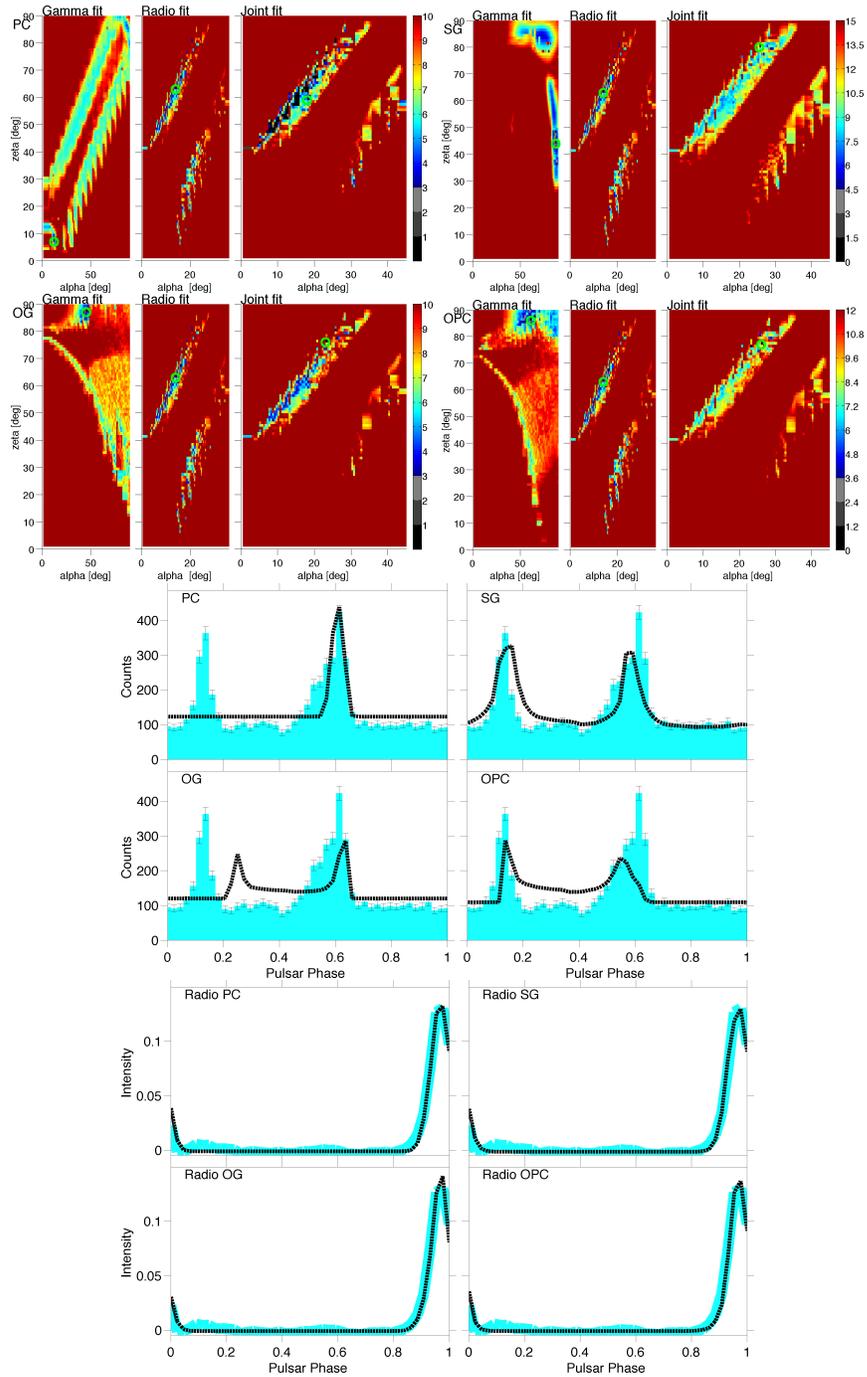

Figure 6.68: PSR J1952+3252. *Top*: for each model is shown the $\alpha$-$\zeta$ likelihood map for the $\gamma$-ray $\chi^2$ fit, the radio $\chi^2$ fit, each one optimised variance, and the sum of the maps. The color-bar is in $\sigma$ units, zero corresponds to the best fit solution. *Middle*: the LAT light-curve (in blue) is compared to the $\gamma$-ray light curve obtained, for each model, by maximising the joint likelihood map. *Bottom*: the radio profile (thick blue line) is compared to the radio light curve obtained, for each model, by maximising the joint likelihood map. The radio model is unique, but the $(\alpha, \zeta)$ solutions vary for each $\gamma$-ray model.

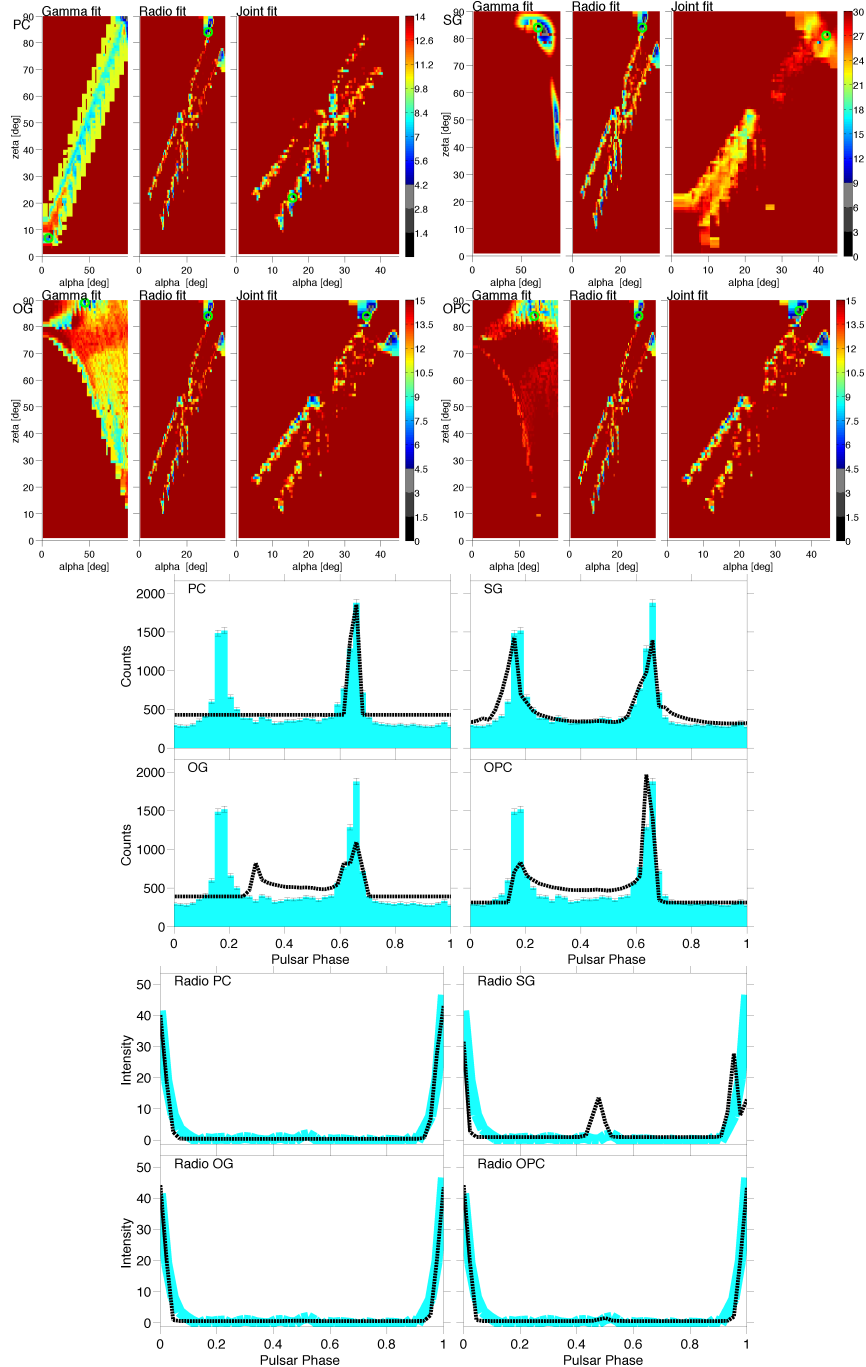

Figure 6.69: PSR J2021+3651. *Top*: for each model is shown the $\alpha$-$\zeta$ likelihood map for the $\gamma$-ray $\chi^2$ fit, the radio $\chi^2$ fit, each one optimised variance, and the sum of the maps. The color-bar is in $\sigma$ units, zero corresponds to the best fit solution. *Middle*: the LAT light-curve (in blue) is compared to the $\gamma$-ray light curve obtained, for each model, by maximising the joint likelihood map. *Bottom*: the radio profile (thick blue line) is compared to the radio light curve obtained, for each model, by maximising the joint likelihood map. The radio model is unique, but the $(\alpha, \zeta)$ solutions vary for each $\gamma$-ray model.

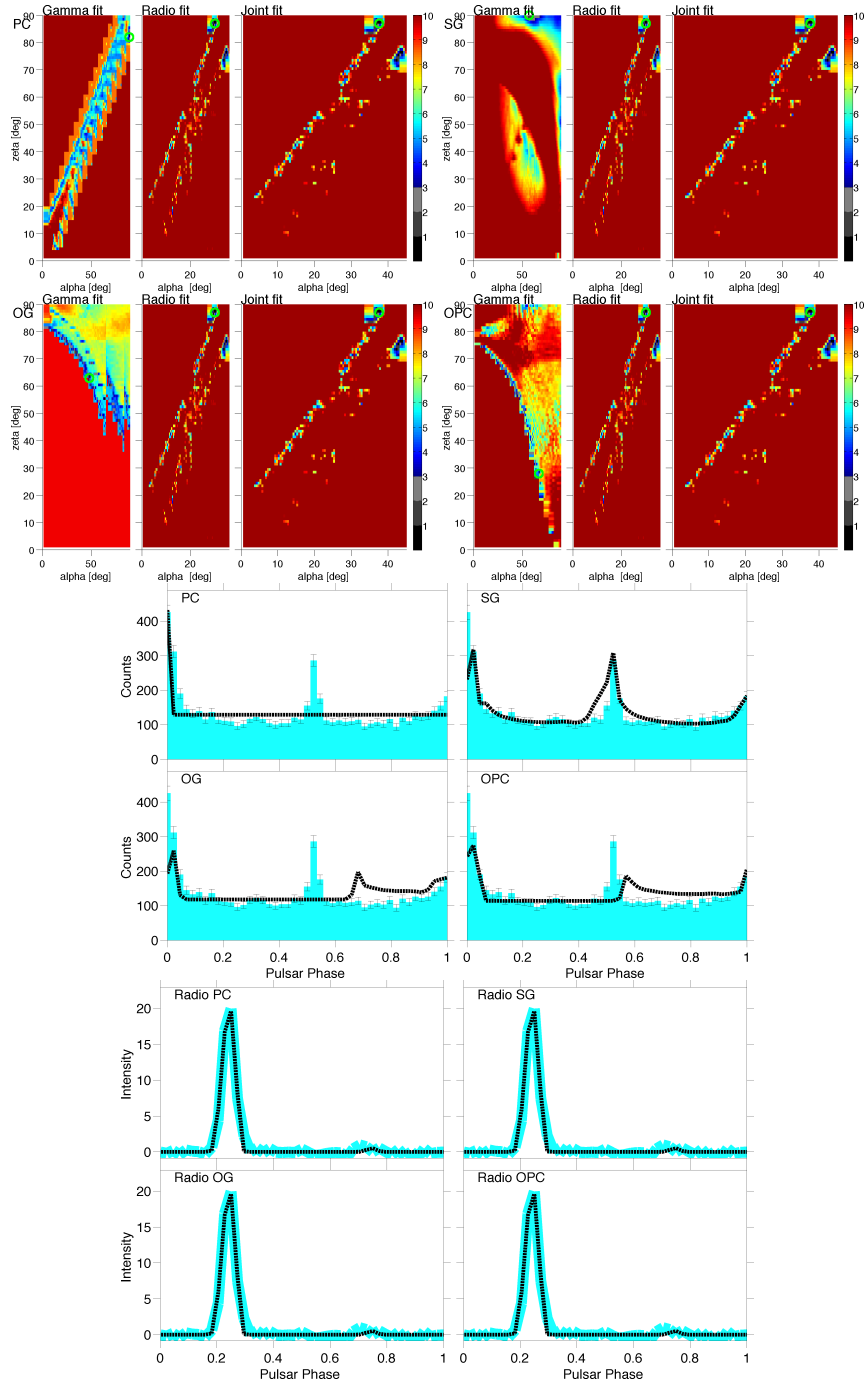

Figure 6.70: PSR J2032+4127. *Top*: for each model is shown the $\alpha$-$\zeta$ likelihood map for the $\gamma$-ray $\chi^2$ fit, the radio $\chi^2$ fit, each one optimised variance, and the sum of the maps. The color-bar is in $\sigma$ units, zero corresponds to the best fit solution. *Middle*: the LAT light-curve (in blue) is compared to the $\gamma$-ray light curve obtained, for each model, by maximising the joint likelihood map. *Bottom*: the radio profile (thick blue line) is compared to the radio light curve obtained, for each model, by maximising the joint likelihood map. The radio model is unique, but the $(\alpha, \zeta)$ solutions vary for each $\gamma$-ray model.

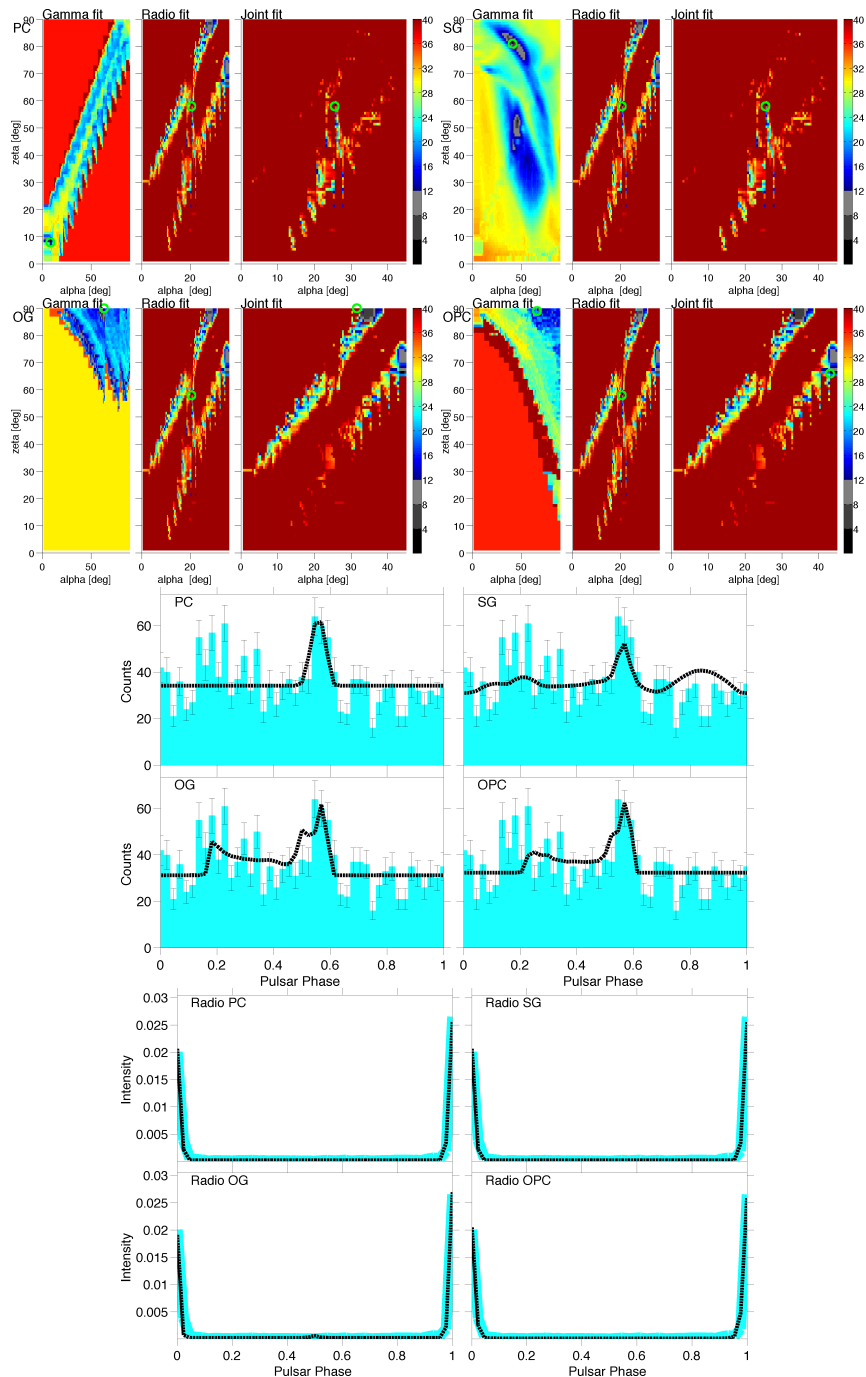

Figure 6.71: PSR J2043+2740. *Top*: for each model is shown the α-ζ likelihood map for the γ-ray χ² fit, the radio χ² fit, each one optimised variance, and the sum of the maps. The color-bar is in σ units, zero corresponds to the best fit solution. *Middle*: the LAT light-curve (in blue) is compared to the γ-ray light curve obtained, for each model, by maximising the joint likelihood map. *Bottom*: the radio profile (thick blue line) is compared to the radio light curve obtained, for each model, by maximising the joint likelihood map. The radio model is unique, but the (α, ζ) solutions vary for each γ-ray model.

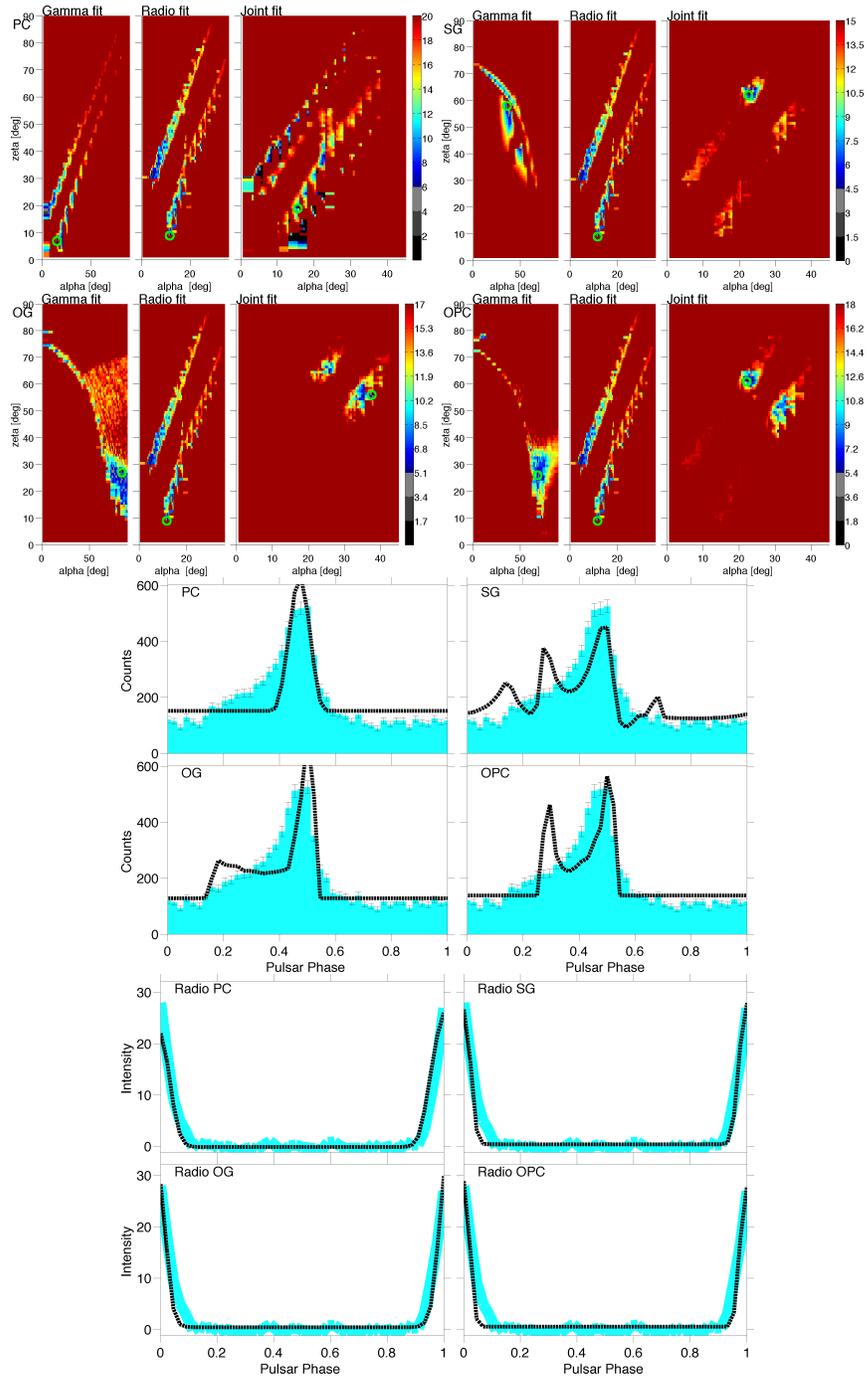

Figure 6.72: PSR J2229+6114. *Top*: for each model is shown the α-ζ likelihood map for the γ-ray χ² fit, the radio χ² fit, each one optimised variance, and the sum of the maps. The color-bar is in σ units, zero corresponds to the best fit solution. *Middle*: the LAT light-curve (in blue) is compared to the γ-ray light curve obtained, for each model, by maximising the joint likelihood map. *Bottom*: the radio profile (thick blue line) is compared to the radio light curve obtained, for each model, by maximising the joint likelihood map. The radio model is unique, but the (α, ζ) solutions vary for each γ-ray model.

## 6.3   Results

This section is dedicated to the illustration and discussion of the results
obtained by fitting the studied LAT pulsar sample with an individual γ-ray
model fit and by implementing a joint γ-radio fit.

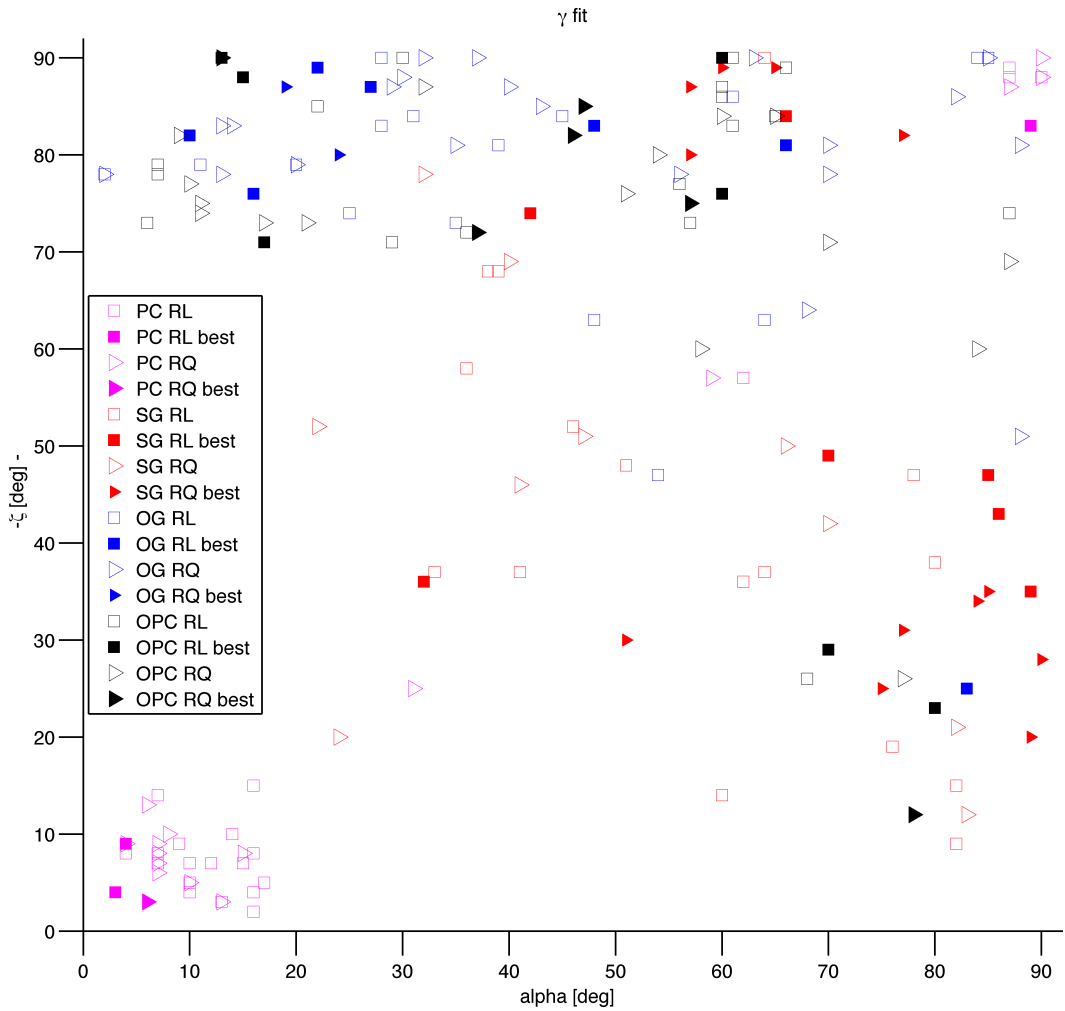

Figure 6.73: Distribution, on the α-ζ plane, of the γ-fit solutions. Triangles and
squares respectively refer to radio-quiet and radio-loud solutions. For each pulsar,
the model that gives the highest likelihood value (best solution) is plotted as filled
marker while the others as empty markers.

The α and ζ estimates obtained as a result of both the individual γ-ray
fit and the joint γ-radio one, have been used to evaluate all the of the studied
pulsars.

### 6.3.1   $\alpha$-$\zeta$ best solutions plane

The first result I will show is the distribution, in the $\alpha$-$\zeta$ plane, of the best fit solutions from both the individual FCB $\gamma$-ray fit (section, 6.1, hereafter *$\gamma$ fit*) and from the joint $\gamma$ plus radio fit (section 6.2, hereafter *joint fit*). Both results are shown, respectively, in figures 6.73 & 6.74.

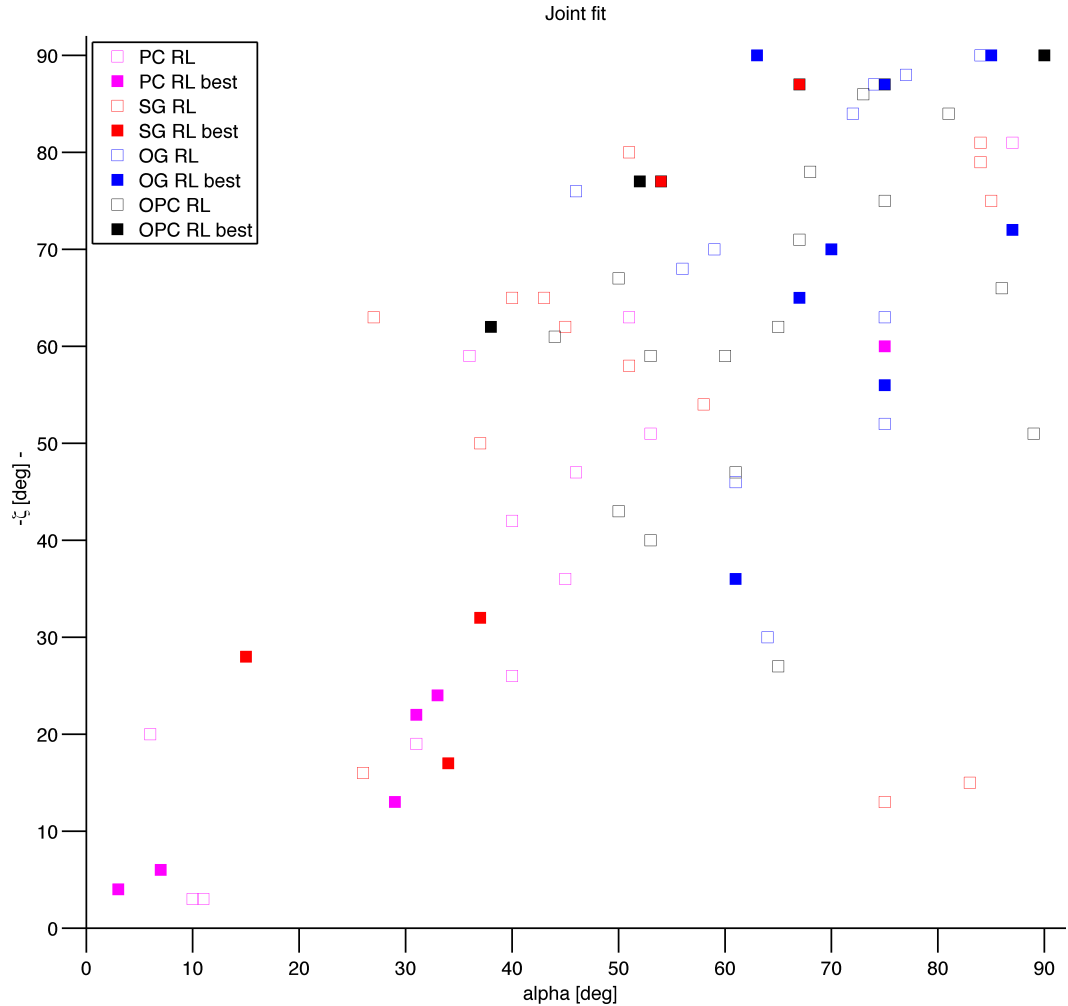

Figure 6.74: Distribution, on the $\alpha$-$\zeta$ plane, of the joint fit solutions. For each pulsar, the model that gives the highest likelihood value (best solution) is plotted as filled marker while the others as empty markers.

Since the radio and PC $\gamma$-ray emission are concentrated in narrow conical beams, coaxial with the magnetic axis, the observer will see the pulsed emission just if its line of sight is close to the magnetic axis. It is required $\alpha$ should be near to $\zeta$. Hereafter, the $\alpha - \zeta$ plane diagonal connecting the points $(0, 0)$, $(90, 90)$, will be called *radio diagonal*. The solution found for the PC model $\gamma$ fit and the solution found for all the models, by fitting jointly $\gamma$ and radio emissions, will lie near the radio diagonal of the $\alpha - \zeta$ plane.

Since the $\gamma$-fit does not take into account the orientation that the pulsar
should have to allow us to see the radio emission, the likelihood maximisation
can find solutions in regions that are far away from the radio diagonal. So,
all the solutions found from the $\gamma$ fit far away from the radio diagonal for
radio-loud pulsars are not trustable. To better understand the differences in
the best solutions found by the $\gamma$ and joint fit, in figure 6.75 is shown, for each
model, the migration of the solution found by the two methods, for the same
object.

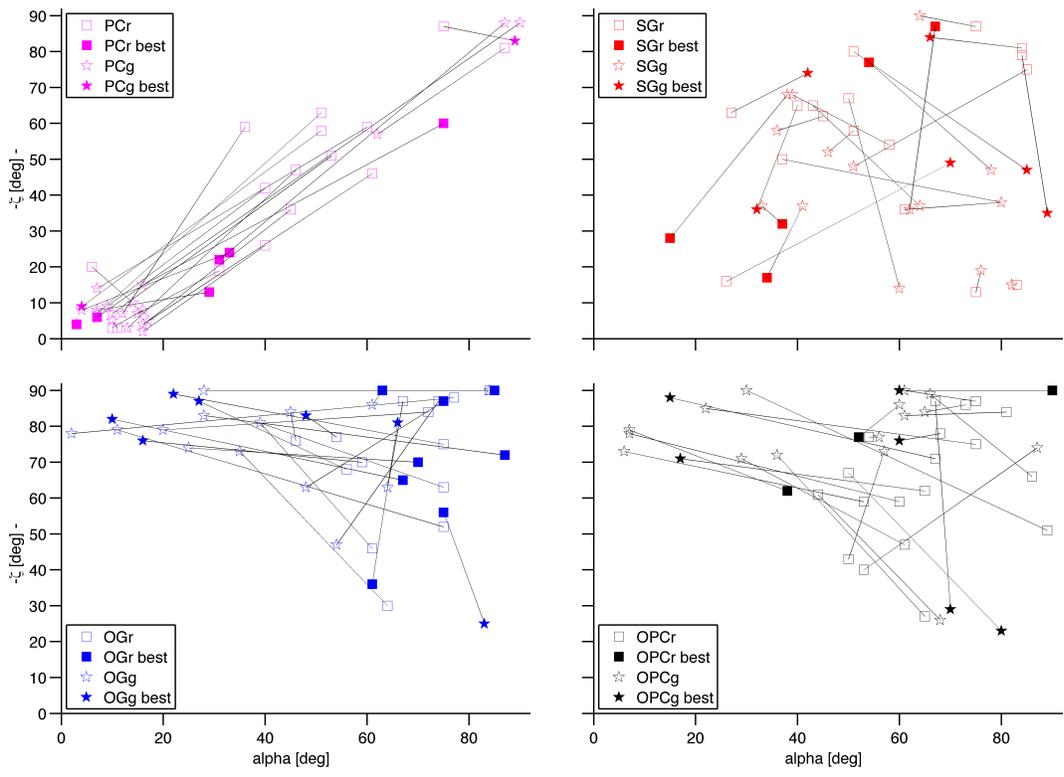

Figure 6.75: Distribution, on the $\alpha$-$\zeta$ plane, of the $\gamma$-fit and joint-fit solutions for
the analysed radio-loud pulsar sample. To show how the introduction of the radio
component in the fit affects the best $\alpha$-$\zeta$ estimate, a black line connects the $\gamma$ and
joint solution of each pulsar. The lines describe how the solution "migrates" when
the radio emission pattern is taken into account. Stars and squares respectively refer
to the $\gamma$ and joint fit solutions. Filled markers note the highest-likelihood solutions.

For both the $\gamma$ and joint fit results, the PC solutions are concentrated
along the radio diagonal in particular at low $\alpha$ and $\zeta$. The concentration of
the PC solutions in the low part of the radio diagonal is due to the size of the
emission cone beam that strongly decreases when $\alpha$ increases, thus decreasing
the probability for the observer line-of-sight to cross the beam. In the PC
case, the migration obviously occurs along the radio diagonal but it shows a
double trend. After the migration, the new best joint fit solutions tend to remain in the bottom left corner of the $\alpha$-$\zeta$ plane but a lot of non best fit solution migrate from this corner to the centre of the radio diagonal, around $\alpha$-$\zeta$=45-45. As I mentioned before, low $\alpha$-$\zeta$ PC values imply a bigger emission beam so it is normal to have the best solutions concentrated in that region. On the other side, the concentration of solutions around $\alpha$-$\zeta$=45-45 is due to the fact that all the real fitted radio profiles are single peak light-curves that are more probable for a pulsar orientation $\alpha$=$\zeta$=45.

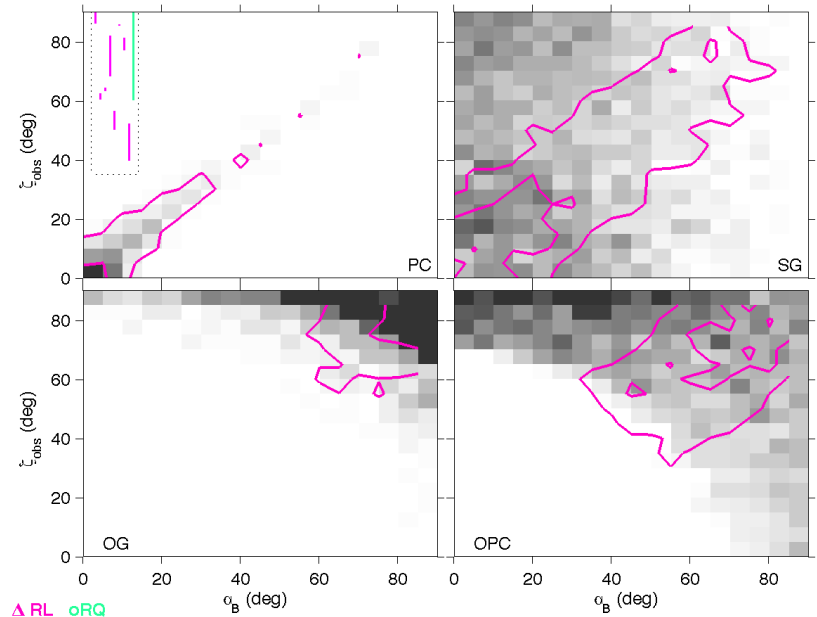

Figure 6.76: Number density of the visible gamma-ray pulsars obtained for each model as a function of magnetic obliquity (alpha) and observer aspect angle (zeta). The linear gray scale saturates at 1.5 star/bin. The pink contours outline the density obtained for the radio-loud gamma-ray sub-sample (at 5% and 50% of the maximum density). The insert gives the set of zeta values measured by (Ng & Romani, 2008) from the orientation of the wind torus seen in X rays (pink lines) and by Caraveo et al. (ref) from the orientation of the Geminga X-ray tails (green line). The separation in alpha in the insert is meaningless.

For the SG case, exception made for some non best fit solutions (other models yield better fits), the results of both the $\gamma$ and joint fits are able to explain a simultaneous radio and gamma emission. All the $\gamma$ radio-loud best fit solutions are close enough to the radio diagonal to explain the radio emission from these pulsars and the radio-quiet ones are far enough to justify the absence of radio detection. One notes an absence of very best joint solutions in the central part of the radio diagonal.

For the OG and OPC models the $\gamma$ fit shows several radio-loud pulsars in $(\alpha, \zeta)$ plane regions where radio emission is impossible, the top left and bottom right corners of the plane. All the non trustable solutions (radio-loud pulsars with $(\alpha, \zeta)$ far away from the radio diagonal) migrate horizontally, toward
the radio diagonal. Exception made for some solutions, this implies that for
the OG & OPC, going from the $\gamma$ to the joint fit, the $\alpha$ distribution changes
completely its shape while the $\zeta$ one keeps constant around the $\gamma$ values.

In figure 6.76 is shown the distribution in the $\alpha$-$\zeta$ plane of the $\gamma$-visible
component obtained from the population synthesis work described in previous
chapters. The comparison with the $\gamma$ and joint-fit best solutions of figures 6.73
& 6.74, shows a good consistence between the LAT pulsars and the simulated
population. It suggests that there is no additional, unaccounted for, selection
effect that would bias the LAT sample. We will now use the knowledge of the
$(\alpha, \zeta)$ solutions to study various properties of the LAT pulsar sample.

### 6.3.2 Beaming factor $f_\Omega$

The beaming factor, evaluated from the $(\alpha, \zeta)$ results of the $\gamma$-ray as well as
joint fit, is plotted, for each model, as a function of the pulsar spin-down power
in figure 6.77 & 6.78. The $f_\Omega$ computation has been done by using equation
5.40, and the emission pattern phase-plot described in section 5.2.1.

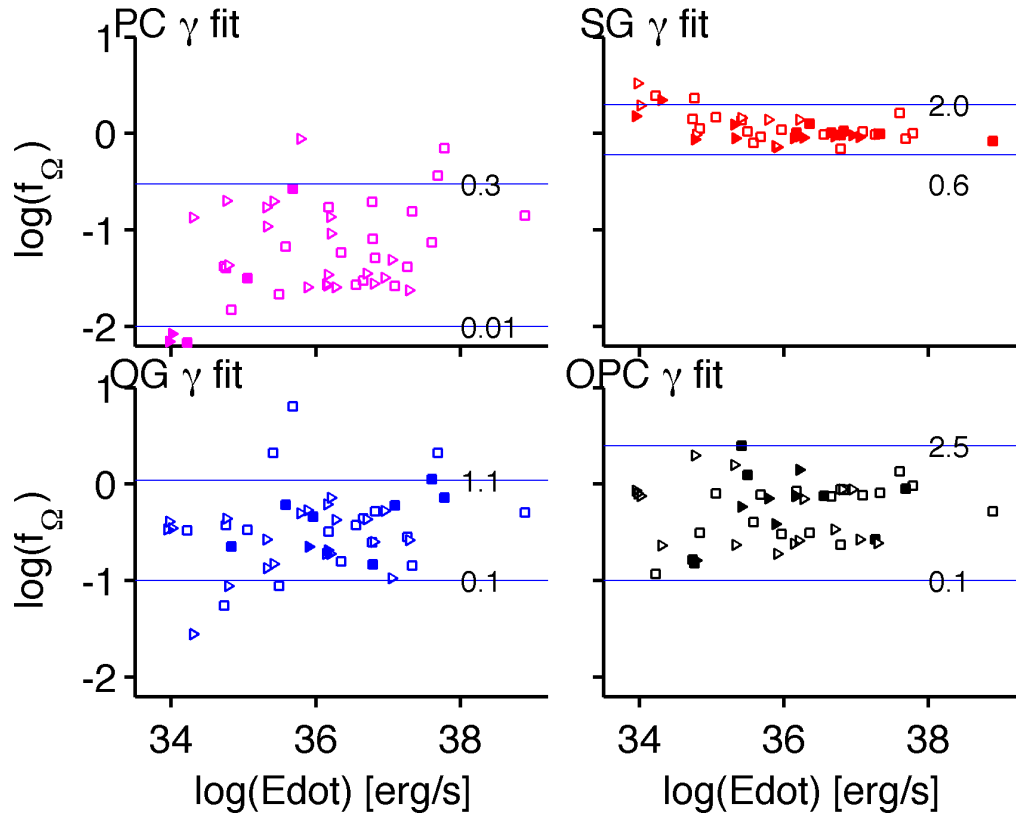

Figure 6.77: Beaming fraction $f_\Omega$ evaluated for the $(\alpha, \zeta)$ solution for the gamma-fit
versus the LAT pulsar spin-down power. Triangles and squares respectively refer
to solutions for radio-quiet and radio-loud pulsars. Filled markers note the highest
likelihood case between the different models.

In the PC case, $f_\Omega$ is low, as expected from the small hollow cone beam used to describe the radio phase-plot (figure 5.6). Since the PC beam is coming from the polar cap region, we expect that $f_\Omega$ decreases as the period increases, thus as $\dot{E}$ decreases. This trend present in the joint best fit solution and can be understood if we consider that the light cylinder radius increases when $\dot{E}$ decreases.

Since in the SG model there is pulsed emission in nearly all directions it is normal to observe a concentration of values around $f_\Omega = 1$. Moreover, the SG $f_\Omega$ is less dispersed for the joint fit solution values. Since the emission is not strongly beamed, $f_\Omega$ does not show any strong trend with $\dot{E}$.

In the OG and OPC cases, the $f_\Omega$ values are much more dispersed with the joint-fit solutions than with the single gamma solutions. This corresponds to the behaviour expected for radio-loud and radio-quiet objects in the population synthesis results (Figure 5.21). It is interesting to note that the beaming fractions of radio-loud pulsars are often larger than the radio-quiet ones.

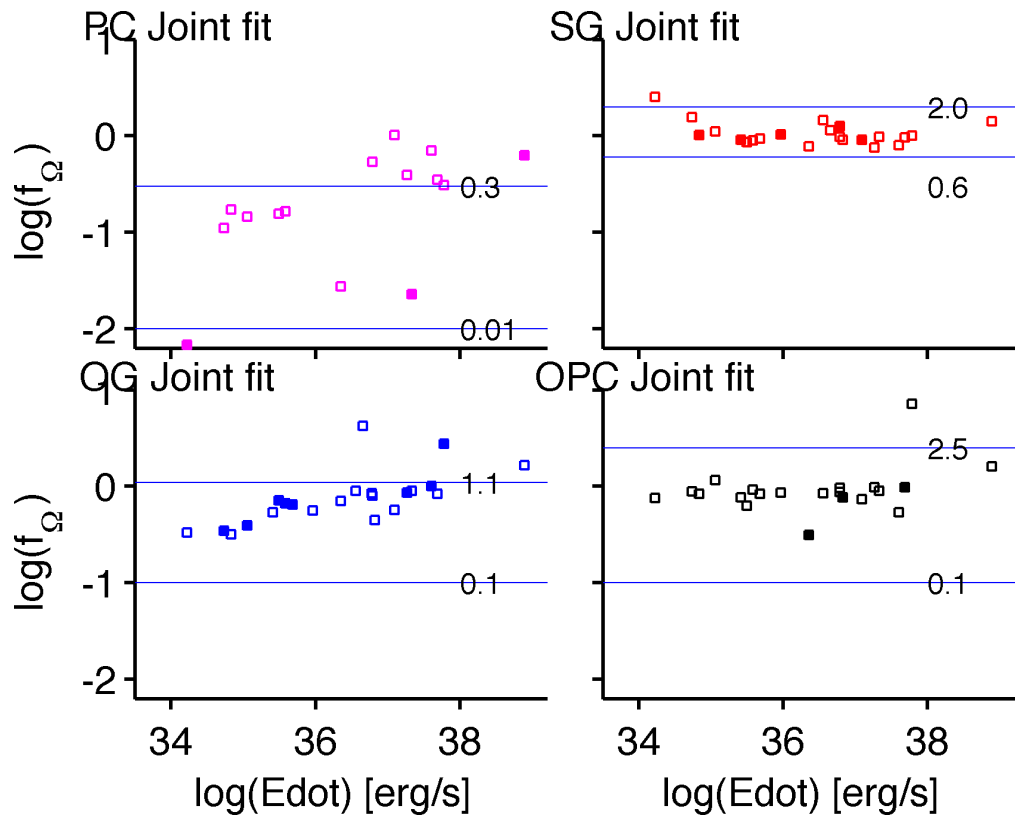

Figure 6.78: Beaming fraction $f_\Omega$ evaluated for the $(\alpha, \zeta)$ solution from the joint-fit versus the LAT pulsar spin-down power. Filled markers note the highest likelihood case between the different models.

The slightly more pronounced shape of $f_\Omega(\dot{E})$ in the OG case compared to the OPC reflect the trend found in the population synthesis results and suggest a different evolution of the beam geometry in the models.

### 6.3.3 Orientation evolution and magnetic alignment

I tried to verify if there is any evidence of an α evolution with pulsar age. The magnetic alignment is a scenario that has been recently analysed by (Young et al., 2010). In addition to a progressive narrowing of the emission cone, pulsar evolution could be characterised by a progressive alignment of the spin and magnetic axes.

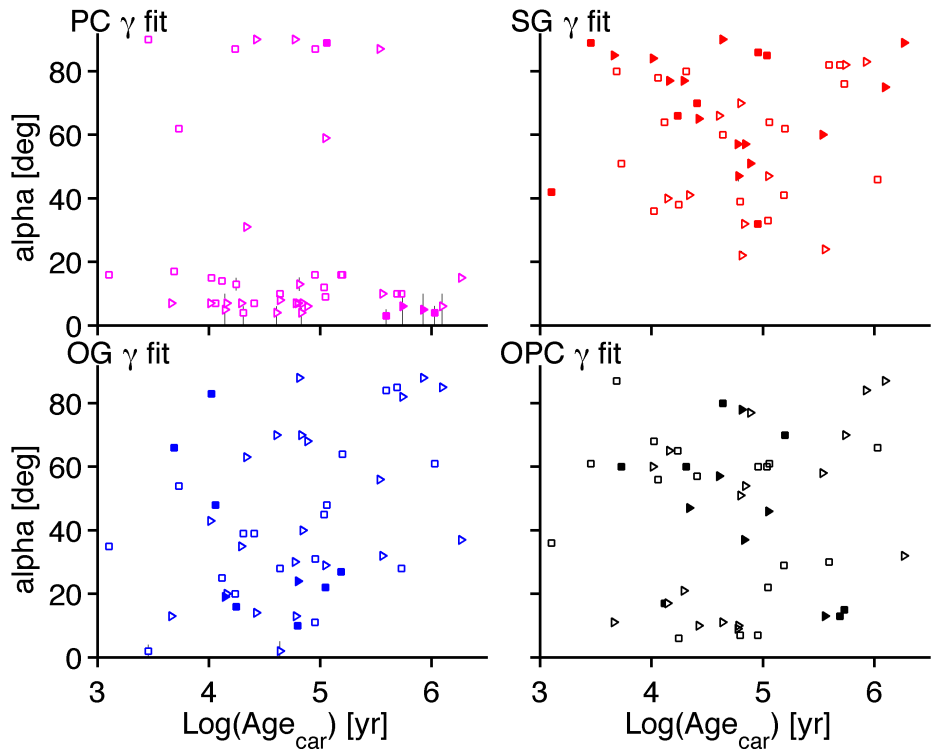

Figure 6.79: Magnetic obliquity α for the (α, ζ) solution for the gamma-fit versus the LAT pulsar characteristic age. Triangles and squares respectively refer to radio-quiet and radio-loud solutions. Filled markers note the highest likelihood case between the different models.

In Figures 6.79 & 6.80 are plotted, for each model, the magnetic obliquity versus the characteristic age for the solutions of the γ and joint fits.

The results are very dispersed for the PC and SG cases and above 40 degrees for the OG and OPC cases. In the OG/OPC phase-plots there is indeed no visible emission for 0 ≤ α ≤ 40. In the OG case, the very mild increase of α with age is not significant and in contrast with the hypothesis of a magnetic alignment as the pulsar ages. Figure 6.81 shows the α evolution with respect to the characteristic age for the γ visible component of the emission models assumed in the population synthesis described in chapter 5. Both the γ-fit and joint fit results are consistent with the alpha evolution obtained from the population study.

In figures 6.82 and 6.83 is shown the evolution of the $|\alpha - \zeta|$ with respect to the pulsar period, for both the $\gamma$ and joint fits. It is evident how the solutions change from the $\gamma$ to the joint fit, and how the latter shows a well defined decreasing trend with pulsar period to capture a radio-loud object. As a pulsar ages, its spin period and polar cap size decrease. So, the observed trend is due to a selection effect for which we have a higher probability to see a young, wide-beam pulsar for high value of $|\alpha - \zeta|$. When the pulsar gets older and the beam shrinks, we can see emission just if the line of sight gets gradually closer to $\alpha$, and so for smaller $|\alpha - \zeta|$ values.

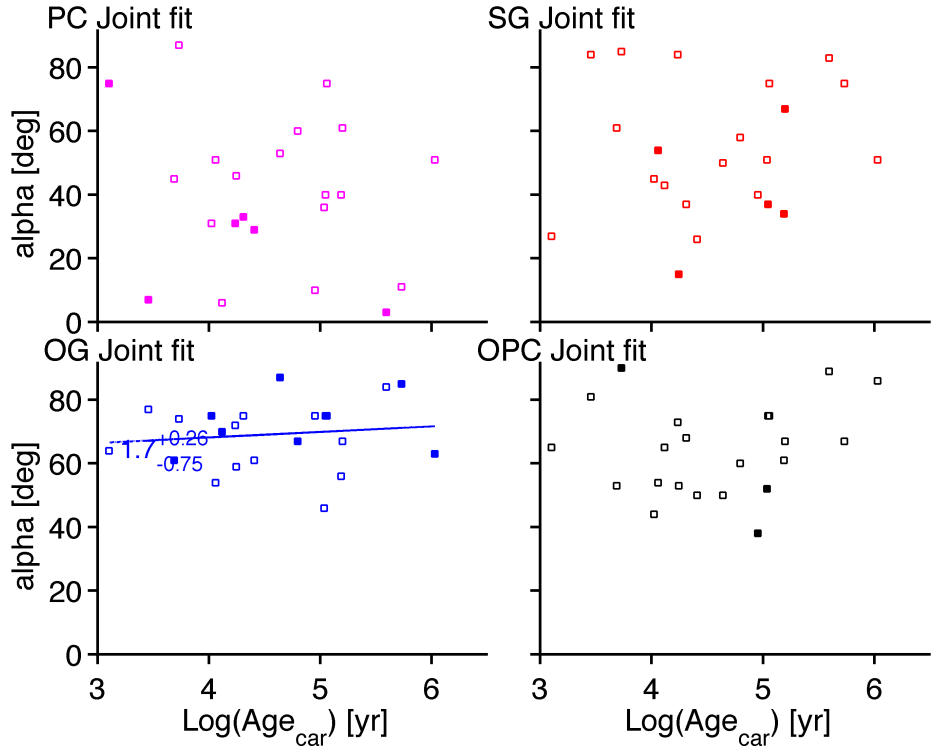

Figure 6.80: Magnetic obliquity $\alpha$ for the $(\alpha, \zeta)$ solution for the joint-fit versus the LAT pulsar characteristic age. Filled markers note the highest likelihood case between the different models.

There are no young pulsars observed with low $|\alpha - \zeta|$ values. This characteristic is not explained by the size reduction of the beam that, on the contrary, implies a young pulsar detection for all the $|\alpha - \zeta|$ values. However, this lack of detections could be explained as selection effect.

In fact, the large majority of the fitted radio profiles shows a single peak structure or two peaks separated by 0.5 in phase. This implies that the observer line of sight $\zeta$ crosses the edge of the radio beam, so the $|\alpha - \zeta|$ angle cannot be lower than $\rho_{cone}/2$ ($\rho$ is the beam opening angle). Another explanation concerns the few radio profiles fitted by a close double peak light curve. Going toward small $\alpha$ values, the emission beam smears and assumes the shape of a bump. The bump pulsar light curves are hardly detectable because they do not have sharp peaks and are characterised by a low signal to noise $(S/N)$ ratio. Moreover, if the young pulsar has $\alpha \simeq \rho_{cone}/2$, it is more difficult to detect the pulsation even if the observer line of sight crosses the emission beam. On the other side, old objects with the same $\alpha$ angle have a smaller and still conical beam, observed with a high $S/N$, and far enough from the magnetic axis direction $\alpha$ to generate a clear pulsation.

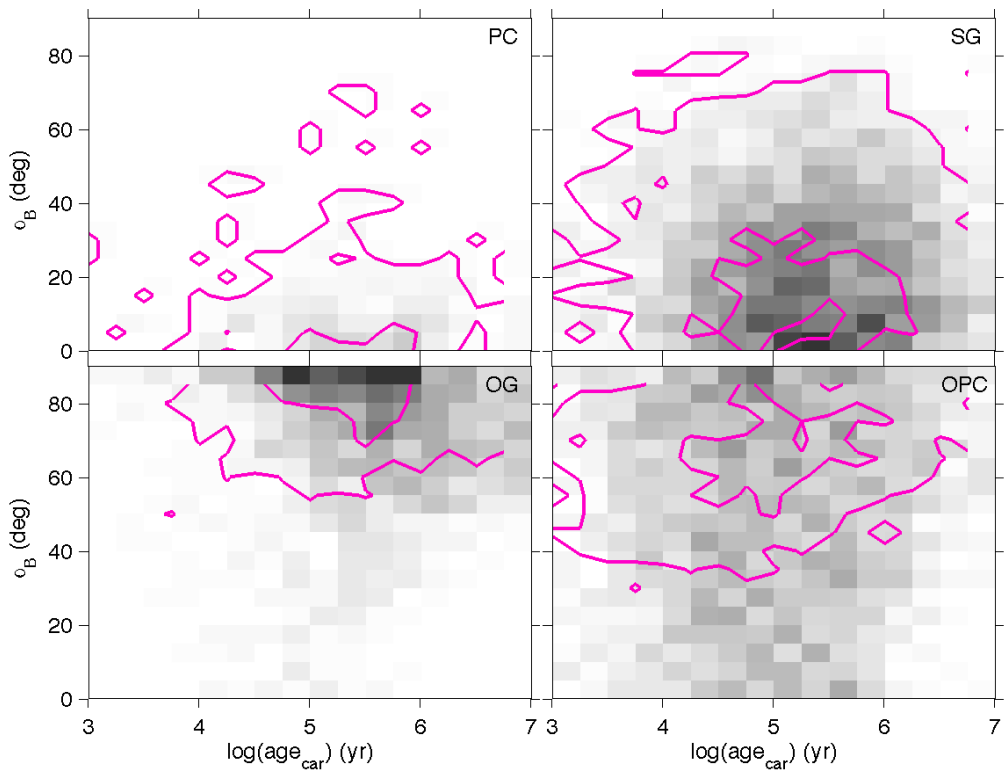

Figure 6.81: Number density of the visible gamma-ray pulsars obtained for each model as a function of magnetic obliquity (alpha) and characteristic age. The linear grey scale saturates at 2 stars/bin. The pink contours outline the density obtained for the radio-loud gamma-ray sub-sample (at 5% and 50% of the maximum density).

In figure 6.84 is shown the evolution of the $|\alpha - \zeta|$ angle with the spin period, obtained, for each emission model, from the population synthesis described in chapter 5. Both the $\gamma$ and joint-fit solutions found for the LAT pulsars are consistent with, respectively, the total and radio loud components of the simulated populations shown in figure 6.84. The decreasing $|\alpha - \zeta|$ angle trend is present but not tight in the population synthesis.

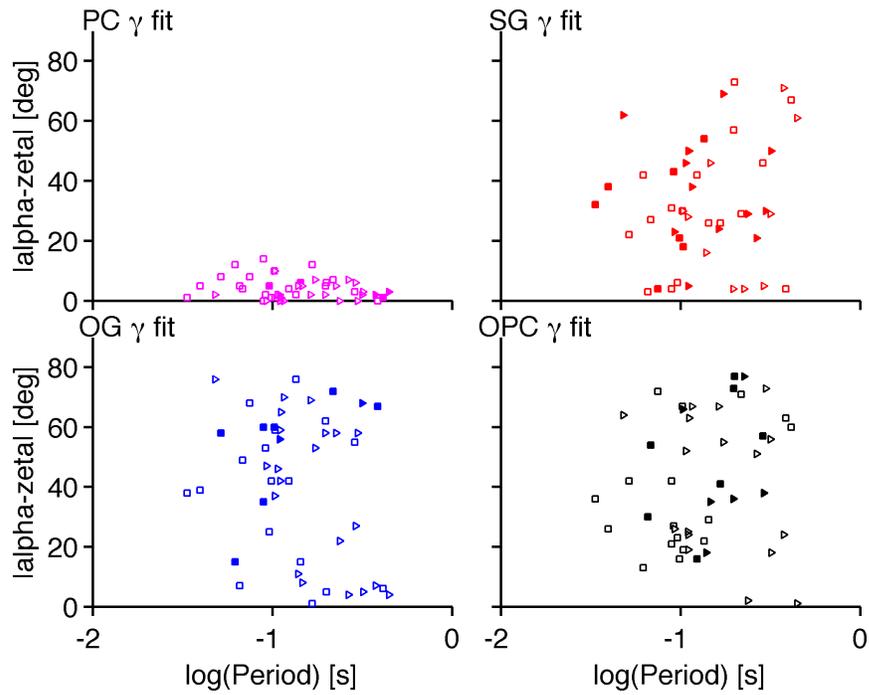

Figure 6.82: Fit solution for the $|\alpha - \zeta|$ angle vs period. Triangles and squares respectively refer to radio-quiet and radio-loud solutions. Filled markers note the highest likelihood case between the different models.

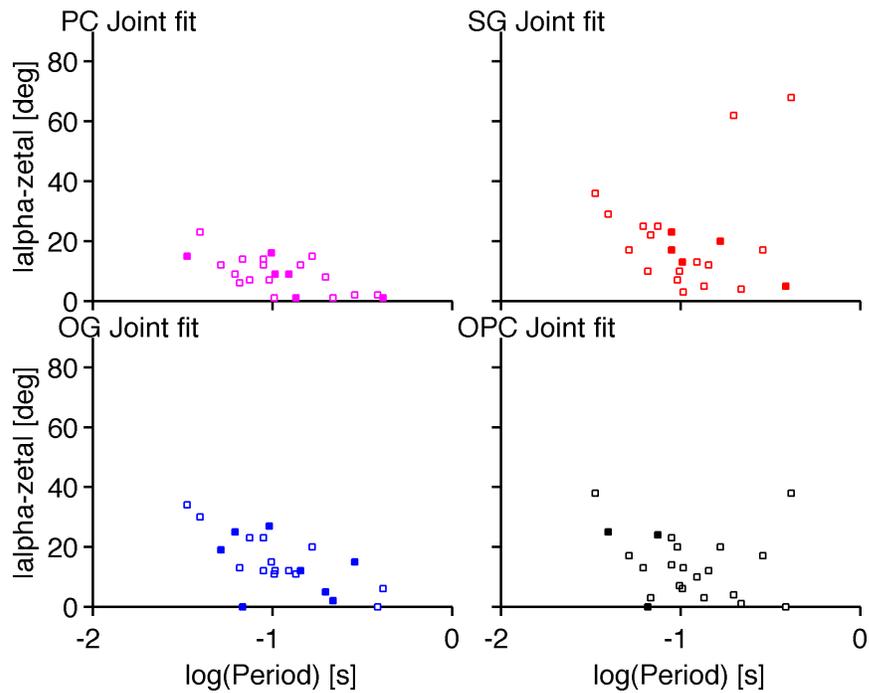

Figure 6.83: Joint best fit solution for the $|\alpha - \zeta|$ angle versus the pulsar period. Filled markers note the highest likelihood case between the different models.

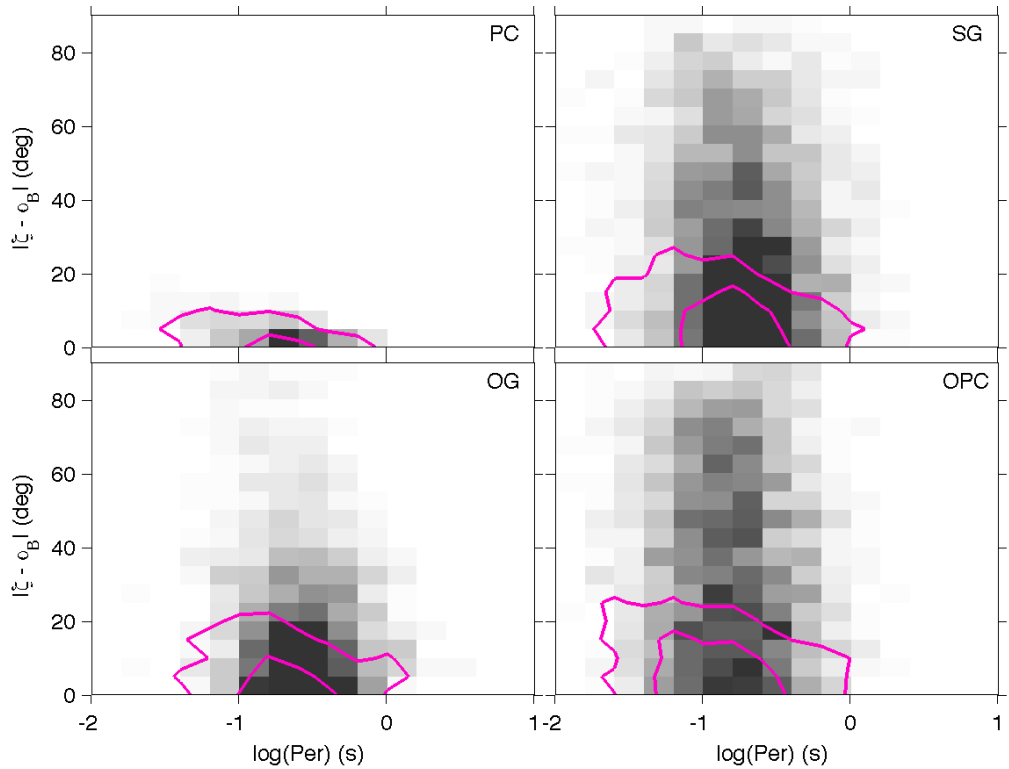

Figure 6.84: Number density of the visible gamma-ray pulsars obtained for each model as a function pulsar period. The linear gray scale saturates at 2 star/bin. The pink contours outline the density obtained for the radio-loud gamma-ray sub-sample (at 5% and 50% of the maximum density).

### 6.3.4  Luminosity

By using the observed LAT pulsar fluxes and equation 5.40, I have evaluated their luminosity for each $(\alpha, \zeta)$ solution found with the two implemented fit methods.

In figures 6.85 & 6.86 are plotted the luminosity versus $\dot{E}$ for the $\gamma$ fit and joint-fit solutions. The first important aspect to be noted is that, for both the joint and gamma fits and for each fitted model (exception made for the PC joint results), the luminosity distribution is consistent with the theoretical behaviour $L_\gamma \propto \dot{E}^{0.5}$.

The model that, in both the $\gamma$ and joint fits, reproduces the expected 0.5 exponent within errors, is the SG. For the other models, the fit results are more distant from the predicted value and this could suggest a further investigation on the theoretical relation or/and on the accuracy of the fit and computations. The power law index becomes higher for all the models going from the $\gamma$-fit to the joint fit. Particularly interesting is the comparison of the fit results, both for the $\gamma$ and joint ones, with the population synthesis results shown in figure 5.22. The large dispersion observed in the simulations for the radio-quiet components is not observed in the $\gamma$ fit results. Both the $\gamma$ and joint fits show much narrower distributions around the expected power law trend, in particular in the OG and OPC cases.

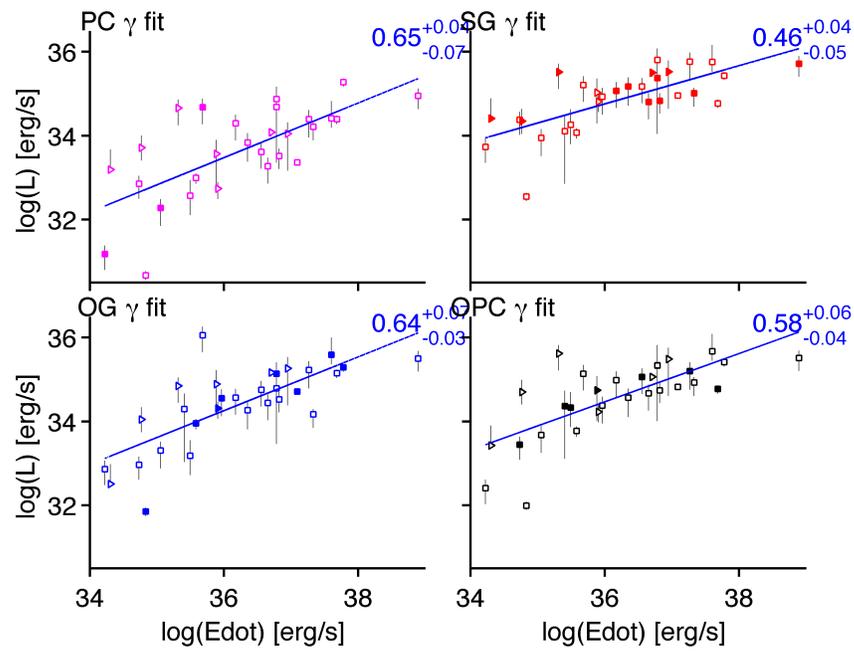

Figure 6.85: γ fit solution for the pulsar luminosity versus $\dot{E}$. Triangles and squares respectively refer to radio-quiet and radio-loud solutions. Filled markers note the highest likelihood case between the different models.

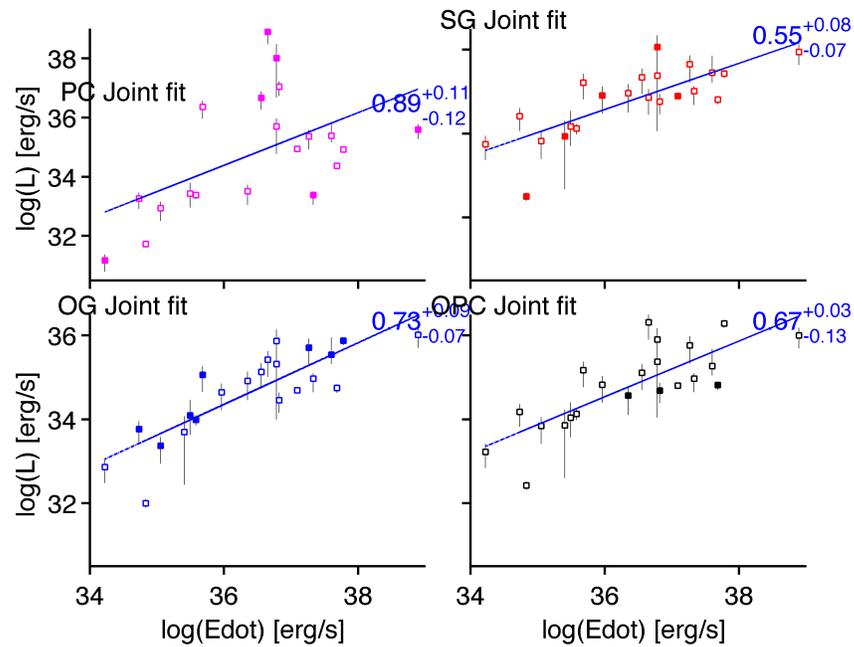

Figure 6.86: Joint best fit solution for the pulsar luminosity versus $\dot{E}$. Filled markers note the highest likelihood case between the different models.

In figure 6.87 is plotted the luminosity versus the pulsar period just for the best joint-fit solutions. What I would like to discuss is the possibility that all four models I used to fit the LAT observations could act at different stages of the pulsar life. Since to search for such a trend requires the most accurate estimates, I choose to plot just the joint fit solutions for the models that gave the highest likelihood value. Figure 6.87 shows that the $\gamma$-ray luminosity decreases with period notwithstanding the emission model considered.

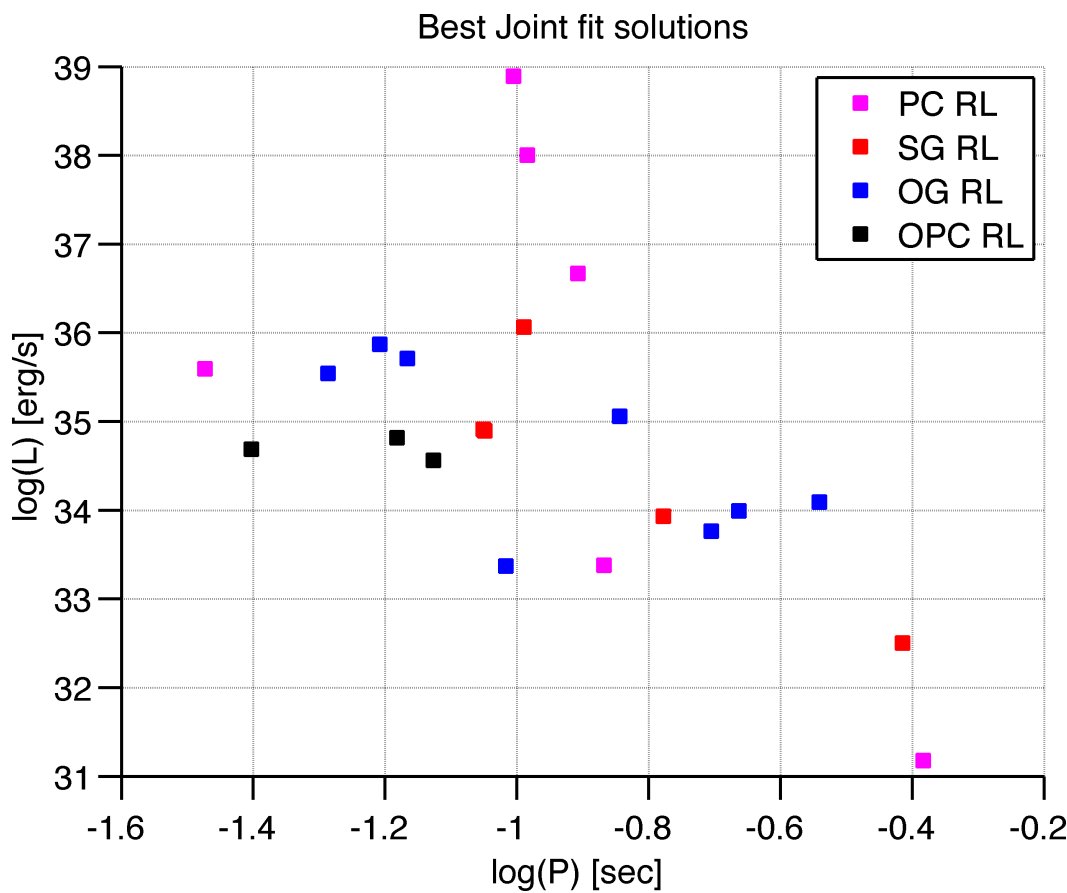

Figure 6.87: Luminosity versus Period for the joint best fit solutions for all the implemented models. For each pulsar it has been plotted just the solution corresponding to the maximum likelihood over all the model solutions. Each pulsar is best fitted by a specific model.

This study can be considered as a first approach to the study of the pulsar luminosity function as generated by different emission mechanisms and requires a more accurate investigation.

### 6.3.5   High energy cut-off & gap width

As a last result of this chapter, Figures 6.88 & 6.89 show the relation between an observable spectral characteristic, the cut-off energy $E_{cut}$, and the width of the emission gaps deduced from the knowledge of $\alpha$ and $\zeta$. We find a pendency for $E_{cut}$ to decrease when the gaps widen. This dependency is particularly important because it defines a relation between the observable $E_{cut}$ and the intrinsic, non directly observable, pulsar gap width, that defines the pulsar electrodynamics. A power law dependency between $E_{cut}$ and $\Delta\xi$ for the SG model can be theoretically obtained as follows. From (Abdo et al., 2010a), the $E_{cut}$ dependency is defined as

$$E_{cut} \propto E_{\parallel}^{3/4} \rho_c^{1/2} \tag{6.13}$$

where $E_{\parallel}$ is the electric field parallel to $B$ lines, and $\rho_c \sim (0.1 - 0.6)R_{LC}$ the radius of curvature of the $B$ lines. Since for all the implemented emission models $E_{\parallel} \propto w^2 B_{LC}$, we have

$$E_{cut} \propto [w^2 B_{LC}]^{3/4} \rho^{1/2} \tag{6.14}$$

where $w$ is the width of the emission gap. From equation 1.26, the light cylinder magnetic field dependency, can be written as

$$B_{LC} = B_S \left( \frac{\Omega R}{c} \right)^3 \propto B_S P^{-3} \qquad (6.15)$$

where $B_S$ is the surface magnetic field, and $P$ the spin period. Since, for all the implemented emission models $\rho \propto R_{LC}^{1/2} \propto P^{1/2}$, the $E_{cut}$ proportionality can be expressed as

$$E_{cut} \propto w^{3/2}[B_S^{-3/8}P]^{-8/4}. \qquad (6.16)$$

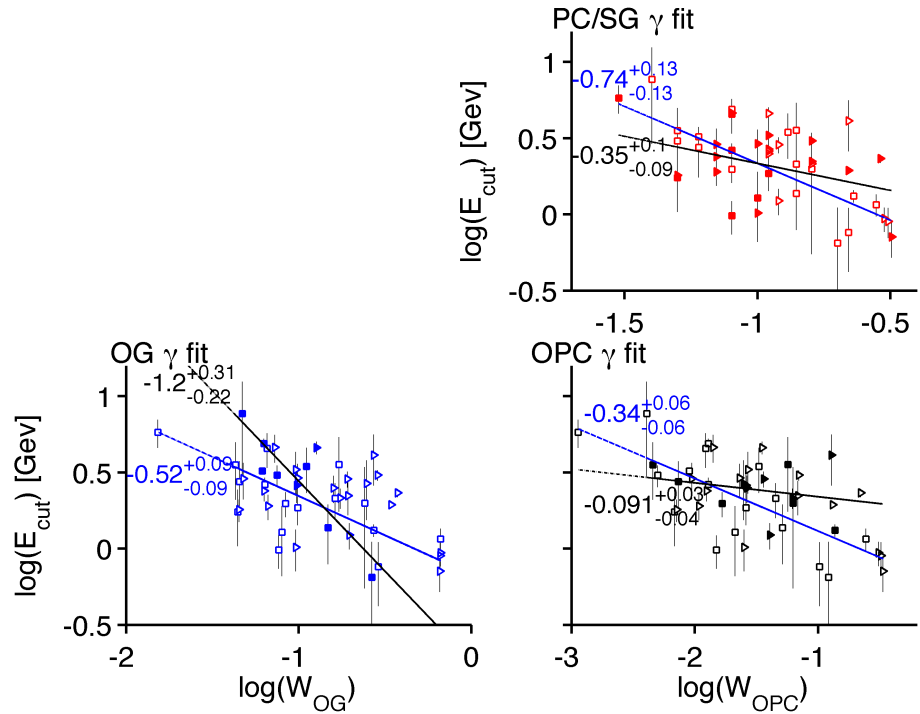

Figure 6.88: Energy cutoff versus gap width for the $\gamma$ fit solutions for all the models. Triangles and squares respectively refer to radio-quiet and radio-loud solutions. Filled markers note the highest likelihood case between the different models. The whole and best fit data set are respectively indicated as blue and black lines.

Now, since the slot gap width dependency follows approximately

$$w \propto PB_S^{-3/7}, \quad B_S > 0.1 \times 10^{12} \ Gauss \qquad (6.17)$$
$$w \propto PB_S^{-4/7}, \quad B_S < 0.1 \times 10^{12} \ Gauss \qquad (6.18)$$

the final $E_{cut}$ dependency can be written as

$$E_{cut} \propto w^{3/2}w^{-8/4} = w^{-0.5}. \qquad (6.19)$$

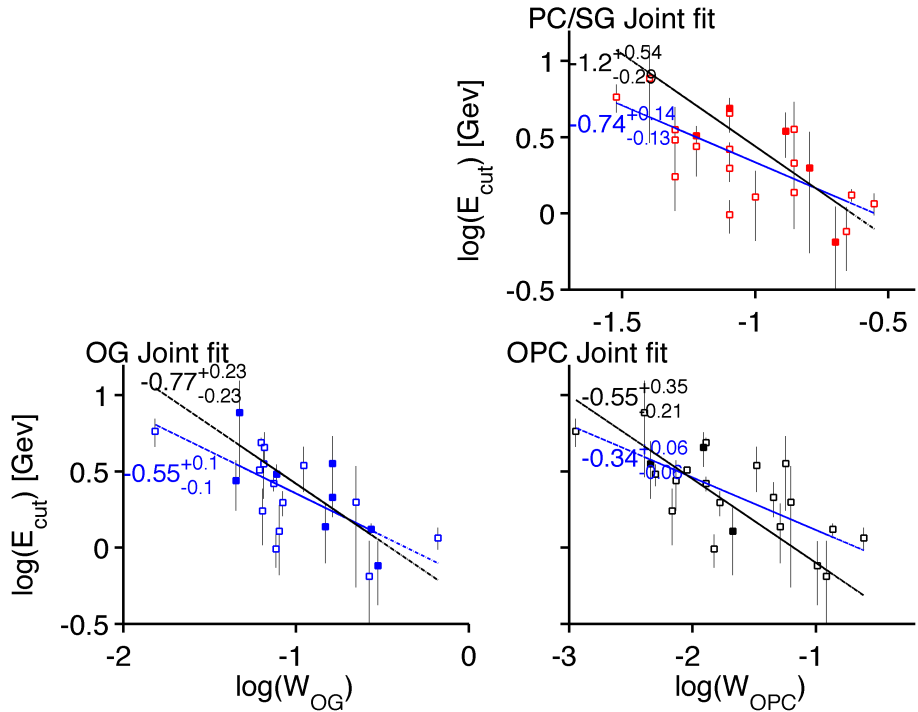

Figure 6.89: Energetic cutoff versus gap width for the joint best fit solutions of all the implemented models. Filled markers note the highest likelihood case between the different models. The whole and best fit data set are respectively indicated as blue and black lines.

In figure 6.88 and 6.89 power-law fits to the data points are given for the whole set and just for the best fit model solutions (respectively fitted in blue and black).

In figure 6.90 is shown the behaviour of $E_{cut}$ with respect to the gap widths $\Delta\xi_{PC,SG}$ & $w_{OG,OPC}$ for the population synthesis results for each models. The fact that no trend is present is due to the definition of the spectral characteristics that have been assigned to each simulated pulsar (section 5.4.1). $E_{cut}$ and the spectral index have been randomly assigned by choosing the double gaussian distribution that better describes the observed values in the LAT sample. By fitting the observed $\gamma$-ray light-curves with different emission patterns, it is possible to highlight relations not predicted by the population modelling. The fact that the results in figure 6.88 and 6.89 show a trend that can be predicted theoretically supports the reliability of the implemented fitting strategy. Moreover, since in the geometric phase-plot modelling there is no $E_{cut} - w_{gap}$ relation, our results imply a real physical relation between the cutoff energy and gap width that can be used to discriminate between the proposed models. A more precise $E_{cut} - w_{gap}$ relation drawn from the analysis of a large LAT sample (soon to be available) should be implemented in the future population synthesis studies.

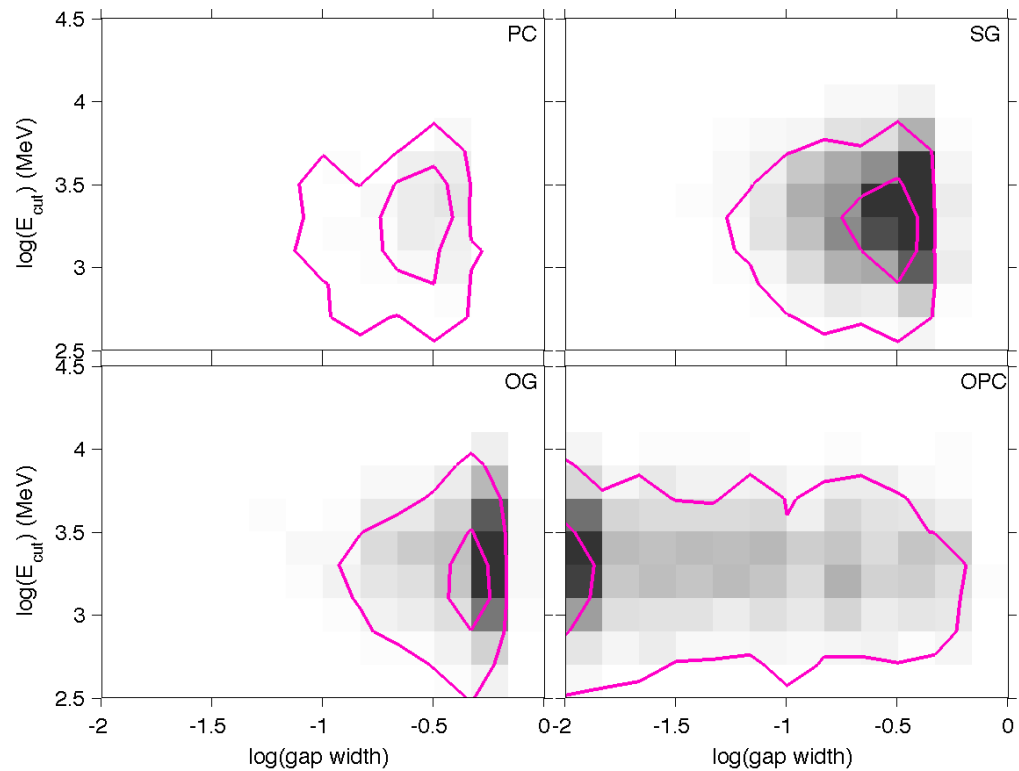

Figure 6.90: Number density of the visible gamma-ray pulsars obtained for each model as a function of gap width and cut-off energy. The linear grey scale saturates at 8 star/bin. The pink contours outline the density obtained for the radio-loud gamma-ray sub-sample (at 5% and 50% of the maximum density).

# Chapter 7

# Light curve structure analysis and shape classification

This chapter is dedicated to the description and discussion of the last part of my PhD project: the shape analysis and classification of the simulated pulsar light curves for the implemented $\gamma$-ray emission models. In the last section of the chapter will also be discussed a comparison with the shapes and morphological characteristics of the observed LAT pulsar light curves.

The light curve structure analysis and shape classification described in this chapter is at the basis of the paper in preparation: *'Gamma-ray pulsars light curve analysis: comparing Fermi data with different emission regions in the magnetosphere*, by Marco Pierbattista, Isabelle Grenier, Alice Harding, Peter Gonthier.

## 7.1 Classification criteria and method

### 7.1.1 Model light curves classification

The model light curve classification that I have implemented is based on 3 curve characteristics: the number of emission bands, the number of curve maxima, and the number of curve minima.

Since the implemented emission patterns (phase-plots, section 5.2) do not include any background emission, in all the implemented models except for the SG, the pulsar emission is concentrated in phase regions here defined as emission bands (figure 7.1). For the SG case, the light curves are characterised by emission at all phases. For the SG analysis, the curve classification is based just on the curve maxima and minima. Figures 7.1, 7.2, and 7.3 show 3 examples of the implemented classification method. Figure 7.1 shows a 2-band, 2-maxima, and 0-minima, class 3, PC light curve; in figure 7.2 is plotted a 3-maxima and 2-minima, class 5, SG light curve; in figure 7.3 is shown a 1-band, 2-maxima, and 1-low minimum, class 5, OG/OPC light curve.

The method adopted to evaluate the maxima and minima in the curve consisted in searching for the zero derivative of the curves and analysing the derivative behaviour to distinguish between maxima and minima. Before the curve derivative evaluation it was necessary to apply a smoothing to get rid of the small curve fluctuations that would have been detected as derivative changes and so as maxima and minima. The smoothing technique I have applied consisted in a convolution of the light curve with a *Gaussian filter*. The method requires evaluating a Fast Fourier Transform (FFT) of both the light curve and gaussian profile, convolve the results, and apply an inverse FFT to obtain the smoothed light curve.

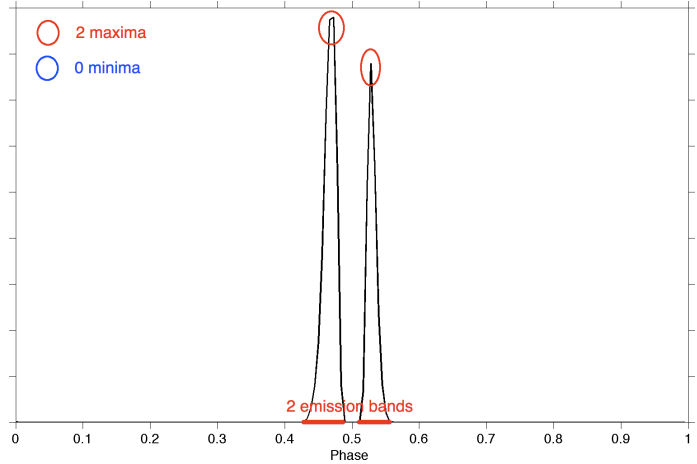

Figure 7.1: Example of a PC class 3 light curve with 2 emission bands, 2 curve maxima and 0 curve minima.

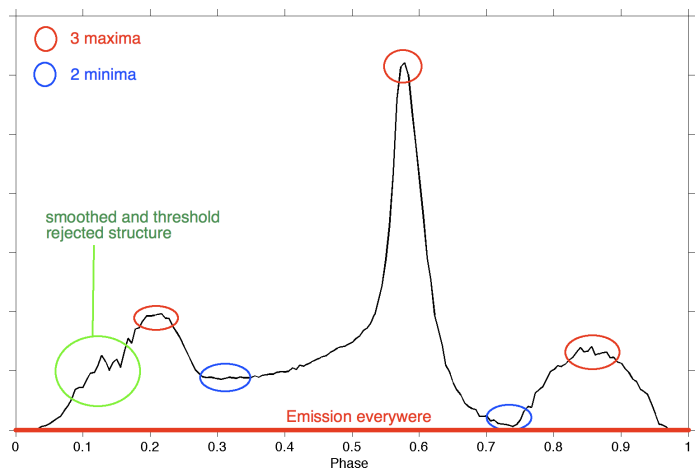

Figure 7.2: Example of a SG class 5 light curve with 1 single band, 3 maxima, and 2 minima. It is also possible to see how the smoothing routine jointly with the definition of a threshold level rejected the structure outlined in green on the left of the curve.

A Fast Fourier Transform is an algorithm based on a Discrete Fourier Transform, DFT, that allows to reduce the computation step numbers from $2N^2$ to $2N \log_2 N$. A DFT and its inverse are defined as

$$F_k = \sum_{j=1}^{N} f_j \omega_N^{(j-1)(k-1)} \tag{7.1}$$

$$f_j = \frac{1}{N} \sum_{k=1}^{N} F_k \omega_N^{-(j-1)(k-1)} \tag{7.2}$$

$$\omega_N = e^{2\pi i/N} \tag{7.3}$$

To apply an FFT to a 2-D signal implies to write it as the sum of harmonic components, with each harmonic characterised by its own frequency. On the other hand, since a Gaussian function is a smooth distribution with $FWHM = f(\sigma)$, its FFT is characterised by one harmonic with frequency $\nu = f(\sigma)$. So, by convolving the light-curve and Gaussian FFT, all the oscillation of frequency $\nu$ will be cancelled from the FFT of the original curve. By applying an inverse FFT to the convolved distribution, one obtains the original curve smoothed for the oscillation of frequency $\nu$. The strength of the smoothing depends on the chosen variance $\sigma$ of the Gaussian. By choosing a small $\sigma$ value, just the high frequency oscillations will be smoothed while by choosing large $\sigma$ values low frequency oscillations will be smoothed.

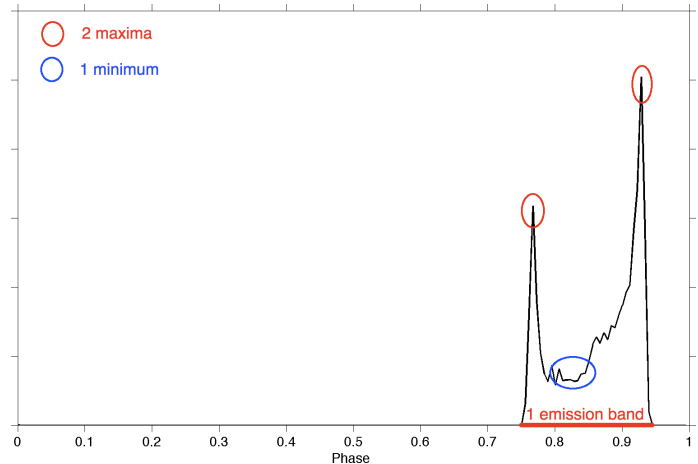

Figure 7.3: Example of an OG/OPC class 5 light curve with 1 emission band, 2 curve maxima and 1 curve minimum.

Since each model is characterised by different irregularities and fluctuations, I had to define an optimised $\sigma$ value to smooth the light-curves of each model. Often the smoothing was not sufficient to get rid of the small irregularities and a strong smoothing changed too much the whole curve structure, erasing also the real peaks.

The solution was to apply a second selection criterion after the curve smoothing to establish if the detected structure was as a peak. An oscillation threshold has been defined for each model and applied to each detected maximum: if the relative height was higher than the threshold it was classified as a maximum, otherwise it was rejected. An example of the rejection of small-scale structures by the joint action of the Gaussian smoothing and thresholding is indicated, in green, in figure 7.2.

In the PC and OG/OPC cases, there are particular geometric configurations for which the pulsar faintly radiates at all phases. In these cases, to be able to define the *number of emission bands*, another threshold criterion has been defined. For each model, all the light curve values lower than a certain percentage of the curve maximum have been set to zero. In this way it was possible to define emission bands also in the few cases for which the faint emission was spread across all phases. The same method could not be applied to the SG case because the all-phase emission is not a negligible component of the total light curve flux.

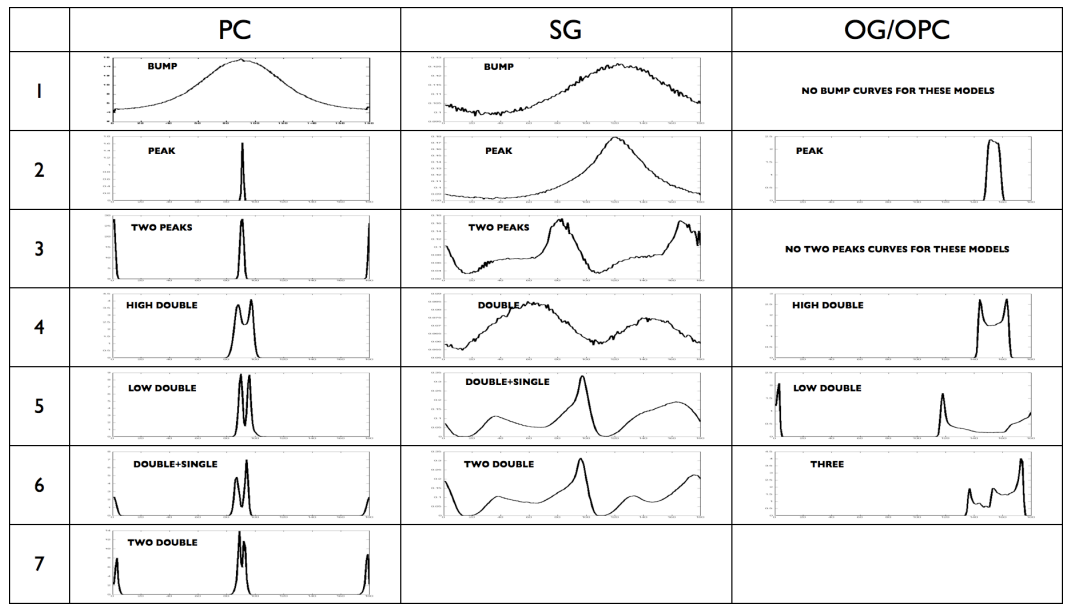

Figure 7.4: Light curve shape classification for each γ-ray emission pattern model.

To make the classification more accurate, after the detection of the number of emission bands, the number of maxima, and the number of minima, other classification criteria have been introduced. They concern a more accurate description of the first four classes of each model classification (figure 7.4). A first, we estimated the FWHM of the peak: if it were more than 33% of the whole phase, the curve was classified as a bump, otherwise as a narrow peak. In the double peak classification we estimated the inter-peak bridge height: if it were lower than half the absolute curve maximum the peak was classified as low double, otherwise as a high double peak. As will be described in section 7.2, these criteria have been used to fit each observed peak profile and evaluate its intrinsic characteristics.

The application of the classification method to the pulsar light curves for each $\gamma$-ray model led to the classification illustrated in figure 7.4.

### 7.1.2   FERMI LAT observed light curves: classification

To be able to compare each model shape classification with the observed population, the same classification criteria have been applied to the LAT curves obtained as described in section 4.2.2, Figures 4.2 to 4.9.

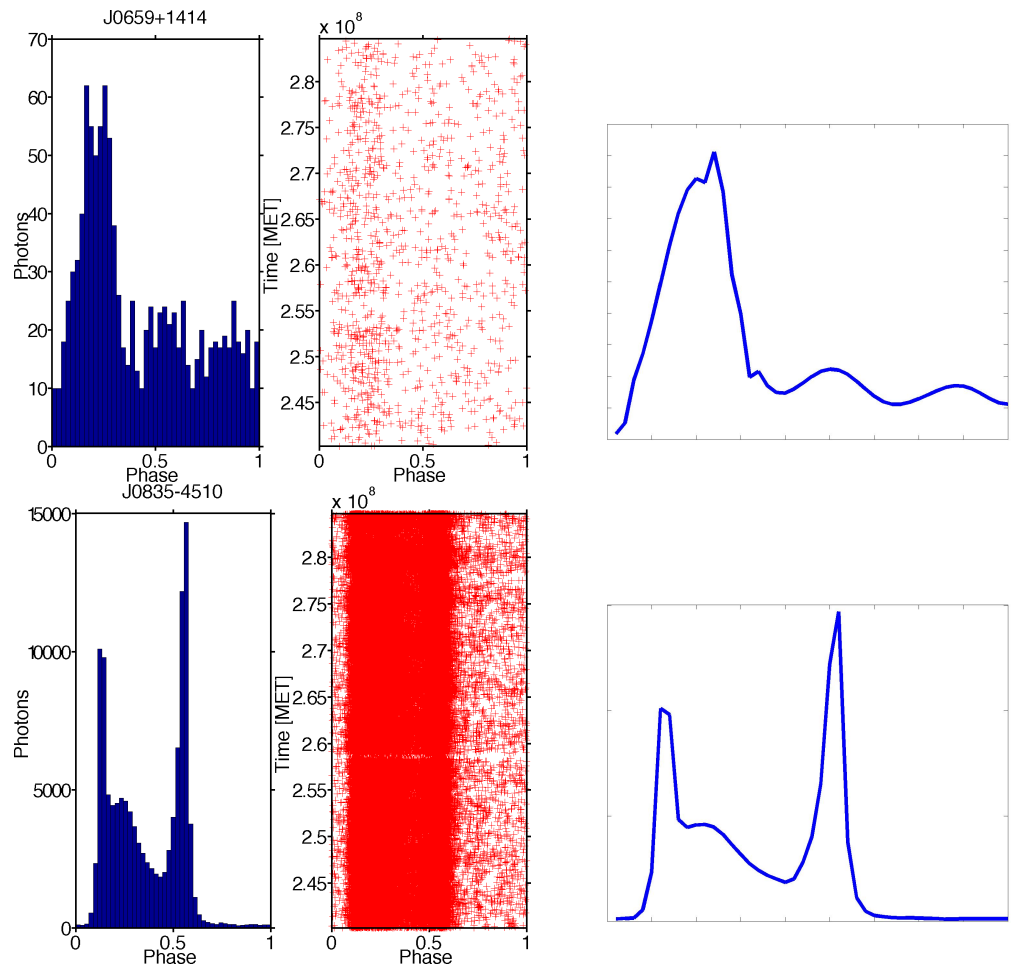

Figure 7.5: Application of a wavelet gaussian de-noising method for two pulsar light curves, with low and high count statistics.

Since the observed light curves are characterised by high noise and a significant background level, and often by low count statistics, before running the classification routine, the LAT profiles have been de-noised by applying wavelet transforms plus a gaussian de-noising method. It has been assumed that the real curve noise is described by Poisson statistics.

To apply a wavelet transform to a light curve means to decompose it in different components characterised by different resolution scales. For each resolution component, the standard deviation $\sigma$ corresponding to each bin of the curve is evaluated. The next step consisted in applying a threshold: for all the resolution components, all the curve values lower than the corresponding $\sigma$ were put to 0. The final step consisted in applying an inverse wavelet transform to recompose the original, de-noised curve. The results of the applied de-noising method is illustrated, for a low and high count statistic pulsar, in figure 7.5. Figure 7.5 shows how, for the noisy low statistic J0659+1414 light curves, the de-noising is efficient both on the pulsed profile and in the background. In the high statistics Vela pulsed profile, the method hardly smooth the light-curve.

The classification method described in the previous section has been applied to the de-noised LAT profiles. Since the PC classification is the one that contains the bigger sample of possible shapes, it has been used to classify the LAT profiles. In the model-LAT comparison discussed in the results section, it has to be taken into account that the LAT classes do not correspond to standard templates as shown in figure 7.4, but they just represent peak multiplicity indices.

## 7.2 Fitting the light curve structures

When classifying the light curve shapes as described in the previous section, we recorded the position of each peak. This gave me the possibility to fit the exact peak phase, the peak FWHM, the peak intensity using a Gaussian or Lorentzian for the peak, and to evaluate the phase separation between double and multiple peaks.

**Polar cap profile**

Since the PC light curve profile is better described by a normal distribution, the peak fit has been implemented by using a Gaussian function with 4 free parameters

$$G(\phi) = C + Ie^{-\frac{(\phi-m)^2}{2s^2}} \tag{7.4}$$

where $I$ defines the peak intensity, $m$ defines the peak position, and $s$ the $RMS$ width.

From the single peak curve (PC classes1 & 2) to the 4 peak one (PC class 7), the PC profiles have been respectively fitted with 1, 2, 3, and 4 Gaussian. For each fit, the best fit gaussian peak position and the FWHM have been stored.

**Slot gap profile**

In the SG case sharp peaks are generally better described by a Lorentz distribution and broad peaks by using a Normal distribution. For the single peak SG classes (classes 1 & 2 of figure 7.4) a double fit strategy has been adopted. The class 2 bump structure has been fitted by using the Gaussian distribution described in equation 7.4. The class 1 SG narrow peak, and all the other complex structures, have been fitted by using the 4 free parameter Lorentz distribution

$$L(\phi) = C + I \frac{s^2}{(\phi - m)^2 + s^2} \tag{7.5}$$

where $I$ defines the peak intensity (combined with $C$), $m$ defines the peak position, and $s$ the FWHM.

As in the PC fit, a single or multiple Lorentzian fit has been applied to the SG light curves to evaluate position, FWHM, and separation of the fitted structures.

**Outer gap & one pole caustic profile**

The OG/OPC light curves are characterised by very thin peaks that are well fitted by a Gaussian distribution. To fit the OG/OPC structures, I have used the Gauss distribution described in equation 7.4, to implement a single or multi-gaussian fit, with respect to the light curve class.

For each implemented fit on each light curve, the values of the peak position and FWHM, and peaks separation have ben stored.

**The LAT pulsar profiles**

The very same characteristics, like peak position, FWHM, peak intensity, and peak separation, have been evaluated also for the de-noised LAT light curves. The observed profiles have been fitted by using the Gauss distribution (Equation 7.4). Since the LAT profiles classification has been made by using the PC's shape classes, all the profiles have been fitted with the same number of gaussians as described in the previous PC section.

## 7.3   Light curves morphological characteristics

The morphological characteristics study has been implemented by evaluating a set of morphological parameters for each light curve of each $\gamma$-ray emission model. The intent was to study all the light curves generated by the same emission mechanism as value distributions, characterised by the same geometrical properties and so grouped in different populations.

W evaluated other morphological parameters such as: Kurtosis and Skewness. The kurtosis is a measure of how flat the top of a symmetric distribution is when compared to a normal distribution of the same variance. Usually a more flat-topped distribution is called *platykurtic*, a less flat-topped distribution, *leptokurtic*, and an equally flat-topped distribution *mesokurtic*. Kurtosis is actually more influenced by the tails of the distribution than its the centre. The mathematical expression I used to evaluate the kurtosis is

$$k = \frac{E(x - \mu)^4}{\sigma^4} \tag{7.6}$$

where $x$ define the values distribution, $\mu$ is the mean of $x$, $\sigma$ is the standard deviation of $x$, and $E(x - \mu)$ is the expected value of $(x - \mu)$.

Skewness refers to the asymmetry of a distribution. A distribution with an asymmetric tail extending out to the right is referred to as positively skewed or skewed to the right, while a distribution with an asymmetric tail extending out to the left is referred to as negatively skewed or skewed to the left. Skewness can range from minus infinity to positive infinity. The mathematical expression I used to evaluate the skewness is

$$s = \frac{E(x - \mu)^3}{\sigma^3} \tag{7.7}$$

with the same notations.

The skewness and kurtosis computation has been made for both the simulated and LAT light curves. In the simulated case, since many curves are fall to zero over large phase intervals, we added a flat background equal to 20% of the curve maximum value. In the LAT profiles cases, the computation has been done directly on the de-noised profiles obtained with the procedure described in section 7.1.2. Since the skewness and kurtosis are parameters that estimate the curve symmetry and sharpness with respect to the centre of a distribution, the barycenter of both the simulated and observed profiles, evaluated as

$$\langle \phi \rangle = \frac{\int_0^{2\pi} \phi \, ltc(\phi) d\phi}{\int_0^{2\pi} ltc(\phi) d\phi}, \tag{7.8}$$

has been shifted to the central bin of the light curve. This shift is equivalent to evaluating the symmetry and sharpness with respect to the same criterion of equal light curve flux to the left and to the right part of the central bin.

## 7.4   Results

In this section I will show the results obtained by implementing the light curve classification and morphological analysis described in the previous sections. Each model result will also be compared with the same analysis performed for the LAT pulsars to try to discriminate which assumed emission pattern better describes the observed population.

### 7.4.1  Light curves shape classification: comparison between the implemented population and the LAT observations.

The analysed curve characteristics are: *shape classification*, *single and double peak FWHM*, *peak separation*, and *radio lag*. To study the peak FWHM, separation and radio lags, we have used only single peak, two peaks, and double peak classes for the LAT and simulated light-curves and compared with the equivalent model classes.

**Shape classes**

In figure 7.6 is shown the recurrence of the shape classes defined in figure 7.4, for the whole simulated sample and for the γ-ray visible component of each implemented emission model. The biggest discrepancy between the whole population and its γ visible component is observed in the PC case. Here, sharp peak class is clearly the most recurrent in the whole parent population, in contrast with the γ-visible profiles for which the narrow peak is the less recurrent shape. The high peak multiplicity profiles (classes 6 & 7) represent a negligible component of the whole PC population and its γ-visible component. In all the other cases, the classes recurrence in the whole population and its γ-visible component, are quite comparable. This is an important result to emphasise since it suggests that there is no evident visibility selection connected to the light curve shape, e.g. the pulsars are visible or not independently of the shape of their light curves.

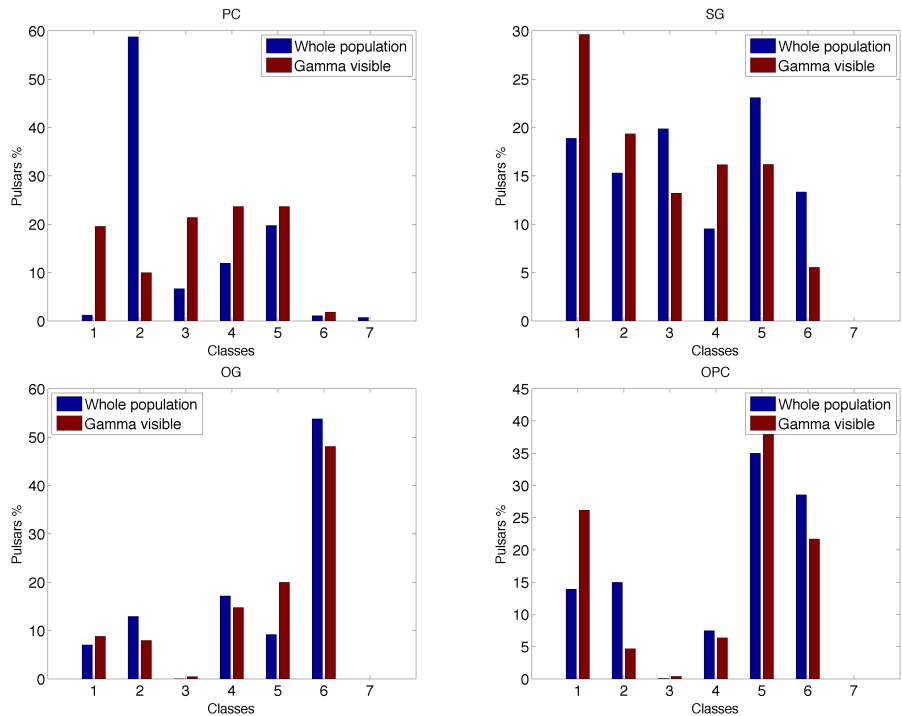

Figure 7.6: Recurrence of the shape classes defined in figure 7.4 for the whole and visible γ sample of each implemented emission model.

In the SG models, the whole population classification is the most heterogeneous. All the possible shapes and multiplicities, have equivalent detection probability. It has to be taken into account that the SG bump excess is probably biased by the necessity to strongly smooth the curves to be able to classify their often complex shapes. Because of this, most of the SG bumps could have been classified as double component or two peaks, with the second peak or component very low and so erased by the smoothing.

The OG and OPC models show the highest agreement between the recurrence in the whole population and the γ-visible sample. From a visibility point of view, the outer magnetosphere emission models are the less affected by visibility selection due to the shape of the light curve.

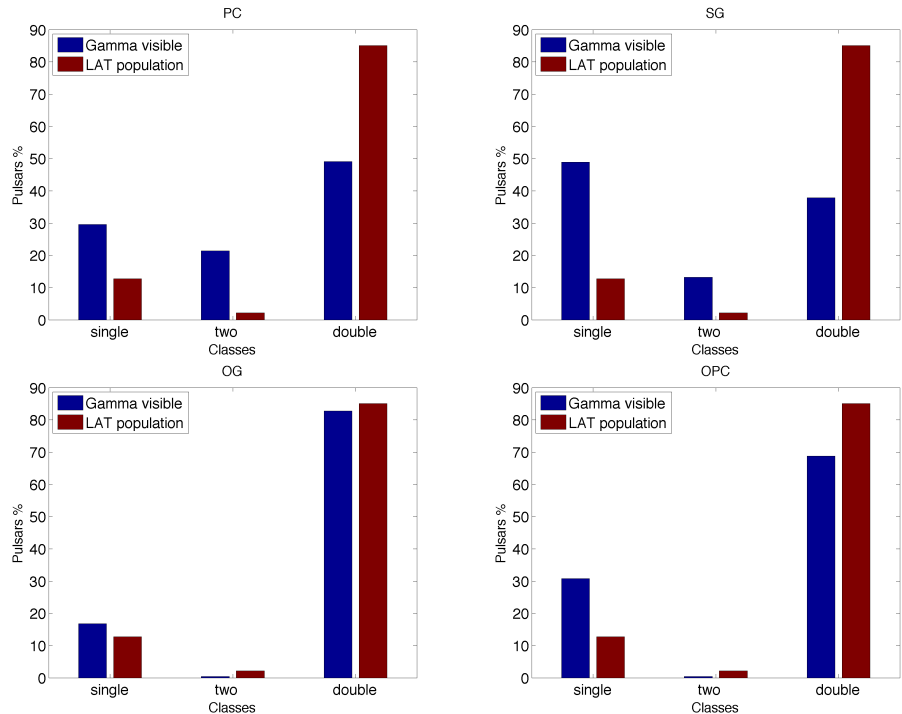

Figure 7.7: Recurrence of the single-peak, two-peak, and double-peak light curves, for the visible γ sample of each emission model and for the LAT pulsar sample.

Concerning the LAT pulsars classification, the peak multiplicity of the LAT profiles obtained by this analysis is never bigger than 2. In figure 7.7 is shown the comparison between each model γ-visible component and the LAT observation. The single peak class has been obtained by merging the bump and narrow peaks, the two peak one represents the two separate peak light curves, and the double class has been obtained by merging the low and high double structure classes (where they exist). The LAT sample classification shows a clear excess of double component structure compared to the two and single peak ones. The single narrow peak structure is completely absent in the LAT data. Single peak emission is observed just in large bump structures. The model that best reproduces the LAT shape recurrence is the OG. It is the only model in which the contribution of the high peak multiplicity classes to both the whole and $\gamma$-visible population is consistent with the LAT observations. It is also able to nicely reproduce the same number of observed bump structures. Even though the OG and OPC share the same emission pattern phase-plot and classification, they show interesting differences in the class recurrence. The assumption of a gap width that does not dependent on the pulsar orientation in the OPC model does not match the LAT distribution as well as the OG model.

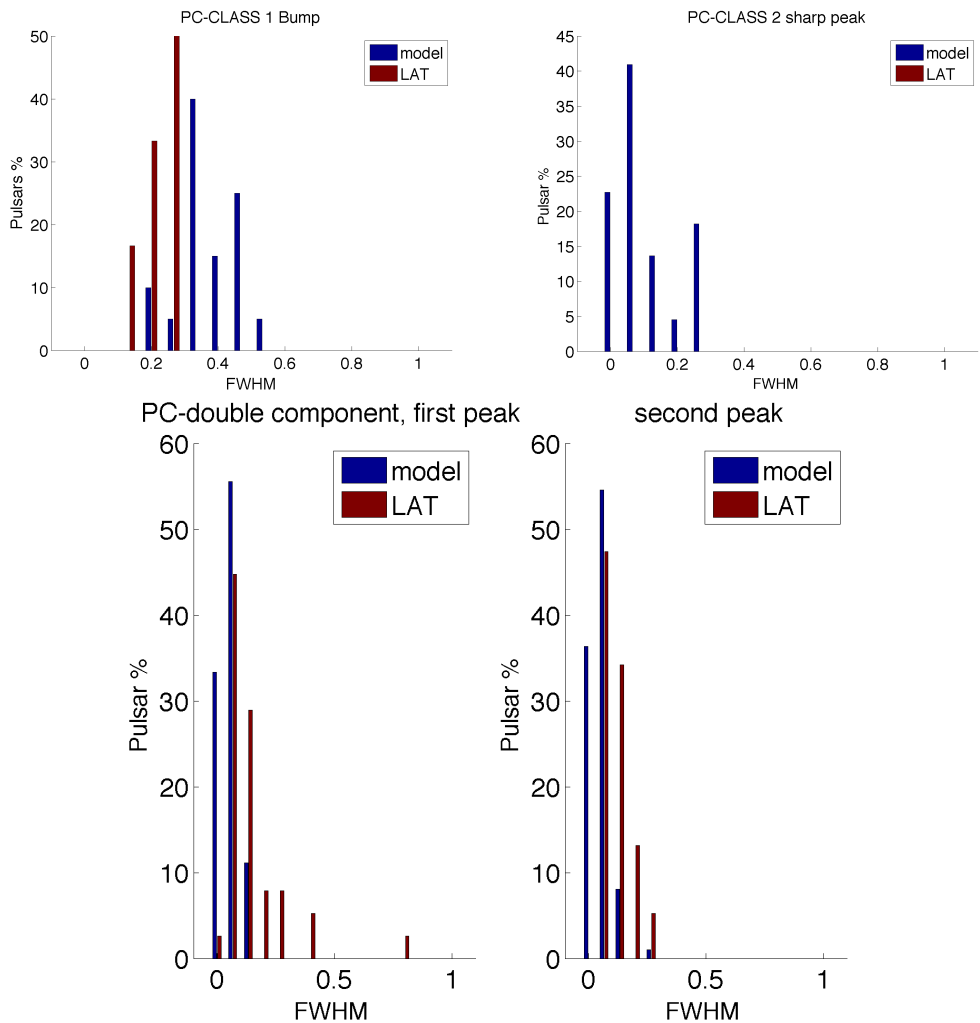

Figure 7.8: Peak FWHM in the light-curves of the PC $\gamma$-visible and LAT population, for the single peak classes and double peak ones.

**Single peak & double component FWHM**

Here I will compare the FWHMs of the peaks found in the LAT and model light-curves for single peak and double peak structures.

In figure 7.8 is shown the comparison between the peak FWHM of the PC $\gamma$-visible sample structures and the LAT ones. Even though the PC model has a narrow emission beam, it is not able to reproduce the narrower LAT bump light curves. On the other hand the LAT double structures show very sharp peaks that are partially reproduced by the narrow rim of the PC beam. The inconsistency in the prediction of double narrow structures plus the too narrow single bump width suggests that the PC emission pattern is not able to describe the LAT much wider structures.

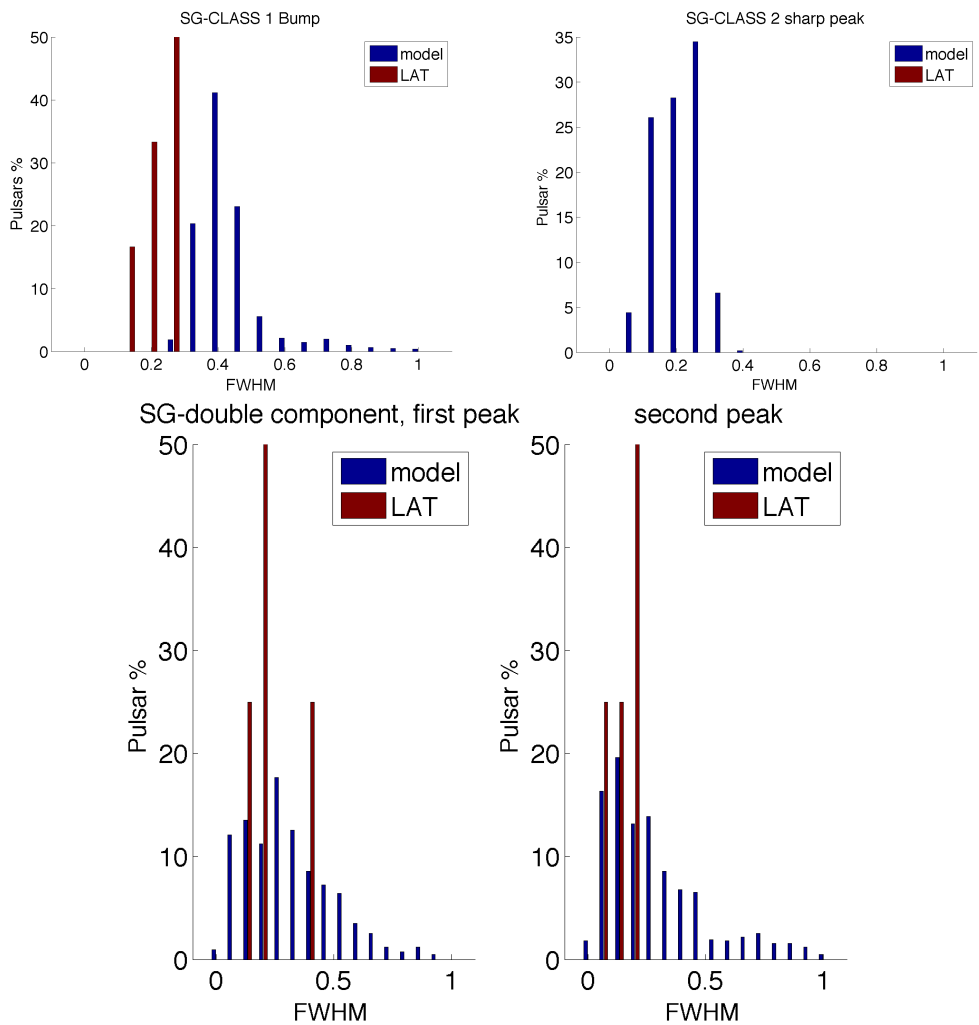

Figure 7.9: Peak FWHM in the light-curves of the SG $\gamma$-visible and LAT population, for the single peak classes and double peak ones.

The SG case, shown in figure 7.9, is analogous to the PC one but in this case the width of the double structure peaks are better described. The same lack of narrow bump is observed in the SG $\gamma$-visible pulsars. Yet the SG single bump is never described by a Gaussian and shows an large number of small structures that are usually erased by the smoothing classification method (section 7.1.1). This implies that the FWHM of the SG bump is often overestimated ( 10% in phase). By taking this overestimate into account, the SG emission model reasonably describes the observed width of the light curves peaks for both the bump and double structures.

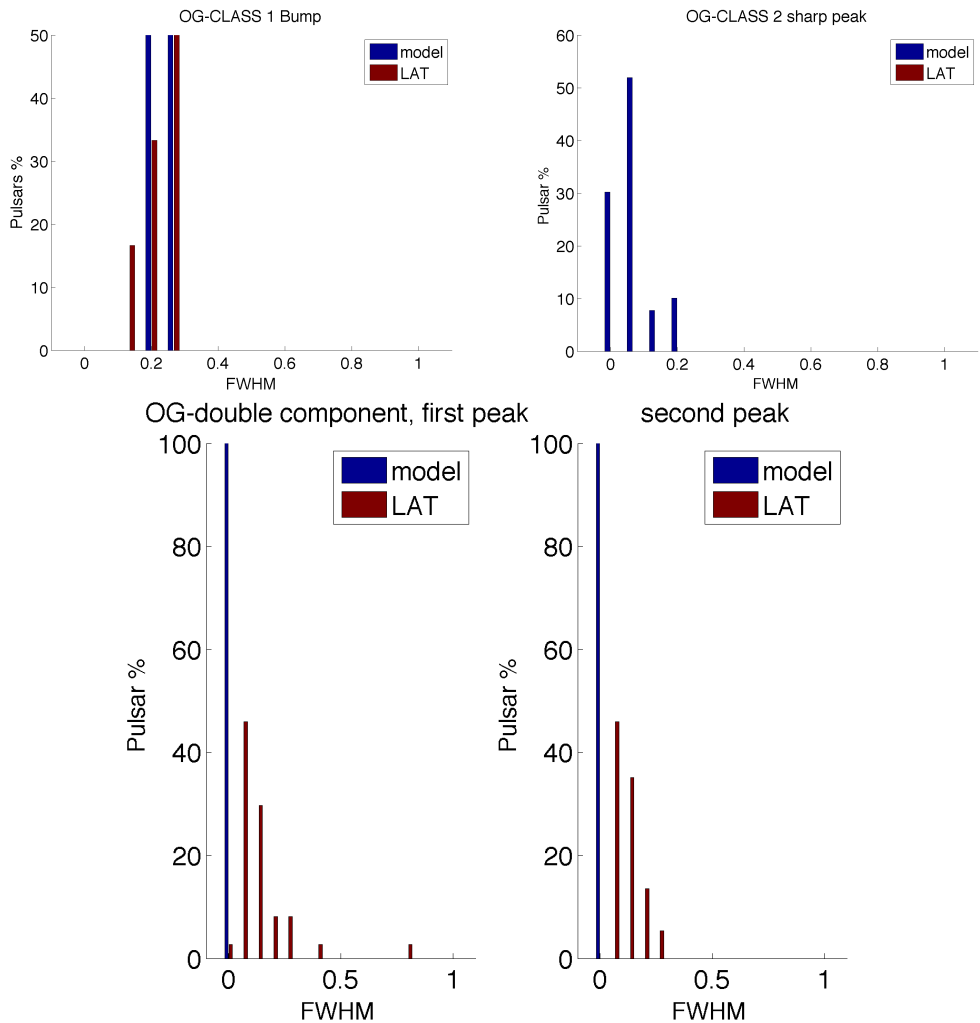

Figure 7.10: Peak FWHM in the light-curves of the PC $\gamma$-visible and LAT population, for the single peak classes and double peak ones.

In figure 7.10 is shown the OG model results. The OG phase-plot is characterised by a large set of wide structures. Even though with a low statistic, the LAT bumps are well described by the $\gamma$-visible component of the OG model but the width of the double component peaks is clearly too small compared to the LAT one. The phase-plot OG double structure is characterised by very narrow peaks and this is probably due to the assumption of an infinitely thin emission layer along the OG gap.

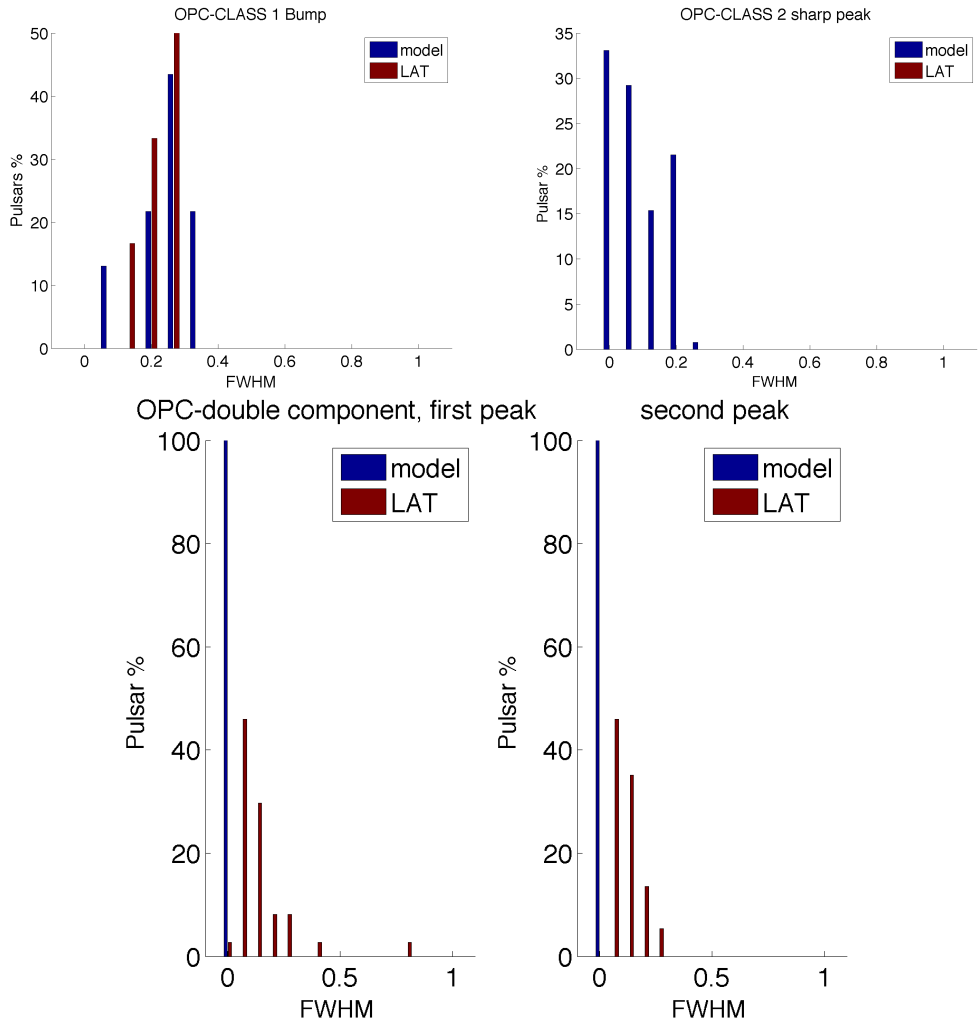

Figure 7.11: Peak FWHM in the light-curves of the PC $\gamma$-visible and LAT population, for the single peak classes and double peak ones.

The OPC case is shown in figure 7.11. As for the OG model the predicted components of the double structure are too thin to represent the LAT data (same assumption of an infinitely thin emission layer along the gap) and the OPC bump structures can reproduce the LAT with an higher statistic.

The light curves analyses in the literature studied the double peak structure in term of separation between the peaks. By implementing a FWHM study it is possible to interpret the observed emission patterns from a more complete point of view, by analysing a characteristic connected with a light curve shape generated by a different pulsar orientation.

**Peak separation**

One important observable characteristic is the distance between the double structures. To be able to define a distance between the peaks it was necessary to define a peak position criterium for all the shape classes of each emission model.

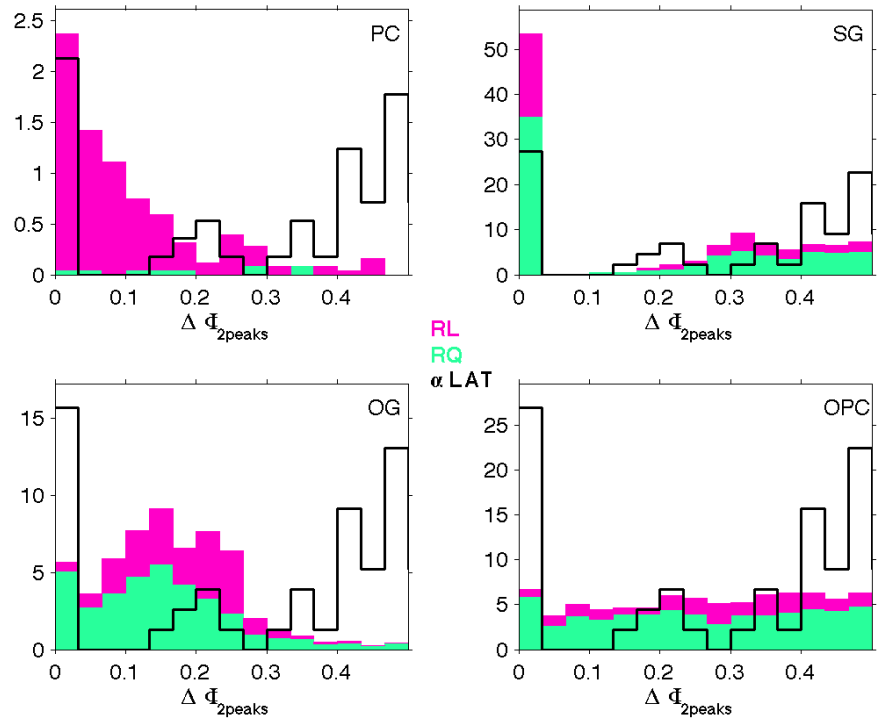

Figure 7.12: Distribution of the peak separation for the $\gamma$-visible component of the simulated population of each emission model. The radio quiet and loud objects are respectively indicated in green and pink, while the LAT observations are plotted as a black contour and scaled to the total number of visible simulated objects for comparison purposes.

- **PC case**: for the double component structure light curves, the definition of peak separation is directly obtained by the peak positions. For the double plus single peak structure, the distance has been evaluated from the barycenter of the double structure to the peak position, and for the two double peak profile, between the barycenters of the two double structures.

- **SG case**: for the double component structure light curves, the definition of peak separation is directly obtained by the peak positions. For the profiles characterised by higher multiplicity the distance between the two
higher peaks has been taken into account.

- **OG/OPC case**: for the double component structure light curves, the
  definition of peak separation is directly obtained by the peak positions.
  For the triple peak structure, the distance between the outer peaks has
  been taken into account.

The peak separation is directly connected with the structure of the
emission region and directly connected to the changes in the magnetosphere
structure. Figure 7.12 shows the peak separation histograms of the $\gamma$-visible
component of each emission model, compared with the LAT sample one. The
LAT observations show two distinct trends: the radio-loud objects show a
decreasing peak separation with decreasing $\dot{E}$ e.g. the distance between the
peaks shrinks with age while the peak separations in the radio-quiet population
shows no age dependency and a majority of large separation values. The first
decreasing trend is predicted by the PC, OG, and OPC model but not by the
SG model. All the models fail to reproduce the dominant number of widely
spaced peaks in the LAT data.

All the emission pattern are based on the same two assumptions: an
emission direction parallel to the magnetic field line and a dipole magnetic field
structure. Since the PC, SG, and OG/OPC emission patterns cover emission
from all the possible regions of the open magnetosphere, the discrepancy
between the observed and simulated data in peak separation may suggest that
the dipole geometry is not quite adequate. As it is also shown in figure 7.13,
only the 2-pole geometry of the SG is consistent with the observed lack of
close peaks ($\Delta\phi < 0.2$). The other models exhibit a continuous distribution
in the $\Delta\phi < 0.2$ region, notably so for the old pulsars from the outer gaps,
because of the shrinking of the polar cap with age. This is quite visible when
comparing the OG and OPC. The large fraction of older, wider gaps from OG
pile up below 0.2. The uniform fraction of OPC objects of all $\dot{E}$ values spreads
uniformly at all separations.

Concerning LAT data, the spread in $\dot{E}$ seen in figure 7.13 at large
separations ($\Delta\phi > 0.2$), is the same as for $\Delta\phi \sim 0.2$. Because of the
polar cap shrinking with age, the "one-pole" models (PC, OG, OPC) show
a concentration of objects along the diagonal that is not very consistent with
the apparent lack of evolution in the LAT peak separations. The lack of
widely spaced peaks observed for all the implemented models is probably
due to the lack of high $\dot{E}$ objects discussed in section 5.7.1. Young objects
would have systematically large peak separations. Concerning the $\gamma$ & radio
emission, except in the PC case, the radio-loud and radio-quiet objects are
rather uniformly distributed across the plot. There is no trend to gain or
loose one peak when one intercepts the radio beam or not. That can be seen in figure 7.12, where the ratios of pink to green does not vary much across $\Delta\phi$. Another important difference between the low and outer magnetosphere emission models could be done by looking at the 0 peak separation trend within the models (single peak). The beams that reach to the inner magnetosphere (PC and SG), tend to over predict the relative number of single-peak light curves compared to the beams born outside the null surface (OG and OPC).

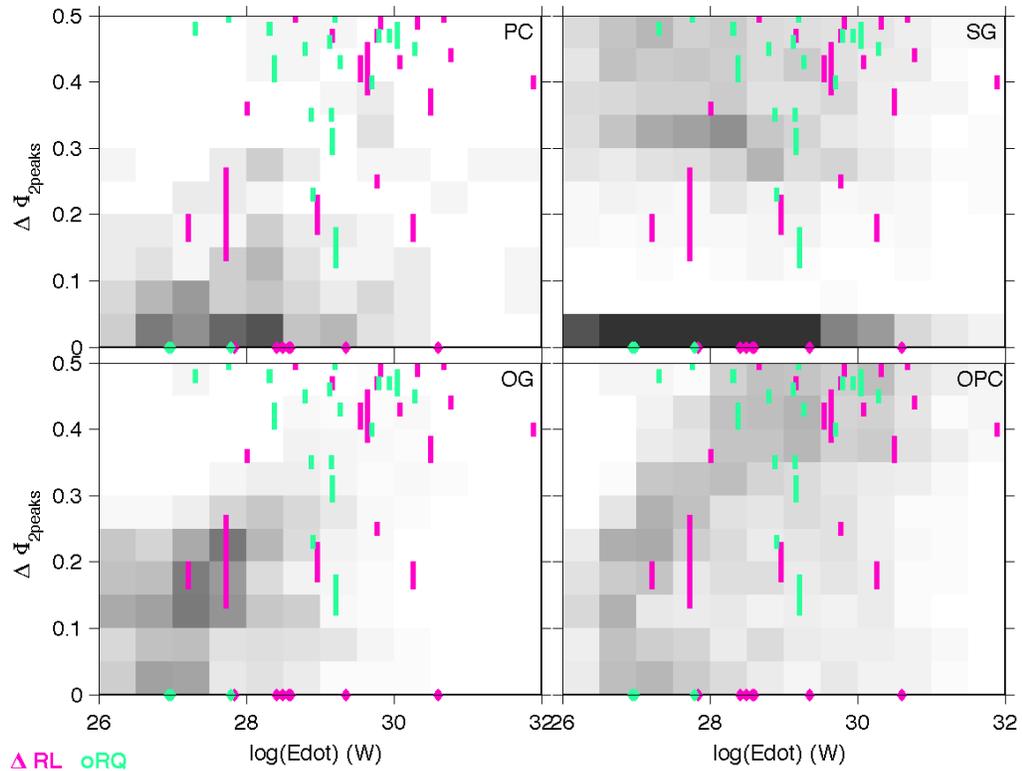

Figure 7.13: Number density of the visible gamma-ray pulsars obtained for each model as a function of peak separation and spin-down power. The PC linear gray scale saturates at 2 star/bin while the other model ones at 7 star/bin. The pink and green bars show the radio-loud and radio-quiet LAT pulsars, respectively.

**Radio lag**

The radio lag between the $\gamma$-ray and radio emission profiles from the same pulsar is a standard measure of the phase interval between the intersection of the observer line of sight $\zeta$ with the radio and $\gamma$-ray beams. In this analysis, the radio lag has been evaluated between a standard $\gamma$ position and a radio phase evaluated on the basis of the following criteria for the different classes shown in Figure 7.14.

- *double, double plus single, and triple structure, triple plus single*: the

barycenter of the double or triple structure has been taken into account.

- *Two or three separate peaks*: the peak nearest to 0 phase has been taken into account.

- *Two double or two triple structure*: the multiple structure (double or triple) barycenter closer to 0 has been taken into account.

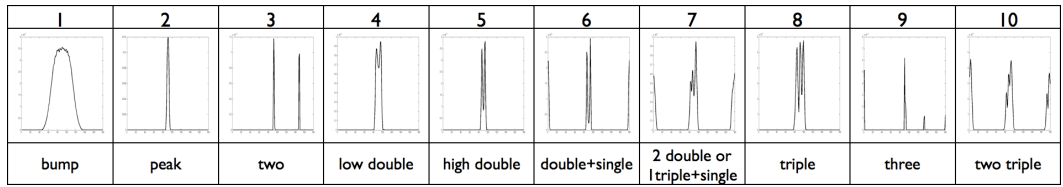

Figure 7.14: Light curve shape classification for the implemented radio core plus cone emission pattern models.

The radio lag plotted in figures 7.15, 7.16, and 7.17, has been evaluated as the radio peak phase as defined above, minus the nearest to 0 phase of the two brightest gamma-ray peaks. In figure 7.15 & 7.16 are respectively plotted the distribution of the radio lag and its evolution with $\dot{E}$ for the $\gamma$-visible component of each implemented emission model. In figure 7.16 a subset of the radio lags measured for the LAT radio-loud pulsars increase with decreasing spin-down power and so with increasing age. Figure 7.17 shows clear increase in radio lag with decreasing peak separation.

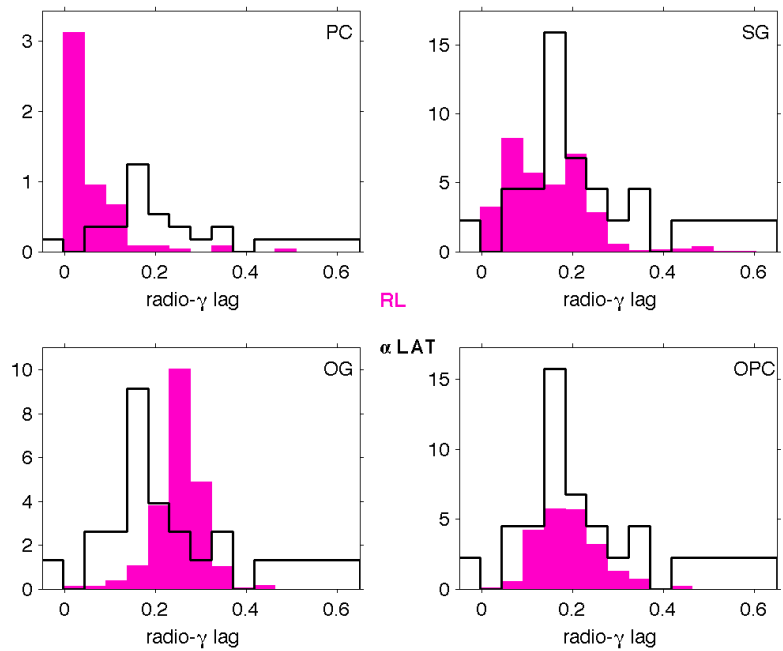

Figure 7.15: Histogram of the radio lag for the radio-loud $\gamma$-ray visible sample of each emission model.

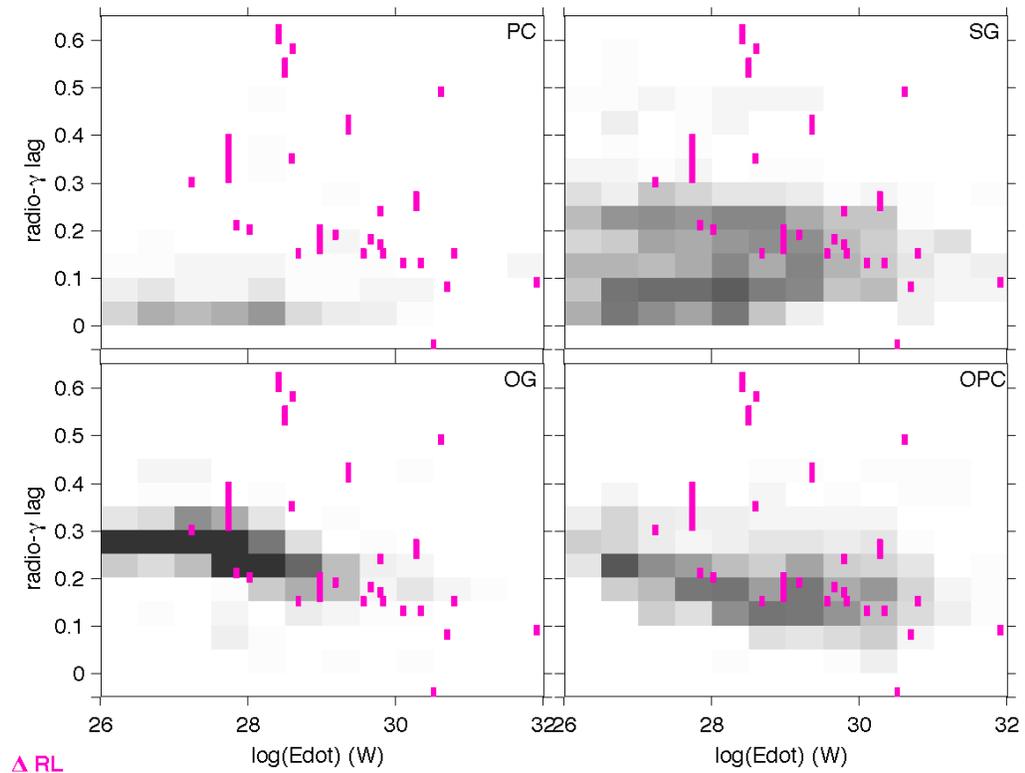

Figure 7.16: Evolution with $\dot{E}$ of the radio phase lag for the radio-loud $\gamma$-ray visible sample of each emission model. Each model prediction is plotted as a grey scale saturation levels while the LAT observation are plotted as pink bars.

The radio lag is better explained by the outer gaps than by the PC and SG that predict too many light curves with the first peak appearing very close ($< 0.1$) after the radio one. This is particularly true for old pulsars, where the brightest peak (which is also the first one and often the single one) comes from the trailing caustic, just after the radio pole. This caustic becomes dominant when the gap opens with age. The outer gaps better fit the radio lag and lack of lags $< 0.1$ because one does not see any emission from the radio pole side of the magnetosphere, but only from the other side. This radio-side trailing caustic is not visible and the first peak is predominantly $> 0.1$ in phase after the radio. So, the LAT data seems to indicate that the first peak comes from the leading side of the pole opposite to the radio pole. This statement is based on the assumption that the vacuum dipole field configuration, that has been used here for all models, applies to the real situation. An alternative to the trailing caustic on the radio pole side that takes into account the first γ-ray peak from
the leading side of the other pole, is to consider that the field configuration,
because of the large currents pervading the outer magnetosphere, becomes less
curved, as in the force-free situation (Bai & Spitkovsky, 2010) or as for a split-
monopole configuration. The OPC vacuum field predicts a convincing trend
in peak separation versus radio lag (Watters & Romani, 2010). The shrinking
with age of the peak separation is a bit too pronounced to match the LAT
data, and outer gaps clearly under predict the number of single-peak pulsars
detected by the LAT.

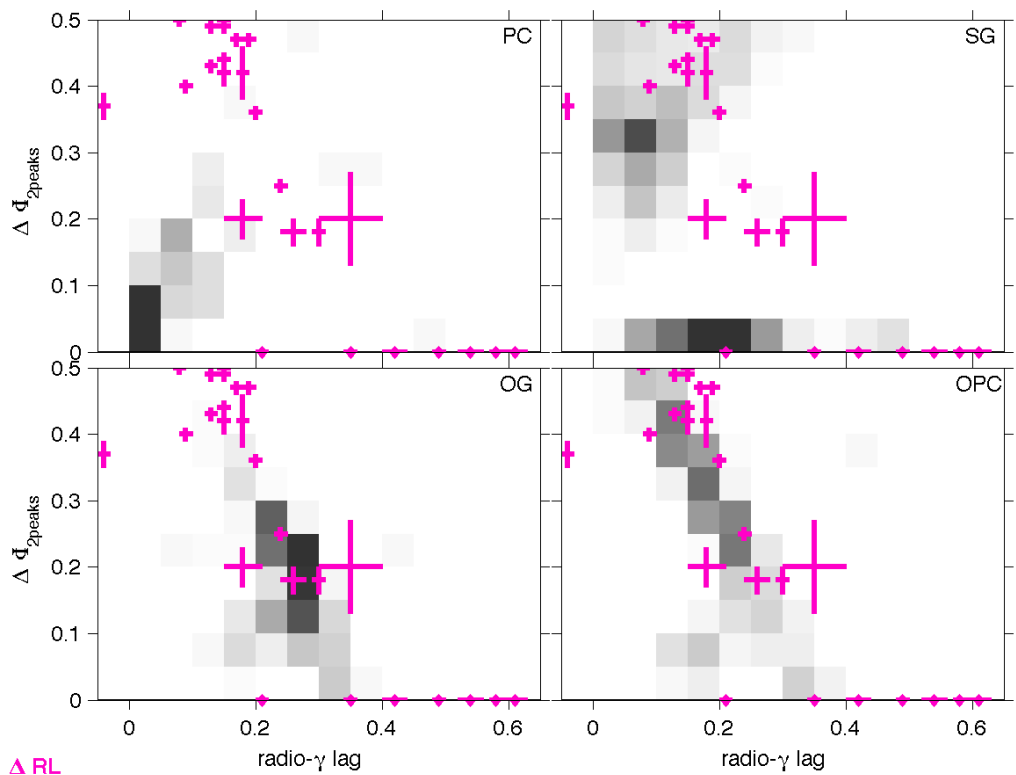

Figure 7.17: The γ-ray peak separation as a function of the radio lag. The model
predictions are plotted as a grey scale saturation levels while the LAT pulsars are
plotted as pink bars.

Figure 7.17 shows the evolution of the γ-ray peak separation with the
radio lag. As it has already been previously described the outer magnetosphere
emission models (OG/OPC) better explain the observed relation. The OPC
is the model that better predicts the observed relation, but the radio lag is
slightly over predicted at all $\Delta\Phi_{peak}$ values. The PC model shows an inverse
trend with respect to the observations. The SG model is not able to explain
the radio lag of the pulsars with small γ-ray peak separation.

### 7.4.2 Light curve morphological characteristics: comparison with the LAT pulsar profiles

In figures 7.18, 7.19, & 7.20 is shown how the light curves generated by the same emission mechanism (PC or SG or OG, or OPC), belong to distinct morphological population. Figure 7.18 shows the behaviour of the light-curve kurtosis with respect to the pulsar period. Exception made for the PC model, the inner and outer magnetosphere emission mechanisms are mostly segregated in the low and high region of the plot. This implies respectively less and more peaked structures in the emission profiles. A comparable behaviour is observed in figure 7.19, which shows the light curve skewness versus spin period. Most $\gamma$-ray visible light curves exhibit an asymmetric tail extending out to high phase, characterised by a positive skewness. The opposite trend is observed in few SG and OPC cases. The abundance of solution at 0 skewness corresponds to symmetric light curves. Each emission model, occupies a well defined region of the P-skewness or P-kurtosis plane, and this offers the possibility to discriminate which model better describes the LAT shapes. In both the figures, the LAT pulsars, indicated as green stars, occupy the low plot region, mainly overlapping the SG population.

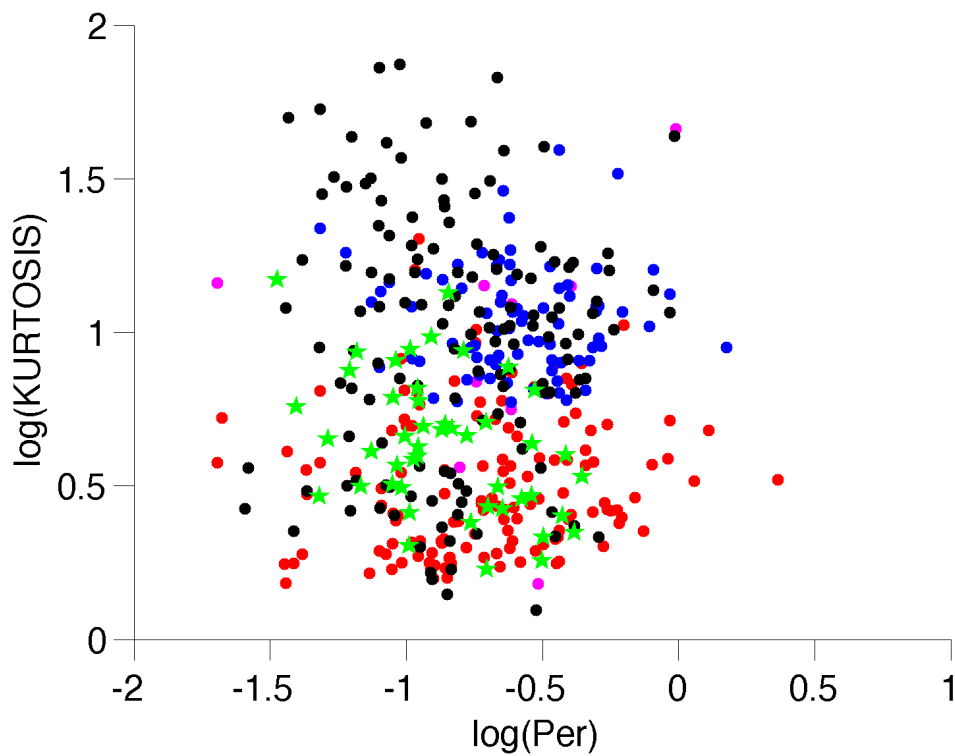

Figure 7.18: Light curve kurtosis as a function of the pulsar spin period. The pink, red, blue, and black points respectively represent the PC, SG, OG, and OPC model. The LAT pulsar light curves are represented by green stars.

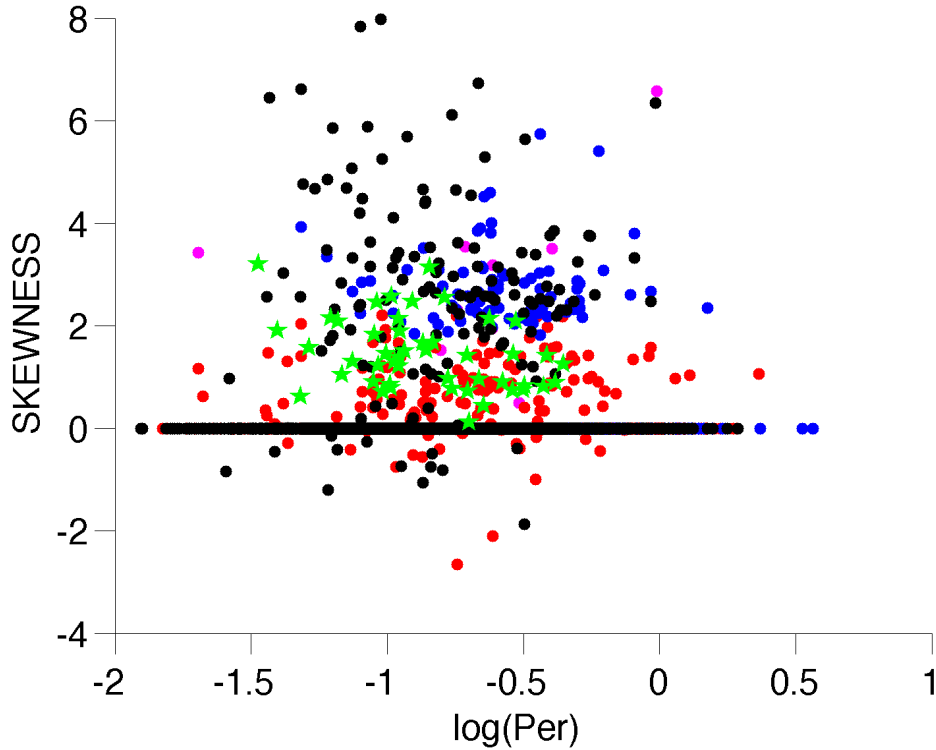

Figure 7.19: Light curve skewness as a function of the pulsar spin period. The pink, red, blue, and black points respectively represent the PC, SG, OG, and OPC model. The LAT pulsar light curves are represented by green stars.

Even though the kurtosis is a sharpness index, it does not generate a segregation of the simulated population. We define another sharpness index as

$$SHARPNESS = \log_{10}\left(\frac{max(ltc)}{mean(ltc)}\right) \times 100.$$  (7.9)

This index complements the kurtosis because it refers mainly to the highest peak than to the whole curve. The results of the efficiency of the new index is shown in figure 7.20 where it is plotted as a function of the skewness. The discrimination between the inner and outer magnetosphere profiles is well marked. The SG population is concentrated in a specific region of the $SHARPNESS - SKEWNESS$ plane as well the OG & OPC ones. The PC appears to overlap the SG region but with few points in the OG/OPC population zone.

The first conclusion about the discrimination of the model that better describes the observed population is in favour of the SG model. The green star markers that correspond to the LAT population in the $SHARPNESS -$

*SKEWNESS* plane are associated with the SG population. There is a clear extension of the LAT population toward higher sharpness indexes that the SG is not able to reproduce. This lack of objects at high sharpness could be due to the lack of high $\dot{E}$ objects that more likely have sharper structures (e.g. Vela, Crab, Geminga).

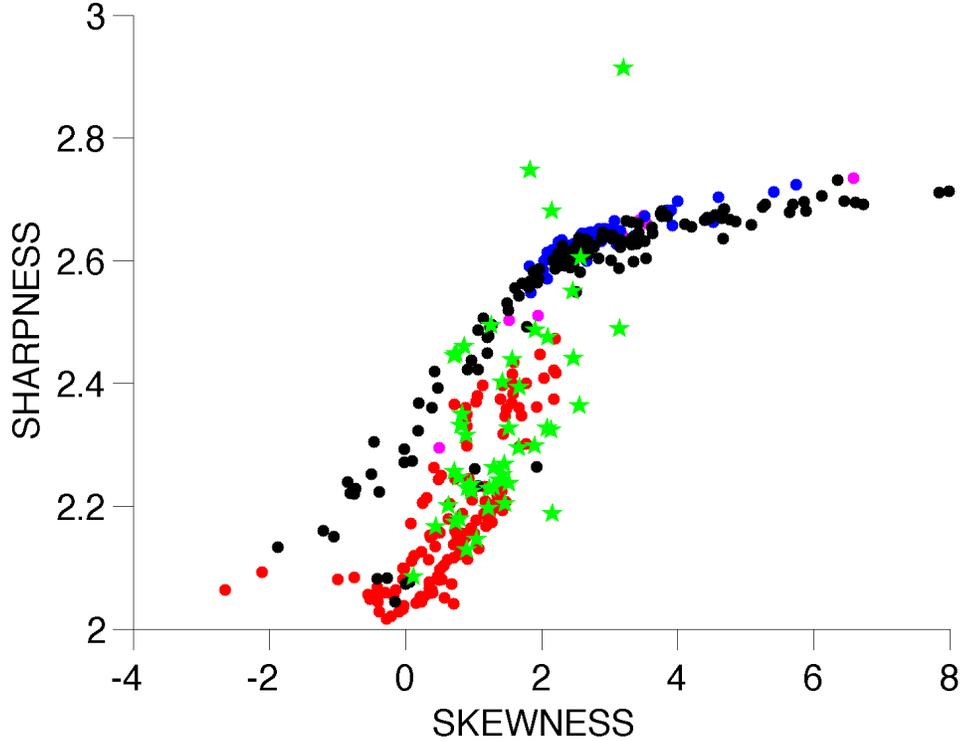

Figure 7.20: Light curve sharpness index as a function of the skewness. The pink, red, blue, and black points respectively represent the PC, SG, OG, and OPC model. The LAT pulsar light curves are represented by green stars.

In figure 7.21 is plotted the 3-D representation of the relation described in figure 7.20, with the addition of the observer line of sight $\zeta$. The trend found between the light curve sharpness and symmetry, can be used to help constrain the $\zeta$ angle. The PC population occupies a very narrow region of this space, narrow in skewness and concentrated at low $\zeta$ values. Consistently with the fact the OG and the OPC $\gamma$-visible population represents respectively the narrow and wide structure of the same phase-plot, they show a short and wide sharpness extension. In the OG case it is possible to define a small range of possible $\zeta$ estimates. For the SG case that best represents the real shapes, the $\zeta$ estimating $\zeta$ could be more complicated. The SG occupies a wide zeta interval, in a narrow skewness range, at low sharpness. The $\zeta$ interval looks to shrink in zeta by increasing the peak sharpness. Since most of the joint fit alpha zeta estimation found in chapter 6 are found for intermediate $\zeta$, their SG estimation could easier.

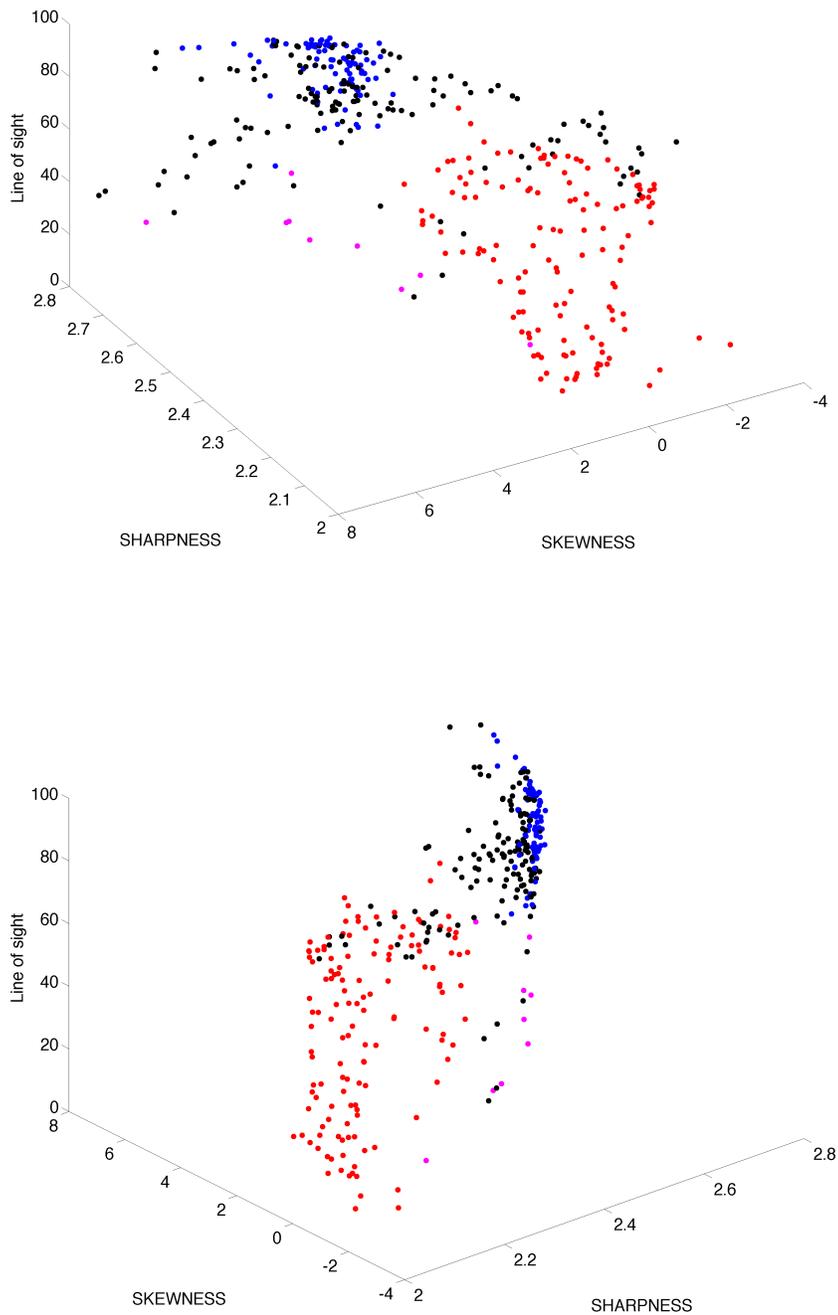

Figure 7.21: 3-D plots showing the relation between the light curve sharpness index, skewness, and pulsar line of sight. The top and bottom panels show the same 3-D plot from a different orientation. In pink, blue, red, and black is respectively indicated the PC, SG, OG, and OPC points.

As a last result, in figures 7.22 & 7.23 is shown the same 3-D plot as in figure 7.20, but with the LAT observed radio-loud pulsar sample for which I have obtained a $\zeta$ estimation through the joint $\gamma$-radio fit.

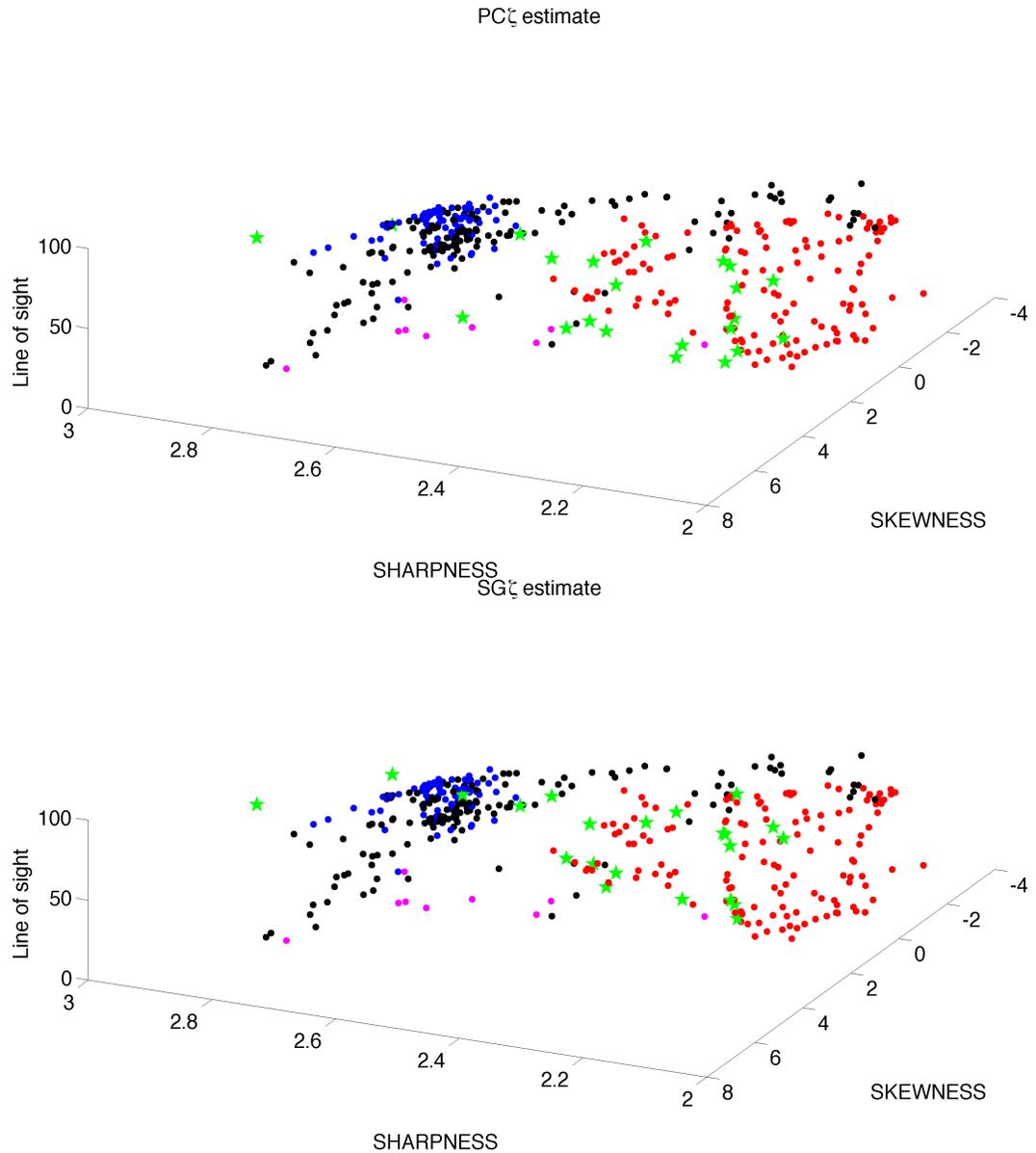

Figure 7.22: 3-D plots showing the relation between the light curve sharpness index, skewness, and pulsar line of sight. In pink, blue, red, black, and green stars are respectively indicated the PC, SG, OG, OPC simulation points, and the OG & OPC estimations of $\zeta$ for the radio-loud LAT pulsars.

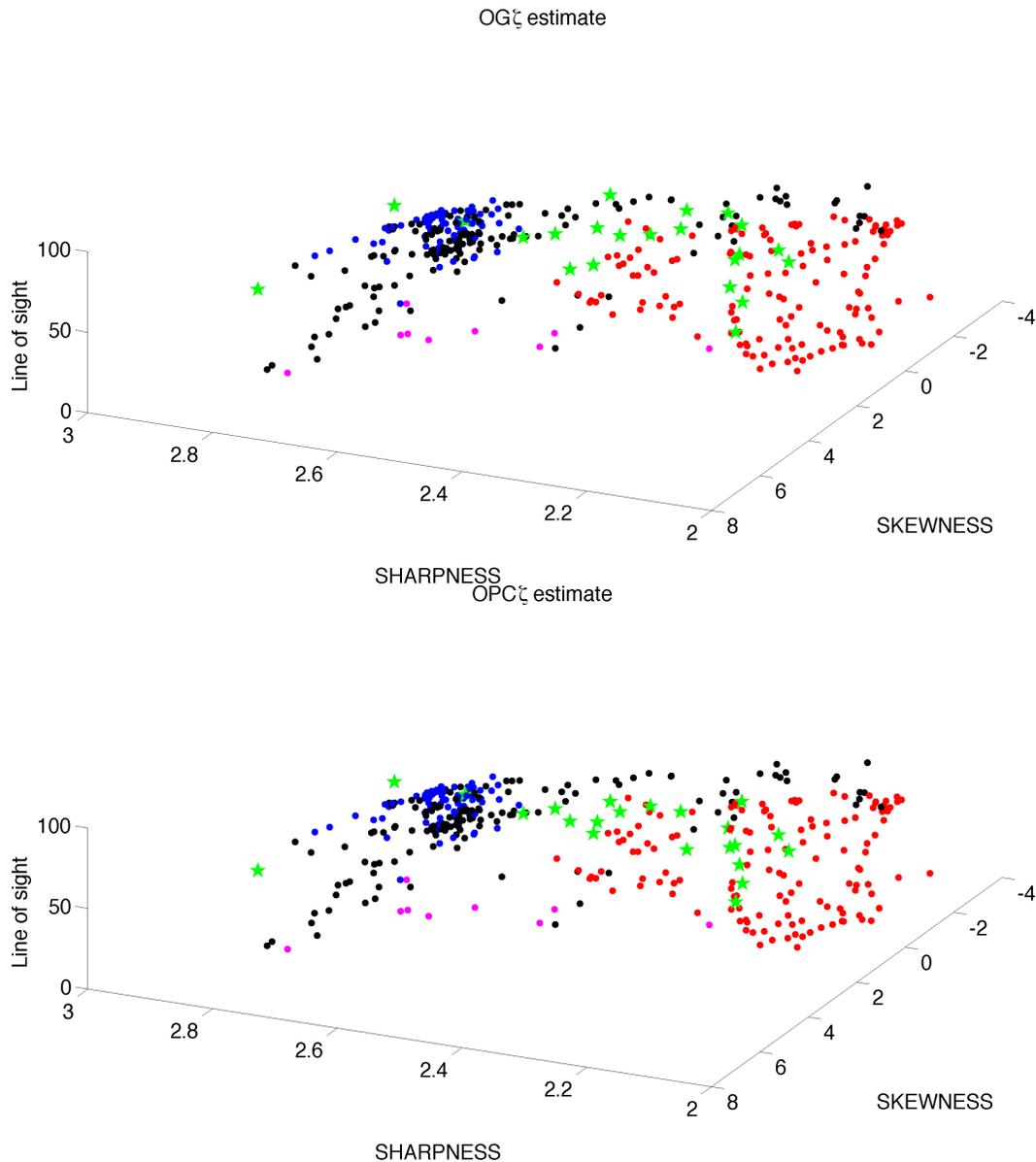

Figure 7.23: 3-D plots showing the relation between the light curve sharpness index, skewness, and pulsar line of sight. In pink, blue, red, black, and green stars are respectively indicated the PC, SG, OG, OPC simulation points, and the OG & OPC estimations of $\zeta$ for the radio-loud LAT pulsars.

Since I have obtained a $\zeta$ estimation for each emission model, the results are shown in the 4 panels of figures 7.22 & 7.23. The first things that has to be noted is that each model $\zeta$ estimation is consistent with each model simulation result. But the most interesting behaviour is that the core of LAT population best corresponds to the SG prediction. The sharpness and symmetry light curve analysis appears to be a powerful test to discriminate between the proposed emission models. They can also be used to help constrain some pulsar orientation parameters. Anyway, the complex and high multiplicity structure of the observed profiles suggest a more accurate investigation of the relation between the skewness and kurtosis, and the real profile shapes.

# Chapter 8

# Overview and joined results

In this last chapter I will discuss and merge the major results obtained in this thesis. By assuming each implemented luminosity model (chapter 2) for the synthesised pulsar population (chapter 5), it was possible to study the pulsar emission mechanism by comparison with the LAT observed sample. I have obtained important constraints about the pulsar energetics, spin period, and radio loud and quiet population ratio that can be used to compare the relative merits of each model.

Moreover, through the implementation of the geometric model that describe the emission pattern across the sky (chapter 5) for each assumed luminosity model, it was possible to compare the expected and observed emission light curves. The beam structure study, led to the definition of light curve characteristics, such as classes of shapes, peak separation and radio lag, that are directly comparable with the observations.

Always based on the light curve structure, a moment analysis has been performed on the simulated and observed light curves. Their comparison shows that there is a precise correspondence between the emission mechanism and light curve symmetry and sharpness that offers another discrimination test between the proposed emission pattern models.

## 8.1   The Polar Cap model

From the population synthesis results concerning the $\gamma$-ray visible pulsars, the PC emission mechanism does not provide sufficient energy to describe the observations. Even with a radiative efficiency of 100% ( figure 5.20) the PC is not energetic enough to reproduce the observed luminosity (figure 5.22). The $P - \dot{P}$ diagram of figure 5.25 shows that the bulk of the observed population is centred in the radio population region, describing a visible population much older than the LAT observed one. The luminosity and age problem can be more directly seen by looking at the spin-down power distribution of figure 5.27. The PC distribution is clearly peaked at low $\dot{E}$ values and the lack of energetic objects affects all the population distributions. Figures 5.28 & 5.31 describe a PC $\gamma$-visible population too old and near the Sun, at odds with of the observed sample. The fraction of radio-loud objects among the total radio and total gamma visible sample as a function of $\dot{E}$, is in strong disagreement with the observed trend (figures 5.33 & 5.34). This disagreement, jointly with the low number of predicted RQ objects, indicate that the emission beam from the polar cap is too faint and too narrow to explain the observations.

Concerning the light curve structure, the most recurrent PC light curve shapes are not consistent with the observed ones. In figure 7.7 is shown that the PC model does not have enough visible double structures to reproduce the observed behaviour. The FWHM of the single and double structure is shown in figure 7.8. The PC emission profiles are not narrow enough to reproduce the observed bump and the double peak profiles are too narrow to fit the observations. The peak separation behaviour, plotted in figures 7.12 & 7.13 is even worse. The PC solutions are concentrated between 0 and 0.2 in phase, exactly the region where there are no detections. The lack of wide peak separation objects is probably due to the lack of high $\dot{E}$ objects. The radio lag model expectations and its evolution with $\dot{E}$ and the peak separation plotted in figures figures 7.15, 7.16, and 7.17, show a strong disagreement with the LAT data. The PC predicts a too small radio lag, $< 0.1$ in phase, against the observed $\sim 0.2$. The observed decreasing trend of the radio lag with $\dot{E}$ and with the $\gamma$ peak separation is not present in the simulation that shows, rather, an inverse trend. In figures 7.18, 7.19 & 7.20 are shown the comparisons between symmetry and sharpness of observed and simulated light curves. In all the cases, the simulated sample is characterised by different moment parameters than the population.

Based on the PC model prediction about the number of visible pulsars, emission beam geometry, and light curve morphology, it is possible to conclude that the current PC model fails to reproduce most of the observed pulsar characteristics, and that the polar cap pulsars, if any, are rare in the LAT sample.

## 8.2   The Slot Gap model

Figure 5.20 shows the best match found between the flux of the $\gamma$-ray visible component of the SG pulsar population and the LAT observations. It has been found for a radiative efficiency of the 700%. This very high $\epsilon_\gamma$ is justified by the fact that a super Goldreich-Julian current and a particularly high accelerating electric field is aloud in the SG model. For this $\epsilon_\gamma$ value, the evolution of the luminosity as a function of $\dot{E}$ shown in figure 5.22, it is in nice agreement with the LAT observations. The SG $P - \dot{P}$ diagram of figure 5.25 shows a visible population that nicely fits both the RL and RQ component of the observed sample. Nevertheless the bulk of the model distribution is slightly shifted toward the older radio population region. By looking at the spin-down power distribution of figure 5.27, it is evident that, even though with a better agreement compared to the PC distribution, the SG simulated population also lacks young energetic objects with high $\dot{E}$. Figures 5.28 & 5.31 are two examples of how the $\dot{E}$ problem can affect the $\gamma$-ray visibility of the whole population. They show a SG visible component composed of older and closer objects, that does not match the observations. Figure 5.34, shows that the SG is the only model that predicts an increase of the fraction of gamma-ray loud objects among the radio visible ones when $\dot{E}$ increases. By looking at figure 5.34, If one labels 'RG' for a radio-loud gamma pulsar, 'G' for a radio-quiet gamma one, 'R' for a gamma-quiet radio one, the fact that the $N_R G/(N_R G + N_G)$ trend with $\dot{E}$ fails against the data and that the $N_R G/(N_R G + N_R)$ trend does better suggests that the high $\dot{E}$ problem affects more the radio quiet gamma sample. We loose too much luminosity or visibility away from the radio polar beams.

The profile shapes study described in figure 7.7 show an excess of single bump structures and a lack of double ones. As it has already been discussed in the *Shape classes* paragraph of section 7.4.1, the SG bump classified profiles are always characterised by an higher complexity that could represent a classification bias in favour of more double structures. The SG single and double profiles FWHM, shown in figure 7.9 , reasonably represent the observed pulse width. Concerning the bump FWHM there could be the same fitting bias described for the bump classification. Often, the bump complexity cause an overestimation of the measured width and so the real SG bump FWHM has to be taken as an upper limit of the real width. The SG peak separation is the one that best reproduce the observed cases. In figures 7.12 & 7.13, the SG simulated profiles predict peak separations in the same interval covered by the observations. Particularly important is the absence of solution in the range $\sim 0 \sim 0.13$, in agreement with the data. From figures 7.15, 7.16, and 7.17, the model radio lag expectations do not reproduce the observed trend. The SG over predicts small radio lags for old pulsars because of the bright caustic on the trailing side of the radio visible pole. This calls for thinner gaps with old age than in the present calculation, but at the expense of lower luminosities, thus visibility, at old age. Concerning the curve symmetry and sharpness, the SG profiles give the best agreement in the sharpness versus skewness plane.

Based on the SG model prediction for the total number of visible pulsar, the pulsar luminosity, emission beam geometry, and light curve morphology, it is possible to conclude that the current SG model is able to reproduce many of the studied pulsar characteristics. The use of a large radiative efficiency, suggests that the electric field in the slot gap is a few times larger than in the current calculation. The SG light curves can often match the LAT data and their statistical level of sharpness and asymmetry. The evolution of the peak separation is correct. The discrepancy in the radio lag evolution indicates either less widening of the gap with age, or the lack of emission from one pole, or a straightening of the field lines at high altitude because of the current feedback.

## 8.3   The Outer Gap and One pole Caustic models

The match between the predicted and observed fluxes plotted in figure 5.20, is obtained for OG and OPC radiative efficiencies of 100% and 50% of the prime particle energy. The luminosities, evaluated from this efficiencies and plotted in figure 5.22, fall within the observed range, but the OG model expectations are too dispersed to say that there is a correspondence between the model and observed luminosity evolution with $\dot{E}$. The OPC power law has been set to match the data and has no predictive value. Concerning the $P - \dot{P}$ distribution of figure 5.25, the OG population appears too shifted toward the older radio population region and not enough extended in magnetic field to fit the LAT observations. The OPC population is the one that better represents the observed sample. It shows the same SG characteristics but with a more extended tail toward energetic objects. The OG model seriously lacks high $\dot{E}$ objects. Between all the implemented emission models, the OPC is the one for which the $\dot{E}$ discrepancy is minimised. As in the SG case, the predicted distribution does not peak and in the OPC case covers the the whole observed distribution with the highest efficiency. This nicer agreement comes from the non physical gap width assumption that, optimising the OPC visibility, implies a higher detection of powerful pulsars. Anyway it does not solve the still present lack of high $\dot{E}$ stars. Both the discussed $\dot{E}$ trend are confirmed by the pulsar age trends (figure 5.28). The OG age distribution is wrongly peaked at old age. Figure 5.31 shows that the OG and OPC visibility and flux predict the right distance range for detection. The evolution with $\dot{E}$ of the RL/$\gamma_{num}$ and RL/$Radio_{num}$ ratios, plotted in figures 5.33 & 5.34, show a strong disagreement with the observations. The relative lack of young radio-loud pulsars in the gamma visible sample is due, as for the SG, to orientation effects. The gamma-ray signal is seen within only 20 degrees of the magnetic axis at old age, therefore with a reasonable chance to intercept both the gamma and radio beams. For young pulsars, a much larger variety of inclinations about the magnetic axis give rise to a gamma-ray signal, thus with a reduced chance to see the radio too. The situation is not symmetryc between the SG and outer gaps for the fraction of gamma-loud objects in the radio visible sample. Once we detect the radio beam, there is a high chance to detect the SG caustic trailing the radio pole and the fraction evolution with $\dot{E}$ is governed primarily by the flux fading with age. In the outer gap geometry, there is an additional evolutionary factor due to the shrinking with age of the emission cusp in the phase-plot. This is why the OG and OPC trends in figure 5.34 are worse than the SG one.

From the light curve shapes recurrence point of view, the high magnetosphere emission models are strongly favoured with respect to the other ones. Figure 7.7 shows that the OG and OPC models are able to reproduce the large proportion of double structures. The FWHM behaviours shown in figures 7.10 and 7.11 for both the OG and OPC models, show that the predicted double structure are clearly too thin compared to the observation. This is due to the assumption of an infinitely thin emission layer along the inner edge of the gap that cannot describe the reality. The OG and OPC expectation for the peak separation is not very consistent with the data. In figures 7.12, the OG model predicts numerous double narrowly peaked light curves that are not observed in the LAT data. The OPC peak separation is completely flat at all the separation values, in particular in the region without any observation ($\sim 0.1$). The observed peak separation evolution with $\dot{E}$ is roughly consistent with both models, and provides the lack of high $\dot{E}$ objects. The observed radio lag, shown in figures 7.15, 7.16, and 7.17, is generally better explained by the OPC model. The OG solutions are anyway consistent with the LAT observations. The light curve sharpness and symmetry analysis suggest a real light curve structure that is significantly different from the ones generated by the OG/OPC phase-plot. The outer magnetosphere models appear to generate too sharp and asymmetric curves compared to the observed ones.

So, the location of the outer gap emission models can explain a number of the light curve characteristics recorded by the LAT, such as the large fraction of double peaks and the radio lag, but it predicts too many close peaks at old age, too few single peaks, and the wrong evolution of sharpness with skewness. From a luminosity point of view, only the physical OG model can be tested against the data and it predicts a population of visible objects too skewed to old ages.

## 8.4   Discrimination

For the comparison between the implemented emission models and the LAT observation, within the vacuum dipole assumption, it is difficult to decide between the two-pole (SG) or one-pole (OG/OPC) versions of the outer emission zone, or on the actual extent of the emission below the null surface (SG) or above (OG/OPC). Both types of models can match some aspect of the data and fail on others. Only an origin of the gamma rays deep down in the inner magnetosphere is systematically rejected on both flux and light curve morphology grounds. To improve on the future discrimination power of the tests implemented here, we need to improve the models in several ways:

- Explore a slot gap width evolution that preserves thinner gaps up to a few hundred kyr.

- Spread the OG and OPC emission in a more realistic distribution across the gap

- Test the population properties of the force-free magnetic field configuration

- Test the influence of a potential magnetic alignment with age, although the obliquities found from the light curve fits to the LAT data do not show evidence of such an alignment over the small age range of the gamma-ray pulsars.

# Conclusions

In the course of my PhD, I have obtained several interesting results on pulsars. We find a significant discrepancy between the number of young and energetic gamma-ray pulsars detected by Fermi and the number we can reasonably expect in the Milky Way for all the current emission models. This discrepancy relates to the evolution of the spin-down power of the neutron star over its first 100 kyr. Because all models under predict young, bright objects, well above the visibility threshold, this discrepancy is linked either to the $P - \dot{P}$ evolution at young age, or magnetic field axis alignment, or beam apertures. All models also fail to reproduce the high probability of observing both the radio and gamma-ray beams from the most energetic pulsars. This relates to the evolution (if any) of the pulsar obliquity over the same time span, or to a higher location of the radio emitting region, or an azimuthal asymmetry in the radio and $\gamma$-ray beam. The simulation shows that the beam correction factor that is commonly used to infer the pulsed luminosity from the observed flux actually evolves with the star spin-down power. It convincingly shows that the relation found in the observations between the $\gamma$-ray luminosity and the spin down power is robust and not plagued by evolving beam apertures or magnetic obliquity and orientation effects.

I have obtained new constraints on the magnetic obliquity angle $\alpha$ and the observer orientation $\zeta$ for 22 pulsars by fitting models to the LAT and radio light curves. The comparison of the pulsar orientation estimated by fitting individually the $\gamma$-ray profiles, and jointly the radio and $\gamma$ ones, leads to the conclusion that a joint $\gamma$-radio fit is the only acceptable way to give reliable estimates of the pulsar orientation. The results in $\alpha$ and $\zeta$ have been used to compare the LAT data with the model characteristics that depend on the estimated angles. We find a suggestive relation between the cutoff energy of the $\gamma$-rays and the accelerator gap width in the magnetosphere. The relation is consistent with the SG predictions $E_{cut} \propto w_{gap}^{-0.5}$. This $E_{cut}$ gap width proportionality has a particular importance because it connects the observed spectral information and the size of the gap region based only on the light curve morphology.

The study of the light curve geometry performed during my PhD, gives important hints towards a possibly different structure and geometry from the assumed dipolar magnetic field. In fact, to explain the observed radio lag and γ-ray peak separation in the outer magnetosphere models, a force-free situation (Bai & Spitkovsky, 2010) or a split-monopole geometry have to be taken into account as alternative magnetic field configurations. Through the light curve shape classification I have investigated the recurrence of the light curve structures in the simulated samples and in their γ-ray visible sub-samples. It emerged that there is no visibility selection connected with the light curve shapes: the pulsars are visible or not independently on their light curve shape. The radio lag and peak separation behaviour of the observed population are better explained by the outer magnetosphere models. The one pole caustic emission geometry assumed in the outer magnetosphere models gives the best explanation of the observed radio lag. The light curve morphology analysis, performed by the study of the light curve sharpness and symmetry, favours the 2 pole emission geometry of SG model in explaining the observation.

**Future projects**

As a follow up of my PhD, one of the first projects will be to extend the implemented population synthesis study. One project will consist in testing the new suggested scenario for the pulsar magnetic field configuration, to define whether a non-dipole spin-down power evolution is able to explain the observed excess of high $\dot{E}$ objects. In this direction, to test the population properties by assuming a force free model magnetic field configuration will help to find out whether a different magnetic configuration could better explain the behaviour of young and energetic objects. The outer magnetosphere models have been implemented under the unreliable assumption of an infinite emission layer across the gap. As a future OG/OPC models improvement, a more realistic distribution of the emission across the gap has to be taken into account. Concerning the SG model, most of the inconsistencies between model expectations and observations are due to an excessive broadening of the slot gap size with age. It suggests to explore the possibility of a different gap width evolution to maintain a thinner SG width up to few hundred kyrs. Since the lack of high $\dot{E}$ objects in the model predictions is shown to be particularly sensitive to the neutron star birth distribution within the galaxy, another study will concern the star birth rate and position. By using the LAT objects joint fit results, the assumptions used to synthesise the population characteristics can be improved. A good example of this is the relation $E_{cut} \propto w_{gap}^{-0.5}$, that has not been taken into account for the population synthesis of the pulsar energy cutoffs and spectral indexes. Another possible scenario that has to be tested is a possible magnetic alignment with age. Although the obliquities found from the light curve fits to the LAT data do not show evidence of such an alignment, it can occur on a short timescale and not be detectable. Anyway, its effect on the population characteristics would be evident and could be responsible for the lack of energetic objects in our simulations.

Another future objective is to extend our models to include the X-ray emission component. From the energetic point of view, a correlation has been found between the pulsar radio and X-ray emission, both in flux and in the light curve structure in terms of timing arrival-times (Lommen et al., 2008, 2007). One project to be implemented would be to test for possible relations between the X-ray and $\gamma$-ray detected fluxes.

A possible project would be to further constrain the pulsar orientation. The measurement of the LAT pulsar parameters I have implemented during my PhD is based on the joint fits of the $\gamma$-ray and radio light curves. They give an estimate of the pulsar obliquity $\alpha$ and line of sight $\zeta$. To include the X-ray components in the fit will provide additional constraints. For instance, one can make use of the thermal emission from the polar cap and to couple the morphological study of the X-ray emission from the pulsar wind nebula (PWN) torus that surrounds many objects (Lipunov et al., 1981). A joint $\alpha$ & $\zeta$ fit including the X-rays to constrain the torus orientation could represent the most accurate $\zeta$ estimation ever done. Moreover, since all the pulsar $\gamma$-ray physical parameters, such as the gap width or voltage, depend on the pulsar orientation, an accurate $\alpha$ & $\zeta$ estimation will lead to improved knowledge of the whole pulsar electrodynamics. In particular, the most debated topic about pulsar physics is the existence, structure and location of the emission gap regions. Since the gap width depends on $\alpha$, and the luminosity scales as $L_\gamma \propto w_{gap}^{-3}$, a best determined $\alpha$ is fundamental to constrain the intrinsic pulsar $L_\gamma$ and its evolution with age.

Another approach to the pulsar orientation study, and more focussed on the $\alpha$ estimation, is based on radio pulsar observation. As an object of a future research, it could be interesting to study the radio polarisation profile as a tracker of the magnetosphere emission region (Gangadhara, 2009) jointly with the multi-wavelength light curves. This method could be particularly interesting since it does not depend on the light curve shape but on an independent emission characteristic. The high accuracy of the radio observations will improve significantly the reliability on the $\alpha$ measurements.

The improvement of the light curve shape morphological study surely represents one of the future projects I would like to carry out. I have shown that the light curve morphological analysis is a powerful tool to deduce the emission gap region structure. Therefore it allows us to discriminate between different emission mechanisms. A future joint analysis between the curve moments study and the non light curve shape dependent radio polarisation will give the possibility to further constrain the emission region structure from two independent points of view. The main goal of such a study will be to better understand the physical nature of the magnetic field (dipole, force-free situation (Bai & Spitkovsky, 2010), or a split-monopole) and its layout. Since the details of this layout depend critically on the pulsar orientation; and the latter may change with potential magnetic alignment; a future research project based on multi-wavelength pulsar light curve and flux analysis would shed new light on pulsar physics.

# List of Figures

# List of Tables

# Acknowledgements

Oh, let me start with some spaghetti English. Well, let me continue. In my train compartment my travel companions sleep. One is from Bangladesh, one from Egypt, one is Rom from Romania, one from Peru, and one from... how to say it in English... Cote d'Ivoire. And I'm going back to Montedinove. I'm not even sure where I come from, not any more. We spoke Italian and Spanish and French. But not English. They were happy to be back in Italy. I don't know whether I'm happy.

I spent almost four years in Paris. I met a lot of people. And I'd like to thanks them first. All of them, all the man, women, guys, and girls I liked to spend time with. Life is not a job, to get a position, or a thesis, or a paper. It is more to find a way to be able to go in a place and to feel home. We found that way and I felt home. Thanks for that.

Relationships are much more rich and interesting between people with a strong personality. Isabelle, I've always admired the huge amount of things you know, that you are able to do, and the impressive number of hours you are able to work and do not sleep (and I think I managed to adapt myself). Thanks Isabelle, for what you taught me, for the patience, the comprehension, the hospitality and the precious suggestions. Sometimes it was quite hard, we argue, we fought, we discussed. But we managed to keep it stable. And possible. Thanks a lot.

A special thank goes to Alice Harding. She was my second advisor. I think there is nobody who knows Pulsars and is able to explain them as well as she does. There is just to learn from her, in life and in science, and I tried to do it as much as possible. Thanks Alice.

And thanks also to Peter Gonthier. He has helped me often to understand about papers and work. It was incredibly nice to know him and his wife in Sardinia.

I would like to acknowledge the Jury of my thesis defence. And assimilate. Andrea Possenti e Marta Burgay, responsible to have "created" me and always present when I needed them. Again Isabelle, for what she did during and said about me the day of my defence. Marcello Fulchignoni has always been very understand and available for suggestions. Gilles Thureau, I found my PhD position thank to him. And I'd like to thank in a special way my two Referees,

David Smith and Patrizia Caraveo. The situation was not so simple. A series of bad circumstances led to a weird "state of the art". I have appreciated a lot your professionalism, comprehension of the problem, and ability in handle it. With two other random referees I would be still there. Thank a lot.

I'd like to thank Jean Marc for the help. He was always available when I needed help. And it is not important the act of helping somebody. What is important is to know there is somebody one can ask.

And thanks to all the people of the FERMI collaboration. I learned a lot from you all, a little bit hided but I learned a lot.

The friends. A lot, really. Already thanked. But of course with each one I had a different way to interact. There was one of them, a tall and slim guy from Amsterdam that I have to thank in a special way. Sacha. I will never understand most of your jokes but it's okay, sometimes the important was to see you having fun 1000 times by telling the same joke that nobody understand (japanese people excluded). You helped me in all the aspects you thought I needed help. And my thesis, a big part of my work is surely due to you. Thanks a lot Sacha.

It was the first year, long time ago but I clearly remember this dirty and crashed car with the radio always tuned on FIP. Chiaretta, you left soon and I missed you. But I remember this year as it was 10 times longer. Thanks for everything.

I like to drive, and I don't know why Anais let me do that. Probably because she didn't know me. Thanks Anais, we talked a lot and it was very nice and helpful to "argue" and to have suggestions from you about thesis, job, and... discriminations.

Thanks to Jenny. She was the first to read my first chapter. And she had nice comments. She was encouraging, and this is always important when one starts a thesis. And thanks also for the suggestions about how to write an applications, how to search for jobs, and how the American system works.

La familia. How to forget la familia, una vera famiglia. Juan, Giova', Anto'. Office mate, flat mate, friend. Thanks Juan, you were always there and I'm sure there is nobody that knows my PhD trajectory and complaining as well as you. You were the first I met, the first friend. Never angry, always available and with a nice suggestion. But it's your fault that I never learned French as well as I could have (but your english got worst, haha). Sometimes it was hard and you were there. I appreciated what you did for me. And how do not cite the other components of the familia. I met them together. Fabio, it is a pity we met so late but now it looks like I know you since forever. Thanks for the help with the thesis, the corrections, suggestions, and helpful discussions. It was important to have somebody like you with which talk about everything. I cannot even imagine something better than a friend like you in that moment of my life. Isabel, Isa, it was a pleasure to know that somebody like you exist. Always positive, willing to talk and to do nice things. Free. Thanks for the help with the thesis. Was it a couple of day before the defence? In the Menagerie talking about my talk and meeting strange retired satellite builder. You cannot imagine how important was it. And thanks for talking several times, to be "apuesta segura", for the friendship and for the beers together. You were the only one able to follow me. Guys, I destroyed your places, I carried you on my shoulders, I risked to kill you more than once. Why you didn't try to kill me (Maybe you tried but didn't manage and I'll never know that, *il luuuuuupo!*)?

And my Parents, always behind the corner, willing to do everything for me. And encouraging, and (daily) present. Supporting me and tolerating my strange and often not correct way to interact with them. And a special, but special thank to my brother (with his wife and my beautiful niece and nephew). He shown to be incredibly present, nice, brother in the proper sense of the word. He helped me economically and personally, always with (sometime) discreet subliminal suggestions I systematically got and made mine. You all crossed Europe to come to see me and do not understand a word. But you were there and I didn't feel alone. Mamma, Papá, Giorgio, Marco, what I made is the projection of the life and education we had together. Congratulations.

The train, this time a different one, is getting closer to Montedinove. Tonight Campesinos, the Peru guy, was controlled twice by the police, the French one before, the Italian one after. They were searching some hided space in his suitcase. The two women, the Rom one and the Cote d'Ivoire one were talking in a strange language that looked like italian with some invented words unexplainable understood from both of them. The Bangladesh man was eating and the Egypt guy sometimes sleeping and sometimes laughing about the strange night of Campesinos.

And me I was thinking that it's amazing how different and far away life can cross and generate situations like that. It's amazing how "far away" is a meaningless concept. You go in a place, you know people and live a life with them. You call this home. You do it several times and you can call home several places and persons. But you don't do it because of the places, not because of the persons, not for the job and not even for the "glory". At the end you want just to know yourself. You want just to be able to call yourself home.